\documentclass[a4paper]{book}

\usepackage[english]{babel}
\usepackage{amsmath,amssymb,graphicx,enumerate,latexsym,calc,capt-of,xcolor}
\usepackage{epigraph}
\usepackage{dsdshorthand}
\usepackage{xcolor}
\usepackage{booktabs}
\usepackage[utf8]{inputenc}
\usepackage{float}
\usepackage{tikz}
\usetikzlibrary{circuits.logic.US,circuits.logic.IEC,fit,positioning,arrows,patterns,decorations.markings,shapes.misc}
\tikzset{cross/.style={cross out, draw=black, fill=none, minimum size=2*(#1-\pgflinewidth), inner sep=0pt, outer sep=0pt}, cross/.default={2pt}}

\usepackage{pifont}
\usepackage{braket}
\usepackage{mathtools}
\usepackage{relsize}
\usepackage{cite}
\usepackage[final]{pdfpages}

\usepackage{amsthm}
\newtheorem  {theorem} {Theorem} [section]
\newtheorem  {lemma}       [theorem] {Lemma}
\newtheorem  {corollary}   [theorem] {Corollary}
\newtheorem  {proposition} [theorem] {Proposition}

\newtheoremstyle{myremark}
{3pt}
{3pt}
{}
{}
{\bf}
{.}
{.5em}
{}
\theoremstyle{myremark}
\newtheorem{remark}{Remark}[section]

\definecolor{darkblue}{rgb}{0.1,0.1,.7}
\usepackage[colorlinks, linkcolor=darkblue, citecolor=darkblue, urlcolor=darkblue, linktocpage,backref=page]{hyperref} 
\renewcommand*{\backref}[1]{}
\renewcommand*{\backrefalt}[4]{%
	\ifcase #1 (Not cited.)%
	\or        (Cited on p.~#2.)%
	\else      (Cited on pp.~#2.)%
	\fi}

\usepackage[]{latexsym}
\usepackage{geometry}
\usepackage{marginnote}

\newcommand{\eps}{}

\def\eps{\epsilon}
\newcommand{\beq}{\begin{equation}} 
	\newcommand{\eeq}{\end{equation}}
 
\def\nn{\nonumber}

\def\half{{\textstyle\frac 12}}

\def\ge{\geqslant}
\def\le{\leqslant}
\def\geq{\geqslant}
\def\leq{\leqslant}
\def\<{\langle}
\def\>{\rangle}

\newcommand{\assign}{:=}
\newcommand{\asterisk}{\mathord{*}}
\newcommand{\cdummy}{\cdot}
\newcommand{\comma}{{,}}
\newcommand{\nin}{\not\in}
\newcommand{\nobracket}{}

\newcommand{\tmem}[1]{{\em #1\/}}
\newcommand{\tmmathbf}[1]{\ensuremath{\boldsymbol{#1}}}
\newcommand{\tmop}[1]{\ensuremath{\operatorname{#1}}}
\newcommand{\tmtextbf}[1]{{\bfseries{#1}}}
\newcommand{\tmtextit}[1]{{\itshape{#1}}}

\newcommand{\infixor}{\mathrm{or}}
\newcommand{\infixand}{\mathrm{and}}

\numberwithin{equation}{section}

\usepackage{chngcntr}
\counterwithin{figure}{section}

\usepackage{etoolbox}
\apptocmd{\thebibliography}{\setlength{\itemsep}{0em}}{}{}

\newcommand{\cmark}{\ding{51}}
\newcommand{\xmark}{\ding{55}}

\newcommand{\bbR}[1]{\mathbb{R}^{#1}}
\newcommand{\bbZ}{\mathbb{Z}}

\newcommand{\fr}[1]{\left(#1\right)}
\newcommand{\abs}[1]{\left\lvert #1 \right\rvert}
\newcommand{\sumlim}[1]{\sum\limits_{#1}}
\newcommand{\limlim}[1]{\lim\limits_{#1}}

\newcommand{\nospace}{}

\newcommand{\tmxspace}{\hspace{1em}}

\usepackage{fancyhdr}

\newcommand{\nnchapter}[1]{
\chapter*{#1}
\addcontentsline{toc}{chapter}{#1}
\markboth{\MakeUppercase{#1}}{\MakeUppercase{#1}}
}

\fancypagestyle{bibliography}{%
  \fancyhead[RO,LE]{\slshape \leftmark}%
}

\newcommand{\figuresandtables}{
\listoffigures
\begingroup
\let\clearpage\relax
\listoftables
\endgroup
}

\makeatletter
\renewcommand\part{%
  \if@openright
    \cleardoublepage
  \else
    \clearpage
  \fi
  \thispagestyle{empty} 
  \if@twocolumn
    \onecolumn
    \@tempswatrue
  \else
    \@tempswafalse
  \fi
  \null\vfil
  \secdef\@part\@spart}
\makeatother

\begin{document}

\frontmatter
\includepdf[pages={1,2}]{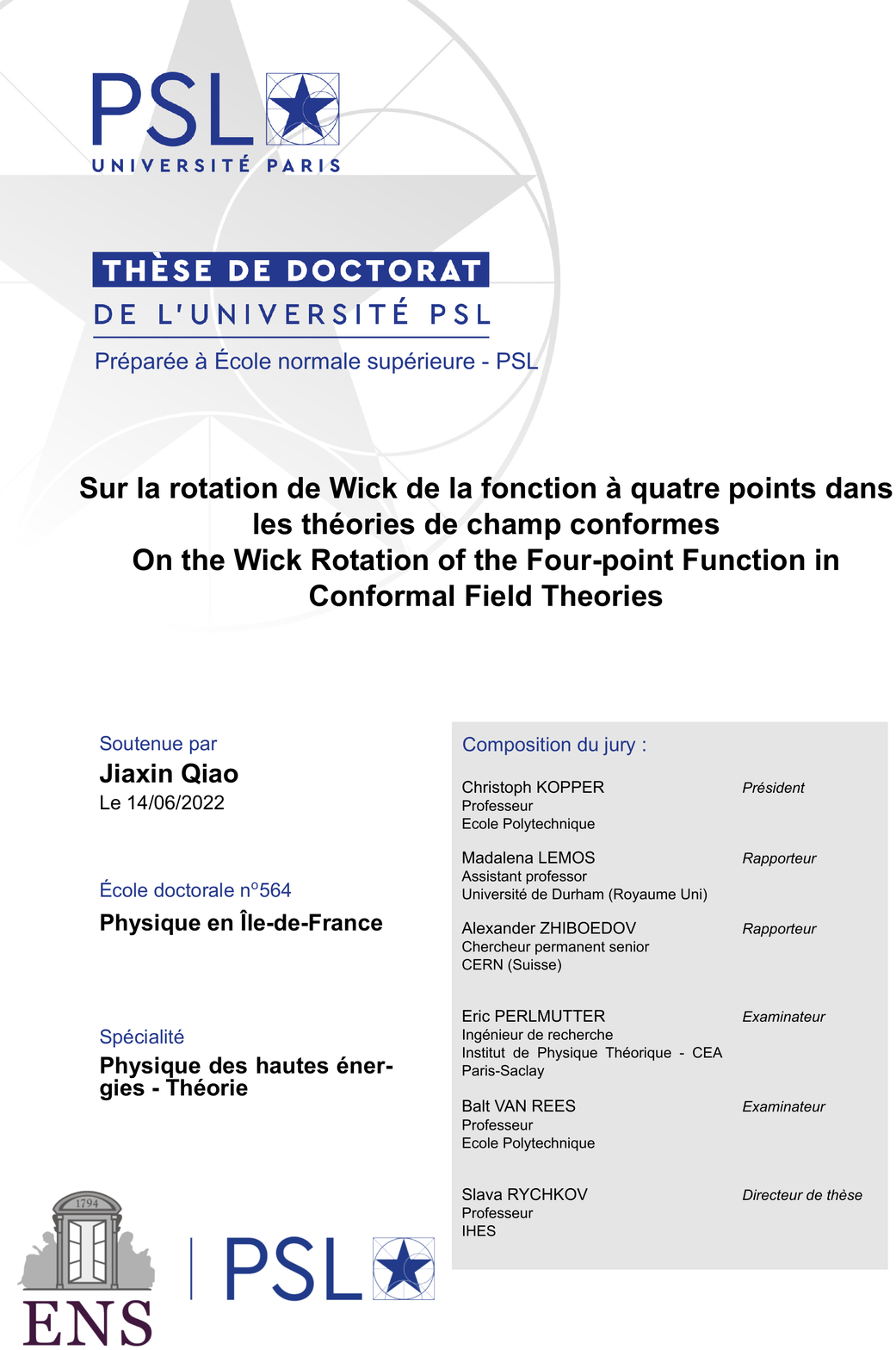}
\tableofcontents
\thispagestyle{empty}
\addtocontents{toc}{\protect\thispagestyle{empty}}
\nnchapter{Abstract}
\thispagestyle{empty}

Conformal field theories (CFTs) in Euclidean signature satisfy well-accepted rules, such as conformal invariance and the convergent Euclidean operator product expansion (OPE). Nowadays, it is common to assume that CFT correlators exist and have various properties in the Lorentzian signature. Some of these properties may represent extra assumptions, and it is an open question if they hold for familiar statistical-physics CFTs such as the critical 3d Ising model. In this thesis, we clarify that at the level of four-point functions, the Euclidean CFT axioms imply the standard quantum field theory axioms such as Osterwalder-Schrader axioms (in Euclidean) and Wightman axioms (in Lorentzian).

In part I, we study the Euclidean CFT four-point functions in the cross-ratio space. We show that the four-point functions in conformal field theory are defined as tempered distributions on the boundary of the region of convergence of the conformal block expansion. The conformal block expansion converges in the sense of distributions on this boundary, i.e.~it can be integrated term by term against appropriate test functions. This result can be interpreted as giving a new class of functionals that commute with the summation of the conformal blocks when applied to the crossing equation, and we comment on the relation of our construction to other types of functionals. Our language is useful in all considerations involving the boundary of the region of convergence, e.g.~for deriving the dispersion relations. We establish our results by elementary methods, relying only on crossing symmetry and the standard convergence properties of the conformal block expansion. 

In part II, we consider CFT four-point functions of
scalar primary operators in Lorentzian signature. We derive a minimal set of their
properties solely from the Euclidean unitary CFT axioms. We establish all Wightman axioms (temperedness, Poincaré invariance, unitarity, spectral
property, local commutativity, clustering), Lorentzian conformal invariance,
and distributional convergence of the s-channel Lorentzian OPE. This is done
constructively by analytically continuing the 4-point functions using
the s-channel OPE expansion in the radial cross-ratios $\rho, \bar{\rho}$.
We prove a crucial fact that $| \rho |, | \bar{\rho} | < 1$ inside the forward
tube, and set bounds on how fast $| \rho |, | \bar{\rho} |$ may tend to 1
when the four-point configuration approaches the Minkowski space.

We also provide a guide to the axiomatic QFT literature for the modern CFT audience. We review the Wightman and Osterwalder-Schrader (OS) axioms for Lorentzian and Euclidean QFTs, and the celebrated OS theorem connecting them. We also review a classic result of Mack about the distributional OPE convergence. Some of the classic arguments turn out useful in our setup. Others fall short of our needs due to Lorentzian assumptions (Mack) or unverifiable Euclidean assumptions (OS theorem).

In part III, we study the OPE convergence properties of the CFT scalar four-point correlation functions in the sense of functions. We establish the criteria on OPE convergence in s-, t- and u-channels. We give a complete classification of four-point configurations in the Lorentzian signature. We show that all configurations in each class have the same OPE convergence properties in s-, t- and u-channels. We give tables including the information of OPE convergence for all classes. By taking the union of all Lorentzian configurations where at least one OPE channel converges, we get a minimal domain of analyticity of Lorentzian CFT four-point functions.

In part IV, we give a preview on two possible generalizations: one is the CFT four-point functions of operators with general SO$(d)$ spins, the other is the CFT four-point functions in the Minkowski cylinder.

\nnchapter{Résumé}
\thispagestyle{empty}

Les théories des champs conformes (\textit{CFTs}) en signature euclidienne satisfont plusieurs règles bien acceptées, telles que l'invariance conforme et la convergence de l'expansion du produit d'opérateurs (\textit{OPE}) en signature euclidienne. De nos jours, il est courant de supposer l'existence des fonctions de corrélation d'une CFT et d'assumer diverses propriétés en signature lorentzienne. Certaines de ces propriétés peuvent représenter des hypothèses supplémentaires, et leur validité reste incertaine dans les CFT de physique statistique familières telles que le modèle d'Ising critique en trois dimensions. Dans cette thèse, nous clarifions qu'au niveau des fonctions de corrélation à quatre points, les axiomes CFT euclidiens impliquent les axiomes standards de la théorie quantique des champs tels que les axiomes d'Osterwalder-Schrader (en signature euclidienne) et les axiomes de Wightman (en signature lorentzienne).

Dans la partie I, nous étudions les fonctions à quatre points de CFT euclidienne dans l'espace des birapports. Nous montrons que les fonctions à quatre points dans les théories des champs conformes sont définies comme des distributions tempérées sur la frontière de la région de convergence de l'expansion en blocs conformes. L'expansion en blocs conformes converge dans le sens des distributions sur cette frontière, c'est-à-dire qu'elle peut être intégrée terme à terme contre des fonctions tests appropriées. Ce résultat peut être interprété comme donnant une nouvelle classe de fonctionnelles qui commutent avec la sommation des blocs conformes lorsqu'elle est appliquée à l'équation de croisement, et nous commentons la relation de notre construction avec d'autres types de fonctionnelles. Notre langage est utile pour toutes questions impliquant la frontière de la région de convergence de l'OPE, par exemple, pour déduire les relations de dispersion. Nous établissons nos résultats par des méthodes élémentaires, en nous appuyant uniquement sur la symétrie de croisement et les propriétés de convergence standards de l'expansion en blocs conformes.

Dans la partie II, nous considérons les fonctions à quatre points des opérateurs primaires scalaires dans une CFT en signature lorentzienne. Nous dérivons un ensemble minimal de leurs propriétés uniquement à partir des axiomes CFT unitaires euclidiens. Nous établissons tous les axiomes de Wightman (caractère tempéré, invariance de Poincaré, unitarité, propriété spectrale, commutativité locale, \textit{clustering}), l'invariance conforme lorentzienne et la convergence distributionnelle de l'OPE lorentzien dans le \textit{canal s}. Ceci est fait de manière constructive en continuant analytiquement les fonctions à 4 points en utilisant l'expansion OPE du canal s dans les birapports radiaux $\rho, \bar{\rho}$. Nous prouvons un fait crucial que $| \rho |, | \bar{\rho} | < 1$ à l'intérieur du tube avant, et fixons des limites sur la vitesse à laquelle $| \rho |, | \bar{\rho} |$ tendent vers 1 lorsque la configuration à quatre points se rapproche de l'espace de Minkowski.

Nous fournissons également un guide de la littérature axiomatique QFT pour le public de CFT. Nous passons en revue les axiomes de Wightman et Osterwalder-Schrader (OS) pour les QFT lorentziennes et euclidiennes, ainsi que le célèbre théorème d'OS qui les relie. Nous passons également en revue un résultat classique de Mack sur la convergence OPE distributionnelle. Certains des arguments classiques s'avèrent utiles dans notre configuration. D'autres ne répondent pas à nos besoins en raison d'hypothèses lorentziennes (Mack) ou d'hypothèses euclidiennes invérifiables (théorème d'OS).

Dans la partie III, nous étudions les propriétés de convergence au sens des fonctions de l'OPE des fonctions de corrélation scalaires à quatre points dans une CFT. Nous établissons les critères de convergence de l'OPE dans les canaux s, t et u. Nous donnons une classification complète des configurations à quatre points possibles dans la signature lorentzienne. Nous montrons que toutes les configurations de chaque classe ont les mêmes propriétes de convergence de l'OPE dans les canaux s, t et u. Nous donnons des tableaux incluant les informations de convergence de l'OPE pour toutes les classes. En prenant l'union de toutes les configurations lorentziennes où au moins un canal de l'OPE converge, nous obtenons un domaine minimal d'analyticité des fonctions à quatre points lorentziennes CFT.

Dans la partie IV, nous donnons un aperçu de deux généralisations possibles : l'une, les fonctions à quatre points des opérateurs avec des spins généraux SO$(d)$; l'autre, les fonctions à quatre points dans le cylindre de Minkowski.

\nnchapter{Foreword}
This thesis is based on the following materials.

Part \ref{part:crossratio} is based on:

P.~Kravchuk, J.~Qiao and S.~Rychkov,
``Distributions in CFT. Part I. Cross-ratio space,''
\href{https://link.springer.com/article/10.1007/JHEP05(2020)137}{JHEP \textbf{05}, 137 (2020)}, 
\href{https://arxiv.org/abs/2001.08778}{arXiv:2001.08778 [hep-th]}.

Part \ref{part:minkowski} is based on:

P.~Kravchuk, J.~Qiao and S.~Rychkov,
``Distributions in CFT. Part II. Minkowski space,''
\href{https://link.springer.com/article/10.1007/JHEP08(2021)094}{JHEP \textbf{08}, 094 (2021)}, \href{https://arxiv.org/abs/2104.02090}{arXiv:2104.02090 [hep-th]}.

Part \ref{part:ope} is based on

J.~Qiao,
``Classification of Convergent OPE Channels for Lorentzian CFT Four-Point Functions,''
\href{https://arxiv.org/abs/2005.09105}{arXiv:2005.09105 [hep-th]}.

Part \ref{part:generalization} is based on some upcoming works:

P.~Kravchuk, J.~Qiao and S.~Rychkov,
``Distributions in CFT III. Spinning Fields in Minkowski space," work in progress.

P.~Kravchuk, J.~Qiao and S.~Rychkov,
``Distributions in CFT IV. Lorentzian Cylinder," work in progress.

\thispagestyle{empty}

\mainmatter

\clearpage
\nnchapter{Acknowledgements}
\thispagestyle{empty}

I'm grateful to too many people who interact with me during my years as a graduate student. More than half of my Ph.D. career has been through the pandemic, and life becomes more isolated than usual. I'm thankful to all people who made my life easier and kept me happy during this supposedly difficult time.  

First of all, I would like to express my deepest gratitude to my Ph.D. supervisor, Slava Rychkov. Our collaboration started when I was a master's student. It's my fortune to have a good supervisor early in my career who guides me into the field.  From him, I learned a huge amount of physics and mathematics, as well as how to be a clear thinker and a good collaborator (although I'm still on my way to this goal). It is always easy to find him when I need help and advice, even on weekends. I enjoyed chatting with him, not only about science but also about all aspects of everyday life and culture. Without him, the goal of my Ph.D. would be difficult to achieve.

I would like to thank my parents. They are always proud of me and give me all the support they could. Because of the Covid pandemic, it's been three years since my last stay in Shanghai, my hometown. Even though we can make frequent video calls, it makes me feel homesick that they care about their son 9,000 kilometers away all the time. When I was halfway through this thesis, Shanghai started the lockdown which goes very bad now. I hope they all stay well.

I am thankful to my collaborator Petr Kravchuk for the tremendous amount of help in carrying out the research projects.  I'm always amazed at his wealth of knowledge. It's a big pleasure to work with and learn from him.

I would like to acknowledge some members of the ENS theory group and other institutes: Gregory Korchemsky, Miguel Paulos, Balt van Rees, Eric Perlmutter, Piotr Tourkine, Jan Troost and Alexander Zhiboedov. They are always happy to help and give me useful advice. I benefited a lot from them during the discussions, journal clubs, seminars, lectures, administrative things, etc. Also thanks to Christoph Kopper, Madalena Lemos, Eric Perlmutter, Balt van Rees, Alexander Zhiboedov and my supervisor Slava Rychkov for being the jury members of my thesis defense and for pointing out the typos in my thesis.

I would like to thank my colleagues. It's always fun for me to chat with them about physics, mathematics, or just daily life. The huge list includes (but is not limited to) Amadou Bah, Gourab Bhattacharya, Andrea Caputo, Corina Ciobotaru, Lucía Córdova, Guilherme Ost de Aguiar, Ya Deng, Aaron Fenyes, Yifei He, Mikhail Isachenkov, François Jacopin, Apratim Kaviraj, Mendes Oulamara, Arthur Parzygnat, Mengxi Ren, Junchen Rong, Fidel Ivan Schaposnik, Benoit Sirois, Emilio Trevisani, Bernardo Zan, and Peng Zhou.

I would like to thank my friends Zhiying Chen, Jianfei He, Yuhang Hou, Tianhao Le, Yi Liu, Dan Mao, and Yujing Zhou. Although we have hardly met in person during these years, being able to keep in touch from time to time makes me feel good. Special thanks to my buddy Yichen Qin, who lives very close to me, so that we always have the opportunity to share good or shitty experiences. 

I would like to thank my former and current roommates in Bures: Jinyu Chen, Botao Dai, Yingyue Luo, and Jing Wang. Living with people from different fields and different cities gave me a lot of fun. I appreciate that they sometimes left a meal when I got home late. They may not know how valuable this is to someone like me who is stuck in research and needs some uplifting. 

Finally, I would like to thank all the friends that I have spent time with during my graduate life, including Shichao Du, Hao Fu, Zhaoheng Guo, Hongye Hu, Meng Kou, Songyuan Li, Chao Li, Mou Li, Zhengying Liu, Yi Pan, Zicheng Qian, Yichen Qin, Yijun Wan, Ao Wang, Yunzhi Wu, Mingchen Xia, Ying Zhang, Haowen Zhang, Yi Zhang, Meiyi Zhang, Xiang Zhao, Zechuan Zheng, Deliang Zhong, Zhaoxuan Zhu$\ldots$ The list can be endless and I have no way of naming them all. They are the ones who made my entire Ph.D. less tedious and more enriching, and our interactions have become some of the best memories of my life.

\nnchapter{Introduction}
\thispagestyle{empty}

The study of conformal field theories (CFTs) is essential for understanding strongly-coupled physics such as thermodynamic phase transition, confinement, quantum gravity, etc. In the framework of quantum field theories (QFTs), CFTs are the fixed points of the renormalization group (RG) flows. They are relatively easier to study than general QFTs because of the additional conformal symmetry and the algebraic structure.   

In the past, there were fruitful results and applications in two-dimensional CFTs \cite{belavin1984infinite,francesco1997conformal}. Due to the infinite-dimensional conformal algebra, namely the Virasoro algebra, one can make many predictions solely from studying the representations of the Virasoro algebra. With the extra self-consistency conditions such as crossing symmetry and modular invariance, the two-dimensional CFTs are even more constrained. In higher-dimensional CFTs, there are much fewer constraints than in $d=2$, mainly because the conformal group there is finite-dimensional. For this reason, few predictions were made in higher-dimensional CFTs.

Things have changed since a decade ago, thanks to the revival of bootstrap philosophy \cite{Rattazzi:2008pe}. The conformal bootstrap approach is based on several Euclidean CFT assumptions (which we call the \emph{Euclidean CFT axioms}), and it analyzes the constraints of self-consistency conditions on the CFT correlation functions. This approach makes precise numerical predictions of experimentally measurable quantities, such as the critical exponents of the 3d Ising model \cite{ElShowk:2012ht,El-Showk:2014dwa,Kos:2014bka,Simmons-Duffin:2015qma}, $O(N)$ model \cite{Kos:2016ysd,Kos:2013tga,Kos:2015mba,Chester:2019ifh}, and other critical systems (see the review \cite{Poland:2018epd}). 

While the basic CFT assumptions are made in Euclidean, many attempts have been made to study the bootstrap equations in the Lorentzian CFT. Due to the causal structure and richer singularities, many constraints are packaged into more visible forms in the Lorentzian signature. These studies include (but are not limited to) the conformal collider physics \cite{Hofman:2008ar}, the light-cone bootstrap \cite{Komargodski:2012ek,Fitzpatrick:2012yx}, the causality constraints and the averaged null energy condition (ANEC) \cite{Hartman:2015lfa,Hartman:2016dxc,Hartman:2016lgu,Kologlu:2019bco}, and the Lorentzian inversion formula \cite{Caron-Huot:2017vep,Simmons-Duffin:2017nub}. It would be interesting to know whether these Lorentzian constraints are available for Euclidean CFTs such as the critical Ising model. 

For general QFTs, the standard set of Lorentzian constraints are the Wightman axioms \cite{Streater:1989vi}:
\begin{itemize}
	\item (W0) Temperedness.
	\item (W1) Poincaré invariance.
	\item (W2) Unitarity.
	\item (W3) Mircocausality.
	\item (W4) Cluster property.
	\item (W5) Spectral condition.
\end{itemize} 
In the early days, these axioms and some extra CFT assumptions (weak conformal invariance, asymptotic operator product expansion) were considered to be the defining properties of Lorentzian CFTs \cite{Luscher:1974ez,Mack:1976pa}. However, it is not known yet whether these Lorentzian axioms follow from Euclidean CFT axioms. In particular, it was not known whether a generic Euclidean CFT correlator becomes a tempered distribution (i.e., satisfying (W0)) in the Lorentzian signature. Although not fully realized in this thesis, our dream goal is to derive Wightman axioms from Euclidean CFT axioms. \\

\textbf{Osterwalder-Schrader reconstruction theorem.}

One motivation of this thesis project comes from the celebrated \emph{Osterwalder-Schrader reconstruction theorem} \cite{osterwalder1973,osterwalder1975}. The theorem describes the relation between Euclidean and Lorentzian QFT axioms. The Euclidean version of Wightman axioms, namely \emph{Osterwalder-Schrader axioms}, is listed as follows
\begin{itemize}
	\item (OS0) Euclidean temperedness.
	\item (OS1) Euclidean invariance.
	\item (OS2) Reflection positivity.
	\item (OS3) Permutation symmetry.
	\item (OS4) Cluster property.
\end{itemize}
Axioms (OS1) - (OS4) are the natural Euclidean analog of Wightman axioms (W1)-(W4). The two main differences between OS and Wightman axioms are the temperedness properties and the spectral condition. In Wightman axioms, temperedness means that correlators are tempered distributions, i.e., continuous linear functionals acting on Schwartz test functions. The Fourier transform of a tempered distribution is still a tempered distribution, so we can consider the momentum-space correlators and formulate the spectral condition (W5). While in OS axioms, the Euclidean temperedness means that correlators are continuous linear functionals acting on a more restricted function space, where the test functions vanish rapidly at coincident points. For this reason, it is not clear a priori whether the Fourier transform of a Euclidean correlator exists in a proper distribution space where we can formulate the spectral condition.

The OS theorem says that under (OS0) - (OS4) plus some extra assumptions, one can Wick rotate the Euclidean correlation functions to the Lorentzian signature and get the Lorentzian correlators that satisfy Wightman axioms. We will describe the exact procedure of Wick rotation in chapter \ref{strategy}.

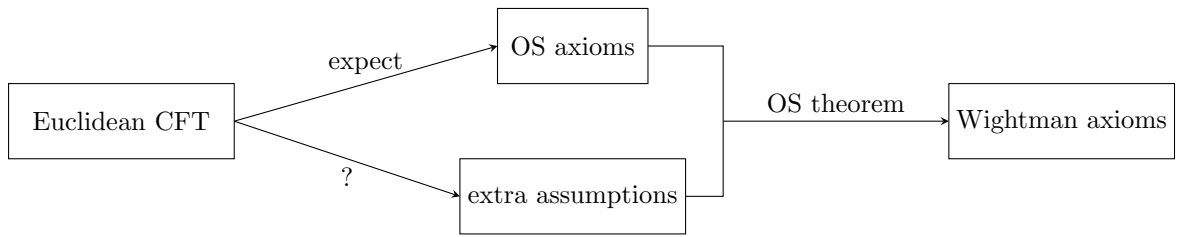
\begin{figure}[H]
	\centering
	\begin{tikzpicture}
		\begin{scope}
			\node[draw, fit={(0,0) (2,1)}, inner sep=0pt, label=center:OS axioms] (A) {};
			\node[draw, fit={(0,0) (3,1)}, xshift=6cm, yshift=-1cm, inner sep=0pt, label=center:Wightman axioms] (B) {};
			\node[draw, fit={(0,0) (3,1)}, xshift=-0.5cm, yshift=-2cm, inner sep=0pt, label=center:extra assumptions] (C) {};
			\node[draw, fit={(0,0) (3,1)}, xshift=-6.5cm, yshift=-1cm, inner sep=0pt, label=center:Euclidean CFT] (D) {};
			\draw (2,0.5) -- (3,0.5) -- (3,-1.5) -- (2.5,-1.5);
			\draw[-stealth] (3,-0.5) -- (6,-0.5) node[pos=0.5, above]{OS theorem};
			\draw[-stealth] (-3.5,-0.5) -- (-0.5,-1.5) node[pos=0.5, below]{?};
			\draw[-stealth] (-3.5,-0.5) -- (0,0.5) node[pos=0.5, above]{expect};
		\end{scope}
	\end{tikzpicture}	 
    \caption{\label{fig:OStheorem} The current status of the (partial) relationship between Euclidean CFT, OS axioms, and Wightman axioms.}
\end{figure}

We expect that the OS axioms are satisfied in Euclidean CFTs since they are also Euclidean QFTs. There are two versions of the extra assumptions: one assumes a stronger form of Euclidean temperedness; another assumes the growth of $n$-point function as a function of $n$, which is called the \emph{linear growth condition}. In practice, both of them are rather difficult to verify. We can check these technical conditions only when there is good control of all the correlators in a QFT, e.g., Gaussian free field, 2d critical Ising model, etc. We summarize the current status in figure \ref{fig:OStheorem}.
In the proof of OS theorem, the main use for introducing extra conditions is to prove power-law bounds on the analytically-continued Euclidean CFT correlators. These bounds imply that correlators are tempered distributions ((W0)) in the Lorentzian signature. Proving (W0) is the most technical part of the theorem: once this step is done, the other Wightman axioms can be obtained from OS axioms almost for free. For this reason, in this thesis we will discuss a lot on tempered distributions.

In CFT, the two- and three-point functions are kinematically fixed by conformal invariance. So it is not hard to prove Wightman axioms at the level of two- and three-point functions. As a modest step, we analyze the four-point functions, which is the first non-trivial CFT correlator. In this thesis, we will derive Wightman axioms for CFT four-point functions, starting from Euclidean CFT axioms. 

\textbf{Distributional properties in cross-ratio space.} \\ \\
As a warm-up exercise, we study the conformally invariant part of the CFT four-point function, denoted as $g(\rho,\bar{\rho})$. Our setup is the series expansion in radial cross-ratios $\rho,\bar{\rho}$ (in 1d there is only one independent cross-ratio variable):
\begin{equation}\label{intro:cbexpansion}
	\begin{split}
		g(\rho)=&\sum_{h}a_{h}\rho^{h}\qquad(d=1), \\
		g(\rho,\bar{\rho})=&\sum_{h,\bar{h}}a_{h,\bar{h}}\rho^{h}\bar{\rho}^{\bar{h}}\qquad(d\geqslant2).
	\end{split}
\end{equation}
It is well known that in the unitarity CFT, this expansion converges in the sense of functions inside the unit disk $\abs{\rho},\abs{\bar{\rho}}<1$ for the radial variable. This setup is more accessible than the position space because we do not have to study the complicated relation between cross-ratios and position-space coordinates. At this point, it is not necessary to introduce the whole Euclidean CFT axioms. 

We would like to show that the above expansion is convergent in the sense of tempered distributions on the boundary of the unit disk: $\abs{\rho}=\abs{\bar{\rho}}=1$. To prove this, we introduce Vladimirov's theorem \cite{Vladimirov}, which tells us that if we have a function $f(z)$ that is holomorphic in the upper half plane, then the limit $\lim\limits_{y\rightarrow0^+}f(x+iy)$ exists in the sense of distributions if the function does not blow up faster than power laws $(1+\abs{x})^\alpha y^{-\beta}$. The same conclusion holds for functions of several complex variables, if we let the imaginary parts of all variables go to zero simultaneously. 

Using Vladimirov's theorem and the fact that $g(\rho)$ has a power-law bound $(1-\abs{\rho})^{-\alpha}$, we show that the limit $\lim\limits_{r\rightarrow1^-}g(re^{i\theta})$ exists in the sense of tempered distributions in the variable $\theta$. The same is true for $g(\rho,\bar{\rho})$, using the two-variable version of Vladimirov's theorem. 

Furthermore, we will show that the above expansion converges in the sense of distributions on the boundary, which means that we can integrate (\ref{intro:cbexpansion}) term by term with any smooth test function $\varphi(\theta)$:
\begin{equation}\label{intro:distrconverge}
	\begin{split}
		\lim\limits_{r\rightarrow1^-}\int d\theta\, g(r e^{i\theta})\varphi(\theta)=&\sum_{h}a_{h}\int d\theta\,e^{ih\theta}\varphi(\theta)\qquad(d=1),  \\
		\lim\limits_{r,\bar{r}\rightarrow1^-}\int d\theta d\bar{\theta}\,g\fr{re^{i\theta},\bar{r}e^{i\bar{\theta}}}\varphi(\theta,\bar{\theta})=&\sum_{h,\bar{h}}a_{h,\bar{h}}\int d\theta d\bar{\theta}\,e^{ih\theta}e^{i\bar{h}\,\bar{\theta}}\varphi(\theta,\bar{\theta})\qquad(d\geqslant2).
	\end{split}
\end{equation}
Our results provide a new point of view about the \emph{bootstrap functionals} that are used in the conformal bootstrap. These functionals satisfy the swapping property \cite{Rychkov:2017tpc} when they are applied to the crossing equation. We will show that under proper coordinates, many bootstrap functionals can be interpreted as integrals against test functions, and the swapping property means distributional convergence, like eq.\,(\ref{intro:distrconverge}).

\textbf{Osterwalder-Schrader axioms from Euclidean CFT.} 

We will postulate the well-accepted Euclidean CFT rules that were used in numerical conformal bootstrap and call them \emph{Euclidean CFT axioms} (see section \ref{ECFTax} for details):
\begin{itemize}
	\item (ECFT0) Real analyticity. 
	\item (ECFT1) Conformal invariance.
	\item (ECFT2) Reflection positivity of two-point functions.
	\item (ECFT3) Hermiticity of two-point and three-point functions.
	\item (ECFT4) Permutation symmetry.
	\item (ECFT5) Convergence of operator product expansion (OPE).
\end{itemize}
Since OS axioms are the properties of Euclidean QFT correlators, it may not be surprising that they follow from Euclidean CFT axioms. In chapter \ref{CFTtoOS}, we will derive OS axioms from Euclidean CFT axioms at the level of two-, three- and four-point functions. Since in the Euclidean CFT axioms, we only postulate a weak form of OPE convergence, extra assumptions are required for deriving OS axioms of higher-point functions. In appendix \ref{OShigher}, we will discuss how CFT axioms with a stronger form of OPE convergence imply OS axioms of higher-point functions.

\textbf{Wightman axioms from Euclidean CFT four-point functions.} \\ \\
As mentioned above, the OS reconstruction theorem is not applicable here because we cannot check whether the extra assumption in the OS theorem is satisfied.

In chapter \ref{strategy}, we will introduce the basic strategy for deriving Wightman axioms from the Euclidean CFT axioms, following the spirit of OS reconstruction theorem. The most crucial step is to show that the analytically continued Euclidean correlators $G_n^E(x_1,\ldots,x_n)$ become tempered distributions in the Lorentzian signature. For this step, we rely on Vladimirov's theorem, which requires two properties of $G_n^E$
\begin{enumerate}
	\item $G_n^E$ has analytic continuation to a regime of complex coordinates, which is called the \emph{forward tube}.
	\item When approaching the Lorentzian regime from the forward tube (which roughly means $x_k=(\epsilon_k+it_k,\mathbf{x}_k+i\mathbf{y}_k)$ with $\epsilon_k,\mathbf{y}_k\rightarrow0$), the correlator should satisfy some power-law bound.
\end{enumerate}
We first consider the cases of two- and three-point functions in chapter \ref{23-point}. One can easily prove the analyticity and power-law bound since they are fixed (up to constant factors) by conformal invariance. Therefore, the analytically continued Euclidean two- and three-point functions become tempered distributions in the Lorentzian signature.

In chapter \ref{sec:4-point}, we consider the scalar CFT four-point functions (which we will simply call four-point functions). The four-point function can be factorized into a scaling prefactor and a conformally invariant factor. The scaling prefactor is the product of two-point functions, so its analyticity and power-law bound are easy to verify. The conformally invariant factor, $g(\rho,\bar{\rho})$, was already mentioned above. It has a series expansion (\ref{intro:cbexpansion}) in $\rho,\bar{\rho}$, convergent when $\abs{\rho},\abs{\bar{\rho}}<1$. We will prove two key facts on the variables $\rho,\bar{\rho}$. The first fact is that $\abs{\rho},\abs{\bar{\rho}}<1$ when the four-point configuration is in the forward tube. The second fact is that when approaching the Lorentzian regime (which is on the boundary of the forward tube), the inverse distances between $\rho,\bar{\rho}$ and the unit circle (i.e., $(1-\abs{\rho})^{-1},(1-\abs{\bar{\rho}})^{-1}$) are bounded by some power laws. Using these geometrical facts and the series expansion in $\rho,\bar{\rho}$, we prove the analyticity and power-law bound of the four-point function in the forward tube. 

Then we apply Vladimirov's theorem to show that the four-point function is a tempered distribution in the Lorentzian regime. Once temperedness is proved, the other Wightman axioms (Lorentzian conformal invariance, clustering, microcausality) follow from Euclidean assumptions by some standard argument in the OS paper \cite{osterwalder1973}. 

Since the Lorentzian four-point function is a tempered distribution instead of a genuine function, it is interesting to understand how its regularity depends on the scaling dimensions of the operators. In chapter \ref{secondpass}, we will show that the radial cross-ratios $\rho,\bar{\rho}$ satisfy a Cauchy-Schwarz type inequality. Using this inequality, we derive an optimal power-law bound of $(1-r)^{-1}$, where $r=\max\{\abs{\rho},\abs{\bar{\rho}}\}$.\footnote{By optimal we mean that the power indices are optimal.} The optimal bound of $\rho,\bar{\rho}$ implies an optimal bound of the four-point function, which is saturated in the generalized free theory.

In chapter \ref{OPEconvMink}, we will discuss the OPE convergence properties, similarly to the discussion in part \ref{part:crossratio}. We show that the s-channel OPE converges in the sense of distributions in Minkowski space. We estimate the OPE convergence rate in the case when the test functions are compactly supported. In Euclidean space, it is well-known that the OPE converges exponentially fast when the scaling-dimension cutoff goes to infinity. However, in Minkowski space, the error term of the truncated OPE decays in a power law in the scaling-dimension cutoff. We also rephrase the OPE convergence in the forward tube, from the point of view of the two-operator states $|\mathcal{O}(x_1)\mathcal{O}(x_2)\rangle$.  We show that in the forward tube region,\footnote{Here we mean that the three-point configuration $(0,x_1,x_2)$ is in the forward tube.} $|\mathcal{O}(x_1)\mathcal{O}(x_2)\rangle$ is an Hilbert-space-valued analytic function, and its OPE converges in the sense of the CFT Hilbert space.

\textbf{Domain of analyticity of the CFT four-point functions.}

In Minkowski space, one can ask: in which regions the CFT four-point function is analytic, not just a distribution. In the framework of Wightman QFTs, the correlators are known to be analytic functions when the operators are pairwise space-like separated \cite{ruelle1959}. However, the correlators involving time-like separated operators are not thoroughly studied. Since there are more constraints in CFT, coming from conformal invariance and OPE convergence, we expect that the CFT correlators are analytic functions in bigger domains than general QFT correlators. In part \ref{part:ope}, we study this problem at the level of four-point functions. Since our Lorentzian four-point function is constructed by the OPE of analytically-continued Euclidean four-point functions, our analysis will be based on the convergence properties of OPE in the Lorentzian regime.

In chapter \ref{section:lorentz4pt}, we will establish some criteria for OPE convergence in s-, t- and u-channels. In the previous part, we have already shown that $\abs{\rho},\abs{\bar{\rho}}<1$ for four-point configurations in the forward tube, which implies the s-channel OPE convergence. Since the Lorentzian configurations lie on the boundary of the forward tube, by continuity $\abs{\rho},\abs{\bar{\rho}}<1$ is equivalent to $\abs{\rho},\abs{\bar{\rho}}\neq1$ there. Thus the s-channel OPE at a Lorentzian configuration is convergent if $\abs{\rho},\abs{\bar{\rho}}\neq1$, or equivalently $z,\bar{z}\notin[1,+\infty)$ in terms of $z,\bar{z}$ variables. This is the criterion of s-channel convergence. 

There are two other ways of expanding the CFT four-point functions: t-channel and u-channel expansions. The three expansions satisfy the \emph{crossing symmetry}, meaning that they agree in their common domain of convergence in the Euclidean signature. By uniqueness of analytic continuation, they also agree in the complex regime. In the Lorentzian signature, it is possible to have some configuration where the four-point function has convergent expansion in only one OPE channel. Therefore, it is essential to know the OPE convergence properties in all channels. We establish criteria for OPE convergence in the t- and u-channels, based on $\abs{\rho_t},\abs{\bar{\rho}_t}<1$ and $\abs{\rho_u},\abs{\bar{\rho}_u}<1$ for radial variables in t- and u-channels. A big difference between the s-channel and other channels is that deciding the s-channel convergence only relies on the values of the cross-ratio variables at the Lorentzian configuration. In contrast, for the t- and u-channel convergence, one must consider the behaviors of cross-ratio variables along the analytic continuation path. 

We will use the above-established criteria to check OPE convergence for all possible Lorentzian four-point configurations. Chapter \ref{section:classifylorentzconfig} gives a classification of Lorentzian four-point configuration, which reduces the OPE check from the whole Lorentzian configuration space to finitely many cases. The classification is basically according to the causal ordering and the range of the cross-ratio variables $z,\bar{z}$. Then we check whether each configuration class satisfies the OPE convergence criteria. As there are still 81 cases to check after reduction, we show the results in terms of some tables in appendix \ref{appendix:tableopeconvergence}.

\textbf{Four-point functions of spinning operators}

In chapter \ref{chap:spinning4pt}, we will discuss the generalization to the four-point functions of operators with SO(d)-spins (which exclude the fermionic operators). For the spinning case, the main extra difficulty is that the tensor structures appear. In $d=2$, the rotation group, which is SO(2), is abelian. As a result, there is only one tensor structure for each spinning four-point function. Then the treatment of the spinning four-point functions in $d=2$ is quite similar to the scalar case. However, in $d\geq3$, a four-point function may contain several tensor structures. It is hard to show that each tensor structure has analytic continuation to the forward tube. Another point of view on this difficulty is that in $d\geq3$, a four-point configuration has a SO(d-2) or SO(d-1) little group, which can be nontrivial. The little group ambiguity will cause non-analyticity of the rotation matrices in the formulation of conformal invariance.

To bypass the above difficulty, we will take another strategy for the analytic continuation of the spinning four-point function. In the Euclidean signature, the four-point function has a conformal partial wave expansion, which is convergent by the OPE convergence assumption in the Euclidean CFT axioms. We would like to show directly that each conformal partial wave, denoted as $G^{a_1a_2a_3a_4}_{1234,\mathcal{O}}$, has analytic continuation to the forward tube, and satisfies a Cauchy-Schwarz type inequality. Using these two properties, we will show that the conformal partial wave expansion is convergent in the forward tube, which performs the analytic continuation of the four-point function. The remaining derivation of the Wightman axioms is similar to the scalar case.

We will show the analyticity and the Cauchy-Schwarz type inequality of the conformal partial wave $G^{a_1a_2a_3a_4}_{1234,\mathcal{O}}$, for generic internal scaling dimension $\Delta_{\mathcal{O}}$. There exists a discrete set of exceptional $\Delta_{\mathcal{O}}$'s, where our proof does not apply, but we hope that the same results still hold by the existence of $G^{a_1a_2a_3a_4}_{1234,\mathcal{O}}$ in the Euclidean signature and its continuity in $\Delta_{\mathcal{O}}$ in the generic case.

\textbf{Scalar four-point functions in the Minkowski cylinder}

In chapter \ref{chap:Minkcylinder}, we will discuss the generalization to the scalar four-point functions in the Minkowski cylinder $\mathbb{R}\times S^{d-1}$, which is another important container of Lorentzian CFTs. In the framework of AdS${}_{d+1}/$CFT$_d$-correspondence, it is well-known that the Minkowski cylinder is the boundary of the Lorentzian AdS${}_{d+1}$ \cite{Aharony:1999ti}.

The CFT correlators in the Minkowski cylinder are defined by analytically continuing the CFT correlators in the Euclidean flat space, using the cylinder coordinates. This Wick rotation procedure has been studied in the early days by Lüscher and Mack \cite{Luscher:1974ez}. Their basic assumptions are the Wightman axioms and the so called \emph{weak conformal invariance}, which basically means conformal invariance in the Euclidean signature. Using the same argument as Glaser, Osterwalder and Schrader  \cite{Glaser1974,osterwalder1975}, they showed that the CFT correlation function has analytic continuation to the cylinder forward tube, where the real parts of the cylinder temporal variables are ordered: $\mathrm{Re}(\tau_1)>\mathrm{Re}(\tau_2)>\ldots>\mathrm{Re}(\tau_n)$. The Minkowski cylinder points correspond to $\mathrm{Re}(\tau)=0$. However, without extra assumptions, they did not manage to show that the limit exists when all the cylinder temporal variables become imaginary. Therefore, deriving Wightman axioms for CFT correlators in the Minkowski cylinder is still an open problem.

We would like to show that in the Minkowski cylinder, the CFT four-point functions are tempered distributions in the cylinder temporal variables and continuous functions in the spherical variables. Our strategy is quite similar to the case of the scalar four-point function in Minkowski flat space. We will show that $\abs{\rho},\abs{\bar{\rho}}<1$ for four-point configurations in the forward tube. We will also derive power-law bounds of $(1-\abs{\rho})^{-1},(1-\abs{\bar{\rho}})^{-1}$ in the forward tube. Then the remaining arguments are the same as the case of Minkowski flat space. 

We will briefly demonstrate the Wightman axioms in the Minkowski cylinder. There are three main differences in the Minkowski cylinder. First, the global conformal invariance holds in the cylinder. In contrast, conformal invariance only holds infinitesimally in the Minkowski flat space. Second, there is no cluster property in the cylinder, because the spatial directions are compactified. Last, the spectral condition in the cylinder case holds only for the Fourier transform of temporal variables.

The Minkowski cylinder is an $\infty$-covering of the (compactified) Minkowski flat space. Since people usually identify the Minkowski space to the Poincaré patch of the Minkowski cylinder \cite{Aharony:1999ti}, it is natural to ask whether the CFT correlators in the Poincaré patch are the same as in the Minkowski space, up to the scaling prefactors. We claim that the answer is yes, for general CFT $n$-point functions. The key observation for the proof is that each configuration in the Poincaré patch is connected to the Euclidean cylinder by a path in the cylinder forward tube, which is mapped to a path in the flat-space forward tube. Then the uniqueness of analytic continuation tells us that the correlators on two sides must match. This result is true not only for four-point functions but also for any $n$-point functions.

The CFT four-point functions have richer singularities in the Minkowski cylinder than in the Minkowski space. One may expect a more complicated classification of the four-point cylinder configurations for the OPE convergence properties in the sense of functions. However, we will show that in the cylinder case, the classification of the four-point cylinder configurations can be easily reduced to the case when all the points are in the Poincaré patch, with the help of the action of a $\mathbb{Z}$-group, which is the center of the cylinder conformal group.

\begin{part}{Distributions in CFT I: cross-ratio space}\label{part:crossratio}
\thispagestyle{fancy}

\chapter{Introduction}

Historically, distributions played a big role in axiomatic approaches to
quantum field theory (QFT), via Wightman axioms~\cite{Streater:1989vi} or Osterwalder-Schrader
axioms~\cite{osterwalder1973,osterwalder1975}. In particular, the language of tempered distributions allows clean
treatment of correlation functions singularities at $x^2 = 0$ in a UV-complete
QFT, where $x^2$ may be Euclidean or Lorentzian distance.

In recent years, a new axiomatic approach---the conformal bootstrap---has
emerged in the study of conformal field theories (CFTs) in dimension $d
\geqslant 2$, i.e.\ quantum field theories invariant under the action of
conformal group (see review~\cite{Poland:2018epd}). This approach is both rigorous and calculable. On the
numerical side, it has allowed precise determinations of many experimentally
measurable quantities, such as the critical exponents of the 3d Ising \cite{ElShowk:2012ht,El-Showk:2014dwa, Kos:2014bka,Simmons-Duffin:2015qma,Kos:2016ysd}, $O (N)$ \cite{Kos:2013tga,Kos:2015mba, Kos:2016ysd, Chester:2019ifh}
and other critical points. On the analytic side, it also led to many insights
into the structure of operator spectrum of general CFTs, in particular
concerning how operators organize themselves in infinite families (Regge
trajectories)~\cite{Komargodski:2012ek,Fitzpatrick:2012yx,Caron-Huot:2017vep}. Numerical bootstrap studies typically take place deep in the
Euclidean region, staying away from the contact term singularities of
correlation functions at short distances. In this regime, the rules of the game
are well-understood and comprise the Euclidean bootstrap axioms. 

On the other hand, analytical
bootstrap studies often boldly go into the Lorentzian space, probe light-cone or other types of singularities. In this regime the most common set of assumptions for correlation functions are the Wightman axioms~\cite{Streater:1989vi}, but it has never been shown how these assumptions follow from the
well-understood Euclidean bootstrap axioms. To achieve this is the goal of
this series of papers. The uniting theme of this work will be tempered
distributions, hence the title.

In this first paper of the series (part \ref{part:crossratio})we will study convergence of the conformal
block decomposition. As is well known, it converges in the sense of functions
inside the unit disk $| \rho |, | \bar{\rho} | < 1$ for the radial variable.
We will show that it converges in the sense of distributions also on the
boundary of this unit disk. This is done using Vladimirov's theorem~\cite{Vladimirov}---a key
result in the theory of functions of several complex variables that we will
carefully introduce.

Vladimirov's theorem provides the answer to the following question: if we have a function
$g(\rho)$ that is holomorphic in the open unit disc $|\rho|<1$, what can we say about
its values for $|\rho|=1$? If $g(\rho)$ were bounded, then the limit $\lim_{r\to 1} g(re^{i\theta})$
would be guaranteed to exist for almost every $\theta$ and give rise to a bounded function $g(e^{i\theta})$. However, the functions of cross-ratios that
we encounter in conformal field theory are not bounded and instead can blow up near the boundary.
Crucially though, it is easy to show (as we do in this part of the thesis) that they blow up only as power laws
$(1-r)^{-K}$. In this case, Vladimirov's theorem guarantees that the limit $\lim_{r\to 1} g(re^{i\theta})$
exists in the space of distributions in the variable $\theta$. We will explain that this conclusion
holds both for the correlation function itself as well as for the individual terms in the conformal 
block expansion, which will allow us to prove convergence of conformal block expansion in the space of distributions. A simple yet illustrative example of distributional convergence is the sum
\be
\sum_{n=-\oo}^{+\oo} e^{in\theta}=2\pi \de(\theta),
\ee
where $\theta\in(-\pi,\pi]$ is the coordinate on the unit circle. This sum doesn't converge
in the usual sense because every term is of absolute value 1, but it does converge after being 
smeared with a smooth test function $f(\theta)$. We will study a more realistic toy example in section~\ref{sec:toy}.

Our results can be interpreted as introducing a new class of functionals which satisfy the swapping property~\cite{Rychkov:2017tpc}
when applied to the crossing equation. This point of view might be helpful for readers with interest in analytic functional bootstrap~\cite{Mazac:2016qev,Mazac:2018mdx,Mazac:2018ycv,Kaviraj:2018tfd,Mazac:2018biw,Hartman:2019pcd,Paulos:2019gtx,Mazac:2019shk}.
Specifically, we show that integration (appropriately defined) of the crossing equation with a test function over the boundary of the crossing region\footnote{By crossing region we mean the region in cross-ratio space where both $s$- and $t$-channel conformal block expansions converge. In the standard $z$-cross ratio it is given by $\C$ minus the cuts along $[1,+\oo)$ and $(-\oo,0]$.} can be exchanged
with the sum over conformal blocks. We prove this result for infinitely smooth test functions, and argue that it likely can be strengthened to enlarge the class
of test functions sufficiently so that our new class of functionals will include all functionals currently known to satisfy the swapping property.

In our second paper {\cite{Kravchuk:2021kwe}} (part \ref{part:minkowski}), CFT Wightman four-point functions in Lorentzian
space will be shown to be tempered distributions, thus establishing Wightman
axioms. In the third paper {\cite{paper3}} (chapter \ref{chap:Minkcylinder}), we will study analytic
continuations of CFT correlation functions to the Lorentzian cylinder (also
known as the boundary of the AdS space). Our goal is to establish everything
from Euclidean bootstrap axioms, without any extra assumptions. When the time
comes, we will explain that the existing classic results in the literature,
like the Osterwalder-Schrader theorem~\cite{osterwalder1973,osterwalder1975} or the construction of L{\"u}scher and
Mack~\cite{Luscher:1974ez}, all require additional assumptions. So our conclusions cannot be recovered from the classic papers.
Fortunately, we found a different way of reasoning which recovers all the results commonly assumed to be true, for the most important in applications case of four-point functions.\footnote{It's an interesting open problem how to extend our arguments to higher point functions.} The good news is that our alternative arguments are really easy, and the main idea can really be summarized in one sentence: ``Look for a powerlaw bound.'' 
This should be contrasted with the classic papers which are quite intricate.

This part of the thesis is organized as follows. In chapter~\ref{sec:cbexpansion} we discuss the motivation for our work from the 
point of view of computing Euclidean and Lorentzian correlation functions. In chapter~\ref{sec:1dcase} we consider the simplified
case of one cross-ratio, starting with a toy example of power series. We also use this simplified setting to discuss possible applications
of our results to analytic functional bootstrap (section~\ref{sec:functionals}) and to proper definition of discontinuities (section~\ref{sec:dispersion}).
In chapter~\ref{sec:higherd} we consider the case of two cross-ratios in scalar correlators in general number of dimensions. We comment on
applications and generalization to spinning correlators. 
In section \ref{sec:Bissi} we discuss an application in the context of a single-variable dispersion relation recently proposed by Bissi, Dey and Hansen \cite{Bissi:2019kkx}. We conclude in chapter~\ref{sec:conclusions}.

\chapter{Conformal block expansion}
\label{sec:cbexpansion}

In this chapter we will state our basic problem, and the main idea how to solve it.
Let us consider the conformal block expansion of a four-point function of identical scalar operators 
(we will consider more general four-point functions later)
\begin{equation}
	\label{eq:4pt}
	\<\f(x_1)\f(x_2)\f(x_3)\f(x_4)\>=\frac{1}{(x_{12}^2)^{\De_\f}(x_{34}^2)^{\De_\f}}g(u,v),
\end{equation}
where, as usual
\begin{equation}
	u=\frac{x_{12}^2x_{34}^2}{x_{13}^2x_{24}^2},\quad v=\frac{x_{14}^2x_{23}^2}{x_{13}^2x_{24}^2}.
\end{equation}
We will mostly be working with the radial coordinates $\rho,\bar\rho$~\cite{Pappadopulo:2012jk,Hogervorst:2013sma} defined as
\begin{equation}
	\label{eq:rho}
	\rho=\frac{z}{(1+\sqrt{1-z})^2},\quad 	\bar\rho=\frac{\bar z}{(1+\sqrt{1-\bar z})^2},
\end{equation}
where $z,\bar z$ are determined by
\begin{equation}
	\label{eq:zzbar0}
	z\bar z=u, \quad (1-z)(1-\bar z)=v.
\end{equation}
We will abuse the notation a bit by writing $g(u,v), g(z,\bar z)$, or $g(\rho,\bar\rho)$ depending
on which set of cross-ratios we want to use.

The function $g(\rho,\bar \rho)$ can be expanded in conformal blocks in $\f(x_1)\times\f(x_2)$ OPE channel as follows,
\be
\label{eq:cbexpansion}
g(\rho,\bar \rho) = \sum_{\De,J} p_{\De,J} g_{\De,J} (\rho,\bar \rho),
\ee
where $p_{\De,J}\geq 0$ are the OPE coefficients squared, and $g_{\De,J}(\rho,\bar \rho)$ 
are the conformal blocks. This expansion is known to be absolutely convergent in the region $|\rho|<1,|\bar\rho|<1$, which we will denote by $\cC$ in what follows.

We will only use the global conformal invariance $SO(d+1,1)$. Under these assumptions, the region $\cC$ is the largest region of convergence of the conformal block decomposition of a general CFT four-point function (we are not aware of any results to the contrary). In 2d CFT, using Virasoro, the region of convergence can be extended further in terms of Al.~Zamolodchikov's uniformizing $q$ variable, being given by $|q|,|\bar q|<1$ which is a strictly larger region than $\cC$~\cite{Hogervorst:2013sma,Maldacena:2015iua}. So our results should be best possible in $d>2$ but not in $d=2$.

Above we focused on the 12 OPE channel ($s$-channel) but the same discussion can be made for the $t$-channel 23 and $u$-channel 13, whose convergence is characterized by the conditions $|\rho_t|,|\bar\rho_t|<1$ and $|\rho_u|,|\bar\rho_u|<1$.

Let us briefly describe what the region $\cC$ corresponds to in the physical space of $x_i$.
In Euclidean signature, this region includes all configurations when the four points $x_i$ do not lie on a circle, which is the generic case. 
If $x_i$ do lie on a circle, the cross-ratios belong to $\cC$ if $x_1$ and $x_2$ are next to each other on the circle. If the points instead fall on the 
circle in the ordering $x_1, x_3, x_2, x_4$ (read in some direction), then we find $|\rho|=|\bar\rho|=1$. Therefore, only a measure zero set of Euclidean configurations does not 
belong to $\cC$ and is instead on its boundary $\ptl \cC$. For these configurations the $s$-channel expansion does not converge. However, it does converge for $t$- and $u$-channels, and so
the value of the Euclidean four-point function can be determined from the OPE for any configuration of the four points.

Our basic problem is to make sense of the four-point function~\eqref{eq:4pt} in Lorentzian signature. 
In order to talk about a Lorentzian four-point function, we need to specify which operator ordering we are interested in. We will only consider here the Wightman functions, i.e.~we fix operator ordering:
\begin{equation}
	\label{eq:bestordering}
	W(x_1,x_2,x_3,x_4)= \<0|\f(x_1)\f(x_2)\f(x_3)\f(x_4)|0\>\,.
\end{equation}
However, we wish to consider all possible time and causal ordering of the points $x_i$.\footnote{Other often considered Lorentzian correlators (retarded, advanced, time-ordered) can be obtained by multiplying Wightman functions with appropriate factors enforcing the needed ordering. The Wightman functions being distributions, and the time-ordering factors being singular, this procedure introduces extra singularities and requires care in a rigorous treatment.} Once we have fixed the operator ordering, the Lorentzian
four-point function can be obtained from the Euclidean one by an appropriate analytic continuation. While in Euclidean we always have
$\bar\rho=\rho^*$, this property is generally lost after the analytic continuation. Furthermore, there are open regions in the Lorentzian configuration space of $x_i$ where $|\rho|$ and/or $|\bar \rho|$ end up $\ge 1$ after the analytic continuation. Then the corresponding conformal block expansion~\eqref{eq:cbexpansion} diverges and thus cannot be used to determine the correlator.

One such well known case is the Regge regime~\cite{Cornalba:2007fs,Cornalba:2008qf,Costa:2012cb}, when $x_1,x_4$ and $x_2,x_3$ pairs are timelike separated, while all other intervals are spacelike (see Fig.~\ref{fig-regge}).
\begin{figure}[h!]
	\centering
	\includegraphics[scale=0.8]{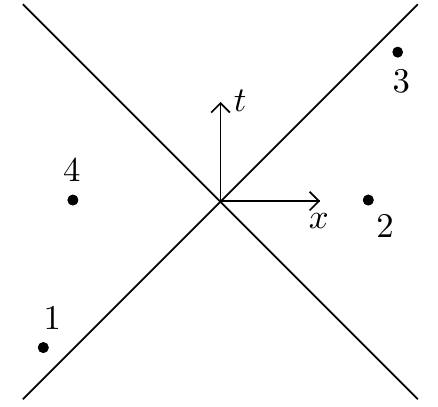}
	\caption{\label{fig-regge} Regge kinematics.}
\end{figure}
One may be tempted to use the 13 OPE for this Lorentzian correlator, because this channel is the most symmetric with respect to the origin, and also because one may be interested in the limit $x_{13}^2\to 0$. However, although this channel converges when points $x_1$ and $x_3$ stay close to the origin, it starts diverging when they cross the lightcones of $x_2$ and $x_4$ and move into the Regge regime, because the corresponding $\rho_u$ variable become larger than 1.\footnote{Conformal Regge theory~\cite{Cornalba:2007fs,Cornalba:2008qf,Costa:2012cb} provides a way to resum the expansion~\eqref{eq:cbexpansion} in the limit $\bar\rho_u\to0$, $\rho_u\to\oo$ with $\rho_u\bar\rho_u$ fixed. We will not consider such resummations in this part of the thesis since they rely on analytically-continued OPE data that we have little control over.} In this particular case, one can switch to the 23 OPE for which $|\rho_t|,|\bar \rho_t|<1$ is less than 1, and so this channel converges. However, this is not always possible: there exist kinematic configurations when no channel converges (appendix~\ref{app:lorentz}).

In this series of works we will propose a different way to solve this problem, and recover the Wightman function in all kinematic configurations. In our construction the key role will be played by the 12 OPE-channel. We call it the ``vacuum channel'', because it involves the two leftmost operators  in the Wightman ordering~\eqref{eq:bestordering}, i.e.~the ones acting on the vacuum. While the vacuum channel OPE does not always converge, it \emph{almost converges} for all possible configurations. What this means is that $|\rho|\leq 1$ and $|\bar\rho|\leq 1$ for all values of $x_i$. This crucial fact will be shown in~\cite{Kravchuk:2021kwe}. It is only true for the vacuum OPE channel, but would not be true for the 23 or 13 channels, for which sometimes $|\rho|$ and/or $|\bar\rho|$ will be strictly greater than 1. In particular, as we show in appendix \ref{app:lorentz}, there exist configurations for which both 23 and 13 channels diverge with $|\rho|,|\bar \rho|>1$, while for 12 channel $|\rho|=|\bar \rho|=1$.


In other words, all possible Lorentzian configurations belong to the closure $\bar \cC$. One can ask how large are the regions in configuration space of $x_i$ which belong in $\ptl \cC$ but not in $\cC$.
In Euclidean, we have seen that these configurations were measure zero, but in Lorentzian this is no longer true: extended regions with non-empty interior have $|\rho|=1$, $|\bar\rho|=1$. So, a fraction of configurations are in $\ptl \cC$ and not in $\cC$. 

If the conformal block expansion converged in $\bar \cC$ and not $\cC$, we would be able to use it to compute any Lorentzian correlator
in any configuration of the points $x_i$. Of course, this is not the case, and the conformal block expansion converges in the usual sense only in $\cC$. However, our goal in this part of the thesis
will be to extend the notion of convergence so that it will become valid in $\bar\cC$. Specifically, we will show that the expansion~\eqref{eq:cbexpansion}
converges in the sense (to be clarified below) of distributions on the boundary $\ptl\cC$ in the cross-ratio space. In the forthcoming work~\cite{Kravchuk:2021kwe,paper3}
we will extend this result to convergence in the sense of distributions in the physical space of $x_i$, either in Minkowski space, or on the Lorentzian cylinder.

One may be wondering what is special about the vacuum channel compared to other OPE channels. Intuitively, the distinguishing feature of vacuum channel is that we can understand it as inserting a complete set of states in the Wightman four-point function. Since Wightman four-point functions are distributions, we cannot generally expect this sum to make sense in terms of functions, but only in terms of distributions. Mack~\cite{Mack:1976pa} understood the vacuum channel OPE expansion in distributional sense in position space. Mack's reasoning is rather nontrivial, and it crucially relies on assuming from the start that Wightman axioms hold in Lorentzian signature---an assumption that we are here not willing to accept. Although our results in cross-ratio space are inspired by Mack's considerations in position space, they do not follow from his results, since we rely on a different and simpler set of assumptions, natural from the modern bootstrap perspective. Also, we are only using rather elementary methods. Position space will be discussed in~\cite{Kravchuk:2021kwe}.

\chapter{One-dimensional case}
\label{sec:1dcase}

First, let us simplify the problem by considering the one-dimensional case where there is a single cross-ratio. The conformal block expansion takes the form
\begin{equation}
	\label{eq:1dcbexpansion}
	g(\rho) = \sum_{\De} p_\De g_\De (\rho),
\end{equation}
where the conformal blocks are given by\footnote{This equation follows from the more familiar one in the $z$ coordinate $g_\De(z)=z^\De {}_2 F_{1}(\De,\De;2\De;z)$ by a hypergeometric identity~\cite{Hogervorst:2013sma}.}
\begin{equation}
	\label{eq:1dblock}
	g_\De(\rho) = (4\rho)^\Delta {}_2 F_{1}(1/2,\De;\De+1/2;\rho^2).
\end{equation}
Furthermore, the sum is over $\De\geq 0$ and $p_\De\geq 0$. This expansion converges, in the usual sense, in the interior of the unit disk $|\rho|<1$, and our goal is to understand whether it can be made convergent, in some generalized sense, on the boundary $|\rho|=1$. 

If we look at~\eqref{eq:1dcbexpansion} more closely, we will notice that the conformal blocks~\eqref{eq:1dblock} are not single-valued in the unit disk of $\rho$. Therefore, we
are really interested in the behavior of this sum on the universal cover of the unit disk branched at $0$, which can be conveniently parametrized by writing
\begin{equation}
	\rho=e^{i\tau}.
\end{equation}
The expansion~\eqref{eq:1dcbexpansion} is then absolutely convergent in the upper-half plane of $\tau$, and we are interested in its convergence for real $\tau$.

\begin{figure}[h!]
	\centering
	\includegraphics[scale=0.7]{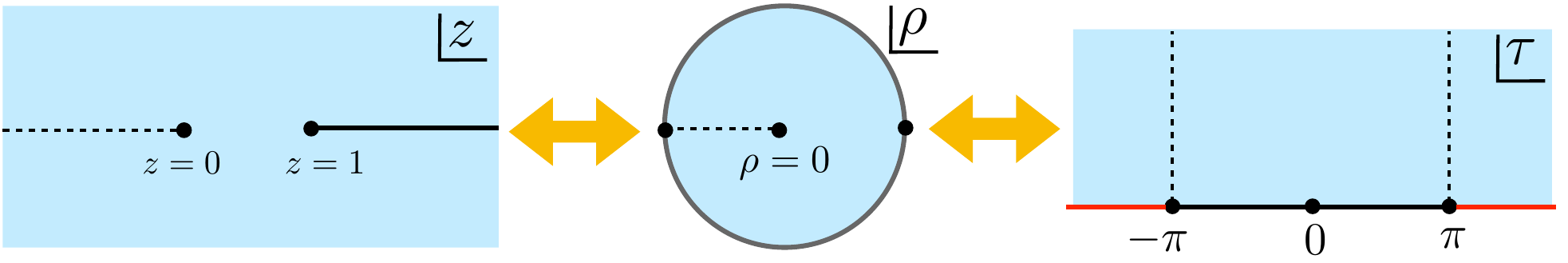}
	\caption{\label{fig:zrhotau} Transformation from the $z$ cut plane to the $\rho$ disk to the $\tau$ upper-half plane, see the text.}
\end{figure}

In Fig.~\ref{fig:zrhotau} we show the transformation from the $z$ cut plane to the $\rho$ disk to the $\tau$ upper-half plane. The two sides of the cut $z\in[1,+\infty)$ are mapped on the boundary of the unit disk $|\rho|=1$, and then to the black part ($\tau\in[-\pi,\pi]$) of the upper-half plane boundary. The rest of the $\tau$ boundary (marked in red) can be accessed in the $\rho$ variable by first going through the cut $\rho\in[-1,0]$ (dashed) and then approaching $|\rho|=1$. 

On the black interval $\tau\in[-\pi,\pi]$ (except at $\tau=0$) the four-point function is actually analytic, as can be shown using the $t$-channel expansion. On the rest of the boundary (red part), the $t$-channel expansion does not converge and provides no information. Below we will show, using the $s$-channel, that the four-point function is a tempered distribution on the whole boundary. We will also show that the $s$-channel conformal block expansion converges in the sense of distributions. When using the $s$-channel, we have to use distributional convergence even on the black part of the boundary, although the function itself is analytic there as explained above. 

\section{A toy problem}
\label{sec:toy}

In order to gain some intuition, it is useful to consider the following toy problem. Let us study the power series
\begin{equation}
	\frac{1}{1-\rho}=\sum_{n=0}^\oo \rho^n.
\end{equation}
It has the similar feature that it converges absolutely for $|\rho|<1$ and that the resulting function has a power-like singularity at $\rho=1$, much like the physical four-point functions do.

In terms of $\tau$ variable we find the sum
\begin{equation}\label{eq:tausum}
	\sum_{n=0}^\oo e^{in\tau},
\end{equation}
which clearly does not converge for any real $\tau$. We claim that it does converge as a tempered distribution.\footnote{A tempered distribution is a distribution that can be paired with Schwartz test functions (see below).} For example, let us compute its real part using the standard formulas of Fourier analysis
\begin{equation}
	\mathrm{Re}\sum_{n=0}^\oo e^{in\tau}=\half\sum_{n=-\oo}^\oo e^{in\tau}+\half=\half+\pi \sum_{k=-\oo}^\oo \de(\tau-2\pi k).
\end{equation}
It is a bit harder to compute the imaginary part, but we can run the following simple argument for the full sum~\eqref{eq:tausum}. Let $f(\tau)$ be a Schwartz test function, i.e. a smooth\footnote{In this part of the thesis ``smooth'' means $C^\infty$ and the two terms are used interchangeably.} function which, together with its derivatives, decays at infinity faster than any power. In order
to show that~\eqref{eq:tausum} converges as a tempered distribution, we need to show, by definition, that the partial sums
\begin{equation}\label{eq:toypartialsums}
	\int d\tau f(\tau) \sum_{n=0}^N e^{in\tau}= \sum_{n=0}^N \int d\tau f(\tau) e^{in\tau}=\sum_{n=0}^N \tl f(n)
\end{equation}
converge to a finite limit as $N\to \oo$. Here, $\tl f(n)$ is the Fourier transform of $f$. Since $f(\tau)$ is a Schwartz test function, so is $\tl f(n)$ (where $n$ is understood as a real parameter) and thus
$\tl f(n)$ decays faster than any power of $n$ as $n\to \oo$. This implies that the partial sums~\eqref{eq:toypartialsums} indeed converge. Strictly speaking, we also need to show that the limit is continuous with respect to $f$ in an appropriate topology. We will delay this question until later. Here the important message is that even though~\eqref{eq:tausum} does not converge in the usual sense, it starts to converge after being smeared with a nice test function.

So far we have learned two things. First, the sum~\eqref{eq:tausum} converges in distributional sense for real $\tau$. Second, the value of this sum is a genuine distribution, since we computed its real part and it is a sum of $\de$-functions. Now, we also know that in the upper-half plane of $\tau$ the sum converges to 
\begin{equation}
	g(\tau)=\frac{1}{1-\rho}=\frac{1}{1-e^{i\tau}}.
\end{equation}
This suggests that on the real line $g(\tau)$ should have a limit that is the tempered distribution computed by~\eqref{eq:tausum}. So we can conjecture that, for real $\tau$,
\begin{equation}
	\label{eq:conjecture}
	\sum_{n=0}^\oo e^{in\tau} = \lim_{\epsilon \to +0} g(\tau+i\e)\equiv \lim_{\e\to +0}\frac{1}{1-e^{-\e+i\tau}},
\end{equation}
where everything is understood in the sense of tempered distributions. 

How can we guarantee that the limit in the right-hand side exists? In the sense of functions, it clearly exists for $\tau\neq 2\pi k$ and is given by $g(\tau)$. However, $g(\tau)$ for real $\tau$ is not obviously a distribution, since it involves non-integrable singularities near $\tau=2\pi k$ that we need to regulate. Specifically, we need to prove that for any Schwartz function $f(\tau)$ the limit
\begin{equation}
	\label{eq:limit-toy}
	\lim_{\epsilon \to +0} \int d\tau g(\tau+i\e) f(\tau)
\end{equation}
exists and depends continuously on $f$ in an appropriate topology. Notice that if $f(\tau)$ were a holomorphic function, for example $f(\tau)=e^{-\tau^2}$, then the existence of the limit would be simple to show.\footnote{We will just give an idea. Expand $f(\tau)$ in Taylor series around $f(\tau+i\eps)$ as $f(\tau)=f(\tau+i\eps)+ (-i\eps) f'(\tau+i\eps)+\ldots+O(\eps^m)$. The terms involving $f^{(k)}(\tau+\eps)$ are easy to analyze: the integrals don't depend on $\eps$ at all because by analyticity we can shift the contour. So only the first of these terms survives. The error term goes to zero provided that $\eps^m g(\tau+i\eps)\to 0$ as $\eps\to0$. This will hold for $m>M$ if $g$ satisfies the slow-growth condition~\eqref{eq:slow-growth} below. This shows that one could equivalently define the pairing between $g$ and holomorphic $f$ by shifting the integration contour for both, as $\int d\tau g(\tau+i\e) f(\tau+i\e)$. This is independent of $\e$ and there is no limit to talk about.} However, the class of holomorphic test functions is too restricted for many purposes. It is more customary to develop the theory of distributions using compactly supported $C^\infty$ test functions, or the even larger class of Schwartz test functions.\footnote{This is not just for the reasons of generality. Compactly supported test functions are needed if one wants to define a very basic notion of support of the distribution. This notion allows as to make statements such as ``distributions $f(x)$ and $g(x)$ agree for $x\in [0,1]$ but disagree outside of this interval.'' Then the class of Schwartz test functions, being invariant under the Fourier transform, plays an important role in all questions involving the Fourier transform of distributions.} For a general Schwartz $f(\tau)$, existence of the limit~\eqref{eq:limit-toy} requires an argument which will be explained in the next section.

We would like to emphasize that the existence of the limits~\eqref{eq:conjecture},~\eqref{eq:limit-toy} is not just some abstract nonsense, but a very concrete prediction. Integrating both parts of~\eqref{eq:conjecture} against an arbitrary Schwartz test function $f(\tau)$, we obtain:
\beq
\label{eq:test-limit}
\lim_{\e\to +0}\int d\tau\,f(\tau)\frac{1}{1-e^{-\e+i\tau}} = \sum_{n=0}^\infty \tl f(n)\,.
\eeq
Let us test this prediction. We pick a function $f(\tau)$ given by $\exp(-1/(1-\tau^2))$ for $\tau\in(-1,1)$, extended by zero outside this interval. It is a compactly supported $C^\infty$ function (in particular Schwartz, but not analytic). We evaluate both sides of the previous equation numerically for $0<\e<1$, and check the limit (see Fig.~\ref{fig-limit}).

\begin{figure}[h!]
	\centering
	\includegraphics[scale=0.5]{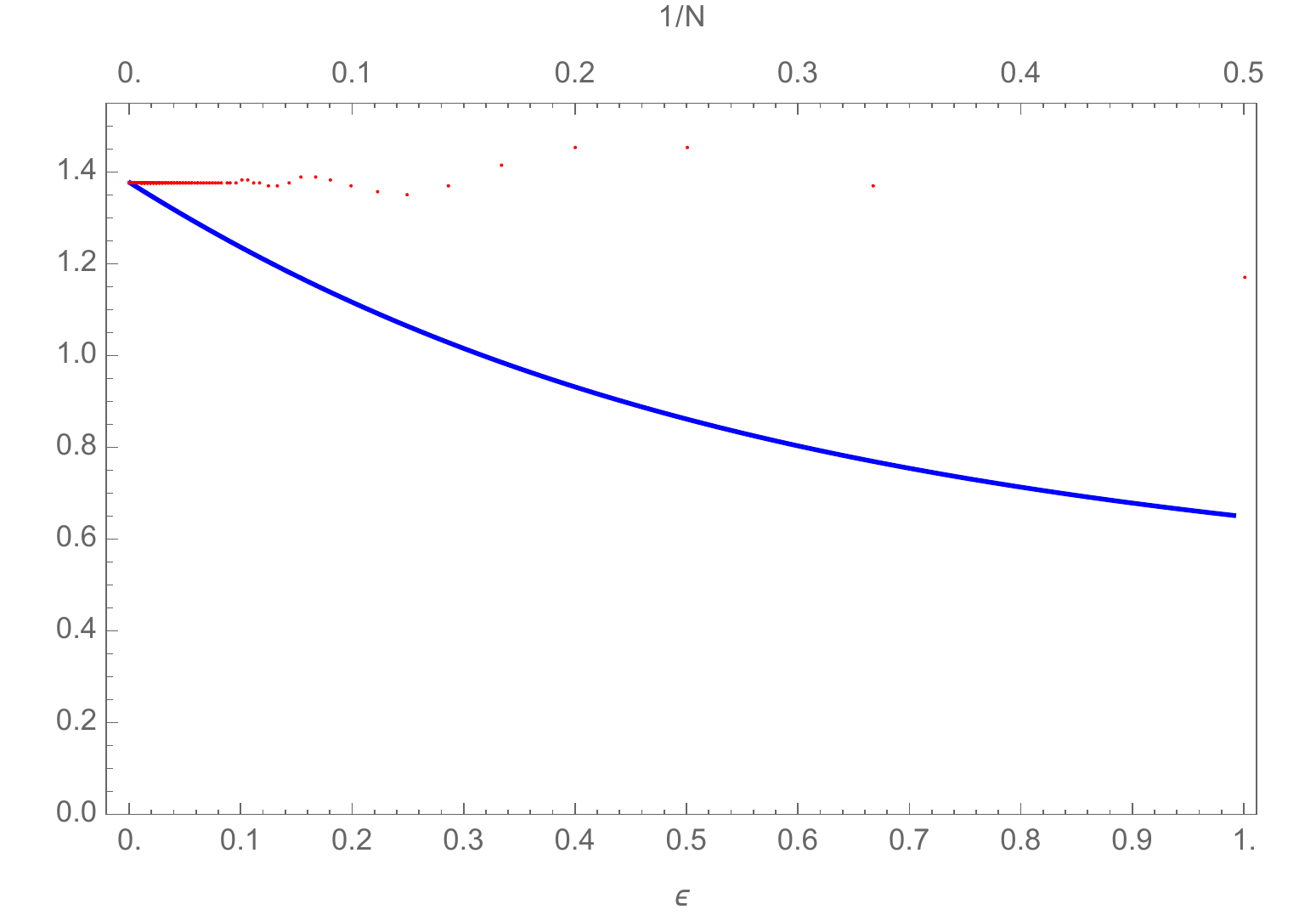}
	\caption{\label{fig-limit} A numerical check of the existence of the limit~\eqref{eq:test-limit}, for $f(\tau)$ given in the text. The curve is the integral under the limit sign, and the red dots are the partials sums of Fourier coefficients in the r.h.s.\ of~\eqref{eq:test-limit} up to $n=N$.}
\end{figure}

\section{Vladimirov's theorem}
\label{sec:proof1}

Fortunately, there is a general result that immediately establishes that~\eqref{eq:conjecture} is valid, i.e.\ that both left and right hand sides converge as tempered distributions and are indeed equal. Before stating this result, let us first clean up some formal definitions. 

For a smooth function $f(x)$ define the semi-norms
\begin{equation}
	\label{eq:semi-norm}
	\|f\|_{m,n}=\sup_{x \in \R} |(1+|x|^m) \ptl_x^n f(x)|.
\end{equation}
The Schwartz space $\cS(\R)$ consists of smooth functions $f$ for which $\|f\|_{m,n}$ is finite for all non-negative integer $m$ and $n$. 
This is a vector space which is given a topology where a sequence $f_k$ is said to converge to $g$ if $h_k=f_k-g$ converges to $0$. In turn,
$h_k$ converges to $0$ iff for all $m,n$ the sequence $\|h_k\|_{m,n}$ converges to $0$.

The space $\cS'(\R)$ of tempered distributions is defined as the space of continuous linear functionals on $\cS(\R)$. We say that
a linear functional $\a$ is continuous if $\a(h_k)\to 0$ for any sequence $h_k\in \cS(\R)$ for which $h_k\to 0$.
We say that a sequence of tempered distributions $\a_k$ converges to a tempered distribution $\beta$ if for any $f\in \cS(\R)$ we have
$\a_k(f)\to \beta(f)$.

Now, let $a>0$ and $g(\tau)$ be a function holomorphic in the strip $0<\mathrm{Im}\,\tau < a$. Suppose there exist $N,M\in \Z_{\geq 0}$ and $C>0$ such that in the strip
\begin{equation}
	\label{eq:slow-growth}
	|g(x+iy)|\leq C(1+|x|^N)y^{-M}
\end{equation}
for all $x\in \R$ and $y\in (0,a)$. We then say that $g$ satisfies a slow-growth condition near $\R$. What this means is that for any $y$ the function $g(x+iy)$ is bounded by a polynomial of fixed degree, and the overall size of this polynomial grows at most as a fixed powerlaw when $y\to0$. Note that thanks to this condition for any $y$, $0<y<a$, the function $g_y(x) \equiv g(x+iy)$
is a tempered distribution in $\cS'(\R)$. We can ask whether the limit
$
\lim_{y\to +0} g_y
$
exists in $\cS'(\R)$. If it does, we say that boundary value of $g$ on $\R$ exists in $\cS'(\R)$ and denote it by $\mathrm{bv}\, g$,
\begin{equation}
	\mathrm{bv}\, g \equiv \lim_{y\to +0} g_y.
\end{equation}

We can now state the theorem
\begin{theorem}
	\label{thm:vlad}
	Let $g(\tau)$ be a function holomorphic for $0<\mathrm{Im}\,\tau<a$ for some $a>0$, satisfying the slow-growth condition near $\R$ as defined above.
	Then the boundary value $\mathrm{bv}\,g$ of $g$ on $\R$ exists in $\cS'(\R)$. Furthermore, if a sequence of functions $g_n$, holomorphic in the same region, satisfies the slow-growth condition with the same constants $C,M,N$ for all $n$ (uniform slow-growth condition), and 
	converges pointwise to $g$ for $0<\mathrm{Im}\,\tau<a$, then 
	$g$ satisfies the same slow-growth condition and
	\begin{equation}
		\lim_{n\to \oo} \mathrm{bv}\, g_n = \mathrm{bv}\, g\quad\text{in }\cS'(\R).
	\end{equation}
\end{theorem}

Such results are rather standard in the theory of distributions (an early mathematics reference is~\cite{Tillmann1961}). In mathematical physics they are very useful in the study of QFT Wightman functions. The standard reference is the book of Vladimirov~\cite{Vladimirov} (section 26), and we will therefore refer to such results as ``Vladimirov's theorems''. A self-contained proof of Theorem~\ref{thm:vlad} will be given below. A more general Vladimirov's theorem will be stated and used in~\cite{Kravchuk:2021kwe}. 

Let us see how this result applies to our toy problem. We have
\begin{equation}
	g_n(\tau)=\sum_{k=0}^n e^{ik\tau}, \quad g(\tau)=\frac{1}{1-e^{i\tau}}.
\end{equation}
Let us check the slow growth condition for $g_n$ on $0<\mathrm{Im}\,\tau <1$:
\begin{equation}
	|g_n(x+iy)|\leq \sum_{k=0}^n |e^{ikx-ky}|\leq \sum_{k=0}^\oo e^{-ky}=\frac{1}{1-e^{-y}}\leq Cy^{-1}
\end{equation}
for some $C>0$. So we see that the slow growth condition is satisfied with $N=0, M=1$. The same condition is then
true for $g(\tau)$, as is easy to check. Then theorem~\ref{thm:vlad} immediately implies our conjecture~\eqref{eq:conjecture}.

\section{Proof of Vladimirov's theorem~\ref{thm:vlad}}
\def\L{L}
\label{sec:proof}

We first prove that $\mathrm{bv}\,g$  exists and is a tempered distribution. So we pick a Schwartz test function $f(x)$ and study the integral
\begin{equation}
	\L(y):= \int dx\,  g (x + i y) f (x)   . 
	\label{hgen}
\end{equation}
We need to show that this has a limit as $y \rightarrow +0$. This looks a bit
magic: estimating naively by absolute value one would conclude that
the integral may blow up as $y^{-M}$. It won't blow up only because of
cancellations, not captured by the naive estimate. In other words, when an analytic function tends somewhere to infinity, it will tend to minus infinity nearby, so that the integral will remain finite. For intuition, recall the
Sochocki formula:
\be 
\lim_{y \rightarrow +0} \overline{} \frac{1}{x + i y}
= \text{PV} \frac{1}{x} - i \pi \delta (x)  
\ee
Principal value PV represents a kind of cancellations whose existence we need to exhibit in
general.

Going back to {\eqref{hgen}},\footnote{We follow the proof in
	{\cite{Streater:1989vi}}, Theorem 2-10.} the first key idea is that we can
estimate not just $\L$ but any its derivative. 
By the
Cauchy-Riemann equations, $y$-derivatives of $\L (y)$ can be transformed into
$x$ derivatives acting on $g$ which then can be integrated by parts to act on
$f$:
\be
L^{(j)}(y)=i^j\int dx g^{(j)}(x+iy) f(x)=(-i)^j \int dx g(x+iy) f^{(j)}(x).
\ee 
Using then the slow-growth condition {\eqref{eq:slow-growth}} we get an estimate of
\emph{any} $y$-derivative $\L^{(j)} (y)$ by $y^{-M}$ times a constant:
\be
\label{eq:jbound}
| L^{(j)} (y) | \leqslant C y^{-M}\,.
\ee
The constant here is proportional to the semi-norm $\|f\|_{N+2,j}$, see~\eqref{eq:semi-norm}; order $N+2$ is needed to make the integral convergent, while derivative order $j$ appears because of integrating by parts. 

This is still growing as $y \rightarrow 0$. Here comes the second key idea:
since we have this bound on any derivative, we can strengthen it recursively
using the Newton-Leibnitz formula:
\be \label{eq:NLformula}
\L^{(j - 1)} (y) = - \int_{y^{}}^{y_0} dy\,\L^{(j)} (y) + \L^{(j - 1)} (y_0) \,.
\ee
Here $y_0$ can be any fixed number in the strip of analyticity, e.g. $y_0=a/2$ will do.

Every time we use this, we obtain a bound on $\L^{(j - 1)}$ of the same type as in~\eqref{eq:jbound} but with the order of
singularity in $y$ reduced by 1 w.r.t.~$\L^{(j)}$. Let us do this repeatedly, starting from $j=M+2$.\footnote{Exercise: once you understand the proof below, show that $j=M+1$ will do as well. Hint: the key requirement is that $\L'(y)$ end up bounded by some integrable function.} Then doing this $M$ times we will prove that $\L''(y)$ has an at most $\log(y)$ singularity, and doing this once more we prove that $\L'(y)$ has no singularity at all, i.e.~it is bounded by a constant, call it $C_1$.

Now we can finally prove that $\L(y)$ has a limit. From the $j=1$ case of~\eqref{eq:NLformula} we can write
\beq
(\text{bv}\, g)(f) = \lim_{y\to+0} \L(y)=-\int_{0}^{y_0} dy\,\L'(y)+L(y_0)\,.
\eeq
The limit exists, since by $|\L'(y)|\le C_1$ the integral in the r.h.s.~converges absolutely at the lower limit of integration.
Thus $\text{bv} \,g$ exists as a linear functional on $\cS(\mathbb{R})$. All constants in the above argument are bounded by some semi-norms of $f$. This proves that $\text{bv}\,g$ is a \emph{continuous} linear functional on $\cS(\mathbb{R})$, i.e.~a tempered distribution.

%
%

Now let us prove the second part of the theorem, about convergence. Replacing $g_n$ by $g_n-g$, it's enough to consider the case $g=0$. We pick an arbitrary Schwartz function $f$ and consider
\beq
\label{eq:bvgn}
(\text{bv}\,g_n)(f)=\lim_{y\to +0} L_n(y)\,.
\eeq
Here $L_n(y)$ is defined by the integral~\eqref{hgen} with $g$ replaced by $g_n$. The existence of the limit for each $n$ is guaranteed by the above argument. As a byproduct of the argument, we have also seen that  $|L'_n(y)|\le C_1$ uniformly in $n$ and $y$, where $C_1$ is bounded by some semi-norm of $f$. 

Furthermore, we claim that $L_n(y)$ tends to zero as $n\to\infty$ for any fixed $y\in(0,a)$. Indeed the integrand in~\eqref{hgen} satisfies two conditions: (a) it tends to zero as $n\to \infty$ because $g_n(x+iy)$ goes pointwise to zero; (b) it is bounded in absolute value by an integrable function which does not depend on $n$:
\beq
|g_n(x+iy)f(x)|\leq \|f\|_{N+2,0}\frac{|g_n(x+iy)|}{1+|x|^{N+2}} \le C\|f\|_{N+2,0}\frac{1+|x|^N}{y^M(1+|x|^{N+2})}\,,
\eeq
where we bounded $f$ by its semi-norm, and then used the slow-growth condition~\eqref{eq:slow-growth}.
So the claim follows by Lebesgue's dominated convergence theorem.

Finally we wish to prove that $(\text{bv}\,g_n)(f)$ tends to zero as $n\to\infty$, as this is what is meant by $\text{bv}\, g_n\to 0$ in $\cS'(\mathbb{R})$. From definition~\eqref{eq:bvgn}, we can bound this quantity as:
\be
|(\text{bv}\,g_n)(f)|\le \sup_{y\in(0,\eps)} | L_n(y)| \le |L_n(\eps)|+C_1\eps\,,
\ee
where in the second inequality we used $|L'_n(y)|\le C_1$. We proved above that $L_n(\eps)$ goes to zero for any $\eps$. So by picking first $\eps$ small enough, and then $n$ large enough, the sum of the two terms in the r.h.s.~is arbitrarily small. This implies that $\limsup_{n\to\infty} |(\text{bv}\,g_n)(f)|$ is arbitrarily small. Thus it is zero.

The attentive reader may notice that the last steps of the proof are not constructive, i.e.~they do not provide a bound on how fast $(\text{bv}\,g_n)(f)$ tends to zero. This is because the used assumption, that $g_n$ converges to zero pointwise, is very general. It allows to conclude, via dominated convergence, that $L_n(y)$ tends to zero pointwise as $n\to\infty$, but it does not tell us how fast this limit is reached. If more detailed information about the rate of the limit $g_n\to 0$ is available, as it usually is in practical applications, then a simple modification of the above argument makes the conclusion $\text{bv}\,g_n \to 0$ in $\cS'(\mathbb{R})$ constructive.

\section{Distributional convergence of conformal block expansion}
\label{sec:distributionalconvergence1d}


Let us now turn back to the 1-dimensional conformal block expansion~\eqref{eq:1dcbexpansion}. We would like to claim that it converges as a tempered distribution for real $\tau$ (recall $\rho=e^{i\tau}$). To prove this,
we will use Vladimirov's theorem~\ref{thm:vlad}, for which we need to establish a uniform slow-growth condition on the partial sums in the left-hand side of~\eqref{eq:1dcbexpansion}.

As a first step, let us derive a slow-growth condition for the four-point function $g(\rho)$ itself. First, note that for $|\rho|<1$ we have
\be\label{eq:1dscalingexpansion}
g(\rho)=\sum_\De \tl p_\De \rho^\De.
\ee
with some positive coefficients $\tl p_\De$. This follows from radial quantization in an appropriate conformal frame~\cite{Hogervorst:2013sma}. Equivalently, we can expand the conformal blocks~\eqref{eq:1dblock} in the right-hand side of~\eqref{eq:1dcbexpansion} in powers of $\rho$ and use the fact that these expansions have positive coefficients. In particular, the sum~\eqref{eq:1dscalingexpansion} can be turned back into the sum~\eqref{eq:1dcbexpansion} by appropriately grouping the terms. Now, we can write
\be
|g(\rho)|=\left|\sum_\De \tl p_\De \rho^\De\right|\leq \sum_\De \tl p_\De |\rho|^\De = g(|\rho|),
\ee
so it suffices to bound $g(\rho)$ for real $\rho\in (0,1)$. This maps to $z\in(0,1)$, and in terms of $z$ variable we know that $g(z)$ satisfies the crossing equation
\be
\label{eq:crossing1d}
z^{-2\De_\f}g(z)=(1-z)^{-2\De_\f}g(1-z).
\ee
When $z\to 1$, we have $g(1-z)=O(1)$, which implies for $z\in (0,1)$ the bound
\be
|g(z)|\leq C(1-z)^{-2\De_\f}
\ee
for some $C>0$. Using the fact that $1-z\sim (1-\rho)^2/4$ as $z\to 1$, we find
\be
\label{eq:b1}
|g(\rho)|\leq g(|\rho|) \leq C' (1-|\rho|)^{-4\De_\f}
\ee
for some $C'>0$. In terms of $\tau = x+iy$ this implies a powerlaw bound
\be
\label{eq:b2}
|g(\tau)|\leq C''y^{-4\De_\f},
\ee
near $y=0$ for a $C''>0$, which is the required slow-growth condition. Therefore, by theorem~\ref{thm:vlad}, $\text{bv}\, g$ exists and is a tempered distribution.

An easy modification establishes the slow-growth condition for the partial sums in~\eqref{eq:1dcbexpansion}. Let $I$ be any (possibly infinite) subset of the terms in~\eqref{eq:1dscalingexpansion} and write
\be
\label{eq:b3}
\left|\sum_{\De\in I}\tl p_\De \rho^\De\right| \leq \sum_{\De\in I}\tl p_\De |\rho|^\De \leq \sum_\De \tl p_\De  |\rho|^\De=g(|\rho|)\leq C''y^{-4\De_\f}.
\ee
Taking $I=I_{\De_*}=\{\De|\De<\De_*\}$ we get a uniform slow-growth condition for partial sums of~\eqref{eq:1dscalingexpansion}. Similarly, by allowing $I=I_n$ to contain the terms corresponding to the first $n$ conformal blocks in~\eqref{eq:1dcbexpansion} we get a uniform slow-growth condition on partial sums of~\eqref{eq:1dcbexpansion}. Therefore, by theorem~\ref{thm:vlad}, we conclude that the expansion~\eqref{eq:1dcbexpansion} converges for the boundary values,
\be
\mathrm{bv}\, g = \sum_{\De} p_\De \mathrm{bv}\, g_\De\quad\text{in }\cS'(\R). 
\ee

Let us unpack this equation a bit. Notice that in the case at hand, $\mathrm{bv}\, g_\De$ is an ordinary locally integrable function which is the easiest kind of distribution. This is because the conformal blocks~\eqref{eq:1dblock} only have a logarithmic singularity at $\rho=1$. Written in full, this equation says that for any Schwartz function $f(\tau)$
\be
\lim_{\eps\to +0} \int d\tau\, g(\rho=e^{-\eps+i\tau}) f(\tau) = \sum_{\De} p_\De \int d\tau\, (\mathrm{bv}\, g_\De)(\rho=e^{i\tau}) f(\tau)\,,
\ee
in the sense that the ${\eps\to +0}$ limit in the l.h.s.~exists (it defines $(\mathrm{bv}\, g)(f)$), the series in the r.h.s.~made of ordinary integrals converges, and that the two sides independently defined in this way are equal.


\section{Convergence for other normalizations and on other boundaries}
\label{sec:1dvariants}

We have proven that the conformal block expansion~\eqref{eq:1dcbexpansion} converges as a distribution on the boundary $\ptl \cC$ of the normal function-like domain of convergence $\cC$. We motivated this question in chapter~\ref{sec:cbexpansion} from the point of view of computing the Wightman functions. However, in other applications the domain $\cC$ may not be the most natural one to consider. For example, 
one of the main objects of study in CFT is the crossing equation 
\be
\label{eq:1dcrossing}
z^{-2\De_\f}g(z)=(1-z)^{-2\De_\f}g(1-z)\,,
\ee
where both left- and right-hand side are expanded in conformal blocks. The two expansions are conventionally referred to as the $s$- and $t$-channel expansions. It is then natural to consider the domain $\cC^{st}=\cC^s\cap \cC^t$ in which both expansions converge in the sense of functions, as well as distributional convergence on its boundary $\ptl \cC^{st}$. Additionally, the function $g(z)$ is multiplied by a factor $z^{-2\De_\f}$ in the above equation, so we should also ask whether such modifications alter our result.

It is easy enough to address the latter question. Indeed, if a function $q(\rho)$ satisfies a slow-growth condition near $|\rho|=1$, so does the function $q(\rho)g(\rho)$ and the partial sums of conformal block expansion~\eqref{eq:1dcbexpansion} multiplied by $q(\rho)$. So we can state the straightforward corollary to theorem~\ref{thm:vlad}:
\begin{corollary}
	\label{cor:factors}
	If function $q(\rho)$ is holomorphic in the branched unit $\rho$-disc and satisfies a slow-growth condition near $\tau\in \R$ (recall $\rho=e^{i\tau}$), then we have
	\be
	\label{eq:1dnormlizedcbexpansion}
	\mathrm{bv}\, (q\cdot g) = \sum_{\De} p_\De \mathrm{bv}\, (q\cdot g_\De)\quad\text{in }\cS'(\R). 
	\ee
\end{corollary}
In the example~\eqref{eq:1dcrossing} we have $q(\rho)=z^{-2\De_\f}$ and it satisfies the assumptions of this theorem as can be seen from the identity $z=\frac{4\rho}{(1+\rho)^2}$.

In order to address the questions related to restricting the domain $\cC$ to smaller domains such as $\cC^{st}$, we can prove the following theorem (see Fig.~\ref{fig:DS}).
\def\SS{S}
\begin{theorem}
	\label{thm:restrictedboundary}
	Let $\mathbb{D}$ be the open unit disk parametrized by $w$ and let $\varphi:w\mapsto\varphi(w)$ be a holomorphic map which maps $\mathbb{D}$ one-to-one onto a domain $ \SS$ inside the cut unit disk of the $\rho$ variable, $ \SS \subset \mathbb{D}\setminus(-1,0]$. Replacing $\rho=\varphi(w)$ in the conformal block expansion~\eqref{eq:1dcbexpansion}, we pull it back to $w\in \mathbb{D}$. Then this pulled-back conformal block expansion in $w$ variable converges on the boundary $|w|=1$ in the sense of distributions (i.e.~when integrated against an arbitrary smooth function on the circle). Furthermore, the same conclusion holds for~\eqref{eq:1dnormlizedcbexpansion} with $q(\rho)=z^{-2\De_\f}$.
\end{theorem}

\begin{figure}[h!]
	\centering
	\includegraphics[scale=0.8]{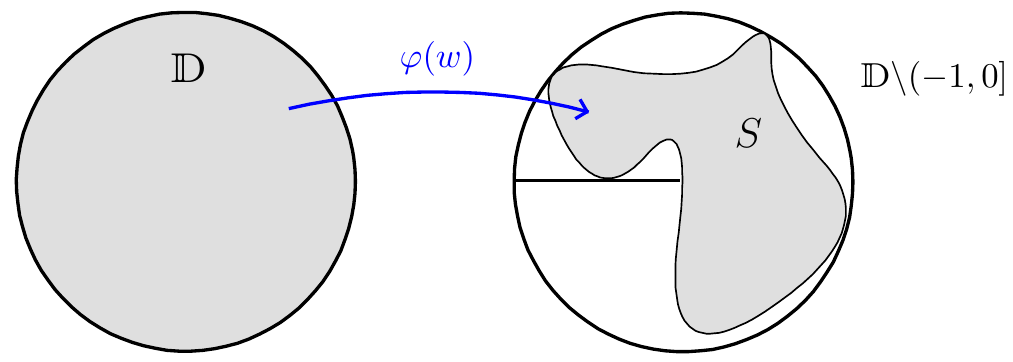}
	\caption{\label{fig:DS} The setting of theorem~\ref{thm:restrictedboundary}. We give one particular example of a possible region $\SS$. In practical applications discussed below $ \SS$ will be either all of $\mathbb{D}\setminus(-1,0]$ or an upper or lower half.}
\end{figure}

The proof will be based on a simple
\begin{lemma}
	\label{lemma:functions}
	For any one-to-one holomorphic function $\varphi$ from $\mathbb{D}$ onto $ \SS\subset \mathbb{D}\setminus(-1,0]$ there are lower bounds
	\be
	\label{eq:Lind-bound}
	1-|\varphi(w)|&\ge C(1-|w|),\nn\\
	|\varphi(w)|&\ge C'(1-|w|)^2,
	\ee 
	with some $C,C'>0$, and for any $w\in \mathbb{D}$. In other words, the first bound says that $|\varphi(w)|$ cannot approach 1 near the boundary faster than linearly in $w$. Similarly, $|\varphi(w)|$ cannot approach $0$ near the boundary faster than quadratically in $w$. 
\end{lemma}

To see why this is intuitively reasonable, consider some model situations. For the first bound, suppose that $\varphi(w)$ has the leading behavior $\varphi_0+ const.(w-w_0)^\alpha$, $|\varphi_0|=1$, near some boundary point $|w_0|=1$. This asymptotics is consistent with~\eqref{eq:Lind-bound} as long as $\alpha\le 1$. The latter condition is implied by the assumption that $\varphi:\mathbb{D}\to \mathbb{D}$: the argument of $w-w_0$ is multiplied by $\alpha$, and for $\alpha>1$ some points will end up outside of the unit circle. A similar check works also for the second bound.

It should be noted that in practical applications the domain $ \SS$ will typically be either the whole of $\mathbb{D}\setminus(-1,0]$ or its upper or lower half. In these cases the functions $\varphi(w)$ will be explicitly known, and bounds~\eqref{eq:Lind-bound} can be verified by an explicit computation. For completeness, a rigorous general proof of this lemma is given in appendix~\ref{app:lemma}.

By the first inequality of the lemma, we have the bound $(1-|\rho|)^{-4\De_\f}\leq C'(1-|w|)^{-4\De_\f}$ for some $C'>0$. So the conformal block expansion pulled back to the unit disk $w\in \mathbb{D}$ satisfies the same bounds throughout the disk as the $\rho$-expansion bounds~\eqref{eq:b1}-\eqref{eq:b3}. Recall in particular that $g(\rho)$ is bounded near $\rho=0$ so whatever happens if the boundary of $ \SS$ touches $\rho=0$, as in figure~\ref{fig:DS}, is not important for this part of the argument. Therefore, the first claim of the theorem follows by the same arguments as in section~\ref{sec:distributionalconvergence1d}. There is even one simplification: since the circle is compact, temperedness of distributions having to do with behavior of infinity is of no importance in the case at hand. The space of test functions are $C^\infty$ functions on the unit circle. 

The second claim does not follow immediately because $z^{-2\De_\f}$ blows up near $\rho=0$. However, thanks to the second bound in~\eqref{eq:Lind-bound}, this does not spoil the slow-growth conditions near $|w|=1$. This finishes the proof of the theorem.

Note that we can replace the unit $\rho$-disk by unit $\rho^{1/n}$-disk for some $n$ if we wish to allow the domain parametrized by $w$ to go under the cut. Similarly, the same result can be proven for a wider class of functions $q(\rho)$ than just $z^{-2\De_\f}$. We won't need these generalizations in this part of the thesis.

\begin{figure}[h!]
	\centering
	\includegraphics[scale=0.7]{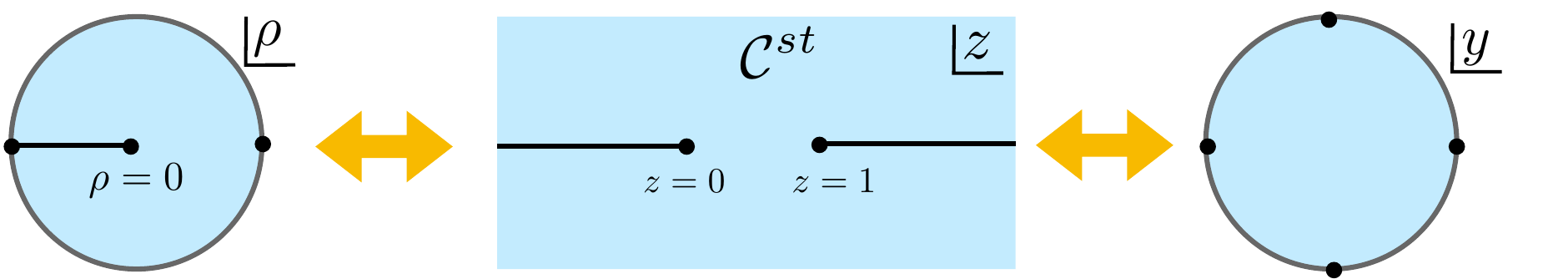}
	\caption{\label{fig:crossing_region} The crossing region $\mathcal{C}^{st}$ and its parametrization using the $\rho$-coordinate and the Zhukovsky $y$-coordinate.}
\end{figure}

\section{Analytic functionals}
\label{sec:functionals}
For the first application of theorem~\ref{thm:restrictedboundary}, consider the common region of convergence $\cC^{st}$ of the two OPE channels for the crossing equation given in $z$-coordinate by the cut plane
\be
\cC^{st}=\C\setminus\p{(-\oo,0]\cup [1,\oo)},
\ee
see figure~\ref{fig:crossing_region}. In $\rho$-variable for either channel it becomes precisely the cut unit disk $\mathbb{D}\setminus(-1,0]$. Following~\cite{Mazac:2016qev}, it is convenient to parametrize domain $\cC^{st}$ via the Zhukovsky map\footnote{The original Zhukovsky (Joukowsky) map $\zeta = y+1/y$ maps the unit circle onto the interval $(-2,2)$. We have $z=1/2+1/\zeta$ so that the unit circle is mapped onto the two cuts $(-\oo,0]\cup [1,\oo)$. The Zhukovsky map is famous in aerodynamics: applying it to offcentric circles one can parametrize airfoil shapes and compute the lift force analytically by conformal invariance of incompressible 2d flows.}
\be
z(y)= \frac{(1+y)^2}{2(1+y^2)}\,.
\ee
This is a holomorphic one-to-one mapping of the unit disk $\mathbb{D}$ onto $\cC^{st}$. Using the function $\varphi(y)=\rho(z(y))$, the $s$-channel conformal block expansion is pulled back to the unit disk of the Zhukovsky variable. Since the region $\cC^{st}$ is symmetric under $z\to 1-z$, the same statement is true for the $t$-channel block expansion (the crossing $z\to 1-z$ corresponds to $y\to -y$). 

We will now apply theorem~\ref{thm:restrictedboundary} with $ \SS=\mathbb{D}\setminus(-1,0]$.
The first conclusion is that the four-point function (both with and without the factor $z^{-2\Delta_\f}$) is a distribution on the boundary of the unit $y$-disk. This statement is only interesting near the points $y=\pm 1$, $y=\pm i$ where the four-point function is singular: on the rest of the boundary it is analytic, as can be shown using the $s$- and $t$-channel expansions.

The second conclusion is that both $s$- and $t$-channel conformal block expansions converge as a distribution on $|y|=1$. This statement is interesting, because in the usual sense each channel converges only on one half of the boundary (the left half for the $s$-channel and the right half for the $t$-channel). 

Distributional convergence has an interesting consequence for the study of the crossing equation using the method of linear functionals~\cite{Rattazzi:2008pe} and in particular for constructing a wide class of functionals satisfying the swapping property of~\cite{Rychkov:2017tpc}. We write the crossing in the usual sum rule form
\be
\label{eq:crossing}
\sum p_\Delta F_\Delta(z)=0,\qquad F_\Delta(z)= z^{-2\De_\f}g_\Delta (z)-(1-z)^{-2\De_\f}g_\Delta(1-z)\,.
\ee
Denote by $F_\Delta(y)$ the same functions pulled back to the unit Zhukovsky disk. They are analytic in the interior and have boundary values $(\text{bv}\,F_\Delta)$ at $|y|=1$. By theorem~\ref{thm:restrictedboundary} we know that~\eqref{eq:crossing} converges on the $|y|=1$ boundary to zero in the sense of distributions. This means that we can integrate it term by term with a smooth function $f(\theta)$:
\be
\label{eq:tbyt}
\sum p_\Delta \int_0^{2\pi} d\theta (\text{bv}\,F_\Delta)(y=e^{i\theta}) f(\theta)=0\,.
\ee
The 1d conformal blocks having only logarithmic singularities, the nature of their boundary values is determined by the singularity of prefactors $z^{-2\De_\f}$ and $(1-z)^{-2\De_\f}$. Thus they are ordinary locally integrable functions for $2\Delta_\f<1$, and distributions otherwise. 

Now, let us fix an infinitely smooth $f(\theta)$ on the boundary of the unit $y$-disk. The support of this function may include points in both halfs of the circle, including the points where the four-point function is singular. Consider a linear functional $\a_f$ defined by the formula
\be\label{eq:functional}
g(y)\mapsto \a_f[g]\equiv\int_0^{2\pi} d\theta (\mathrm{bv}\,g)(y=e^{i\theta}) f(\theta)\,.
\ee
We can write~\eqref{eq:tbyt} equivalently as
\be
\sum p_\De \a_f[F_\De] = 0\,.
\ee
This means, in the terminology of~\cite{Rychkov:2017tpc}, that the functional~\eqref{eq:functional} satisfies swapping property. 

Note that many simple functionals can be rewritten in the form~\eqref{eq:functional}. For example, the derivative evaluation functional $\a_{n,y_0}$
\be
\label{eq:evaluationfunctional}
g_{n,y_0}(y)\mapsto \a_{n,y_0}[g]\equiv g^{(n)}(y_0)
\ee
for integer $n\geq 0$ and $|y_0|<1$ can be written using Cauchy theorem as\footnote{In Cauchy theorem we integrate over the contour at $|y|=1-\eps$, where the function is analytic. As $\eps\to 0$, the Cauchy kernel tends to $\frac{1}{(e^{i\theta}-y_0)^{n+1}}$ in the $C^\infty$ topology of test functions on the circle, while $g(y)$ tends to $\mathrm{bv}\,g$ in the sense of distributions by theorem \ref{thm:restrictedboundary}. This justifies pushing the contour all the way to the boundary $|y|=1$. }
\be
\a_{n,y_0}[g]=\frac{n!}{2\pi i}\int_0^{2\pi} d\theta \frac{ie^{i\theta}}{(e^{i\theta}-y_0)^{n+1}}(\mathrm{bv}\,g)(y=e^{i\theta}).
\ee
This coincides with $\a_{f_{n,y_0}}$ with $f_{n,y_0}$ given by
\be\label{eq:evaluationf}
f_{n,y_0}(\theta)=\frac{n!}{2\pi i}\frac{ie^{i\theta}}{(e^{i\theta}-y_0)^{n+1}}.
\ee

A type of functionals commonly used in analytic functional conformal bootstrap~\cite{Mazac:2016qev,Mazac:2018mdx,Mazac:2018ycv,Kaviraj:2018tfd,Mazac:2018biw,Hartman:2019pcd,Paulos:2019gtx,Mazac:2019shk} can be described as
\be\label{eq:originalcuttouching}
g(y)\mapsto \a_{h,\G}[g]\equiv \int_{\G} dy h(y)g(y),
\ee
where $h(y)$ is some holomorphic function and $\G$ is a contour in $\mathbb{D}$ which is allowed to have end points on the boundary $|y|=1$. Conditions on $h(y)$ that guarantee the swapping property for $\a_{h,\G}$ were studied in~\cite{Rychkov:2017tpc}. We can try to identify $\a_{h,\G}$ with $\a_{f_{h,\G}}$ where
\be\label{eq:cuttouchingf}
f_{h,\G}(\theta) = \int_\G dy h(y)f_{0,y}(\theta),
\ee
with $f_{n,y}$ defined in~\eqref{eq:evaluationf}. Unfortunately, if $\G$ ends or starts on $|y|=1$ then for generic $h(y)$ the function $f_{h,\G}(\theta)$ will not be smooth (and so will not be a test function) and thus we have not proven that $\a_{f_{h,\G}}$ is well defined and satisfies the swapping property for this class of functionals. In other words, so far the class of functionals~\eqref{eq:functional} is too small to accommodate the modern results in analytic functional bootstrap. 

However, the swapping conditions of~\cite{Rychkov:2017tpc} require $h(y)$ to decay sufficiently quickly (as some power-law) near the end points of $\G$ that are on $|y|=1$. In this case $f_{h,\G}(\theta)$ is still generically not infinitely smooth, but it will have some finite number of derivatives, i.e.\ we will have $f_{h,\G}(\theta)\in C^k(S^1)$ for some $k>0$. In particular, under the swapping conditions on $h(y)$ derived in~\cite{Rychkov:2017tpc} $k$ is proportional to $\De_\f$. On the other hand, by examining the proof of Vladimirov's theorem~\ref{thm:vlad} given in section~\ref{sec:proof}, we can see that we only use a finite number of semi-norms of the test functions, corresponding to derivatives of order related to the power $M$ in the slow-growth condition~\eqref{eq:slow-growth}, which in turn is related to the dimension $\De_\f$. This implies that in order for the functionals~\eqref{eq:functional} to be well-defined and satisfy the swapping property, we only need $f$ to have $k'$ derivatives with $k'$ proportional to $\De_\f$. 

We thus see that if the functional~\eqref{eq:originalcuttouching} satisfies the swapping conditions derived in~\cite{Rychkov:2017tpc}, then the function~\eqref{eq:cuttouchingf} has $k\propto \De_\f$ derivatives. Similarly, we concluded that our results can be strengthened so that the functional~\eqref{eq:functional} is well defined and satisfies the swapping property if $f$ has $k'\propto \De_\f$ derivatives. This suggests that it is possible to define a space $\cB_{\De_\f}$  of functions on $S^1$ with the following properties. First, we would like $\a_f$ to be well defined and satisfy the swapping property for all $f\in \cB_{\De_\f}$. Furthermore, all functionals used in analytic functional bootstrap should be representable by $\a_f$ with $f\in \cB_{\De_\f}$, i.e.\ we want $f_{h,\G}\in \cB_{\De_\f}$ for all $h$ and $\G$ which satisfy the swapping conditions of~\cite{Rychkov:2017tpc}.

As alluded to above, the first approximation to the space $\cB_{\De_\f}$ is $C^k(S^1)$ with appropriately chosen $k$. However, this seems too coarse, since $k$ is a discrete parameter, while $\De_\f$ is continuous. Moreover, not all the points $y$ with $|y|=1$ are equal---there are special points $y=\pm 1,\pm i$, where the correlator might have a singularity that needs to be controlled, but at all other points we know from crossing that the correlator is smooth (but this does not imply that the conformal block expansion converges there pointwise). It would be interesting to find the appropriate definition for $\cB_{\De_\f}$ since it would provide a uniform description of all functionals suitable for analyzing the crossing equation. We leave these questions for future work.
%
%

\section{Dispersion relation in cross-ratio space and the discontinuity}
\label{sec:dispersion}

For a second application, we consider the upper half-plane in $z$ variable. This region is a subset of $\cC^{st}$ and thus we can again use theorem~\ref{thm:restrictedboundary} (this time with $ \SS$ being the upper half of $\mathbb{D}$) to conclude that both $s$- and $t$- conformal block expansions converge as distributions on the boundary of unit disk in the variable $w=\frac{z-i}{z+i}$. 
This boundary minus one point is smoothly mapped to the real line in $z$-plane, and so both $s$- and $t$-channels also converge as distributions on the real line $\R$ in $z$-plane when approached from above.  
By repeating the same arguments for the lower half-plane mapped to the unit disk via $\tl w=\frac{z+i}{z-i}$, we find that both channels converge as distributions on the real line in $z$-plane when approached from below. 

Let us now see how this kind of arguments can be used to write rigorous dispersion relations and give a proper definition of discontinuity (including the point at infinity). Let $z_0$ be a point in the upper half-plane, and $C$ and $\tl C$ be contours in the upper and lower half-planes, with $C$ surrounding $z_0$. Then we have  
\begin{align}
	\label{eq:disp0}
	g(z_0)&=\frac 1{2\pi i }\oint_C \frac{dz}{z-z_0} g(z)\,,\nn\\
	0&=\frac 1{2\pi i }\oint_{\tl C} \frac{dz}{z-z_0} g(z)\,.
\end{align}
Intuitively, to derive the dispersion relation we push $C$ and $\tl C$ to the real axis and infinity, and take the difference of the two equations, which gives a dispersion relation
\begin{equation}
	\label{eq:disp}
	\quad g(z_0)=\frac 1{2\pi i }\int_{-\infty}^\infty \frac{dx}{x-z_0} {\rm Disc}\, g(x)+\text{contribution at infinity}\,,
\end{equation}
where ${\rm Disc}\, g(x)$ is the difference in two limits of $g(z)$. Contribution at infinity cannot be generally computed in this approach, unless one has some information about the asymptotics of $g(z)$ as $z\to\infty$.

Let us now turn this reasoning into a rigorous dispersion relation, including the contribution at infinity. First of all we pull Eqs. \eqref{eq:disp0} to the unit discs of $w$ and $\tl w$ variable, which gives: 
\begin{align}
	\label{eq:disp1}
	g(w_0)&=\frac 1{2\pi i } (w_0-1)\oint_C \frac{dw}{w-w_0} \frac{g_+(w)}{w-1},\nn\\
	0&=\frac 1{2\pi i } (w_0-1) \oint_{\tl C} \frac{d\tl w/\tl w^2}{\tl w^{-1}-w_0} \frac{g_-(\tl w)}{\tl w^{-1}-1} \,.
\end{align}
where we denoted by $g_+(w)$, $g_-(\tl w)$ the function $g(z)$ from the upper/lower half-plane pulled to the corresponding unit disk. Then we push the contours $C$, $\tl C$ to $|w|=1$, $|\tl w|=1$ and get:
\begin{align}
	\label{eq:disp2}
	g(w_0)&=\frac 1{2\pi i } (w_0-1)\oint_{|w|=1} \frac{dw}{w-w_0} {\rm bv}\frac{g_+(w)}{w-1},\nn\\
	0&=\frac 1{2\pi i } (w_0-1) \oint_{|\tl w|=1}  \frac{d\tl w/\tl w^2}{\tl w^{-1}-w_0} {\rm bv} \frac{g_-(\tl w)}{\tl w^{-1}-1} \,.
\end{align}
Notice that we have to include the singular factors $1/(w-1)$ and $1/(\tl w-1)$, arising due to the transformation of the measure $dz$, under the ``bv'' sign. Since these factors are power-like, the resulting limiting boundary values exist as distributions also in presence of these factors. Finally we take the difference of the two equations and we get:
\begin{align}
	\label{eq:disp3}
	g(w_0)&=\frac 1{2\pi } (w_0-1)\int_{0}^{2\pi} \frac{d\theta\,e^{i\theta} }{e^{i\theta}-w_0} D(\theta),\\
	D(\theta)&=\left.{\rm bv}\frac{g_+(w)}{w-1} \right|_{w=e^{i\theta}} -  \left.{\rm bv} \frac{g_-(\tl w)}{\tl w^{-1}-1}  \right|_{\tl w=e^{-i\theta}}\,.
\end{align}
Here $D(\theta)$ is a distribution on the unit circle, which plays the role of a rigorously defined discontinuity, including the point $z=\infty$ mapped to $\theta=0$. For points away from $\theta=0$ and $\theta=2\pi$ we can pull the factors $1/(w-1)$ and $1/(\tl w-1)$ from under $\mathrm{bv}$ and $D(\theta)$ becomes just
\be
D(\theta)=\frac{1}{e^{i\theta}-1}\mathrm{Disc}\, g(x=-\cot \tfrac{\theta}{2}), \qquad \theta\neq 0,2\pi.
\ee
Here $\mathrm{Disc}\, g(x=-\cot \tfrac{\theta}{2})$ is defined as $\left.{\rm bv}{g_+(w)} \right|_{w=e^{i\theta}} -  \left.{\rm bv} {g_-(\tl w)} \right|_{\tl w=e^{-i\theta}}$, which is equivalent to taking the boundary values in $z$-space from above and below the real axis, which is simply the intuitive definition of discontinuity. Using this value of $D(\theta)$ in~\eqref{eq:disp3} and changing back to $x$ variable, we recover~\eqref{eq:disp}. So we see that~\eqref{eq:disp3} is indeed an analogue of~\eqref{eq:disp}. However, using $D(\theta)$ allows us to rigorously include the contribution at $x=\oo$.

An intuitive way to think about this construction is that it defines $\mathrm{Disc}\,g(x)$ as a distribution on a class of test functions $\cS_0(\R)$ larger than $\cS(\R)$. The space $\cS_0(\R)$ consists of smooth functions $f(x)$ such that $f(1/x')$ is smooth and vanishing at $x'=0$. Pairing with $\mathrm{Disc}\,g(x)$ is defined by the formula
\be
\int dx f(x)\mathrm{Disc}\,g(x)\equiv-2\int d\theta e^{i\theta} \tl f(\theta) D(\theta),
\ee
where 
\be\label{eq:iso}
\tl f(\theta) \equiv \frac{1}{e^{i\theta}-1}f(-\cot \tfrac{\theta}{2})
\ee
is a smooth function on the circle parametrized by $\theta$.\footnote{This equation established isomorphism between $\cS_0(\R)$ and $C^\oo(S^1)$ in the sense that $f\in \cS_0(\R)$ if and only if $\tilde f\in C^\oo(S^1)$.} With this definition we can write the dispersion relation~\eqref{eq:disp3} as
\be\label{eq:disp4}
g(z_0)=\frac 1{2\pi i }\int_{-\infty}^\infty \frac{dx}{x-z_0} {\rm Disc}\, g(x),
\ee
since the Cauchy kernel $\frac{1}{x-z_0}$ belongs to our new class of test functions. In this language our results imply that both $\mathrm{Disc}$ of the four-point function and $\mathrm{Disc}$ of partial sums of the conformal block expansion are distributions in $\cS'_0(\R)$, and the partial sums converge to the four-point function in this space (i.e.\ discontinuity can be computed term-by-term). 

Let us consider an example. First take $g(z)=\log z$. This is not a good four-point function since it does not satisfy crossing, but it will allow us to clarify the notion of the discontinuity as a distribution and how it can be concretely computed. Going to the $\rho$ variable we easily see that the slow-growth condition is satisfied. For finite $x<0$ we have $\mathrm{Disc}\,g(x)=2\pi i$. This is a distribution in $\cS'(\R)$, but not obviously in $\cS'_0(\R)$. To extend it to $\cS'_0(\R)$ let us write
\be
\log z = -\lim_{\a\to +0} \ptl_\a z^{-\a}.
\ee
The point here is that $z^{-\a}$ also satisfies a power-law bound and for $\a>0$ the discontinuity
\be
\mathrm{Disc} \,x^{-\a}=-2 i\sin \pi\a |x|^{-\a}
\ee
is in $\cS'_0(\R)$. We can then obtain $\mathrm{Disc}\,g(x)$ by taking derivative and limit $\a\to +0$.\footnote{Justification for this comes from the limit part of the statement of theorem~\ref{thm:vlad} and (for the derivative) from arguments as in appendix~\ref{app:proofVlad2}.} Pairing $\mathrm{Disc}\,g(x)$ with functions that vanish as $1/x^2$ or faster we get integrals that converge in the usual sense. So we only need to use the limiting construction to define the pairing with $1/x$. We have:
\be
\int_{-\oo}^{-1}dx \frac{1}{x}\mathrm{Disc}\,g(x) &=-\lim_{\a\to +0} \ptl_\a\int_{-\oo}^{-1}dx\frac{1}{x}\p{-2 i\sin \pi\a |x|^{-\a}}\nn\\
&=-2 i\lim_{\a\to +0} \ptl_\a\frac{\sin \pi\a}{\a}=0\,,
\ee
where the choice of the integral's upper limit $-1$ is just for convenience since it leads to a simple answer (zero). We can therefore define the distribution $\mathrm{Disc}\,g(x)$ by
\be
\int dx f(x) \mathrm{Disc}\,g(x) = \int_{-\oo}^{-1} dx (f(x)-f_1x^{-1})2\pi i+\int_{-1}^0 dx f(x)2\pi i.
\ee
where $f_1\equiv\lim\limits_{x\rightarrow\infty}xf(x)$. The dispersion relation~\eqref{eq:disp4} then becomes
\be
\log z_0 = \int_{-\oo}^{-1} dx \p{\frac{1}{x-z_0}-\frac{1}{x}}+\int_{-1}^0 dx \frac{1}{x-z_0}\,.
\ee
This is easy to verify.

Another example, which we will find useful in section~\ref{sec:Bissi}, is $\mathrm{Disc}\,1$. Naively, this discontinuity must be zero. This is indeed correct, except at $x=\oo$. Indeed, analogously to the above, we have
\be
1=\lim_{\a\to+0}z^{-\a},
\ee
so
\be
\int_{-\oo}^{-1}dx \frac{1}{x}(\mathrm{Disc}\,1)(x) &=\lim_{\a\to +0} \int_{-\oo}^{-1}dx\frac{1}{x}\p{-2 i\sin \pi\a |x|^{-\a}}\nn\\
&=2 i\lim_{\a\to +0} \frac{\sin \pi\a}{\a}=2\pi i\,,
\ee
and thus 
\be
\int dx f(x) (\mathrm{Disc}\,1)(x) = 2\pi i f_1,
\ee
where as before $f_1=\lim_{x\to \oo} xf(x)$.

\chapter{Scalar four-point functions in higher dimensions}
\label{sec:higherd}

We will now generalize our results to general scalar four-point functions in any number of dimensions $d$.  This generalization is mostly technical, and all the conceptual points were already explained in chapter~\ref{sec:1dcase}. Our strategy is therefore very similar: first we will introduce analogues of the expansions~\eqref{eq:1dcbexpansion} and~\eqref{eq:1dscalingexpansion}, and then use these expansions to prove bounds on the correlation function and partial sums of the conformal block expansion. Finally, we will apply a higher-dimensional version of Vladimirov's theorem~\ref{thm:vlad} to conclude that the conformal block expansion converges in the sense of distributions on the boundary of the region $|\rho|,|\bar\rho|<1$.

\section{Conformal block expansion}
\label{sec:generalcb}
We consider a correlation function of four not necessarily identical scalar operators $\f_i$ with scaling dimensions $\De_i$,
\be
\label{eq:general4ptfunction}
\<\f_1(x_1)\f_2(x_2)\f_3(x_3)\f_4(x_4)\> =  
\frac{1}
{
	(x^2_{12})^{\frac{\De_1+\De_2}2}
	(x_{34}^2)^{\frac{\De_3+\De_4}2}
}
\p{\frac{x^2_{24}}{x^2_{14}}}^{
	\frac{\De_1-\De_2}2}\p{\frac{x^2_{14}}{x^2_{13}}}^{\frac{\De_3-\De_4}2}g_{1234}(\rho,\bar \rho),
\ee
which is a simple generalization of~\eqref{eq:4pt}. The subscript $1234$ on $g_{1234}$ indicates that it relates to the four-point function of $\phi_1,\ldots,\phi_4$. The function $g_{1234}(\rho,\bar\rho)$ has a conformal block expansion of the form
\be\label{eq:generalscalarcbexpansion}
g_{1234}(\rho,\bar\rho)=\sum_{\cO} \lambda_{12\bar\cO}\lambda_{43\cO} g_{\De,J}(\rho,\bar\rho),
\ee
where we sum over primaries $\cO$ in $\f_1\times\f_2$ OPE, $\lambda$'s are the three-point coefficients, $\De, J$ are the spin and dimension of $\cO$, and $g_{\De,J}(\rho,\bar\rho)$ are the conformal blocks. The conformal blocks also depend implicitly on $\De_{12}=\De_1-\De_2$ and $\De_{34}=\De_3-\De_4$.

We would like to show that the function $g_{1234}(\rho,\bar\rho)$ satisfies a powerlaw bound as $\rho$ and $\bar\rho$ approach the boundaries of their respective unit disks. We would also like to show that partial sums of the conformal block expansion~\eqref{eq:generalscalarcbexpansion} satisfy a uniform powerlaw bound. We will prove this by relating $g_{1234}(\rho,\bar\rho)$ to the four-point function where operators are inserted symmetrically with respect to the origin~\cite{Pappadopulo:2012jk}.

Let us focus on configurations when all points $x_i$ lie in the 2-plane $P$ defined by $x^\mu=0$ for $\mu>2$. It is convenient to introduce complex coordinates $y,\bar y$ in this plane
\be
y=x^1+ix^2,\quad \bar y=x^1-ix^2.
\ee
Notice that in Euclidean configurations (i.e.\ when $x^\mu$ are real) we have $\bar y=y^*$. Using the notation $\f_i(y,\bar y)$ for operator insertions in $P$ parametrized by $y,\bar y$, we consider for $\bar\rho=\rho^*$
a symmetrically-inserted four-point function
\be\label{eq:rhocf}
\tl g_{1234}(\rho,\bar \rho) = (\rho\bar\rho)^{\frac{\De_1+\De_2}{2}} \<\f_1(-\rho,-\bar\rho)\f_2(\rho,\bar\rho)\f_3(1,1)\f_4(-1,-1)\>\,,
\ee
where the factor $ (\rho\bar\rho)^{\frac{\De_1+\De_2}{2}}$ is inserted for further convenience (basically to make Eq.~\eqref{eq:higherdscalingexpansion} look maximally nice)\,.
For operators inserted as shown, the meaning of $\rho$ in~\eqref{eq:rhocf} and~\eqref{eq:general4ptfunction} is the same, justifying the notation. Evaluating also the prefactor in~\eqref{eq:general4ptfunction}, we find the following relation between $\tl g_{1234}$ and $g_{1234}$:\footnote{Both $\tl g_{1234}$ and $g_{1234}$ depend on $\rho,\bar\rho$ and both can pretend to be called the conformally invariant part of the general four-point function. One could switch from one convention to the other by changing the prefactor in~\eqref{eq:general4ptfunction}. We will still express our final results in terms of $g_{1234}$, since Eq.~\eqref{eq:general4ptfunction} is the most standard convention.}
\be
\label{eq:gtlgrelation}
\tl g_{1234}(\rho,\bar\rho)={2^{-\De_1-\De_2-\De_3-\De_4}}\p{\frac{(1+\rho)(1+\bar\rho)}{(1-\rho)(1-\bar\rho)}}^{\half(\De_{12}-\De_{34})} g_{1234}(\rho,\bar\rho).
\ee
For $\bar\rho=\rho^*$~\eqref{eq:rhocf} is a Euclidean configuration, radial quantization of which~\cite{Pappadopulo:2012jk,Fitzpatrick:2012yx,Hogervorst:2013sma,Hartman:2015lfa} gives the following absolutely convergent expansion for $|\rho|=|\bar\rho|<1$
\be
\label{eq:higherdscalingexpansion}
\tl g_{1234}(\rho,\bar\rho) = \sum_{\psi} \tl \lambda_{12\psi}\tl\lambda_{43\bar\psi} \rho^{h}{\bar\rho}^{\bar h},
\ee
where we sum over eigenstates $\psi$ of dilatations and planar rotations in radial quantization, and $h,\bar h$ are appropriate combinations of the corresponding eigenvalues. Since it converges absolutely for $|\rho|=|\bar\rho|<1$ when $\bar\rho=\rho^*$,
it also does so for independent $\rho$ and $\bar\rho$ when $|\rho|,|\bar\rho|<1$. Furthermore, the conformal block expansion~\eqref{eq:generalscalarcbexpansion} can be understood as a reorganization of expansion~\eqref{eq:higherdscalingexpansion} by
grouping $\psi$ into conformal families.

\section{Bounds on $g(\rho,\bar\rho)$ and partial sums of the conformal block expansion}
\label{sec:generalbounds}
\def\mm{\hspace{0.05em}}
Consider the following analogues of~\eqref{eq:rhocf},\eqref{eq:higherdscalingexpansion} where two pairs of operators are hermitean conjugates of each other:
\be
\label{eq:an1}
\tl g_{12\bar 2\mm\bar 1}(\rho,\bar\rho) &
=  \sum_{\psi} \tl \lambda_{12\psi}\tl\lambda_{\bar 1\mm\bar 2\mm\bar\psi} \rho^{h}{\bar\rho}^{\bar h}
=\sum_{\psi} |\tl \lambda_{12\psi}|^2\rho^{h}{\bar\rho}^{\bar h},\\
\label{eq:an2}
\tl g_{\bar 4\mm\bar 334}(\rho,\bar\rho) &=  \sum_{\psi} \tl \lambda_{\bar 4\mm\bar 3\psi}\tl\lambda_{43\bar\psi} \rho^{h}{\bar\rho}^{\bar h}
=  \sum_{\psi} |\tl\lambda_{43\bar\psi}|^2 \rho^{h}{\bar\rho}^{\bar h},
\ee
where we use $\bar 1$, etc., to denote three-point coefficients of hermitian conjugates $\phi_1^\dagger$, etc.. As shown, because of $\tl\lambda_{\bar 1\mm\bar 2\bar\psi}=(\tl\lambda_{12\psi})^*$ and $\tl\lambda_{\bar 4\bar 3\psi}=(\tl\lambda_{43\bar\psi})^*$, these two expansions have non-negative real coefficients. Furthermore, estimating by absolute value and applying Cauchy-Schwarz, we can bound~\eqref{eq:higherdscalingexpansion} in terms of~\eqref{eq:an1},~\eqref{eq:an2}:
\be
\label{eq:scalarbound1}
|\tl g_{1234}(\rho,\bar \rho)| \leq 
\sum_{\psi} |\tl \lambda_{12\psi}||\tl\lambda_{43\bar\psi}| r^{h+\bar h}\le \bigl [\tl g_{12\bar 2\bar 1}(r,r)\tl g_{\bar 4\bar 334}(r,r)\bigr]^{1/2}
\ee
where $r=\max(|\rho|,|\bar\rho|)$.\footnote{We also have a more nuanced bound by $\bigl[\tl g_{12\bar 2\bar 1}(|\rho|,|\bar\rho|)\tl g_{\bar 4\bar 334}(|\rho|,|\bar\rho|)\bigr]^{1/2}$ but we won't need it.} Note that the same bound holds if we replace the sum over $\psi$ by a sum over a subset of all allowed $\psi$'s. This, similarly to the argument in section~\ref{sec:distributionalconvergence1d}, implies that the partial sums of expansions~\eqref{eq:generalscalarcbexpansion} and~\eqref{eq:higherdscalingexpansion} satisfy the same bound~\eqref{eq:scalarbound1} (with $\tl g_{1234}$ related to $g_{1234}$ via~\eqref{eq:gtlgrelation} where needed).

To proceed we need a bound on $\tl g_{12\bar 2\bar 1}(r,r)$ and $\tl g_{\bar 4\bar 334}(r,r)$. 
This bound is easy to obtain from the corresponding definition~\eqref{eq:rhocf}. In the limit $r\to1$ two pairs of hermitean conjugate operators approach each other. Using OPE between the approaching pairs, we get a leading asymptotics for the correlator.\footnote{This can be equivalently formulated via crossing symmetry in $z$ space and then transforming to the $\rho$ space, as in section~\ref{sec:distributionalconvergence1d}.} This implies a bound of the same functional form as the leading asymptotics times a constant. The resulting bounds have the form:
\be
\tl g_{12\bar2\bar 1}(r, r)&\leq C(1-r)^{-2\De_1-2\De_2},\\
\tl g_{\bar 4\bar 334}(r, r)&\leq C(1-r)^{-2\De_3-2\De_4},
\ee
with some $C>0$. Notice that there is no blowup as $r\to 0$ since it's overcome by the prefactor in~\eqref{eq:rhocf}. Combining these with~\eqref{eq:scalarbound1} we find
\be
|\tl g_{1234}(\rho,\bar \rho)|\leq C (1-r)^{-\De_1-\De_2-\De_3-\De_4},
\ee
and finally via~\eqref{eq:gtlgrelation} we get a bound for $g_{1234}$
\be
\label{eq:generalpowerbound}
|g_{1234}(\rho,\bar\rho)|\leq C'(1-r)^{-\De_1-\De_2-\De_3-\De_4-|\De_{12}-\De_{34}|},\qquad r=\max(|\rho|,|\bar\rho|),
\ee
for some $C'>0$. Again, the same bound with the same $C'$ holds for the partial sums of expansions~\eqref{eq:generalscalarcbexpansion} and~\eqref{eq:higherdscalingexpansion}.

We repeat the logic of this argument. The key idea is to use OPE in the cross channel to infer the leading singularity of the correlator and then to argue that a similar bound holds throughout the range $|\rho|,|\bar \rho|<1$. This does not work directly for $g_{1234}$, but only for 4pt functions with non-negative $\rho$,$\bar\rho$ expansion coefficients, such as $g_{12\bar 2\bar 1}$ and $g_{\bar 4\bar 3 34}$. So we run the argument for those, and recover the general case by Cauchy-Schwarz.

\section{Vladimirov's theorem}

Now that we have the bound~\eqref{eq:generalpowerbound} we would like to use a higher-dimensional version of Vladimirov's theorem~\ref{thm:vlad} to argue for the distributional convergence of conformal block expansion~\eqref{eq:generalscalarcbexpansion}.

\begin{theorem}
	\label{thm:vlad2}
	Consider $\C^{N}=\C^n\times \C^d$ with coordinates $w_k$ on $\C^n$ and $u_k=x_k+i y_k$ on $\C^d$. Let $U$ be an open subset of $\C^n$ and let $M=U\times \R^d$ be the manifold defined by $w\in U, \, y_k=0,\, k=1\ldots d$. Let $V$ be a convex open cone in $\R^d$ with vertex at $y=0$ that doesn't contain $y=0$.  Let $\cW$ be the subset of $\C^{N}$ for which $y\in V$, $|y_k|<a$ for some $a>0$, and $w\in U$. Let $g(w,u)$ be a function holomorphic in $\cW$ that satisfies the slow-growth condition near $M$\footnote{More precisely, we'd like to have this condition satisfied uniformly on compact subsets $w\in \cK\subset U$ with $C,L,K$ allowed to depend on $\cK$.\label{ft:compactfootnote}}
	\be
	\label{eq:slow-growth2}
	|g(w,u)|\leq C\p{1+\sum_k x_k^2}^L \p{\sum_k y_k^2}^{-K}.
	\ee
	Finally, let $v$ be a vector in $V$. Then for fixed $w$ the boundary value
	\be
	(\mathrm{bv}\,g)(w,x) = \lim_{\e\to+0} g(w,x+iv\e)
	\ee
	exists in $\cS'(\R^d)$ and is independent of the choice of $v$. Furthermore, this boundary value depends on $w$ holomorphically, which means that for any $f\in \cS(\R^d)$ the function $h(w)$ defined by\footnote{Here the integral of course just means the pairing of the distribution with the test function.}
	\be
	h(w)\equiv \int d^dx\, f(x) (\mathrm{bv}\,g)(w,x)
	\ee
	is holomorphic for $w\in U$. Furthermore, suppose that sequence of functions $g_n$ holomorphic in $\cW$ converges to $g$ in $\cW$ pointwise and satisfies the slow-growth condition near $M$ uniformly in $n$. Then for all $w\in U$
	\be\label{eq:generalfnconvergence}
	(\mathrm{bv}\,g_n)(w,x) \to (\mathrm{bv}\,g)(w,x)\quad\text{in }\cS'(\R^d).
	\ee
\end{theorem}
The proof of this theorem is very similar to the proof of theorem~\ref{thm:vlad} given in section~\ref{sec:proof}, and we summarize it in appendix~\ref{app:proofVlad2}. For more general results in this direction see, for example,~\cite{Vladimirov} and~\cite{RealSubmanifolds}.

Let us now apply theorem~\ref{thm:vlad2} to the conformal block expansion~\eqref{eq:generalscalarcbexpansion}. As a first step, we introduce the coordinates $\tau$ and $\bar\tau$ via
\be
\rho=e^{i\tau},\quad\bar\rho=e^{i\bar\tau}.
\ee
Note that in Euclidean configurations we have $\bar\tau=-\tau^*$. The function $g_{1234}(\tau,\bar\tau)$ as well as the partial sums of~\eqref{eq:generalscalarcbexpansion} are holomorphic functions in the region
\be
\cW_0=\{(\tau,\bar\tau)|\mathrm{Im}\,\tau,\mathrm{Im}\,\bar\tau>0\},
\ee
which is the universal cover of the product of open unit discs of $\rho$ and $\bar\rho$. Furthermore, the expansion~\eqref{eq:generalscalarcbexpansion} converges absolutely in $\cW_0$. We can apply theorem~\ref{thm:vlad2} in two essentially different ways.

Firstly, we can take $\mathrm{Im}\,\bar\tau$ to zero while keeping $\tau$ fixed. This corresponds to $n=d=1$ case of theorem~\ref{thm:vlad2}, in which $\C^n$ is parametrized by $\tau$ and $\C^d$ by $\bar\tau$. The open set $U$ is then given by $\mathrm{Im}\,\tau>0$ and the cone $V$ is given by $y_1=\mathrm{Im}\,\bar\tau>0$. The set $\cW$ is then
\be\label{eq:simplecW}
\cW=\{(\tau,\bar\tau)|\mathrm{Im}\,\tau>0,a>\mathrm{Im}\,\bar\tau>0\},
\ee
for some $a>0$, say $a=1$.
The slow-growth condition for $g_{1234}(\tau,\bar\tau)$ and the partial sums of~\eqref{eq:generalscalarcbexpansion} follows from~\eqref{eq:generalpowerbound}. In this way, for each $\tau$ we get a distribution
\be
(\mathrm{bv}\,g_{1234})(\tau,\mathrm{Re}\,\bar\tau)=\sum_{\cO} \lambda_{12\bar\cO}\lambda_{43\cO} (\mathrm{bv}\,g_{\De,J})(\tau,\mathrm{Re}\,\bar\tau)\quad\text{in }\cS'(\R)
\ee
that is holomorphic in $\tau$. Similarly, we can send $\mathrm{Im}\,\tau$ to $0$ while keeping $\tau$ fixed to get
\be
(\mathrm{bv}\,g_{1234})(\mathrm{Re}\,\tau,\bar\tau)=\sum_{\cO} \lambda_{12\bar\cO}\lambda_{43\cO} (\mathrm{bv}\,g_{\De,J})(\mathrm{Re}\,\tau,\bar\tau)\quad\text{in }\cS'(\R),
\ee
holomorphic in $\bar\tau$.

Secondly, we can take the simultaneous limit $\mathrm{Im}\,\tau,\mathrm{Im}\,\bar\tau\to 0$. This corresponds to $n=0$ and $d=2$ in theorem~\ref{thm:vlad2}. A small subtlety is that with $\cW$ as in~\eqref{eq:simplecW} the slow-growth condition doesn't follow from~\eqref{eq:generalpowerbound}, since $r$ in~\eqref{eq:generalpowerbound} can approach $1$ even if only one of $\mathrm{Im}\,\tau,\mathrm{Im}\,\bar\tau$ is small. To fix this, choose any $\a<1$ and define
\be
V=\{(\mathrm{Im}\,\tau,\mathrm{Im}\,\bar\tau)\,|\, \mathrm{Im}\,\tau,\mathrm{Im}\,\bar\tau>0,\, \a^{-1}<\mathrm{Im}\,\tau/\mathrm{Im}\,\bar\tau<\a\}.
\ee
The corresponding $\cW$ has form as in figure~\ref{fig:W}. In this new $\cW$ we have $1-r>C[(\mathrm{Im}\,\tau)^2+(\mathrm{Im}\,\bar\tau)^2]^{1/2}$ for some $C>0$ and the slow-growth condition follows from~\eqref{eq:generalpowerbound}. We therefore conclude the existence of the boundary values and the distributional convergence of the boundary value series:
\be
(\mathrm{bv}\,g_{1234})(\mathrm{Re}\,\tau,\mathrm{Re}\,\bar\tau)=\sum_{\cO} \lambda_{12\bar\cO}\lambda_{43\cO} (\mathrm{bv}\,g_{\De,J})(\mathrm{Re}\,\tau,\mathrm{Re}\,\bar\tau)\quad\text{in }\cS'(\R^2).
\ee

\begin{figure}[h!]
	\centering
	\includegraphics[scale=0.5]{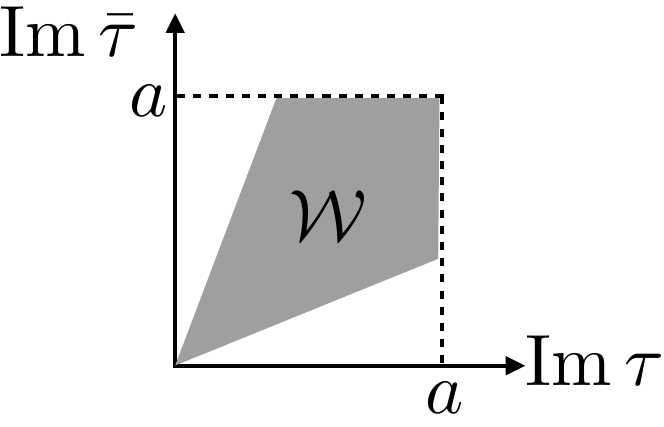}
	\caption{\label{fig:W} The region $\cW$ relevant for the second application of Vladimirov's theorem.}
\end{figure}

\section{Analytic functionals}

Similarly to the one-dimensional case, we can consider various generalizations. In particular, we have the obvious generalizations of corollary~\ref{cor:factors} and theorem~\ref{thm:restrictedboundary}.

\begin{corollary}
	\label{cor:factors2}
	If function $q(\rho,\bar\rho)$ is holomorphic in the branched unit $\rho,\bar\rho$-polydisc and satisfies the appropriate slow-growth conditions near $\tau,\bar \tau\in \R$ (recall $\rho=e^{i\tau},\bar\rho=e^{i\bar\tau}$), then we have
	\be
	(\mathrm{bv}\,q\cdot g_{1234})(\tau,\mathrm{Re}\,\bar\tau)&=\sum_{\cO} \lambda_{12\bar\cO}\lambda_{43\cO} (\mathrm{bv}\,q\cdot g_{\De,J})(\tau,\mathrm{Re}\,\bar\tau)\quad\text{in }\cS'(\R),\\
	(\mathrm{bv}\,q\cdot g_{1234})(\mathrm{Re}\,\tau,\bar\tau)&=\sum_{\cO} \lambda_{12\bar\cO}\lambda_{43\cO} (\mathrm{bv}\,q\cdot g_{\De,J})(\mathrm{Re}\,\tau,\bar\tau)\quad\text{in }\cS'(\R),\\
	(\mathrm{bv}\,q\cdot g_{1234})(\mathrm{Re}\,\tau,\mathrm{Re}\,\bar\tau)&=\sum_{\cO} \lambda_{12\bar\cO}\lambda_{43\cO} (\mathrm{bv}\,q\cdot g_{\De,J})(\mathrm{Re}\,\tau,\mathrm{Re}\,\bar\tau)\quad\text{in }\cS'(\R^2).
	\ee
\end{corollary}

\begin{theorem}
	\label{thm:restrictedboundary2}
	Let $\mathbb{D}$ be the open unit disk parametrized by $w$ and let $\varphi:w\mapsto\varphi(w)$ be a holomorphic map which maps $\mathbb{D}$ one-to-one onto a domain $ \SS$ inside the cut unit disk of the $\rho$ variable, $ \SS \subset \mathbb{D}\setminus(-1,0]$. Let $\bar\phi$ be a map of the same kind with $ \SS$ replaced by $\bar{ \SS}\subset \mathbb{D}\setminus(-1,0]$. Replacing $\rho=\phi(w),\bar\rho=\bar\phi(\bar w)$ in the conformal block expansion~\eqref{eq:generalscalarcbexpansion}, we pull it back to $w,\bar w\in \mathbb{D}\times\mathbb{D}$. Then this pulled-back conformal block expansion in $w,\bar{w}$ variables converges on the boundaries $|w|=1$, $|\bar w|=1$, or $|w|=|\bar w|=1$ in the sense of distributions. Furthermore, the same conclusion holds if expansion~\eqref{eq:generalscalarcbexpansion} is multiplied by  $q(\rho,\bar \rho)=(z\bar z)^{-\frac{\De_1+\De_2}{2}}$.
\end{theorem}

For example, the discussion of analytic bootstrap functionals in section~\ref{sec:functionals} can be extended to the two-variable case as follows. In Zhukovsky variables $y,\bar y$ the crossing domain $\cC^{st}$ is given by $\mathbb{D}\times \mathbb{D}$. The boundary $\ptl(\mathbb{D}\times \mathbb{D})$ is topologically a 3-sphere $S^3$. This $S^3$ is a disjoint union
\be
S^3 = (\mathbb{D}\times S^1) \sqcup (S^1 \times \mathbb{D}) \sqcup \mathbb{T}^2,
\ee
where the first solid torus $\mathbb{D}\times S^1$ corresponds to $|y|=1$ and $|\bar y|<1$, the second solid torus corresponds to $|\bar y|=1$ and $|y|<1$, while the torus $\mathbb{T}^2=S^1\times S^1$ corresponds to $|y|=|\bar y|=1$. We have shown that the conformal block expansion in either $s$- or $t$- channel converges in the sense of distributions on each of these boundary components. 

Let us focus on the component $\mathbb{T}^2=S^1\times S^1$. Our results imply that the functionals  $\a_f$ of the form
\be
g(y,\bar y)\mapsto \a_f[g]\equiv \int_0^{2\pi}\int_0^{2\pi} d\theta d\bar\theta f(\theta,\bar\theta) g(y=e^{i\theta},\bar y=e^{i\bar\theta})
\ee
where $f(\theta,\bar\theta)$ is a smooth function, satisfy the swapping property. As in section~\ref{sec:functionals}, by taking $f$ to be the Cauchy kernel
\be
f_{m,n;y_0,\bar y_0}(\theta,\bar\theta) = \frac{m!n!}{(2\pi)^2} \frac{e^{i\theta}}{(e^{i\theta}-y_0)^{m+1}}\frac{e^{i\bar\theta}}{(e^{i\bar\theta}-\bar y_0)^{n+1}},
\ee
we can reproduce the evaluation functionals $\a_{m,n;y_0,\bar y_0}$ 
\be
g(y,\bar y)\mapsto \a_{m,n;y_0,\bar y_0}[g]\equiv \ptl_y^m\ptl_{\bar y}^n g(y_0,\bar y_0).
\ee
We can again ask about the space of functions $f$ for which the functional $\a_f$ satisfies the swapping property and try to see if this space is large enough to incorporate the functionals that are useful in analytic conformal bootstrap. Just as in section~\ref{sec:functionals}, we leave these questions for future work.

\section{Spinning operators}

Another natural generalization available in higher dimensions is to operators with spin. In cross-ratio space this question is somewhat non-canonical due to the freedom of choosing the tensor structures for spin indices, which is similar to the freedom of selecting the prefactor in~\eqref{eq:general4ptfunction}. Nevertheless, it is clear that for reasonable choices of the basis of tensor structures, the four-point functions of spinning operators should satisfy similar power-law bounds in cross-ratio space. For example, one could use equation~\eqref{eq:rhocf} with $\phi_i$ replaced by plane-rotation eigencomponents of some spinning operators $\cO_i$, and the arguments of sections~\ref{sec:generalcb} and~\ref{sec:generalbounds} would still go through. This would correspond to using the ``conformal frame'' basis of four-point structures~\cite{Kravchuk:2016qvl}, which is related to all reasonable choices of tensor structures by matrices which themselves satisfy power-law bounds.\footnote{There is a subtlety for $z=\bar z$, in which case the transition matrices to/from conformal frame basis become singular. These singularities are canceled by special conditions satisfied by four-point functions in conformal frame basis near this locus (see appendix A of~\cite{Kravchuk:2016qvl} and appendix D of~\cite{Karateev:2019pvw}).} While it would be a good exercise to explicitly repeat our arguments in the case of spinning correlators, we do not do it in this part of the thesis for the sake of space.

\section{Single-variable dispersion relation for the four-point function in $d\ge 2$}
\label{sec:Bissi}
This section generalizes section \ref{sec:dispersion} to the case of two cross-ratios $z,\bar{z}$. Ref.~\cite{Bissi:2019kkx} presented a single-variable dispersion relation recovering the four-point function in terms of its discontinuities. We will state their story in our language, clarifying some issues. Consider the four-point function satisfying the crossing equation
\be
\label{eq:crossing_disp}
F(z,\bar z)=F(1-z,1-\bar z) = (z \bar z)^{-\De_\f} F(1/z,1/\bar{z})\,,
\ee
where $F(z,\bar z)= (z \bar z)^{-\De_\f}g(z,\bar z)$ and the third equation corresponds to the $u$-channel. This channel representation does not exist for a general 1d four-point function considered in section \ref{sec:1dvariants}.

Ref.~\cite{Bissi:2019kkx} considers a dispersion relation for the function $F(z,\bar z)$ using the discontinuity w.r.t.~$z$ and keeping $\bar z$ fixed. In our language this dispersion relation would be written in the form 
\begin{equation}
	\label{eq:dispgen}
	\quad F(z,\bar z)=\frac 1{2\pi i }\int_{-\infty}^\infty \frac{dx}{z'-z} \mathop{{\rm Disc}}\limits_{z'}\, F(z',\bar z)
\end{equation}
where the discontinuity has to be understood in a distributional sense, including the contribution at infinity, as discussed in section \ref{sec:dispersion}.

Then the question arises how to compute the discontinuity. There are three cases: $-\infty<z'\le 0$, $1\le z'< +\infty$, and $z'=\infty$. In the first case we can use the $s$-channel conformal block decomposition, which converges in the sense of distributions (in fact in ordinary sense for $z'<0$). The discontinuity at $z'\ge 1$ is reduced to the one at $z'\le 0$ via the first crossing equation in \eqref{eq:crossing_disp}.\footnote{It is also possible to compute the discontinuity at $z\ge 1$ by summing the $s$-channel conformal block expansion since by our results it converges on this cut in the sense of distributions. Ref.~\cite{Bissi:2019kkx} mentions this result in footnote 1, attributing it to Mack~\cite{Mack:1976pa}. This is not correct: Mack's paper studies distributional convergence of OPE expansion in position space, not in the cross-ratio space as needed here. In due fairness, footnote 1 is not central for \cite{Bissi:2019kkx}, being only used in section 4.2.2.
	
	We note in passing that Mack \cite{Mack:1976pa} relied on validity of Wightman axioms and rather non-trivial representation theory. It is only in~\cite{Kravchuk:2021kwe,paper3} that we will show, for the first time, how some of Mack's assumptions follow from more mundane Euclidean CFT rules. In comparison, our arguments here are very elementary and rely only on the well-established properties of the conformal block expansion.}

One can try to fix the contribution at infinity using the $u$-channel conformal block expansion, which determines the behavior of the correlator at $z'=\infty$. Let us assume that 
\be
\label{eq:asu}
F(z',\bar z) = 1+O((z')^{-\tau/2})\,.
\ee 
Ref.~\cite{Bissi:2019kkx} argued this by appealing to the second crossing relation in \eqref{eq:crossing_disp}, expanding $F(1/z,1/\bar z)$ in conformal blocks, keeping only the unit operator and dropping all the other operators which seem to be naively suppressed by $(1/z)^{\tau/2}$ where $\tau=\min(\Delta-\ell)$ is the minimal twist, assumed positive. This reasoning includes a subtlety, see below. But assuming \eqref{eq:asu} we can argue that, in the language of section~\ref{sec:dispersion},
\be
\mathop{{\rm Disc}}\limits_{z}\,F(z,\bar z) = (\mathrm{Disc}\, 1)(z) + \mathop{{\rm Disc}'}\limits_{z}\,F(z,\bar z),
\ee
where $\mathrm{Disc}\, 1$ was computed in section~\ref{sec:dispersion} and $\mathop{{\rm Disc}'}\limits_{z}\,F(z,\bar z)$ is a distribution that is represented near $z=\oo$ by an ordinary function. In other words, ${\rm Disc}'$ is the discontinuity ``without the contribution at $\oo$.''\footnote{Note, however, that we can only unambiguously define such discontinuity because of~\eqref{eq:asu}. For example, this is not possible for $\log z$ example from section~\ref{sec:dispersion}.}

Using this decomposition of $\mathop{\mathrm{Disc}}\limits_{z}\,F(z,\bar z)$, one obtains from~\eqref{eq:dispgen} a dispersion relation in the form given by \cite{Bissi:2019kkx}
\begin{equation}
	\label{eq:dispgen1}
	\quad F(z,\bar z)=1+ \Bigl(\frac 1{2\pi i }\int_{-\infty}^0 \frac{dz'}{z'-z} \mathop{{\rm Disc}'}\limits_{z'}\, F(z',\bar z)+(z,\bar z\to 1-z,1-\bar z)\Bigr)\,,
\end{equation}
where, as mentioned above, the discontinuity $\mathrm{Disc}'$ does not include the contribution at infinity that is instead explicitly included as ``$1+$'', and we used crossing symmetry to account for discontinuity on the cut $[1,+\oo)$.

Note that independently of the assumption~\eqref{eq:asu}, our results imply that $\mathop{\mathrm{Disc}}\limits_{z} F(z,\bar z)$ can be computed term-by term in conformal block expansion (including the contribution at infinity), and then used in~\eqref{eq:dispgen}, although it is not guaranteed that the decomposition~\eqref{eq:dispgen1} exists in that case.

Let us now discuss the subtlety in the asymptotics \eqref{eq:asu}. Upon a closer look, this asymptotic is only justified provided that $z$ and $\bar z$ 
belong to the different halfplanes of the region $\cC^{st}$, i.e.~if ${\rm Im}\,z$ and ${\rm Im}\,\bar z$ have opposite sign. This is because the $u$-channel conformal block expansion stops converging when $z$ crosses the cut $(0,1)$ and moves into the same half-plane as $\bar z$. Thus, if $\bar z$ is fixed, asymptotics \eqref{eq:asu} is rigorously true only on one of the two arcs at infinity $z$. The asymptotics on the second arc is somewhat similar to the Regge limit asymptotics, in the sense that $1/z$ goes through the $s$-channel cut and then is sent to zero (while, unlike in the Regge limit, $\bar z$ stays fixed). 

There are two ways around this difficulty. One way is to take $\bar z\in (0,1)$ real. Then, by our results, the $u$-channel OPE expansion converges in the sense of distributions on both arcs. In this case the asymptotics \eqref{eq:asu} is true provided that the error term is understood in the sense of distributions, and it goes to zero as $z\to \infty$. Since a zero distribution is a zero function, we recover the dispersion relation \eqref{eq:dispgen1}.

The second way around the difficulty is to apply the dispersion relation in perturbation theory around a mean field theory, which was in fact the main focus of \cite{Bissi:2019kkx}. In their case the zeroth order term satisfies the asymptotics \eqref{eq:asu} by inspection, while perturbative corrections have an even better behavior. The use of dispersion relation in such a limited context is justified.

\chapter{Conclusions}
\label{sec:conclusions}

In this part of the thesis we studied the properties of the conformal block expansion on the boundary of its region of convergence. We showed that both the correlation functions and conformal blocks can be interpreted as distributions on this boundary, and that the conformal 
block expansion converges in the space of distributions. We have proven these results in one- and higher-dimensional cases for correlators of scalar operators, but the extension to general spinning four-point functions is straightforward.

An important feature of our analysis is that we did not rely on anything but the modern Euclidean bootstrap axioms. Specifically, we essentially only used the reality properties of OPE coefficients and the usual convergence properties of the conformal block expansion. There is a growing consensus that the Euclidean bootstrap axioms provide a good conceptual and practical definition for CFTs. Their conceptual appeal is due to them being rooted in cutting-and-gluing properties of Euclidean path integrals, which is a natural expected consequence of locality. The practical utility of these axioms has been demonstrated by the numerical conformal bootstrap studies, which have yielded extremely precise values of critical exponents and other parameters in various strongly-coupled CFTs such as the 3d Ising CFT and the $O(2)$ model (see \cite{Kos:2016ysd, Chester:2019ifh} for the most precise determination to date). These values are in agreement with a plethora of other completely independent methods (most notably Monte Carlo simulations and the $\eps$-expansion).

Our results are important for understanding the nature of conformal correlation functions in Lorentzian signature. Indeed, as we show in appendix~\ref{app:lorentz}, the best one can guarantee in general configurations in Lorentzian signature is that the conformal cross-ratios are on the boundary of the region of convergence for one of the OPE channels. It is thus important to understand the value of CFT four-point functions on this boundary. We have shown that the conformal block expansion converges there in distributional sense, which gives a practical way for computing correlation functions. For example, we can now imagine collecting numerical OPE data for 3d Ising CFT as in~\cite{Simmons-Duffin:2016wlq} and using it to compute pairings of the boundary value of $\<\s\s\s\s\>$ four-point function with various tests functions.

One important byproduct of our results, which we discuss in section~\ref{sec:functionals}, is a hint at a uniform description of the space of functionals with which we can probe the crossing equation. Starting with numerical conformal bootstrap~\cite{Rattazzi:2008pe}, it has become standard to disprove the existence (under certain spectral assumptions) of solutions to the crossing equation by exhibiting functionals that separate the left-hand side of the crossing equation from the right-hand side. In numerical bootstrap (see~\cite{Poland:2018epd} for review) these functionals are finite combinations of evaluation functionals $\a_{n,y}$~\eqref{eq:evaluationfunctional}, while in more recent analytical functional bootstrap~\cite{Mazac:2016qev,Mazac:2018mdx,Mazac:2018ycv,Kaviraj:2018tfd,Mazac:2018biw,Hartman:2019pcd,Paulos:2019gtx,Mazac:2019shk} the appropriate functionals are given by contour integrals $\a_{h,\G}$~\eqref{eq:originalcuttouching}. Having a uniform description of a sufficiently large class $\cB_{\De_\f}$ of functionals (that in particular would include $\a_{n,y}$ and $\a_{h,\G}$) would allow us to formulate and hopefully answer some interesting conceptual questions. For example, 
\begin{itemize}
	\item is it true that for any spectral assumption for which there is no solution to crossing equation there exists a functional in $\cB_{\De_\f}$ that disproves the existence of a solution? 
	\item Is it true that when the spectral assumption is not ``extremal,'' this functional can be taken as a finite linear combination of evaluation functionals? (In other words, is numerical conformal bootstrap complete?) 
	\item When the spectral assumption is extremal, is it true that there exists a unique extremal functional? 
\end{itemize}

Most practitioners would probably guess that the answer to these three questions should be ``yes'', ``yes'' and ``generically yes''. To put this intuition on firm footing we need first of all understand better the space $\cB_{\De_\f}$ and the appropriate topology on this space. 
Answering these questions will be important for advancing our analytical understanding of conformal bootstrap.

\end{part}

\begin{part}{Distributions in CFT II: Minkowski space}\label{part:minkowski}
\thispagestyle{fancy}

\chapter{Introduction}

Quantum Field Theory (QFT) can be studied via constructive or axiomatic
approaches. The former use microscopic formulations, while the latter rely on
axioms. There are many constructive approaches, e.g.\ using Hamiltonian, path
integral, lattice, etc. There are also many axiomatic approaches, corresponding
to various sets of axioms (Wightman {\cite{Streater:1989vi}},
Osterwalder-Schrader {\cite{osterwalder1973,osterwalder1975}}, Haag-Kastler
{\cite{Haag:1963dh,Haag:1992hx}}, etc.). Historically, axiomatic approaches
played an important role in clarifying general QFT properties, but they did
not have a tremendous success in making predictions about concrete theories in $d>2$ dimensions.\footnote{In $d=2$ significant progress has been achieved axiomatically for massive integrable models using the S-matrix bootstrap \cite{Zamolodchikov:1978xm} as well as for rational CFTs \cite{belavin1984infinite}.}
This started to change recently, with the revival of the bootstrap philosophy
{\cite{Rattazzi:2008pe}}. Our focus here will be on conformal field theories
(CFTs) in dimension $d \geqslant 2$, i.e.\ QFTs invariant under the action of
conformal group, which are nowadays studied via the \tmtextit{conformal
	bootstrap}. This axiomatic approach led to precise determinations of many
experimentally measurable quantities, such as the critical exponents of the 3d
Ising
{\cite{ElShowk:2012ht,El-Showk:2014dwa,Kos:2014bka,Simmons-Duffin:2015qma,Kos:2016ysd}},
$O (N)$ {\cite{Kos:2013tga,Kos:2015mba,Kos:2016ysd,Chester:2019ifh,Chester:2020iyt}} and other
critical points (see review~{\cite{Poland:2018epd}}).\footnote{There is also an ongoing
	revival of the S-matrix bootstrap applicable to nonintegrable massive QFTs in $d\ge 2$
	{\cite{Paulos:2016fap,Paulos:2016but,Paulos:2017fhb,Cordova:2018uop,Guerrieri:2019rwp,EliasMiro:2019kyf,Cordova:2019lot,Karateev:2019ymz,Correia:2020xtr,Guerrieri:2020bto,Hebbar:2020ukp,Tourkine:2021fqh,Sinha:2020win,Haldar:2021rri,He:2021eqn}}.}

The numerical conformal bootstrap relies on the ``Euclidean CFT
axioms'',\footnote{The term ``Euclidean bootstrap axioms'' is also sometimes
	used.} which specify properties of correlation functions in any unitary CFT in
$\mathbb{R}^d$ via a set of simple and commonly accepted rules, such as the
unitarity bounds on primary operator dimensions, conformally invariant and
convergent Operator Product Expansion (OPE), and reality constraints on OPE
coefficients.

On the other hand, correlation functions in a general unitary QFT (and in
particular in a CFT) should satisfy Osterwalder-Schrader (OS) and Wightman
axioms. It is then interesting and important to inquire what is the relation
of Euclidean CFT axioms to these other sets of axioms.\footnote{Clarifying the
	relation to the Haag-Kastler axioms appears more challenging as those axioms
	do not deal with correlation functions but with operator algebras.} To carry
out this analysis will be the main goal of this thesis. Our main conclusion will
be that the Euclidean CFT axioms imply OS axioms and Wightman axioms for $2,
3$ and $4$-point functions. In this thesis we only consider bosonic operators.

The relation of Euclidean CFT and OS axioms is perhaps not so surprising since
they both deal with the Euclidean correlation functions. It is more
interesting that we are able to construct Minkowski $n$-point functions (for
$n = 2, 3, 4$), and show that they satisfy Wightman axioms, such as
temperedness, spectral condition, and unitarity. Temperedness (being a
tempered distribution) is a crucial property of Minkowski correlation
functions: it shows that in a certain averaged sense they are meaningful
everywhere including the light-cone and double light-cone singularities. One
might be tempted to think that in CFT this question is relatively trivial due
to OPE. However, as discussed in part \ref{part:crossratio} and part \ref{part:ope}, already for
4-point functions there exist causal configurations of points in Minkowski
space, away from the null cones, for which no OPE channel is convergent in the
conventional sense. We briefly discuss one such example in the conclusions
(chapter \ref{chap:conclusions}).

A theorem of Osterwalder and Schrader
{\cite{osterwalder1973,osterwalder1975}} says that, under some extra
assumption, OS axioms imply Wightman axioms. Unfortunately this extra
assumption, the so called ``linear growth condition'', which involves the
Euclidean $n$-point functions with arbitrarily high $n$, appears impossible to
verify from the Euclidean CFT axioms (see chapter \ref{OS}). For this reason
we cannot appeal to the OS theorem, and we will give an independent derivation
of Wightman axioms for CFT correlators.

The study of distributional properties of CFT correlators started in part \ref{part:crossratio}. There, we considered expansions of the CFT 4-point
function $g (\rho, \bar{\rho})$ in terms of conformally invariant cross-ratios
$\rho$, $\bar{\rho}$. While such expansions converge in the usual sense for $|
\rho |, | \bar{\rho} | < 1$, in part \ref{part:crossratio} we showed that they also
converge for $| \rho |, | \bar{\rho} | = 1$ in the sense of distributions. As
explained in part \ref{part:crossratio}, results of this sort follow naturally from the
theory of functions of several complex variables (namely Vladimirov's
theorem), given some apriori information about the growth of the analytically
continued correlator. That key insight of part \ref{part:crossratio}, \emph{``Look for a
	powerlaw bound!''}, will be transported here from the cross-ratio to the
position space.

The readers interested in our main technical result---analytic continuation of
a scalar Euclidean CFT 4-point function to the forward tube and showing that the
Minkowski 4-point function is a tempered distribution---may follow the
\tmtextbf{fast track:} start with the executive summary in Sec.\ \ref{executive}, proceed to Secs.\ \ref{strategy} and \ref{23-point} (skipping
\ref{MinkFromEucl} and \ref{comparison}), then continue with Secs.\ \ref{sec:informal}-\ref{power4-point} (optionally including chapter \ref{secondpass}) and finish with the discussion in chapter \ref{chap:conclusions}.
This is only about 20-25 pages.

On the other hand, we made an effort to make the exposition self-contained
and to review main ideas and results of the axiomatic quantum field
literature, directly or tangentially related to our discussion. This explains
the great total length of our work. The reader will find here:
\begin{itemize}
	\item A review of classic QFT axioms: Wightman (Sec.\ \ref{Waxioms}), OS
	(Sec.\ \ref{OSaxioms}). A review of main implications among these axioms:
	how OS reflection positivity robustly implies Wightman positivity (Sec.\ \ref{sec:Wpos}). A review of the Osterwalder-Schrader theorem about how OS
	axioms imply Wightman axioms under the additional assumptions of the linear
	growth condition (chapter \ref{OS}).
	
	\item A \ formulation of `Euclidean CFT axioms' for unitary CFT in
	Euclidean space $\mathbb{R}^d$ (Sec.\ \ref{ECFTax}). We consider bosonic
	fields in arbitrary tensor representations. Our axioms encode in a
	consistent and non-redundant manner the main properties used in the
	numerical conformal bootstrap calculations.\footnote{See also
		{\cite{Rychkov:2020rcd}} for a recent informal exposition of Euclidean CFT
		axioms (incomplete as it omits tensor fields) for mathematical physics
		audience. Ref.\ {\cite{Schwarz:2015fva}} attempted the axiomatization of
		Euclidean CFT in $d > 2$ dimensions similar to Segal's axioms in $d = 2$
		{\cite{tillmann2004}}. It is not immediately obvious if the axioms of Ref.\ {\cite{Schwarz:2015fva}} are equivalent to ours, or how to connect them to
		practical CFT calculations.} They are applicable to any globally conformally
	invariant theory in $d \geqslant 2$. We do not include the axioms involving
	the local stress tensor and the conserved currents. In particular our axioms
	would be too weak (but valid) when applied to local 2d CFTs, as they know
	nothing about the Virasoro algebra.\footnote{Recall that while in $d=2$
		assuming the existence of a local stress tensor immediately implies Virasoro
		symmetry, no such dramatic statements are currently known in $d>2$.} See Remark \ref{ECFTax-comp} for a
	comparison between our axioms and the CFT rules gathered in the conformal
	bootstrap review {\cite{Poland:2018epd}}.
	
	\item A derivation of OS axioms from Euclidean CFT axioms for 4-point function
	(chapter \ref{CFTtoOS}). A notable result is a rigorous proof that the state
	produced by two operators in lower Euclidean half-space belongs to the CFT
	Hilbert space generated by single-point operator insertions. The higher-point
	case is discussed in App.\ \ref{OShigher}, where we need a somewhat
	stronger form of the OPE axiom than in Sec.\ \ref{ECFTax}.
	
	\item A derivation of Wightman axioms from Euclidean CFT axioms for scalar
	4-point functions (chapter \ref{sec:4-point}). As mentioned above, this is the main
	technical result of this part of the thesis. The key observation is that the analytic
	continuation from Euclidean to Minkowski can be done in a way which keeps
	the s-channel $\rho, \bar{\rho}$ less than 1 in absolute value along the
	continuation contour. When we take the Minkowski limit $| \rho |, |
	\bar{\rho} |$ stay less than 1 for some causal orderings and approaches 1
	for others (see part \ref{part:ope} for a classification) but even if $|
	\rho |, | \bar{\rho} | \rightarrow 1$ they approach this limit sufficiently
	slowly (``powerlaw bound'') which guarantees that the Minkowski 4-point
	function is a tempered distribution everywhere. E.g.\ using this argument we
	can show for the first time that the CFT 4-point function is a tempered
	distribution on the double light-cone singularity.
	
	\item We include also a derivation of other expected properties of Minkowski
	4-point functions, such as conformal invariance, unitarity, clustering, and
	local commutativity (Secs.\ \ref{ConfMink}-\ref{local-comm}). The reader
	may find it curious how some of the steps do not use conformal invariance at
	all but follow simply from analyticity and/or OS positivity.
	
	\item chapter \ref{secondpass} proves a curious geometric ``Cauchy-Schwarz''
	inequality for $\rho, \bar{\rho}$ variables which provides an alternative
	way of understanding why $| \rho |, | \bar{\rho} | < 1$ in the forward tube.
	It bounds $\rho, \bar{\rho}$ for a generic configuration by $\rho,
	\bar{\rho}$ for reflection-symmetric configurations. It would be interesting
	to find an elementary proof of this inequality (our proof uses some facts
	about conformal blocks).
	
	\item chapter \ref{OPEconvMink} shows that the (s-channel) conformal block
	expansion of 4-point Wightman functions converges in the sense of
	distributions for all configurations of points in Minkowski space. It is
	also shown that the OPE for the state-valued distributions $| \mathcal{O}
	(x_1) \mathcal{O} (x_2) \rangle$ with $x_1, x_2 \in \mathbb{R}^{d - 1, 1}$
	converges in the sense of distributions. We discuss the relationship of
	these results to the classic work of Mack {\cite{Mack:1976pa}} and prove
	estimates for the convergence rates of these expansions.
	
	\item chapter \ref{OS} contains a review of the papers
	{\cite{osterwalder1973,osterwalder1975}} by Osterwalder and Schrader. In
	particular, we discuss the gap in the arguments of {\cite{osterwalder1973}}
	which precludes the derivation of Wightman axioms from the OS axioms of
	{\cite{osterwalder1973}}, and explain in detail how this gap is filled in
	{\cite{osterwalder1975}} with the addition of new axioms. 
	
	\item {App.~\ref{literature} is a guide to the modern Lorentzian
		CFT literature: conformal collider bounds, light-cone bootstrap, causality constraints, the Lorentzian OPE inversion formula, light-ray operators, etc. Our results help put some of these considerations on a firmer footing. We indicate the most critical remaining steps, which still wait to be rigorously derived from the Euclidean CFT axioms.}
	
\end{itemize}
We conclude in Sec.~\ref{chap:conclusions}. Some additional technical details are given in
Apps.~\ref{OShigher}-\ref{IntLem1}. 

\section{Executive summary of results for CFT experts}\label{executive}

This part of the thesis is rather lengthy as a result of our attempt to make it
self-contained. In this section we give a brief summary of the main technical
results, aimed at the more expert readers who may not wish to read the
expository parts of this part. Note, however, that here we omit many
secondary results, some of which are mentioned above.

The basic question we address in this part is the question of the
distributional properties of Wightman 4-point functions in CFTs. As is
well-known, Wightman $n$-point distributions are recovered from the boundary
values of functions holomorphic in the forward tube $\mathcal{T}_n$. {For an $n$-point
	function 
	\be
	\<0|\cO_1(x_1)\cdots \cO_n(x_n)|0\>
	\ee
	the forward tube is defined as the set of $x_i\in \C^{1,d-1}$ subject to 
	\be
	\mathrm{Im}\, x_1\prec \mathrm{Im}\, x_2\prec\cdots \prec \mathrm{Im}\, x_n,
	\label{forward-tube-Mink}
	\ee
	where $\prec$ denotes the causal ordering in $\R^{1,d-1}$. Analyticity in $\cT_n$ and existence of
	the boundary value as $\mathrm{Im}\, x_i\to 0$ is usually derived 
	from Wightman or OS axioms (with extra assumptions in the latter case). 
	In this part of the thesis we want to understand this question from the point of view
	of CFT axioms.} 

With the
cases $n = 2, 3$ being relatively trivial in a CFT, our main observation is
that a particular OPE channel for 4-point functions converges everywhere in
the interior of $\mathcal{T}_4$. Specifically, we take the OPE $\mathcal{O}
(x_1) \times \mathcal{O} (x_2)$ in the Wightman function
\begin{equation}
	\langle 0 | \mathcal{O} (x_1) \mathcal{O} (x_2) \mathcal{O} (x_3)
	\mathcal{O} (x_4) | 0 \rangle . \label{introcrr}
\end{equation}
This OPE is expected to converge, at least distributionally for real $x_i$,
from the results of Mack {\cite{Mack:1976pa}}. However, his work assumes
Wightman axioms from the beginning, and our goal here is to clarify the
implications of Euclidean CFT axioms, which only assume convergence of the
Euclidean OPE.

To see that this OPE channel converges, we show in Lemma \ref{bound} that for
any configuration of $x_i$ in $\mathcal{T}_4$ the radial cross-ratios $\rho$
and $\bar{\rho}$ for this OPE belong to the open unit disk,
\begin{equation}
	| \rho |, | \bar{\rho} | < 1.
\end{equation}
This implies convergence of the conformal block expansion in $\cO (x_1)
\times \cO (x_2)$ channel in the interior of $\mathcal{T}_4$. A technical
way to see this is to note that the expansion in descendants
\begin{equation}\label{eq:introexpansion}
	g (\rho, \bar{\rho}) = \sum_{h, \bar{h} > 0} p_{h, \bar{h}} \rho^h
	\bar{\rho}^{\bar{h}},
\end{equation}
where $g$ is the conformally-invariant part of the 4-point function and
$p_{h, \bar{h}} > 0$, can be bounded term-by-term by
\begin{equation}
	| p_{h, \bar{h}} \rho^h \bar{\rho}^{\bar{h}} | \leqslant p_{h, \bar{h}} r^h
	r^{\bar{h}}, \label{introtermbyterm}
\end{equation}
where $r = \max (| \rho |, | \bar{\rho} |) < 1$. The right-hand side of this
inequality is a term in the expansion of $g (r, r)$, a Euclidean configuration
in which the OPE is known to converge, so~\eqref{eq:introexpansion} is dominated
by a convergent series. Therefore, \eqref{eq:introexpansion} is convergent for $r<1$, and moreover
uniformly so on compact subsets, since each term $p_{h, \bar{h}} r^h
r^{\bar{h}}$ is monotonic. We can then conclude that the sum $g(\rho,\bar \rho)$ is a holomorphic function.

This reasoning also gives us the inequality
\begin{equation}
	| g (\rho, \bar{\rho}) | \leqslant g (r, r) .
\end{equation}
So, we find that the correlator can be recovered inside of $\mathcal{T}_4$ from
the $\cO (x_1) \times \cO (x_2)$ OPE, is analytic there, and is bounded
by a Euclidean configuration.

In Sec.\ \ref{section:1-r} we establish a stronger form of Lemma
\ref{bound}, schematically,
\begin{equation}
	1 - r (c) \geqslant \tmop{dist} (c, \partial \mathcal{T}_4)^a \label{rbound}
\end{equation}
for some $a > 0$, where $c \in \mathcal{T}_4$ is a configuration of 4 points
in $\mathcal{T}_4$. (The more precise form also bounds $1 - r (c)$ as $c$ goes
to infinity.) This immediately implies a powerlaw bound on $g (r, r)$ near the
boundary of $\mathcal{T}_4$. Indeed, near $r \rightarrow 1$ the correlator is
dominated by the identity in the crossed channel, and so
\begin{equation}
	g (r (c), r (c)) \leqslant C (1 - r (c))^{- 4 \Delta_{\varphi}},
\end{equation}
and thus
\begin{equation}
	| g (c) | \leqslant C \tmop{dist} (c, \partial \mathcal{T}_4)^b
\end{equation}
for some real $b$. This allows us to use Vladimirov's Theorem \ref{ThVlad},
which implies that the boundary limit (as $x_i$ approach real Minkowski
values) of {\eqref{introcrr}} exists in the space of tempered distribution.
(We establish a more refined bound for $x_i \rightarrow \infty$ to claim
temperedness.)

The above bounds hold just as well for the truncated conformal block expansion
as for the full correlator. A variant of Theorem \ref{ThVlad} then allows us
to conclude that the conformal block expansion, while converging in the sense
of functions in the interior of $\mathcal{T}_4$, converges in the space of
tempered distributions on the Minkowski boundary.

We extend the above results to correlators of non-identical scalars by
replacing the term-by-term bound {\eqref{introtermbyterm}} with a standard
Cauchy-Schwartz argument, bounding the correlator in terms of a product of two
reflection-symmetric Euclidean correlators. While it is intuitively obvious
that similar arguments should also work for operators with spin, we found that
the extension to spinning operators, due to the complexity of tensor
structures, requires enough additional work to warrant a separate paper
{\cite{paper2a}}.

Finally, in chapter \ref{secondpass} we prove Theorem \ref{boundThm}, which
gives an optimal bound of the form {\eqref{rbound}}. Specifically, it is
\begin{equation}
	r (c)^2 \leqslant r (c_{12}) r (c_{34}), \label{introoptimal}
\end{equation}
together with a bound for the right-hand side. Here, if $c = (x_1, x_2, x_3,
x_4)$ (where $x_i$ are real in Minkowski space), then $c_{12} \equiv (x_1,
x_2, x_2^{\ast}, x_1^{\ast})$ and $c_{34} \equiv (x_4^{\ast}, x_3^{\ast}, x_3,
x_4)$. The bound for the right-hand side is easier to obtain because the
configurations $c_{12}$ and $c_{34}$ are reflection symmetric. This is done in
Sec.\ \ref{rhsbound}. The bound {\eqref{introoptimal}} looks like a
Cauchy-Schwartz-type inequality, and is indeed derived from the
Cauchy-Schwartz inequality for unitary conformal blocks {\eqref{CB-CS}}. The
latter is true because of the unitarity of conformal representations
corresponding to these blocks. In the limit $\Delta + \ell \rightarrow
\infty,$ $\Delta - \ell$ fixed, conformal blocks are dominated by $r^{(\Delta
	+ \ell) / 2}$, which reduces the conformal block Cauchy-Schwartz inequality to
{\eqref{introoptimal}}.

\chapter{Axioms}\label{chap:axioms}

\section{Wightman axioms}\label{Waxioms}

In this section we will state the properties of Wightman correlation functions in
a unitary QFT, to which we will refer here as ``Wightman axioms.'' These
axioms appear as as ``properties of
the vacuum expectation values'' in {\cite{Streater:1989vi}}, Sec.\ 3-3, and as (W1)-(W5) in \cite{simon1974}, Theorem II.1. Refs.\ {\cite{Streater:1989vi,simon1974}} give in
addition another set of axioms (called G{\aa}rding-Wightman axioms in \cite{simon1974}) saying that fields are operator-valued
distributions in the Hilbert space on which the Lorentz group acts, etc. This
other set of axioms will not be used in this thesis. In any case, the Wightman
reconstruction theorem {\cite{Streater:1989vi}} says that the two sets of
axioms are equivalent.

A unitary QFT in Minkowski space studies $n$-point correlators
\begin{equation}
	\langle \varphi_1 (x_1) \ldots \varphi_n (x_n) \rangle, \label{nptMink}
\end{equation}
(Wightman functions) of local operators $\varphi_i (x)$, $x \in
\mathbb{R}^{1, d - 1}$. For simplicity in this part of the thesis we will only consider
bosonic operators, although more generally one should allow fermionic
operators and spinor representations. Wightman functions are translation and
$\tmop{SO} (1, d - 1)$ invariant. We will choose a basis of local operators
$\mathcal{O}_i$ transforming in irreducible $\tmop{SO} (1, d - 1)$
representations $\rho_i$. Then, Wightman functions remain invariant when
\begin{equation}
	\mathcal{O}_i^{\alpha} (x) \rightarrow {\rho_i (g)^{\alpha}_{\ \beta}}
	\mathcal{O}_i^{\beta} (g^{- 1} x), \label{transform}
\end{equation}
where $g \in \tmop{SO} (1, d - 1)$, and $\rho_i (g)$ are finite-dimensional
matrices of the representation $\rho_i$ $(\alpha, \beta = 1 \ldots \dim
\rho_i)$. Let $\mathcal{C}$ be the complex vector space whose elements are
arbitrary components of $\mathcal{O}_i$'s, and their \tmtextit{finite} linear
combinations with constant complex coefficients. Operators $\varphi
\mathcal{}_i$ in {\eqref{nptMink}} can be arbitrary elements of $\mathcal{C}$,
and Wightman function {\eqref{nptMink}} is multi-linear in $\varphi
\mathcal{}_i$. Note that in this and the next section derivatives of local
operators (of any order) are counted as independent operators, while in the
CFT Sec.\ \ref{ECFTax} we will start making distinction between primaries
and their derivatives.

Wightman functions {\eqref{nptMink}} are required to be tempered
distributions, i.e.\ can be paired with Schwartz class test functions $f (x_1,
\ldots, x_n)$. For this reason they are sometimes referred to as ``Wightman
distributions''. Note that the test functions $f (x_1, \ldots, x_n)$ with
which Wightman functions are paired do not have to vanish at coincident points
(unlike for the Schwinger functions discussion in the next section). This
means that, in a distributional sense, Wightman functions have meaning for all
configurations, including coincident points and light-cone singularities.
Translation and Lorentz invariance of Wightman functions are also understood
not pointwise but in the sense of distributions (i.e.\ that the pairing should
remain invariant if the test function is transformed in the dual
way).\footnote{Although Wightman functions can be shown to be real-analytic at
	some totally spacelike-separated configurations (Jost points), in general they
	may be singular even away from light cones (in particular when there are
	timelike separations).}

We will not consider here other Minkowski correlators, such as retarded,
advanced, or time ordered, which are obtained from Wightman functions
multiplying by theta-functions of time coordinate differences, and whose
distributional properties require a separate discussion.

Limiting to the bosonic case as we are, \tmtextbf{local commutativity} (also
called microcausality) morally says that operators commute at spacelike
separation. Wightman axioms impose this as a constraint on Wightman functions:
\begin{equation}
	\langle \varphi_1 (x_1) \ldots \mathcal{} \varphi_p (x_p) \mathcal{}
	\varphi_{p + 1} (x_{p + 1}) \ldots \varphi_n (x_n) \rangle = \langle
	\varphi_1 (x_1) \ldots \mathcal{} \varphi_{p + 1} (x_{p + 1}) \mathcal{}
	\varphi_p (x_p) \ldots \varphi_n (x_n) \rangle \label{Wightman:causality}
\end{equation}
whenever $x_p - x_{p + 1}$ is spacelike: $(x_p - x_{p + 1})^2 > 0.$ Since we
are talking about distributions, this constraint means that
{\eqref{Wightman:causality}} holds when paired with any test function whose
support is contained in $(x_p - x_{p + 1})^2 > 0$.

\tmtextbf{Clustering} says that correlators should factorize if two groups of
points are far separated in a spacelike direction:
\begin{equation}
	\langle \varphi_1 (x_1) \ldots \mathcal{} \varphi_p (x_p) \mathcal{}
	\varphi_{p + 1} (x_{p + 1} + \lambda a) \ldots \varphi_n (x_n + \lambda a)
	\rangle \rightarrow \langle \varphi_1 (x_1) \ldots \mathcal{} \varphi_p
	(x_p) \rangle \langle \varphi_{p + 1} (x_{p + 1}) \ldots \varphi_n (x_n)
	\rangle \label{Wightman:cluster}
\end{equation}
as $\lambda \rightarrow \infty$ for any spacelike vector $a$, limit
understood in the sense of distributions.

We next discuss the spectral condition. By translation invariance we can
write
\begin{equation}
	\langle \varphi_1 (x_1) \ldots \varphi_n (x_n) \rangle = W (\xi_1,
	\ldots, \xi_{n - 1}), \hspace{3em} \xi_k = x_k - x_{k + 1},
\end{equation}
where $W$ is a tempered distribution in one less variable. Consider its
Fourier transform:
\begin{eqnarray}
	\hat{W} (q_1, \ldots, q_{n - 1}) & = & \int W (\xi_1, \ldots,
	\xi_{n - 1}) e^{i \underset{k = 1}{\overset{n - 1}{\sum}} q_k \cdummy \xi_k}\,
	d \xi_1 \ldots d \xi_n, 
\end{eqnarray}
where $q_k = (E_k, \mathbf{q}_k)$, $\xi_k = (t_k, \boldsymbol{\xi}_k)$, $q_k
\cdummy \xi_k = - E_k t_k +\mathbf{q}_k \cdummy \boldsymbol{\xi}_k$. Since $W$
is a tempered distribution, the Fourier transform $\hat{W}$ is well
defined and is also a tempered distribution. The \tmtextbf{spectral condition}
then says that $\hat{W}$ must be supported in the product of closed
forward light cones, i.e.\ in the region
\begin{equation}
	E_k \geqslant \mathbf{q}_k, \hspace{3em} k = 1, 2, \ldots, n - 1.
	\label{Wightman:spectral}
\end{equation}
For the two remaining conditions we need to discuss conjugation. Physically,
each operator $\varphi$ should have a conjugate $\varphi^{\dagger}$. In the
discussed framework we cannot define $\varphi^{\dagger}$ as an adjoint of an
operator acting on a Hilbert space, since we do not have a Hilbert space.
Instead, we will simply assume that there is a rule which associates
$\varphi^{\dagger}$ to $\varphi$, and impose the expected relations at the
level of correlation functions (Eq.\ {\eqref{Hermiticity}} below). This rule,
conjugation map $\dagger : \mathcal{C} \rightarrow \mathcal{C}$, associates to
each independent component $\mathcal{O}_i^{\alpha}  (\alpha = 1 \ldots \dim
\rho_i)$ of the above-mentioned basis of $\mathcal{C}$ a conjugate operator
$(\mathcal{O}_i^{\alpha})^{\dagger}$. This map is required to be an
involution, i.e.\ $\dagger \dagger = 1$. Furthermore, it is extended to
the whole of $\mathcal{C}$ by anti-linearity, i.e.\ $(c_1 \varphi_1 + c_2
\varphi_2 \mathcal{})^{\dagger} = c_1^{\ast} \varphi_1^{\dagger} + c_2^{\ast}
\varphi_2^{\dagger}$.\footnote{The $\dagger$ operation is denoted by $*$ in \cite{Streater:1989vi}.}  

Let us group operators $(\mathcal{O}_i^{\alpha})^{\dagger}$ in a multiplet
which we denote by $\mathcal{O}_i^{\dagger}$, i.e.\ $(\mathcal{O}_i^{\dagger})^{\alpha} =(\mathcal{O}_i^{\alpha})^{\dagger}$.\quad We will see below that
$(\mathcal{O}_i^{\dagger})^{\alpha}$ transform under $g \in \tmop{SO} (1, d -
1)$ with matrices complex-conjugate to those of $\mathcal{O}_i^{\alpha}$:
\begin{equation}
	\mathcal{O}_i^{\alpha} \rightarrow \rho_i (g)^{\alpha}_{\ \beta} \,
	\mathcal{O}_i^{\beta} \quad \Rightarrow \quad
	(\mathcal{O}^{\dagger}_i)^{\alpha} \rightarrow \overline{\rho_i
		(g)^{\alpha}_{\ \beta}} \, (\mathcal{O}^{\dagger}_i)^{\beta} .
	\label{R*}
\end{equation}
In other words, $\mathcal{O}^{\dagger}_i$ transforms in the conjugate
representation $\overline{\rho_i}$.

Since we are considering only bosonic operators, the relevant representations
$\rho_i$ are tensors $T^{\mu_1 \ldots \mu_l}$, on which $g \in \tmop{SO} (1, d
- 1)$ act as:
\begin{equation}
	\begin{array}{lll}
		T^{\mu_1 \ldots \mu_l} & \rightarrow & (\rho_i (g) T)^{\mu_1 \ldots
			\mu_l} = {g^{\mu_1}_{\ \nu_1} \ldots g^{\mu_l}_{\ \nu_l}}
		T^{\nu_1 \ldots \nu_l} .
	\end{array} \label{rhoi}
\end{equation}
Depending on $\rho_i$, these tensors have some fixed rank and mixed symmetry
properties. In addition, in even $d$, for tensors with $d / 2$ antisymmetric
indices, (anti-)chirality\footnote{Chiral and anti-chiral representations are sometimes also called ``self-dual'' and ``anti-self-dual''. We use
	``chiral'' and ``anti-chiral'' to avoid the clash with ``dual representation'' in mathematician's sense.} constraints must be imposed. All tensor
representations of $\tmop{SO} (1, d - 1)$ are real (i.e.\ matrices $\rho_i
(g)^{\alpha}_{\ \beta}$ in {\eqref{R*}} can be chosen real), except for
(anti-)chiral representations in $d = 0 \tmop{mod} 4$ which are
complex-conjugate to each other. For operators in real representations we can
choose a basis such that $\mathcal{O}_i =\mathcal{O}^{\dagger}_i$.

After this intermezzo we are ready to formulate hermiticity and positivity
conditions. \tmtextbf{Hermiticity} says that complex conjugate correlators
equal correlators of conjugated operators in inverted order:
\begin{equation}
	\overline{\langle \varphi \nobracket_1 (x_1) \ldots \varphi_n (x_n)
		\rangle} = \langle \varphi_n^{\dagger} (x_n) \ldots \varphi_1^{\dagger}
	(x_1) \rangle . \label{Hermiticity}
\end{equation}
This would be true of course if $\varphi$'s were operators acting on a Hilbert
space, with $\varphi^{\dagger}$'s their adjoints. In the present framework
without Hilbert space it is imposed as an axiom. This axiom implies in
particular {\eqref{R*}}, i.e.\ that $\mathcal{O}^{\dagger}_i$ transforms in
the conjugate irrep $\overline{\rho_i}$.

The last Wightman axiom, \tmtextbf{positivity}, is most conveniently written
down using the language of states. One considers basic ket states $| \psi (f,
\varphi_1, \ldots, \varphi_n) \rangle$, associated with $n$ local operators
$\varphi_1, \ldots, \varphi_n \in V$ and a complex Schwartz test function of
$n$ variables $f$. One defines the inner product on basic ket states by
\begin{eqnarray}\label{Wightman:inner}
	\langle \psi (g, \chi_1, \ldots, \chi_m) | \psi (f, \varphi_1, \ldots,
	\varphi_n) \rangle & \assign & \int d x\, d y\, \overline{g (x_1, \ldots,
		x_m)} f (y_1, \ldots, y_n) \\
	&  & \qquad \times \langle \chi^{\dagger}_m (x_m) \ldots
	\chi_1^{\dagger} (x_1) \varphi_1 (y_1) \ldots \varphi_n (y_n) \rangle .
	\nonumber
\end{eqnarray}
The vector space of ket states $\mathcal{H}_0$ consists of \tmtextit{finite}
linear combinations $| \Psi \rangle$ of basic ket states, with the inner
product extended to it by (anti)linearity. \tmtextbf{Positivity} then says
that the so defined inner product is positive semidefinite:
\begin{equation}
	\langle \Psi | \Psi \rangle \geqslant 0 \quad \forall | \Psi \rangle \in
	\mathcal{H}_0 . \label{Wightman:positivity}
\end{equation}
\begin{remark}
	\label{rem:states}A comment is in order concerning the meaning of these
	states. They may be seen as just a convenient notation, since Eq.
	{\eqref{Wightman:positivity}} can be rewritten without ever using the word
	``state'' (see {\cite{Streater:1989vi}}, Eq.\ (3-35)). But they are more than
	that: \ the vector space of states $\mathcal{H}_0$ is ``almost'' the Hilbert
	space $\mathcal{H}$ of our QFT. The only difference between $\mathcal{H}_0$
	and $\mathcal{H}$ is that $\mathcal{H}_0$ is not complete and may contain
	some states of zero norm. However, since $\mathcal{H}_0$ has a positive
	semidefinite inner product, as expressed by Eq.
	{\eqref{Wightman:positivity}}, we can obtain from it a Hilbert space
	$\mathcal{H}$ via a standard procedure of completion and
	modding out by states of zero norm. This is the first step of the Wightman
	reconstruction theorem {\cite{Streater:1989vi}}, and the resulting Hilbert
	space $\mathcal{H}$ turns out to be (possibly a superselection sector of)
	\tmtextit{the} Hilbert space of the QFT, on which fields can then be
	realized as operator valued distributions.
\end{remark}

\begin{remark}
	\label{hermfrompos}Although we included hermiticity as a separate axiom
	because of its suggestive form, it can be derived from positivity,
	considering the states of the form $| \Psi \rangle = | \psi (f_0, 1) \rangle
	\nobracket + | \psi (f, \varphi_1, \ldots, \varphi_n) \rangle$ where $f_0
	\in \mathbb{C}$ and $1$ is the unit operator.
\end{remark}

\begin{remark}
	Another interesting positivity property of Wightman functions is called Rindler Reflection positivity,
	or Wedge Reflection positivity~\cite{Casini:2010bf}. A restricted version of this property (with wedge-ordered points)
	can be derived from Wightman axioms, while a stronger version (no wedge-ordering)
	follows from Tomita-Takesaki theory which relies on Haag-Kastler axioms~\cite{Casini:2010bf}. In CFT context this property has been discussed, e.g.,
	in~\cite{Hartman:2016lgu}. We will not discuss these properties in this part of the thesis. However, it would be interesting to
	see whether the stronger form of Rindler positivity (including distributional information) can be derived from CFT axioms without the appeal to Tomita-Takesaki theory (the weaker version following from our results on Wightman axioms and~\cite{Casini:2010bf}).{We believe this can be done, and it could be a nice exercise for someone wishing to master our techniques.} 
\end{remark}

\section{Osterwalder-Schrader axioms}\label{OSaxioms}

We next describe a version of the Osterwalder and Schrader axioms
{\cite{osterwalder1973,osterwalder1975}} of Euclidean unitary QFT (see the end
of the section about the relation to the original OS axioms). The setup is
similar to Wightman axioms with $\tmop{SO} (d)$ replacing $\tmop{SO} (1, d -
1)$. We consider a basis of local bosonic operators $\mathcal{O}_i$
transforming\footnote{In the sense of Eq.\ {\eqref{transform}} where now $g \in
	\tmop{SO} (d)$.} in $\tmop{SO} (d)$ irreps $\rho_i$, counting derivatives as
independent operators. Finite linear combinations of their components span a
complex vector space $\mathcal{C}$ of local operators. The axioms specify
properties of translation and $\tmop{SO} (d)$ invariant $n$-point correlators {(often called Schwinger functions)}
\begin{equation}
	\langle \varphi_1 (x_1) \ldots \varphi_n (x_n) \rangle, \quad \varphi_i \in
	\mathcal{C}, \quad x_i \in \mathbb{R}^d \label{nptEucl} .
\end{equation}
These correlators are defined away from coincident points (i.e.\ whenever $x_i
\neq x_j$ for each $i, j$). We will assume
that\footnote{\label{real-anal}Recall that a $C^{\infty}$ function of $m$ real
	variables is called real-analytic in a domain $D \subset \mathbb{R}^m$ if it
	has a convergent Taylor series expansion in a small ball around every point of
	this domain. Equivalently, such a function has an analytic extension to a
	small open neighborhood of this domain inside $\mathbb{C}^m$.}
\begin{equation}
	\text{correlators are real-analytic,} \label{real-anal1}
\end{equation}
and grow not faster than some power when some points approach each other or go
to infinity, i.e.
\begin{equation}
	| \langle \varphi_1 (x_1) \ldots \varphi_n (x_n) \rangle | \leqslant C
	\left( 1 + \max_{i \neq j} \left( \frac{1}{| x_i - x_j |}, | x_i - x_j |
	\right) \right)^p \label{OSmod}
\end{equation}
with some correlator-dependent positive constants $C, p$. Unlike Wightman
axioms, OS axioms do not bother what happens \tmtextit{precisely} at
coincident points (not even in the sense of distributions).

As we are limiting to the bosonic case, correlators remain invariant when
operators are permuted:\footnote{In particular one can sort all operators so
	that the Euclidean time coordinates are ordered $x^0_1 \geqslant x_2^0
	\geqslant \cdots \geqslant x_n^0$, and Euclidean correlator for any other
	ordering can be obtained by trivially permuting field labels.}
\begin{equation}
	\langle \varphi_1 (x_1) \ldots \varphi_n (x_n) \rangle = \langle
	\varphi_{\pi (1)} (x_{\pi (1)}) \ldots \varphi_{\pi (n)} (x_{\pi (n)})
	\rangle . \label{perminv}
\end{equation}
To formulate the Euclidean version of hermiticity and positivity, we will need
some simple facts about $\tmop{SO} (d)$ representations. Abstractly, for any
irrep $\rho$ acting $T^{\alpha} \rightarrow \rho (g)^{\alpha}_{\ \beta}
T^{\beta}$, the conjugate representation $\bar{\rho}$ acts with complex
conjugate matrices $\overline{\rho (g)^{\alpha}_{\ \beta}}$. Since
$\tmop{SO} (d)$ is compact, we have $\bar{\rho} \simeq \rho^{\ast}$, the dual
representation. The $\tmop{SO} (d)$ irreps $\rho$ are again tensors
$T^{\mu_1 \ldots \mu_l}$ like in {\eqref{rhoi}}, of in general mixed symmetry,
and with (anti)-chirality constraints if having $d / 2$ antisymmetric
indices in even $d$. All of them are real, except for chiral
representations in $d = 2 \tmop{mod} 4$ which are complex-conjugate to the
anti-chiral ones.\footnote{This well-known shift from $d = 0 \tmop{mod} 4$
	for $\tmop{SO} (1, d - 1)$ is induced by raising the indices of the
	$\varepsilon$-tensor. E.g.\ $\varepsilon^{01} \varepsilon_{10} = - 1$ for
	$\tmop{SO} (2)$, while it is 1 for $\tmop{SO} (1, 1)$.}

We will also need the reflected representation $\rho^R$ with matrices $\rho^R
(g) = \rho (g^R)$, where $g \rightarrow g^R = \Theta g \Theta,$ $\Theta =
\tmop{diag} (- 1, 1, \ldots, 1),$ is an automorphism of $\tmop{SO} (d)$. For
tensor representations, we can consider the map
\begin{equation}
	T^{\mu_1 \ldots \mu_l} \rightarrow {\Theta^{\mu_1}_{\ \nu_1} \ldots
		\Theta^{\mu_l}_{\ \nu_l}} T^{\nu_1 \ldots \nu_l},
	\label{intertwiner:theta}
\end{equation}
which preserves rank and mixed symmetry properties. It also maps chiral to
anti-chiral tensors in any even $d$. Whenever the representation space is
preserved, this map serves as an intertwiner between $\rho^R$ and $\rho$. This
means that $\rho^R \simeq \rho$ for all tensor representations without
chirality constraints, while this operation interchanges chiral and
antichiral irreps in any even $d$.\footnote{In odd $d$, $\Theta$ is a
	product of $- 1$ and an $\tmop{SO} (d)$ matrix, so that $g \rightarrow g^R$ is
	an inner automorphism. This provides another argument why $\rho^R \simeq \rho$
	for all irreducible $\tmop{SO} (d)$ representations in odd $d$.}

Applying both conjugation and reflection we get the conjugate reflected
representation $\bar{\rho}^R$ (isomorphic to dual reflected). From the above
it follows that $\bar{\rho}^R \simeq \rho$ for all $\tmop{SO} (d)$ irrreps,
except for (anti-)chiral tensors in $d = 0 \tmop{mod} 4$ which are
interchanged.

Just as for Wightman axioms, we will need a conjugation operation $\dagger :
\mathcal{C} \rightarrow \mathcal{C}$ on the vector space of local operators,
which is involutive, anti-linear, and associates to each independent component
$\mathcal{O}_i^{\alpha}  (\alpha = 1 \ldots \dim \rho_i)$ a conjugate operator
$(\mathcal{O}_i^{\dagger})^{\alpha} \assign
(\mathcal{O}_i^{\alpha})^{\dagger}$. Then the \tmtextbf{hermiticity} axiom
takes the form\footnote{Although we write the operators in the r.h.s.\ in the
	inverted order like in {\eqref{Hermiticity}}, permutation invariance renders
	this detail unimportant for the OS axioms.}
\begin{equation}
	\overline{\langle \varphi \nobracket_1 (x_1) \ldots \varphi_n (x_n)
		\rangle} = \langle \varphi_n^{\dagger} (x^{\theta}_n) \ldots
	\varphi_1^{\dagger} (x^{\theta}_1) \rangle, \label{HermiticityOS}
\end{equation}
similar to the Minkowski counterpart {\eqref{Hermiticity}} but with an
important difference that the operators in the r.h.s.\ are put at reflected
positions
\begin{equation}
	x^{\theta} \assign \Theta x.
\end{equation}
This change has a consequence that $\mathcal{O}^{\dagger}_i$ transforms in the
conjugate \tmtextit{reflected} representation $\overline{\rho_i}^R$,
explaining why we introduced this concept in the first place.\footnote{Indeed
	we have ${\langle (\mathcal{O}^{\dagger}_i)^{\alpha} (x) \ldots \rangle =
		\overline{\langle \mathcal{O}^{\alpha}_i (x^{\theta}) \ldots \rangle}
		=\overline{\rho (g)^{\alpha}_{\ \beta}} \langle
		\mathcal{O}^{\beta}_i (g^{- 1} x^{\theta}) \ldots \rangle = \overline{\rho
			(g)^{\alpha}_{\ \beta}} \langle (\mathcal{O}^{\dagger}_i)^{\beta}
		((g^R)^{- 1} x) \ldots \rangle}$.} For self-conjugate-reflected
representations we may choose a basis such that
\begin{equation}
	(\mathcal{O}^{\dagger}_i)^{(\mu)} = \Theta^{(\mu)}_{(\nu)}
	\mathcal{O}_i^{(\nu)}, \label{realOS}
\end{equation}
where $\Theta^{(\mu)}_{(\nu)} : = {\Theta^{\mu_1}_{\ \nu_1} \ldots
	\Theta^{\mu_l}_{\ \nu_l}}$ is the intertwiner {\eqref{intertwiner:theta}}.

To write positivity, basic ket states $| \psi (f, \varphi_1, \ldots,
\varphi_n) \rangle$ are associated with $n$ local operators $\varphi_1,
\ldots, \varphi_n \in \mathcal{C}$ and a complex compactly supported
Schwartz test function of $n$ variables $f (x_1, \ldots, x_n)$ which
vanishes unless all points are in the lower half space and have time variables
ordered: $0 > x_1^0 > x_2^0 > \cdots > x_n^0$. These support requirements
were absent in the Wightman case. The inner product on the basic ket states is
defined by
\begin{eqnarray}
	\langle \psi (g, \chi_1, \ldots, \chi_m) | \psi (f, \varphi_1, \ldots,
	\varphi_n) \rangle & \assign & \int d x\, d y\, \overline{g (y_1^{\theta},
		\ldots, y_m^{\theta})} f (x_1, \ldots, x_n) \nonumber\\
	&  & \times \langle \chi^{\dagger}_m (y_m) \ldots \chi_1^{\dagger} (y_1)
	\varphi_1 (x_1) \ldots \varphi_n (x_n) \rangle,  \label{inner:OS}
\end{eqnarray}
and is extended by (anti)linearity to the vector space
$\mathcal{H}_0^{\tmop{OS}}$ of \tmtextit{finite} linear combinations $| \Psi
\rangle$ of basic ket states. In this notation, \tmtextbf{positivity} takes
the same form as {\eqref{Wightman:positivity}}, i.e.\ that the so defined inner
product must be positive semidefinite:
\begin{equation}
	\langle \Psi | \Psi \rangle \geqslant 0 \quad \forall | \Psi \rangle \in
	\mathcal{H}_0^{\tmop{OS}} . \label{OS:positivity}
\end{equation}
This is referred to as ``OS reflection positivity'' because of the reflected
$g$ arguments in {\eqref{inner:OS}}, differently from the Wightman case.
Because of this reflection and the above test function support requirements,
all operators in {\eqref{inner:OS}} sit at separated positions. This is one
reason why the OS axioms involve ordinary functions, without worrying about
coincident points. In contrast, Wightman positivity integrates operator
insertions over coincident points and makes sense only for distributions.

Just as in the Wightman case (Remark \ref{rem:states}), we can complete the
vector space $\mathcal{H}_0^{\tmop{OS}}$, mod out by states of zero norm, and
obtain a Hilbert space $\mathcal{H}^{\tmop{OS}}$ of the Euclidean theory.

Although we included hermiticity as an independent axiom, it can be derived
from positivity, just as in Remark \ref{hermfrompos} in the Wightman case.

One simple consequence of OS reflection positivity is pointwise positivity of
$2 n$-point functions at reflection invariant configurations of points:
\begin{equation}
	\langle \varphi^{\dagger}_n (x^{\theta}_n) \ldots \varphi_1^{\dagger}
	(x^{\theta}_1) \varphi_1 (x_1) \ldots \varphi_n (x_n) \rangle \geqslant 0
	\label{OSnaive1}
\end{equation}
for any $x_1, \ldots, x_n$ in the lower half space.\footnote{For tensor
	operator in self-conjugate-reflected representations, choosing the real basis
	{\eqref{realOS}}, this becomes $\langle \ldots \Theta^{(\mu)}_{(\nu)}
	\mathcal{O}^{(\nu)} (x^{\theta}) \mathcal{O}^{(\mu)} (x) \ldots
	\rangle \geqslant 0$ (no sum on $\mu$), i.e.\ tensor indices are also
	reflected.} This follows from {\eqref{OS:positivity}} by taking $| \Psi
\rangle = | \psi (f, \varphi_1, \ldots, \varphi_n) \rangle$ and localizing $f$
near one configuration of points. In general, imposing {\eqref{OSnaive1}} for
all $\varphi$'s and $x$'s would be clearly weaker than full OS reflection
positivity. E.g.\ {\eqref{OS:positivity}}, but not {\eqref{OSnaive1}}, can be
used to bound 3-point functions in terms of 2- and 4-point functions, or
non-reflection-invariant 4-point functions by reflection-invariant ones.
However for CFTs we will see below that OS reflection positivity can be
reduced to a form of {\eqref{OSnaive1}} for 2-point functions plus a form of
{\eqref{HermiticityOS}} for 3-point functions.

Finally, the OS \tmtextbf{clustering} asserts that
\begin{eqnarray}
	& \lim_{\lambda \rightarrow \infty} \int d x\, d y\, \overline{g
		(y_1^{\theta}, \ldots, y_m^{\theta})} f (x_1, \ldots, x_n) \langle
	\chi^{\dagger}_m (y_m) \ldots \chi_1^{\dagger} (y_1) \varphi_1 (x_1 +
	\lambda a) \ldots \varphi_n (x_n + \lambda a) \rangle &  \nonumber\\
	& = \int d x\, d y\, \overline{g (y_1^{\theta}, \ldots, y_m^{\theta})} f
	(x_1, \ldots, x_n) \langle \chi^{\dagger}_m (y_m) \ldots \chi_1^{\dagger}
	\nobracket (y_1) \rangle \langle \nobracket \varphi_1 (x_1) \ldots \varphi_n
	(x_n) \rangle &  \label{clusterint}
\end{eqnarray}
for any Schwartz test functions $f (x_1, \ldots, x_n)$ and $g (y_1, \ldots,
y_m)$ supported for $0 > x_1^0 > x_2^0 > \ldots > x_n^0$ and $0 > y_1^0 >
y_2^0 > \ldots > y_n^0$, for any local fields $\varphi_1, \ldots, \varphi_n$
and $\chi_1, \ldots, \chi_m$, and for any $a \in \mathbb{R}^d$ which is
parallel to the $x^0$ plane ($a^0 = 0$). The latter requirement is somewhat
analogous to having the Wightman cluster property {\eqref{Wightman:cluster}}
to be satisfied only for spacelike $a$.\footnote{This is axiom E4 in
	{\cite{osterwalder1973}}. Ref.\ {\cite{osterwalder1973}} also mentions a
	stronger axiom E4', but we will be content here with checking the easier axiom
	E4.}

Note that the Minkowski operators can be mapped to Euclidean operators. In
particular any $\tmop{SO} (1, d - 1)$ irrep can be mapped to an $\tmop{SO}
(d)$ irrep. This map of irreps originates from the map between the two Lie
algebras which have the same complexification. It can then be shown that a
pair of conjugate $\tmop{SO} (1, d - 1)$ irreps is mapped to a pair of
$\tmop{SO} (d)$ irreps which are conjugate-reflected to each other. This gives
another rationale for the appearance of reflected irreps in the OS axioms.

\begin{remark}
	\label{OSnewVSold}The stated version of OS axioms includes the assumption of
	real analyticity {\eqref{real-anal1}} and the bound {\eqref{OSmod}}. These
	assumptions are natural from physics perspective; they also easily follow
	from Wightman axioms. The original OS axioms did not include
	{\eqref{real-anal1}} nor {\eqref{OSmod}}, but included instead a differently
	stated assumption:
	\begin{equation}
		\text{correlators are distributions on $^0 \mathcal{S}$,} \label{OSorig}
	\end{equation}
	where $^0 \mathcal{S}$ is the space of Schwartz test function vanishing at
	coincident points with all their \ derivatives.
	
	We would like to discuss here the relation between
	{\eqref{real-anal1}}+{\eqref{OSmod}} and {\eqref{OSorig}}. In one direction
	this is easy: clearly {\eqref{OSmod}} implies {\eqref{OSorig}}. In the other
	direction it can be shown that {\eqref{OSorig}} and other OS axioms (in
	particular OS positivity and rotation invariance) imply real analyticity
	{\eqref{real-anal1}}. This is a result of \
	{\cite{osterwalder1973,osterwalder1975}} and {\cite{Glaser1974}}. It is also
	possible to derive {\eqref{OSmod}} from {\eqref{OSorig}} and other OS axioms
	{\cite{osterwalder1975}}. These issues will be reviewed further in chapter \ref{OS}.
\end{remark}

\section{Euclidean CFT axioms}\label{ECFTax}

Wightman and OS axioms stated in the previous two sections are standard. We
took care to present them for general operator representations and in general
$d$. We will now present axioms for Euclidean unitary CFT. Just as OS axioms,
these concern correlators in Euclidean signature, but there is an extra
assumption of conformal invariance. Another feature of the CFT axioms is that
assumptions are imposed on simple building blocks (2- and 3-point functions)
from which more complicated correlators can be constructed. Properties of
these complicated correlators then follow. The point of this part is how one
can recover OS axioms and (after Wick rotation) Wightman axioms in this setup.

A Euclidean unitary CFT in $\mathbb{R}^d$ ($d \geqslant 2$) deals with local
\tmtextit{primary} operators $\mathcal{O}_i (x)$ and with their $n$-point
correlation functions $\langle \mathcal{O}_{i_1} (x_1) \ldots
\mathcal{O}_{i_n} (x_n) \rangle$. Correlators are real-analytic functions
defined away from coincident points, which are permutation-invariant as in
{\eqref{perminv}}. Each primary is characterized by its scaling dimension
$\Delta_i$ and is an $\tmop{SO} (d)$ tensor transforming in an irreducible
representation $\rho_i$.\footnote{Operators can also be grouped into
	multiplets of the global symmetry group $G$ which a CFT might have, but we
	will not discuss global symmetry here. For simplicity we will only consider
	bosonic operators. More generally one should allow fermionic operators and
	spinor representations.} The scaling dimensions are real and nonnegative, with
the unit operator having dimension zero. The set of scaling dimensions
(``spectrum'') is assumed to be discrete, by which we mean that there are finitely
many $\Delta_i$'s in any finite interval $[a, b] \subset
\mathbb{R}$.\footnote{There exist 2d unitary CFTs, such as the Liouville
	theory, with a continuous spectrum of scaling dimensions. In this case axioms
	need to be modified. All known unitary CFTs in $d \geqslant 3$ have a discrete
	spectrum.}

The set of all local operators of a CFT consists of primaries $\mathcal{O}_i
(x)$ and their space-time derivatives $\partial_{\mu_1} \ldots
\partial_{\mu_n} \mathcal{O}_i (x)$, often refered to as descendants. The
correlation functions of the descendant operators are simply the derivatives
of the correlation functions of primary operators. They are well-defined since
the correlators of primaries are assumed to be real-analytic.

Parameters $\Delta_i$ and $\rho_i$ determine transformation properties of
$\mathcal{O}_i (x)$ under the conformal group $\tmop{SO} (d + 1, 1)$, and
correlators remain invariant under these standard transformations which we
will not write down. These constraints determine the functional form of
1,2,3-point functions. In particular, the unit operator is the only one with a
nonzero 1-point function. {See, e.g.,~\cite{Poland:2018epd} for a review of these
	facts.} 

{
	An important fact that follows from the conformal invariance of correlation functions
	is that one is allowed to insert an operator at spatial infinity. This is defined
	as
	\be\label{eq:inftydefn}
	\<\cO_i(\oo)\cdots\>\equiv\lim_{L\to +\oo}L^{2\De_i}\<\cO_i(L\hat e_0)\cdots\>.
	\ee
	To see that this limit exists one can use a conformal map that takes
	$\oo$ to a finite point and moves no other operators to infinity. After applying this 
	map the limit~\eqref{eq:inftydefn} turns into a limit in 
	which all points approach finite values. We conclude that~\eqref{eq:inftydefn}
	exists, and is then of course independent of the concrete conformal map that we chose.
	In the definition~\eqref{eq:inftydefn} we have chosen a particular direction ($\hat e_0$) for
	the limit. Using conformal symmetry it is easy to show that~\eqref{eq:inftydefn} is independent
	of this direction, up to a rotation on the indices of $\cO_i$. In what follows we will
	always allow for Euclidean CFT correlators to have one of the operators to be at $\oo$.
}

For every primary $\mathcal{O}_i$ there is a unique conjugate primary
$\mathcal{O}^{\dagger}_i$ (where $\dagger$ is involutive) such that the
2-point function $\langle \mathcal{O}_i^{\dagger} \mathcal{O}_i \rangle$ does
not vanish. The $\mathcal{O}_i$ and $\mathcal{O}^{\dagger}_i$ have equal
scaling dimensions, and transform in the conjugate-reflected irreps. Recall
that in Sec.\ \ref{OSaxioms} we saw that most $\tmop{SO} (d)$ irreps are
self-conjugate-reflected, $\rho_i \simeq \bar{\rho}_i^R$, the only exception
being (anti-)chiral tensors in $d = 0 \tmop{mod} 4$ which are exchanged by
this operation. For operators in self-conjugate-reflected irreps we may choose
operator basis such that Eq.\ {\eqref{realOS}} holds, which we copy here:
\begin{equation}
	(\mathcal{O}^{\dagger}_i)^{(\mu)} = \Theta^{(\mu)}_{(\nu)}
	\mathcal{O}_i^{(\nu)} . \label{realOS1}
\end{equation}
The functional form of the $\langle \mathcal{O}_i^{\dagger} (x) \mathcal{O}_i
(y) \rangle$ 2-point function is fixed by conformal symmetry:
\begin{equation}
	\langle (\mathcal{O}^{\dagger}_i)^{(\mu)} (x) \mathcal{O}_i^{(\nu)} (y)
	\rangle = \cN_i I^{(\mu), (\nu)} (x - y), \label{2-pointfixed}
\end{equation}
where $(\mu), (\nu)$ are collections of tensor indices (of equal length),
$I^{(\mu), (\nu)} (x - y)$ is a tensor function depending only on $\Delta_i,
\rho_i$, {and $\cN_i$ is a constant whose phase is determined by the positivity condition discussed below. The remaining freedom to rescale $\cN_i$ by a positive real number is fixed in some arbitrary unimportant way, e.g.\ so that some component of the 2-point function
	is one at unit separation.}

Positivity is imposed in Euclidean CFT axioms only on 2-point functions. We
write it again using the language of states. Basic ket states are $|
\partial^{(\beta)} \mathcal{O}_i^{(\nu)} \rangle \nobracket$ where
$\mathcal{O}_i^{(\nu)}$ is a primary component and $\partial^{(\beta)}$ an
arbitrary derivative. The inner product is defined as
\begin{equation}
	\langle \partial^{(\alpha)} \mathcal{O}_i^{(\mu)} | \partial^{(\beta)}
	\mathcal{O}_i^{(\nu)} \rangle = \Theta^{(\alpha)}_{(\alpha')} \langle
	\partial^{(\alpha')} (\mathcal{O}^{\dagger}_i) \nobracket^{(\mu)} (x_N)
	\partial^{(\beta)} \mathcal{O}_i^{(\nu)} (x_S) \rangle, \label{innerCFT}
\end{equation}
i.e.\ as the value of the shown 2-point function inserting the operators at
$x_S = (- 1, 0, \ldots, 0)$ and $x_N = (1, 0, \ldots, 0) = (x_S)^{\theta}$ \
(where N,S stands for north, south). For ket states with $i \neq j$ the inner
product vanishes since the 2-point function is zero. This inner product is
extended by (anti)linearity to the vector space $\mathcal{H}_0^{\tmop{CFT}}$
of finite complex linear combinations of basic ket states. In this language,
Euclidean CFT \tmtextbf{positivity} reads exactly as the Wightman and OS
positivity: $\langle \Psi | \Psi \rangle \geqslant 0$ for all states of this
restricted form. More prosaically, this can also be stated that the infinite
matrices $M^{(\alpha) (\mu), (\beta) (\nu)}_i$ built out of 2-point functions
in the r.h.s.\ of {\eqref{innerCFT}} are all positive semidefinite when
restricted to finite subspaces. 

{CFT positivity can be analyzed primary by primary, and it depends only on
	the primary 2-point function, Eq.\ {\eqref{2-pointfixed}} which determines the full matrix $M^{(\alpha) (\mu), (\beta)
		(\nu)}_i$. Clearly, only one phase of the normalization constant $\cN_i$ in Eq.\ {\eqref{2-pointfixed}} can give rise to a positive definite matrix, so that phase is uniquely fixed. Once the phase of $\cN_i$ is fixed, positivity for a given primary depends only on its $\Delta$, $\rho$.} It then can be shown that CFT positivity holds if and only if every $\Delta$, in addition to being real and non-negative, lies above a certain
minimal $\rho$-dependent value (``unitarity bound''):
\begin{equation}
	\Delta \geqslant \Delta_{\min} (\rho) .
\end{equation}
These unitarity bounds are documented in the literature, e.g.\ we have
$\Delta_{\min} = d / 2 - 1$ for scalars, and $d + \ell - 2$ for spin-$\ell$,
$\ell \geqslant 1$. For arbitrary $\tmop{SO} (d)$ representations see
{\cite{Minwalla:1997ka}}.\footnote{We chose to express CFT positivity
	inserting operators at the points $(\pm 1, 0, \ldots, 0)$ which corresponds to
	the N-S quantization (see {\cite{EPFL}}) and will facilitate the comparison
	with the Osterwalder-Schrader reflection positivity. Equivalently, one could
	go via a conformal transformation to the more familiar radial quantization corresponding to inserting the
	operators at $0$ and $\infty$. CFT positivity is then equivalent to radial
	quantization states having positive norm on every level, which is how the
	unitarity bounds are usually worked out in Euclidean CFTs
	{\cite{Minwalla:1997ka}}. In mathematical language, this latter condition
	corresponds to having a positive-definite Shapovalov form on the parabolic
	Verma module. Recent work {\cite{Yamazaki:2016vqi,Penedones:2015aga}}
	explained how the determinant formulas by Jantzen
	{\cite{jantzen_kontravariante_1977}} provide a rigorous justification of the
	Euclidean unitarity bounds (both in the necessary and sufficient directions).}

For future uses, we wish to define the CFT Hilbert space
$\mathcal{H}^{\tmop{CFT}}$ via completion of $\mathcal{H}_0^{\tmop{CFT}}$,
after modding out by zero norm states (for operators saturating the unitarity
bounds, some descendants have zero norm). This can be done abstractly, or
explicitly using a basis as we now describe. Throwing out zero-norm
descendants, the remaining states can be organized choosing an orthonormal
basis. We may choose such a basis independently among descendants of each
primary, and then combine all these bases, e.g.\ in the order of non-decreasing
scaling dimensions. \ The elements of $\mathcal{H}^{\tmop{CFT}}$ are then
formal linear combinations $\sum_n c_n | n \rangle,$where $| n \rangle$ are
orthonormal basis elements, and $c_n$ is an arbitrary complex $\ell_2$
sequence. The norm on $\mathcal{H}^{\tmop{CFT}}$ is the $\ell_2$ norm of
the sequence $c_n$. Restricting to sequences $c_n$ which have only a finite
number of nonzero elements, we get elements of $\mathcal{H}_0^{\tmop{CFT}}$
(modulo the zero-norm states).

Let us continue with the axioms. CFT \tmtextbf{hermiticity} condition is
imposed only on the 2-point and 3-point functions, namely:
\begin{equation}
	\langle (\mathcal{O}^{\dagger}_i)^{(\mu)} (x_1) \mathcal{O}_i^{(\nu)}
	(x_2) \rangle =  \overline{\langle \mathcal{O}_i^{(\mu)} (x_1^{\theta})
		(\mathcal{O}_i^{\dagger})^{(\nu)} (x_2^{\theta}) \rangle}, 
	\label{hermicity2-point}
\end{equation}
{which is also a consequence CFT
	positivity and in particular fixed the phase of $\cN_i$ up to a sign},\footnote{For the special case $x_1 = x_N$, $x_2 = x_S$, Eq.\ {\eqref{hermicity2-point}} is nothing but hermiticity of the matrix $M_i^{(\mu),
		(\nu)}$, a consequence of positive-semidefiniteness. The general case reduces
	to the special one mapping $x_1, x_2$, to $x_N$, $x_S$ by a conformal
	transformation (both sides of {\eqref{hermicity2-point}} have the same conformal
	transformation properties).} and
\begin{eqnarray}
	\langle \mathcal{O}_i^{(\mu)} (x_1) \mathcal{O}_j^{(\nu)} (x_2)
	\mathcal{O}_k^{(\lambda)} (x_3) \rangle & = & \overline{\langle
		(\mathcal{O}_i^{\dagger})^{(\mu)} (x^{\theta}_1)
		(\mathcal{O}_j^{\dagger})^{(\nu)} (x^{\theta}_2)
		(\mathcal{O}^{\dagger}_k)^{(\lambda)} (x^{\theta}_3) \rangle} . 
	\label{CFThermiticity}
\end{eqnarray}
for any 3 primaries $\mathcal{O}_i$, $\mathcal{O}_j$,
$\mathcal{O}_k$.\footnote{Note that since $\dagger$ is involutive, this
	covers the case when an operator and its conjugate are interchanged between
	the two sides of this equation.} Similarly to the 2-point function case, this
condition can be simplified using the conformally invariant tensor structures,
with an important difference that the normalization of operators has already
been fixed. Conformal invariance constrains the 3-point functions to take the
form
\begin{equation}
	\langle \mathcal{O}_i^{(\mu)} (x_1) \mathcal{O}_j^{(\nu)} (x_2)
	\mathcal{O}_k^{(\lambda)} (x_3) \rangle = \sum_{a = 1}^{N_{i j k}} f^a_{i j
		k}  \langle \mathcal{O}_i^{(\mu)} (x_1) \mathcal{O}_j^{(\nu)} (x_2)
	\mathcal{O}_k^{(\lambda)} (x_3) \rangle_a, \label{3-pointGeneral}
\end{equation}
where $\langle \mathcal{O}_i^{(\mu)} (x_1) \mathcal{O}_j^{(\nu)} (x_2)
\mathcal{O}_k^{(\lambda)} (x_3) \rangle_a$ span the finite-dimensional space
(of dimension $N_{i j k}$) of solutions of conformal invariance constraints on
the 3-point functions of the operators with given $\Delta_s, \rho_s$ ($s = i,
j, k$). On the other hand the coefficients $f_{i j k}^a \in \mathbb{C}$ are
not fixed by conformal symmetry (no sum on $i, j, k$ in the r.h.s.\ of
{\eqref{3-pointGeneral}}). We often refer to $f^a_{i j k}$ as the ``OPE
coefficients.'' It is always possible to choose the basis structures $\langle
\mathcal{O}_i^{(\mu)} (x_1) \mathcal{O}_j^{(\nu)} (x_2)
\mathcal{O}_k^{(\lambda)} (x_3) \rangle_a$ to satisfy the hermiticity
constraint {\eqref{CFThermiticity}} individually, in which case the OPE
coefficients must satisfy
\begin{equation}
	(f^a_{i j k})^{\ast} = f_{\bar{i}  \bar{j}  \bar{k} }^a,
	\label{OPEcoeff:conjugation}
\end{equation}
where the barred indices refer to the conjugate operators
$\mathcal{O}_i^{\dagger}, \mathcal{O}_j^{\dagger}, \mathcal{O}^{\dagger}_k .$
In particular, when all three operators are self-conjugate-reflected i.e.\ satisfy {\eqref{realOS1}}, the OPE coefficients $f^a_{i j k}$ must be real.

Finally, unitary CFTs enjoy a \tmtextbf{convergent operator product
	expansion} (OPE). This means that any correlation function\footnote{Importantly, we allow here for one of the operators to
	be inserted at spatial infinity.} satisfies
\begin{equation}
	\langle \mathcal{O}_i^{(\mu)} (x_1) \mathcal{O}_j^{(\nu)} (x_2)
	\mathcal{O}_m^{(\rho)} (x_3) \cdots \rangle = \sum_k \sum_{a = 1}^{N_{i j
			k}} f_{i j k}^a C_{a, (\lambda)}^{(\mu) (\nu)} (x_1, x_2, x_0, \partial_0)
	\langle (\mathcal{O}^{\dagger}_k)^{(\lambda)} (x_0)
	\mathcal{O}_m^{(\rho)} (x_3) \cdots \rangle, \label{OPEgeneral}
\end{equation}
where the first sum runs over all primary operators $\mathcal{O}_k$ in the
theory, and $C_{a, (\lambda)}^{(\mu) (\nu)} (x_1, x_2, x_{0,} \partial_0)$ is
a formal sum of the form
\begin{equation}
	C_{a, (\lambda)}^{(\mu) (\nu)} (x_1, x_2, x_0, \partial_0) =
	\sum_{\alpha} C_{a, (\lambda), \alpha}^{(\mu) (\nu)} (x_1, x_2, x_0)
	(\partial / \partial x_0)^{\alpha} . \label{Cexp}
\end{equation}
This differential operator is determined by conformal symmetry\footnote{There is an ambiguity when $\cO_k$ is in a short conformal
	representation (in unitary theories this happens only if $\cO_k$ is a conserved current or a free field). This subtlety will not play
	any role in this part.}
and depends only on
\ $\Delta_s, \rho_s$ ($s = i, j, k$). Here $(\partial / \partial
x)^{\alpha} = (\partial / \partial x^0)^{\alpha_0} \ldots (\partial /
\partial x^{d - 1})^{\alpha_{d - 1}}$ with $\alpha = (\alpha_0, \ldots,
\alpha_{d - 1}) \in (\mathbb{Z}_{\geqslant 0})^d$ a multiindex. Convergence of
OPE means that the sum {\eqref{OPEgeneral}}, with $C$ expanded as in
{\eqref{Cexp}}, converges whenever \
\begin{equation}
	x_1, x_2 \in B (x_0, R), \quad R = \min (| x_3 - x_0 |, | x_4 - x_0 |,
	\ldots) \label{ball}
\end{equation}
where $B (x_0, R)$ is an open ball centered at $x_0$ and of radius $R$. In
other words, OPE converges whenever $x_1$ and $x_2$ are the two closest
operator insertions to $x_0$ (in Euclidean distance). Convergence should be
understood carefully as follows. For each $\mathcal{O}^{\dagger}_k$ in the
r.h.s.\ of {\eqref{OPEgeneral}}, and for each $n \in \mathbb{Z}_+$, we perform
finite summation over $a, \lambda$, and all multiindices $\alpha$ with $|
\alpha | = n$. We are left then with the doubly infinite sum
\begin{equation}
	\sum_k \sum_{n = 0}^{\infty} g_{k, n} (\{ x_i \}) \label{doublyinf} .
\end{equation}
This doubly infinite sum has to converge absolutely for every $x_1, x_2$ as in
{\eqref{ball}}.\footnote{The requirement of absolute convergence can be
	somewhat relaxed, see Sec.\ \ref{OSfromCFT}.}

That the same coefficients $f_{i j k}^a$ appear in the OPE
{\eqref{OPEgeneral}} and the 3-point function {\eqref{3-pointGeneral}} follows
immediately by using the OPE inside the latter 3-point function.

Local Euclidean CFTs contain the conserved stress tensor operator $T_{\mu
	\nu}$ of dimension $d$, and in case of continuous global symmetry, conserved
global symmetry currents $J_{\mu}$ of dimensions $d - 1$. We will not discuss
here additional axioms involving 3-point functions and OPE coefficients of
these operators, related to their conservation and Ward identities, see e.g.\ {\cite{Dymarsky:2017xzb,Dymarsky:2017yzx}}.

\begin{remark}
	\label{ECFTax-comp}The just given Euclidean CFT axioms are more careful in
	what concerns reality constraints than the set of CFT rules gathered in the
	conformal bootstrap review {\cite{Poland:2018epd}}. They are also more economical: e.g.\ Ref.\ {\cite{Poland:2018epd}} assumed OS reflection positivity and clustering for
	$n$-point functions, which for us will be theorems to prove, not
	assumptions. Ref.\ {\cite{Poland:2018epd}} also included some constraints on the CFT
	data which emerge when considering CFT in Minkowski signature, most notably
	the Averaged Null Energy Condition (ANEC). In this part of the thesis we will establish
	all Wightman axioms for scalar Minkowski CFT 4-point functions from the
	Euclidean CFT axioms, but we will not discuss ANEC. A proof of ANEC
	{\cite{Faulkner:2016mzt}} has been given using the Haag-Kastler axioms for
	general QFT. CFT arguments have also been given in
	{\cite{Hartman:2016lgu,Kravchuk:2018htv}}, but they rely on some assumptions
	which have not been rigorously proven from axioms. It would be interesting
	to fill these gaps and establish ANEC as a theorem from Euclidean CFT
	axioms.\footnote{\label{caveats} {The argument in
			{\cite{Hartman:2016lgu}} uses an OPE asymptotic expansion on the second
			sheet, outside of the range of convergence of the OPE rigorously implied by
			the Euclidean CFT axioms (see App.\ \ref{Tom} where we review this method
			going back to {\cite{Hartman:2015lfa}}).} In {\cite{Kravchuk:2018htv}} ANEC
		is derived using manipulations with a generalization of the Lorentzian
		inversion formula of {\cite{Caron-Huot:2017vep}}, of which some have not
		been rigorously justified. For example, the derivation starts with the
		Euclidean inversion formula, which is readily justified from Harmonic
		analysis only for external scaling dimensions on principal series and
		square-integrable correlators, none of which is generically the case in the
		required setup.}
\end{remark}

\chapter{Euclidean CFT $\Rightarrow$ Osterwalder-Schrader}\label{CFTtoOS}
In this section we will discuss some simple consequences of CFT axioms and in
particular will show that they imply OS axioms for 4-point functions (the case of
higher-point functions is more subtle and is discussed in App.\ \ref{OShigher}). Here we will prove only the OS reflection-positivity and the
cluster property. The ``Euclidean temperedness'' bound {\eqref{OSmod}} will
follow from our arguments in the following sections, where we establish
power-law bounds on CFT correlation functions. The remaining OS axioms are a
subset of the CFT axioms.

\section{OS reflection positivity}\label{OSfromCFT}

{In this section we prove OS positivity
	for compactly-supported test functions. The extension to Schwartz functions is easy once 
	we establish~\eqref{OSmod} for CFT correlators, see Remark \ref{lastOS}.}

First let us slightly reformulate the OPE convergence property. Consider an
$n$-point correlation function with operators inserted at $x_1 \ldots x_n .$
Let $S$ be a hyperplane and $x_0$ be a point such that and $x_1, x_2, x_0$ are
on one side of $S$ while all the other points $x_i$, $i>2$, are on the other side.
Using a conformal transformation we can map $S$ to a sphere $S'$ so that $x_0$
is mapped to the center of $S'$ which we denote by $x_0'$. Let $x_i'$ denote
the positions of all the other points $x_i$ after this map. We can then use
the OPE {\eqref{OPEgeneral}} for the correlation function evaluated at $x_i'$,
\begin{equation}
	\langle \mathcal{O}_i^{(\mu)} (x'_1) \mathcal{O}_j^{(\nu)} (x'_2)
	\mathcal{O}_m^{(\rho)} (x'_3) \cdots \rangle = \sum_k \sum_{a = 1}^{N_{i j
			k}} f_{i j k}^a C_{a, (\lambda)}^{(\mu) (\nu)} (x'_1, x'_2, x_0',
	\partial_{0'}) \langle (\mathcal{O}^{\dagger}_k)^{(\lambda)} (x'_0)
	\mathcal{O}_m^{(\rho)} (x'_3) \cdots \rangle .
\end{equation}
Transforming this expansion term-by term to the original coordinates $x_i$ we
find the convergent expansion (with convergence understood in the same sense
as in the previous section\footnote{Careful reading of the argument below
	shows that, in the 4-point case, the requirement of absolute convergence of
	{\eqref{doublyinf}} could be replaced by a weaker requirement that we can find
	any subsequence of partial sums of {\eqref{doublyinf}} which approximates the
	correlator pointwise.})
\begin{equation}
	\langle \mathcal{O}_i^{(\mu)} (x_1) \mathcal{O}_j^{(\nu)} (x_2)
	\mathcal{O}_m^{(\rho)} (x_3) \cdots \rangle = \sum_k \sum_{a = 1}^{N_{i j
			k}} f_{i j k}^a \tilde{C}_{a, (\lambda)}^{(\mu) (\nu)} (x_1, x_2, x_0,
	\mathcal{D}_0) \langle (\mathcal{O}^{\dagger}_k)^{(\lambda)} (x_0)
	\mathcal{O}_m^{(\rho)} (x_3) \cdots \rangle . \label{OPEplanar}
\end{equation}
Here the differential operators $\mathcal{D}^{(\alpha)}$ are simply the
derivatives $\left( \frac{\partial}{\partial x'} \right)^{(\alpha)}$ expressed
in the original coordinates $x$, and conjugated by the conformal
transformation factor of $\mathcal{O}_k^{\dagger}$. The functions
$\tilde{C}_{a, (\lambda)}^{(\mu) (\nu)}$ are obtained from $C_{a,
	(\lambda)}^{(\mu) (\nu)}$ by our conformal transformation. The important point
is that truncation of $C_{a, (\lambda)}^{(\mu) (\nu)}$ in order of derivatives
$\partial^{(\alpha)}$ corresponds to truncation of $\tilde{C}_{a,
	(\lambda)}^{(\mu) (\nu)}$ in order of operators $\mathcal{D}^{(\alpha)}$.

We now specialize to $S$ being the $x^0 = 0$ plane, $x_0 = x_S = (- 1, 0,
\ldots),$ and take $S'$ to be the unit sphere with the center $x_0' = 0$. Then
the derivatives $\left( \frac{\partial}{\partial x'} \right)^{(\alpha)}
\mathcal{O}_k^{(\mu)} (x_0')$ are eigenstates of the standard dilatation
generator $D$ with eigenvalues $\Delta_k + | \alpha |$. Note that $D$ has two
fixed points: $x_0' = 0$ and infinity. Applying our conformal map, we find
that the derivatives $\mathcal{D}^{(\alpha)} \mathcal{O}_k^{(\mu)} (x_S)$ are
in turn eigenstates, with the same eigenvalues, of the conformal generator $D'
= (K^0 - P^0) / 2$ that preserves $x_S$ and $x_N = x_S^{\theta}$ (which is the
image of infinity under our conformal map) and acts by dilatations near these
two points. This, together with the conformal invariance and diagonality of
2-point functions, implies
\begin{equation}
	\langle (\mathcal{D}^{(\alpha)})^{\theta} \mathcal{O}_j^{\dagger (\mu)}
	(x_N) \mathcal{D}^{(\beta)} \mathcal{O}_k^{(\nu)} (x_S) \rangle \propto
	\delta_{| \alpha |, | \beta |} \delta_{j, k}, \label{descOrth}
\end{equation}
where $(\mathcal{D}^{(\alpha)})^{\theta}$ is obtained from
$\mathcal{D}^{(\alpha)}$ by replacing $x \rightarrow x^{\theta}$.

The OPE {\eqref{OPEplanar}} gives an expansion for ket states $| \Psi \rangle
\in \mathcal{H}_0^{\tmop{OS}}$ created by two local operators in terms of ket
states created by a single operator, which are elements of
$\mathcal{H}_0^{\tmop{CFT}}$. We would like to have a dual version of this
expansion for bra states $\langle \Psi |$. For this we need to understand how
the OPE transforms under the conjugation. Note that the formal differential
operators $\tilde{C}_{a, (\lambda)}^{(\mu) (\nu)} (x_1, x_2, x_0,
\mathcal{D}_0)$ can be uniquely determined by the equation
\begin{equation}
	\langle \mathcal{O}_1^{(\mu)} (x_1) \mathcal{O}_2^{(\nu)} (x_2)
	\mathcal{O}_3^{(\rho)} (x_3) \rangle_a = \tilde{C}_{a, (\lambda)}^{(\mu)
		(\nu)} (x_1, x_2, x_0, \mathcal{D}_0) \langle
	(\mathcal{O}^{\dagger}_3)^{(\lambda)} (x_0) \mathcal{O}_3^{(\rho)} (x_3)
	\rangle,
\end{equation}
where it is understood that the points are arranged as above, so that the
formal sum defined by $\tilde{C}_{a, (\lambda)}^{(\mu) (\nu)} (x_1, x_2, x_0,
\mathcal{D}_0)$ actually converges. By applying complex conjugation on both
sides and using the 2-point and 3-point hermiticity constraints
{\eqref{hermicity2-point}} and {\eqref{CFThermiticity}} we find
\begin{equation}
	\langle (\mathcal{O}^{\dag}_1)^{(\mu)} (x_1^{\theta})
	(\mathcal{O}^{\dag}_2)^{(\nu)} (x_2^{\theta})
	(\mathcal{O}^{\dag}_3)^{(\rho)} (x_3^{\theta}) \rangle_a = [\tilde{C}_{a,
		(\lambda)}^{(\mu) (\nu)} (x_1, x_2, x_0, \mathcal{D}_0)]^{\asterisk}
	\langle \mathcal{O}_3^{(\lambda)} (x_0^{\theta})
	(\mathcal{O}^{\dag}_3)^{(\rho)} (x_3^{\theta}) \rangle,
\end{equation}
which implies that
\begin{equation}
	[\tilde{C}_{a, (\lambda)}^{(\mu) (\nu)} (x_1^{\theta}, x_2^{\theta},
	x_0^{\theta}, \mathcal{D}_0^{\theta})]^{\asterisk} =
	\widetilde{\tilde{C}}_{a, (\lambda)}^{(\mu) (\nu)} (x_1, x_2, x_0,
	\mathcal{D}_0), \label{OPEkernelDagger}
\end{equation}
where $\widetilde{\tilde{C}}_{a, (\lambda)}^{(\mu) (\nu)} (x_1, x_2, x_0,
\mathcal{D}_0)$ is the formal sum that appears in the OPE for operators
with conjugate-reflected quantum numbers.

We are now ready to prove OS positivity for 4-point functions. Let $|
\Psi_0 \rangle$ be an OS ket state involving at most two local operators, i.e.
\begin{equation}
	| \Psi_0 \rangle = \sum_{i, j, \alpha, \beta} \int dx_1\, dx_2\, f_{i, j, (\alpha)
		(\beta)} (x_1, x_2) \mathcal{O}_i^{(\alpha)} (x_1) \mathcal{O}_j^{(\beta)}
	(x_2), \label{H2}
\end{equation}
where $f (x_{1,} x_2)$ is a compactly supported test function vanishing unless
$0 > x_1^0 > x_2^0$. (Terms with one or no operators are realized by setting
one or both operators to the identity.) Since by Euclidean CFT axioms the
correlation functions are real-analytic, the integrals that appear in the
expression for $\langle \Psi_0 | \nobracket \Psi_0 \rangle$ can be
approximated by finite Riemann sums, reflection-symmetric if necessary. This
implies that, for any $\varepsilon > 0$, we can pass from $| \Psi_0 \rangle$
to a ket state $| \Psi \rangle$ which is created by a \tmtextit{finite}
linear combination of insertions of up to two local operators with $x_1^0,
x_2^0 < 0$:
\begin{equation}
	| \Psi \rangle = \sum_{i, j, x_1, x_2, \alpha, \beta} c_{i, j, (\alpha)
		(\beta)} \mathcal{O}_i^{(\alpha)} (x_1) \mathcal{O}_j^{(\beta)} (x_2),
	\label{H2fin}
\end{equation}
and has the property that
\begin{equation}
	| \langle \Psi_0 | \Psi_0 \rangle - \nobracket \langle \Psi_0 | \Psi
	\rangle \nobracket | < \varepsilon, \qquad | \langle \Psi_0 | \Psi
	\rangle - \nobracket \langle \Psi | \Psi \rangle \nobracket | <
	\varepsilon, \label{PsiPsi0prop}
\end{equation}
so that as a result
\begin{equation}
	| \langle \Psi_0 | \Psi_0 \rangle - \nobracket \langle \Psi | \Psi
	\rangle \nobracket | < 2 \varepsilon \label{stp}
\end{equation}
We are therefore reduced to proving the nonnegativity of $\langle \Psi |
\Psi \rangle \nobracket$.

Now, the OPE convergence axiom implies that, for any $\varepsilon > 0$,
starting from $| \Psi \rangle$ and using the OPE {\eqref{OPEplanar}} in the
half-space $x^0 < 0$, we can construct a state $| \psi \rangle \nobracket = |
\psi_{ \Lambda} \rangle \in \mathcal{H}_0^{\tmop{CFT}}$ created by a finite
linear combination of local operators at $x_S$ such that
\begin{equation}
	| \langle \Psi | \psi_{} \rangle - \langle \Psi | \Psi \rangle |
	< \varepsilon . \label{appr1}
\end{equation}
Here $\Lambda$ is an OPE truncation cutoff which we need to increase
appropriately as $\varepsilon$ gets smaller. Namely, we will obtain $|
\psi_{} \rangle$ by keeping in the OPE all terms with $\Delta_k + | \alpha
| < \Lambda$, where $\Delta_k$ is the dimension of a primary
$\mathcal{O}_k$ appearing in the OPE, and $\alpha$ is the order of the
descendant $\mathcal{D}^{(\alpha)} \mathcal{O}_k$.

We can then repeat this procedure in the upper half-plane $x^0 > 0$ and
construct a state $\langle \psi' | = \langle \psi_{\Lambda'} |
\nobracket$ of local operators inserted at $x_N$ such that
\begin{equation}
	| \langle \Psi | \psi \rangle - \nobracket \langle \psi' | \psi
	\rangle \nobracket | < \varepsilon . \label{psiprime}
\end{equation}
Eq.\ {\eqref{OPEkernelDagger}} and the reality constraint
{\eqref{OPEcoeff:conjugation}} for the OPE coefficients imply that the state
$\langle \psi' |$ differs from $\langle \psi |$ at most by where the OPE
expansion was truncated. Furthermore, we can always assume that $\langle
\psi' |$ contains at least all the terms that $\langle \psi |$ does (i.e.\ $\Lambda' \geqslant \Lambda$), since adding more OPE terms to $\langle \psi'
|$ can only improve {\eqref{psiprime}}.

Eq.\ {\eqref{descOrth}} then implies that $\langle \psi' | \psi \rangle
\nobracket = \langle \psi | \psi \rangle \nobracket$ and therefore
\begin{equation}
	| \langle \Psi | \psi \rangle - \nobracket \langle \psi | \psi \rangle
	\nobracket | < \varepsilon . \label{appr2}
\end{equation}
Combining this with {\eqref{appr1}} we conclude:
\begin{equation}
	| \langle \Psi | \Psi \rangle - \nobracket \langle \psi | \psi \rangle
	\nobracket | < 2 \varepsilon, \label{PsiPsi}
\end{equation}
Since by the CFT positivity axiom $\langle \psi | \psi \rangle \nobracket$ is
non-negative, we conclude that $\langle \Psi | \Psi \rangle \nobracket$ is
also non-negative. This completes the proof of OS positivity for states
created by up to two operator insertions.

\section{Denseness and Hilbert space implications}\label{Hilbert}

Here we will describe some useful byproducts of the just given argument. Note
that {\eqref{appr1}} and {\eqref{appr2}}, in addition to {\eqref{PsiPsi}},
also implies
\begin{equation}
	\| \Psi - \psi_{\Lambda} \| \equiv \langle \Psi - \psi_{\Lambda} | \Psi -
	\psi_{\Lambda} \rangle = \langle \Psi | \Psi \rangle - \langle \Psi |
	\psi_{\Lambda} \rangle + \langle \psi_{\Lambda} | \psi_{\Lambda} \nobracket
	\rangle - \langle \psi_{\Lambda} | \Psi \rangle < 2 \varepsilon .
	\label{PsiPsiDense}
\end{equation}
This means that any $| \Psi \rangle$ can be approximated arbitrarily well
by a $| \psi_{\Lambda} \rangle \nobracket \in \mathcal{H}_0^{\tmop{CFT}}$. In
other words, $\mathcal{H}_0^{\tmop{CFT}}$ is a dense subspace of the vector
space of $\Psi$'s.

This fact has a simple but quite powerful consequence involving the CFT
Hilbert space $\mathcal{H}^{\tmop{CFT}}$, defined in Sec.\ \ref{ECFTax}
as the completion of $\mathcal{H}_0^{\tmop{CFT}}$. Eq.\ {\eqref{PsiPsiDense}}
implies, using the triangle inequality $\| \psi_{\Lambda_1} - \psi_{\Lambda_2}
\| \leqslant \| \Psi - \psi_{\Lambda_1} \| + \| \Psi - \psi_{\Lambda_2} \|$,
that the states $| \psi_{\Lambda} \rangle \nobracket$ corresponding to smaller
and smaller $\varepsilon$ form a Cauchy sequence. Therefore, these states have
a limit in $\mathcal{H}^{\tmop{CFT}}$ as $\Lambda \rightarrow \infty$,
which we call $| \psi_{\infty} \rangle \nobracket$. This $\psi_{\infty}$ is
nothing but the full, untruncated, OPE expansion of the state $\Psi$. We claim
that the map mapping $\Psi$'s to the corresponding $\psi_{\infty}$'s is
isometric, i.e.\ it preserves the inner products:
\begin{equation}
	\langle \Phi | \Psi \rangle = \langle \varphi_{\infty} | \psi_{\infty}
	\rangle . \label{OPEhilbert}
\end{equation}
Here the inner product on the l.h.s.\ is the OS inner product, computed using
CFT 4-point functions with operators inserted in the lower and upper
half-spaces, while the inner product in the r.h.s.\ is the \
$\mathcal{H}^{\tmop{CFT}}$ inner product, defined as the limit of
$\mathcal{H}_0^{\tmop{CFT}}$ inner product removing the cutoff:
\begin{equation}
	\langle \varphi_{\infty} | \psi_{\infty} \rangle \assign \lim_{\Lambda
		\rightarrow \infty} \langle \varphi_{\Lambda} | \psi_{\Lambda} \rangle .
	\label{HCFTinner}
\end{equation}
The proof of {\eqref{OPEhilbert}} is straightforward. We write:
\begin{eqnarray}
	& \langle \varphi_{\Lambda} | \psi_{\Lambda} \rangle = \langle \Phi +
	(\varphi_{\Lambda} - \Phi) | \Psi + (\psi_{\Lambda} - \Psi) \rangle =
	\langle \Phi | \Psi \rangle + \tmop{err} (\Lambda), & \\
	& \tmop{err} (\Lambda) = \langle \varphi_{\Lambda} - \Phi | \Psi \rangle +
	\langle \Phi | \psi_{\Lambda} - \Psi \rangle + \langle \varphi_{\Lambda} -
	\Phi | \psi_{\Lambda} - \Psi \rangle . & 
\end{eqnarray}
By Eq.\ {\eqref{PsiPsiDense}} we know that $\| \Psi - \psi_{\Lambda} \|$, $\|
\Phi - \varphi_{\Lambda} \|$ go to zero as $\Lambda \rightarrow \infty$.
Hence, $\tmop{err} (\Lambda) \rightarrow 0$ and {\eqref{OPEhilbert}} is
proved.

Eqs.\ {\eqref{OPEhilbert}} and {\eqref{HCFTinner}} mean that OPE converges in
the sense of the CFT Hilbert space. This property is often used in the CFT
literature (see Sec.\ \ref{Eucl4-point}). Note that CFT axioms in Sec.\ \ref{ECFTax} only assume pointwise OPE convergence, which is a weaker
statement. Curiously, by the given arguments, pointwise OPE convergence plus
CFT positivity imply Hilbert space convergence, at least for the 4-point
functions.

In the above argument we used the Hilbert space $\mathcal{H}^{\tmop{CFT}}$,
the completion of $\mathcal{H}_0^{\tmop{CFT}}$. We may introduce a second
Hilbert space as the completion of the space of $\Psi$'s, call it
$\mathcal{H}^{(2)}$. This Hilbert space contains e.g.\ all $\Psi_0$ states \
{\eqref{H2}}. (Similarly to {\eqref{PsiPsiDense}}, Eqs.\ {\eqref{PsiPsi0prop}}
and {\eqref{stp}} imply that $\Psi$ states are dense in the $\Psi_0$ states.)
Although $\mathcal{H}^{(2)}$ may look like a ``bigger'' space than
$\mathcal{H}^{\tmop{CFT}}$, actually it's not. Indeed, the map from $\Psi$
to $\psi_{\infty}$ extends to an isometric map from $\mathcal{H}^{(2)}$
to $\mathcal{H}^{\tmop{CFT}}$. In other words, Eq.\ {\eqref{OPEhilbert}}
remains true for any $\Phi, \Psi \in \mathcal{H}^{(2)}$. Since we can view $\cH^\text{CFT}_0$ as a
subspace of the space of $\Psi$'s, we also have that this map is surjective. The Hilbert spaces
$\mathcal{H}^{(2)}$ and $\mathcal{H}^{\tmop{CFT}}$ are thus unitarily
equivalent and may be identified. 

\section{OS clustering}\label{OSclustering}

Here we will derive the OS clustering {\eqref{clusterint}} from CFT axioms. We
will consider only $m + n = 4$, i.e.\ when the left-hand side of
{\eqref{clusterint}} can be written in terms of a 4-point function (this
also covers $m + n < 4$ since we can choose some of $\varphi$'s or $\chi$'s to
be the trivial identity field). We can assume that all $\chi$'s and
$\varphi$'s in {\eqref{clusterint}} are primary fields, since any derivatives
can be integrated by parts.

First as a general remark, assuming OS positivity, clustering (for any $m, n$)
only needs to be established point-wise, i.e.
\begin{equation}
	\lim_{\lambda \rightarrow \infty} \langle \chi^{\dagger}_m (y_m) \ldots
	\chi_1^{\dagger} (y_1) \varphi_1 (x_1 + \lambda a) \ldots \varphi_n (x_n +
	\lambda a) \rangle = \langle \chi^{\dagger}_m (y_m) \ldots
	\chi_1^{\dagger} \nobracket (y_1) \rangle \langle \nobracket \varphi_1 (x_1)
	\ldots \varphi_n (x_n) \rangle .
\end{equation}
This follows from the dominated convergence theorem. Indeed, OS positivity and
translation invariance (recall that we only consider $a^0 = 0$!) implies a
uniform in $\lambda$ bound\footnote{This follows, similarly to
	{\eqref{OSnaive1}}, by applying OS positivity to the state $| \Psi \rangle = |
	\psi (F, \varphi_1 \ldots \varphi_n) \rangle + e^{i \alpha} | \psi (G, \chi_1
	\ldots \chi_n) \rangle \nobracket$ where $F, G$ tend to delta functions
	localizing the operators at points $x_1 + \lambda a, \ldots, x_n + \lambda a$
	and $y_1^{\theta}, \ldots, y_n^{\theta}$ respectively, and choosing the phase
	$\alpha$ appropriately.}
\begin{eqnarray}
	& | \langle \chi^{\dagger}_m (y_m) \ldots \chi_1^{\dagger} (y_1)
	\varphi_1 (x_1 + \lambda a) \ldots \varphi_n (x_n + \lambda a) \rangle |
	\leqslant & \\
	&  \langle \chi^{\dagger}_m (y_m) \ldots \chi_1^{\dagger} (y_1) \chi_1
	(y_1^{\theta}) \ldots \chi_n (y_n^{\theta}) \rangle \times \langle
	\varphi^{\dagger}_n (x_n^{\theta}) \ldots \varphi_1^{\dagger}
	(x_1^{\theta}) \varphi_1 (x_1) \ldots \varphi_n (x_n) \rangle . &  \nonumber
\end{eqnarray}
It then follows that the integrand in {\eqref{clusterint}} is bounded by a
$\lambda$-independent integrable function, and the dominated convergence theorem is
applicable.

Going back to the 4-point function case which is our focus in this section,
let us start with $m = n = 2$. Since we already proved OS positivity for
states created by at most two operators (Sec.\ \ref{OSfromCFT}), in this
case we can rely on the above observation and we only need to check the
point-wise limit:
\begin{equation}
	\lim_{\lambda \rightarrow \infty} \langle \chi^{\dagger}_2 (y_2)
	\chi_1^{\dagger} (y_1) \varphi_1 (x_1 + \lambda a) \varphi_2 (x_2 + \lambda
	a) \rangle = \langle \chi^{\dagger}_2 (y_2) \chi_1^{\dagger} \nobracket
	(y_1) \rangle \langle \nobracket \varphi_1 (x_1) \varphi_2 (x_2) \rangle .
\end{equation}
To see this, we apply the OPE {\eqref{OPEplanar}} to $\chi^{\dagger}_2
(y_2) \chi_1^{\dagger} (y_1)$ in the left-hand side. The results of the
previous section imply that this OPE can be interpreted as expanding the state
in the Hilbert space $\mathcal{H}$ created by $\chi^{\dagger}_2 (y_2)
\chi_1^{\dagger} (y_1)$ in terms of eigenstates of $(K^0 - P^0) / 2$. This
implies that the OPE converges uniformly in $\lambda$ since the norm of $|
\varphi_1 (x_1 + \lambda a) \varphi_2 (x_2 + \lambda a) \rangle$ is
independent of $\lambda$ due to translation invariance ($a^0 = 0$!). We can
thus use the OPE to approximate $\langle \chi^{\dagger}_2 (y_2)
\chi_1^{\dagger} (y_1) \varphi_1 (x_1 + \lambda a) \varphi_2 (x_2 + \lambda a)
\rangle$ for any $\lambda$ to within any $\varepsilon > 0$ by a finite sum of
3-point functions of the form $\langle (\mathcal{D}^{(\alpha)})^{\theta}
\mathcal{O}_i^{(\nu)} (x_N) \varphi_1 (x_1 + \lambda a) \varphi_2 (x_2 +
\lambda a) \rangle$ times some $\lambda$-independent coefficients. Of these,
only the term corresponding to the identity operator, i.e.\ the one with
$(\mathcal{D}^{(\alpha)})^{\theta} \mathcal{O}_i^{(\nu)} = 1$, does not decay
with $\lambda$. It is easily verified that the contribution of this term is
precisely equal to $\langle \chi^{\dagger}_2 (y_2) \chi_1^{\dagger}
\nobracket (y_1) \rangle \langle \nobracket \varphi_1 (x_1) \varphi_2 (x_2)
\rangle$. This finishes the proof of clustering for $m = n = 2.$

In the remaining case $m = 3, n = 1$, we will consider the limit for the
integral (since we have not yet proved OS positivity for states involving 3
operators). Note that $\langle \varphi_1 (x) \rangle = 0$ unless $\varphi_1
\propto 1$, in which case the cluster property becomes trivial. This means
that we only need to prove
\begin{equation}
	\lim_{\lambda \rightarrow \infty} \int d x\, d y\, \overline{g
		(y_1^{\theta}, y_2^{\theta}, y_3^{\theta})} f (x_1) \langle
	\chi_3^{\dagger} (y_3) \chi_2^{\dagger} (y_2) \chi_1^{\dagger} (y_1)
	\varphi_1 (x_1 + \lambda a) \rangle = 0 \label{m3n1}
\end{equation}
for $\varphi \neq 1.$ This in turn is a very simple consequence of conformal
invariance, and of the fact that $\Delta_{\varphi} > 0$ for all operators but
the identity. We will be somewhat schematic. The main point is that the
configuration $(y_3, y_2, y_1, \infty)$ is nonsingular from the conformal
kinematics point of view (for which the conformal compactification $S_d$ of
the Euclidean space $\mathbb{R}^d$ is the appropriate arena). One way to see
it is that the cross ratios are finite in this limit. Thinking in a pedestrian
way, we can find a conformal transformation $g_{\lambda}$ which will move
points $(y_3, y_2, y_1, x_1 + \lambda a)$ to some points which have finite
limits as $\lambda \rightarrow \infty$. Transforming the integral
{\eqref{m3n1}} to this coordinate system, the only singular behavior at large
$\lambda$ comes from the Weyl transformation factor as the operator
$\varphi_1$ is moved from near infinity to a finite position. This factor
implies that the integral {\eqref{m3n1}} will go to zero as $\lambda^{- 2
	\Delta_{\varphi}}$, proving clustering in this particular case. See Sec.\ \ref{3+1MinkCluster} for additional details. More generally, the same argument
will also work for arbitrary $m$ as long as $n = 1$.

\chapter{Euclidean CFT $\Rightarrow$ Wightman: Basic
	strategy}\label{strategy}

We will now pass to the main task of this part of the thesis: given a Euclidean unitary CFT,
recover Minkowski correlators and show that that they satisfy Wightman axioms.

Let us first discuss this problem without assuming conformal invariance.
Suppose we know correlators $G^E_n (x_1, \ldots, x_n)$ of a scalar field in a
Euclidean QFT which is translationally and rotationally invariant, but not
necessarily conformally invariant. We are assuming, as discussed above, that
the correlators $G^E_n$ are defined and real-analytic (see footnote
\ref{real-anal}) for non-coincident Euclidean points ($x_k \in \mathbb{R}^d,
x_i \neq x_j$).

We would like to recover correlators in Minkowski signature. We are only
interested here in Wightman correlation functions, where the operator ordering
is fixed while the Minkowski time coordinates vary independently. We will call
them simply ``Minkowski correlators''. Starting from this section we will
focus on correlators of scalar primaries; correlators of fields in general
$\tmop{SO} (d)$ representations will be considered in our future publication
{\cite{paper2a}} (chapter \ref{chap:spinning4pt}).

To understand the equations below, it helps to keep in mind the basic
heuristic. If we had a Hilbert space, field operators $\phi$, and a
Hamiltonian $H$, then the Minkowski correlators would be given by
\begin{equation}
	G_n^M (x^M_1, \ldots, x^M_n) = \langle 0 | \phi \left( 0,
	\mathbf{x}_1 \right) e^{- i H (t_1 - t_2)} \phi \left( 0, \mathbf{x}_2
	\right) e^{- i H (t_2 - t_3)} \ldots | 0 \rangle, \qquad x_k^M
	= \left( t_k, \mathbf{x}_k \right) . 
\end{equation}
while the Euclidean correlators by
\begin{equation}
	G^E_n (x_1, \ldots, x_n) = \langle 0 | \phi \left( 0,
	\mathbf{x}_1 \right) e^{- H (\epsilon_1 - \epsilon_2)} \phi \left( 0,
	\mathbf{x}_2 \right) e^{- H (\epsilon_2 - \epsilon_3)} \ldots | 0
	\rangle, \qquad x_k = \left( \epsilon_k, \mathbf{x}_k \right),
	\quad \epsilon_k > \epsilon_{k + 1}, 
\end{equation}
We stress that the r.h.s.\ of these two equations will never be used in this
part of the thesis. We just use them to illustrate the intuitive property that $G_n^M$ can
be recovered from $G^E_n$ by analytic continuation $\epsilon_k \rightarrow
\epsilon_k + i t_k$ and sending $\epsilon_k \rightarrow 0$ while respecting
$\epsilon_k > \epsilon_{k + 1}$. In other words, there is a holomorphic function
$G_n$ which reduces to $G_n^M$ in one limit and to $G^E_n$ in another. The
precise domain of analyticity of $G_n$ can be clarified from the Wightman
axioms {\cite{Streater:1989vi}}. Their basic consequence is that Minkowski
correlators can be analytically continued to the ``forward tube'' (see below),
which contains the Euclidean space as a section. In this part of the thesis we will derive
Wightman axioms, rather than assume them. In particular, we will carry out
analytic continuation to the forward tube just from the properties of the
Euclidean correlators.

Let us put these observations into a definition of what it means to recover
$G_n^M$ from $G^E_n$. We consider $n$-point configurations with
complexified coordinates:
\begin{equation}
	c = (x_1, \ldots, x_n), \qquad x_k = \left( x_k^0, \mathbf{x}_k \right) \in
	\mathbb{C}^d . \label{ccompl}
\end{equation}
The ``forward tube'' $\mathcal{T}_n$ is defined as the set of all such
configurations for which the differences $y_k = x_k - x_{k + 1} = (y_k^0,
\mathbf{y}_k) \in \mathbb{C}^d$ satisfy the constraint:
\begin{equation}
	\tmop{Re} y_k^0 > | \tmop{Im} \mathbf{y}_k | \label{forward}, \qquad k = 1,
	\ldots, n - 1.
\end{equation}
Equivalently, this means that vectors $\tmop{Im} (i y_k^0, \mathbf{y}_k)$
belong to the open forward light cone of $\mathbb{R}^{1, d - 1}$, explaining
the name ``forward tube''.\footnote{The just given definition of the forward tube is adapted to the Euclidean coordinates. In Sec.\ \ref{executive}, Eq.\ \eqref{forward-tube-Mink}, we wrote the same definition in terms of Minkowski coordinates $x_k^M=(-i x_k^0, \mathbf{x}_k)$.}

Let $\mathcal{D}_n$ be the subset of the forward tube consisting of the
configurations with real spatial parts $\mathbf{x}_k$. Equivalently, we have:
\begin{equation}
	\label{def:Dn} \mathcal{D}_n = \{\, c\ |\ x_k^0 = \epsilon_k + i
	t_k,\ \mathbf{x}_k \in \mathbb{R}^{d - 1},\ \epsilon_1 > \epsilon_2 > \ldots >
	\epsilon_n \} .
\end{equation}
Finally, we denote by $\mathcal{D}^E_{n}$ the Euclidean part of
$\mathcal{D}_n$ obtained by setting all $t_k = 0$.

Minkowski correlators are then defined by the following two-step procedure:

\tmtextbf{Step 1.} One finds an extension $G_n^E$ from $\mathcal{D}^E_n$ to a
function $G_n (x_1, \ldots, x_n)$ such that one of the two conditions is
satisfied:
\begin{equation}
	\text{$G_n$ is defined on $\mathcal{T}_n$, and holomorphic in all variables
		$x_k^0$, $\mathbf{x}_k$,} \label{Gass1}
\end{equation}
or
\begin{equation}
	\text{$G_n$ is defined on $\mathcal{D}_n$, is holomorphic in variables $x_k^0$
		and is real-analytic in $\mathbf{x}_k$.} \label{Gass}
\end{equation}
Real analyticity in $\mathbf{x}_k$ means that $G_n$ can be extended from
$\mathcal{D}_n$ to a holomorphic function defined on a neighborhood of
$\mathcal{D}_n$ which allows small imaginary parts for $\mathbf{x}_k$. This
neighborhood can be arbitrarily small. Condition {\eqref{Gass}} is thus weaker
than {\eqref{Gass1}} and may be easier to check, although Theorem \ref{ThVlad}
below shows that the two conditions are equivalent under the ``powerlaw
bound'' assumption.

\tmtextbf{Step 2.} Minkowski correlators are defined as the limits of $G_n$
from inside $\mathcal{D}_n$ by sending $\epsilon_i \rightarrow 0$:\footnote{We will see in Theorem \ref{ThVlad} that this limit has to be taken
	along a fixed direction and is independent of direction. If the stronger
	condition {\eqref{Gass1}} holds, the limit can in fact be taken along any
	direction in the forward null cone.}
\begin{equation}
	G_n^M (x_1^M, \ldots, x_n^M) = \lim_{\epsilon_i \rightarrow 0} G_n (x_1,
	\ldots, x_n), \qquad x_k^M = \left( t_k, \mathbf{x}_k \right), \quad k = 1
	\ldots n \label{limit} .
\end{equation}
As mentioned several times, Minkowski correlators are expected to be tempered
distributions, and therefore this limit has to be understood in the
distributional sense. To show that the limit exists and has properties
required by Wightman axioms, one relies on the following powerful theorem of
several complex variables: 

\begin{theorem}[Vladimirov's theorem]
	\label{ThVlad}Suppose that the function $G_n$ is translation- and rotation-invariant, satisfies
	{\tmem{{\eqref{Gass}}}} and in addition satisfies everywhere on
	$\mathcal{D}_n$ the following `powerlaw bound' with some positive constants
	$C_n, A_n, B_n$:
	\begin{gather}
		|G_n (x_1, \ldots, x_n)| \leqslant C_{n}  \left( 1 + \max_k 
		\dfrac{1}{\epsilon_k - \epsilon_{k + 1}} \right)^{A_n}  (1 + \max_i  | x_i
		- x_{i + 1} |)^{B_n},  \label{powerlawbound}\\
		|x_i - x_j |^2 \equiv  | \epsilon_i + it_i - \epsilon_j - it_j |^2 +
		| \mathbf{x}_i - \mathbf{x}_j|^2.  \label{absdef} 
	\end{gather}
	Then:
	\begin{enumerate}
		\item Limit {\tmem{{\eqref{limit}}}} exists in the sense of tempered
		distributions. The limiting value $G_n^M$ is a tempered distribution.\footnote{This part of the theorem does not need translation- and rotation-invariance of $G_n$.}
		
		\item The distribution $G_n^M$ is Poincar\'e-invariant and satisfies the
		Wightman spectral condition. I.e.\ its Fourier transform $W (p_1, \ldots,
		p_{n - 1})$ with respect to the differences $x_k^M - x_{k + 1}^M$ has
		support in the product of closed forward light cones, which is the region
		$E_k \geqslant | \mathbf{p}_k |$, $p_k = (E_k, \mathbf{p}_k)$.
		
		\item The function $G_n$ can be extended from $\mathcal{D}_n$ to an
		holomorphic function on the whole forward tube $\mathcal{T}_n$. The limit
		{\eqref{limit}} exist also from the forward tube, i.e.\ when $\tmop{Re}
		y_k^0 \rightarrow 0, | \tmop{Im} \mathbf{y}_k | \rightarrow 0$, satisfying
		{\eqref{forward}}.
	\end{enumerate}
\end{theorem}

See App.\ \ref{Vlad} for the proof of Vladimirov's theorem and a reminder
of what the limit in the sense of distributions means. In the process of the
proof, it will be established that the holomorphic function $G_n$ on
$\mathcal{T}_n$ can be written as a ``Fourier-Laplace'' transform
\begin{equation}
	G_n (x_1, \ldots, x_n) = \int d p_1 \ldots d p_{n - 1}\, W (p_1, \ldots, p_{n
		- 1}) e^{\underset{k = 1}{\overset{n - 1}{\sum}} (- E_k (x^0_k - x^0_{k +
			1}) + i\mathbf{p}_k \cdummy (\mathbf{x}_k -\mathbf{x}_{k + 1}))},
	\label{FL1}
\end{equation}
where $W$ is a tempered distribution which is the Fourier transform of the
tempered distribution $G_n^M$, mentioned in Part 2 of the theorem.

To use Theorem \ref{ThVlad}, one needs to verify the powerlaw bound
{\eqref{powerlawbound}}. This strategy was first developed by Osterwalder and
Schrader (OS) {\cite{osterwalder1973,osterwalder1975}}.\footnote{OS used a
	slightly stronger version of Theorem \ref{ThVlad} with real analyticity in
	$\mathbf{x}_k$ replaced by the weaker assumption of continuity in these
	variables, but this difference is not essential.} Their full list of
assumptions included, in addition to reflection positivity and other OS axioms
listed in Sec.\ \ref{OSaxioms}, a less widely known \tmtextit{linear growth
	condition}, which roughly says that $G_n^E$ (appropriately integrated) grows
with $n$ not faster than a power of $n!$ and the degree of its singularities
grows not faster than linearly in $n$. The proof of the powerlaw bound was the
most technical part of the OS construction, and it crucially relied on the
linear growth condition. See App.\ \ref{OS} for the review.

In this part of the thesis we aim to define Minkowski correlators of a \tmtextit{conformal}
field theory, given Euclidean correlators satisfying the CFT axioms of Sec.\ \ref{ECFTax}. As seen in Sec.\ \ref{OSfromCFT}, reflection positivity is
robustly encoded in CFT via the positivity requirements for 2-point functions
and reality constraints on the OPE coefficients. On the other hand, not much
is known about how CFT $n$-point functions grow with $n$. In particular, we
are unable to justify the OS growth condition in our setup, hence we cannot
appeal to the OS theorem.

In this part of the thesis we will be able to circumvent this difficulty, by giving an
alternative proof of the powerlaw bound for the most important in applications
cases of 2, 3 and 4-point functions. Then, by \ Theorem \ref{ThVlad},
these correlators exist in Minkowski space and are tempered distributions. Our
proof of the powerlaw bound uses only the Euclidean CFT axioms. In fact, the
two- and 3-point function case is almost trivial, these correlators being
fixed by conformal invariance. The 4-point function case is much deeper and
is one of our main results. Remaining Wightman axioms not mentioned in Theorem
\ref{ThVlad} (positivity and clustering) will also be shown to hold.

\begin{remark}
	In practice, to compute the Minkowski correlator function one may connect a
	Minkowski configuration to a Euclidean configuration by a curve $c (s), 0
	\leqslant s \leqslant 1$, where $c (0)$ is Euclidean, $c (1)$ Minkowskian,
	and $c (s)$, $0 < s < 1$, belong to the forward tube. In general, the curve
	should remain in the forward tube except for the endpoint $c (1)$. This
	means that we must have strict inequalities:
	\begin{equation}
		\tmop{Re} x_1^0 (s) > \tmop{Re} x_2^0 (s) > \cdots > \tmop{Re} x_n^0 (s)
		\label{ftineq}
	\end{equation}
	everywhere along the analytic continuation contour, except for $s = 1$. See
	Fig.\ \ref{illustr}.
	
	In the literature, one sometimes encounters a different prescription for
	computing the Minkowski correlators (see e.g.\ {\cite{Hartman:2015lfa}},
	Sec.\ 3.1), where one puts all points but one at Minkowski positions, and
	considers correlators as a holomorphic function of the complexified coordinate
	of the remaining point. One then imagines that Wightman functions are
	holomorphic functions branching at light-cone separation, and that one can
	access different operator orderings by going around branch points. We would
	like to warn the reader that this prescription has to be taken with a grain
	of salt. To our knowledge there is no general result that the only Wightman
	functions singularities are branch cuts on the light cones. This is known to
	be true only in some special cases, e.g.\ for CFT 2-point and 3-point functions, as
	well as for CFT 4-point functions in $d = 2$ {\cite{Maldacena:2015iua}}. While
	some analytic continuation beyond the forward tube can be done in a general
	QFT (to the so called permuted extended tube), it does not suffice to
	justify the analytic continuation prescription of {\cite{Hartman:2015lfa}}
	in a general QFT. {In CFTs in higher dimensions, the
		prescription of Ref.\ {\cite{Hartman:2015lfa}} has some applicability, with
		the understanding that the correlator is analytic along the continuation
		contour but may stop being analytic at the endpoint (see App.\ \ref{Tom}).}
\end{remark}

\begin{figure}[h]\centering
	{\includegraphics[width=0.33\textwidth]{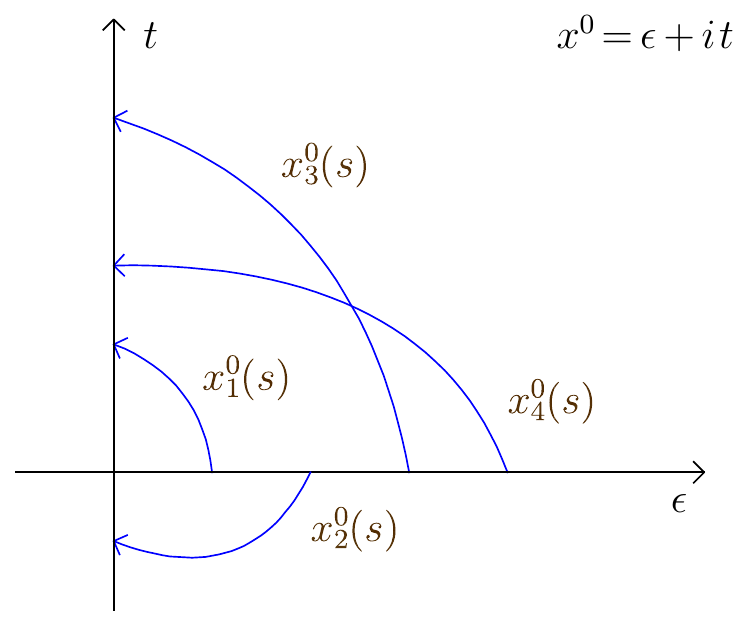}}
	\caption{Inequalities {\eqref{ftineq}} should be satisfied along the
		analytic continuation contour.\label{illustr}}
\end{figure}

\begin{remark}
	\label{lastOS}
	A powerlaw bound in the forward tube \eqref{powerlawbound} of course implies in a powerlaw bound for the Euclidean 4pt function itself. Together with rotation invariance, this will imply the remaining OS axiom, the ``Euclidean temperedness bound'' \eqref{OSmod}. Indeed, by rotation invariance, we can choose the direction of the $x^0$ axis before applying the Euclidean powerlaw bound. Let us choose the $x^0$ direction so that, after ordering the operators according to $\eps_1>\eps_2>\eps_3>\eps_4$, we have $\eps_k-\eps_{k+1}\geqslant \alpha |x_k-x_{k+1}|$ for each $k$. Such a direction exists for a sufficiently small positive $\alpha$, depending on $d$ and the number of points but not on $x_i$.\footnote{For each pair of points $(x_i,x_j)$ we consider the set of direction $\hat{e}_0$ such that $|(x_i-x_j)\cdot\hat{e}_0|\leqslant|x_i-x_j|\sin \delta$. This gives a subset $\mathcal{U}_\delta$ of the sphere $S^{d-1}$ with ${\rm Vol}(\mathcal{U}_\delta)\leqslant2\delta\, {\rm Vol}(S^{d-2})$. If we choose 
		$\delta_*=\frac{{\rm Vol}(S^{d-1})}{2 n(n-1)\,\text{Vol}(S^{d-2})}$, then the total volume excluded by considering all possible $(x_i,x_j)$ pairs is less than ${\rm Vol}(S^{d-1})/2$. Therefore, we can find a direction $\hat{e}_0$ such that the opposite inequality $|(x_i-x_j)\cdot\hat{e}_0|\geqslant  |x_i-x_j| \sin \delta_*$ holds for all pairs. Then, renumbering the points in the order of decreasing $x_i^0$, we obtain $x^0_k-x^0_{k+1}\geqslant  |x_k-x_{k+1}| \sin \delta_*$.} Applying the Euclidean case of the powerlaw bound \eqref{powerlawbound} in this frame we obtain \eqref{OSmod}.
\end{remark}

\section{Recovering Minkowski averages from Euclidean
	averages}\label{MinkFromEucl}

Minkowski correlators provided by Theorem \ref{ThVlad}, being tempered distributions, can be paired with a
Schwartz test function $F$:
\begin{equation}
	(G_n^M, F) = \int d^d x_1\ldots d^dx_n\, G_n^M (x_1 \ldots x_n) F (x_1 \ldots x_n) .
	\label{GMav}
\end{equation}
Here we will discuss how these pairings can be computed given the Euclidean correlators (compare {\cite{osterwalder1973}}, Sec.\ 4.3). This discussion will be needed in Sec.\ \ref{sec:Wpos} below and may be skipped on the first reading. 

Eq.\ \eqref{GMav} can equivalently be expressed via the Fourier transform $W$ of $G_n^M$
with respect to $x_k - x_{k + 1}$:
\begin{equation}
	(G_n^M, F) = \int d^dp_1\ldots d^dp_{n-1}\, W (p_1, \ldots, p_{n - 1}) f (p_1 \ldots p_{n - 1}),
	\label{GMavFT}
\end{equation}
where $f (p_1 \ldots p_{n - 1}) = \hat{F} (- p_1, p_1 - p_2, \ldots, p_{n - 2}
- p_{n - 1}, p_{n - 1})$.
Natural pairings for Euclidean
correlators are
\begin{equation}
	(G_n^E, \varphi) = \int d^dx_1\ldots d^dx_n\, G_n^E (x_1 \ldots x_n) \Phi (x_1 \ldots x_n) .
	\label{GEav}
\end{equation}
where $\Phi$ is a $C^{\infty}$ test function compactly supported in $x_1^0 >
\cdots > x_n^0$. We wish to discuss how pairings {\eqref{GMav}} or
{\eqref{GMavFT}} can be found given {\eqref{GEav}}. 

By the Fourier-Laplace
representation {\eqref{FL1}}, we can write {\eqref{GEav}} as
\begin{equation}
	(G_n^E, \varphi) = \int d^dp_1\ldots d^dp_{n-1}\, W (p_1, \ldots, p_{n - 1}) g (p_1, \ldots, p_{n
		- 1}), \label{GEavFT}
\end{equation}
where $g$ is any Schwartz class function which agrees inside the forward light
cones with $\tilde{\varphi}$, the Fourier-Laplace transform of $\varphi (y_1,
\ldots, y_{n - 1}) = \int d^d x_n\, \Phi \left( x_n + \sum_{i = 1}^{n - 1} y_i,
x_n + \sum_{i = 2}^{n - 1} y_i, \ldots, x_n \right)$:
\begin{gather}
	g \in \mathcal{S}, \qquad g (p) = \tilde{\varphi} (p) \qquad (p_k \in
	\overline{V_+}) ,  \label{gphi}\\
	\tilde{\varphi} (p_1, \ldots, p_{n - 1}) =  \int d^d y_1\ldots d^dy_{n-1}\, \varphi (y_1 \ldots
	y_{n - 1}) e^{\underset{k = 1}{\overset{n - 1}{\sum}} (- p^0_k y^0 +
		i\mathbf{p}_k \cdummy \mathbf{y}_k)}. \nn
\end{gather}
Note that we cannot just put $g = \tilde{\varphi}$ because $\tilde{\varphi}$
is by itself not a Schwartz function (it may grow exponentially in the
negative $p^0_{k}$ directions, although it will decrease exponentially in
the positive one, since $\varphi$ is supported at $y_k^0 > 0$). On the other
hand the values of $g$ outside the light cones, where $W$ is supported, are
unimportant. We can for example take
\begin{equation}
	g (p_1, \ldots, p_{n - 1}) = \chi (p^0_1) \ldots \chi (p^0_{n - 1})
	\tilde{\varphi} (p_1, \ldots, p_{n - 1}), \label{gex}
\end{equation}
where $\chi (s)$ is a $C^{\infty}$ function which equals identically 1 for $s
\geqslant 0$ and 0 for $s < - 1$.

Suppose then that we find a sequence of $C^{\infty}$ functions $\{ \varphi_r
\}_{r = 1}^{\infty}$ compactly supported at $y^0_k > 0$, the corresponding
functions $g_r \in \mathcal{S}$ such that $g_r = \tilde{\varphi}_r$ inside the
light cones, and in addition that $g_r \rightarrow f$ in the sense of the
Schwartz space (i.e.\ that all Schwartz space seminorms of the difference go to
zero), where $f$ is the function in {\eqref{GMavFT}}. Let us put $\Phi_r (x_1,
\ldots, x_n) = \varphi (x_1 - x_2, x_2 - x_3, \ldots, x_{n - 1} - x_n) \omega
(x_n)$ where $\omega$ is any $C_0^{\infty}$ function of integral one. Then we will
have
\begin{equation}
	(G_n^E, \Phi_r) = (W, g_r) \longrightarrow (W, f) = (G_n^M, F) \qquad (r
	\rightarrow \infty),
\end{equation}
and so we will solve the problem of computing Minkowski averages given
Euclidean averages. The following lemma, loosely
related to Lemma 8.2 in {\cite{osterwalder1973}}, shows that it is indeed possible to find such sequences
$\varphi_r$ and $g_r$ for any Schwartz class $f$. 

\begin{lemma}
	\label{lemma:fcheckdense}The set of functions $g \in \mathcal{S}
	(\mathbb{R}^{d (n - 1)})$ which satisfy {\eqref{gphi}} for some $\varphi$ a
	$C^{\infty}$ test function compactly supported in $y_k^0 > 0$ is dense in
	the Schwartz space.
\end{lemma}

We will give a formal proof; see App.\ \ref{IntLem1} for some intuition. We
will consider the case $n = 2$ as $n > 2$ is no more complicated. For each
$\varphi$ we consider the set $A_{\varphi}$ of Schwartz functions $g$ which satisfy
{\eqref{gphi}}:
\begin{equation}
	A_{\varphi} \assign \left\{ g \in \mathcal{S} (\mathbb{R}^d) \middle| 
	\nobracket g |_{\overline{V_+}} = \tilde{\varphi} \right\} .
\end{equation}
We know that $A_{\varphi}$ is non-empty, e.g.\ we can take $g$ from
{\eqref{gex}} (it is not hard to show that this is a Schwartz function). Our
lemma says that $A \equiv$ ``the union of $A_{\varphi}$ over all $\varphi$''
is dense in $\mathcal{S} (\mathbb{R}^d)$. The proof will be by contradiction.
Note that $A$ is a linear subspace of $\mathcal{S} (\mathbb{R}^d)$. If
$\overline{A} \neq \mathcal{S} (\mathbb{R}^{d})$, then there exists a
tempered distribution $T \in \mathcal{S}' (\mathbb{R}^d)$ such that $T$
vanishes on all test functions from $A$ but does not vanish
identically.\footnote{The corresponding statement for normed spaces is
	standard, being a well-known consequence of the Hahn-Banach theorem (see e.g.\ {\cite{brezis}}, Corollary 1.8). For the Schwartz space, we can first find a
	Schwartz norm $| \cdot |_n$, such that $\overline{A}$ is not everywhere dense
	with respect to this norm, and then apply the standard statement with the norm
	$| \cdot |_n$. This gives a linear functional $T$ on $\mathcal{S}
	(\mathbb{R}^d)$ continuous with respect to $| \cdot |_n$, hence $T \in
	\mathcal{S}' (\mathbb{R}^d)$.}

So $T$ in particular vanishes on $A_0 = \left\{ g \in \mathcal{S}
(\mathbb{R}^{d}) \middle|  \nobracket g |_{\overline{V_+}} = 0 \right\}$
(take $\varphi = 0$). This means that the support of the distribution $T$ is
contained inside $\overline{V_+}$. Consider the Fourier transform of $T$,
\begin{equation}
	\hat{T} (x) \assign \int \frac{d^d p}{(2 \pi)^{d}}\, T (p) e^{i p^0 x^0 -
		i\mathbf{p} \cdummy \mathbf{x}} .
\end{equation}
We can consider $\hat{T} (x)$ for real $x$ where it is a distribution. Since
$\tmop{supp} (T) \subseteq \overline{V_+}$ it is also natural to consider
$\hat{T} (\xi + i \eta)$ where $\xi, \eta$ are real and $\eta$ is in the
forward cone. We know that $\hat{T} (x)$ is a holomorphic function for such $x
= \xi + i \eta$. We also know that the distribution $\hat{T} (x)$ for real $x$
can be obtained as a limit of the holomorphic function $\hat{T} (\xi + i
\eta)$ as $\eta \rightarrow 0$.

Let us now come back to the assumption that $(T, g) = 0$ for any $g \in
A_{\varphi}$. We will apply this to a function $g$ of the form $g = X (p) \tilde{\varphi}$ where $X
(p)$ is a $C^{\infty}$ function identically 1 on the forward
light cone and such that $X (p) e^{- p^0 x^0 + i\mathbf{p} \cdot
	\mathbf{x}}$ is in Schwartz class for any $x^0 > 0$. It is easy to see that
such functions $X(p)$ exist. Writing $(T, g)$ in full we get:
\begin{eqnarray}
	0 = (T, g) & = & \int d p\, T (p) X (p) \int d x\, e^{- p^0 x^0 +
		i\mathbf{p} \cdummy \mathbf{x}} \varphi (x) \nonumber\\
	& = & \int d x\, \varphi (x) \int d p\, T (p) X (p) e^{- p^0 x^0 +
		i\mathbf{p} \cdummy \mathbf{x}} \nonumber\\
	& = & \int d x\, \varphi (x) \hat{T} (i x^0, \mathbf{x}) . 
	\label{int:Tf}
\end{eqnarray}
The swap of the order of integration between the first and the second line can
be justified as follows. Since $T (p)$ is a tempered distribution, we can
write it as a finite sum of derivatives of continuous functions of power
growth: $T (p) = \underset{}{} \sum_{\alpha} \partial_p^{\alpha} F_{\alpha}
(p)$. Using distributional integration by parts, we can the rewrite the first
line of {\eqref{int:Tf}} as a sum of ordinary integrals, apply Fubini's
theorem to swap the integration order, and integrate by parts back to express
the answer in terms of $T (p)$.

Because $\varphi (x)$ has compact support in the region $x^0 > 0$, the
argument of $\hat{T}$ in the last line of (\ref{int:Tf}) is of the form $\xi +
i \eta$ with $\eta = (x^0, \tmmathbf{0})$ in the forward light cone, where
we know $\hat{T}$ is analytic. So, from the fact that the last line of
(\ref{int:Tf}) vanishes for any $\varphi$ we conclude that
\begin{equation}
	\hat{T} (\xi + i \eta) = 0, \qquad \xi = (0, \mathbf{x}), \quad \eta = (i
	x^0, \tmmathbf{0}), \quad x^0 > 0.
\end{equation}
The set of these points is a totally real submanifold, and so by analyticity
we conclude that $\hat{T} (\xi + i \eta)$ is identically zero for any $\xi
\in \mathbb{R}^d$ and any $\eta$ in the forward cone. Furthermore, as
mentioned above, $\hat{T} (x)$ for real $x$ is a boundary value of $\hat{T}
(\xi + i \eta)$. Therefore, $\hat{T} = 0$ in the sense of distributions.
However we assumed above that $T$ was not identically zero. The reached
contradiction shows that $A$ is dense in $\mathcal{S}$.

\chapter{Two- and 3-point functions}\label{23-point}

Let us see how the strategy from chapter \ref{strategy} works for the CFT 2-point
and 3-point functions. The Euclidean 2-point and 3-point correlators of scalar primaries
are given by [$x_{i j}^2 = (x_i - x_j)^2$]
\begin{eqnarray}
	G^E_2 (x_1, x_2) & = & \frac{1}{(x^2_{12})^{\De}},  \label{G2E}\\
	G^E_3 (x_1, x_2, x_3) & = & \frac{c_{123}}{(x^2_{12})^{h_{123}}
		(x^2_{13})^{h_{132}} (x_{23}^2)^{h_{23}}}, \qquad h_{i j k} = (\Delta_i +
	\Delta_j - \Delta_k) / 2.  \label{G3E}
\end{eqnarray}

In this case, the standard way to obtain the Wightman functions is to write
these Euclidean correlators in terms of $x^2_{i j}$ with $i < j$ (as we did).
Substituting the analytic continuation of $x_{i j}^2$,
\begin{equation}
	x^2_{i j} = (x_i - x_j)^{\mu} (x_i - x_j)_{\mu} = (x^0_i -
	x^0_j)^2 + (\mathbf{x}_i - \mathbf{x}_j)^2, \qquad x_i = \left( x_i^0,
	\mathbf{x}_i \right) \in \mathbb{C}^d \label{xij-cont} .
\end{equation}
into the Euclidean 2-point and 3-point functions expressions, we obtain their analytic
continuations. Suppose further that $x^2_{i j} \neq 0$, $i < j$, in the
forward tube (this will be shown below). Then the functions
\begin{equation}
	c \mapsto x_{i j}^2 \quad (i < j) \label{cxij2}
\end{equation}
are holomorphic functions from the forward tube to $\widetilde{\mathbb{C}
	\backslash \{ 0 \}}$, the universal covering of the complex plane minus the
origin. On the other hand $z \mapsto z^h$ is holomorphic from this universal
covering to $\mathbb{C}$. Composing these two holomorphic functions, we conclude
that $(x_{i j}^2)^h$, $i < j$, are holomorphic on the forward tube. Hence this
procedure analytically extends the Euclidean 2-point and 3-point functions to the
whole forward tube $\mathcal{T}_n$ $(n = 2, 3)$.

We will now give a simple lemma which proves that indeed $x^2_{i j} \neq 0$,
$i < j$, in the forward tube. Actually the lemma says that a bit more is true,
namely $x^2_{i j} \in \mathbb{C} \backslash (- \infty, 0]$. This has the
following practical consequence. In general, to compute the analytic
continuation of $(F (c))^h$, where $F (c)$ is a nonzero holomorphic function, we
need to know the phase of $F (c)$, i.e.\ to which sheet of the Riemann surface
$\widetilde{\mathbb{C} \backslash \{ 0 \}}$ it belongs. To compute the phase
we need to connect $c$ to a $c_E$ by a curve and analytically continue along
this curve, following the phase. However, this is unnecessary for $(x_{i
	j}^2)^h$. Indeed, by the lemma below $x_{i j}^2$ always belongs to the
principal sheet. So there is no need to use a curve to compute the phase: it
can be computed unambiguously just by plugging the coordinates into
{\eqref{xij-cont}}.

\begin{lemma}
	\label{xij2h}Let $y = (y^0, \mathbf{y}) \in \mathbb{C}^d$ satisfy $\tmop{Re}
	y^0 > | \tmop{Im} \mathbf{y} |$. Then
	\begin{equation}
		y^2 \equiv (y^0)^2 +\mathbf{y}^2 \in \mathbb{C} \backslash (- \infty, 0] .
		\text{} \label{xij2}
	\end{equation}
\end{lemma}

\begin{proof}
	We will denote by Greek letters $\xi, \eta$, etc., vectors of Minkowski space
	$\mathbb{R}^{1, d - 1}$ with the Minkowski inner product $\xi^2 = -
	(\xi^0)^2 + \tmmathbf{\xi}^2$.\footnote{\label{signature}Everywhere in this
		part of the thesis we are using $-, +, \ldots, +$ Minkowski signature.} Decomposing the
	vector $(i y^0, \mathbf{y})$ into its real and imaginary parts:
	\begin{equation}
		(i y^0, \mathbf{y}) = \xi + i \eta,
	\end{equation}
	condition $\tmop{Re} y^0 > | \tmop{Im} \mathbf{y} |$ means that $\eta^0 > 0$
	and $- \eta^2 > 0$, i.e.\ $\eta$ is in the open forward light cone, which we
	will denote by $\eta \succ 0$. In this notation, we have to prove that
	\begin{equation}
		(\xi + i \eta)^2 = \xi^2 - \eta^2 + 2 i (\xi \eta) \nin \mathbb{C}
		\backslash (- \infty, 0] . \label{toprove}
	\end{equation}
	where by our conventions all inner products involving $\xi, \eta$ are
	Minkowski. Suppose this is violated, i.e.
	\begin{equation}
		(\xi \eta) = 0, \quad \xi^2 - \eta^2 < 0, \label{incomp}
	\end{equation}
	for some $\xi, \eta$. Since $\eta$ is timelike, $(\xi \eta) = 0$ implies
	that $\xi$ is spacelike. But then $\xi^2 - \eta^2 = \xi^2 + (- \eta^2) > 0$.
	Thus the two conditions in {\eqref{incomp}} cannot both be true, and
	{\eqref{toprove}} is proved.
\end{proof}

As the next step of implementing the strategy from chapter \ref{strategy}, we
need to check that the constructed analytic continuations satisfy a powerlaw
bound so that we can apply Theorem \ref{ThVlad}. Although we already
constructed analytic continuation to the whole $\mathcal{T}_n$, we only need
to check the bound on $\mathcal{D}_n$ which is somewhat easier. The powerlaw
bound follows from the following lemma.

\begin{lemma}
	\label{x2bnd}
	(a) Let $y = (\varepsilon + i s, \mathbf{y})$, $\varepsilon, s
	\in \mathbb{R}$, $\mathbf{y} \in \mathbb{R}^{d - 1}$. Then $y^2$ is bounded
	above and below in the absolute value, as follows:
	\begin{equation}
		\varepsilon^2 \leqslant | y^2 | \leqslant | y |^2 \equiv | \varepsilon + i
		s |^2 +\mathbf{y}^2 ;
	\end{equation}
	(b) On $\mathcal{D}_n$ ($n = 2, 3$), each $1 / (x^2_{i j})^h$ factor in
	{\eqref{G2E}}, {\eqref{G3E}} ($h \in \mathbb{R}$) satisfies a powerlaw
	bound:
	\begin{equation}
		\Bigl| \frac{1}{(x^2_{i j})^h} \Bigr| \leqslant \frac{| x_i - x_j
			|^B}{(\epsilon_i - \epsilon_j)^A} \qquad (i > j),
	\end{equation}
	where $A = 2h$, $B = 0$ for $h$ positive and $A = 0$, $B = - 2h$ for $h$
	negative.
\end{lemma}

\begin{proof}
	(a) The upper bound is obvious. Let us show the lower bound by an explicit
	computation (see Lemma \ref{zeta2bnd} below for an alternative proof). We
	have:
	\begin{equation}
		| y^2 |^2 \equiv | (\varepsilon + i s)^2 +\mathbf{y}^2 |^2 =
		(\varepsilon^2 - s^2 +\mathbf{y}^2)^2 + 4 \varepsilon^2 s^2,
	\end{equation}
	Minimizing this in $\mathbf{y}$, we get
	\[ \min_{\mathbf{y}} | y^2 |^2 = \left\{\begin{array}{l}
		(\varepsilon^2 - s^2)^2 + 4 \varepsilon^2 s^2 = (\varepsilon^2 +
		s^2)^2, \qquad | s | \leqslant \varepsilon,\\
		4 \varepsilon^2 s^2, \qquad | s | \geqslant \varepsilon .
	\end{array}\right. \]
	Minimizing this next in $s$, we find that the absolute minimum is located at
	$\mathbf{y}= 0$,  $s = 0$, and is equal to $\varepsilon^4$. Part (b) follows
	from (a).
\end{proof}

Now that we have the powerlaw bound, we can apply Theorem \ref{ThVlad}. We
conclude that the Minkowski 2-point and 3-point functions, defined as $\epsilon_i
\rightarrow 0$ limits of the analytically continued Euclidean correlators,
exist, are Lorentz-invariant tempered distributions, and satisfy the spectral
condition.\footnote{\label{noteMarc1}Since these are tempered distributions,
	their Fourier transforms are well defined. Explicit expressions for these
	Fourier transforms are known in many cases. See {\cite{Gillioz:2018mto}} for
	$\langle \mathcal{O}_{\Delta, l} (p) \mathcal{O}_{\Delta, l} (- p) \rangle$
	and {\cite{Gillioz:2019lgs}} for $\langle \mathcal{O}_{\Delta_1} (p_1)
	\mathcal{O}_{\Delta_2} (p_2) \mathcal{O}_{\Delta, l} (p_3) \rangle$.}

\section{Comparison with the $i
	\varepsilon$-prescription}\label{comparison}

Here we will comment on the ``$i \varepsilon$-prescription'' often used in the
literature to define Minkowski 2-point and 3-point correlators, and how it compares
with our definition. We will focus on the 2-point case for definiteness (same
remarks hold for the 3-point case).

The $i \varepsilon$-prescription defines the Minkowski 2-point correlator $G_2^M
(x_1^M, x_2^M)$ as
\begin{equation}
	\frac{1}{(- (s - i \varepsilon)^2 +\mathbf{y}^2)^{\De}}, \label{GMnaive}
\end{equation}
with $s = t_1 - t_2$, $\mathbf{y}=\mathbf{x}_1 -\mathbf{x}_2$ and taking the
$\varepsilon \rightarrow 0^+$ limit. The precise meaning of the limit is often
left implicit in the physics literature. Away from the light cone the 2-point correlator is
an ordinary function, the limit can be understood pointwise and it agrees with
our definition. Clearly, on the light cone the limit must be understood in
distributional sense, integrating against a test function $f (s, \mathbf{y})$. {That is what we showed above: Vladimirov's theorem guarantees that the limit $\e\to 0$ exists as a tempered distribution and can be therefore integrated against any Schwartz test function. In physics literature, one instead often hears that such integrals should be defined by ``shifting the
	integration contour''. Note however that this alternative way of understanding the $\e\to 0$ limit would only work for analytic test functions.} Let us discuss the consequences of this limitation.

It is helpful to recall that the theory of distributions commonly uses three
classes of test functions, denoted $\mathcal{S}, \mathcal{K}, \mathcal{Z}$
{\cite{gelfandshilov}}. Here $\mathcal{S}$ is the space of Schwartz functions,
$\mathcal{K}$ (denoted sometimes by $\mathcal{D}$) is the space of compactly
supported $C^{\infty}$ functions, and $\mathcal{Z}$ consists of entire
holomorphic functions decreasing faster than any power in the real directions
and bounded by some fixed exponential in the imaginary directions. Note that
$\mathcal{K}, \mathcal{Z} \subset \mathcal{S}$. The corresponding distribution
spaces thus satisfy the opposite inclusion: $\mathcal{S}' \subset
\mathcal{K}', \mathcal{Z}'$. The elements of $\mathcal{S}'$ are precisely the
tempered distributions discussed above, $\mathcal{K}'$ are distributions on
the compactly supported test functions\footnote{They are briefly mentioned in
	the proof of Theorem \ref{ThVlad}, App.\ \ref{Proof1}, after Eq.
	{\eqref{gge}}.}, while $\mathcal{Z}'$ is yet another distributional class.

Importantly, the Fourier transform $\mathcal{F}$ leaves $\mathcal{S}$
invariant. Since the Fourier transform is defined in the theory of
distributions by duality, we also have $\mathcal{F} (\mathcal{S}')
=\mathcal{S}'$: the Fourier transform of a tempered distribution is also a
tempered distribution. On the other hand, one can show (see
{\cite{gelfandshilov}}) that $\mathcal{F} (\mathcal{K}) =\mathcal{Z}$. This is
the rationale behind introducing the space $\mathcal{Z}$, and this also
implies that $\mathcal{F}$ maps $\mathcal{K}'$ to $\mathcal{Z}'$ and vice
versa.

Coming back to {\eqref{GMnaive}}, shifting the integration contour defines
this distribution as an element of $\mathcal{Z}'$. The pairing with a test
function $f \in \mathcal{Z}$ is thus defined by
\begin{equation}
	\int_C d z\, \int d\mathbf{y}\, \frac{1}{(- z^2 +\mathbf{y}^2)^{\De}} f (z,
	\mathbf{y})
\end{equation}
with the contour $C$ running parallel to the real axis in the lower half
plane.\footnote{We can somewhat relax the condition $f \in \mathcal{Z}$. At
	the very least, $f$ must be holomorphic in the lower half-plane close to the real
	axis and decrease sufficiently fast at infinity for the integral to be
	convergent.} \ By the previous paragraph, this is then sufficient to define
the Fourier transform of the 2-point function as an element of $\mathcal{K}'$. By
moving the contour far away from the real axis, one shows that the Fourier
transform vanishes for negative energies, and by Lorentz invariance one
concludes that the support must belong to the forward light cone. These
arguments have parallels in the proof of Part 2 of Theorem \ref{ThVlad} (see
App.\ \ref{Proof1}).

Compared to this simple and almost elementary discussion, Theorem
\ref{ThVlad} proves a stronger statement that the 2-point distribution
{\eqref{GMnaive}} can be extended to test functions of Schwartz class and,
furthermore, to functions which have only a finite number of derivatives as
expressed by Eq.\ {\eqref{GMcont}}. This can be seen as a finer
characterization of the singularity structure at short distances. The Fourier
transform is then also a tempered distribution, thus bounded by some power,
which is a stronger statement than it being an element of $\mathcal{K}'$ since
those can grow arbitrarily fast at infinity.

Since the 2-point and 3-point correlators are known in closed form, one can in
principle verify that their Fourier transform does not grow too fast at
infinity by an explicit computation. This would provide an alternative proof
of temperedness. Our point here is that Theorem \ref{ThVlad} reaches this
conclusion without any computations. For the 4-point correlators considered below,
the Fourier transform cannot be evaluated easily, and Theorem \ref{ThVlad}
appears to be the only realistic way to show temperedness.

It is instructive to discuss why we insist so much on temperedness. In other
words, why Wightman axioms require that the Minkowski $n$-point correlators
must be tempered distributions, and not of some other class? There is a simple
reason why temperedness is a natural requirement, while $\mathcal{K}'$ or
$\mathcal{Z}'$ would not suffice. The point is that Wightman axioms include
\tmtextbf{both} commutativity at spacelike separation and the spectral
condition (the Fourier transform supported in the forward tube). Both these
conditions need compactly supported test functions: the former in position
space, the latter in momentum space. The space $\mathcal{S}$ is large enough
to write both these conditions, while $\mathcal{K}$ or $\mathcal{Z}$ are
inadequate as we would lose one of them.\footnote{For completeness it should
	be noted that one can reduce $\mathcal{S}$ a bit and still be able to
	formulate both these axioms, as for Jaffe fields {\cite{Jaffe}}, which may
	have stronger-than-powerlaw singularities at short distances. For CFTs and for
	any theory which asymptotes to a CFT at short distances, there is no reasons
	to consider such modifications, and $\mathcal{S}$ remains the natural choice.}

Finally, sometimes by the $i \varepsilon$-prescription one
means the following simplified form of {\eqref{GMnaive}}:
\begin{eqnarray}
	&  & \frac{1}{(- s^2 + i 0^+ \tmop{sign} (s) +\mathbf{y}^2)^{\De}}, 
	\label{Expr1}
\end{eqnarray}
which agrees with {\eqref{GMnaive}} away from the light cone. {By Vladimirov's theorem, this defines a distribution for $s>0$ (including the light cone) and another distribution for $s<0$, but it is not an adequate starting point for defining the distribution around $(s,\mathbf{y})=(0,0)$.} 

\chapter{Scalar 4-point function}\label{sec:4-point}

This section is the heart of this part of the thesis. In it we will show how to define
Minkowski 4-point functions starting from Euclidean 4-point functions of a unitary
CFT. \ We will follow the strategy of chapter \ref{strategy} and in particular
will rely on Theorem \ref{ThVlad}. To avoid inessential details, we will focus
on the case of four identical scalars. Non-identical scalars can be treated by
the same argument (see Sec.\ \ref{nonId}). Additional complications arise
for spinning operators; this case is postponed to a future publication
{\cite{paper2a}}.

So, we consider the Euclidean CFT 4-point function of four identical scalar
Hermitian primaries, which by conformal invariance can be written as:
\begin{equation}
	G^E_4 (c_E) \equiv \langle
	\mathcal{O}(x_1)\mathcal{O}(x_2)\mathcal{O}(x_3)\mathcal{O}(x_4) \rangle =
	\frac{1}{(x_{12}^2 x_{34}^2)^{\Delta_{\mathcal{O}}}} g (c_E) .
	\label{def:Euclidean4-point}
\end{equation}
Here $c_E = (x_1, x_2, x_3, x_4)$ denotes an ordered configuration of four
Euclidean non-coincident points ($x_k \in \mathbb{R}^d, \hspace{1em} x_i \neq
x_j$), $\Delta_{\mathcal{O}}$ is the scaling dimension of $\mathcal{O}$, and
$g (c_E)$ is a real function which depends only on the conformal class of
$c_E$. It can be written as a function of two conformally invariant
cross-ratios $u, v$:
\begin{equation}
	g (c_E) = g (u, v), \qquad u = \dfrac{x_{12}^2 x_{34}^2}{x_{13}^2 x_{24}^2},
	\quad v = \dfrac{x_{14}^2 x_{23}^2}{x_{13}^2 x_{24}^2} . \label{uv}
\end{equation}
Our plan is as follows. After a discussion of the basic issues involved in the
analytic continuation of the 4-point function (Sec.\ \ref{sec:informal}), we
will introduce a representation in terms of the radial coordinates $\rho,
\bar{\rho}$ (Sec.\ \ref{Eucl4-point}), and use it to construct the analytic
continuation to the whole forward tube $\mathcal{T}_4$ (Sec.\ \ref{anal4-point}). This construction works because $\rho, \bar{\rho}$ remain
strictly inside the unit disc everywhere in the forward tube (Lemma
\ref{bound} and Eq.\ {\eqref{3b}}), a fundamental fact proved in Sec.\ \ref{PetrProof}. We then briefly review the well-established powerlaw bound on
$g (\rho, \bar{\rho})$ with respect to $\rho, \bar{\rho}$, and prove a
powerlaw bound on $| \rho (c) |, | \bar{\rho} (c) |$ with respect to $c \in
\mathcal{T}_4$. Combining these powerlaw bounds together, we will get a
powerlaw bound on the analytically continued 4-point function $G_4 (c)$, which by
Theorem \ref{ThVlad} implies (as $c$ approaches the Minkowski region) the
existence of the boundary limit of $G_4 (c)$ as a tempered distribution
(Sec.\ \ref{power4-point}). After establishing temperedness, we will derive the
Minkowski conformal invariance (Sec.\ \ref{ConfMink}), Wightman positivity
(Sec.\ \ref{sec:Wpos}), Wightman clustering (Sec.\ \ref{clusterWightman})
and local commutativity (Sec.\ \ref{local-comm}). Some of them do not rely
on conformal properties: for these we will use the standard arguments given by
Osterwalder and Schrader {\cite{osterwalder1973}}. In Sec.\ \ref{nonId}, we
will generalize the above results to non-identical scalars by using
Cauchy-Schwarz arguments.

\section{Informal introduction to basic issues}\label{sec:informal}

Here we wish to outline a few basic difficulties which must be dealt with when
analytically continuing the 4-point function. We will be using $u, v$ coordinates as
an example, although we will see below that other coordinates will be more
suitable for our task. Readers uninterested in philosophical discussions may
skip directly to Sec.\ \ref{Eucl4-point}.

\begin{figure}[h]\centering
	\includegraphics[width=0.5\textwidth]{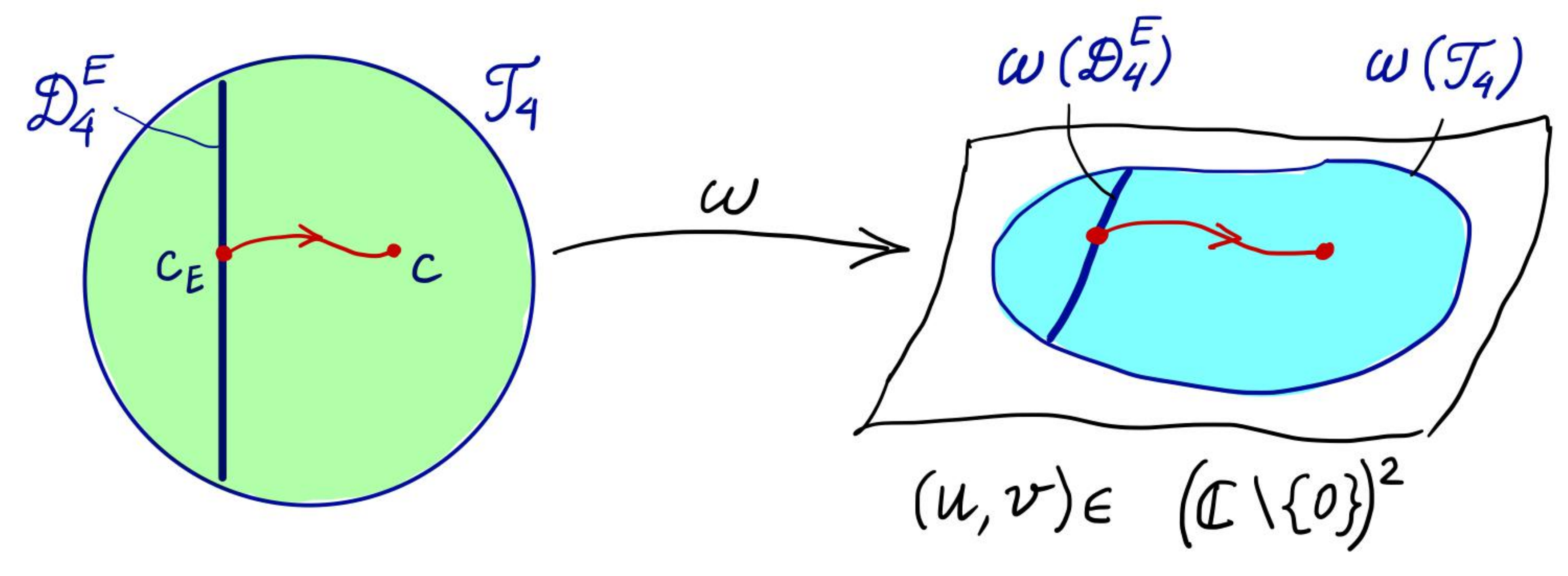}
	\caption{\label{fig:uv}Illustration of the discussion in Sec.\ \ref{sec:informal}.}
\end{figure}

Given any point $c$ of the forward tube, we can connect it to a Euclidean
point $c_E$ by a curve, and analytically continue the 4-point function along the
curve (see Fig.\ \ref{fig:uv}, left). The forward tube being simply connected,
the analytic continuation (if it exists) does not depend on the curve.
Furthermore, let us take into account that our conformal 4-point function
effectively only depends on two variables $u, v$. Applying Lemma \ref{xij2h},
we see that $u, v$ are both nonzero holomorphic functions on the forward tube.
Consider the map:
\begin{equation}
	\omega : c \rightarrow (u, v) \label{omegauv} .
\end{equation}
Since the forward tube is simply connected, we can consider this map as acting
from $\mathcal{T}_4$ to $(\widetilde{\mathbb{C} \backslash \{ 0 \}})^2$, where
tilde denotes the universal cover. Denote by $\omega (\mathcal{T}_4)$ and
$\omega (\mathcal{D}^E_4)$ the images of the forward tube and of its Euclidean
part under this map (see Fig.\ \ref{fig:uv}, right).

Now suppose that we found an analytic continuation of $g (u, v)$ from $\omega
(\mathcal{D}^E_4)$, where it is initially defined, to the whole of $\omega
(\mathcal{T}_4)$. Then we could immediately write down the analytic
continuation of the 4-point function to the forward tube as follows:\footnote{The
	prefactor analytically continues just as the 2-point and 3-point functions in chapter \ref{23-point}.}
\begin{equation}
	G_4 (c) = \frac{1}{(x_{12}^2 x_{34}^2)^{\Delta_{\mathcal{O}}}} g (u (c), v
	(c)) . \label{simple}
\end{equation}
This formula defines the analytic continuation to the forward tube as a
composition of two holomorphic functions:
\begin{equation}
	\mathcal{T}_4 \xrightarrow{\omega} \omega (\mathcal{T}_4)
	\xrightarrow{g} \mathbb{C}.
\end{equation}
We would like to use this strategy, but its direct implementation is hindered
by a couple of difficulties:
\begin{itemize}
	\item We don't know much about the shape or even topology of $\omega
	(\mathcal{T}_4)$. E.g.\ we don't know if this set is simply connected. The
	continuous image of a simply connected set, such as the forward tube, does
	not have to be simply connected (Fig.\ \ref{fig:simply}). If $\omega
	(\mathcal{T}_4)$ is not simply connected, there is no guarantee that $g (u,
	v)$ will be single-valued on it. And if $g (u, v)$ has branch cuts, then a
	simple formula like {\eqref{simple}} using only the endpoint values $(u (c),
	v (c))$ will not work; we will need to know in addition ``from which side of
	the cut'' we got to this point along the analytic continuation contour (see
	Fig.\ \ref{fig:simply}).
	
	\item To be sure, we don't know if the above difficulty is actually
	realized. Perhaps the set $\omega (\mathcal{T}_4)$ is, after all, simply
	connected, and $g (u, v)$ has a single-valued analytic continuation to it.
	Even if this is the case, how can we construct this extension starting from
	$g (u, v)$ in the Euclidean region?
\end{itemize}
\begin{figure}[h]\centering
	\includegraphics[width=0.5\textwidth]{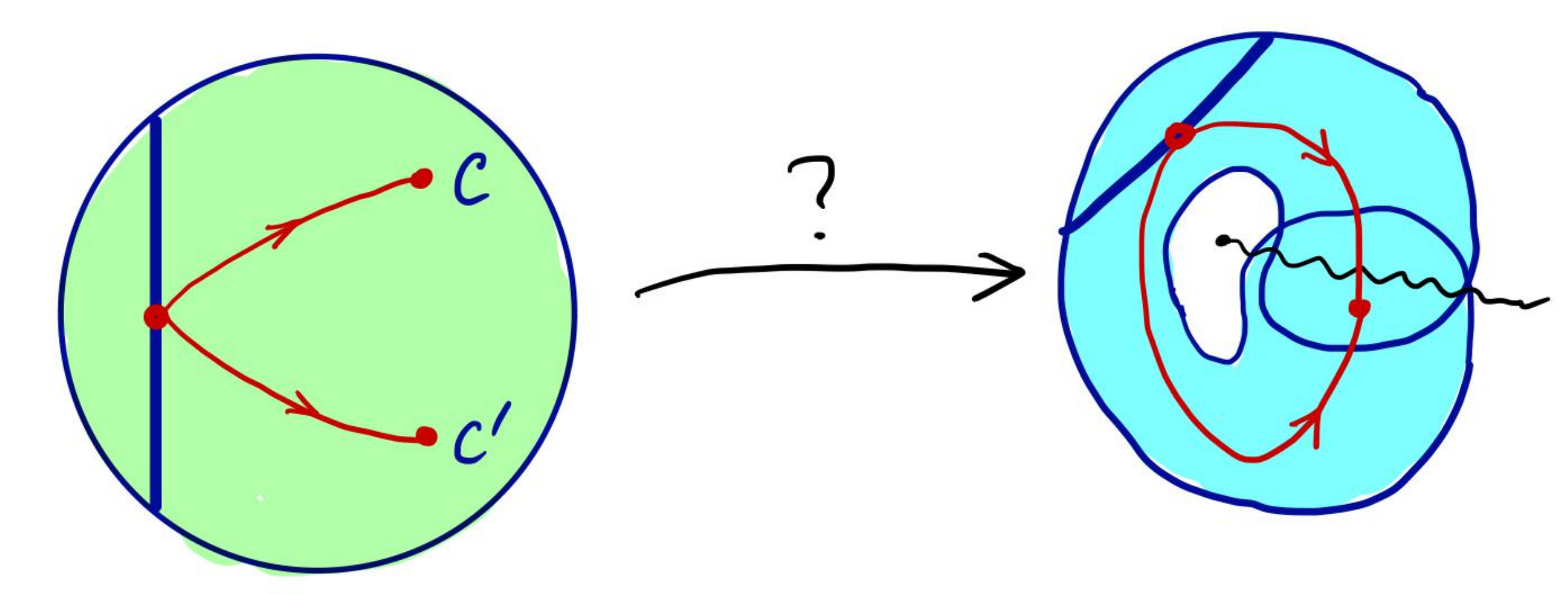}
	\caption{\label{fig:simply}Illustration of a potential difficulty if the set
		$\omega (\mathcal{T}_4)$ were not simply connected (see Sec.\ \ref{sec:informal}).}
\end{figure}

In this part of the thesis we will circumvent these difficulties rather than attacking them
head-on. In the Euclidean region, one often uses different variables to
parametrize the cross-ratios $u, v$, such as the Dolan-Osborn variables $z,
\bar{z}$, or the radial variables $\rho, \bar{\rho}$. As one can imagine, a
smart choice of Euclidean variables can greatly simplify the analytic
continuation. We will see that the radial variables are ideally suited for
this task, allowing a natural resolution of the above-mentioned difficulties.

\section{Euclidean 4-point function in radial coordinates}\label{Eucl4-point}

We first recall the well-known Dolan-Osborn variables $z, \bar{z}$
{\cite{Dolan:2000ut,Dolan:2003hv}}, which are two complex variables related to
$u, v$ by
\begin{gather}
	u = z \bar{z}, \quad v = (1 - z)  (1 - \bar{z}), \qquad\infixor 
	\label{zzbar}\\
	z, \bar{z} = \frac{1}{2} \left( 1 + u - v \pm \sqrt{(1 + u - v)^2 - 4
		u} \right) .  \label{zzbarsolved}
\end{gather}
Since in the Euclidean case we only consider non-coincident points, we have $u, v \neq 0$, and hence
$z, \bar{z} \neq 0, 1$. It is possible to fix a Euclidean conformal frame by
setting the four points to positions
\begin{equation}
	x_1 = 0, \quad x_2 = a \hat{e}_0 + b \hat{e}_1, \qquad x_3 = \hat{e}_0,
	\qquad x_4 = \infty \hat{e}_0, \label{frameE}
\end{equation}
where $\hat{e}_{\mu}$ is the standard orthonormal basis of $\mathbb{R}^d$.
Using this frame, we obtain $z, \bar{z} = a \pm i b$. This shows that in the
Euclidean, the variables $z, \bar{z}$ are complex-conjugate $(\bar{z} =
z^{\ast})$.

Euclidean configurations with real $z = \bar{z}$ correspond to four points
lying on a circle, which maps in the frame {\eqref{frameE}} to four points on
a line. The three possibilities $z < 0$, $z \in (0, 1)$, $z \in (1, + \infty)$
are then realized for different cycling orderings. \

The radial variables $\rho, \bar{\rho} \in \mathbb{C}$
{\cite{Pappadopulo:2012jk,Hogervorst:2013sma}} are defined in terms of the
Dolan-Osborn variables by the formula:
\begin{equation}
	\label{def:rho} \rho = f (z), \quad \bar{\rho} = f (\bar{z}), \quad f (w) :
	= \dfrac{w}{(1 + \sqrt{1 - w})^2} .
\end{equation}
The function $f (w)$ in this definition\footnote{The definition assumes the
	standard branch of the square root function.} is the uniformization map for
the complex plane minus the cut $(1, + \infty),$ i.e.\ it is a one-to-one map
of $\mathbb{C} \backslash [1, + \infty)$ onto the unit disk. Eq.
{\eqref{def:rho}} thus associates with any Euclidean configuration a pair of
complex conjugate $\rho, \bar{\rho}$ $(\bar{\rho} = \rho^{\ast})$ belonging to
the unit disk: $| \rho | \leqslant 1$. Moreover we have $| \rho | < 1$ except
for the Euclidean configurations with $z = \bar{z} \in (1, + \infty)$. As
explained above, this happens when four points lie on a circle in the cyclic
order $1324$. For such exceptional configurations one may define $\rho,
\bar{\rho}$ by continuity so that $| \rho | = 1$, $\bar{\rho} = \rho^{\ast}$.

The meaning of the coordinate $\rho$ is clarified by mapping the 4-point
configuration to a conformal frame (compare {\eqref{frameE}})
\begin{equation}
	x_1 = - \alpha \hat{e}_0 - \beta \hat{e}_1, \quad x_2 = \alpha \hat{e}_0 +
	\beta \hat{e}_1, \qquad x_3 = \hat{e}_0, \qquad x_4 = - \hat{e}_0,
	\label{frameErho}
\end{equation}
Using this frame, we obtain $\rho, \bar{\rho} = \alpha \pm i \beta$.

There is a small difference between $d = 2$ and $d \geqslant 3$ dimensions.
In $d \geqslant 3$, conformal frames {\eqref{frameE}} and {\eqref{frameErho}}
are unique only up to a sign of $b$ and $\beta$ (flipped rotating by $\pi$ in
the $12$ plane), which implies that pairs $(z, \bar{z})$ and $(\rho,
\bar{\rho})$ are defined only up to permutation. On the other hand in $d = 2$
flipping the sign of $b$ or $\beta$ is a parity transformation, which is not
in the identity component of the conformal group. Hence the conformal frames
are unique and $z, \bar{z}$ as well as $\rho, \bar{\rho}$ are individually
meaningful.

In a unitary Euclidean CFT, the 4-point function admits a power-series
expansion in the $\rho$ coordinate, absolutely convergent whenever $| \rho | <
1$ {\cite{Pappadopulo:2012jk}}. Specifically, \tmtextit{the function $g
	(c_E)$ appearing in the 4-point function {\eqref{def:Euclidean4-point}} of four
	identical scalar Hermitean primaries has a series expansion of the form}
\begin{equation}
	g (c_E) = \sum_{\delta, m} p_{\delta, m} r^{\delta} e^{i m \theta},
	\label{g:rhoexpansion}
\end{equation}
\tmtextit{where the sum runs over a discrete set of pairs $(\delta, m)$ with
	$\delta \geqslant 0$, $m \in 2\mathbb{Z}$, and the variables $r$, $\theta \in
	\mathbb{R}$ are the modulus and the phase of $\rho (c_E) = r e^{i \theta}$.
	The sum is absolutely convergent when $r = | \rho (c_E) | < 1$. In addition,
	we know that $| m | \leqslant \delta$ and $p_{\delta, m} \geqslant 0$ for all
	terms in {\eqref{g:rhoexpansion}}. Finally, when $d \geqslant 3$ we have
	$p_{\delta, - m} = p_{\delta, m}$, so that the r.h.s.\ of
	{\eqref{g:rhoexpansion}} is uniquely defined in spite of $\rho (c_E)$ being
	defined only up to complex conjugation.}

The readers familiar with this fact may skip to Sec.\ \ref{anal4-point} where we
will use it to perform analytic continuation. In the rest of this section we
recall how it follows from the CFT axioms
{\cite{Pappadopulo:2012jk,Fitzpatrick:2012yx}}.

We consider the 4-point function in the conformal frame configuration
{\eqref{frameErho}} and write it as the inner product of two states created by
the operators outside and inside a unit sphere $S$ centered at the origin:
\begin{equation}
	\langle \mathcal{O} (1, 0, \tmmathbf{0}) \mathcal{O} (- 1, 0, \tmmathbf{0}
	\tmmathbf{}) | \mathcal{O} (\alpha, \beta, \tmmathbf{0}) \mathcal{O} (-
	\alpha, - \beta, \tmmathbf{0}) \rangle \label{frame0}
\end{equation}
We can find a conformal transformation which maps the sphere $S$ to $x^0 = 0$
plane, its center 0 to $x_S$ and the infinity to $x_N$. This is the setup in
which we developed the CFT Hilbert space picture in Sec.\ \ref{Hilbert}.
Applying the inverse transformation, we are allowed to use the Hilbert space
language in the frame {\eqref{frame0}}, which is the familiar setting of
radial quantization. We decompose the radial quantization Hilbert space,
produced by local operators inserted at the origin, in orthonormalized
eigenstates $| \delta, m \rangle$ of the dilatation $D$ and the planar
rotation $M_{01}$. The ket state is expanded in this basis as
\begin{equation}
	\nobracket | \mathcal{O} (\alpha, \beta, \tmmathbf{0}) \mathcal{O} (-
	\alpha, - \beta, \tmmathbf{0}) \rangle = \sum_{\delta, m} c_{\delta, m}
	r^{\delta - 2 \Delta_{\varphi}} e^{i m \theta} | \delta, m \rangle .
	\label{ketexpr}
\end{equation}
The dependence of the expansion coefficients in this formula on $r$ and
$\theta$ is fixed by knowing how the state in the l.h.s.\ transforms under
rotations and dilatations. The transformation $\theta \rightarrow \theta +
\pi$ swaps the two operators leaving the state invariant for the considered
case of identical operators. Hence the state in the r.h.s.\ also must remain
invariant, proving that $m$ must be even.

Setting $r = 1, \theta = 0$ in {\eqref{ketexpr}}, we get
\begin{equation}
	\nobracket | \mathcal{O} (1, 0, \tmmathbf{0}) \mathcal{O} (- 1, 0,
	\tmmathbf{0}) \rangle = \sum_{\delta, m} c_{\delta, m} | \delta, m \rangle .
	\label{ketexpr1}
\end{equation}
In the considered frame the OS reflection is the inversion with respect to the
sphere $S$: $x^{\mu} \rightarrow x^{\mu} / x^2 .$ In particular, this leaves
$x_3$ and $x_4$ invariant. Applying this transformation to {\eqref{ketexpr1}},
we get
\begin{equation}
	\langle \mathcal{O} (1, 0, \tmmathbf{0}) \mathcal{O} (- 1, 0, \tmmathbf{0}
	\tmmathbf{}) | = \sum_{\delta, m} c^{\ast}_{\delta, m} \langle \delta, m | .
	\label{braexpr}
\end{equation}
Taking the inner product of {\eqref{ketexpr}} and {\eqref{braexpr}}, we get
\begin{equation}
	\langle \mathcal{O} (1, 0, \tmmathbf{0}) \mathcal{O} (- 1, 0, \tmmathbf{0}
	\tmmathbf{}) | \mathcal{O} (\alpha, \beta, \tmmathbf{0}) \mathcal{O} (-
	\alpha, - \beta, \tmmathbf{0}) \rangle = \sum_{\delta, m} | c_{\delta, m}
	|^2 r^{\delta - 2 \Delta_{\varphi}} e^{i m \theta} .
\end{equation}
Comparing this with Eq.\ {\eqref{def:Euclidean4-point}}, and using that $x_{12}^2 =
4 r^2$, $x_{34}^2 = 4$ in the considered conformal frame, we obtain
{\eqref{g:rhoexpansion}} with $p_{\delta, m} = 16^{\Delta_{\mathcal{O}}} |
c_{\delta, m} |^2 \geqslant 0$.

In the above argument we chose for simplicity the sphere of radius 1, but any
sphere of radius $r < r_0 < 1$ would work equally well and give rise to the
same expression. Absolute convergence for $r < 1$ follows, because both the
bra and the ket states are normalizable for such $r_0$ (while for $r_0 = 1$ as
above the bra state $\langle \varphi (x_3) \varphi (x_4 \tmmathbf{}) |$ is not
normalizable).

The restriction $| m | \leqslant \delta$ follows from the 2d unitarity bounds.
The 2d unitarity bound applies, as any $d$-dimensional CFT restricted to a
plane can be seen as a unitary 2d CFT. For 2d primaries of spin $J$ and
dimension $\Delta$, the 2d unitarity bound says $| J | \leqslant \Delta$. The
descendants at level $n \in \mathbb{Z}_{\geqslant 0}$ have $\delta = \Delta +
n,$ $| m - J | \leqslant n$, hence $| m | \leqslant \delta$ follows.

Finally, let us prove that $p_{\delta, m} = p_{\delta, - m}$ in $d \geqslant
3$. We consider Eq.\ {\eqref{ketexpr1}} and perform a $\pi$ rotation in the 12
plane. In the r.h.s. $| \delta, m \rangle \rightarrow | \delta, - m \rangle$
because $M_{01} \rightarrow - M_{01}$ under such a rotation. On the other hand
the l.h.s.\ does not change. This implies that we must have $c_{\delta, m} =
c_{\delta, - m}$, and hence $p_{\delta, m} = p_{\delta, - m}$. (In $d = 2$,
these properties also hold under the additional assumption of parity
invariance.)

\section{Analytic continuation}\label{anal4-point}

In this section we will construct the analytic continuation of the Euclidean
4-point function {\eqref{def:Euclidean4-point}} to the forward tube
$\mathcal{T}_4$ (recall the forward tube definition {\eqref{forward}}).
Analytic continuation to $\mathcal{D}_4 \subset \mathcal{T}_4$ has already
been given in {\cite{Qiao:2020bcs}}, Sec.\ 3.4,\footnote{Since the analytic continuation result is similar, we do not present that section of \cite{Qiao:2020bcs} in this thesis.} and we will use a somewhat
streamlined version of that construction. We will analytically continue to the
full forward tube $\mathcal{T}_4$, since this does not lead to additional
complications.

The analytic continuation will be given by the formula
\begin{equation}
	G_4 (c) = \frac{1}{(x_{12}^2 x_{34}^2)^{\Delta_{\mathcal{O}}}} g (c), \qquad
	c \in \mathcal{T}_4 . \label{G4c}
\end{equation}
Here the prefactor trivially analytically continues to $\mathcal{T}_4$
similarly to the 2-point and 3-point functions discussed in chapter \ref{23-point}. We
will construct $g (c)$, analytic continuation of $g (c_E)$, starting from Eq.
{\eqref{g:rhoexpansion}}.

First we have to define the variables $z (c)$, $\bar{z} (c)$ on the forward
tube, which is naturally done as follows. Given a configuration $c \in
\mathcal{T}_4$, we evaluate $u = u (c),$ $v = v (c)$ via {\eqref{uv}}. By
Lemma \ref{xij2h}, $u (c)$ and $v (c)$ are nonzero holomorphic functions on the
forward tube. We then define $z (c)$, $\bar{z} (c)$ via {\eqref{zzbarsolved}}:
\begin{equation}
	z (c), \bar{z} (c) = \frac{1}{2} \left( 1 + u (c) - v (c) \pm \sqrt{[1 + u
		(c) - v (c)]^2 - 4 u (c)} \right) . \label{zuv}
\end{equation}
Unlike for Euclidean configurations, for a general configuration $c \in
\mathcal{T}_4$ these are two complex numbers unrelated by conjugation. Since
$u (c)$ and $v (c)$ are nonzero, Eq.\ {\eqref{zzbar}} implies $z (c), \bar{z}
(c) \in \mathbb{C} \backslash \{ 0, 1 \}$.

Since Eq.\ {\eqref{zuv}} only defines $z (c), \bar{z} (c)$ up to permutation,
we view it as a map from the forward tube to $\mathbb{C}^2 /\mathbb{Z}_2$,
the set of unordered pairs of complex numbers. This map is continuous, and is
analytic everywhere except on $\Gamma \subset \mathcal{T}_4$ where the
expression under the square root vanishes:
\begin{equation}
	\Gamma = \{ c \in \mathcal{T}_4 : [1 + u (c) - v (c)]^2 - 4 u (c) = 0 \} .
	\label{Gamma}
\end{equation}
Actually, it turns out that in $d = 2$ one can resolve the ambiguity inherent
in Eq.\ {\eqref{zuv}} and define $z (c)$, $\bar{z} (c)$ as individually
globally holomorphic functions on $\mathcal{T}_4$. We will bring up this fact
below when we need it. Ref.\ {\cite{Qiao:2020bcs}}, App.\ A, showed that such an
improvement is impossible in $d \geqslant 3$.

The following result is fundamental for our construction. The proof is
elementary but a bit tricky and is postponed to Sec.\ \ref{PetrProof}.

\begin{lemma}
	\label{bound}For any $c \in \mathcal{T}_4$ we have $z (c)$, $\bar{z} (c)
	\nin [1, + \infty)$.
\end{lemma}

We next define $\rho (c)$, $\bar{\rho} (c)$ on $\mathcal{T}_4$, via
\begin{equation}
	\rho (c) = f (z (c)), \quad \bar{\rho} (c) = f (\bar{z} (c)), \label{rc}
\end{equation}
where $f$ is the same function as in {\eqref{def:rho}}, mapping $\mathbb{C}
\backslash [1, + \infty)$ onto the unit disk. By Lemma \ref{bound}, we then
have
\begin{equation}
	0 < | \rho (c) |, | \bar{\rho} (c) | < 1 \text{\qquad for any\qquad} c \in
	\mathcal{T}_4 .\footnote{Note that the converse is not true: the region
		in which $0<|\rho|,|\bar\rho|<1$ is larger than the forward tube. For example, it includes
		the extended forward tube (see Sec.~\ref{localCFT}).} \label{3b}
\end{equation}
{Moreover, $\rho(c)$ and $\bar\rho(c)$ are locally holomorphic away from $\G$.}
Because of this, and since Eq.\ {\eqref{g:rhoexpansion}} for the 4-point
function converges in the Euclidean for any $| \rho | < 1$, we may hope to use
Eq.\ {\eqref{g:rhoexpansion}} to analytically continue $g (c)$ to the whole
forward tube. We will now carry out this strategy. Note that some extra care
is needed, because $\rho (c)$ and $\bar{\rho} (c)$ are, just as $z (c)$ and
$\bar{z} (c)$, not globally holomorphic and are defined only up to permutation
(except in $d = 2$, see below), and because {\eqref{g:rhoexpansion}} contains
in general non-integer powers.

To begin with, we rewrite Eq.\ {\eqref{g:rhoexpansion}} equivalently as
\begin{equation}
	\label{g:rhoexpansion2} g (c_E) = \sum_{\delta, 0 \leqslant m \leqslant
		\delta} (\rho \bar{\rho})^{\delta / 2 - m / 2}  (p_{\delta, m} \rho^m +
	p_{\delta, - m} \bar{\rho}^m) \hspace{0.17em},
\end{equation}
Various pieces of this formula need to be analytically continued to the
forward tube. Consider first
\begin{equation}
	R (c) = \rho (c)  \bar{\rho} (c),
\end{equation}
which is a candidate for the analytic continuation of $\rho \bar{\rho}$ from
the Euclidean region. We can view it as a composition of two functions: $c
\mapsto (\rho (c), \bar{\rho} (c))$ which is a continuous function from
$\mathcal{T}_4$ to $\mathbb{C}^2 /\mathbb{Z}_2$ analytic away from $\Gamma$,
followed by $(\rho, \bar{\rho}) \mapsto \rho \bar{\rho}$ which is a continuous
holomorphic function from $\mathbb{C}^2 /\mathbb{Z}_2$ to $\mathbb{C}$. Hence $R
(c)$ is a continuous function on the forward tube, analytic everywhere except
perhaps on $\Gamma$. However, manifold $\Gamma$ has complex codimension one,
and by an analogue of Riemann's theorem about removable singularities we
conclude that $R (c)$ is in fact analytic also on $\Gamma$, and thus on the
whole $\mathcal{T}_4$.\footnote{The precise argument is as follows. Let us
	keep all complex coordinates fixed and vary just one, say $x_{1}^{0}$.
	There are two cases: either {\eqref{Gamma}} is identically zero as a function
	of $x_{1}^{0}$, or it is a nonzero polynomial of $x_{1}^{0}$. In
	the first case $R (c)$ is trivially holomorphic in $x_{1}^{0}$. In the
	second case {\eqref{Gamma}} vanishes at most for a few isolated values of
	$x_1^0$. We can then apply 1d Riemann's theorem to say that $R (c)$ is also
	analytic at those isolated points. By these arguments, we conclude that $R
	(c)$ is holomorphic in each variable separately. Finally, a continuous function
	of several complex variables holomorphic in each variable separately is jointly
	holomorphic {\cite{Osgood}}.}

In addition, $R (c)$ is nonzero in the forward tube. Thus we can lift $R (c)$
to a holomorphic function $\tilde{R} (c)$ from the forward tube to the universal
cover {$\widetilde{\mathbb{C} \backslash \{ 0 \}}$}. Composing this function with $z^h :
\widetilde{\mathbb{C} \backslash \{ 0 \}} \rightarrow \mathbb{C}$, we obtain
an analytic continuation of $(\rho \bar{\rho})^h$ for any $h \in \mathbb{R}$,
which we denote by $R_h (c)$. This discussion mirrors the one around Eq.
{\eqref{cxij2}} in chapter \ref{23-point}. However, unlike $x_{i j}^2$ in that
discussion, it is not true that $\tilde{R} (c)$ always belongs to the
principal sheet of {$\widetilde{\mathbb{C} \backslash \{ 0 \}}$}. So, in general, to
compute the phase of the analytically continued function, one should follow
the phase of $\rho \bar{\rho}$ along a curve joining $c_E$ to $c$.

Following a curve is perfectly fine as a theoretical device. For practical
computations of the phase, one may wish to use instead the following trick
which avoids having to look at the curve. (The reader happy to follow the curve may skip the trick and go directly to Eq.\ {\eqref{Phi}}.) Consider the
identity:
\begin{equation}
	\rho \bar{\rho} = \frac{1}{16} u (1 + \rho)^2 (1 + \bar{\rho})^2 =
	\frac{1}{16} \frac{x^2_{12} x^2_{34}}{x^2_{13} x^2_{24}} Y^2, \quad Y = (1 +
	\rho) (1 + \bar{\rho})^{}, \label{rhotrick}
\end{equation}
which follows by using $z = \frac{4 \rho}{(1 + \rho)^2}$, the inverse of the
relation {\eqref{def:rho}} between $\rho$ and $z$, as well as $u = z \bar{z}$
and the expression for $u$. The function $Y (c) = (1 + \rho (c)) (1 +
\bar{\rho} (c))$ is holomorphic on $\mathcal{T}_4$ by the same ``analyticity
on $\mathcal{T}_4 \backslash \Gamma$ plus Riemann's theorem'' argument as used
above for $\rho \bar{\rho}$. In addition, and this is the key point, because
$| \rho (c) |, | \bar{\rho} (c) | < 1$, we know that $Y (c) \in \mathbb{C}
\backslash (- \infty, 0]$. The upshot of the trick is that Eq.
{\eqref{rhotrick}} expresses $\rho \bar{\rho}$ as a product of factors which
all remain on the principal sheet of $z^h$ upon the analytic continuation.
Hence we can compute the analytic continuation of $(\rho \bar{\rho})^h$ by
\begin{equation}
	R_h (c) = \frac{1}{16^h} \frac{(x^2_{12})^h (x^2_{34})^h}{(x^2_{13})^h
		(x^2_{24})^h} Y (c)^{2 h} \qquad (h \in \mathbb{R}), \label{Rh}
\end{equation}
This determines the phase of $R_h (c)$ unambiguously without having to
look at the curve joining $c_E$ to $c$.

Next, we consider for an integer $m$ a function
\begin{equation}
	\Phi_m (c) = \rho (c)^m + \bar{\rho} (c)^m . \label{Phi}
\end{equation}
Just as $\rho \bar{\rho}$ and $Y$, it is continuous on $\mathcal{T}_4$ and
holomorphic on $\mathcal{T}_4 \backslash \Gamma$, and thus holomorphic on the
whole $\mathcal{T}_4$.

We can now define the analytic continuation of {\eqref{g:rhoexpansion2}}.
Consider first $d \geqslant 3$, when $p_{\delta, - m} = p_{\delta, m}$. In
this case the analytic continuation is given by the formula
\begin{equation}
	g (c) = \sum_{m, \delta, 0 \leqslant m \leqslant \delta} p_{\delta, m}
	R_{\delta / 2 - m / 2} (c) \Phi_m (c) . \label{eq:gtilde}
\end{equation}
This series consists of holomorphic functions, and it reduces to
{\eqref{g:rhoexpansion2}} in the Euclidean region. Furthermore, every term in
the series can be bounded in absolute value by:
\begin{eqnarray}
	| p_{\delta, m} R^{\delta / 2 - m / 2} (c) \Phi_m (c)  | & \leqslant &
	p_{\delta, m}  | \rho (c) \bar{\rho} (c) |^{\delta / 2 - m / 2} (| \rho (c)
	|^m + | \bar{\rho} (c) |^m) \nonumber\\
	& \leqslant & p_{\delta, m} r^{\delta - m} (r^m + r^m),  \label{maj}
\end{eqnarray}
where $r = r (c) = \max (| \rho (c) |, | \bar{\rho} (c) |)$, which is $< 1$ by
Eq.\ {\eqref{3b}}. Here we used $p_{\delta, m} \geqslant 0$ in the first line,
and $\delta - m \geqslant 0$ in the second line. The terms in the r.h.s.\ of
{\eqref{maj}} comprise a positive convergent series whose sum is the Euclidean
4-point function {\eqref{g:rhoexpansion2}} evaluated at $\rho = \bar{\rho} = r
(c)$. This proves that {\eqref{eq:gtilde}} converges uniformly on compact
subsets of $\mathcal{T}_4$, and hence defines a holomorphic function in
$\mathcal{T}_4$.

It remains to consider $d = 2$. As anticipated above, in this case the
functions $z (c), \bar{z} (c)$ are individually globally holomorphic on
$\mathcal{T}_4$. This can be seen introducing coordinates (see
{\cite{Qiao:2020bcs}}, Sec.\ 3.5)
\begin{equation}\label{def:coord2d}
	z_k = x_k^0 + i x_k^1, \qquad \bar{z}_k = x_k^0 - i x_k^1, \qquad k = 1, 2,
	3, 4.
\end{equation}
Then the explicit formulas for $z (c), \bar{z} (c)$ are given by:
\begin{equation}
	z (c) = \frac{(z_1 - z_2) (z_3 - z_4)}{(z_1 - z_3) (z_2 - z_4)}, \qquad
	\bar{z} (c) = \frac{(\bar{z}_1 - \bar{z}_2) (\bar{z}_3 -
		\bar{z}_4)}{(\bar{z}_1 - z_3) (\bar{z}_2 - \bar{z}_4)} . \label{zzbarglobal}
\end{equation}
The functions $\rho (c), \bar{\rho} (c)$ defined by {\eqref{rc}} are also
individually globally holomorphic on $\mathcal{T}_4$. As a consequence, the
functions $\rho (c)^m$ and $\bar{\rho} (c)^m$ are individually holomorphic in $d
= 2$, and not just their sum {\eqref{Phi}}. We can therefore define the
analytic continuation of $g (c)$ by the formula (compare {\eqref{eq:gtilde}}):
\begin{equation}
	g (c) = \sum_{m, \delta, 0 \leqslant m \leqslant \delta} R_{\delta / 2 -
		m / 2} (c)  [p_{\delta, m} \rho (c)^m + p_{\delta, - m} \bar{\rho} (c)^m]
	. \label{eq:gtilde2d}
\end{equation}
This formula would be appropriate for non-parity invariant 2d CFTs which may
have $p_{\delta, m} \neq p_{\delta, - m}$. Analyticity follows from the
uniform convergence on compact subsets, by the same argument as for $d
\geqslant 3$.

Finally, we wish to explain how the above construction may be translated into
the language of Sec.\ \ref{sec:informal}, to see how the issues raised there
are resolved. This is instructive but not strictly speaking necessary, so we
will be schematic. In the 2d case, when $\rho (c)$, $\bar{\rho} (c)$ are
individually defined, the translation is in terms of the map
\begin{equation}
	\Omega : c \mapsto (\rho (c), \bar{\rho} (c)) \in (\widetilde{\mathbb{D}
		\backslash \{ 0 \}})^2,
\end{equation}
where $\mathbb{D}$ is the open unit disk, and we lifted each of the maps $\rho
(c)$, $\overline{\rho (c)}$ to the universal cover of $\mathbb{D} \backslash
\{ 0 \}$. This map is the present analogue of $\omega$ in {\eqref{omegauv}}.
The function $g (\rho, \bar{\rho})$ extends analytically to the whole
$(\widetilde{\mathbb{D} \backslash \{ 0 \}})^2$, which makes it unnecessary to
understand the precise shape of $\Omega (\mathcal{T}_4)$.

For $d \geqslant 3$, $\rho (c)$, $\bar{\rho} (c)$ are defined only up to
permutation. Translation can then be done in terms of their symmetric
combinations $\rho \bar{\rho}, \rho + \bar{\rho}$. Any symmetric polynomial in
$\rho$, $\bar{\rho}$, such as the r.h.s.\ of {\eqref{Phi}}, can \ be expressed
as a polynomial in these coordinates. Let then $X$ be the image of
$(\mathbb{D} \backslash \{ 0 \})^2$ under the map $(\rho, \bar{\rho}) \mapsto
(\rho \bar{\rho}, \rho + \bar{\rho})$. The following map is holomorphic on
$\mathcal{T}_4$:
\begin{equation}
	\Omega : c \mapsto (\rho (c) \bar{\rho} (c), \rho (c) + \bar{\rho} (c)) \in
	\tilde{X},
\end{equation}
where we lifted to the universal cover. The above argument can be interpreted
as showing that the function $g (\rho, \bar{\rho})$ extends analytically to
the whole $\tilde{X}$. Understanding the precise shape of $\Omega
(\mathcal{T}_4)$ is once again unnecessary.

\section{Proof of $z, \bar{z} \nin [1, + \infty)$}\label{PetrProof}

Here we will prove Lemma \ref{bound} which played such a fundamental role in
the previous section. Just as for Lemma \ref{xij2h}, it will be helpful to use
the Minkowski metric. Thus we pass from Euclidean complex coordinates $x_k \in
\mathbb{C}^d$ to Minkowski complex coordinates $\zeta_k = (i x^0_k,
\mathbf{x}_k) \in \mathbb{C}^{1, d - 1}$. Definitions of $u, v$ are then
rewritten equivalently as
\begin{equation}
	u = \frac{\zeta_{12}^2 \zeta_{34}^2}{\zeta_{13}^2 \zeta_{24}^2}, \quad v =
	\frac{\zeta_{23}^2 \zeta_{14}^2}{\zeta_{13}^2 \zeta_{24}^2},
\end{equation}
where $\zeta_{i j} = \zeta_i - \zeta_j$ and $\zeta^2 = - (\zeta^0)^2
+\tmmathbf{\zeta}^2 .$ We denote
\begin{equation}
	\zeta_k = \xi_k + i \eta_k, \quad \xi_k, \eta_k \in \mathbb{R}^{1, d - 1} .
\end{equation}
We will thus use Minkowski norm for $\xi$'s, $\eta$'s and their differences.
The forward tube condition on $x_k$ is rewritten as $\eta_k - \eta_{k + 1}
\succ 0$ which is the notation for
\begin{equation}
	\eta^0_k - \eta^0_{k + 1} > 0 \quad \infixand \quad - (\eta_k - \eta_{k +
		1})^2 > 0. \label{FTc}
\end{equation}
We will need the following lemma which is related to Lemma \ref{xij2h} (see
the proof at the end of the section).

\begin{lemma}
	\label{Petr}Let $\zeta = \xi + i \eta$ and $\eta^2 < 0$. Then
	
	(a) $\zeta^2 \neq 0$;
	
	(b) Define $\zeta' = \xi' + i \eta'$ by
	\begin{equation}
		\zeta' = \zeta / \zeta^2,
	\end{equation}
	which is finite by Part (a). Then $\eta'$ belongs to the same causal part of
	the light cone (future or past) as $\eta$. I.e.\ $\eta \succ 0 \Rightarrow
	\eta' \succ 0$. Analogously, $\eta \prec 0 \Rightarrow \eta' \prec 0$.
\end{lemma}

Let us start the proof of Lemma \ref{bound}. The $z, \bar{z}$ are defined from
$u, v$ via {\eqref{zzbar}}. It is not hard to see from the first line of
{\eqref{zzbar}} that $z, \bar{z}$ are precisely the two solutions of the
quadratic equation
\begin{equation}
	z^2 - (1 + u - v) z + u = 0. \label{quadeq}
\end{equation}
We thus have to show that, assuming {\eqref{FTc}}, this equation has no
solutions which are real and belong to the interval $[1, + \infty)$.

Without loss of generality, we can assume that $\zeta_3 = 0$.\footnote{It is
	important to move $\zeta_3$ (or $\zeta_2$) to zero rather than $\zeta_1$ or
	$\zeta_4$, because only then, after applying the inversion, one gets causal
	information not only on $\eta'_k$'s but also on some of their differences.}
Then we have $\eta_1, \eta_2 \succ 0$ while $\eta_4 \prec 0$. Then we apply
Lemma \ref{Petr} and map the configuration $(\zeta_1, \zeta_2, 0, \zeta_4)$ to the
configuration $(\zeta_1', \zeta_2', \infty, \zeta_4')$ with $\eta_1', \eta_2'
\succ 0$ while $\eta_4' \prec 0$. These relations imply $\eta'_{14} \succ 0,
\eta'_{24} \succ 0$ which will be used below.

Since $u, v$ are invariant under the inversion, we have {(this can be checked by a direct computation)}
\begin{equation}
	u = \frac{(\zeta'_{12})^2}{(\zeta'_{24})^2}, \quad v =
	\frac{(\zeta'_{14})^2}{(\zeta'_{24})^2},
\end{equation}
and Eq.\ {\eqref{quadeq}} reduces to
\begin{equation}
	(\zeta'_{24})^2 z^2 - [(\zeta'_{24})^2 + (\zeta'_{12})^2 -
	(\zeta'_{14})^2] z + (\zeta'_{12})^2 = 0 .
\end{equation}
Using that $\zeta'_{12} = \zeta_{14}' - \zeta_{24}'$, this equation can be
written equivalently as
\begin{equation}
	(\zeta_{14}' + (z - 1) \zeta_{24}')^2 = 0 . \label{equivquad}
\end{equation}
Now let us suppose that $z \in [1, + \infty)$. Then
\begin{equation}
	\tmop{Im} [\zeta_{14}' + (z - 1) \zeta_{24}'] = \eta_{14}' + (z - 1)
	\eta_{24}' \succ 0 .
\end{equation}
Then Eq.\ {\eqref{equivquad}} is in contradiction with Lemma \ref{xij2h}. Lemma
\ref{bound} is demonstrated.

\tmtextbf{Proof of Lemma \ref{Petr}.} This was shown in
{\cite{Kravchuk:2018htv}}, footnote 74, and we reproduce the argument here for
completeness. Part (a) is a partial case of Lemma \ref{xij2h} (for $\eta \prec
0$ we should apply it to the complex conjugate vector $\zeta^{\ast} = \xi - i
\eta$). Let us show Part (b). To show that $\eta \succ 0 \Rightarrow \eta'
\succ 0$, we write
\begin{equation}
	\zeta' = \frac{\xi + i \eta}{\xi^2 - \eta^2 + 2 i (\xi, \eta)} = \frac{(\xi
		+ i \eta)  (\xi^2 - \eta^2 - 2 i (\xi, \eta))}{(\xi^2 - \eta^2)^2 + 4 (\xi,
		\eta)^2}  .
\end{equation}
So, up to a positive factor, $\eta'$ is given by
\begin{equation}
	(\xi^2 - \eta^2) \eta - 2 (\xi, \eta) \xi .
\end{equation}
For $\xi = 0$ this is given by $(- \eta^2) \eta \succ 0$. More generally, this
squares to
\begin{equation}
	(\xi^2 - \eta^2)^2 \eta^2 + 4 (\xi, \eta)^2 \xi^2 - 4 (\xi, \eta)^2  (\xi^2
	- \eta^2) = \eta^2  ((\xi^2 - \eta^2)^2 + 4 (\xi, \eta)^2) < 0 .
	\label{eta1sq}
\end{equation}
Therefore, for all $\xi$, we have that $\eta'$ is timelike. Since we have
shown that $\eta' \succ 0$ for $\xi = 0$, by continuity it follows that $\eta'
\succ 0$ for all $\xi$.

Finally, the implication $\eta \prec 0 \Rightarrow \eta' \prec 0$ follows by
complex conjugation.

\section{4-point function powerlaw bound}\label{power4-point}

We wish to show next that the analytically continued 4-point function satisfies a
powerlaw bound, so that we can apply Theorem \ref{ThVlad}. The prefactor in
Eq.\ {\eqref{G4c}} satisfies a powerlaw bound by Lemma \ref{x2bnd}.
Furthermore, Eq.\ {\eqref{maj}} implies that the analytic continuation $g (c)$
constructed in Sec.\ \ref{anal4-point} is bounded by a Euclidean 4-point function,
namely:
\begin{equation}
	| g (c) | \leqslant g_E (c_{\ast}), \label{ggE}
\end{equation}
where $c_{\ast}$ is any Euclidean 4-point function configuration having $\rho
(c_{\ast}) = \bar{\rho} (c_{\ast}) = r = r (c) = \max (| \rho (c) |, |
\bar{\rho} (c) |)$. We choose the conformal frame {\eqref{frameErho}}:
\begin{equation}
	c_{\ast} : \qquad x_1 = - r \hat{e}_0,\ x_2 = r \hat{e}_0,\ x_3 = \hat{e}_0,\
	x_4 = - \hat{e}_0 .
\end{equation}
Using the convergent OPE in the $x_2 \rightarrow x_3$, $x_1 \rightarrow x_4$
channel, we have the asymptotics
\begin{equation}
	G_4^E (c_{\ast}) \sim \frac{1}{(1 - r)^{4 \Delta_{\varphi}}} \qquad (r
	\rightarrow 1) . \label{g4cstar}
\end{equation}
The function $g_E (c_{\ast})$ satisfies the same asymptotics up to a constant,
being related to $G_4^E (c_{\ast})$ via Eq.\ {\eqref{def:Euclidean4-point}} by a
factor which is non-singular in the $r \rightarrow 1$ limit. Since $g_E
(c_{\ast})$ is a positive monotonically increasing function for $0 \leqslant r
< 1$ (see Eq.\ {\eqref{g:rhoexpansion}}), we conclude that it has a bound
\begin{equation}
	g_E (c_{\ast}) \leqslant \frac{\tmop{const} .}{(1 - r (c))^{4
			\Delta_{\varphi}}}, \label{gEcstar}
\end{equation}
and $| g (c) |$ by {\eqref{ggE}} satisfies the same bound.

The upshot of this discussion is that we will have a powerlaw bound on $G_4
(c)$ if we manage to get a powerlaw bound on $\frac{1}{1 - r (c)}$. We will
next state and prove such a bound.

{Before launching into the technical discussion, let us discuss
	intuitively why a result like this is expected to be true. We know (Lemma
	\ref{bound}) that $| \rho (c) |, | \bar{\rho} (c) | < 1$ and now we wish to
	prove that $| \rho (c) |, | \bar{\rho} (c) |$ do not approach 1 too quickly as
	$c$ goes to the Minkowski boundary of the forward tube. This may remind the
	reader of the Schwarz-Pick lemma, which says that if $f (w)$ is a function
	from a unit disk to itself and $f (0) = 0$, then $| f (w) | \leqslant | w |$,
	hence providing a bound on how fast $| f (w) |$ can approach 1 as $| w |
	\rightarrow 1$. In the 2d case, when $\rho (c)$ and $\bar{\rho} (c)$ are
	individually defined holomorphic functions in the forward tube, it is indeed
	possible to use the Schwarz-Pick lemma to prove a powerlaw bound on $\max (|
	\rho (c) |, | \bar{\rho} (c) |)$ {\cite{lecturesSaclay}}. It should be
	possible to generalize the Schwarz-Pick argument to any $d$, although we have
	not worked it out in full details.\footnote{For any $d$, the Schwarz-Pick
		lemma allows a natural generalization to holomorphic functions in the forward
		tube {\cite{KravchukSchwarz-Pick}}.} The proof below will be different and
	more direct: it will simply mimic the proof of Lemma \ref{bound}, replacing
	all ``$> 0$'' inequalities by ``$\geqslant \varepsilon$'' with an explicit
	positive $\varepsilon$.}

\subsection{A powerlaw bound on $\frac{1}{1 - r (c)}$}\label{section:1-r}

Let us introduce some notation. We will measure the size of a complex vector
$\zeta \in \mathbb{C}^{1, d - 1}$ by $| \zeta |$,
\begin{equation}
	| \zeta |^2 = | \zeta^0 |^2 + | \zeta^1 |^2 + \cdots + | \zeta^{d -
		1} |^2 .
\end{equation}
Clearly $| (\zeta_1, \zeta_2) | \leqslant | \zeta_1 | | \zeta_2 |$. We also
define for $\zeta = \xi + i \eta$, $\xi, \eta \in \mathbb{R}^{1, d - 1}$, and
$\eta^2 < 0$ (i.e.\ timelike)
\begin{equation}
	S (\zeta) = \max \Bigl( \frac{1}{\sqrt{- \eta^2}}, | \zeta | \Bigr) .
\end{equation}
Thus $S (\zeta)$ is large either if some component of $\zeta$ (real or
imaginary) is large or if $\eta$ approaches the light cone. Note that $S
(\zeta) \geqslant 1$ for any $\zeta$. We will never need $S (\xi + i \eta)$
for spacelike $\eta$.

Finally we consider an analogous function on $\mathcal{T}_4$:
\begin{equation}
	S (c) = \max_{i < j} S (\zeta_{i j}),
\end{equation}
which becomes large if any of $S (\zeta_{i j})$ become large. We claim that
there is the following bound (recall $r (c) = \max (| \rho (c) |, | \bar{\rho}
(c) |)$)
\begin{equation}
	\frac{1}{1 - r (c)} \leqslant 720 S (c)^{12} \qquad (c \in \mathcal{T}_4)
	\label{rhobound} .
\end{equation}
This bound will be shown for any $c$ in the forward tube, which is the natural
setting. When we specify to $c \in \mathcal{D}_4 \subset \mathcal{T}_4$ [see
Eq.\ {\eqref{def:Dn}}], we have
\begin{equation}
	S (c) = \max_{i < j} \max \left\{  \frac{1}{| \epsilon_i - \epsilon_j
		|}, | x_i - x_j | \right\} . \label{Scbound}
\end{equation}
Eq.\ {\eqref{rhobound}} then becomes a powerlaw bound for $\frac{1}{1 - r (c)}$
on $\mathcal{D}_4$ of the form {\eqref{powerlawbound}}, precisely as needed
for applying Theorem \ref{ThVlad}.

The proof of the bound {\eqref{rhobound}} will build upon the proof of $z,
\bar{z} \nin [1, + \infty)$ given in Sec.\ \ref{PetrProof}. There we showed
that $z$ solves Eq.\ {\eqref{equivquad}}, which however is inconsistent
for $z \in [1, + \infty)$ and $c$ in the forward tube. Here we will use the
same Eq.\ {\eqref{equivquad}}, but make the rest of the argument
quantitative, by showing that if $c$ stays away from the boundary or infinity
of the forward tube, so that $S (c)$ is bounded, then both $z (c)$ and
$\bar{z} (c)$ must stay a finite distance away from $[1, + \infty)$, as
measured by an upper bound on $\frac{1}{1 - r (c)}$ expressed by Eq.
{\eqref{rhobound}}. The proof is straightforward but somewhat technical and we
split it into a series of lemmas.

\begin{lemma}
	\label{zeta2bnd}Let $\zeta = \xi + i \eta$, $\eta^2 < 0$. Then for any $\xi$
	\begin{equation}
		| \zeta^2 | \geqslant (- \eta^2) . \label{zeta2bnd0}
	\end{equation}
\end{lemma}

\begin{proof}
	This is a generalization of Lemma \ref{x2bnd}(a) and could be proven
	analogously. We give a slightly different proof for a change. We have
	\begin{equation}
		| \zeta^2 |^2 = (\xi^2 - \eta^2)^2 + 4 (\xi, \eta)^2 = (\xi^2)^2 +
		(\eta^2)^2 + 2 [2 (\xi, \eta)^2 - \xi^2 \eta^2] .
	\end{equation}
	The lemma now follows from the inequality:
	\begin{equation}
		2 (\xi, \eta)^2 - \xi^2 \eta^2 \geqslant 0 . \label{ineqtoprove}
	\end{equation}
	Eq.\ {\eqref{ineqtoprove}} is obvious for $\xi^2 \geqslant 0$, so let us
	consider $\xi^2 < 0$. By Lorentz invariance and homogeneity it's enough to
	consider $\xi = (\pm 1, 0, \ldots, 0)$ in which case the l.h.s.\ of
	{\eqref{ineqtoprove}} becomes $(\eta^0)^2 +\tmmathbf{\eta}^2 \geqslant 0$.
\end{proof}

Then we have the following strengthening of Lemma \ref{Petr}(b):

\begin{lemma}
	\label{Sinv}Let $\zeta = \xi + i \eta$, $\eta^2 < 0$, and $\zeta' = \zeta /
	\zeta^2$. Then
	\begin{equation}
		S (\zeta') \leqslant [S (\zeta)]^3 .
	\end{equation}
\end{lemma}

\begin{proof}
	We have
	\begin{equation}
		| \zeta' | = \frac{| \zeta |}{| \zeta^2 |} \leqslant \text{[by Lemma
			\ref{zeta2bnd}]}  \frac{| \zeta |}{- \eta^2} \leqslant S (\zeta)^3 .
	\end{equation}
	We also have (see the proof of Lemma \ref{Petr}, in particular Eq.
	{\eqref{eta1sq}}) that $\eta^{\prime 2} < 0$ and
	\[ \frac{1}{- \eta^{\prime 2}} = \frac{| \zeta^2 |^2}{- \eta^2} \leqslant
	\text{[by Lemma \ref{zeta2bnd}]} S (\zeta)^6 . \]
\end{proof}

\begin{lemma}
	\label{SSS}Let $\zeta_i \in \mathbb{C}^{1, d - 1}, \eta_i \succ 0$ ($i =
	1, 2$). Then
	\begin{equation}
		S (\zeta_1 + \zeta_2) \leqslant S (\zeta_1) + S (\zeta_2) .
	\end{equation}
\end{lemma}

\begin{proof}
	We have $| \zeta_1 + \zeta_2 | \leqslant | \zeta_1 | + | \zeta_2 |$ and $-
	(\eta_1 + \eta_2)^2 \geqslant - \eta^2_1 - \eta_2^2 $ {(since $\eta_1\cdot\eta_2 <0$).}
\end{proof}

\begin{lemma}
	\label{lemma8}Let $\Upsilon_i = \Phi_i + i \Psi_i \in \mathbb{C}^{1, d - 1},
	\Phi_i, \Psi_i \in \mathbb{R}^{1, d - 1}$, $\Psi_i \succ 0$ (i=1,2), and $z$
	solves the equation
	\begin{equation}
		(\Upsilon_1 + (z - 1) \Upsilon_2)^2 = 0 . \label{Yeq}
	\end{equation}
	Then
	\begin{equation}
		1 - | \rho (z) | \geqslant \delta_0 : = \frac{1}{45 S^4}, \qquad S = \max
		(S (\Upsilon_1), S (\Upsilon_2)) . \label{rhoclose}
	\end{equation}
\end{lemma}

\begin{proof}
	Note that $z = 4 \rho / (1 + \rho)^2$, and so Eq.\ {\eqref{Yeq}} can be
	rewritten as
	\begin{equation}
		((\rho + 1)^2 \Upsilon_1 - (\rho - 1)^2 \Upsilon_2)^2 = 0. \label{eqrew1}
	\end{equation}
	For $\rho = e^{i \alpha}$, multiplying this equation by $e^{- 2 i \alpha}$,
	it becomes
	\begin{equation}
		(\Upsilon)^2 = 0, \qquad \Upsilon \equiv \left( 2 \cos \frac{\alpha}{2}
		\right)^2 \Upsilon_1 + \left( 2 \sin \frac{\alpha}{2} \right)^2
		\Upsilon_2,
	\end{equation}
	which contradicts Lemma \ref{Petr}(a), since $\tmop{Im} \Upsilon \succ 0$.
	So $\rho$ cannot lie precisely on the unit circle (as we already knew). It
	should then not be surprising that it also cannot get too close to the unit
	circle, which is what {\eqref{rhoclose}} says. This can be shown by a
	straightforward although somewhat technical generalization of the above
	argument.
	
	Denoting $\rho = r e^{i \alpha} = e^{i \alpha} - \delta e^{i \alpha}$,
	$\delta = 1 - r > 0$, and multiplying {\eqref{eqrew1}} by $e^{- 2 i
		\alpha}$, it becomes
	\begin{equation}
		\left( \left( 2 \cos \frac{\alpha}{2} - \delta e^{i \alpha / 2} \right)^2
		\Upsilon_1 + \left( 2 \sin \frac{\alpha}{2} + i \delta e^{i \alpha / 2}
		\right)^2 \Upsilon_2 \right)^2 = 0,
	\end{equation}
	or
	\begin{equation}
		(\Upsilon + \Upsilon')^2 = 0 \label{Y+Y}
	\end{equation}
	with
	\begin{eqnarray}
		& \Upsilon = 4 \cos^2 \frac{\alpha}{2} \Upsilon_1 + 4 \sin^2
		\frac{\alpha}{2} \Upsilon_2, & \\
		& \Upsilon' = \kappa_1 \Upsilon_1 + \kappa_2 \Upsilon_2, & \\
		& \kappa_1 = - 4 \cos \frac{\alpha}{2} \delta e^{i \alpha / 2} + \delta^2
		e^{i \alpha}, \quad \kappa_2 = 4 i \sin \frac{\alpha}{2} \delta e^{i
			\alpha / 2} - \delta^2 e^{i \alpha} . & 
	\end{eqnarray}
	So for $\delta$ small, $\Upsilon'$ is a small correction to $\Upsilon$. We
	write $\tmop{Im} (\Upsilon + \Upsilon') = \Psi + \Psi'$, where
	\[ \Psi = \tmop{Im} \Upsilon = 4 \cos^2 \frac{\alpha}{2} \Psi_1 + 4 \sin^2
	\frac{\alpha}{2} \Psi_2, \quad \Psi' = \tmop{Im} \Upsilon' . \]
	We know that $\Psi \succ 0$. In addition we also have a lower bound on $-
	\Psi^2$:
	\begin{equation}
		- \Psi^2 \geqslant 16 \cos^4 \frac{\alpha}{2} (- \Psi^2_1) + 16 \sin^4
		\frac{\alpha}{2} (- \Psi^2_2) \geqslant \frac{1}{S^2} \times 16 \min
		\left\{ \cos^4 \frac{\alpha}{2}, \sin^4 \frac{\alpha}{2} \right\} =
		\frac{4}{S^2} . \label{lowerbnd}
	\end{equation}
	We will now show that $- (\Psi + \Psi')^2$ remains strictly positive if
	$\delta < \delta_0$. This will imply, by Lemma \ref{Petr}(a), that Eq.
	{\eqref{Y+Y}} cannot hold, and hence we must have $\delta \geqslant
	\delta_0$, i.e.\ Eq.\ {\eqref{rhoclose}}, proving the lemma.
	
	To implement this natural strategy, we will need only crude estimates of
	the size of various terms. Note that $\delta_0 < 1$ since $S \geqslant 1$,
	so in particular we have $\delta^2 \leqslant \delta$. Using this we have the
	bounds $| \kappa_i | \leqslant 5 \delta$, and hence an upper bound
	\begin{equation}
		| \Psi' | \leqslant | \Upsilon' | \leqslant 10 \delta S.
	\end{equation}
	We also have an upper bound $| \Psi | \leqslant 4 S$. Using these,
	{\eqref{lowerbnd}}, and $\delta^2 \leqslant \delta$, we have:
	\begin{eqnarray}
		- (\Psi + \Psi')^2 = - \Psi^2 - 2 (\Psi, \Psi') - (\Psi')^2 & \geqslant &
		\frac{4}{S^2} - 2 | \Psi | | \Psi' | - | \Psi' |^2 \nonumber\\
		& \geqslant & \frac{4}{S^2} - 80 \delta S^2 - 100 \delta^2 S^2
		\nonumber\\
		& \geqslant & \frac{4}{S^2} - 180 \delta S^2 = \frac{4 (1 - \delta /
			\delta_0)}{S^2} \,,
	\end{eqnarray}
	which is strictly positive for $\delta < \delta_0$. As explained above this
	proves the lemma.
\end{proof}

Finally we can prove {\eqref{rhobound}}. We repeat the proof of Lemma
\ref{bound} given in Sec.\ \ref{PetrProof}. As there, we reduce to
configuration having $\zeta_3 = 0$ and obtain that $z$ (as well as $\bar{z}$)
is a solution of Eq.\ {\eqref{equivquad}}, which has the form
{\eqref{Yeq}} with
\begin{equation}
	\Upsilon_1 = \zeta_{14}' = \zeta_1' - \zeta_4', \quad \Upsilon_2 =
	\zeta_{24}' = \zeta_2' - \zeta_4', \quad \zeta_i' = \zeta_i / \zeta^2 \quad
	(i = 1, 2, 4) . \label{def:Upsilon12}
\end{equation}
Let us write $\Upsilon_i = \Phi_i + i \Psi_i \in \mathbb{C}^{1, d - 1},
\Phi_i, \Psi_i \in \mathbb{R}^{1, d - 1}$. As was already pointed out in
Sec.\ \ref{PetrProof}, we have $\Psi_i \succ 0$ ($i = 1, 2$). Furthermore,
by Lemma \ref{Sinv} we know that $S (\zeta_i') \leqslant S (c)^3$, and then
applying Lemma \ref{SSS} that $S (\Upsilon_i) \leqslant 2 S (c)^3$. Thus Lemma
\ref{lemma8} implies {\eqref{rhobound}} (note that $720 = 45 \times 16$).

\begin{remark}
	The bound {\eqref{rhobound}} is not optimal. We will prove a better bound in
	chapter \ref{secondpass}, by a different argument. 
\end{remark}

Let us recap. In Sec.\ \ref{anal4-point} we have analytically continued the
Euclidean 4-point function to the forward tube, and here we showed that this
analytic continuation satisfies a powerlaw bound. Then by Theorem
\ref{ThVlad}, the Minkowski 4-point function defined as the limit {\eqref{limit}}
exists, is a Lorentz-invariant tempered distribution, and satisfies Wightman
spectral condition. In the remainder of this section we will show that this
distribution is also conformally invariant (Sec.~\ref{ConfMink}), that it satisfies the remaining
Wightman axioms (positivity in Sec.~\ref{sec:Wpos}, clustering in Sec.~\ref{clusterWightman}, and local commutativity in Sec.~\ref{local-comm}). {Later in Sec.~\ref{OPEconvMink} we will also show that it
	can be computed by a convergent (in the sense of distributions) OPE.}

Now that we know that the Minkowski 4-point function is a distribution everywhere,
one may inquire about the regularity of this distribution. {E.g.\ for some
	configurations the 4-point function is actually real-analytic (see part \ref{part:ope}). We will come back to
	this question in the conclusion section.}

\section{Conformal invariance}\label{ConfMink}

Conformal invariance of Euclidean 4-point function {\eqref{def:Euclidean4-point}} can
be described as invariance under finite conformal transformations $x
\rightarrow x' = f (x)$,
\begin{equation}
	\Omega_1 \Omega_2 \Omega_3 \Omega_4 G^E_4 (x'_1, x'_2, x'_3, x'_4)
	= G^E_4 (x_1, x_2, x_3, x_4), \label{EfinInv}
\end{equation}
where $\Omega_i = J (x_i)^{\Delta_{\mathcal{O}}}$ and $J (x) = \det
(\partial f^{\mu} / \partial x^{\nu})^{1 / d}$ is the local scale factor.
Alternatively, and equivalently, this can be expressed as invariance under
infinitesimal conformal transformations, a conformal Ward identity, which says
that the Euclidean correlator is annihilated by a sum of differential
operators, one per point:
\begin{equation}
	\sum_{i = 1}^4 \mathcal{D} (x_i, \partial_{x_i}) G^E_4 (x_1, x_2, x_3, x_4)
	= 0 . \label{Ward}
\end{equation}
There is a differential operator per conformal group generator
($\partial^{\mu}$ for $P_{\mu}$, $x^{\mu} \partial^{\nu} - x^{\nu}
\partial^{\mu}$ for $M_{\mu \nu}$, $x \cdot \partial +
\Delta_{\mathcal{O}}$ for $D$, $x^2 \partial^{\mu} - 2 x^{\mu} (x \cdot
\partial) - 2 x^{\mu} \Delta_{\mathcal{O}}$ for $K_{\mu}$).

Since all these differential operators have polynomial coefficients, Ward
identities {\eqref{Ward}} continue to hold in the forward tube for the
function $G (x_1, x_2, x_3, x_4)$. Taking the limit to the Minkowski boundary,
we obtain that the Minkowski 4-point function satisfies infinitesimal Minkowski
conformal invariance expressed by the Ward identities.

The possibility to take the limit is guaranteed by the standard result that
distributional limits commute with derivatives. Indeed, suppose that we have,
in the sense of distributions, $\lim_{\varepsilon \rightarrow 0}
f_{\varepsilon} = g$. This means that for any test function $\varphi$, we have
$\lim_{\varepsilon \rightarrow 0} (f_{\varepsilon}, \varphi) = (g, \varphi)$.
But then for any derivative $\partial,$
\begin{equation}
	(\partial g, \varphi) = - (g, \partial \varphi) = - \lim_{\varepsilon
		\rightarrow 0} (f_{\varepsilon}, \partial \varphi) = \lim_{\varepsilon
		\rightarrow 0} (\partial f_{\varepsilon}, \varphi), \label{derLimit}
\end{equation}
which implies that $\lim_{\varepsilon \rightarrow 0} \partial f_{\varepsilon}
= \partial g$. A similar argument shows that the limit commutes with
multiplication of distributions by polynomials. All this is analogous to how
we prove Lorentz invariance of the Minkowski correlator in App.\ \ref{Vlad}.

So we have shown that the Minkowski 4-point function satisfies Lorentzian
conformal Ward identities. This means that
\begin{equation}
	\sum_{i = 1}^4 (\mathcal{D}_i G^M_4, \varphi) = 0, \label{MinkConfWard}
\end{equation}
where $\mathcal{D}_i$ are the analytic continuations of the Euclidean
differential operators to Minkowski space, and the pairing with the Schwartz
test functions is defined by integration by parts. Note that the conformal
Ward identities in Minkowski space hold also at coincident points (i.e.\ the
test function $\varphi$ does not have to be zero at coincident points).

Now let us discuss invariance of Minkowski 4-point function under
\tmtextit{finite} Lorentzian conformal transformations. Since $G_4^M$ is a
distribution, the appropriate form of writing is to transform the test
function:
\begin{equation}
	(G_4^M, \varphi) = (G_4^M, \varphi^f), \label{MfinInv}
\end{equation}
where $\varphi^f (x_1, \ldots, x_4) = \varphi (f^{- 1} (x_1), \ldots, f^{- 1}
(x_4)) \prod_{i = 1}^4 J (f^{- 1} (x_i))^{\Delta_{\mathcal{O}} - d}$. However
we have to be careful. This invariance is true not for every test function
$\varphi$ but only for an $f$-dependent subset of test functions.

Let $f_t$ be a smooth family of Lorentzian conformal transformations
connecting $f$ to the identity: $f_0 = \tmop{id}$, $f_1 = f$. Suppose that
\begin{equation}
	\text{$\varphi^{f_t}$ is a Schwartz function for any $f_t$ in the family} .
	\label{req1}
\end{equation}
Then we can integrate infinitesimal conformal invariance and prove that
{\eqref{MfinInv}} is true. For translations, Lorentz transformations and
dilatations, Eq.\ {\eqref{req1}} is clearly satisfied and Eq.\ {\eqref{MfinInv}}
holds for any $\varphi$. However, for general conformal transformations,
{\eqref{req1}} may not necessarily be true. The problems will appear if $f$
is singular on the support of $\varphi$, as $\varphi^f$ may then not be a
Schwartz function. As a concrete example, consider the Lorentzian special
conformal transformation:
\begin{equation}
	f (x) = \frac{x^{\mu} + x^2 b^{\mu}}{1 + 2 x \cdummy b + x^2 b^2} .
	\label{fb}
\end{equation}
The corresponding scale factor is $J (x) = \frac{1}{1 + 2 x \cdummy b + x^2
	b^2}$. Take for definiteness spacelike $b = \beta \hat{e}_1$, where $\beta >
0$ and $\hat{e}_1$ is the unit vector in the $x^1$ direction. The
transformation {\eqref{fb}} is then singular for $x^0 = \pm | \mathbf{x}+
\beta^{- 1} \hat{e}_1 |$, where the scale factor blows up, i.e.\ on the
light cone whose vertex is at $x = - \beta^{- 1} \hat{e}_1$.\footnote{Recall
	that we are using $-, + \cdots +$ Minkowski signature.} Scaling $\beta$ to zero
we can connect the transformation {\eqref{fb}} to the identity. Under this
scaling the light cone of singularities moves away to infinity along the
negative $x^1$ direction. Requirement {\eqref{req1}}, and hence finite
invariance {\eqref{MfinInv}}, will hold if the light cone of singularities,
while moving away, does not touch the support of $\varphi$ (see Fig.\ \ref{figure:supp0} for the 2d case).

\begin{figure}[h]\centering
	\includegraphics[width=8.24452315361406cm,height=4.28230355503083cm]{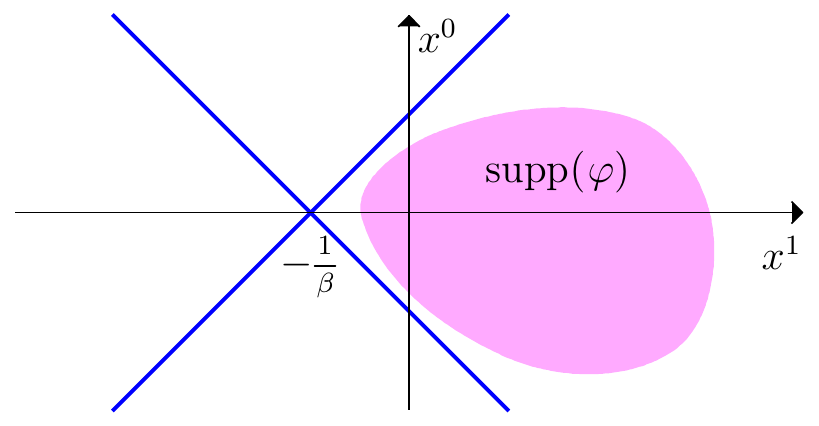}
	\caption{\label{figure:supp0}In
		the 2d case, the special conformal transformation {\eqref{fb}} is singular on
		the blue light cone $x^0 = \pm | x^1 + \beta^{- 1} |$. Suppose $\varphi$ is
		supported as shown on the right of the light cone. As $\beta \rightarrow 0$,
		the light cones moves towards the left infinity and does not touch $\tmop{supp}
		(\varphi)$. Therefore, such a $\varphi$ satisfies the condition for the 
		invariance under a finite special conformal transformation {\eqref{fb}}. }
\end{figure}

Note that such a support requirement still leaves possibility for both
spacelike and timelike separations among the points $x_i$ in the support of
$\varphi$. For $x_i \in \tmop{supp} \varphi$, the points $f (x_i)$ will have
the same causal structure as the points $x_i$, i.e.~$(f (x_i) - f (x_j))^2$
will have the same sign as $(x_i - x_j)^2$. This follows from the fact that $J
(x) > 0$ on $\tmop{supp} (\varphi)$, as guaranteed by being able to
continuously connect to the identity without singularities.

In the early CFT days, it was considered puzzling that Lorentzian special
conformal transformations may change the causal structure of a point
configuration. As we see here, the puzzle can be avoided by either limiting to
infinitesimal conformal invariance, or by restricting the class of test
functions so that the causal structure is preserved. A third way to deal with
the puzzle is to consider the Lorentzian conformal transformations acting on
the Lorentzian cylinder as opposed to the Minkowski space
{\cite{Luscher:1974ez}}. We will revisit the Lorentzian cylinder in our future
publication {\cite{paper3}}.

\begin{remark}
	We would like to contrast the Minkowski conformal Ward identities
	{\eqref{MinkConfWard}} with conformal Ward identities valid for Euclidean
	correlators. Euclidean correlators are real-analytic away from coincident
	points and naturally satisfy conformal Ward identities for such
	configurations. Although in this part of the thesis we don't need it, in some questions
	it might be useful to extend Euclidean correlators, in the sense of
	distributions, also to coincident points. One may ask if such an extension
	can be done in a way so that the resulting distributional correlators
	satisfy conformal Ward identities analogous to~\eqref{MinkConfWard}. In general the answer is no, already for
	2-point functions. Namely 2-point functions of primaries of dimension
	$\Delta$ such that $2 \Delta - d \in \mathbb{Z}$ will in general not allow
	even a scale invariant extension at coincident points, let alone
	conformally invariant one. E.g.\ this feature will always be present for the
	stress tensor 2-point function. 
\end{remark}

\subsection{Conformal invariance in terms of cross
	ratios}\label{conf-cross}

So as we have just seen, Minkowski correlator $G_4^M$ is conformally
invariant. If it were a function, conformal invariance would imply that we
could write it as the usual prefactor times a function of the cross-ratios.
Since it is a distribution, one might hope that it can be written as the
prefactor times a {\tmem{distribution}} of the cross-ratios. We will now
develop this point of view and show that it indeed works, at least 
locally.\footnote{One does not expect a very nice global statement. Indeed, 
	the cross-ratio space is morally the moduli space of four points on Minkowski cylinder
	$\cM$. (We ignore the fact that we have not constructed the distributions on Minkowski
	cylinder yet in favor of having an actual action of the conformal group.) This is a quotient space $(\cM^4)/G$
	where $G$ is the universal cover of Lorentzian conformal group. This quotient space is rather singular,
	which has to do with different configurations in $\cM^4$ having different stability subgroups (light-cones, $z=\bar z$).
	The quotient space $(\cM^4)/G$ is not only not smooth, it is not even Hausdorff. So away from some regular regions
	of $(\cM^4)/G$ one shouldn't expect a simple statement of the form~\eqref{G4g}, unless one finds a smoother model 
	of this moduli space.}

Our goal will be to make sense of the formula:
\begin{equation}
	G_4 (c) = \frac{g (z (c), \bar{z} (c))}{(x_{12}^2
		x_{34}^2)^{\Delta_{\mathcal{O}}}}, \label{G4g}
\end{equation}
where $g (z, \bar{z})$ will be in general a distribution in two variables, and
$g (z (c), \bar{z} (c))$ its pullback to the space $\mathbb{R}^{4 d}$ of
Minkowski 4-point configurations $c$. This equation will be understood in the
sense of integrating both parts with a test function. We will only consider
compactly supported $C^{\infty}$ test functions $\varphi (c)$, with the
additional requirement that all $c \in \tmop{supp} \varphi$ have the same
causal ordering. In particular, this implies that $\tmop{supp} \varphi$
contains no $c$'s with lightlike separated pairs. The causal ordering of a
configuration $c = (x_1, x_2, x_3, x_4)$ is encoded by the directed graph with
vertices $1, 2, 3, 4$ and edges $i \rightarrow j$ if $x_j$ belongs to the open
future light cone of $x_i$ (no edge if two points are spacelike).

Since $u, v$ are real in Minkowski space, $z, \bar{z}$ are either both real
(excluding $0, 1$) or complex conjugate. Part \ref{part:ope} divided
all causal orderings into 4 classes according to possible values of $(z,
\bar{z})$:
\begin{itemize}
	\item Class S: $(z, \bar{z}) \in (0, 1) \times (- \infty, 0)$ or the other
	way around
	
	\item Class T: $(z, \bar{z}) \in (0, 1) \times (1, + \infty)$ or the other
	way around
	
	\item Class U: $(z, \bar{z}) \in (- \infty, 0) \times (1, + \infty)$ or the
	other way around
	
	\item Class E causal orderings which contain configurations realizing the
	remaining possibilities:
	\begin{itemize}
		\item E$_{\tmop{su}}$: $(z, \bar{z}) \in (- \infty, 0) \times (- \infty,
		0)$
		
		\item E$_{\tmop{st}}$: $(z, \bar{z}) \in (0, 1) \times (0, 1)$
		
		\item E$_{\tmop{tu}}$: $(z, \bar{z}) \in (1, + \infty) \times (1, +
		\infty)$
		
		\item E$_{\tmop{stu}}$: $z, \bar{z}$ are complex-conjugate and not real
	\end{itemize}
\end{itemize}
Some class E causal orderings realize only one of the four subclasses, while
others contain configurations in each \ subclass. In the latter case different
subclasses are connected along configurations with $z = \bar{z}$ (see Fig.\ \ref{fig:zzbarclass}).

To simplify the discussion, we will assume that $\tmop{supp} \varphi$ does
not include any configurations with $z = \bar{z}$. In particular, this implies
that all configurations from $\tmop{supp} \varphi$ fall into a single class S,
T, U or a single subclass E$_{\tmop{su}}$, E$_{\tmop{st}}$, E$_{\tmop{tu}}$,
E$_{\tmop{stu}}$. Below we will comment how one can add the $z = \bar{z}$
configurations.

If $\tmop{supp} \varphi$ falls into class S, E$_{\tmop{su}}$, E$_{\tmop{st}}$,
E$_{\tmop{stu}}$, we will have $| \rho |, | \bar{\rho} | < 1$. These cases do
not require special treatment, since the correlator is a function, and Eq.
{\eqref{G4g}} is true in the ordinary sense of functions.

If $\tmop{supp} \varphi$ falls into class T or U, we will have $| \rho | < 1,
| \bar{\rho} | = 1$ or the other way around. Then $g (z, \bar{z})$ will be a
function in $z$ and a distribution in $\bar{z}$.\footnote{For some (but not
	all) of these causal orderings, it can be shown using another OPE channel that
	$g (z, \bar{z})$ is actually a function of both variables. See part \ref{part:ope}.} This case can be treated analogously, and simpler,
than the $| \rho |, | \bar{\rho} | = 1$ case discussed below.

Finally, if $\tmop{supp} \varphi$ falls into class E$_{\tmop{tu}}$, we will
have $| \rho |, | \bar{\rho} | = 1$. Then $g (z, \bar{z})$ will generally be a
distribution in two variables. This is the case we will focus on. E.g.\ it is
realized for the causal ordering $1 \rightarrow 3 \rightarrow 2 \rightarrow
4$.

Let us define the distribution $g (z, \bar{z})$ for $z, \bar{z} \in (1, +
\infty)$. We first define the distribution $g (\rho, \bar{\rho})$ with $|
\rho |, | \bar{\rho} | = 1$. This is done using the series in the r.h.s.\ of
Eq.\ {\eqref{g:rhoexpansion2}}, which we now consider as a function of two
independent variables $\rho, \bar{\rho}$. To be precise we consider the
series:
\begin{equation}
	g (\rho, \bar{\rho}) = \sum_{\delta, 0 \leqslant m \leqslant \delta} e^{i
		\Phi (\delta / 2 - m / 2)} (\rho \bar{\rho})^{\delta / 2 - m / 2} 
	(p_{\delta, m} \rho^m + p_{\delta, - m} \bar{\rho}^m) \hspace{0.17em},
	\label{gseries}
\end{equation}
which we view as a holomorphic function on $(\mathbb{D} \backslash (- 1,
0])^2$. Here $e^{i \Phi}$, $\Phi \in \{ 0, \pm 2 \pi, \pm 4 \pi \}$, is the
phase acquired by $\rho (c) \bar{\rho} (c)$ upon analytic continuation from
Euclidean space (as discussed in Sec.\ \ref{anal4-point} this phase is the same
as for $u (c)$). This phase is constant for each causal ordering and it may be
determined by following a path from $c_E$ to $c$ for any particular $c$.
Alternatively, the phase can also be determined from {\eqref{rhotrick}}. E.g.\ the causal ordering $1 \rightarrow 3 \rightarrow 2 \rightarrow 4$ has $\Phi =
0$.\footnote{We have that $x_{12}^2$, $x^2_{34}$, $x^2_{13}$, $x^2_{24}$ all
	acquire phase $- \pi$, hence $u = \frac{x_{12}^2 x^2_{34}}{x_{13}^2 x^2_{24}}$
	acquires phase 0.}

It's easy to see that function {\eqref{gseries}} satisfies a powerlaw bound as
$| \rho |, | \bar{\rho} | \rightarrow 1$. This is a baby version of the
problems studied in this part of the thesis, which was considered in part \ref{part:crossratio}. The
limit of $g (r e^{i \theta_1}, r e^{i \theta_2})$ as $r \rightarrow 1$
defines a tempered distribution on the boundary of the domain of analyticity,
parametrized by the two angles $\theta_1, \theta_2$. We can write it as $g
(\rho, \bar{\rho})$, with $\rho, \bar{\rho} \in S^1$.\footnote{In part \ref{part:crossratio} we also discussed a more general distribution
	defined on the product of universal covers of two circles. Here Eq.
	{\eqref{gseries}} with fixed $\alpha$ will be sufficient for our purposes.}

In fact we are interested only in a part of this distribution, {because $\rho,\bar{\rho}\neq\pm1$ for each fixed causal ordering.} The
points $- 1, 1$ divide the circle into two open arcs, and within $\tmop{supp}
\varphi$, $\rho$ and $\bar{\rho}$ will each live in one or the other arc. Each
arc is mapped smoothly and one-to-one to $(1, + \infty)$ by the $\rho \mapsto
z$ map. Thus we obtain the distribution $g (z, \bar{z})$ defined for $z,
\bar{z} > 1$. Although in general $z (c), \bar{z} (c)$ are defined only up to
permutation, let us define them in the case at hand, with real $z \neq
\bar{z}$, so that $\bar{z} (c) > z (c)$.

Now let us go back to making sense of {\eqref{G4g}}. Suppose first $g (z (c),
\bar{z} (c))$ were a function. Integrating {\eqref{G4g}} against a test
function we have:
\begin{equation}
	\int d^{4 d} c \, G_4 (c) \varphi (c) = \int d^4 c\, g (z (c), \bar{z} (c))
	\tilde{\varphi} (c), \qquad \tilde{\varphi} (c) = \frac{\varphi
		(c)}{(x_{12}^2 x_{34}^2)^{\Delta_{\mathcal{O}}}} .
\end{equation}
Note that $\tilde{\varphi} (c)$ is still $C^{\infty}$ since we are away from
light cones. We would like to continue by expressing the r.h.s. of the previous equation as an integral of $g
(z, \bar{z})$ against a two-dimensional test function:
\begin{equation}
	\begin{gathered}
		\int d^4 c\, g (z (c), \bar{z} (c)) \tilde{\varphi} (c) = \int d z\, d \bar{z}\, g
		(z, \bar{z}) \psi (z, \bar{z}), \\
		\psi (x_1, x_2) = \int d^4 c \,\delta (x_1 - z (c)) \delta (x_2 - \bar{z} (c)) \tilde{\varphi} (c) .
		\label{psidef}
	\end{gathered}
\end{equation}
We would like to know if $\psi (x_1, x_2)$ is a smooth function. By our
assumptions, $\tilde{\varphi} (c)$ is supported away from $\bar{z} (c) = z
(c)$. In this region the map $c \rightarrow (z (c), \bar{z} (c))$ is a
submersion, which means that the Jacobian has maximal rank (i.e.\ 2).
Alternatively, this means that the form $d z \wedge d \bar{z}$ does not vanish
anywhere away from $z = \bar{z}$. Showing this is a matter of an easy
computation.\footnote{Start by noting that, away from $z = \bar{z}$, we have
	$d z \wedge d \bar{z} \propto d u \wedge d v$ with a nonvanishing prefactor.
	We need to understand where $\nabla u$ can become proportional to $\nabla v$.
	Using the embedding space formalism {\cite{Costa:2011mg}} we write $u =
	\frac{(X_1 X_2) (X_3 X_4)}{(X_1 X_3) (X_2 X_4)}$, $v = \frac{(X_1 X_2) (X_3
		X_4)}{(X_1 X_3) (X_2 X_4)}$ where $X_i$ are null cone $d + 2$ dimensional
	vectors. For any $X_i, X_j, X_k$ the direction $R_{i, j k} = X_j (X_i X_k) -
	X_k (X_i, X_j)$ is tangent to the null cone at $X_i$. Imposing $R_{i, j k}
	\cdot \nabla_{X_i}  (u - \alpha v) = 0$ for all unequal $i, j, k$ where
	$\alpha$ is a constant, one gets a set of simple algebraic constraints on $u,
	v$. These constraints can be easily solved to show that $\alpha = \frac{2 u}{-
		1 + u + v}$ while $(1 + u - v)^2 - 4 u = 0$. The latter is precisely the
	constraint characterizing $z = \bar{z}$.}

Using the fact that $c \rightarrow (z (c), \bar{z} (c))$ is an submersion,
it's easy to show that $\psi (x_1, x_2)$ is smooth for $\tilde{\varphi} (c)$
supported away from $\bar{z} (c) = z (c)$ (see Chapter III.1 of
{\cite{gelfandshilov}} for such arguments). To summarize, for every smooth
function $\varphi (c)$ compactly supported away from $\bar{z} (c) = z (c)$ and
from the light cones, we constructed a smooth function $\psi (z, \bar{z})$
compactly supported in $1 < z < \bar{z}$ such that
\begin{equation}
	\int d^{4 d} c \, G_4 (c) \varphi (c) = \int d z\, d \bar{z}\, g (z, \bar{z})
	\psi (z, \bar{z}) \label{Gg1}
\end{equation}
holds in case $g (z, \bar{z})$ is a function. We now claim that this equation
continues to hold, with the same $\psi$, in case $g (z, \bar{z})$ is a
distribution. The point is that we can find a sequence of functions $g_n (z,
\bar{z})$ which tend to $g (z, \bar{z})$ in the sense of distributions, so
that the corresponding $\frac{g_n (z (c), \bar{z} (c))}{(x_{12}^2
	x_{34}^2)^{\Delta_{\mathcal{O}}}}$ tend to $G_4 (c)$ in the sense of
distributions on $\mathbb{R}^{4 d}$. Since both $\varphi$ and $\psi$ are
smooth, we are allowed to pass to the limit on both sides of the equation,
proving the claim. The functions $g_n (z, \bar{z})$ are given e.g.\ by the
partial sums of the series {\eqref{gseries}}, transformed from the $\rho$ to
the $z$ coordinates.

Let us now discuss how configurations where $z = \bar{z}$ can be included into
this discussion. The basic difficulty is that the map $c \mapsto (z, \bar{z})$
fails to be a submersion near such configurations. So in general the function
$\psi (z, \bar{z})$ will not be smooth. Consider e.g.\ the causal ordering $1
\rightarrow 3 \rightarrow 2 \rightarrow 4$. In this case it's possible to show
(we omit the proof) that the function $\psi (z, \bar{z})$ behaves like
\begin{equation}
	\text{$| z - \bar{z} |^{d - 2}$ times a smooth function near $z = \bar{z}$},
	\label{psizzbar}
\end{equation}
which in general is not smooth unless $d$ is even.

We need to be able to make sense of the r.h.s.\ of {\eqref{Gg1}} for such
non-fully-smooth test functions. This is possible due to the following
observation. Above we explained, following the arguments first presented in
part \ref{part:crossratio}, that $g (\rho, \bar{\rho})$ is a distribution for $| \rho |,
| \bar{\rho} | = 1$. But in fact it's a bit better than that (the fact not
mentioned in part \ref{part:crossratio}): it is a distribution in $\rho$ for each fixed
value of $\bar{\rho} / \rho = e^{i \alpha}$! Indeed if we substitute
$\bar{\rho} = e^{i \alpha} \rho$ with a fixed $\alpha$ into {\eqref{gseries}},
we get a holomorphic function in the unit disk of $\rho$, which satisfies a
powerlaw bound, hence its boundary value is a distribution. This can be
generalized to holomorphic maps $\bar{\rho} = f (\rho)$ which maps the unit disk
into itself (or at least a portion of the unit disk near $\rho = \rho_0$ into
the unit disk). Translating to $z, \bar{z}$, this implies in particular that
$g (z, z + t)$ is a distribution for any fixed $t$. In fact, the map $\bar{z}
= z + t$ corresponds to a map $\bar{\rho} = f_t (\rho)$ to which the previous
argument is applicable. So $g (z, \bar{z})$ is by no means the most general
distribution in two variables, as it allows the restriction to the submanifold
$\bar{z} = z + t$ for any $t$. E.g.\ $\delta (z - \bar{z})$ is not allowed by
this property, while $\delta (z + \bar{z})$ is allowed. Following this logic a
bit more carefully, it can be shown (we omit the proof) that $g (z, \bar{z})$
can be paired with test functions $\psi (z, \bar{z})$ which, when expressed in
terms of $s = z + \bar{z}$, $t = \bar{z} - z$, have the following property:
$\psi (s, t)$ is $C^{\infty}$ with respect to $s$ for any fixed $t$, with
bounds on derivatives in the $s$ direction which are integrable in the $t$
direction. Eq.\ {\eqref{psizzbar}} is compatible with this requirement.

A further complications arises near the $z = \bar{z} > 1$ locus for the
causal orderings which include configurations in both E$_{\tmop{tu}}$ and
E$_{\tmop{stu}}$ subclasses. In this case the function $\psi (z, \bar{z})$
defined in {\eqref{psidef}} will consist of two functions $\psi_1 (z,
\bar{z})$ and $\psi_2 (z, \bar{z})$: one defined for real $z, \bar{z}$,
another for complex-conjugate $z, \bar{z}$. The two functions $\psi_i$ will be
glued along the $z = \bar{z} > 1$ line. The resulting glued function will not
in generally be smooth on the $z = \bar{z} > 1$ line (while it will be smooth
away from it). However the directions orthogonal to the line turn out
analogous to the $t$ direction in the previous paragraph, i.e.\ the test
function is actually not required to be smooth in these directions for the
pairing to be defined. This allows to make sense of the formula {\eqref{Gg1}}
also in this case. We omit the details.

\subsection{Fixing points}

We would like to put the results of the previous section in the context of a
general question of ``fixing points'' in a distribution. E.g.\ we know that the
Minkowski 4-point function is a translationally invariant distribution. Using
translation invariance we can always fix one of the 4 points to a given
position, e.g.\ zero, and consider it as a distribution with respect to the
remaining 3 positions. One could ask if one can do better than that, i.e.\ to
fix $n$ points to given positions and consider the 4-point function as a
distribution with respect to the remaining $4 - n$ positions. Where the 4-point
function is real-analytic we can of course consider all four points as fixed.

Now, results of Sec.\ \ref{conf-cross} show that, if one excludes lightlike
separations limiting to configurations having some fixed causal ordering, one
can fix a conformal frame, i.e.\ fix three points to some fixed positions, and
the fourth point to a position characterized by two conformal cross ratios,
and consider the distribution as a distribution in only two variables (cross
ratios). It is not clear if results of Sec.\ \ref{conf-cross} can be
generalized to cover lightlike separations.

In some cases it is possible to argue that one can fix more than one point
without using conformal invariance. E.g.\ we may always fix a consecutive pair
of points, i.e.\ $(x_k, x_{k + 1})$, where $k = 1, 2$ or $3$, to
spacelike-separated positions in Minkowski space, while allowing the remaining
two points to approach Minkowski limit from the forward tube. The proof of
Lemma \ref{bound} can be slightly modified to show that $| \rho |, |
\bar{\rho} | < 1$ for such configurations (see Sec.\ \ref{localCFT} below).
Moreover, a powerlaw bound also holds, by a slight modification of the
argument after Eq.\ (\ref{def:Upsilon12}).\footnote{Since $S (c) = \infty$ in
	these cases, we cannot rely on (\ref{rhobound}). Instead we directly show
	powerlaw bounds on $S (\Upsilon_1), S (\Upsilon_2)$ defined in Eq.
	(\ref{def:Upsilon12}). Then the powerlaw bound on $| \rho |, | \bar{\rho} |$
	holds by Lemma \ref{lemma8}. For $k = 1$, by fixing $\zeta_3 = 0$, and using
	Lemmas \ref{Sinv} and \ref{SSS}, we have $S (\Upsilon_i) \leqslant S
	(\zeta_i') + S (- \zeta_4') \leqslant S (x_{i 3})^3 + S (x_{34})^3$ $(i = 1,
	2)$. This is the desired powerlaw bound with respect to $x_3$ and $x_4$. For
	$k = 2$, $S (\Upsilon_1)$ is bounded as for $k = 1$, while for $S
	(\Upsilon_2)$ we argue as follows. Since $x_2$ and $x_3$ are spacelike
	separated, after fixing $\zeta_3 = 0$, $\zeta_2$ is a spacelike Minkowski
	point, hence so is $\zeta_2'$, i.e.\ $\tmop{Im} (\zeta_4' + \zeta_2') =
	\tmop{Im} (\zeta_4')$. Then by Lemma \ref{Sinv}, \ $S (\Upsilon_2) \leqslant S
	(\zeta_4') + | \zeta_2' | \leqslant S (\zeta_4)^3 + | \zeta_2' |$, which is
	the needed bound. Case $k = 3$ follows by similar arguments or by mapping it
	to $k = 1$ via $(x_1, x_2, x_3, x_4) \rightarrow (x_1' = x_4^{\theta}, x_2' =
	x_3^{\theta}, x_3' = x_2^{\theta}, x_4' = x_1^{\theta})$ which maps $\rho$ and
	$\bar{\rho}$ are to their complex conjugates.} Then our arguments show that
the Minkowski 4-point function is a distribution with respect to the two unfixed
coordinates, which depends analytically on the fixed coordinates. In this case
the unfixed coordinates may have any causal orderings and also lightlike
separation.

One interesting case is that of the double light cone (DLC) singularity, i.e.\ the region close to $x_1 = 0$, $x_3 = \hat{e}_1$, $x_4 = \infty$, while $x_2$
on the light cones of $x_1, x_3$. Our results are the first ones which establish
the existence of the Wightman 4-point function in a neighborhood of DLC.
However, there is a difference between restricting to one causal ordering near
DLC or studying an open neighborhood of DLC which includes several causal
orderings (see Fig.\ \ref{134} in Conclusions). In the former case we can use
directly the results of Sec.\ \ref{conf-cross} and represents the 4-point
function as a distribution in two variables $z, \bar{z}$. In the latter case
we can fix, by the above argument, two successive spacelike points $x_3$ and
$x_4$. We are left with a distribution depending on $x_1, x_2$, i.e.\ $2\times d$
coordinates. This distribution still satisfies conformal invariance Ward
identities w.r.t.\ infinitesimal conformal transformations preserving $x_2$. It
would be interesting to understand how this constrains the distribution at the
DLC.

{Although it is not directly related, }we would also like to
mention here the classic result of Borchers {\cite{Borchers1964}} which says
that it is enough to smear Wightman functions $G_M (x_1, \ldots, x_n = 0)$
with respect to the time variables only, i.e.\ integrating with respect to $h_1
(x_1^0) \ldots h_{n - 1} (x_{n - 1}^0)$ where $h_i \in \mathcal{S}
(\mathbb{R})$, after which they become $C^{\infty}$ functions in the remaining
spatial variables $\mathbf{x}_i$. This result is valid in any QFT satisfying
Wightman axioms. It holds because smearing in time, which acts as an energy
cutoff, is effectively also a momentum cutoff because $| \mathbf{p} |
\leqslant E$.

\section{Wightman positivity}\label{sec:Wpos}

Recall that in Sec.\ \ref{OSfromCFT} we showed that CFT axioms imply OS
reflection positivity for 4-point functions. That discussion gives us access to OS
states $| \mathcal{O} (x) \mathcal{O} (y) \rangle \nobracket$ with $0 > x^0
> y^0$, with finite norm, and inner products measured by the Euclidean 4-point
function. We know that these states belong to the CFT Hilbert space, i.e.\ can
be arbitrarily well approximated in norm by states produced by inserting
finite linear combinations of CFT local operators at one point in the
half-space $x^0 < 0$, e.g.\ the south pole $x_S = (- 1, \tmmathbf{0})$.

Now that we analytically continued the 4-point function, we can consider other
states involving operators at complexified coordinates. We wish to prove that
those states belong to the CFT Hilbert space and have a positive definite
inner product. This can be shown by a robust argument, going back to
Osterwalder and Schrader {\cite{osterwalder1973}}, Sec.\ 4.3. The argument
uses only OS positivity and the Fourier-Laplace representation, but not
directly the CFT axioms.

We will consider two new kinds of states. First, states generated by a pair of
Minkowski operators smeared with respect to an arbitrary Schwartz test
function:
\begin{equation}
	| \Psi_M (F) \rangle \nobracket = \int d x\, d y\, F (x, y) | \mathcal{O} (i
	x^0, \mathbf{x}) \mathcal{O} (i y^0, \mathbf{y}) \rangle \nobracket,
	\label{PsiF}
\end{equation}
and second, states generated by a pair of Euclidean operators at complexified
time positions:
\begin{equation}
	\left| \mathcal{O} (x_3) \mathcal{O} (x_4) \rangle, \quad x_i = (\epsilon_i
	+ i t_i, \mathbf{x}_i) \right., \quad 0 > \epsilon_3 > \epsilon_4 .
	\label{Ocompl}
\end{equation}
The inner products of states {\eqref{PsiF}} are given by integrals of the
Minkowski 4-point function
\begin{equation}
	\langle \Psi_M (F_1) | \Psi_M (F_2) \rangle \nobracket = \int d x\, G_4^M
	(x_1, x_2, x_3, x_4) \overline{F_1 (x_2, x_1)} F_2 (x_3,
	x_4), \label{PsiF1PsiF2}
\end{equation}
while the natural inner product on the states {\eqref{Ocompl}} is:
\begin{equation}
	\langle \mathcal{O} (x_1) \mathcal{O} (x_2) | \nobracket \mathcal{O} (x_3)
	\mathcal{O} (x_4) \rangle = G_4 (x^{\theta}_2, x^{\theta}_1, x_3, x_4),
	\label{complinner}
\end{equation}
where the OS reflection operation extends to points with complex time
coordinates by:
\begin{equation}
	x = (\varepsilon + i t, \mathbf{x}) \mapsto x^{\theta} = (- \varepsilon + i
	t, \mathbf{x}) . \label{reflCompl}
\end{equation}
The states {\eqref{PsiF}} also have a natural inner product $\langle
\mathcal{O} (x_1) \mathcal{O} (x_2) | \Psi (F) \rangle \nobracket$ with the OS
states.\footnote{We start from the analytically continued Euclidean 4-point
	function $G_4 (x_1, x_2, x_3, x_4)$ and take the limit where $x_1, x_2$ are
	kept at fixed Euclidean positions, while $x_3, x_4$ approach the Minkowski
	space. By Theorem \ref{ThVlad}, the limit is a distribution in $x_3, x_4$, and
	the inner product is its pairing with the test function $F$.}

We wish to show that all these new inner products are positive definite and,
moreover, that the new states can be approximated in norm by the smeared OS
2-operator states at Euclidean positions. Note that the positive definiteness
of {\eqref{PsiF1PsiF2}} is precisely Wightman positivity for the 4-point case.

\subsection{Wightman states}

Let us start with {\eqref{PsiF1PsiF2}}. Rewriting the inner product in terms
of $W (p_1, p_2, p_3)$, the (distributional) Fourier transform of $G_4^M$
with respect to $y_k = x_k - x_{k + 1}$, we obtain
\begin{equation}
	\langle \Psi_M (F_1) | \Psi_M (F_2) \rangle \nobracket = \int d p\, W (p_1,
	p_2, p_3) [\widehat{F_1} (p_2 - p_1, p_1)]^{\ast}  \hat{F}_2 (p_2 - p_3,
	p_3) \label{pos4-pointFT} .
\end{equation}
We will also need the inner products of the (smeared) OS states
\begin{equation}
	| \nobracket \Psi (H) \rangle = \int d x\, d y\, H (x, y) | \nobracket
	\mathcal{O} (x) \mathcal{O} (y) \rangle \label{PsiH}
\end{equation}
where $H$ is any $C^{\infty}$ function compactly supported at $0 > x^0 >
y^0$. Their inner products are given by
\begin{equation}
	\langle \Psi (H_1) | \Psi (H_2) \rangle \nobracket = \int d x\, G^E_4
	(x_1, x_2, x_3, x_4) \overline{H_1 (x^{\theta}_2, x^{\theta}_1)} H_2
	(x_3, x_4) .
\end{equation}
This can be expressed using the Fourier-Laplace representation {\eqref{FL1}}.
We obtain
\begin{equation}
	\langle \Psi (H_1) | \Psi (H_2) \rangle \nobracket = \int d p\, W
	(p_1, p_2, p_3) \overline{g (H_1) (p_2, p_1)} g (H_2) (p_2, p_3),
	\label{fphipos}
\end{equation}
where $g (H) (p, q)$ is a Schwartz class function related to $H (x, y)$ as
follows. First we form the function $h (y_1, y_2) = H (- y_1, - y_1 - y_2)$
which has support at $y_1^0, y_2^0 > 0$. Next we consider $\tilde{h}$, the
Fourier-Laplace transform of $h (y_1, y_2)$:
\begin{eqnarray}
	& \tilde{h} (p_1, p_2) = \int d y_1\, d y_2\, e^{- p_1^0 y_1^0 + i\mathbf{p}_1
		\cdummy \mathbf{y}_1 - p_2^0 y_2^0 + i\mathbf{p}_2 \cdummy \mathbf{y}_2} h
	(y_1, y_2) . &  \label{fphi}
\end{eqnarray}
Finally, $g (H)$ is an arbitrary Schwartz class function which coincides
with $\tilde{h}$ inside the forward light cones. We also have an analogous
formula for the inner product between states of two types:
\begin{equation}
	\langle \Psi (H) | \Psi_M (F) \rangle \nobracket = \int d p\, W
	(p_1, p_2, p_3) \overline{g (H) (p_2, p_1)} \hat{F}_2 (p_2 - p_3,
	p_3) . \label{PsiHPsiF}
\end{equation}
At this point we recall Lemma \ref{lemma:fcheckdense} from Sec.\ \ref{MinkFromEucl}. That lemma implies that Schwartz functions of the form $g
(H)$ are dense in the Schwartz space. In particular, for any Schwartz $F$, we
can find a sequence of functions $\{ H_r \}_{r = 1}^{\infty}$ such that $g
(H_r) (p_2, p_3) \rightarrow \hat{F} (p_2 - p_3, p_3)$ in the Schwartz space.
Then it follows from {\eqref{pos4-pointFT}}, {\eqref{fphipos}}, {\eqref{PsiHPsiF}}
that
\begin{eqnarray}
	& \langle \Psi (H_r) | \Psi (H_r) \rangle \nobracket \rightarrow
	\langle \Psi_M (F) | \Psi_M (F) \rangle \nobracket, & \\
	& \langle \Psi (H_r) | \Psi_M (F) \rangle \nobracket \rightarrow
	\langle \Psi_M (F) | \Psi_M (F) \rangle \nobracket . &  \nonumber
\end{eqnarray}
From the first equation we conclude that $\langle \Psi_M (F) | \Psi_M
(F) \rangle \nobracket \geqslant 0$, proving Wightman positivity. The two
equations taken together imply that
\begin{equation}
	\langle \Psi (H_r) - \Psi_M (F) | \Psi (H_r) - \Psi_M (F) \rangle
	\nobracket \rightarrow 0,
\end{equation}
i.e.\ OS states can approximate Wightman states in norm.

\subsection{OS states for complexified times}\label{CScompl}

Let us discuss next the states {\eqref{Ocompl}} obtained by putting operators
at complexified time positions. In these states we don't take the limit to
Minkowski space, so they are defined without smearing. Using the
Fourier-Laplace representation, their inner product {\eqref{complinner}} is
expressed as
\begin{equation}
	\langle \mathcal{O} (x_1) \mathcal{O} (x_2) | \nobracket \mathcal{O} (x_3)
	\mathcal{O} (x_4) \rangle = G_4 (x^{\theta}_2, x^{\theta}_1, x_3, x_4) =
	\int d p\, W (p_1, p_2, p_3) \overline{f_{x_1, x_2} (p_2, p_1)}
	f_{x_3, x_4} (p_2, p_3),
\end{equation}
where $f_{x, y} (p, q)$, where $0 > \tmop{Re} (x^0) > \tmop{Re} (y^0)$, is
any Schwartz function which agrees with
\begin{equation}
	e^{p^0 x^0 - i\mathbf{p} \cdummy \mathbf{x} - q^0 (x^0 - y^0) +
		i\mathbf{q} \cdummy (\mathbf{x}-\mathbf{y})} .
\end{equation}
for $p, q$ in the forward light cone (where this function is exponentially
decreasing) and extends it somehow outside the light cones (it does not matter
how because $W$ has support in the forward light cones).

Since $f_{x, y}$ is a Schwartz function, it can be approximated by Schwartz
functions of the form $g (H)$. This implies that non-smeared complexified OS
states can be approximated in norm by Euclidean OS states smeared with
compactly supported test functions. In particular, the inner product
{\eqref{complinner}} is positive definite, providing an extension of pointwise
OS positivity to complexified times:
\begin{equation}
	G_4 (y^{\theta}, x^{\theta}, x, y) \geqslant 0, \qquad (0 > \tmop{Re} x^0
	> \tmop{Re} y^0) .
\end{equation}
As usual, positive-definite inner product implies a Cauchy-Schwarz inequality
for the complexified times:
\begin{equation}
	| G_4 (x_1, x_2, x_3, x_4) |^2 \leqslant G_4 (x_1, x_2,
	x^{\theta}_2, x^{\theta}_1) G_4 (x^{\theta}_4, x^{\theta}_3, x_3, x_4),
	\label{CS4-point}
\end{equation}
valid for $\tmop{Re} x^0_1 > \tmop{Re} x^0_2 > 0 > \tmop{Re} x_3^0 > \tmop{Re}
x^0_4$. The analogues of these properties for conformal blocks will be useful
in chapter \ref{secondpass}.

\begin{remark}
	\label{genrefl}We can extend the reflection operation further for points
	with complexified both time and space coordinates, as
	\begin{equation}
		x = (\varepsilon + i t, \mathbf{x}+ i\mathbf{y}) \mapsto x^{\theta} = (-
		\varepsilon + i t, \mathbf{x}- i\mathbf{y}) . \label{reflCompl1}
	\end{equation}
	With this definition, we can show by the same arguments as above that
	$G_4 (y^{\theta}, x^{\theta}, x, y) \geqslant 0$ (pointwise OS positivity) remains true for $0 \succ (\tmop{Re} x^0, \tmop{Im} \mathbf{x}) \succ
	(\tmop{Re} y^0, \tmop{Im} \mathbf{y})$ where $\eta_1 \succ \eta_2$ means
	$\eta_1 - \eta_2 \succ 0$ (i.e.\ in the forward light cone). We can then show
	that the states $| \nobracket \mathcal{O} (x) \mathcal{O} (y) \rangle$
	make sense for such $x, y$ and can be approximated in norm by integrated
	Euclidean OS states.
\end{remark}

\section{Wightman clustering}\label{clusterWightman}

\subsection{$2 + 2$ split}

In this section we will derive clustering {\eqref{Wightman:cluster}} for
Wightman 4-point functions (see {\cite{osterwalder1973}}, Sec.\ 4.4). As in
Sec.\ \ref{OSclustering} for the OS case, we will consider 2+2 and 3+1
splits separately. The property we need to prove in the 2+2 case can be
written conveniently in the language of Wightman states $| \nobracket \Psi_M
(F) \rangle$, at our disposal by the discussion in Sec.\ \ref{sec:Wpos}:
\begin{equation}
	\langle \Psi_M (F_1) | \Psi_M \nobracket (U_{\lambda a} F_2) \rangle
	\rightarrow \langle \Psi_M (F_1) | \Omega \nobracket \rangle \langle \Omega
	| \Psi_M \nobracket (F_2) \rangle \label{WCshow}
\end{equation}
as $\lambda \rightarrow \infty$ for any spacelike vector $a$ and any Schwartz
test functions $F_1, F_2$, where $U_{\lambda a}$ is translation: $(U_{\lambda
	a} F_2) (x, y) = F_2 (x - \lambda a, y - \lambda a)$, and $\Omega$ is the
vacuum state corresponding to inserting the unit operator. By Lorentz
invariance it's enough to prove this for $a = (0, \mathbf{a})$, purely spatial
vector. In Sec.\ \ref{OSclustering} we showed the OS clustering, which we
can also write using the integrated OS states {\eqref{PsiH}}, as
\begin{equation}
	\langle \Psi (H_1) | \Psi \nobracket (U_{\lambda a} H_2) \rangle
	\rightarrow \langle \Psi (H_1) | \Omega \nobracket \rangle \langle \Omega
	| \Psi \nobracket (H_2) \rangle \label{OSCknow} .
\end{equation}
As explained in Sec.\ \ref{sec:Wpos}, we can find states $| \Psi (H_1)
\rangle \nobracket$ and $| \Psi (H_2) \rangle \nobracket$ which approximate
$| \Psi_M \nobracket (F_1) \rangle$ and $| \Psi_M \nobracket (F_2) \rangle$ in
norm within any $\varepsilon > 0$. Moreover it's obvious from that
construction that the norm is invariant under shifts in purely spatial
direction (i.e.\ the operator $U_{\lambda a}$ is unitary). Hence we have $\|
\Psi (U_{\lambda a} H_2) - \Psi_M (U_{\lambda a} F_2) \| = \| \Psi (H_2)
- \Psi_M (F_2) \| \leqslant \varepsilon$ for any $\lambda$. By these
properties, {\eqref{OSCknow}} implies {\eqref{WCshow}}.\footnote{Indeed we
	have $| \langle \Psi_M (F_1) | \Psi_M \nobracket (U_{\lambda a} F_2) \rangle -
	\langle \Psi (H_1) | \Psi \nobracket (U_{\lambda a} H_2) \rangle |
	\leqslant C \varepsilon$ with some $C$ independent of $\lambda$. Now passing
	to the limit $\lambda \rightarrow \infty$ and using {\eqref{OSCknow}} we
	obtain $\text{lim sup}_{\lambda \rightarrow \infty} \langle \Psi_M (F_1) |
	\Psi_M \nobracket (U_{\lambda a} F_2) \rangle \leqslant \langle \Psi_M (F_1) |
	\Omega \nobracket \rangle \langle \Omega | \Psi_M \nobracket (F_2) \rangle +
	C' \varepsilon$, and an analogous lower bound on $\text{lim inf}_{\lambda
		\rightarrow \infty}$. Since $\varepsilon > 0$ is arbitrary we obtain
	{\eqref{WCshow}}.}

\subsection{$3 + 1$ split}\label{3+1MinkCluster}

Let us first restate the Euclidean 3+1 clustering argument from Sec.\ \ref{OSclustering} in a somewhat more explicit form, and specializing to
scalars. So let $\varphi (x_1), \chi (x_2, x_3, x_4)$ be two smooth functions
with compact support\footnote{For simplicity, in this section we prove clustering for compactly-supported, as opposed to Schwartz, test functions. We expect that it should be possible to find a proof for Schwartz test functions as well. In any case, the most natural proof would use positivity and the OPE similarly to 2+2 split, provided positivity for higher-point functions is proven (which we don't do in this part of the thesis).}
\begin{equation}
	\tmop{supp} (\varphi) \subset \{ x_1^0 > 0 \}, \quad \tmop{supp} (\chi)
	\subset \{ 0 > x_2^0 > x_3^0 > x_4^0 \} . \label{suppreq}
\end{equation}
We would like to show
\begin{equation}
	\lim_{\lambda \rightarrow \infty} (G, \varphi_{\lambda} \otimes \chi) = 0
	\label{toshow31}, \qquad \varphi_{\lambda} \assign \varphi (\cdot - \lambda
	\hat{e}_1)
\end{equation}
where $G = G_4^E$ is the Euclidean 4-point function of four identical scalars, and
$\hat{e}_1$ is the $x^1$ unit vector. The main idea is that we can find a
conformal transformation which moves the point at infinity as well as all the
other points to some finite positions. The suppression of the integral then
comes from the Jacobian of this transformation. Consider a special conformal
transformation $f (x) = \frac{x^{\mu} + x^2 b^{\mu}}{1 + 2 x \cdummy b + x^2
	b^2} =\mathcal{J} \circ T_b \circ \mathcal{J}$, where $\mathcal{J}$ is
inversion and $T_b$ is a translation by $b = \hat{e}_1$. We have $f (-
\hat{e}_1) = \infty$, while $f$ is non-singular on $\tmop{supp}
(\varphi_{\lambda})$ and $\tmop{supp} (\chi)$. We also have $f (\infty) =
\hat{e}_1 .$ By conformal invariance we have (compare {\eqref{MfinInv}}) $(G,
\Phi) = (G, \Phi^f)$ where $\Phi^f (x_1, \ldots, x_4) = \Phi (f^{- 1} (x_1),
\ldots, f^{- 1} (x_4)) \prod_{i = 1}^4 J (f^{- 1} (x_i))^{\Delta_{\mathcal{O}}
	- d}$, where $J (x) = \frac{1}{1 + 2 x \cdummy b + x^2 b^2}$. We apply this
equation with $\Phi = \varphi_{\lambda} \otimes \chi$. The function $\chi$ is
mapped by this transformation to some smooth function. Suppression of the
integral in the limit $\lambda \rightarrow \infty$ will come from the
transformation of $\varphi_{\lambda}$, which is mapped to
\begin{equation}
	\varphi_{\lambda}^f (x_1) \assign \varphi (f^{- 1} (x_1) - \lambda
	\hat{e}_1) J (f^{- 1} (x_1))^{\Delta_{\mathcal{O}} - d} .
\end{equation}
Namely we have
\begin{equation}
	| (G, \varphi_{\lambda}^f \otimes \chi^f) | \leqslant C (\lambda) I, \quad I
	= \int d x_1\, | \varphi_{\lambda}^f (x_1) |, \quad C (\lambda) = \sup_{x_1
		\in \tmop{supp} \varphi_{\lambda}^f} | (G (x_1, \cdot), \chi^f) | .
\end{equation}
The function $\varphi_{\lambda}^f$ is nonzero for $f^{- 1} (x_1) \in
\tmop{supp} (\varphi) + \lambda \hat{e}_1$, which is a point near infinity for
$\lambda$ large. We conclude that $\varphi_{\lambda}^f$ is supported in a small
neighborhood, order $1 / \lambda$, of $f (\infty) = \hat{e}_1$. Since $G$
is real-analytic at nonzero point separation, this implies that $C
(\lambda)$ is bounded by some constant for $\lambda \geqslant \lambda_0$.
To compute $I$, we do the change of variables $x_1 = f (y)$:
\begin{equation}
	I = \int d y\, | \varphi (y - \lambda \hat{e}_1) | J
	(y)^{\Delta_{\mathcal{O}}} \sim \frac{\tmop{const}}{\lambda^{2
			\Delta_{\mathcal{O}}}} .
\end{equation}
This finishes the proof of Euclidean 3+1 clustering, Eq.\ {\eqref{toshow31}}.

Let us proceed next to show Wightman 3+1 clustering. We will show the same
equation as {\eqref{toshow31}}, namely
\begin{equation}
	\lim_{\lambda \rightarrow + \infty} (G, \varphi_{\lambda} \otimes \chi) = 0,
\end{equation}
where now $G = G_4^M$ is the Minkowski 4-point function, which is a tempered
distribution, and $\varphi (x_1)$ and $\chi (x_2, x_3, x_4)$ are arbitrary
compactly supported test functions (i.e.\ no support requirements analogous to
{\eqref{suppreq}}).\footnote{The method described below cannot be
	straightforwardly generalized to the case of Schwartz test functions.} \ The
proof will be based on the same idea of moving the point at infinity to a
finite position, paying attention to $G$ now being distribution, and to the
requirement {\eqref{req1}} on invariance under finite Minkowski conformal
transformations.

We will use the same transformation $f (x) = \frac{x^{\mu} + x^2 b^{\mu}}{1 +
	2 x \cdummy b + x^2 b^2}$, $b = \hat{e}_1$. By translation invariance, we may
assume that $\tmop{supp} (\chi)$ lies at larger $x^1$ values than of the
singularity light cone $x^0 = \pm | \mathbf{x}+ \hat{e}_1 |$ of this
transformation (see Sec.\ \ref{ConfMink}). For sufficiently large $\lambda$,
$\tmop{supp} (\varphi_{\lambda})$ will also satisfy this condition. As we
scale $b$ to zero to connect $f$ to the identity, the singularity light cone
moves away to infinity along the negative $x^1$ direction, without touching
$\tmop{supp} (\varphi_{\lambda})$ nor $\tmop{supp} (\chi)$, see Fig.\ \ref{3+1Mink}. Hence requirement {\eqref{req1}} is satisfied and we may apply
invariance {\eqref{MfinInv}}, which says $(G, \varphi_{\lambda} \otimes \chi)
= (G, \varphi_{\lambda}^f \otimes \chi^f)$.

\begin{figure}[h]\centering
	\raisebox{-0.504762041315234\height}{\includegraphics[width=10.1973796405615cm,height=4.57459333595697cm]{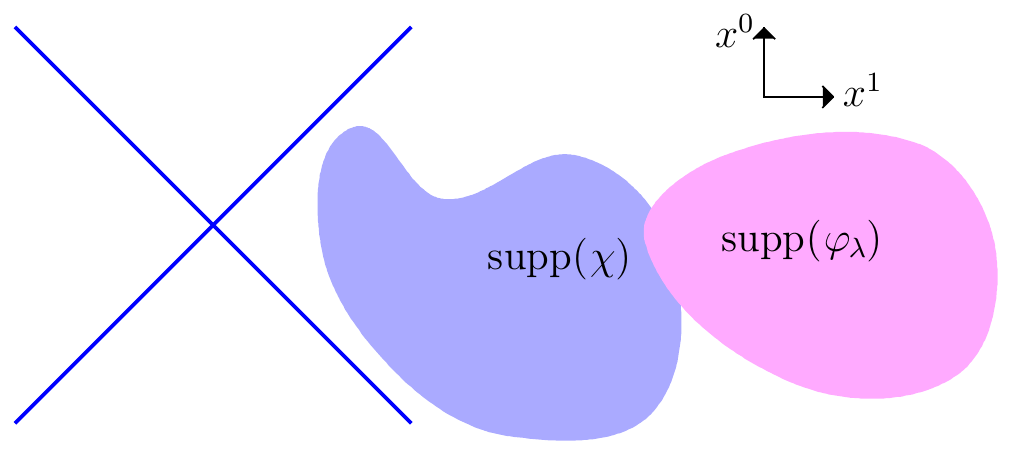}}
	\caption{Location of supports of $\varphi_{\lambda}$ and $\chi$ with respect
		to the singularity light cone of $f$.\label{3+1Mink}}
\end{figure}

Now, using translation invariance of the 4-point function, $G (x_1, x_2, x_3, x_4)
= \tilde{G} (x_2 - x_1, x_3 - x_1, x_4 - x_1)$ we may write
\begin{equation}
	(G, \varphi_{\lambda}^f \otimes \chi^f) = \int d x_1\, \varphi_{\lambda}^f
	(x_1) F (x_1), \label{intMink}
\end{equation}
where $F (x_1) = (\tilde{G}, T_{x_1} \cdot \chi^f)$ and $T_a$ is a
translation. $\tilde{G}$ is a distribution, but since translation is a
continuous operation in the space of test functions, we know that $F (x_1)$ is
a continuous function of $x_1$. When $\lambda$ goes to $+ \infty$, the support
of $\varphi_{\lambda}^f$ shrinks to the point $\hat{e}_1$.\footnote{It is
	important for the argument that, as one can easily check, $\tmop{supp}
	(\varphi_{\lambda}^f)$ shrinks to a compact set (in fact, a point) and not,
	say, spreads out along some light cone.} Hence for $\lambda \geqslant
\lambda_0$ we can bound $| F |$ on $\tmop{supp} (\varphi_{\lambda}^f)$ by a
constant, and estimate {\eqref{intMink}} in absolute value by $\tmop{const}
\times \int d x_1\,  | \varphi_{\lambda}^f (x_1) |$. This remaining integral is
computed via the change of variables as the Euclidean one, and goes to zero as
$\lambda^{- 2 \Delta_{\mathcal{O}}}$, completing the proof.

\section{Local commutativity}\label{local-comm}

Let us show that the constructed Minkowski correlators satisfy local
commutativity. This follows by a robust argument which uses only Lorentz
invariance, analyticity in the forward tube, existence of the boundary
distribution, and real analyticity of the Euclidean correlators away from
coincident points (OS {\cite{osterwalder1973}}, Sec.\ 4.5). Here for
completeness we will provide this argument for $n$-point functions which is
its natural setting. In Sec.\ \ref{localCFT} below we will make some remarks
specific to CFT 4-point functions.

So, we start from the Euclidean correlator $G^E (x_1, \ldots, x_n)$ at $x_1^0
> x^0_2 > \cdots > x_n^0$ and its analytic continuation $G (x_1, \ldots,
x_n)$ to the forward tube $\mathcal{T}_n$ which is the set of points $x_k \in
\mathbb{C}^d$ such that their differences $y_k = x_k - x_{k + 1}$ satisfy
$\tmop{Re} y_k^0 > | \tmop{Im} \mathbf{y}_k |$ or equivalently $\eta_k \succ
0$ in terms of $\zeta_k = (i y_k^0, \mathbf{y}_k) = \xi_k + i \eta_k$, $\xi_k,
\eta_k \in \mathbb{R}^{1, d - 1}$. We will write $G$ instead of $G_n$. We know
by Theorem \ref{ThVlad} that this analytic continuation is invariant under
Lorentz transformations $\zeta_k \rightarrow \Lambda \zeta_k$ where $\Lambda
\in L_+^{\uparrow}$, the identity component of the real Lorentz group. 
Since $G$ is translationally invariant, it depends only on
$\zeta_k$, and we will abuse of notation by sometimes writing $G (\zeta_1,
\ldots, \zeta_{n - 1})$ and $(\zeta_1, \ldots, \zeta_{n - 1}) \in
\mathcal{T}_n$ instead of $G (x_1, \ldots, x_n)$ and $(x_1, \ldots, x_n)
\in \mathcal{T}_n$.

\tmtextbf{Step 1.} We will extend domain of analyticity of $G$ using the
complex Lorentz group $L (\mathbb{C})$, defined as the set of complex matrices
$A$ preserving the Minkowski metric, i.e.\ $A^T g A = g$ where $g = \tmop{diag}
(- 1, 1 \ldots 1)$. We will only need the component of $L (\mathbb{C})$
connected to the identity, denoted $L_+ (\mathbb{C})$. For any $\Lambda \in
L_+ (\mathbb{C})$ consider the equation
\begin{equation}
	G (\zeta_1, \ldots, \zeta_{n - 1}) = G (\Lambda^{- 1} \zeta_1, \ldots,
	\Lambda^{- 1} \zeta_{n - 1}) . \label{BHWeq}
\end{equation}
The two sides of this equation coincide for real $\Lambda \in L_+^{\uparrow}$
(by Lorentz invariance of $G_n$), and hence by analyticity in the components
of $\Lambda$ also for complex $\Lambda \in L_+ (\mathbb{C})$, at least for
$\Lambda$ close to 1. In other words, Eq.\ {\eqref{BHWeq}} is just an identity
if $\Lambda \approx 1$ and the arguments of $G_n$ on both sides are in the
forward tube. But a general $\Lambda \in L_+ (\mathbb{C})$ does not preserve
the forward tube. For such $\Lambda$, Eq.\ {\eqref{BHWeq}} extends analytically
$G$ from the forward tube to the set
\begin{equation}
	\mathcal{T}_n' = \bigcup_{\Lambda \in L_+ (\mathbb{C})} \Lambda \cdot
	\mathcal{T}_n,
\end{equation}
called the extended tube. The Bargmann-Hall-Wightman theorem shows that no
further topological obstructions arise in this analytic continuation; see
{\cite{jost1979general}}, p.78 for details. Call this extension $\tilde{G}$.

\tmtextbf{Step 2.} Let us consider $\tilde{G} (x_1, \ldots, x_n)$ for
\begin{equation}
	\epsilon_1 > \ldots > \epsilon_{k - 1} > 0 > \epsilon_{k + 2} > \ldots >
	\epsilon_n,
\end{equation}
while assuming that $\epsilon_k, \epsilon_{k + 1}$ are near zero and much
smaller than other $\epsilon_i$'s, and $| t_k - t_{k + 1} | < |\mathbf{x}_k
-\mathbf{x}_{k + 1} | \nobracket$, $\mathbf{x}_k$, $\mathbf{x}_{k + 1}$ real,
so that $x_k - x_{k + 1}$ approaches a spacelike separation. For $\epsilon_k >
\epsilon_{k + 1}$ this configuration is in the forward tube, so we know
$\tilde{G}$ is analytic there and agrees with $G (x_1, \ldots, x_n)$. Let us
show that the configurations with $\epsilon_k < \epsilon_{k + 1}$ are in the
extended tube. We may set $x_{k + 1} = 0$ for this argument, so that
\begin{equation}
	\zeta_k = (t_k + i \epsilon_k, \mathbf{x}_k) .
\end{equation}
We may assume without loss of generality that $\mathbf{x}_k = (x^1_k, 0,
\ldots 0)$, $x_k^1 > | t_k |$. Then acting on $\zeta_k$ with the complexified
Lorentz transformation
\begin{equation}
	\Lambda_{\theta} = \left(\begin{array}{cc}
		\cosh (i \theta) & \sinh (i \theta)\\
		\sinh (i \theta) & \cosh (i \theta)
	\end{array}\right) \in L_+ (\mathbb{C}),
\end{equation}
with small $\theta$ we get, using $\Lambda_{\theta} \approx
\left(\begin{array}{cc}
	1 & i \theta\\
	i \theta & 1
\end{array}\right)$, $\zeta_k' = \Lambda_{\theta} \zeta_k \approx (t_k, x^1_k)
+ i (\theta x^1_k + \epsilon_k, \theta t_k)$, and thus $\eta_k' \approx
(\theta x^1_k + \epsilon_k, \theta t_k)$. If $\epsilon_k$ is negative but very
small, we can can achieve $\eta_k' \succ 0$ by choosing an appropriate small
$\theta$. We need $\theta$ small so that all the other $\zeta_i'$ remain in
the forward light cone, and this will work because we are assume that
$\epsilon_k$ is very much smaller than all the other $\epsilon_i$'s.

The bottom line is that the extended tube contains an open set of
configurations as above, with $| t_k - t_{k + 1} | < |\mathbf{x}_k
-\mathbf{x}_{k + 1} | \nobracket$ and $\epsilon_k, \epsilon_{k + 1}$ small,
with $\epsilon_k - \epsilon_{k + 1}$ of any sign. Let us call this set
$\mathcal{Q}_{n, k}$. By restricting this set a bit, we may assume that
$\mathcal{Q}_{n, k}$ is invariant under permutations of $x_k$ and $x_{k + 1}$.
By Step 1 we know that function $\tilde{G}$ is holomorphic in the extended tube
and hence also in $\mathcal{Q}_{n, k}$. In particular, it is analytic if we
set $\epsilon_k = \epsilon_{k + 1} = 0$. This already has an interesting
consequence: Minkowski correlator is analytic with respect to a pair of
spacelike-separated points (while it remains a distribution with respect to
all the other points). Projection of $\mathcal{Q}_{n, k}$ to the plane
$(\epsilon_k - \epsilon_{k + 1}, t_k - t_{k + 1})$ is shown schematically in
Fig.\ \ref{Qnk}.

\begin{figure}[h]\centering
	\raisebox{-0.496957195215242\height}{\includegraphics[width=5.07528204119113cm,height=4.30852354715991cm]{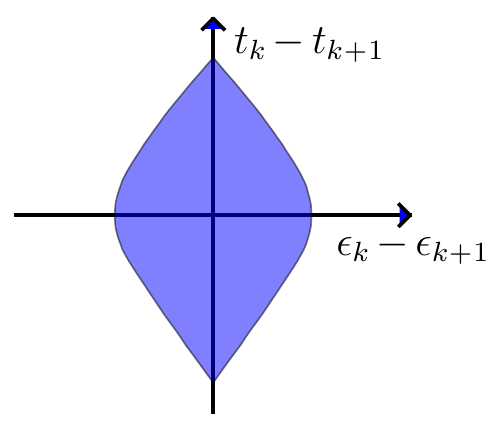}}
	\caption{\label{Qnk}Projection of the set $\mathcal{Q}_{n, k}$, where the
		function $\tilde{G}$ is holomorphic, to the plane $(\epsilon_k - \epsilon_{k +
			1}, t_k - t_{k + 1})$. The vertical extent of this region is determined by
		the condition $| t_k - t_{k + 1} | < | \mathbf{x}_k -\mathbf{x}_{k - 1} |$.
		The horizontal extent is determined, among other things, by the condition
		that $\epsilon_k, \epsilon_{k + 1}$ have to be much smaller that all the
		other $\epsilon_i$'s.}
\end{figure}

The set $\mathcal{Q}_{n, k}$ contains real configurations (horizontal axis in
Fig.\ \ref{Qnk}, setting other $t_i \rightarrow 0$ as well). Restriction of
$\tilde{G}$ to the real part of $\mathcal{Q}_{n, k}$ agrees with the Euclidean
correlator $G^E$. (They agree for $\epsilon_k > \epsilon_{k + 1}$ by
construction and for $\epsilon_k < \epsilon_{k + 1}$ by the uniqueness of
analytic continuation. Recall that the Euclidean correlator $G^E$ is real
analytic everywhere away from coincident points, i.e.\ for $\epsilon_k -
\epsilon_{k + 1}$ of any sign as long as $\mathbf{x}_k \neq \mathbf{x}_{k -
	1}$.) One consequence of this fact is that $\tilde{G}$ restricted to the real
part $\mathcal{Q}_{n, k}$ is permutation invariant w.r.t. $x_k \leftrightarrow
x_{k + 1}$:
\begin{equation}
	\tilde{G} (\ldots x_k, x_{k + 1} \ldots) = \tilde{G} (\ldots
	x_{k + 1}, x_k \ldots), \label{Gtperm}
\end{equation}
because the Euclidean correlator has this property. Finally, since
$\mathcal{Q}_{n, k}$ is connected to the real configurations (see Fig.\ \ref{Qnk}), we conclude that permutation invariance {\eqref{Gtperm}} holds
everywhere in $\mathcal{Q}_{n, k}$.\footnote{In fact, $\tilde{G}$ can be
	extended to a single-valued holomorphic function on the ``permuted extended
	tube'' $\bigcup_{\pi \in S^n} \pi \mathcal{T}_n'$, and satisfied permutation
	invariance {\eqref{Gtperm}} on this large set. See {\cite{jost1979general}},
	App.\ II, {\cite{Tomozawa}} and {\cite{bogolubov2012general}}, Sec.\ 9.D.
	However for our purposes analyticity and permutation invariance on
	$\mathcal{Q}_{n, k}$ will suffice.}

We now see the meaning of $\tilde{G}$ for configurations with $\epsilon_k <
\epsilon_{k + 1}$. Via permutation invariance {\eqref{Gtperm}}, such
configurations are mapped to the forward tube and hence can be evaluated as
$G$ for the permuted configurations.

\tmtextbf{Step 3.} We are now ready to show local commutativity. We have to
prove that boundary value limits of two holomorphic functions agree:
\begin{equation}
	\lim_{\epsilon_i \rightarrow 0} G(\ldots x_k, x_{k + 1} \ldots) =
	\lim_{\epsilon_i \rightarrow 0} G(\ldots x_{k + 1}, x_k
	\ldots) \text{} \label{perm1},
\end{equation}
when approaching a Minkowski configuration in which $x_k - x_{k + 1}$ is
spacelike. Note that, by the original definition, the two limits are from
different forward tubes: the first one must respect the condition $\epsilon_k
> \epsilon_{k + 1}$, while the second $\epsilon_{k + 1} > \epsilon_k$. By
Theorem \ref{ThVlad}, Part 3, we can take the limits $\epsilon_i \rightarrow
0$ in any order, so let us send $\epsilon_k, \epsilon_{k + 1} \rightarrow 0$
first, while keeping other $\epsilon_i$ fixed for the moment. For very small
$\epsilon_k, \epsilon_{k + 1}$, the configurations on both sides will be in
$\mathcal{Q}_{n, k}$ where both sides are restrictions of the function
$\tilde{G}$ analytic around $\epsilon_k, \epsilon_{k + 1} = 0$ and
satisfying permutation invariance {\eqref{Gtperm}}. It follows that the two
sides of {\eqref{perm1}} agree in the limit $\epsilon_k, \epsilon_{k + 1}
\rightarrow 0$. Sending the remaining $\epsilon_i \rightarrow 0$ we recover
the local commutativity.

\subsection{Local commutativity for CFT 4-point functions}\label{localCFT}

In this part of the thesis we analytically continued the CFT 4-point function $\langle
\mathcal{O} (x_1) \mathcal{O} (x_2) \mathcal{O} (x_3) \mathcal{O} (x_4)
\rangle$ to the forward tube using $\rho, \bar{\rho}$ coordinates. We would
like to indicate here that this provides an alternative path to understanding
local commutativity. 
{We have shown previously that $0<|\rho|,|\bar\rho|<1$ in the forward tube.
	Since the extended tube is obtained from the forward tube by complexified Lorentz transformations
	and $\rho,\bar\rho$ are invariant under such transformations, it follows that $0<|\rho|,|\bar\rho|<1$ also in the extended tube. Below we will show this explicitly for the configurations used
	in the proof of local commutativity.}
We consider separately $k = 1$ and $k = 2$ ($k = 3$ being
analogous to $k = 1$).

\tmtextbf{$k = 1$:} Here $x_1, x_2$ approach spacelike-separated Minkowski
points. We know that \ $| \rho |, | \bar{\rho} | < 1$ in $\mathcal{D}_4$,
$\epsilon_1 > \epsilon_2 > \epsilon_3 > \epsilon_4$. Extended tube analyticity
suggests that this must remain true also for $\epsilon_1 = \epsilon_2 >
\epsilon_3 > \epsilon_4$. Indeed, this follows from critical rereading of the
proof of Lemma \ref{bound} (Sec.\ \ref{PetrProof}, Eq.\ {\eqref{quadeq}} and
below). (That proof does not use the condition $\eta_1 \succ \eta_2$ but only
$\eta_1, \eta_2 \succ 0$.\footnote{An alternative argument is as follows. In
	chapter \ref{secondpass} we will show the Cauchy-Schwarz inequality for $\rho,
	\bar{\rho}$, Theorem \ref{boundThm}, which bounds $\rho, \bar{\rho}$ for any
	configuration in the forward tube with $\epsilon_1 > \epsilon_2 > 0 >
	\epsilon_3 > \epsilon_4$ in terms of $\rho, \bar{\rho}$ of
	``reflection-symmetric'' configurations having $\epsilon_3 = - \epsilon_2$,
	$\epsilon_4 = - \epsilon_1$. The proof of Lemma \ref{boundLemma}, Eq.
	{\eqref{z12zbar12}} shows that $\rho, \bar{\rho}$ remain less than 1 for the
	latter configurations in the limit $\epsilon_1 \rightarrow \epsilon_2$.}) It
is also important for analyticity that $\rho, \bar{\rho}$ not vanish. In the
forward tube $\rho, \bar{\rho}$ do not vanish because $x^2_{i j} \neq 0$, $i <
j$ (Lemma \ref{xij2h}). When $\epsilon_1 = \epsilon_2$ we have $x^2_{12} > 0$
(spacelike separation), hence also nonzero. These observations show that the
CFT 4-point function can be analytically extended, using the $\rho, \bar{\rho}$
expansion, to a neighborhood of points with $\epsilon_1 = \epsilon_2 > 0 >
\epsilon_3 > \epsilon_4$, $x^2_{12} > 0$, in agreement with the general QFT
arguments given above.

Let us now permute the first two points: $(\epsilon_1 + i t_1, \mathbf{x}_1)
\leftrightarrow (\epsilon_2 + i t_2, \mathbf{x}_2)$. In the Euclidean region,
this transformation maps $\rho \rightarrow - \rho, \bar{\rho} \rightarrow -
\bar{\rho}$ and leaves the 4-point function of identical scalars invariant because
the expansion {\eqref{g:rhoexpansion}} contains only even $m$. The same
transformation remains true for complexified times for spacelike separation.
Taking the limit $\epsilon_1, \epsilon_2 \rightarrow 0$, we recover local
commutativity very explicitly.

\tmtextbf{$k = 2$:} Now we are interested in the limit $\epsilon_2 \rightarrow
\epsilon_3$ from inside $\epsilon_1 > \epsilon_2 > \epsilon_3 >
\epsilon_4$.\footnote{\label{noteShock}The discussion on the local
	commutativity of this type can also be found in the study of causality in a
	shockwave background (see Sec.\ 5 of {\cite{Hartman:2015lfa}}). In
	{\cite{Hartman:2015lfa}}, the 2-point function in a shockwave background is
	defined by $\langle \mathcal{O} (x) \mathcal{O} (y) \rangle_{\Psi} \assign
	\frac{\langle \Psi (i \delta) \mathcal{O} (x) \mathcal{O} (y) \Psi (- i
		\delta) \rangle}{\langle \Psi (i \delta) \Psi (- i \delta) \rangle}$, where
	$x$, $y$ are Minkowski points and ``$i \delta$'' means the Euclidean point
	$(\delta, 0, \ldots, 0)$. In our language it corresponds to the 4-point
	function $\langle \Psi (x_1) \mathcal{O} (x_2) \mathcal{O} (x_3) \Psi (x_4)
	\rangle$ with $\varepsilon_1 = - \varepsilon_4 = \delta > 0$ and
	$\varepsilon_2 = \varepsilon_3 = 0$. We know that the 4-point function is
	regular analytic at such configurations. So the commutator $[\mathcal{O} (x),
	\mathcal{O} (y)]$ vanishes in the shockwave background when $x$ and $y$ are
	spacelike separated.} As for $k = 1$, critical rereading of the proof of
Lemma \ref{bound} shows that $| \rho |, | \bar{\rho} |$ remain less than 1.
(We put in that proof $\zeta_2 = \xi_2 + i \eta_2$, $\xi_2 = (t_2,
\mathbf{x}_2)$ spacelike, and $\eta_2 = (\epsilon_2, \tmmathbf{0})$,
$\epsilon_2 > 0$. The proof does not use the condition $\eta'_2 \succ 0$ but
only $\eta'_{24} \succ 0$. The latter condition remains true for $\epsilon_2
\rightarrow \epsilon_3 = 0$, as $\zeta_2'$ goes to a finite real vector.)
Hence, the CFT 4-point function can be analytically extended, using the $\rho,
\bar{\rho}$ expansion, to a neighborhood of points with $\epsilon_1 >
\epsilon_2 = \epsilon_3 > \epsilon_4$, $x^2_{23} < 0$.

To finish the proof of local commutativity, we fall back on the general
argument, appealing to the permutation invariance of the (real-analytic) CFT
4-point function under $x_2 \leftrightarrow x_3$. (Unlike for $k = 1$, the
s-channel OPE expansion {\eqref{g:rhoexpansion}} cannot be used to make this
step more explicit, as it does not manifestly have this invariance.)

\section{Generalization to non-identical scalars}\label{nonId}

In the previous subsections we proved that the 4-point function of
identical scalars has analytic continuation to the forward tube
$\mathcal{T}_4$, and its boundary value in the Minkowski region is a tempered
distribution. Then Minkowski conformal invariance, Wightman positivity,
Wightman clustering and local commutativity follow from their Euclidean
analogues.

In this section we will indicate how to generalize analytic continuation and
temperedness to 4-point functions of non-identical scalars. The proof of
the other properties is the same as in the case of identical scalars.

We consider the 4-point function of scalar primary operators
$\mathcal{O}_i$ with scaling dimensions $\Delta_i$,
\begin{eqnarray}
	G^E_{1234} (c_E) & \assign & \langle \mathcal{O}_1 (x_1) \mathcal{O}_2 (x_2)
	\mathcal{O}_3 (x_3) \mathcal{O}_4 (x_4) \rangle \nonumber\\
	& = & \frac{1}{(x_{12}^2)^{\frac{\Delta_1 + \Delta_2}{2}}
		(x_{34}^2)^{\frac{\Delta_3 + \Delta_4}{2}}} \left( \frac{x_{24}^2}{x_{14}^2}
	\right)^{\frac{\Delta_1 - \Delta_2}{2}} \left( \frac{x_{14}^2}{x_{13}^2}
	\right)^{\frac{\Delta_3 - \Delta_4}{2}} g_{1234} (c_E), 
	\label{def:Eucl4-pointgeneral}
\end{eqnarray}
which reduces to (\ref{def:Euclidean4-point}) when $\Delta_i$'s are identical. The
analytic continuation of the prefactor to the forward tube $\mathcal{T}_4$ is
straightforward. The function $g_{1234} (c_E)$ only depends on the conformal
equivalence class of $c_E$, i.e.\ $g_{1234} (c_E) = g_{1234} (\rho (c_E),
\bar{\rho} (c_E))$. By the similar argument to that in Sec.\ \ref{Eucl4-point},
the function $g_{1234} (c_E)$ has the following series expansion
\begin{equation}
	g_{1234} (c_E) = \left[ \frac{(1 - \rho) (1 - \bar{\rho})}{(1 + \rho) (1 +
		\bar{\rho})} \right]^{\frac{\Delta_1 - \Delta_2 - \Delta_3 + \Delta_4}{2}}
	\underset{\delta, m}{\sum} a_{12} (\delta, m) a_{\bar{4} \bar{3}} (\delta,
	m)^{\ast} r^{\delta} e^{i m \theta}, \qquad \rho (c_E) = r e^{i m \theta},
\end{equation}

where the sum runs over a discrete set of pairs $(\delta, m)$ with $\delta
\geqslant 0$, $m \in \mathbb{Z}$ (not necessarily even for non-identical
scalars), $| m | \leqslant \delta$. Analogously to the case of identical
scalars, the sum is absolutely convergent when $| \rho (c_E) | < 1$ (see
below). Also, when $d \geqslant 3$ we have $p_{\delta, - m} = p_{\delta, m}$,
where $p_{\delta, m} = a_{12} (\delta, m) a_{\bar{4} \bar{3}} (\delta,
m)^{\ast}$. Analogously to Sec.\ \ref{anal4-point}, the analytic continuation of
$g_{1234} (c)$ in $d \geqslant 3$ will be given by the formula (compare
(\ref{eq:gtilde}))
\begin{equation}
	g_{1234} (c) = \left( \frac{x_{14}^2 x_{23}^2}{x_{13}^2 x_{24}^2}
	\right)^{\frac{\Delta_1 - \Delta_2 - \Delta_3 + \Delta_4}{4}} \sum_{m,
		\delta, 0 \leqslant m \leqslant \delta} p_{\delta, m} R_{\delta / 2 - m /
		2} (c) \Phi_m (c) . \label{eq:gtildegeneral}
\end{equation}
In $d = 2$, $p_{\delta, m} \neq p_{\delta, - m}$ but the functions $\rho
(c)^m$ and $\bar{\rho} (c)^m$ are individually holomorphic. In this case the
analytic continuation of $g_{1234} (c)$ is given by the formula (compare
{\eqref{eq:gtilde2d}}):
\begin{equation}
	g_{1234} (c) = \left( \frac{x_{14}^2 x_{23}^2}{x_{13}^2 x_{24}^2}
	\right)^{\frac{\Delta_1 - \Delta_2 - \Delta_3 + \Delta_4}{2}} \sum_{m,
		\delta, 0 \leqslant m \leqslant \delta} R_{\delta / 2 - m / 2} (c) 
	[p_{\delta, m} \rho (c)^m + p_{\delta, - m} \bar{\rho} (c)^m] .
	\label{eq:gtilde2dgeneral}
\end{equation}
We would like to show that

(a) when $r = \max \{ | \rho |, | \bar{\rho} | \} < 1$, the series
\begin{equation}
	\tilde{g}_{1234} (\rho, \bar{\rho}) = \underset{\delta, m}{\sum} a_{12}
	(\delta, m) a_{\bar{4} \bar{3}} (\delta, m)^{\ast} \rho^{(\delta + m) / 2}
	\bar{\rho}^{(\delta - m) / 2} \label{gtildeexp}
\end{equation}
is absolutely convergent;

(b) the remainder $\tilde{g}_{1234} (\rho, \bar{\rho} ; \delta_{\ast}) \assign
\underset{\delta \geqslant \delta_{\ast}, m}{\sum} a_{12} (\delta, m)
a_{\bar{4} \bar{3}} (\delta, m)^{\ast} \rho^{(\delta + m) / 2}
\bar{\rho}^{(\delta - m) / 2}$ has a powerlaw bound, uniform in
$\delta_{\ast}$:
\begin{equation}
	| \tilde{g}_{1234} (\rho, \bar{\rho} ; \delta_{\ast}) | \leqslant C (1 -
	r)^{- \Delta_1 - \Delta_2 - \Delta_3 - \Delta_4} .
	\label{gtildegeneral:bound}
\end{equation}
This is done as follows (compare Sec.\,\ref{sec:generalbounds}). Consider the
4-point functions $\langle \mathcal{O}_1 \mathcal{O}_2 \mathcal{O}_2^{\dag}
\mathcal{O}_1^{\dag} \rangle$, $\langle \mathcal{O}_4^{\dag}
\mathcal{O}_3^{\dag} \mathcal{O}_3 \mathcal{O}_4 \rangle$, and let
$\tilde{g}_{12 \bar{2} \bar{1}}$, $\tilde{g}_{\bar{4} \bar{3} 34}$ be the
analogues of (\ref{gtildeexp}):
\begin{eqnarray}
	\tilde{g}_{12 \bar{2} \bar{1}} (\rho, \bar{\rho}) & = & \underset{\delta,
		m}{\sum} | a_{12} (\delta, m) |^2 \rho^{(\delta + m) / 2}
	\bar{\rho}^{(\delta - m) / 2},  \nn\\
	\tilde{g}_{\bar{4} \bar{3} 34} (\rho, \bar{\rho}) & = & \underset{\delta,
		m}{\sum} | a_{\bar{4} \bar{3}} (\delta, m) |^2 \rho^{(\delta + m) / 2}
	\bar{\rho}^{(\delta - m) / 2} .  \label{gtildeexp:34}
\end{eqnarray}
Noticing that $| m | \leqslant \delta$, we estimate (\ref{gtildeexp}) by
absolute value and apply Cauchy-Schwarz inequality:
\begin{equation}
	| \tilde{g}_{1234} (\rho, \bar{\rho} ; \delta_{\ast}) | \leqslant
	\underset{\delta, m}{\sum} | a_{12} (\delta, m) | | a_{\bar{4} \bar{3}}
	(\delta, m) | r^{\delta} \leqslant [\tilde{g}_{12 \bar{2} \bar{1}} (r, r)
	\tilde{g}_{\bar{4} \bar{3} 34} (r, r)]^{1 / 2} . \label{gtilde:CS}
\end{equation}
The functions $\tilde{g}_{12 \bar{2} \bar{1}} (r, r)$ and $\tilde{g}_{\bar{4}
	\bar{3} 34} (r, r)$ correspond to the 4-point functions at the Euclidean
configurations with $\rho = \bar{\rho} = r < 1$, hence their series expansions
(\ref{gtildeexp:34}) are convergent by the Euclidean OPE
axiom. Therefore, (\ref{gtildeexp}) is absolutely convergent when $| \rho |, |
\bar{\rho} | < 1$. This finishes the proof of part (a).

Using the t-channel OPE, we can show that for $0 \leqslant r < 1$,
\begin{eqnarray}
	\tilde{g}_{12 \bar{2} \bar{1}} (r, r) & \leqslant & C (1 - r)^{- 2 \Delta_1
		- 2 \Delta_2}, \nn
	\\
	\tilde{g}_{\bar{4} \bar{3} 34} (r, r) & \leqslant & C (1 - r)^{- 2 \Delta_3
		- 2 \Delta_4},  \label{gtildebound:34}
\end{eqnarray}
with some $C > 0$. Combining 
(\ref{gtildebound:34})
with (\ref{gtilde:CS}) we get (\ref{gtildegeneral:bound}). This finishes the
proof of part (b).

\chapter{Optimal powerlaw bound from Cauchy-Schwarz $\rho, \bar{\rho}$
	inequality}\label{secondpass}

In Sec.\ \ref{power4-point} we provided a powerlaw bound for the 4-point function,
based on the inequality {\eqref{rhobound}} for $\max (| \rho (c) |, |
\bar{\rho} (c) |)$. That did the job of allowing us to apply Theorem
\ref{ThVlad} and prove that the Minkowski 4-point function is a distribution, but
the actual bound {\eqref{rhobound}} is not optimal. It is interesting to get a
better bound on $| \rho (c) |, | \bar{\rho} (c) |$, because this will
translate into a better powerlaw bound for the 4-point function, allowing us to
get a better idea about the regularity of the Minkowski 4-point function as a
distribution, i.e.\ how many derivatives test functions must have. In the proof
of Theorem \ref{ThVlad}, parameters $A_n$ and $B_n$ of the powerlaw bound
enter into Eq.\ {\eqref{GMreg}} which provides an upper bound on the
regularity.

In this section we will provide such an optimal bound on $| \rho (c) |, |
\bar{\rho} (c) |$. The main idea of the bound and of its proof is inspired by
Sec.\ \ref{CScompl}. Let us denote by $\mathcal{D}^{(0)}_4$ the subset of
configurations $c \in \mathcal{D}_4$ satisfying the condition $\tmop{Re} x_1^0
> \tmop{Re} x_2^0 > 0 > \tmop{Re} x_3^0 > \tmop{Re} x_4^0$. We showed that the
4-point functions for complexified times satisfy the Cauchy-Schwarz inequality
{\eqref{CS4-point}} for $c = (x_1, x_2, x_3, x_4) \in \mathcal{D}^{(0)}_4$. For a
general configuration $c \in \mathcal{D}^{(0)}_4$ we define two configurations
\begin{equation}
	\label{def:configC12C34} c_{12} = (x_1, x_2, x_2^{\theta}, x_1^{\theta}),
	\quad c_{34} = (x_4^{\theta}, x_3^{\theta}, x_3, x_4),
\end{equation}
where $\theta$ is the operation in {\eqref{reflCompl}} which generalizes the
OS reflection to complexified times. We will call such configurations, for
obvious reasons, reflection-symmetric. It is clear that both $c_{12}, c_{34}
\in \mathcal{D}^{(0)}_4$. Eq.\ {\eqref{CS4-point}} can now be written as
\begin{equation}
	| G_4 (c) |^2 \leqslant G_4 (c_{12}) G_4 (c_{34}) \qquad (c \in
	\mathcal{D}^{(0)}_4) . \label{Gc12c34}
\end{equation}
Since we know that $G_4$ can be written as a convergent power series in
$\rho$, $\bar{\rho}$, Eq.\ {\eqref{Gc12c34}} suggests that there should be a
corresponding bound for the $\rho$, $\bar{\rho}$ coordinates. This is indeed
the case, as we have the following couple of results:

\begin{lemma}
	\label{boundLemma}Any reflection-symmetric configuration $c \in
	\mathcal{D}^{(0)}_4$ has $\rho (c), \bar{\rho} (c) \in (0, 1)$.
\end{lemma}

\begin{theorem}[Cauchy-Schwarz inequality for $\rho, \bar{\rho}$]
	\label{boundThm}For any configuration $c \in \mathcal{D}^{(0)}_4$ we have
	the inequality:
	\begin{equation}
		\label{maxrhoineq} \max \{ | \rho (c) |, | \bar{\rho} (c) | \}^2
		\leqslant \max \{ \rho (c_{12}), \bar{\rho} (c_{12}) \} \times \max
		\{ \rho (c_{34}), \bar{\rho} (c_{34}) \} .
	\end{equation}
\end{theorem}

We will next prove Lemma \ref{boundLemma}. We will then show how, combined
with Theorem \ref{boundThm}, this implies an optimal bound on $\rho,
\bar{\rho}$. Finally we will present a proof of Theorem \ref{boundThm}, which
is surprisingly subtle.

\section{Proof of Lemma \ref{boundLemma}}

To prove the lemma, consider a reflection-symmetric configuration $c$ as in
{\eqref{def:configC12C34}} with:
\begin{eqnarray}
	&  & x_1 = (\epsilon_1 + it_1, \mathbf{x}_1), \quad x_2 = (\epsilon_2 +
	it_2, \mathbf{x}_2), \quad \epsilon_1 > \epsilon_2 > 0, 
	\label{config:reflectionsym1}\\
	&  & x_3 = x_2^{\theta} = (- \epsilon_2 + it_2, \mathbf{x}_2), \quad x_4 =
	x_1^{\theta} = (- \epsilon_1 + it_1, \mathbf{x}_1) . \nonumber
\end{eqnarray}
We will compute $z (c)$, $\bar{z} (c)$ explicitly. We can use translations in
the $\mathbf{x}$ direction, as well as spatial rotations to simplify these
computations. All these transformations do not change the conformal class of
configuration, hence preserve $u, v$ and $z, \bar{z}$. They also commute with time
reflection, and so map reflection-symmetric configurations to
reflection-symmetric ones. By using this freedom, we get an equivalent
configuration $c'$ with the same $z, \bar{z}$:
\begin{equation}
	\label{config:reflectionsym2} x_1' = (\epsilon_1 + i t_1, \tmmathbf{0}),
	\quad x_2' = (\epsilon_2 + it_2, | \mathbf{x}_2 - \mathbf{x}_1 |, 0, \ldots,
	0), \quad x_3' = (x_2')^{\theta}, \quad x_4' = (x_1')^{\theta} .
\end{equation}
This is an effectively two-dimensional configuration. The $z, \bar{z}$
variables of a two-dimensional 4-point configuration $x_k = (x_k^0, x^1_k)$ are
given by Eq.\ {\eqref{zzbarglobal}}, which we copy here
\begin{equation}
	z = \dfrac{(z_1 - z_2)  (z_3 - z_4)}{(z_1 - z_3)  (z_2 - z_4)}, \quad
	\bar{z} = \dfrac{(\bar{z}_1 - \bar{z}_1)  (\bar{z}_3 -
		\bar{z}_4)}{(\bar{z}_1 - \bar{z}_3)  (\bar{z}_2 - \bar{z}_4)}, \quad z_k =
	x_k^0 + i x^1_k, \quad \bar{z}_k = x_k^0 - i x^1_k .
\end{equation}
Applying this to the configuration $c'$, we get $z, \bar{z}$ for $c'$ (which
are the same as for $c$). It's easy to see that $z_3 - z_4 = (z_1 -
z_2)^{\ast}$, $z_1 - z_3 = (z_2 - z_4)^{\ast}$ as a consequence of
reflection symmetry, and similarly for $\bar{z}$'s. So we get $z (c),
\bar{z} (c)$ both real and positive. Explicit expressions come out to be
\begin{eqnarray}
	z (c) = & \dfrac{(\epsilon_1 - \epsilon_2)^2 + (t_1 - t_2 - |
		\mathbf{x}_1 - \mathbf{x}_2 |)^2}{(\epsilon_1 + \epsilon_2)^2 + (t_1 - t_2 -
		| \mathbf{x}_1 - \mathbf{x}_2 |)^2},  \label{z12zbar12}\\
	\bar{z} (c) = & \dfrac{(\epsilon_1 - \epsilon_2)^2 + (t_1 - t_2 + |
		\mathbf{x}_1 - \mathbf{x}_2 |)^2}{(\epsilon_1 + \epsilon_2)^2 + (t_1 - t_2 +
		| \mathbf{x}_1 - \mathbf{x}_2 |)^2} . \nonumber
\end{eqnarray}
In particular we see that $0 < z (c), \bar{z} (c) < 1$. The function $f
(\zeta)$ in the definition of $\rho$ variables maps the interval $(0, 1)$ to
itself. Hence also $0 < \rho (c), \bar{\rho} (c) < 1$, and the lemma and
proved.

\section{Optimal bound for $\rho, \bar{\rho}$}\label{rhsbound}

We wish to derive a powerlaw bound on $\frac{1}{1 - r}$, $r = \max (| \rho |,
| \bar{\rho} |)$, since by the arguments in Sec.\ \ref{power4-point} this
implies a powerlaw bound for the 4-point function. Our aim here is to improve on
{\eqref{rhobound}}, {\eqref{Scbound}}.

Consider first a configuration $c \in \mathcal{D}^{(0)}_4$. For such a
configuration, by Theorem \ref{boundThm}, we have
\begin{equation}
	r (c) \leqslant \sqrt{r (c_{12}) r (c_{34})} \leqslant \max (r (c_{12}),
	r (c_{34})),
\end{equation}
and hence
\begin{equation}
	\frac{1}{1 - r (c)} \leqslant \max \left( \frac{1}{1 - r (c_{12})},
	\frac{1}{1 - r (c_{34})} \right) . \label{ineq:1-r}
\end{equation}
We are thus reduced to study $r (c)$ for reflection-symmetric configurations,
like in {\eqref{config:reflectionsym1}}. By definition (\ref{def:rho}) of
$\rho$ variables, we have
\begin{equation}
	\dfrac{1}{1 - \rho} = \dfrac{1 + \sqrt{1 - z}}{2 \sqrt{1 - z}} \leqslant
	\dfrac{1}{\sqrt{1 - z}}, \quad z \in [0, 1), \label{rhozest}
\end{equation}
so it suffices to study $1 / (1 - z)$ and $1 / (1 - \bar{z})$. Using $z,
\bar{z}$ for reflection-symmetric configurations computed in Eqs.\ {\eqref{z12zbar12}} we have
\begin{equation}
	\frac{1}{1 - z (c_{12})} = \dfrac{(\epsilon_1 + \epsilon_2)^2 + (t_1 - t_2 -
		| \mathbf{x}_1 - \mathbf{x}_2 |)^2}{4 \epsilon_1 \epsilon_2},
\end{equation}
and an analogous relation for $\frac{1}{1 - \bar{z} (c_{12})}$. From these
equations, using $\epsilon_2 < \epsilon_1$, and estimating $\epsilon_1 -
\epsilon_2, | t_1 - t_2 |, \left| \mathbf{x}_1 - \mathbf{x}_2 \right|$ from
above by $| x_1 - x_2 |$ (see {\eqref{absdef}}), we easily get
\begin{equation}
	\frac{1}{1 - z (c_{12})}, \frac{1}{1 - \bar{z} (c_{12})} \leqslant \left( 1
	+ \frac{1}{\epsilon_2} \right)^2 (1 + | x_1 - x_2 |)^2,
\end{equation}
Putting together this relation, an analogous relation for $z (c_{34})$,
$\bar{z} (c_{34})$, Eqs.\ {\eqref{ineq:1-r}} and {\eqref{rhozest}}, we get
\begin{equation}
	\frac{1}{1 - r (c)} \leqslant \max \left\{ \left( 1 + \frac{1}{\epsilon_2}
	\right) (1 + | x_1 - x_2 |), \left( 1 + \frac{1}{| \epsilon_3 |}
	\right) (1 + | x_3 - x_4 |) \right\} \qquad (c \in \mathcal{D}_4^{(0)})\,.
	\label{1-r0}
\end{equation}
This was for $c \in \mathcal{D}_4^{(0)}$. For a general configuration $c \in
\mathcal{D}_4$, we will shift the coordinates by a translation in time
direction (which of course does not change $\rho, \bar{\rho}$), arranging so
that the shifted configurations $c'$ has $\epsilon_2 > 0 > \epsilon_3$, i.e.\ $c' \in \mathcal{D}_4^{(0)}$. Specifically we will choose
\begin{equation}
	\epsilon_2 (c') = \frac{1}{2} (\epsilon_2 (c) - \epsilon_3 (c)), \quad
	\epsilon_3 (c') = - \frac{1}{2} (\epsilon_2 (c) - \epsilon_3 (c)) . \quad
\end{equation}
Then, using {\eqref{1-r0}} for $c'$, we obtain a bound on $\frac{1}{1 - r
	(c)}$ which for example can be expressed as
\begin{equation}
	\frac{1}{1 - r (c)} \leqslant 2 \left( 1 + \frac{1}{\epsilon_2 - \epsilon_3}
	\right) (1 + \max \{ | x_1 - x_2 |, | x_3 - x_4 | \}) \qquad (c \in
	\mathcal{D}_4) . \label{1-rbnd}
\end{equation}
This is a powerlaw bound of the type we were looking for. {By considering reflection-symmetric configurations, it's easy to see that the exponents in this bound cannot be improved. Eq.~\eqref{1-rbnd} is much stronger than our
	previous suboptimal bound {\eqref{rhobound}}; in fact it implies a bound of the same form as \eqref{rhobound} with the power exponent 12 replaced by 2.} 


\section{Proof of Theorem \ref{boundThm}}\label{rhorhobarProof}

Although {\eqref{maxrhoineq}} looks like a simple-enough geometric inequality,
we do not know an elementary proof of this fact. We essentially guessed this
inequality, checked it numerically, and then looked for a proof. Our guess
started in the Euclidean region, where $\bar{\rho} = \rho^{\ast}$, and
{\eqref{maxrhoineq}} takes the form
\begin{equation}
	| \rho (c) |^2 \leqslant \rho (c_{12}) \rho (c_{34}) \qquad ( c \in
	\mathcal{D}^{(0)}_4 \text{\quad Euclidean} ) . \label{CSrrbarEucl}
\end{equation}
Even in this case we don't know an elementary proof. We guessed that this must
hold, because otherwise it was hard to imagine that the 4-point function itself
would satisfy a Cauchy-Schwarz inequality. Indeed {\eqref{CSrrbarEucl}}
implies the Euclidean version of {\eqref{CS4-point}}. We then guessed
{\eqref{maxrhoineq}} as a generalization of {\eqref{CSrrbarEucl}} for
complexified times.

Our proof of {\eqref{maxrhoineq}} reverses this logic, by deriving it from
{\eqref{CS4-point}}. There exist many explicit CFT 4-point functions, e.g.\ mean field
theories (MFT). One could imagine that by considering {\eqref{CS4-point}} for a
family of such 4-point functions, and passing to some limit (e.g.\ of scaling
dimension of the mean field going to infinity), one could recover
{\eqref{maxrhoineq}}. We haven't managed to make this work using MFTs, but a
closely related strategy does work. Namely we will apply this sort of argument
not to the full 4-point function, but to a single conformal block, since the
latter also satisfy {\eqref{CS4-point}} (as we will explain).

Now that we explained the main idea, let us supply the details. By applying a
translation, we may set $\mathbf{x}_1 = 0$. The remaining spacial component
vectors $\mathbf{x}_2, \mathbf{x}_3, \mathbf{x}_4$ span at most
three-dimensional subspace of $\mathbb{R}^{d - 1}$. This shows that it is
enough to prove the inequality {\eqref{maxrhoineq}} in the case $d = 4$. We
assume that the readers are familiar with the conformal blocks, which encode
contributions of a primary into a 4-point function. In the considered case of 4
identical Hermitean scalar, the relevant OPE has the form (simplifying the
general case considered in Sec.\ \ref{OSfromCFT})
\begin{equation}
	\varphi (x_1) \varphi (x_2) = f_{\varphi \varphi \mathcal{O}} C_{(\lambda)}
	(x_1, x_2, x_0, \mathcal{D}_0) \mathcal{O}^{(\lambda)} (x_0)
\end{equation}
where $\mathcal{O}^{(\lambda)}$ is a dimension $\Delta$, spin $\ell$ symmetric
traceless primary. The conformal block then can be computed by
\begin{equation}
	g_{\Delta, \ell} (c) = C_{(\lambda)} (x_1, x_2, x_0, \mathcal{D}_0)
	C_{(\mu)} (x^{\theta}_3, x^{\theta}_4, x_0, \mathcal{D}^{\theta}_0)
	\langle \mathcal{O}^{(\lambda)} (x_0) \mathcal{O}^{\dagger (\mu)}
	(x^{\theta}_0) \rangle . \label{CBrepr}
\end{equation}
The 4d Euclidean conformal blocks are known explicitly
{\cite{Dolan:2000ut,Dolan:2003hv}}:
\begin{equation}
	\label{cb:dolanosborn} g_{\Delta, \ell} (c) = \frac{z \bar{z}}{z -
		\bar{z}} [k_h (z) k_{\bar{h} - 1} (\bar{z}) - k_h (\bar{z}) k_{\bar{h} - 1}
	(z)],
\end{equation}
where $h, \bar{h} = (\Delta \pm \ell) / 2$, and $k_{\beta} (z) = z^{\beta}{}_2 F_1 (\beta, \beta, 2 \beta ; z)$. (We only cite the result for equal
external dimensions.) We will assume that the exchanged operator satisfies the
4d unitarity bound $\Delta \geqslant \ell + 2$. As Eq.
{\eqref{cb:dolanosborn}} shows, Euclidean conformal blocks are real-analytic
functions whenever $| z | < 1$. We can also use this formula to analytically
continue them to the forward tube. We wish to show that this analytic
continuation satisfies some properties. This is best shown not from the
explicit formula, but by adapting the robust 4-point function arguments from
chapter \ref{sec:4-point}. Indeed, conformal blocks allow an expansion of the same
form as {\eqref{g:rhoexpansion}}, with non-negative coefficients which are
fixed by conformal invariance. This can be shown by arguments similar to those
in Sec.\ \ref{Eucl4-point}. The existence of the representation {\eqref{CBrepr}}
guarantees Hilbert space unitarity. Then, by the arguments of Sec.\ \ref{anal4-point}, conformal blocks admit an analytic extension to the forward
tube (which is of course the same as the one following from the explicit
formula {\eqref{cb:dolanosborn}}). The point of the current construction is
that it shows that the analytic extension satisfies an inequality analogous to
{\eqref{ggE}}:
\begin{equation}
	| g_{\Delta, \ell} (c) | \leqslant g_{\Delta, \ell} (c_{\ast}) \label{ggECB}
\end{equation}
Then, by the arguments in Sec.\ \ref{power4-point}, conformal blocks satisfy the
powerlaw bound in the forward tube. (As is easy to see from
{\eqref{cb:dolanosborn}}, 4d Euclidean conformal blocks grow as $1 / (1 - z)$
as $z \rightarrow 1^-$ along the real axis, which replaces Eq.
{\eqref{g4cstar}}.) Finally, by the arguments analogous to Sec.\ \ref{CScompl} we conclude that the analytically continued conformal blocks
satisfy Cauchy-Schwarz inequality:
\begin{equation}
	| g_{\Delta, l} (c) |^2 \leqslant g_{\Delta, l} (c_{12}) g_{\Delta, l}
	(c_{34}) \quad \text{for any } c \in \mathcal{D}^{(0)}_4 . \label{CB-CS}
\end{equation}
(Euclidean reflection positivity of conformal blocks follows from the
representation {\eqref{CBrepr}}, which we assume to be valid in the Euclidean
region.)

In the rest of the argument we will only need two facts, the Cauchy-Schwarz
inequality {\eqref{CB-CS}} and the explicit Dolan-Osborn formula
{\eqref{cb:dolanosborn}}. We will apply {\eqref{CB-CS}} to the blocks of spin
$\ell \geqslant 1$ at the unitarity bound, i.e.\ with $\bar{h} = 1, h = \ell +
1$. The Cauchy-Schwarz inequality for $\rho, \bar{\rho}$ will follow by
extracting the asymptotics in the limit $h \rightarrow + \infty$. The
asymptotic behavior of $k_h$ is given by the following lemma:

\begin{lemma}
	\label{lemma:khasymp}For any fixed $z \in \mathbb{C} \setminus [1, +
	\infty)$, the function $k_h (z)$ has the following asymptotic behavior in
	terms of the $\rho$ variable defined in (\ref{def:rho}):
	\begin{equation}
		\label{kh:asymptotic} k_h (z) = (4 \rho)^h \left[ \dfrac{1}{\sqrt{1 -
				\rho^2}} + o (1) \right], \quad h \rightarrow + \infty .
	\end{equation}
\end{lemma}

\begin{proof}
	We have the following identity for $k_h (z)$ {\cite{Hogervorst:2013sma}}:
	\begin{equation}
		k_h (z) = (4 \rho)^h _2 F_1 (1 / 2, h ; h + 1 / 2 ; \rho^2) .
	\end{equation}
	The region $z \notin [1, + \infty)$ corresponds to $| \rho | < 1$, where the
	hypergeometric function $_2 F_1$ has the power series representation
	\begin{equation}
		\label{f21:series}_2 F_1 (1 / 2, h ; h + 1 / 2 ; \rho^2) = \sum_{n =
			0}^{\infty} \dfrac{(1 / 2)_n (h)_n}{n! (h + 1 / 2)_n} \rho^{2 n} .
	\end{equation}
	When $h \rightarrow + \infty$, each coefficient of the series increases
	monotonically, and tends to the coefficients of the convergent in the disk
	$| \rho | < 1$ series
	\[ \sum_{n = 0}^{\infty} \dfrac{(1 / 2)_n}{n!} \rho^{2 n} =
	\dfrac{1}{\sqrt{1 - \rho^2}} . \]
	This implies the statement of the lemma.
\end{proof}

Consider now inequality {\eqref{CB-CS}} for the blocks with $\bar{h} = 1$, $h
= \ell + 1$. Since $k_0 \equiv 1$, it reads:
\begin{equation}
	| w \cdot [k_h (z) - k_h (\bar{z})] |^2 \leqslant w_{12} w_{34} \cdot [k_h
	(z_{12}) - k_h (\bar{z}_{12})] [k_h (z_{34}) - k_h (\bar{z}_{34})],
	\label{CBineq}
\end{equation}
where we denoted $w = \frac{z \bar{z}}{z - \bar{z}}$, and similarly $w_{12},
w_{34}$. Let us assume that the configuration $c \in \mathcal{D}^{(0)}_4$ is
such that
\begin{equation}
	\label{condition:rho} | \rho | \neq | \bar{\rho} |, \quad \rho_{12} \neq
	\bar{\rho}_{12}, \quad \rho_{34} \neq \bar{\rho}_{34} .
\end{equation}
Then, using Lemma \ref{lemma:khasymp}, for large $h$ inequality
{\eqref{CBineq}} becomes:
\begin{equation}
	(A + o (1)) \max \{ | \rho |, | \bar{\rho} | \}^{2 h} \leqslant (B + o
	(1)) \max \{ \rho_{12}, \bar{\rho}_{12} \}^h \max \{ \rho_{34},
	\bar{\rho}_{34} \}^h,
\end{equation}
where $A, B$ are some positive $h$-independent quantities. Now, taking the
limit $h \rightarrow + \infty$, we obtain inequality {\eqref{maxrhoineq}}.

It's easy to see that configurations which violate the condition
{\eqref{condition:rho}} are non-generic. They can be approached by
configurations which do satisfy {\eqref{condition:rho}}. Therefore, by
continuity inequality {\eqref{maxrhoineq}} is valid also for such exceptional
configurations.

\chapter{OPE convergence in the forward tube and in Minkowski
	space}\label{OPEconvMink}

We have several OPE convergence statements scattered throughout this part of the thesis. The
Euclidean CFT axioms assume convergence of the OPE series for $\mathcal{O}_1
(x_1) \mathcal{O}_2 (x_2)$ whenever the two points $x_1, x_2$ are closer to
the OPE center than any other point. Then we established OPE convergence in
the Hilbert space sense (Sec.\ \ref{Hilbert}) in the Euclidean region for
states generated by two operators in the half-space. Then in Sec.\ \ref{Eucl4-point} we used Hilbert space language to derive the power series
representation {\eqref{g:rhoexpansion}} for the 4-point function, whose
convergence is thus morally equivalent to OPE convergence (for the 4-point
functions). We then used this power series representation to analytically
continue the 4-point function to the forward tube, and then define the Minkowski
4-point function as a boundary value in the sense of distributions. Finally, in
Sec.\ \ref{sec:Wpos} we showed, by arguments not using conformal invariance,
that the OS states $| \nobracket \mathcal{O}_1 (x_1) \mathcal{O}_2 (x_2)
\rangle$ can be extended to the forward tube and (when integrated against test
functions) to the Minkowski region, and that the so obtained states can be
arbitrarily well approximated by (integrated) OS states. Therefore, OPE
convergence holds for these states, as for the OS states.

In this section we will give a more explicit discussion of the OPE
convergence for the Minkowski 4-point function and for the 2-operator states in
the forward tube and Minkowski space. We will also explain how our approach
and results compare to the classic paper by Mack {\cite{Mack:1976pa}}.

\section{Convergence of conformal block decomposition for 4-point functions}

Let us consider the 4-point function of identical scalars
{\eqref{def:Euclidean4-point}}:
\begin{equation}
	G (x_1, x_2, x_3, x_4) \equiv G (c) = \langle \mathcal{O} (x_1) \mathcal{O}
	(x_2) \mathcal{O} (x_3) \mathcal{O} (x_4) \rangle = \frac{g (\rho,
		\bar{\rho})}{(x_{12}^2 x_{34}^2)^{\Delta_{\mathcal{O}}}} .
\end{equation}
The discussion below can be easily extended to non-identical scalars using the
same ideas as in Sec.\ \ref{nonId}.

We know that in the Euclidean region the function $g (\rho, \bar{\rho})$ has a
convergent conformal block decomposition
\begin{equation}
	g (\rho, \bar{\rho}) = \underset{\Delta, l}{\sum} C_{\Delta, l}^2 g_{\Delta,
		l} (\rho, \bar{\rho}) .
\end{equation}
As in Sec.\ \ref{rhorhobarProof}, we will assume that the reader is familiar
with conformal blocks. The main point is that the conformal block
decomposition is obtained by separating the series {\eqref{g:rhoexpansion}}
into parts corresponding to the conformal multiplets of primary operators
$\mathcal{O}_{\Delta, l}$ occurring in the $\mathcal{O} \times \mathcal{O}$ OPE
with coefficients $C_{\Delta, l}$. Conformal blocks in the Euclidean region by
themselves have power series expansions like {\eqref{g:rhoexpansion}} with
positive coefficients (fixed by conformal symmetry). As in Sec.\ \ref{anal4-point}, we can use this expansion to analytically continue conformal
blocks to the forward tube. By an analogue of the bound {\eqref{maj}} we know
that conformal block expansion remains convergent everywhere in the forward
tube, since $| \rho |, | \bar{\rho} | < 1$ there. Since individual conformal
blocks are smaller than the 4-point function in the Euclidean region, by the
arguments in Sec.\ \ref{power4-point} we know that they satisfy a powerlaw
bound, and hence they become tempered distributions when going to the
Minkowski region.\footnote{This argument shows that any conformal block which
	occurs in a reflection-positive CFT 4-point function satisfies a powerlaw bound.
	E.g.\ conformal blocks for $l \geqslant 0$, $\Delta \geqslant l + d - 2$ occur
	in a 4-point function $\langle \varphi_1 \varphi_2 \varphi_1 \varphi_2 \rangle$
	where $\varphi_1, \varphi_2$ are two GFFs of appropriately chosen equal
	dimension. It should be also possible to show that conformal blocks satisfy a
	powerlaw bound without relying on a fiducial 4-point function. E.g.\ for $d = 4$
	conformal blocks this follows from their explicit Dolan-Osborn expressions.
	For general $d$, powerlaw bound on the diagonal $z = \bar{z}$ can be shown
	using the differential equation found in {\cite{Hogervorst:2013kva}} and
	extended to $z \neq \bar{z}$ by the usual arguments.}${}^{,}$\footnote{It should be
	noted that away from light cones conformal blocks are better than
	distributions: they are real-analytic there (although this fact won't play a
	role for us). In even $d$ this is obvious from their explicit expressions in
	terms of hypergeometric functions. For general $d$ this follows from a
	first-order matrix ODE satisfied by a finite-length vector including the
	conformal block and its low-order derivatives. Such an ODE exists for a
	length-8 vector and can be built using the quadratic and quartic Casimir
	equations {\cite{KravchukCB}}.}${}^{,}$\footnote{\label{noteMarc2}Also ``conformal
	partial waves'' $\frac{g_{\Delta, l} (\rho, \bar{\rho})}{(x_{12}^2
		x_{34}^2)^{\Delta_{\mathcal{O}}}}$ are tempered distributions in the
	Minkowski space. Therefore their Fourier transforms are well defined. Explicit
	expressions for these Fourier transforms were found recently in
	{\cite{Gillioz:2020wgw}}. }

By the arguments like in Sec.\ \ref{power4-point}, $g (\rho, \bar{\rho})$,
the partial sums of the conformal block decomposition $g (\rho, \bar{\rho} ;
\Delta_{\ast}) = \underset{\Delta \leqslant \Delta_{\ast}, l}{\sum} C_{\Delta,
	l}^2 g_{\Delta, l} (\rho, \bar{\rho})$, and the corresponding remainders satisfy in the forward
tube a uniform bound:
\begin{equation}
	| g (\rho, \bar{\rho} ; \Delta_{\ast}) |, | g (\rho, \bar{\rho}) - g (\rho,
	\bar{\rho} ; \Delta_{\ast}) | \leqslant C (1 - r)^{- 4 \Delta}, \qquad r =
	\max \{ | \rho |, | \bar{\rho} | \} . \label{powerlawbound:block}
\end{equation}
Consider the 4-point partial sums including the prefactor $G (c ; \Delta_{\ast}) =
\frac{1}{(x_{12}^2 x_{34}^2)^{\Delta_{\mathcal{O}}}} g (\rho, \bar{\rho} ;
\Delta_{\ast})$. By the powerlaw bound of $(1 - r (c))^{- 1}$, we have the
powerlaw bounds
\begin{eqnarray}
	| G (c ; \Delta_{\ast}) |, | G (c) - G (c ; \Delta_{\ast}) | & \leqslant &
	C_{}  \left( 1 + \max_k  \dfrac{1}{\epsilon_k - \epsilon_{k + 1}}
	\right)^{A}  (1 + \max_i  | x_i - x_{i + 1} |)^B  \label{powerlawbound:G}
\end{eqnarray}
for all $c \in \mathcal{D}_4$ and $\Delta_{\ast} \geqslant 0$. \ Consider the
boundary value of $G (c ; \Delta_{\ast})$, call it $G^M (x_1, x_2, x_3, x_4 ;
\Delta_{\ast})$, where $x_i \in \mathbb{R}^{1, d - 1}$; it is a tempered
distribution by Theorem \ref{ThVlad}. The following theorem establishes
distributional convergence of conformal block decomposition.

\begin{theorem}
	\label{theorem:districonverge}We have $G^M (x ; \Delta_{\ast}) \rightarrow
	G^M (x)$ in the sense of tempered distributions. 
\end{theorem}

\begin{proof}
	Denote $H (c ; \Delta_{\ast}) = G (c) - G (c ; \Delta_{\ast})$. We have to
	show that, as $\Delta_{\ast}$ goes to infinity, the boundary value of $H (c
	; \Delta_{\ast})$ converges to 0 in the sense of tempered distributions,
	i.e, for any Schwartz test function $f \in \mathcal{S} (\mathbb{R}^{4 d})$
	\begin{equation}
		\underset{\Delta_{\ast} \rightarrow \infty}{\lim} \lim_{\lambda
			\rightarrow 0^+}  \int H (\lambda \epsilon + i t, \mathbf{x};
		\Delta_{\ast}) f (t, \mathbf{x})\, d t\, d \mathbf{x}= 0, \qquad
		(\epsilon_1 > \epsilon_2 > \epsilon_3 > \epsilon_4), \label{Hf}
	\end{equation}
	where we write for brevity $t$ instead of $t_1, t_2, t_3, t_4$ etc. The
	proof is the same as in part \ref{part:crossratio}, theorem \ref{thm:vlad}. We will
	retrace here the main steps for completeness and because we will need it to
	establish a stronger result below. Define $L_f (\lambda ; \Delta_{\ast})
	\assign \int H (\lambda \epsilon + i t, \mathbf{x}; \Delta_{\ast}) f (x)\,
	d x$ with $x = (t, \mathbf{x})$. Since $H$ is holomorphic in $\tau = \lambda
	\epsilon + i t$ we have
	\begin{equation}
		\begin{split}
			L_f^{(n)} (\lambda ; \Delta_{\ast}) =& \int \left( \left( \epsilon \cdummy
			\frac{\partial}{i \partial t} \right)^n H (\lambda \epsilon + i t,
			\mathbf{x}; \Delta_{\ast}) \right) f (x)\, d x \\
			=& \int H (\lambda \epsilon
			+ i t, \mathbf{x}; \Delta_{\ast}) \left( \left( i \epsilon \cdummy
			\frac{\partial}{\partial t} \right)^n f (x) \right)\, d x,
		\end{split}    
	\end{equation}
	which by the powerlaw bound {\eqref{powerlawbound:G}} implies
	\begin{equation}
		L_f^{(n)} (\lambda ; \Delta_{\ast}) \leqslant \frac{C_n}{\lambda^A} | f
		|_{p_n}, \qquad \lambda \in (0, 1], \qquad p_n = \max \{ n, \lceil B
		\rceil + 4 d + 1 \} . \label{Lfn:bound1}
	\end{equation}
	These bounds blow up in the $\lambda \rightarrow 0$ limit, but by using the
	Newton-Leibniz repeatedly one can get bounds which do not blow up:
	\begin{equation}
		L_f^{(n)} (\lambda ; \Delta_{\ast}) \leqslant D_n | f |_{p_{n + [A] + 1}},
		\qquad \lambda \in (0, 1] . \label{Lfn:bound2}
	\end{equation}
	Using this for $n = 1$ one proves that the limit $L_f (0 ; \Delta_{\ast})
	= \lim_{\lambda \rightarrow 0^+} L_f (\lambda ; \Delta_{\ast})$ exists
	and
	\begin{equation}
		| L_f (0 ; \Delta_{\ast}) - L_f (\lambda ; \Delta_{\ast}) | \leqslant D_1
		\lambda | f |_{\max \{ [A] + 2, \lceil B \rceil + 4 d + 1 \}} .
	\end{equation}
	By Lebesgue's dominated convergence theorem, for any fixed $\lambda$ in $(0,
	1]$, $L_f (\lambda ; \Delta_{\ast})$ tends to zero as $\Delta_{\ast}
	\rightarrow + \infty$. Thus the previous bound implies
	\begin{eqnarray}
		\underset{\Delta_{\ast} \rightarrow \infty}{\overline{\lim}} | L_f (0,
		\Delta_{\ast}) | & \leqslant & \underset{\Delta_{\ast} \rightarrow
			\infty}{\overline{\lim}} | L_f (0, \Delta_{\ast}) - L_f (\lambda ;
		\Delta_{\ast}) | + \underset{\Delta_{\ast} \rightarrow
			\infty}{\overline{\lim}} | L_f (\lambda ; \Delta_{\ast}) | \nonumber\\
		& \leqslant & D_1 \lambda | f |_{\max \{ [A] + 2, \lceil B \rceil + 4 d +
			1 \}} .  \label{Lf0estimate}
	\end{eqnarray}
	Since $\lambda$ can be arbitrarily small, we get $\underset{\Delta_{\ast}
		\rightarrow \infty}{\overline{\lim}} | L_f (0, \Delta_{\ast}) | = 0$. This
	finishes the proof.
\end{proof}

\subsection{Convergence rate for compactly supported test
	functions}\label{subsection:rateconvergence}

Because of the use of Lebesgue's theorem on dominated convergence, Theorem
\ref{theorem:districonverge} does not give the rate of convergence. We will
now give the rate in an important special case of compactly supported test
functions. This provides an explicit example for the remark in the last paragraph of Sec.\ \ref{sec:proof}.

The idea is that not only $H (c, \Delta_{\ast}) \rightarrow 0$ pointwise but
it does so exponentially fast. We will first derive the exponential
convergence bound, upgrading the Euclidean argument from
{\cite{Pappadopulo:2012jk}}, to the forward tube, and then use it. Let $F (t)$
be the Laplace transform of a positive measure $\mu (E)$ on $E \geqslant 0$:
\begin{equation}
	F (t) = \int_0^{\infty} \mu (E) e^{- E t}\, d E, \qquad \mu (E) \geqslant 0.
	\label{def:ft}
\end{equation}
We assume that the integral is convergent for $t > 0$ and $F (t) \sim t^{-
	\alpha}$ as $t \rightarrow 0^+$, and $\alpha > 0$. Then by the
Hardy-Littlewood tauberian theorem we know that
\begin{equation}
	M (E) = \int_0^E \mu (E')\, d E' \sim \frac{E^{\alpha}}{\Gamma (\alpha + 1)}
	\tmop{as} E \rightarrow + \infty .
\end{equation}
We can now estimate the remainder $F_{E_{\ast}} (t) \assign
\int_{E_{\ast}}^{\infty} \mu (E) e^{- E t}\, d E,$ via
\begin{equation}
	F_{E_{\ast}} (t) = \int_{E_{\ast}}^{\infty} e^{- E t}\, d M (E) = - M
	(E_{\ast}) e^{- E_{\ast} t} + t \int_{E_{\ast}}^{\infty} M (E) e^{- E t}\, d
	E,
\end{equation}
which gives $\begin{array}{lll}
	| F_{E_{\ast}} (t) | & \leqslant & C_1 e^{- E_{\ast} t} E_{\ast}^{\alpha} +
	C_2 e^{- E_{\ast} t} \left( \frac{1 + E_{\ast} t}{t} \right)^{\alpha},
\end{array}$ and finally
\begin{equation}
	| F_{E_{\ast}} (t) | \leqslant \tmop{const} \times e^{- E_{\ast} t} (t^{- 1}
	+ E_{\ast})^{\alpha} . \label{ft:remainderestimate}
\end{equation}
for any $E_{\ast} \geqslant 1$ (say).

Now let's go back to the 4-point function of four identical scalars $G (c) =
\langle \mathcal{O} (x_1) \mathcal{O} (x_2) \mathcal{O} (x_3)
\mathcal{O} (x_4) \rangle$, $c \in \mathcal{D}_4$. The remainder $H (c ;
\Delta_{\ast})$ can clearly be bounded by replacing $\rho, \bar{\rho}$ with $r
= \max \{ | \rho |, | \bar{\rho} | \}$:
\begin{equation}
	| H (c ; \Delta_{\ast}) | \leqslant \frac{1}{| x_{12}^2
		|^{\Delta_{\mathcal{O}}} | x_{34}^2 |^{\Delta_{\mathcal{O}}}} [g (r, r) - g
	(r, r ; \Delta_{\ast})] \label{CS:GN},
\end{equation}
By setting $t = \log (1 / r)$, $g (r, r)$ with its representation
{\eqref{g:rhoexpansion}} is in the same form as (\ref{def:ft}) and by Eq.
{\eqref{g4cstar}} we have that the corresponding $F (t) \sim t^{- \alpha}$
with $\alpha = 4 \Delta_{\mathcal{O}}$. Bounding the remainder $g (r, r) - g
(r, r ; \Delta_{\ast})$ by (\ref{ft:remainderestimate}), and using \
{\eqref{CS:GN}}, we get a bound on the remainder $| H (c ; \Delta_{\ast})
|$ for any $\Delta_{\ast} \geqslant 1$ (say):
\begin{equation}
	| H (c ; \Delta_{\ast}) | \leqslant \frac{\tmop{const}}{| x_{12}^2
		|^{\Delta_{\mathcal{O}}} | x_{34}^2 |^{\Delta_{\mathcal{O}}}} \times r
	(c)^{\Delta_{\ast}} \left( \frac{1}{\log (r (c)^{- 1})} + \Delta_{\ast}
	\right)^{4 \Delta_{\mathcal{O}}} . \label{g1221:bound}
\end{equation}
Let us now convert this into an explicit estimate for the distributional
convergence rate, improving on Theorem \ref{theorem:districonverge} for
compactly supported test functions. Recall that we have an upper bound on $r
(c) = \max \{ |\rho(c)|, |\bar{\rho}(c)| \}$ ($c = (x_1, x_2, x_3,
x_4)$), Eq.\ {\eqref{1-rbnd}}, which we copy here:
\begin{equation}
	\frac{1}{1 - r (c)} \leqslant 2 \left( 1 + \frac{1}{\epsilon_2 - \epsilon_3}
	\right) (1 + \max \{ | x_1 - x_2 |, | x_3 - x_4 | \}) . \label{copybnd}
\end{equation}
This bound tells us how much $r (c)$ is separated from 1. In turn, by
{\eqref{g1221:bound}} this translates into an explicit bound on $| H (c ;
\Delta_{\ast}) |$. Let us retrace the proof of Theorem
\ref{theorem:districonverge}, replacing Eq.\ {\eqref{Lf0estimate}} by
\begin{eqnarray}
	| L_f (0, \Delta_{\ast}) | & \leqslant & | L_f (0, \Delta_{\ast}) - L_f
	(\lambda ; \Delta_{\ast}) | + | L_f (\lambda ; \Delta_{\ast}) | \nonumber\\
	& \leqslant & D_1 \lambda | f |_{\max \{ [A] + 2, \lceil B \rceil + 4 d + 1
		\}} + | L_f (\lambda ; \Delta_{\ast}) | .  \label{Lf0estimate1}
\end{eqnarray}
We will now choose $\lambda$ small, as a function of $\Delta_{\ast}$, so that
the second term in the r.h.s.\ is smaller than the first one. Let us choose and
fix $\epsilon_1 > \epsilon_2 > \epsilon_3 > \epsilon_4$. For $x_k^{\lambda} =
(\lambda \epsilon_k + i t_k, \mathbf{x}_k)$, $x_k^M = (i t_k, \mathbf{x}_k)$,
bound {\eqref{copybnd}} gives
\begin{equation}
	\frac{1}{1 - r (c^{\lambda})} \leqslant \frac{C_{\epsilon}}{\lambda} (1 +
	\max \{ | x^M_1 - x^M_2 |, | x^M_3 - x^M_4 | \}) \qquad (0 < \lambda
	\leqslant 1), \label{rlambda:bound}
\end{equation}
where $C_{\epsilon}$ is a constant which depends only on $\epsilon_i$ but not
on $\lambda$ or $x_k^M$. If $f \in C^{\infty}_0 (\mathbb{R}^{4 d})$, a
compactly supported test function, then (\ref{rlambda:bound}) implies
\begin{equation}
	\frac{1}{1 - r (c^{\lambda})} \leqslant \frac{A_f}{\lambda}
	\label{rlambda:fbound}  \qquad (0 < \lambda \leqslant 1, (x_k^M) \in
	\tmop{supp} (f)),
\end{equation}
where $A_f = 2 C_{\epsilon} \underset{(x_k^M) \in \tmop{supp} (f)}{\sup} (1 +
\max \{ | x^M_1 - x^M_2 |, | x^M_3 - x^M_4 | \})$.

Now by {\eqref{g1221:bound}}, {\eqref{rlambda:fbound}}, and Lemma
\ref{x2bnd}(b) we have a bound for a compactly supported test function $f$:
\begin{equation}
	| L_f (\lambda ; \Delta_{\ast}) | \leqslant \frac{\tmop{const}}{\lambda^{4
			\Delta_{\mathcal{O}}}} e^{- \frac{\lambda}{A_f} \Delta_{\ast}} \left(
	\frac{A_f}{\lambda} + \Delta_{\ast} \right)^{4 \Delta_{\mathcal{O}}} \int |
	f |\, d x.
\end{equation}
By this and {\eqref{Lf0estimate1}} we have
\begin{equation}
	| L_f (0 ; \Delta_{\ast}) | \leqslant A_1 \lambda + \frac{A_2}{\lambda^{4
			\Delta_{\mathcal{O}}}} e^{- \frac{\lambda}{A_f} \Delta_{\ast}} \left(
	\frac{A_f}{\lambda} + \Delta_{\ast} \right)^{4 \Delta_{\mathcal{O}}},
\end{equation}
where all constants $A_1, A_2, A_f$ are $f$-dependent. If we choose $\lambda =
1 / \Delta_{\ast}^{\gamma}$ then for any $\gamma \in (0, 1)$ the first term
dominates (the second term decays exponentially fast for large
$\Delta_{\ast}$). It is easy to see that the first term still dominates for
$\lambda = \kappa \frac{\log \Delta_{\ast}}{\Delta_{\ast}}$ with
sufficiently large $\kappa$. We therefore obtain the following promised
strengthening of Theorem \ref{theorem:districonverge}.

\begin{theorem}
	\label{theorem:districonverge1}For any compactly supported $C^{\infty}$ test
	function $f$, we have $(G^M (\Delta_{\ast}), f) - (G^M, f) \rightarrow 0$ as
	$O \left( \left. \frac{\log \Delta_{\ast}}{\Delta_{\ast}} \right) \right.$.
\end{theorem}

It is somewhat surprising that conformal block decomposition converges so
slowly in the Minkowski region, while it converges exponentially fast in the
Euclidean region.

\section{OPE convergence in the sense of
	$\mathcal{H}^{\tmop{CFT}}$}\label{OPEHCFT}

We will now rephrase the question of OPE convergence from the point of view of
states generated by two operators (rather than 4-point functions which represent
inner products of such states). We already discussed these questions to some
extent in Sec.\ \ref{Hilbert} in the Euclidean region, and Sec.\ \ref{sec:Wpos} in the forward tube and in the Minkowski region. We will now
update that discussion.

Recall that in Sec.\ \ref{sec:Wpos} we defined states (see Eq.
{\eqref{Ocompl}} and Remark \ref{genrefl})
\begin{equation}
	\psi (x_1, x_2) = \left| \mathcal{O} (x_1) \mathcal{O} (x_2) \rangle, \quad
	x_k = (\epsilon_k + i t_k, \mathbf{x}_k + i\mathbf{y}_k) \right., \quad 0
	\succ (\epsilon_1, \mathbf{y}_2) \succ (\epsilon_1, \mathbf{y}_2),
	\label{f.t.states}
\end{equation}
as elements of a vector space with inner products {\eqref{complinner}}:
\begin{equation}
	\langle \mathcal{O} (x_1) \mathcal{O} (x_2) | \nobracket \mathcal{O} (x_3)
	\mathcal{O} (x_4) \rangle = G_4 (x^{\theta}_2, x^{\theta}_1, x_3, x_4),
\end{equation}
where $x^{\theta} = (- \epsilon + i t, \mathbf{x}- i\mathbf{y})$ for $x =
(\epsilon + i t, \mathbf{x}+ i\mathbf{y})$. We then proved that this inner
product was positive definite, and that these states could be approximated in
norm by Euclidean states, and so belong to the same Hilbert space. The
arguments of Sec.\ \ref{sec:Wpos} did not use conformal symmetry.

So, by arguments of Sec.\ \ref{sec:Wpos} we have a map $\psi (x_1, x_2)$
from $x_1, x_2$ as in {\eqref{f.t.states}} into $\mathcal{H}^{\tmop{CFT}}$. {We claim that this map is holomorphic.
	To begin with, this map is continuous with respect to
	the $\mathcal{H}^{\tmop{CFT}}$ norm, and in particular bounded on compact subsets. This follows easily from the continuity of $G_4$. To show holomorphicity, we will use Morera's theorem and Osgood's lemma (which
	remain valid for Hilbert-space-valued functions of complex
	variables). Morera's theorem says that a locally continuous
	function of one complex variable is holomorphic if its integral over any small contour is zero. Let $C$ be a small 1d contour in the region of $\xi =
	(x_1, x_2)$ where $\psi$ is defined (we assume that one of the components of $\xi$ goes around the contour while the others stay fix). We have
	\begin{equation}
		\left\| \int_C d \xi\, \psi (\xi) \right\|^2 = \int_{\xi' \in C} d
		\overline{\xi'}\, \int_{\xi \in C} d \xi\, G_4 (\xi^{\prime \theta}, \xi) = 0\,,
		\label{Morera}
	\end{equation}
	the last integral being zero because $G_4$ is holomorphic in $\xi$.
	Hence $\int_C d \xi\, \psi (\xi)=0$ and by Morera's theorem $\psi(x_1,x_2)$ is holomorphic in each component separately.
	Finally, by Osgood's lemma \cite{Osgood} $\psi(x_1,x_2)$ is holomorphic in all variables jointly.}

Let us connect this discussion with the OPE. In the Euclidean region, OPE says
\begin{eqnarray}
	\psi (x_1, x_2) = | \mathcal{O} (x_1) \mathcal{O} (x_2) \rangle & = &
	\underset{k, \lambda}{\sum} f_{\mathcal{O}\mathcal{O}k} C_{k, \lambda} (x_1,
	x_2, x_S, \mathcal{D}) | (\mathcal{O}^{\dagger}_k)^{(\lambda)} (x_S)
	\rangle,  \label{OPE:Euclidean}\\
	C_{k, \lambda} (x_1, x_2, x_S, \mathcal{D}) & = & \sum_{\alpha}
	f_{\mathcal{O}\mathcal{O}k} C_{k, \lambda, \alpha} (x_1, x_2, x_S)
	\mathcal{D}^{\alpha}, \nonumber
\end{eqnarray}
where $x_1, x_2$ are two Euclidean points in the lower halfspace, $0 > x_1^0,
x_2^0$, $x_S = (- 1, 0, \ldots)$ is the south pole, and $\mathcal{D}
=\mathcal{D}_{x_S}$ is the image of $\partial / \partial x |_{x = 0}
\nobracket$ under a conformal transformation which maps $0, \infty$ to $x_S,
x_N$. We proved in Sec.\ \ref{Hilbert}, from the Euclidean CFT axioms, that
the series in the r.h.s.\ converges in $\mathcal{H}^{\tmop{CFT}}$. As discussed
in Sec.\ \ref{Hilbert}, convergence holds provided that the series is summed
in a certain manner: for each $\Lambda$ we define a partial sum
$\psi_{\Lambda} (x_1, x_2)$ over all terms with $\Delta_k + | \alpha | <
\Lambda$, and then tend $\Lambda \rightarrow \infty$. This procedure is needed
because, although the states $\mathcal{D}^{\alpha} |
(\mathcal{O}^{\dagger}_k)^{(\lambda)} (x_S) \rangle$ are orthogonal for
different $| \alpha |$ (because they are eigenvectors of $\frac{K^0 - P^0}{2}$
with different eigenvalues), they are not orthonormal. This can be corrected
as follows. For each $k$, let us orthonormalize the infinite multiplet of
states $\mathcal{D}^{\alpha} | (\mathcal{O}^{\dagger}_k)^{(\lambda)} (x_S)
\rangle$. Let $e_{k, n}$, $n \in \mathbb{Z}_{\geqslant 0}$, be the
corresponding orthonormal basis of states (there is obviously a lot of
arbitrariness in this basis). We then can write
\begin{equation}
	\psi (x_1, x_2) = | \mathcal{O} (x_1) \mathcal{O} (x_2) \rangle =
	\sum_k f_{\mathcal{O}\mathcal{O}k} \underset{n}{\sum} \tilde{C}_{k, n} (x_1,
	x_2) e_{k, n}, \label{psialtexp}
\end{equation}
where $\tilde{C}_{k, n}$'s are some finite linear combinations of $C_{k,
	\lambda, \alpha} (x_1, x_2, x_S)$. Since $e_{k, n}$'s with different $k$ are
also orthogonal, this equation is an expansion of the state $\psi (x_1, x_2)$
in an orthonormal basis. Convergence of the series is now equivalent to the
finiteness of the norm of $\psi (x_1, x_2)$:
\begin{equation}
	\| \psi (x_1, x_2) \|_{\mathcal{H}^{\tmop{CFT}}} = \sum_{k, n} |
	f_{\mathcal{O}\mathcal{O}k} \tilde{C}_{k, n} (x_1, x_2) |^2 < \infty .
	\label{psinorm}
\end{equation}
Moreover, by definition, this norm is nothing but the 4-point function $\langle
\mathcal{O} (x_2^{\theta}) \mathcal{O} (x_1^{\theta}) \mathcal{O} (x_1)
\mathcal{O} (x_2) \rangle$.

Eqs.\ {\eqref{OPE:Euclidean}}-{\eqref{psinorm}} were all in the Euclidean
region, but we claim that they continue to make sense in the forward tube. The
argument is as follows. We know by the arguments around {\eqref{Morera}} that
$\psi (x_1, x_2)$ have analytic continuation to the region
{\eqref{f.t.states}}. The inner product $\langle e_{k, n} | \psi (x_1, x_2)
\rangle$ is then the analytic continuation of $f_{\mathcal{O}\mathcal{O}k}
\tilde{C}_{k, n} (x_1, x_2)$ from the Euclidean to the same region. (This
inner product is a finite linear combination of $x_N$-derivatives of the 3-point
function $\langle (\mathcal{O}_k)^{(\lambda)} (x_N) \mathcal{O} (x_1)
\mathcal{O} (x_2) \rangle$, hence holomorphic in the forward tube.) We thus
obtain the following fact:

\begin{theorem}
	Expansion {\eqref{psialtexp}}, analytically continued from the Euclidean
	region to the forward tube term by term, converges in the sense of
	$\mathcal{H}^{\tmop{CFT}}$ to the same states $\psi (x_1, x_2)$ in the
	region {\eqref{f.t.states}} that we defined in Sec.\ \ref{sec:Wpos}.
\end{theorem}

For the subsequent discussion, we also define the states
\begin{equation}
	\psi_k (x_1, x_2) = f_{\mathcal{O}\mathcal{O}k} \underset{n}{\sum}
	\tilde{C}_{k, n} (x_1, x_2) e_{k, n} .
\end{equation}
The norms of these states is given by conformal blocks (up to prefactor
$f_{\mathcal{O}\mathcal{O}k}^2 / (x_{12}^2)^{4 \Delta_{\mathcal{O}}}$). Just
as the state $\psi (x_1, x_2)$, each state $\psi_k (x_1, x_2)$ is an
$\mathcal{H}^{\tmop{CFT}}$-valued holomorphic function in the region
{\eqref{f.t.states}}, moreover in this region we have
\begin{equation}
	\psi (x_1, x_2) = \sum_k \psi_k (x_1, x_2), \label{psipsik}
\end{equation}
the series convergent in the sense of $\mathcal{H}^{\tmop{CFT}}$. The norm of
the tail of this series is given by the function $H (c, \Delta_{\ast})$ from
the proof of Theorem \ref{theorem:districonverge}:
\begin{equation}
	\Biggl\| \sum_{\Delta_k > \Delta_{\ast}} \psi_k (x_1, x_2)
	\Biggr\| = H (c, \Delta_{\ast}), \qquad c = (x_2^{\theta}, x_1^{\theta},
	x_1, x_2) . \label{tailbound1}
\end{equation}

Vladimirov's Theorem \ref{ThVlad} remains true for Hilbert-space-valued
holomorphic functions, whose norm satisfies a powerlaw bound in the forward tube.
Applying such a version of Theorem \ref{ThVlad}, as well as arguments from the
proof of Theorem \ref{theorem:districonverge} and from Sec.\ \ref{subsection:rateconvergence} (in particular using the bound
{\eqref{g1221:bound}}), it's easy to obtain the following result (we omit the
proof).

\begin{theorem}
	\label{th:distropeconvergence}(a) The boundary value $\tmop{bv} (\psi) =
	\lim_{\epsilon_i \rightarrow 0} \psi (x_1, x_2)$ exists as
	$\mathcal{H}^{\tmop{CFT}}$-valued tempered distributions, and similarly for
	each $\tmop{bv} (\psi_k)$: $\tmop{bv} (\psi), \tmop{bv} (\psi_k) \in
	\mathcal{S}' (\mathbb{R}^{2 d}) \otimes \mathcal{H}^{\tmop{CFT}}$.
	Explicitly, the limit
	\begin{equation}
		(\tmop{bv} (\psi), f) = \underset{\epsilon \rightarrow 0}{\lim} \int
		\psi (\epsilon + i t, \mathbf{x}) f (t, \mathbf{x})\, d^2 t\, d^{2 (d - 1)}
		\mathbf{x}
	\end{equation}
	exists as a vector in $\mathcal{H}^{\tmop{CFT}}$ for any Schwartz function
	$f \in \mathcal{S} (\mathbb{R}^{2 d})$, and is a continuous linear operator
	from $\mathcal{S} (\mathbb{R}^{2 d})$ to $\mathcal{H}^{\tmop{CFT}}$, and
	analogously for $\tmop{bv} (\psi_k)$;
	
	(b) (Distributional OPE convergence in $\mathcal{H}^{\tmop{CFT}}$) For each
	Schwartz test function $f \in \mathcal{S} (\mathbb{R}^{2 d})$, the series
	$\underset{k}{\sum} (\tmop{bv} (\psi_k), f)$ converges in
	$\mathcal{H}^{\tmop{CFT}}$ norm to $(\tmop{bv} (\psi), f)$;
	
	(c) (Convergence rate) For compactly supported test functions, the series in
	(b) summed over $\Delta_k \leqslant \Delta_{\ast}$ converges with rate $O
	\left( \left. \sqrt{\frac{\log \Delta_{\ast}}{\Delta_{\ast}}} \right)
	\right.$.
\end{theorem}

\begin{remark}
	The following more finegrained version of Theorem
	\ref{th:distropeconvergence}(b) also holds. Denote
	\begin{equation}
		E_{k, n} (x_{1,} x_2) = f_{\mathcal{O}\mathcal{O}k} \underset{}{}
		\tilde{C}_{k, n} (x_1, x_2) e_{k, n} .
	\end{equation}
	As explained above, $\tilde{C}_{k, n} (x_1, x_2)$ is a finite sum of terms
	like $(\mathcal{D}^{\theta})^{\alpha} \langle (\mathcal{O}_k)^{(\lambda)}
	(x_N) \mathcal{O} (x_1) \mathcal{O} (x_2) \rangle$ (all having the same $|
	\alpha |$), so $\tmop{bv} (E_{k, n})$ exists by Vladimirov's theorem. Then
	\tmtextit{for each Schwartz test function $f \in \mathcal{S} (\mathbb{R}^{2
			d})$, the series $\underset{k, n}{\sum} (\tmop{bv} (E_{k, n}), f)$ converges
		in $\mathcal{H}^{\tmop{CFT}}$ norm to $(\tmop{bv} (\psi), f)$.}
	
	It would be interesting to prove a version Theorem
	\ref{th:distropeconvergence}(c), truncating the series $\underset{k,
		n}{\sum} (\tmop{bv} (E_{k, n}), f)$ to $k, n$ such as the corresponding
	$\Delta_k + | \alpha | \leqslant \Delta_{\ast}$. To do so one would have to
	find an analogue of the bounds {\eqref{tailbound1}}, {\eqref{g1221:bound}}
	valid for such a truncation. This is not straightforward because the partial
	sums of the series {\eqref{psinorm}} truncated to $\Delta_k + | \alpha |
	\leqslant \Delta_{\ast}$ do not correspond to a simple truncation of the
	$\rho, \bar{\rho}$ series of the full 4-point function (basically because the
	transformation which maps $x_1, x_2, x_1^{\theta}, x_2^{\theta}$ to their
	$\rho, \bar{\rho}$ conformal frame does not necessarily map $x_S, x_N$ to
	$0, \infty$).
\end{remark}

\section{Comparison to Mack's work on OPE convergence}\label{MackComp}

To assume OPE convergence as an axiom in Euclidean CFT, and to derive
Minkowski physics from it, as we did in this part of the thesis, seems natural from the
modern perspective. On the contrary, in the early days of CFTs it was
considered natural to assume standard Minkowski physics (such as Wightman
axioms). The OPE convergence was not assumed at the time, but was something to
be derived.

This was the underlying philosophy of the works by L\"uscher and Mack
{\cite{Luscher:1974ez}} and of Mack {\cite{Mack:1975je,Mack:1976pa}}. Written
45 years ago, these papers remain widely cited, but not everyone is familiar
with what precisely has been derived there. Here we will present a short
review for the benefit of the modern audience.

These works make two main assumptions. \tmtextbf{First,} that we have a
unitary quantum field theory in the Minkowski signature which satisfies
Wightman axioms (in particular has a Hilbert space $\mathcal{H}$ on which the
Poincar\'e group acts unitarily and quantum fields are operator-valued
distributions). Correlators then have the usual analyticity properties of
Wightman functions, in particular they are real-analytic in the Euclidean. The
\tmtextbf{second} main assumption is that these Euclidean correlators are
invariant under the action of the Euclidean conformal group.

Using these two assumptions, L\"uscher and Mack {\cite{Luscher:1974ez}} proved
that the Hilbert space $\mathcal{H}$ carries a unitary representation not just
of the Poincar\'e but of the group $G^{\ast} =$universal cover of the Lorentzian
conformal group $\tmop{SO} (d, 2)$.\footnote{\label{LMcomplain}One also often
	quotes L\"uscher and Mack {\cite{Luscher:1974ez}} for proving that CFT
	correlation functions may be continued to the Minkowski cylinder $S^{d - 1}
	\times \mathbb{R}$. This is a misquotation as they did not prove this, but
	posed it as a conjecture. What they did prove was that CFT correlation
	functions can be analytically continued to a domain of which $S^{d - 1} \times
	\mathbb{R}$ is a real boundary. One still needs to establish a powerlaw bound
	to take the boundary limit and obtain a distribution. We plan to derive this
	fact in our future work {\cite{paper3}}, for 4-point functions, from the
	Euclidean CFT axioms, using the $\rho, \bar{\rho}$ expansion.} \ Mack
{\cite{Mack:1975je}} then classified all unitary positive energy
representations of $G^{\ast}$. It should be mentioned that Refs.\ {\cite{Luscher:1974ez,Mack:1975je,Mack:1976pa}} only consider $d = 4$
spacetime dimensions. Many explicit group theoretic calculations are done only
for this value of $d$. The results should however generalize to arbitrary $d$
with appropriate modifications.

Continuing this program, Mack {\cite{Mack:1976pa}} studied distributional OPE
convergence in Minkowski CFT. Since we also have results of this kind (Sec.\ \ref{OPEHCFT}), it will be particularly interesting to compare with Mack's
discussion. So let us review his argument. Compared to
{\cite{Luscher:1974ez,Mack:1975je}}, Ref.\ {\cite{Mack:1976pa}} includes one
extra assumption: that the OPE $\varphi_i (x_1) \varphi_j (x_2)$ of two fields
acting on the Minkowski vacuum is valid in an asymptotic sense. Namely that
for some dense set of states $\psi$ we have\footnote{Mack assumes $x_2 = -
	x_1$ but here for simplicity we will assume that this is valid for any $x_1,
	x_2$.}
\begin{equation}
	\langle \psi | \nobracket \varphi_i (x_1) \varphi_j (x_2) \rangle \sim
	\sum_k C_{i j k} (x_1, x_2) \langle \psi | \nobracket \varphi_k (0) \rangle,
	\label{asOPE}
\end{equation}
as an asymptotic expansion for rescaling $x_1, x_2 \rightarrow \lambda x_1,
\lambda x_2$, where $C_{i j k} (x_1, x_2)$ are some $\psi$-independent
homogeneous functions: $C_{i j k} (\lambda x_1, \lambda x_2) =
\lambda^{\Delta_k - \Delta_i - \Delta_j} C_{i j k} (x_1, x_2)$.\footnote{In fact  $C_{i j k}$ is a distribution so homogeneity should be understood in the sense of pairing with a rescaled test function.} Asymptotic
means that if we truncate the expansion at $\Delta_k = \Delta_{\ast}$ and take
$\lambda \rightarrow 0$ limit for any fixed $x_1, x_2$ then the error is $o
(\lambda^{\Delta_{\ast} - \Delta_i - \Delta_j})$. Note that there are both
primaries and descendants among $\varphi_k$'s. The main result of
{\cite{Mack:1976pa}} is to convert this asymptotic expansion to an expansion
convergent in the Hilbert space sense.

The first step is to use a general result that any Hilbert space carrying a
unitary representation of a semisimple Lie group can be decomposed as a direct
integral of unitary irreducible representations {\cite{Mackey1}}. Since, by
the above-mentioned result of {\cite{Luscher:1974ez}}, $\mathcal{H}$ carries a
unitary representation of $G^{\ast}$, we thus have
\begin{equation}
	\mathcal{H}= \int^{\oplus} d \mu_{\chi}\, d \tilde{\mu}_{\nu}\,
	\mathcal{H}^{\chi \nu}, \label{directint}
\end{equation}
where $\chi$ labels different unitary irreps of $G^{\ast}$, $\chi =
(\Delta_{\chi}, \rho_{\chi})$ with $\Delta_{\chi}$ the scaling dimension and
$\rho_{\chi}$ a Lorentz group irrep, and $\nu$ labels different copies of the
same irrep. Only positive energy irreps may occur, since all states of
Wightman theory have positive energy. By definition, Eq.\ {\eqref{directint}}
identifies every vector $\psi \in \mathcal{H}$ with a Hilbert-space-valued
function $(\chi, \nu) \mapsto \psi_{\chi \nu} \in \mathcal{H}^{\chi}$, some
standard realization of the irrep $\chi$. It is assumed that $\langle \psi |
\psi \rangle < \infty$, inner products being given by the following integral:
\begin{equation}
	\langle \psi | \psi' \rangle \nobracket = \int d \mu_{\chi}\, d
	\tilde{\mu}_{\nu}\, \langle \psi_{\chi \nu} | \psi'_{\chi \nu}
	\rangle_{\mathcal{H}^{\chi}}, \label{psipsiprime}
\end{equation}
Also $G^{\ast}$ acts on $\psi$ by acting on each $\psi_{\chi \nu}$.
Furthermore, Ref.\ {\cite{Mack:1975je}} realized $\mathcal{H}^{\chi}$ as a
space of distributions $\psi (x)$ on Minkowski space\footnote{More properly
	$\psi (x)$ is a distribution on the Lorentzian cylinder on which the group
	$G^{\ast}$ acts naturally, but due to a periodicity condition it may be
	reconstructed from its values on the Poincar\'e patch.} taking values in the
representation space of $\rho_{\chi}$, with Fourier transform supported in the
forward light cone, and for which the following inner product (defined
initially on a dense subset of smooth $\psi, \psi'$) is finite:
\begin{equation}
	\langle \psi | \psi' \rangle_{\mathcal{H}^{\chi}} = \int d x\, d y\,
	\overline{\psi (x)} I^{\chi} (x - y) \psi' (y),
	\label{Hchi}
\end{equation}
where $I^{\chi}$ is an ``intertwining kernel''. Physically, $I^{\chi}$
is the Minkowski CFT 2-point function of the primary in the ``shadow irrep'' of
$\chi$.

The above integration measure $d \mu_{\chi} d \tilde{\mu}_{\nu}$ depends on
the theory; from the abstract arguments alone it may be continuous or
discrete. Ref.\ {\cite{Mack:1976pa}} then proceeds to show that (a) this
measure is actually discrete (a sum of delta-functions), so that the direct
integral is a direct sum; (b) that the state $| \nobracket \mathcal{O}_i (x_1)
\mathcal{O}_j (x_2) \rangle$ produced by two Minkowski primary operators
acting on the Minkowski vacuum can be written as
\begin{equation}
	| \nobracket \mathcal{O}_i (x_1) \mathcal{O}_j (x_2) \rangle = \sum_{k, a}
	f^a_{i j k} \int d x\, B^a_{k, i j} (x, x_1, x_2) | \nobracket \mathcal{O}_k
	(x) \rangle, \label{OMack}
\end{equation}
where $B_{i j k}^a (x, x_1, x_2)$, $a = 1 \ldots N_{i j k},$ are some
kinematically determined distributions, the convergence is in the Hilbert
space sense after integrating with an arbitrary test functions $f (x_1, x_2)$,
and the local primary operators $\mathcal{O}_k$ occurring in this sum have
quantum numbers in the discrete set where the integration measure $d
\mu_{\chi} d \tilde{\mu}_{\nu}$ is supported.

To show how this comes about, let us focus on the case of scalar identical
$\mathcal{O}_i =\mathcal{O}_j =\mathcal{O}$ for simplicity. In this case
expansion {\eqref{OMack}} will end up being precisely our expansion
{\eqref{psipsik}} (although derived under very different assumptions), with
$B (x, x_1, x_2)$ related to the OPE kernel in {\eqref{OPE:Euclidean}}.

{Applying \eqref{psipsiprime} with 
	$|\psi'\rangle =| \nobracket \mathcal{O} (x_1) \mathcal{O}(x_2) \rangle$ gives} (Eq.\ (2.6) in {\cite{Mack:1976pa}}):
\begin{equation}
	\langle \psi | \mathcal{O} (x_1) \mathcal{O} (x_2) \rangle \nobracket =
	\int d \mu_{\chi}\, d \tilde{\mu}_{\nu}\, c_{\chi \nu} \int d x\,
	\overline{\psi_{\chi \nu} (x)} B_{\chi} (x, x_1, x_2), \label{psiO}
\end{equation}
where we denoted $\int d y\, I^{\chi} (x-y) \psi_{\chi \nu}' (y) =
c_{\chi \nu} B_{\chi} (x, x_1, x_2)$ where $c_{\chi \nu}$ is a proportionality
factor, and $B_{\chi} (x, x_1, x_2)$ is a kinematically determined
distribution (it is basically the amputated Minkowski 3-point function $\langle
\mathcal{O}_{\chi} (x) \mathcal{O} (x_1) \mathcal{O} (x_2) \rangle$). Mack then undertakes a meticulous study of $B_{\chi} (x, x_1, x_2)$ and of its
Fourier transform with respect to the first argument $\hat{B}_{\chi} (p, x_1,
x_2)$. This actually takes most of his paper, and involves many explicit
nontrivial calculations (e.g.\ it involves the first ever explicit
characterization of the most general 3-point function of CFT primaries in
arbitrary irreps). One of the main results is that $\hat{B}_{\chi} (p, x_1,
x_2)$ are entire functions of $p$:
\begin{equation}
	\hat{B}_{\chi} (p, x_1, x_2) = \sum_{| \alpha | \geqslant 0}
	b^{\alpha}_{\chi} (x_1, x_2) (- i p)_{\alpha}, \label{Bhatexp}
\end{equation}
where $b^{\alpha}_{\chi} (x_1, x_2) = (x^2_{12})^{- \Delta_{\mathcal{O}} +
	\Delta_{\chi} / 2}$ times a polynomial in $x_1, x_2$ of degree $| \alpha |$,
in particular $b^{\alpha}_{\chi} (\lambda x_1, \lambda x_2) =
\lambda^{\Delta_{\chi} + | \alpha | - 2 \Delta_{\mathcal{O}}}
b^{\alpha}_{\chi} (x_1, x_2)$.

Let us now specialize to states $\psi$ for which the function $\psi_{\chi \nu}
(x)$ has Fourier transform of compact support (one can show that such
states are dense in $\mathcal{H}$). Then the previous equations imply the
following convergent expansion for the integrand in {\eqref{psiO}}:
\begin{equation}
	\int d x\, \overline{\psi_{\chi \nu} (x)} B_{\chi} (x, x_1, x_2) =
	\sum_{\alpha} b^{\alpha}_{\chi} (x_1, x_2)  \overline{\partial^{\alpha}
		\psi_{\chi} (0)}, \label{Bhatint:exp}
\end{equation}
Mack then claims (before Eq.\ (2.11$'$)) that if, for each $\chi$, this
convergent expansion is truncated at $\Delta_{\chi} + | \alpha | =
\Delta_{\ast}$, and inserted back into {\eqref{psiO}}, this results in an
asymptotic expansion for the l.h.s.\ of {\eqref{psiO}}. I.e.\ for any
$\Delta_{\ast}$ (Mack does not write this equation explicitly):
\begin{equation}
	\langle \psi | \mathcal{O} (x_1) \mathcal{O} (x_2) \rangle \nobracket =
	\left\{ \int d \mu_{\chi}\, d \tilde{\mu}_{\nu}\, c_{\chi \nu}
	\sum_{\Delta_{\chi} + | \alpha | \leqslant \Delta_{\ast}} b^{\alpha}_{\chi}
	(x_1, x_2)  \overline{\partial^{\alpha} \psi_{\chi} (0)} \right\} + E
	(x_1, x_2 ; \Delta_{\ast}), \label{gap?}
\end{equation}
where the error term $E (\lambda x_1, \lambda x_2 ; \Delta_{\ast}) = O
(\lambda^{\Delta_{\ast} - 2 \Delta_{\mathcal{O}}})$ as $\lambda \rightarrow 0$
for any fixed $x_1$. Unfortunately, Mack does not give any justification of
this claim, which to us does not appear self-evident. The difficulty is that
although for every $\chi, \nu$ the truncated series has error $O
(\lambda^{\Delta_{\ast} - 2 \Delta_{\mathcal{O}}})$, the constant will
certainly depend on $\chi, \nu$. How do we know that the error estimate
survives after the integration in $\chi, \nu$? It might be possible to close
this omission in Mack's reasoning using normalizability of $| \psi \rangle
\nobracket$, but this needs extra arguments compared to what is given in his
paper, and we have not investigated this in detail.\footnote{We also tried, but unfortunately we did not manage, to get feedback from Prof. Gerhard Mack concerning this matter.}
Researchers relying on Mack's result should keep this caveat in mind.

Assuming that {\eqref{gap?}} is true, the argument is completed as follows. We
now have two asymptotic expansions for the l.h.s.\ of $\langle \psi |
\mathcal{O} (x_1) \mathcal{O} (x_2) \rangle \nobracket$, one coming from
{\eqref{gap?}}, and another from {\eqref{asOPE}}. The second one is discrete
(by assumption), so the first one also must be discrete. This establishes that
the measure $d \mu_{\chi} d \tilde{\mu}_{\nu}$ is discrete, a sum of delta
functions, hence we can write {\eqref{psiO}} with the r.h.s.\ as a sum, not an
integral:
\begin{equation}
	\langle \psi | \mathcal{O} (x_1) \mathcal{O} (x_2) \rangle \nobracket =
	\sum_n c_{\chi_n} \int d x\, \overline{\psi_{\chi_n} (x)} B_{\chi_n} (x,
	x_1, x_2) . \label{psiO1}
\end{equation}
A more detailed comparison of this equation with {\eqref{asOPE}} leads us to
conclude that (a)
\begin{equation}
	c_{\chi_n}  \overline{\psi_{\chi_n} (x)} = f_n \langle \psi |
	\mathcal{O}_{\chi_n} (x) \rangle, \label{psiO2}
\end{equation}
where $\mathcal{O}_{\chi_n}$ are primary operators related by rescaling to a
subset of the local operators $\varphi_k$, we choose them unit-normalized
(hence a coefficient $f_n$); (b) that all the other operators $\varphi_k$ are
the descendants $\mathcal{O}_{\chi_n}$'s; and (c) that all coefficients $C_k
(x_1, x_2)$ are basically the expansion coefficients $b^{\alpha}_{\chi_n}
(x_1, x_2)$ in {\eqref{Bhatexp}}. From {\eqref{psiO1}} and {\eqref{psiO2}}, we
have
\begin{equation}
	\langle \psi | \mathcal{O} (x_1) \mathcal{O} (x_2) \rangle \nobracket =
	\sum_n f_n \int d x\, \langle \psi | \mathcal{O}_{\chi_n} (x) \rangle
	B_{\chi_n} (x, x_1, x_2), \label{psiO3}
\end{equation}
for a dense set of states $\psi$. Because of the orthogonality of different $|
\nobracket \mathcal{O}_{\chi_n} (x) \rangle$'s, this implies that
\begin{equation}
	| \mathcal{O} (x_1) \mathcal{O} (x_2) \rangle \nobracket = \sum_n f_n
	\int d x\, B_{\chi_n} (x, x_1, x_2) | \nobracket \mathcal{O}_{\chi_n} (x)
	\rangle, \label{psiO4}
\end{equation}
the sum convergent in the Hilbert space sense after integrating out with any
test function $f (x_1, x_2)$. This is Eq.\ {\eqref{OMack}} in the considered
case $\mathcal{O}_i =\mathcal{O}_j =\mathcal{O}$.

\subsection{Relating Mack's kernel $B$ to the Euclidean OPE kernel $C$}\label{section:relateB}

Now we would like to relate Mack's OPE kernel $B$ to our OPE kernel $C_{a,
	(\lambda)}^{(\mu) (\nu)} (x_1, x_2, x_0, \partial_0)$ defined by Eqs.~(\ref{OPEgeneral}) and (\ref{Cexp}). We only consider the OPE kernel for the
scalar external operators for simplicity, i.e.\ $C_{\chi} (x_1, x_2, x_0,
\partial_0)$, $\chi = (\Delta, \ell = 0)$; similar remarks apply in the
general case. We first give the conclusion:
\begin{equation}
	C_{\chi} (x_1, x_2, x_0, \partial_0) = \underset{\mu}{\sum} b_{\chi}^{\mu}
	(x_1 - x_0, x_2 - x_0) \partial_0^{\mu}, \label{ope:relation}
\end{equation}
where the coefficient functions $b_{\chi}^{\mu}$ are the same as in
(\ref{Bhatexp}). One could ``derive'' this by using (\ref{OMack}) and formally
manipulating the integral in the momentum space:\footnote{Since here we are
	only interested in the OPE kernels $B_{\chi}$ and $C_{\chi}$, we set $f_{\chi}
	= 1$ (the overall coefficient) for convenience.}
\begin{eqnarray}
	| \mathcal{O} (x_1) \mathcal{O} (x_2) \rangle_{\chi} & = & \int d x\, B_{\chi}
	(x, x_1, x_2) | \nobracket \mathcal{O}_{\chi} (x) \rangle \nonumber\\
	& = & \int d p\, \hat{B}_{\chi} (p, x_1, x_2) | \nobracket
	\hat{\mathcal{O}}_{\chi} (p) \rangle = \underset{\mu}{\sum} b_{\chi}^{\mu}
	(x_1, x_2) | \partial^{\mu} \mathcal{O}_{\chi} (0) \rangle, 
\end{eqnarray}
which shows (\ref{ope:relation}) in the case when $x_0 = 0$. The general $x_0$
case follows by translation invariance. This derivation is not rigorous for
various reasons: (a) we did not clarify the meaning of
$\hat{\mathcal{O}}_{\chi} (p)$; (b) why can we swap the order of summation and
integration? (c) the above derivation is done in Minkowski region, how do we
match the coefficients $b_{\chi}^{\mu} (x_1, x_2)$ with the Euclidean
coefficients in (\ref{Cexp})?

Below we will give a rigorous justification of (\ref{ope:relation}), using
only the two- and 3-point functions which are kinematically determined by
conformal invariance. Recall that on the Euclidean side, the formal power
series of $C_{\chi}$ (the scalar version of (\ref{Cexp}))
\begin{equation}
	C_{\chi} (x_1, x_2, x_0, \partial_0) =
	\frac{1}{(x_{12}^2)^{\Delta_{\mathcal{O}} - \Delta_{\chi} / 2}}
	\underset{\mu}{\sum} c_{\chi}^{\mu} (x_{10}, x_{20}) \partial_0^{\mu}
	\label{Cexp:scalar}
\end{equation}
is determined by the Euclidean two- and 3-point functions:
\[ \langle \mathcal{O}_{\chi}^{\dag} (y) \mathcal{O} (x_1) \mathcal{O} (x_2)
\rangle_E = \frac{1}{(x_{12}^2)^{\Delta_{\mathcal{O}} - \Delta_{\chi} / 2}}
\underset{\alpha}{\sum} c_{\chi, \alpha} (x_{10}, x_{20}) \langle
\mathcal{O}_{\chi}^{\dag} (y) \partial^{\alpha} \mathcal{O}_{\chi} (x_0)
\rangle_E . \label{c:Euclidean} \]
Here we already used translation invariance, which implies $C_{a, (\lambda),
	\alpha}^{(\mu) (\nu)} (x_1, x_2, x_0) = c_{a, (\lambda), \alpha}^{(\mu) (\nu)}
(x_{10}, x_{20})$ on the r.h.s.~of (\ref{Cexp}). One can match the coefficients
$c_{\chi}^{\mu} (x_{10}, x_{20})$ with the Taylor expansion of $\langle
\mathcal{O}_{\chi}^{\dag} (y) \mathcal{O} (x_1) \mathcal{O} (x_2) \rangle_E$
around $x_1 = x_2 = x_0$. In Euclidean one can always find a proper region for
the matching: let $y$ be sufficiently far from the $(x_0, x_1, x_2)$ cluster,
so that the Taylor expansions of $[(y - x_1)^2]^{\#}$ around $x_1 = x_0$ and
$[(y - x_2)^2]^{\#}$ around $x_2 = x_0$ are convergent.

On the Minkowski side, the OPE kernel $B_{\chi}$ is kinematically determined
by the equality
\begin{equation}
	\hat{G}_{\mathcal{O}_{\chi} \mathcal{O}\mathcal{O}} (p, x_1, x_2) =
	\hat{B}_{\chi} (p, x_1, x_2) \hat{G}_{\chi} (p), \label{Mackope:3-pointFT}
\end{equation}
where $\hat{G}_{\mathcal{O}_{\chi} \mathcal{O}\mathcal{O}} (p, x_1, x_2) =
\int d y\, \langle \mathcal{O}_{\chi}^{\dag} (y) \nobracket \mathcal{O} (x_1)
\mathcal{O} (x_2) \rangle_M e^{- i p \cdummy y}$ and $\hat{G}_{\chi} (p) =
\int d y\, \langle \mathcal{O}_{\chi}^{\dag} (y) \nobracket
\mathcal{O}_{\chi} (0) \rangle_M e^{- i p \cdummy y}$ (see
{\cite{Mack:1976pa}}, Eq.\ (8.2)).\footnote{In the unitary CFTs,
	$\hat{G}_{\bar{n} \mathcal{O}\mathcal{O}} (p, x_1, x_2)$ and $\hat{G}_{\chi_n}
	(p)$ vanish unless $p \in \overline{V_+}$, so the behavior of
	$\hat{B}_{\chi_n} (p, x_1, x_2)$ outside the forward light cone is not
	important.} All Fourier transforms here are in the sense of distributions. To
get an equation valid in the sense of functions we pick a test function
$\varphi$ with compactly supported Fourier transform, and integrate
{\eqref{Mackope:3-pointFT}} against $\hat{\varphi}$, which gives:
\begin{equation}
	\int \langle \mathcal{O}_{\chi}^{\dag} (x) \nobracket \mathcal{O} (x_1)
	\mathcal{O} (x_2) \rangle_M \varphi (x)\, d x = \int d p\, \hat{\varphi} (p)
	\hat{B}_{\chi} (p, x_1, x_2) \hat{G}_{\chi} (p) . \label{3-point:exp10}
\end{equation}
The variable $x$ ranges over the Minkowski space, while we will pick $x_1,
x_2$ complex, in the forward tube region
\begin{equation}
	\tmop{Im} (x_1), \tmop{Im} (x_2) \prec 0. \label{region:matching}
\end{equation}
Then the 3-point function $\langle \mathcal{O}_{\chi}^{\dag} (x) \nobracket
\mathcal{O} (x_1) \mathcal{O} (x_2) \rangle_M$ is nonsingular as a function of
$x$ and the l.h.s.\ of {\eqref{3-point:exp10}} is a finite number. To transform the
r.h.s.\ of {\eqref{3-point:exp10}} we will use the fact that $\hat{B}_{\chi}$ has
the following form (a more detailed version than (\ref{Bhatexp})):
\begin{equation}
	\hat{B}_{\chi} (p, x_1, x_2) = \frac{e^{- i p \cdummy
			x_1}}{(x_{12}^2)^{\Delta_{\mathcal{O}} - \Delta_{\chi} / 2}} E_{\chi}
	(x_{12} \cdummy p, x_{12}^2 p^2), \label{Bhat:entire}
\end{equation}
where $E_{\chi} (z_1, z_2)$ is some entire function on $\mathbb{C}^2$. Hence
as long as $x_{12}^2 \neq 0$ (not necessarily real), $\hat{B}_{\chi} (p, x_1,
x_2)$ has the following convergent expansion:
\begin{equation}
	\hat{B}_{\chi} (p, x_1, x_2) = \frac{e^{- i p \cdummy
			x_1}}{(x_{12}^2)^{\Delta_{\mathcal{O}} - \Delta_{\chi} / 2}}
	\underset{\alpha}{\sum} a_{\chi}^{\alpha} (x_{21}) (- i p)_{\alpha},
	\label{Bhatexp2}
\end{equation}

where $a_{\chi}^{\alpha} (x)$ is some $\tmop{SO} (1, d - 1)$-covariant,
homogeneous, symmetric polynomial of degree $| \alpha |$. Plugging this into
{\eqref{3-point:exp10}}, using that the expansion {\eqref{Bhatexp2}} converges
uniformly on the support of $\hat{\varphi}$ (assumed compact), and the fact
that $\hat{G}_{\chi}$ is a tempered measure,\footnote{This is a consequence of
	the Bochner-Schwartz theorem: any positive tempered distribution is the
	Fourier transform of some positive tempered measure (see {\cite{Vladimirov2}},
	Sec.\ 8.2).} we obtain:
\begin{equation}
	\int \langle \mathcal{O}_{\chi}^{\dag} (x) \nobracket \mathcal{O} (x_1)
	\mathcal{O} (x_2) \rangle_M \varphi (x)\, d x =
	\frac{1}{(x_{12}^2)^{\Delta_{\mathcal{O}} - \Delta_{\chi} / 2}}
	\underset{\alpha}{\sum} a_{\chi}^{\alpha} (x_{21}) \int \langle
	\mathcal{O}_{\chi}^{\dag} (x) \partial_{\alpha} \mathcal{O}_{\chi} (x_1)
	\rangle_M \varphi (x)\, d x. \label{3-point:exp1}
\end{equation}
At this stage we have established that for any $x_1, x_2$ as in
{\eqref{region:matching}}, and for any test $\varphi$ with compact
$\hat{\varphi}$, the series in the r.h.s.\ converges to the l.h.s.

Now the key point is that the r.h.s.\ of {\eqref{3-point:exp1}} is a convergent
power series in $x_{21}$, while the l.h.s.\ can be expanded in such a
convergent power series. Indeed we know the explicit form of $\langle
\mathcal{O}_{\chi}^{\dag} (x) \nobracket \mathcal{O} (x_1) \mathcal{O} (x_2)
\rangle_M$:
\begin{equation}
	\langle \mathcal{O}_{\chi}^{\dag} (x) \nobracket \mathcal{O} (x_1)
	\mathcal{O} (x_2) \rangle_M = \frac{1}{(x_{12}^2)^{\Delta_{\mathcal{O}} -
			\Delta_{\chi} / 2}}  \frac{1}{[(x - x_1)^2 (x - x_2)^2]^{\Delta_{\chi} / 2}}
	. \label{3-point:explicit}
\end{equation}
For all $x$ in the Minkowski space, the function $[(x - x_1)^2 (x - x_2)^2]^{-
	\Delta_{\chi} / 2}$ is holomorphic in $x_1, x_2$ as long as $\tmop{Im} (x_1),
\tmop{Im} (x_2) \prec 0$. It's easy to show that this remains true after
integration in $\varphi$. At this point we can match the expansions for the
two sides of {\eqref{3-point:exp1}}, and get
\begin{equation}
	\sum_{| \alpha | = n} \int d x\, \varphi (x) \left\{ \frac{
		(x_{21})_{\alpha}}{\alpha !} \partial_{x_2}^{\alpha} [(x - x_1)^2 (x -
	x_2)^2]^{- \Delta_{\chi} / 2}  | \nobracket_{x_2 = x_1} -
	a_{\chi}^{\alpha} (x_{21}) \langle \mathcal{O}_{\chi}^{\dag} (x)
	\partial_{\alpha} \mathcal{O}_{\chi} (x_1) \rangle_M \right\} = 0 .
	\label{phiint}
\end{equation}
Up to this point it was crucial to keep the function $\varphi$ in the game to
keep convergence issues under control, but now we can get rid of it. Indeed
$\underset{| \alpha | = n}{\sum}$ is a finite sum, also $\langle
\mathcal{O}_{\chi}^{\dag} (x) \partial_{\alpha} \mathcal{O}_{\chi} (x_1)
\rangle_M$ is a holomorphic function in the forward tube $\mathcal{T}_2$. For
any Minkowski point $x_0$, we choose a sequence of test functions $\varphi_k$
of compact support $\hat{\varphi}_k$ such that $\varphi_k$ tends to $\delta (x
- x_0)$. Passing to the limit, {\eqref{phiint}} implies the same equality for
the integrand. I.e.\ for any fixed $n \in \mathbb{N}$, and any Minkowski $x$,
\begin{equation}
	\frac{ (x_{21})_{\alpha}}{\alpha !} \partial_{x_2}^{\alpha} [(x - x_1)^2
	(x - x_2)^2]^{- \Delta_{\chi} / 2}  | \nobracket_{x_2 = x_1} -
	a_{\chi}^{\alpha} (x_{21}) \langle \mathcal{O}_{\chi}^{\dag} (x)
	\partial_{\alpha} \mathcal{O}_{\chi} (x_1) \rangle_M = 0. \label{ope:match}
\end{equation}
Now as promised we are reduced to an equation which only involves 2-point
and 3-point functions which are holomorphic. E.g.\ we can take $x = 0$ and
$x_1, x_2$ in Euclidean. Then this equation is the same one as the equation
which determines the Euclidean OPE kernel for $x_0 = x_1$, i.e.\ $C_{\chi}
(x_1, x_2, x_1, \partial)$. For convenience in this discussion we use
Minkowski coordinates for Euclidean correlators (i.e.\ we write the Euclidean
correlators as $\langle \nobracket \langle \mathcal{O} (- i \tau, \mathbf{x})
\ldots \rangle_M$). Under this convention we have
\begin{equation}
	C_{\chi} (x_1, x_2, x_1, \partial) =
	\frac{1}{(x_{12}^2)^{\Delta_{\mathcal{O}} - \Delta_{\chi} / 2}}
	\underset{(\mu)}{\sum} a_{\chi}^{\mu} (x_{21}) \partial_{\mu} =
	\underset{(\mu)}{\sum} b_{\chi}^{\alpha} (0, x_{21}) \partial_{\alpha} .
	\label{ope:kernelmatch1}
\end{equation}

This establishes (\ref{ope:relation}) for $x_0 = x_1$. The general case
reduces to this one by noticing that $c_{\chi}^{\alpha}$ satisfies the
relation:
\begin{equation}
	c_{\chi}^{\alpha} (x_{10}, x_{20}) = \underset{}{\underset{\beta
			\leqslant \alpha}{\sum}} \frac{1}{\beta !} c_{\chi}^{\alpha - \beta} (0,
	x_{21}) x_{10}^{\beta},
\end{equation}
where $\beta \leqslant \alpha$ means $\beta_i \leqslant \alpha_i$ for all $i$;
and $b_{\chi}$ satisfies the identical relation with $c_{\chi} \rightarrow
b_{\chi}$. For $c_{\chi}^{\alpha}$ this follows by translation invariance and
analyticity of CFT two- and 3-point functions, and for $b_{\chi}^{\alpha}$
from $\hat{B}_{\chi} (p, x_{10}, x_{20}) = e^{- i p \cdummy x_{10}}
\hat{B}_{\chi} (p, 0, x_{21})$.

\chapter{Review of Osterwalder-Schrader theorem}\label{OS}

In this section we review the results of
{\cite{osterwalder1973,osterwalder1975}} and, in particular, discuss the
linear growth condition and why it was necessary for establishing Wightman
axioms in {\cite{osterwalder1975}}.

In {\cite{osterwalder1973}} Osterwalder and Schrader formulated an equivalence
theorem which stated that a set of axioms for Euclidean correlation functions
(a version of the Osterwalder-Schrader axioms described in Sec.\ \ref{OSaxioms}) is
equivalent to Wightman axioms for Euclidean correlation functions.
Unfortunately, later a technical error was discovered in their proof, and in
{\cite{osterwalder1975}} Osterwalder and Schrader gave two new results.

The first result of {\cite{osterwalder1975}} is a revised equivalence
theorem, which shows that a stronger version of Euclidean axioms is in fact
equivalent to Wightman axioms. The proof of this theorem is rather simple.
However, as we will review, this is at the expense of the new version of
Euclidean axioms being rather hard to verify.

The second result of {\cite{osterwalder1975}} shows that the original OS
axioms, plus a ``linear growth condition,'' imply Wightman axioms and a growth
condition on Wightman distributions. A partial result in the reverse direction
is also valid. It assumes a stronger growth condition on the Wightman
distributions than follows from the direct result, and it yields a growth
condition on Euclidean correlators which is weaker than the linear growth
condition. Therefore, these latter results do not establish an equivalence of
any two systems of axioms, but they do allow to establish Wightman axioms from
OS axioms in some situations.

In what follows we will review the general structure of the arguments of
{\cite{osterwalder1973,osterwalder1975}}. For our purposes it will suffice to
ignore the space coordinates and focus only on the time arguments of the
fields. We will not completely reproduce all arguments of
{\cite{osterwalder1973,osterwalder1975}}, and in some of the omitted steps the
space arguments and Lorentz symmetry are important. We will also work with
correlation functions involving a single hermitian scalar field $\phi$,
similarly to {\cite{osterwalder1973,osterwalder1975}}. In CFT applications we
are interested in correlation functions of all local operators. It should be
relatively straightforward to adapt the discussion of
{\cite{osterwalder1973,osterwalder1975}} to this more general setup.

Our main goal is to construct an analytic continuation of the Euclidean
correlation functions
\begin{equation}
	G^E_n (t_1, \ldots, t_n) \equiv \langle \phi (t_1) \ldots \phi (t_n) \rangle
\end{equation}
from real to complex $t_k$, and to establish that the Wightman functions
recovered in the limit of pure imaginary $t_k$ (real Lorentzian times) are
tempered distributions. This is the most non-trivial part of the argument.
Other Wightman axioms such as positivity, spectrum condition, etc., follow
relatively easily and have been reviewed in chapter \ref{sec:4-point}.

\section{The argument of {\cite{osterwalder1973}}}

Physically, the analyticity of position-space correlation functions is due to
positivity of energy. More concretely, the Euclidean evolution operator $e^{-
	H t}$ is well-defined and holomorphic in $t$ for $\tmop{Re} t > 0$ due to the
spectrum of $H$ being non-negative. The first step to establishing analyticity
is thus to construct the operator $H$, and for this we first need to construct
a Hilbert space on which it acts.

The Hilbert space $\mathcal{H}^{\tmop{OS}}$ is constructed, as we discussed in
Sec.\ \ref{OSaxioms}, by considering the vector space
$\mathcal{H}^{\tmop{OS}}_0$ of formal linear combinations of
states\footnote{In Sec.\ \ref{OSaxioms} the states are introduced as
	integrals of these quantities. This is also what is done in
	{\cite{osterwalder1973,osterwalder1975}}, since they assume only that the
	Euclidean correlators are distributions. Here, for simplicity of discussion,
	we use the knowledge that correlators are functions and use states evaluated
	at points. The arguments easily generalize to distributions and smeared
	states, but become more technical.}
\begin{equation}
	| \phi (t_1) \phi (t_2) \ldots \phi (t_n) \rangle \label{basisstate}
\end{equation}
with $0 > t_1 > t_2 > \cdots > t_n .$ A Hermitian inner product is introduced
on $\mathcal{H}^{\tmop{OS}}_0$ by
\begin{equation}
	\langle \phi (s_1) \phi (s_2) \ldots \phi (s_m) \nobracket | \phi (t_1) \phi
	(t_2) \ldots \phi (t_n) \rangle \equiv G_n^E (- s_m, \ldots, - s_1, t_1,
	\ldots, t_n) .
\end{equation}
By OS reflection positivity-axiom this inner product is positive-semidefinite.
The Hilbert space $\mathcal{H}^{\tmop{OS}}$ is obtained from
$\mathcal{H}^{\tmop{OS}}_0$ by modding out null states and completing the
resulting quotient space with respect to the above inner product. We can
naturally think of $| \phi (t_1) \phi (t_2) \ldots \phi (t_n) \rangle$ as
states in $\mathcal{H}^{\tmop{OS}}$.

Physically, to construct the Hamiltonian $H$, we first define it by its action
on {\eqref{basisstate}}. Then we note that $H$ has to be positive, otherwise
the correlation functions would grow exponentially at large distances.
Formally, one first defines for $t > 0$ an operator $U_t$ on
$\mathcal{H}^{\tmop{OS}}_0$ by
\begin{equation}
	U_t | \phi (t_1) \phi (t_2) \ldots \phi (t_n) \rangle \equiv
	{| \phi (t_1 - t) \phi (t_2 - t) \ldots \phi (t_n - t)
		\rangle} .
\end{equation}
The usual care must be taken to ensure that this defines an operator on
$\mathcal{H}^{\tmop{OS}}$. For this one notes that for any $\Psi \in
\mathcal{H}^{\tmop{OS}}_0$ we have $| \nobracket \langle \Psi | U_t | \Psi
\rangle | \nobracket \leqslant P (t)$ for some polynomial $P (t)$ since the
Euclidean correlation functions are assumed to be powerlaw-bounded when groups
of points are separated to infinity. Then a simple estimate gives
\begin{equation}
	| \nobracket \langle \Psi | U_t | \Psi \rangle | \nobracket \leqslant \|
	\Psi \| \| U_t \Psi \| = \| \Psi \| | \nobracket \langle \Psi | U_{2 t} |
	\Psi \rangle | \nobracket^{1 / 2} \leqslant \cdots \leqslant \| \Psi \|^{1 +
		1 / 2 + \cdots + 1 / 2^{n - 1}} | \nobracket \langle \Psi | U_{2^n t} | \Psi
	\rangle | \nobracket^{1 / 2^n} .
\end{equation}
Using $| \langle \Psi | U_t | \Psi \rangle | \leqslant P (t)$ we get in the
limit $n \rightarrow \infty$
\begin{equation}
	| \nobracket \langle \Psi | U_t | \Psi \rangle | \nobracket \leqslant \|
	\Psi \|^{1 + 1 / 2 + \cdots + 1 / 2^{n - 1}} (P (2^n t))^{1 / 2^n}
	\rightarrow \| \Psi \|^2 . \label{contractive}
\end{equation}
This shows that $U_t$ maps null states to null states and thus is defined on
(a dense subset of) $\mathcal{H}^{\tmop{OS}}$. By the above, it is also a
bounded operator, so it extends in a unique way to all of
$\mathcal{H}^{\tmop{OS}}$. Furthermore, noting that it is symmetric, of norm
at most 1, and we have the semigroup law $U_t U_s = U_{t + s}$, we find that
$U_t = e^{- H t}$ for a non-negative self-adjoint Hamiltonian $H$ (see, e.g.,
{\cite{funcan}} Sec.\ 141).

Since the domain in which we need to construct the analytic continuation of
$G_n^E$ is awkward to define in $t_k$ variables, we introduce the difference
variables $y_k \equiv t_k - t_{k + 1} .$ Due to translation invariance, $G_n^E
(t_1 \ldots t_n)$ can be rewritten as
\begin{equation}
	G_n^E (t_1 \ldots t_n) = S_{n - 1} (y_1 \ldots y_{n - 1})
\end{equation}
for some functions $S_n$. Similarly, we will use the following notation for
states in terms of $y_k$ variables,
\begin{equation}
	| \Psi_n (- t_1 ; y_1 \ldots y_{n - 1}) \rangle \equiv | \phi (t_1) \phi
	(t_2) \ldots \phi (t_n) \rangle .
\end{equation}
Note that
\begin{equation}
	\langle \nobracket \Psi_m (t' ; y'_1 \ldots y'_{m - 1}) | \Psi_n (t ;
	y_1 \ldots y_{n - 1}) \rangle = S_{m + n - 1} (y_{m - 1}', \ldots, y_1', t'
	+ t, y_1, \ldots, y_{n - 1}) .
\end{equation}
In terms of $S_{n - 1} (y_1 \ldots y_{n - 1})$, our goal is to construct an
analytic continuation to $\tmop{Re} y_k > 0$ and show that the limit of all
$\tmop{Re} y_k \rightarrow 0^+$ exists in the sense of tempered distributions.

With a positive $H$ now constructed, we can define a holomorphic family of
bounded operators $U_{\tau} = e^{- H \tau}$ for $\tmop{Re} \tau > 0$, which
will be our main tool for analytically continuing the correlation functions
$S_n$. In particular, we can now consider the matrix elements
\begin{equation}
	\langle \nobracket \Psi_m (t' ; y'_1 \ldots y'_{m - 1}) | U_{\tau} |
	\nobracket \Psi_n (t ; y_1 \ldots y_{n - 1}) \rangle = S_{m + n - 1}
	(y_{m - 1}', \ldots, y_1', t' + t + \tau, y_1, \ldots, y_{n - 1}),
\end{equation}
which are analytic for $\tmop{Re} \tau>0$. This establishes the desired
analyticity of $S_{n - 1} (y_1 \ldots y_{n - 1})$ in each variable $y_k$
separately. In {\cite{osterwalder1973}} they additionally establish some
growth conditions on these individual holomorphic functions which then imply that
for fixed $y_k, y'_k$ and $\tmop{Re} \tau > 0$ the above function can be
represented as the Fourier-Laplace transform
\begin{equation}
	S_{m + n - 1} (y_{m - 1}', \ldots, y_1', \tau, y_1, \ldots, y_{n - 1}) =
	\int d \alpha\, e^{- \alpha \tau}  \check{S} (\alpha)
\end{equation}
for some tempered distribution $\check{S} (\alpha)$. In other words, $S_{m + n
	- 1}$ can be extended to a holomorphic function in the right-half plane in each
variable separately, and each such holomorphic function can be represented as a
Fourier-Laplace transform of a tempered distribution. The erroneous Lemma 8.8
of {\cite{osterwalder1973}} states that under these conditions, the full
function $S_{m + n - 1}$ is a simultaneous Fourier-Laplace transform in all
its variables of a tempered distribution,
\begin{equation}
	S_{m + n - 1} (\tau_1 \ldots \tau_{m + n - 1}) = \int d \alpha\, e^{- \alpha_1
		\tau_1 - \cdots - \alpha_{m + n - 1} \tau_{m + n - 1}}  \check{S}_{m + n
		- 1} (\alpha_1 \ldots \alpha_{m + n - 1}) . \label{FLtransform}
\end{equation}
From this, the tempered Wightman distributions are obtained immediately by
setting $\tmop{Re} \tau_k \rightarrow 0^+$ in which case the Fourier-Laplace
transform above becomes a Fourier transform of a tempered distribution.
Fourier transform of a tempered distribution is, of course, itself tempered.

\section{The argument of {\cite{osterwalder1975}}}

\subsection{Fixing the equivalence theorem}

Unfortunately, Lemma 8.8 of {\cite{osterwalder1973}} is wrong. As explained in
{\cite{osterwalder1975}}, the function $S_2 (y_1, y_2) = e^{- y_1 y_2}$ gives
a simple counter-example. For fixed $y_2 > 0,$we find that $S_2 (y_1, y_2)$ is
holomorphic for $\tmop{Re} y_1 > 0$ and is there the Fourier-Laplace transform of
the tempered distribution $\delta (\alpha - y_2)$. The same statements hold
with $y_1$ and $y_2$ exchanged. However, $S_2 (y_{1,} y_2)$ is not a
Fourier-Laplace transform of a tempered distribution in both variables
simultaneously. For if this were the case, the corresponding Wightman function
$S_2 (i x_1, i x_2) = e^{x_1 x_2}$ would be a tempered distribution, which it
is not since it grows faster than any power in some directions.

The first result of {\cite{osterwalder1975}} (see also the review in \cite{simon1974}) 
rescues Lemma 8.8 by making a
stronger assumption about $S_n (y_1 \ldots y_n)$ which they denote by
$\check{E 0}$. Concretely, let $\mathbb{R}^n_+$ be the set of points $(y_1,
\ldots, y_n)$ with $y_k > 0.$ Let $\mathcal{S} (\mathbb{R}_+^n)$ be the
subspace of the space of Schwartz functions, consisting of functions supported
on $\mathbb{R}^n_+$ with the induced topology. The functions $S_n (y_1 \ldots
.y_n)$ can be viewed as distributions in the continuous dual space
$\mathcal{S}' (\mathbb{R}_+^n)$ defined by, for $f \in \mathcal{S}
(\mathbb{R}_+^n)$
\begin{equation}
	S_n (f) \equiv \int d y_1 \ldots d y_n\, S_n (y_1 \ldots y_n) f (y_1 \ldots
	y_n) .
\end{equation}
Note that smoothness of $f$ together with its support properties ensures that
$f (y_1 \ldots y_n)$ vanishes with all derivatives whenever $y_k = y_j$ for $k
\neq j$. The assumption that $S_n$ has at most powerlaw singularities at
coincident points and at infinity means that $S_n (f)$ is continuous in $f$ in
the topology of $\mathcal{S} (\mathbb{R}_+^n)$. The additional assumption
$\check{E 0}$ is that it is also continuous in $f$ in a weaker topology. This
weaker topology is defined by the usual Schwartz norms on
$\overline{\mathbb{R}_+^n}$
\begin{equation}
	| g |_{p, +} = \sup_{x \in \overline{\mathbb{R}_+^n}, | \alpha | \leqslant
		p} | (1 + x^2)^{p / 2} \partial^{(\alpha)} g (x) |
\end{equation}
but applied not to $f$ and instead to its Fourier-Laplace\footnote{As written,
	this is a Laplace transform. It is a Fourier transform in the spatial
	variables which we are omitting.} transform $\check{f}$
\begin{equation}
	\check{f} (q_1 \ldots q_n) \equiv \int d y_1 \ldots d y_n\, e^{- q_1 y_1 -
		\cdots - q_n y_n} f (y_1 \ldots y_n) .
\end{equation}
One establishes that $\check{f} = 0$ iff $f = 0$ (injectivity) and that the
set of all images $\check{f}$ is dense in an appropriate space of Schwartz
functions (denseness). The proof of {\eqref{FLtransform}} then becomes
straightforward: one first defines $\check{S}_n$ by $\check{S}_n (\check{f}) =
S_n (f)$. This definition makes sense due to the injectivity property. The
assumption $\check{E 0}$ ensures that $\check{S}_n$ is continuous. The
denseness property just mentioned then allows to extend $\check{S}_n$ to an
appropriate space of Schwartz functions by continuity, establishing
temperedness of $\check{S}_n$ and allowing one to define tempered Wightman
distributions as Fourier transforms of $\check{S}_n$. It is similarly not
difficult to show that Wightman axioms imply $\check{E 0} .$

\subsection{Wightman axioms from linear growth condition}

As we can see, the axiom $\check{E 0}$ is not very different from directly
assuming temperedness of Wightman distributions, even though it is formulated
for Euclidean correlators. It is also unclear how to verify this axiom in
practice.\footnote{We would also like to mention related work by Zinoviev
	{\cite{Zinoviev1995}}. Zinoviev replaces axiom $\check{E 0}$ by an axiom E5
	which imposes that certain limits exist which allow to compute the inverse
	Laplace transform of $S_n$. While E5 may look like a more constructive version
	of $\check{E 0}$, in practice its verification appears just as hard as
	assuming outright that $S_n$ is a Laplace transform (which is what $\check{E
		0}$ essentially does). We are grateful to David Brydges for an enlightening
	explanation of Zinoviev's construction, and in particular for pointing out
	that it represents a generalization of Post's Laplace transform inversion
	formula {\cite{Post}} to the case of distributions.} For this reason,
{\cite{osterwalder1975}} introduced an alternative ``linear growth condition''
on the correlation functions $S_n$ which is easier to verify and has been
established in some models (see below), yet is also sufficient to establish
temperedness of Wightman functions (though this condition is not known to follow
from Wightman axioms). The construction of the analytic continuation of the
functions $S_n$ as well the proof of the temperedness of the resulting
Wightman distributions is much more complicated than using $\check{E 0}$.
Therefore, our review of these arguments will be even more schematic than the
above, and we will only try to illustrate the key ideas and explain why and
how the linear growth condition is used. We are not aware of any previous attempt to review this part of \cite{osterwalder1975}.

First of all, let us state the linear growth condition of
{\cite{osterwalder1975}}. Note that the correlation functions $G_n^E$ can be
viewed as distributions in $({}^0 \mathcal{S})' (\mathbb{R}^{d \cdummy n})$,
where ${}^0 \mathcal{S} (\mathbb{R}^{d \cdummy n})$ is the space of Schwartz
functions of $n$ arguments in $\mathbb{R}^d$ which vanish with all derivatives
at coincident points, by
\begin{equation}
	G_n^E (f) = \int d^d x_1 \ldots d^d x_n\, f (x_1, \ldots, x_n) G_n^E (x_{1,}
	\ldots, x_n) .
\end{equation}
Here we have temporarily restored the spatial coordinates. In fact,
{\cite{osterwalder1973,osterwalder1975}} do not assume that $G_n^E$ are
functions, and only that they are distributions in $({}^0 \mathcal{S})'
(\mathbb{R}^{d \cdummy n})$. It follows, however, from the OS axioms (without
the linear growth condition) that $G_n^E$ are real-analytic functions, as
shown in {\cite{Glaser1974,osterwalder1975}}.

Note that the assumption that $G_n^E \in ({}^0 \mathcal{S})' (\mathbb{R}^{d
	\cdummy n})$ means $G_n^E$ is sufficiently continuous as a linear functional or, equivalently, is sufficiently bounded. That is,
\begin{equation}
	| G_n^E (f) | \leqslant \sigma_n | f |_{q_n} \label{J0distr}
\end{equation}
for all $f \in {}^0 \mathcal{S} (\mathbb{R}^{d \cdummy n})$ and some $\sigma_n
> 0$ and $q_n \in \mathbb{Z}_{\geqslant 0}$, where $| f |_p$ denotes the
Schwartz norms on ${}^0 \mathcal{S} (\mathbb{R}^{d \cdummy n})$. The linear
growth condition requires $q_n$ to grow at most linearly, and $\sigma_n$ at
most as a power of a factorial. In other words, the linear growth condition is
the statement that there exists a positive integer $s$ and a sequence
$\sigma_n$ such that
\begin{equation}
	| G_n^E (f) | \leqslant \sigma_n | f |_{n \cdummy s} \label{lineargrowth}
\end{equation}
for any $n$ and $f \in {}^0 \mathcal{S} (\mathbb{R}^{d \cdummy n})$, and
$\sigma_n \leqslant \alpha (n!)^{\beta}$ for some constants $\alpha \comma
\beta$.

The unusual feature of the linear growth condition is that this is a condition
on all $n$-point correlation functions $G_n^E .$ It has to hold for all $n$ in
order for the result of {\cite{osterwalder1975}} to imply, say, even just the
temperedness of $3$-point Wightman distribution. In order to understand why
this is required, below we will review the basic strategy behind the proof of
{\cite{osterwalder1975}}. There are two steps in the argument. In the first
step, one establishes analyticity of $S_n (y_1, \ldots ., y_n)$ in the region
$\tmop{Re} y_k > 0.$ This does not require the linear growth condition
{\cite{Glaser1974,osterwalder1975}}. In the second step, which does use the
linear growth condition, one proves a bound on $S_n$ in this region, which
allows the application of Vladimirov's theorem and thus the construction of
tempered Wightman distributions.

We conclude this section with additional comments about the linear growth
condition. First of all, Appendix of {\cite{osterwalder1975}} shows that the
linear growth condition follows from requiring that $G_n^E \in \mathcal{S}'
(\mathbb{R}^{d n})$ and imposing
\begin{equation}
	| G_n^E (f_1 \otimes \ldots \otimes f_n) | \leqslant \sigma_n \prod_{i =
		1}^n | f_i |_r, \label{E0pp}
\end{equation}
for any $n$, where $f_i \in \mathcal{S} (\mathbb{R}^d)$, $| \cdot |_r$ is some
fixed Schwartz space norm, and $\sigma_n \leqslant \alpha (n!)^{\beta}$. In
other words, while in {\eqref{lineargrowth}} the $n$-point function variables
are smeared jointly, here each variable is smeared separately. Note that the
total smearing $f_1 \otimes \ldots \otimes f_n$ does not necessarily exclude
coincident points, that's why we need to assume $G_n^E \in \mathcal{S}'
(\mathbb{R}^{d n})$ and not $G_n^E \in ({}^0 \mathcal{S})' (\mathbb{R}^{d
	\cdummy n})$ as above.

Although {\eqref{E0pp}} is stronger than {\eqref{lineargrowth}}, it is easier
to verify in particular models. E.g.\ it holds for any gaussian scalar field
$\mathcal{O}$ with a two point function $G_2$ having a powerlaw asymptotics in
the UV.\footnote{Then $G_n^E (f_1 \otimes \ldots \otimes f_n)$ is a sum of $(n - 1) !!$
	terms, products of Wick contractions $G_2  (f_i \otimes f_j)$, which can be
	bounded by $A | f_i |_r | f_j |_r$ where $r$ depends on the UV dimension of
	$\mathcal{O}$. We thus get {\eqref{E0pp}} with $\sigma_n = (n - 1) !!A^{n/2}$.}
It has been also established in some non-gaussian models.\footnote{See e.g.~\cite{GJS_os_axiom},Theorem 1.1.8, which establishes Eq.~\eqref{E0pp} for Schwinger functions of arbitrarily high normal-ordered powers $:\!\phi^n\!:$ of the fundamental field  $\phi$ in weakly coupled $P(\phi)_2$ theories.}

More generally, bound {\eqref{E0pp}} is natural for field theories realizable
as random distributions.\footnote{See~\cite{Glimm:1981xz}, Sec.\ 6. This book introduced axioms for random distributions, numbered OS0-OS5. This chosen name is a bit unfortunate because these axioms are quite different in spirit from the original Osterwalder-Schrader axioms  described in Sec.~\ref{OSaxioms}, and appear much stronger. E.g.~they make the recovery of Wightman axioms a relatively trivial task. We don't know how to derive the axioms of \cite{Glimm:1981xz} from the Euclidean CFT axioms.} Imagine that there
is a measure $d \mu$ in the space of distributions $\phi \in \mathcal{S}'
(\mathbb{R}^d)$ such that for every test function $f \in \mathcal{S}
(\mathbb{R}^d)$ the following expectation value is finite:
\begin{equation}
	S (f) = \int d \mu \, e^{\phi (f)}\,. \label{genf}
\end{equation}
Such measures make rigorous sense of the Feynman path integral. Eq.
{\eqref{genf}} is a rigorous version of generating functional, and
differentiating with respect to $f$ one defines correlation functions $\langle
\phi (x_1) \ldots \phi (x_n) \rangle$ which are in this framework
automatically distributions in $\mathcal{S}' (\mathbb{R}^{d n})$. Bound
{\eqref{E0pp}} in this case can be reduced to an estimate on the growth of $S
(f)$. The Osterwalder-Schrader and Wightman axioms then follow.

{The field $\phi$ in \eqref{genf} is naturally a ``fundamental field'' of some model, such as $P(\phi)_2$ \cite{Glimm:1981xz} or $(\phi^4)_3$ (see \cite{Glimm:1981xz}, Sec. 23.1 for references). Sometimes this framework can be extended to generating functionals $\int d \mu \, e^{\phi' (f)}$ where $\phi'$ is a composite operator. E.g. $\phi'=\,:\!\phi^n\!:$, $n<\deg P$, in $P(\phi)_2$ is treated in \cite{Glimm:1981xz}. See also 
	\cite{Abdesselam:2016npc} for the general problem to construct $:\!\phi^2\!:$ as a random distribution given $\phi$.}

\subsection{Analytic continuation}

There are three tricks used together to construct the analytic continuation of $S_n$.
The first trick was already used above: it is the observation that if the
states $| \Psi_n (t ; y_{1,} \ldots, y_n) \rangle$ and $| \Psi_m (t' ; y'_{1,}
\ldots, y'_m) \rangle$ are defined for some values of $t, y_k$ and $t'$,
$y'_k$, then we can compute the matrix elements
\begin{equation}
	\langle \nobracket \Psi_m (t' ; y'_1 \ldots y'_{m - 1}) | U_{\tau} |
	\nobracket \Psi_n (t ; y_1 \ldots y_{n - 1}) \rangle = S_{m + n - 1}
	(\overline{y_{m - 1}'}, \ldots, \overline{y_1'}, t' + t + \tau, y_1, \ldots,
	y_{n - 1}) \label{trick1}
\end{equation}
with $\mathrm{Re}\,\tau>0$, thus potentially extending the domain of analyticity of $S_{m + n - 1}$.

The second trick, intuitively, says that we can write
\begin{equation}
	\langle \Psi_n (t ; y_1 \ldots y_{n - 1}) | \nobracket \Psi_n (t ; y_1
	\ldots y_{n - 1}) \rangle = S_{2 n - 1} (\overline{y_{n - 1}}, \ldots,
	\overline{y_1}, 2 t, y_1, \ldots, y_{n - 1}), \label{trick2}
\end{equation}
and so the state $| \Psi_n (t ; y_1 \ldots y_{n - 1}) \rangle$, whose norm
appears in the left-hand side, should be well-defined as long as the
correlation function in the right-hand side is well-defined. That is, while we
start with the states $| \Psi_n (t ; y_1 \ldots y_{n - 1}) \rangle$ defined
for positive real $y_k$, we should be able to analytically continue them in
$y_k$ if we manage to analytically continue the correlators $S_{2 n - 1} .$ Of
course, this is not a proof that $| \Psi_n (t ; y_1 \ldots y_{n - 1})
\rangle$ is well-defined. We will give the proof below, after we get more
information about the domain in which we wish to construct it.

The final trick is the idea of analytic completion for functions of several
complex variables. Recall that for $n > 1$ not every domain in $\mathbb{C}^n$
is the domain of holomorphy of some holomorphic function: there exist domains
$\mathcal{D} \subset \mathbb{C}^n$ such that any $f$ holomorphic in
$\mathcal{D}$ can be extended to a function holomorphic in a strictly larger
domain $\mathcal{D}' \supset \mathcal{D}$. For our applications the relevant
theorem is Bochner's tube theorem, which states that any holomorphic function
in a tube domain of the form $\mathcal{D}=\mathbb{R}^n + i X$, where $X$ is a
connected open subset of $\mathbb{R}^n$, can be extended to a holomorphic
function on $\mathcal{D}' = \tmop{ch} (\mathcal{D}) =\mathbb{R}^n + i
\tmop{ch} (X),$ where $\tmop{ch}$ denotes the convex hull. Note that since
$\mathcal{D}'$ is a convex set, it is a domain of holomorphy\footnote{To see
	this, it suffices to show that for any point $z$ on the boundary of
	$\mathcal{D}'$ there exists a function holomorphic in $\mathcal{D}'$ but
	singular at $z$. In general, such functions might not exist since the set of
	singularities of a holomorphic function cannot be arbitrary. However, it is
	easy to construct a function singular on any complex codimension-1 hyperplane
	in $\mathbb{C}^n$ (take the reciprocal of an affine-linear function). For a
	convex $\mathcal{D}'$ one can always find such a hyperplane passing through a
	given boundary point but staying away from the interior of $\mathcal{D}'$.}
and so $f$ cannot be extended any further by analytic completion alone. The
requirement that $X$ is open is a bit too restrictive and we'll need also a
degenerate case of this theorem, as described below.

These three tricks are applied one by one infinitely many times in order to
construct the full analytic continuation of $S_n .$ Instead of setting up the
procedure in its full glory, we will only follow the first steps to see how it
works in principle. The full analysis is performed in
{\cite{osterwalder1975}}.

First, it helps to introduce new variables $w_i$ by
\begin{equation}
	e^{w_i} = y_i .
\end{equation}
Our domains of analyticity in terms of $w_i$ will always be tubes of the form
$(w_1 \ldots w_n) \in \mathcal{D} (X) \equiv \mathbb{R}^n + i X$ for various
$X \subset \mathbb{R}^n$, and so we'll often just describe $X$. For example,
we start with $S_n$ and $\Psi_n$ defined for real positive $y_i$, which
corresponds to the domain $\mathcal{D} (\{ 0 \})$ in $w_i$.

Consider the 2-point function $S_1 (y_1)$. We start with the domain
$\mathcal{D} (\{ 0 \}) =\mathbb{R}$ in $w_1$, corresponding to real positive
$y_1 .$ Next, we apply the first trick. Specifically, we write
\begin{equation}
	\langle \nobracket \Psi_1 (t') | U_{\tau} | \nobracket \Psi_1 (t)
	\rangle = S_1 (t' + t + \tau),
\end{equation}
and since we are free to choose $t > 0$ and $t' > 0$ arbitrarily small, while
$U_{\tau}$, as discussed above, is a well-defined bounded operator for
$\tmop{Re} \tau > 0$, we obtain an analytic continuation of $S_1 (y_1)$ to the
right half-plane.

We are now done with the analytic continuation of $S_1$, since our goal was to
continue all $y_k$ to the right-half plane. In terms of $w_i$, this
corresponds to the domain $w_1 \in \mathcal{D} \left( \left( - \frac{\pi}{2},
+ \frac{\pi}{2} \right) \right)$, i.e.\ a strip. For higher-point functions, in
terms of $w_i$, we should stop when our domain of analyticity is $\mathcal{D}
\left( \left( - \frac{\pi}{2}, + \frac{\pi}{2} \right) \times \cdots \times
\left( - \frac{\pi}{2}, + \frac{\pi}{2} \right) \right)$.

Consider now the 3-point function $S_2 (y_1, y_2)$. We can again use the
first trick and define it on $(w_1, w_2) \in \mathcal{D} (X_2)$, where $X_2 =
\{ 0 \} \times \left( - \frac{\pi}{2}, + \frac{\pi}{2} \right) \cup \left( -
\frac{\pi}{2}, + \frac{\pi}{2} \right) \times \{ 0 \}$ (see Fig.\ \ref{extend1}, left). In more detail, we write the following two equations for
$S_2 (y_1, y_2)$, representing it as an inner product in two ways, and
inserting a $U_{\tau},$
\begin{equation}
	\langle \Psi_2 (t' ; y'_1) | U_{\tau} | \Psi_1 (t) \rangle = S_2
	(y_1', t' + t + \tau), \label{s2trick11}
\end{equation}
\begin{equation}
	\langle \Psi_1 (t') | U_{\tau} | \Psi_2 (t, y_1) \rangle = S_2 (t' + t
	+ \tau, y_1), \label{s2trick12}
\end{equation}
where the left-hand sides are well-defined (at this point) for $t, t', y_1,
y_1'>0$ and $\tmop{Re} \tau > 0$. We see that the first equation defines $S_2
(y_1, y_2)$ for real $y_1 > 0$ and $\tmop{Re} y_2 > 0$ as a holomorphic function
of $y_2$. The second equation does the same, but with $y_1$ and $y_2$
exchanged. In terms of $(w_1, w_2)$ this corresponds to the ``analyticity
domain'' $\mathcal{D} (X_2)$ described above. We write ``analyticity domain''
in quotes because $\mathcal{D} (X_2)$ is not open (and has empty interior),
and thus is not a domain. Correspondingly, we cannot say that $S_2$ is an
holomorphic function of two variables on $\mathcal{D} (X_2)$. We will deal with
this problem momentarily.

\begin{figure}[h]\centering
	\raisebox{-0.5\height}{\includegraphics[width=4.27990948445494cm,height=4.16486291486291cm]{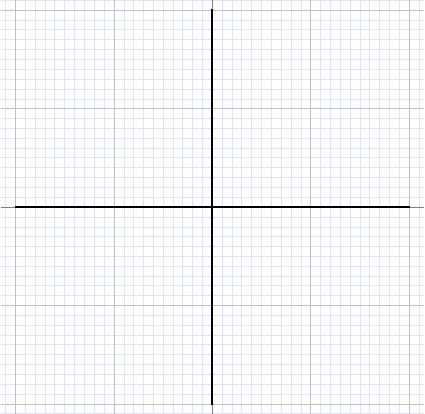}}
	$\Longrightarrow$
	\raisebox{-0.5\height}{\includegraphics[width=4.27990948445494cm,height=4.16486291486291cm]{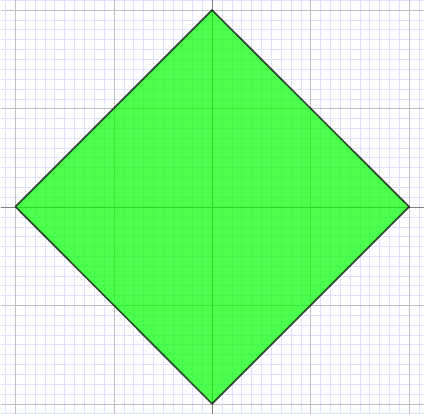}}
	\caption{\label{extend1}Left: set $X_2$. Right: domain $X_2'$ which defines
		the envelope of holomorphy $\mathcal{D} (X_2')$ of $\mathcal{D} (X_2)$.}
\end{figure}

To proceed with the analytic continuation of $S_2 (y_1, y_2)$, we want to use
the third trick, the tube theorem, to extend the analyticity domain from
$\mathcal{D} (X_2)$ to $\mathcal{D} (X_2')$, with $X_2' \equiv \tmop{ch}
(X_2)$ (Fig.\ \ref{extend1}, right).

The problem with this is that $X_2$ is not open, as mentioned above, so the
tube theorem does not apply. Instead, for this step one has to use
Malgrange-Zerner theorem {\cite{Epstein:1966yea}}, which allows $X_2$ to be a
union of intervals, with $S_2 (y_1, y_2)$ separately holomorphic in one variable
on each of these intervals, as is the case in our setup. The conclusion is
still that $S_2 (y_1, y_2)$ can be analytically continued to $\mathcal{D}
(X_2')$.

Note that the domain $\mathcal{D} (X_2')$ is not yet the full analyticity
domain $\mathcal{D} \left( \left( - \frac{\pi}{2}, + \frac{\pi}{2} \right)
\times \left( - \frac{\pi}{2}, + \frac{\pi}{2} \right) \right)$ that we are
aiming for. In particular, $X_2'$ is a proper subset of the square $\left( -
\frac{\pi}{2}, + \frac{\pi}{2} \right) \times \left( - \frac{\pi}{2}, +
\frac{\pi}{2} \right)$, see the right panel of Fig.\ \ref{extend1}.
Importantly, it doesn't approach the corners $\left( \pm \frac{\pi}{2}, \pm
\frac{\pi}{2} \right)$, which correspond to pure imaginary $y_1, y_2$. Pure
imaginary $y_1, y_2$ is, in turn, where we want to recover the Wightman
distributions.

To extend the domain of analyticity of $S_2 (y_1, y_2)$ even further, we need
to first extend the domain of $\Psi_2 (t, y_1)$, which can be done by the
second trick above --- via the equality
\begin{equation}
	\langle \Psi_2 (t, y_1) | \Psi_2 (t, y_1) \rangle = S_3 (\overline{y_1}, 2
	t, y_1) .
\end{equation}
Note that we are not interested in the analytic continuation in $t$ here ---
it is automatic when we act on $\Psi_2$ with $e^{- H t}$ --- so we can assume
$t$ is real. For $S_3$ we can run the same argument as we just did for $S_2$
and conclude that it is holomorphic in $\mathcal{D} (X_3')$, where $X_3'$ is the
convex hull of three intersecting intervals on coordinate axes (an
octahedron). As discussed above, we expect that $\Psi_2 (t, y_1)$ is defined
whenever $t$ and $y_1$ are such that the arguments of $S_3$ above are in its
analyticity domain. This happens whenever
\[ (\overline{w_1}, \log 2 t, w_1) \in \mathcal{D} (X_3'), \]
which is equivalent to
\begin{equation}
	(\tmop{Im} \overline{w_1}, \tmop{Im} \log 2 t, \tmop{Im} w_1) \in X_3' .
\end{equation}
Since we take $t$ to be real and positive, we have $\tmop{Im}
\log 2 t = 0$ and so $t$ is otherwise unconstrained. By construction of $X_3'$
and $X_2'$, $(\tmop{Im} \overline{w_1}, 0, \tmop{Im} w_1) \in X_3'$ is
equivalent to $(\tmop{Im} \overline{w_1}, \tmop{Im} w_1) \in X_2'$. Using
$\tmop{Im} \overline{w_1} = - \tmop{Im} w_1$, we conclude that $w_1$ is
constrained by
\begin{equation}
	(- \tmop{Im} w_1, \tmop{Im} w_1) \in X_2' .
\end{equation}
This is equivalent to $| \tmop{Im} w_1 | < \frac{\pi}{4}$, which is the same
as $w_1 \in \mathcal{D} \left( \left( - \frac{\pi}{4}, + \frac{\pi}{4} \right)
\right) .$ To conclude, we expect $\Psi_2 (t, y_1)$ to be defined and holomorphic
in $y_1$ for $t > 0$ and $w_1 \in \mathcal{D} \left( \left( - \frac{\pi}{4}, +
\frac{\pi}{4} \right) \right)$.

We can now apply the first trick to $S_2 (y_1, y_2)$ again, writing it as
inner product of $\Psi_1$ and $\Psi_2$ in the two ways {\eqref{s2trick11}} and
{\eqref{s2trick12}}. However, this time we can use $\Psi_2 (t, y_1)$ in a
wider domain of $y_1$, as computed above, equivalent to $w_1 \in \mathcal{D}
\left( \left( - \frac{\pi}{4}, + \frac{\pi}{4} \right) \right)$. From
{\eqref{s2trick11}} we conclude that $S_2 (y_1, y_2)$ is analytic for
\begin{equation}
	(w_1, w_2) \in \mathcal{D} \left( \left( - \frac{\pi}{4}, + \frac{\pi}{4}
	\right) \times \left( - \frac{\pi}{2}, + \frac{\pi}{2} \right) \right),
\end{equation}
where the domain of analyticity in $w_1$ comes from that of $\Psi_2 (t, y_1)$,
and in $w_2$ from $e^{- H \tau}$. Similarly, {\eqref{s2trick12}} now implies
analyticity in the domain
\begin{equation}
	(w_1, w_2) \in \mathcal{D} \left( \left( - \frac{\pi}{2}, + \frac{\pi}{2}
	\right) \times \left( - \frac{\pi}{4}, + \frac{\pi}{4} \right) \right) .
\end{equation}
Combining the two together, we find that $S_2 (y_1, y_2)$ is analytic for
$(w_1, w_2) \in \mathcal{D} (X_2'')$, where
\begin{equation}
	X_2'' \equiv \left( - \frac{\pi}{4}, + \frac{\pi}{4} \right) \times \left( -
	\frac{\pi}{2}, + \frac{\pi}{2} \right) \cup \left( - \frac{\pi}{2}, +
	\frac{\pi}{2} \right) \times \left( - \frac{\pi}{4}, + \frac{\pi}{4}
	\right),
\end{equation}
see the left panel of Fig.\ \ref{extend2}.

\begin{figure}[h]\centering
	\centering
	\raisebox{-0.5\height}{\includegraphics[width=4.27990948445494cm,height=4.16486291486291cm]{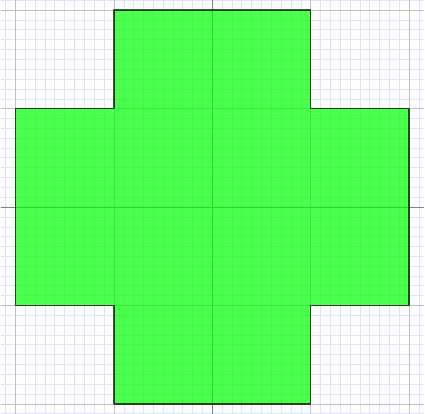}}
	$\Longrightarrow$
	\raisebox{-0.5\height}{\includegraphics[width=4.27990948445494cm,height=4.16486291486291cm]{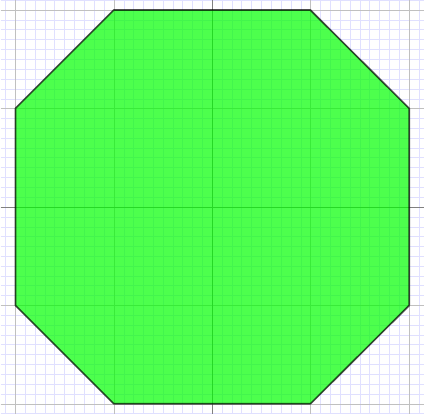}}
	\caption{Left: set $X_2''$. Right: domain $X_2'''$ which defines the
		envelope of holomorphy $\mathcal{D} (X_2''')$ of $\mathcal{D}
		(X_2'')$.\label{extend2}}
\end{figure}

Using the tube theorem, we can now extend the analyticity domain from
$\mathcal{D} (X_2'')$ further to $\mathcal{D} (X_2''')$, where $X_2''' \equiv
\tmop{ch} (X_2'')$ is the convex hull of $X_2''$ shown in the right panel of
Fig.\ \ref{extend2}.

We see that in order to analytically continue the 3-point function $S_2$,
it was useful to split it into an inner product of one-operator and
two-operator states $\Psi_1$ and $\Psi_2$, and use the information about the
latter that is provided by its norm, the 4-point function $S_3 .$ Still, we
have not yet managed to analytically continue $S_2$ to the entire region of
interest (we still have the corners missing in the right panel of Fig.\ \ref{extend2}). The only way to fix this is to extend the region of
analyticity of $S_3 .$ For that, we have to split it into a product of two
states, and extend the region of analyticity of these states. It is useless to
split it as a product of two $\Psi_2$ states, since their norm is computed by
$S_3$ itself and we won't learn anything new in this way. Instead, we have to
split it as a product of $\Psi_1$ and $\Psi_3$. This will lead us to consider
the norm of $\Psi_3$, which is computed by the six-point function $S_5$.
Following this logic, eventually, we will be forced to consider $S_n$ with
arbitrarily high $n$ just in order to construct the analytic continuation of
$S_2$. Fortunately, it can be shown that this procedure converges to the
desired domain $\mathcal{D} \left( \left( - \frac{\pi}{2}, + \frac{\pi}{2}
\right) \times \cdots \times \left( - \frac{\pi}{2}, + \frac{\pi}{2} \right)
\right)$ for all $S_n$, see {\cite{osterwalder1975}} for details.

To finish the discussion of the analytic continuation of $S_n$, let us justify
the second trick, which constructs the states $\Psi_n$ based on analyticity of
their norm $S_{2 n - 1} .$ Let $C$ be the domain of analyticity of $S_{2 n -
	1} (y_1 \ldots y_{2 n - 1})$ known to us, expressed in terms of $w_i$, and let
$D$ be the domain of the arguments $t, w_1 \ldots w_{n - 1}$ of $\Psi_n (t ;
y_1 \ldots y_{n - 1})$ for which the arguments of $S_{2 n - 1}$ in the
right-hand side of
\begin{equation}
	\langle \Psi_n (t ; y_1 \ldots y_{n - 1}) | \nobracket \Psi_n (t ; y_1
	\ldots y_{n - 1}) \rangle = S_{2 n - 1} (\overline{y_{n - 1}}, \ldots,
	\overline{y_1}, 2 t, y_1, \ldots, y_{n - 1}),
\end{equation}
belong to $C$. As is clear from the above discussion, $C$ (expressed in terms
of $w_i$) is always of the form $C =\mathcal{D} (X)$ for some $X$. We
similarly have $D =\mathcal{D} (Y)$ for some $Y$. By the tube theorem (or
Malgrange-Zerner theorem), we can assume that $X$ (and thus also $Y$)
is open, non-empty, and convex. Furthermore, it is easy to convince oneself
that $X$, and thus $Y$, is invariant under reflections along any of the
coordinate real axes (i.e.\ sending $w_i$ to $\overline{w_i}$ for some $i$).

Suppose now that we have a point $\tmmathbf{} (t ; w_1^0 \ldots w_{n - 1}^0)
\in D$. Then by definition of $D$ we have
\begin{equation}
	p \equiv \tmmathbf{} (\overline{w_{n - 1}^0} \ldots \overline{w_1^0}, \log 2
	t, w_1^0 \ldots w_{n - 1}^0) \in C.
\end{equation}
The above properties imply that there are $r_i > 0$ such that the polydisk
\begin{equation}
	P = \{ (w_1 \ldots w_{2 n - 1})  |  | w_i | < r_i \} \nobracket + \tmop{Re}
	p
\end{equation}
is contained in $C$, $P \subset C$, and moreover $p \in P.$ Indeed, since $C
=\mathcal{D} (X),$ this will be true if $\tmop{Im} P \subset X$ and $\tmop{Im}
p \in \tmop{Im} P$.\footnote{The latter is because $\tmop{Im}$ of the section
	of $P$ by $\tmop{Re} x = \tmop{Re} p$ is $\tmop{Im} P$.} By construction,
$\tmop{Im} P$ is a box with sides $2 r_i$ centered at 0. On the other hand,
the properties of $X$ imply that together with any point $x$, $X$ contains
such a box with $x$ being one of its vertices. We can then find an
$\varepsilon > 0$ such that $(1 + \varepsilon) \tmop{Im} p \in X$, and take
$\tmop{Im} P$ to be the box defined by the vertex $x = (1 + \varepsilon)
\tmop{Im} p$. See Fig.\ \ref{CPfig} for an intuitive picture.

\begin{figure}[h]\centering
	\raisebox{-0.5\height}{\includegraphics[width=8.94149285058376cm,height=5.35478158205431cm]{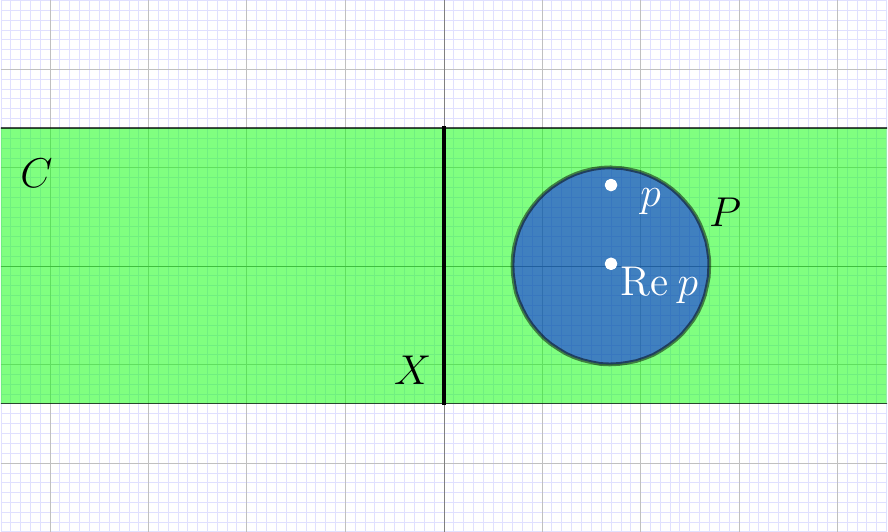}}
	\caption{Schematic picture of the tube $C$ and polydisk $P$.\label{CPfig}}
\end{figure}

Writing temporarily the state $\Psi_n$ as a function of $w_k$ instead of
$y_k$, we define it at $w_k$ by the Taylor series
\begin{equation}
	| \Psi_n (t ; w_1 \ldots w_{n - 1}) \rangle \equiv \sum_{\alpha} \frac{(w -
		\tmop{Re} w^0)^{\alpha}}{\alpha !} \partial^{\alpha} | \Psi_n (t ; \tmop{Re}
	w^0_1 \ldots \tmop{Re} w^0_{n - 1}) \rangle \label{stateext}
\end{equation}
($\alpha$ is a multiindex so $w^{\alpha} = w_1^{\alpha_1} w_2^{\alpha_2}
\ldots$ etc.). Note that the state in the right-hand side is well defined since
the corresponding $y_k = e^{\tmop{Re} w^0_k} > 0$. To check whether this
Taylor series converges, we look at its remainder
\begin{equation}
	\sum_{| \alpha | > N} \frac{(w - \tmop{Re} w^0)^{\alpha}}{\alpha !}
	\partial^{\alpha} | \Psi_n (t ; \tmop{Re} w^0_1 \ldots \tmop{Re} w^0_{n -
		1}) \rangle,
\end{equation}
whose norm squared is
\begin{equation}
	\sum_{| \alpha | > N} \sum_{| \beta | > N} \frac{(w - \tmop{Re}
		w^0)^{\alpha}}{\alpha !} \frac{(\bar{w} - \tmop{Re} w^0)^{\beta}}{\beta !}
	\partial^{\alpha} \partial^{\beta} S_{2 n - 1} (\tmop{Re} w^0_{n - 1} \ldots
	\tmop{Re} w^0_1, \log 2 t, \tmop{Re} w^0_1 \ldots \tmop{Re} w^0_{n - 1}),
\end{equation}
where $\beta$-derivatives act on the first $n - 1$ arguments of $S_{2 n - 1}$,
while $\alpha$-derivatives act on the last $n - 1$ arguments. Here we also
temporarily write $S_{2 n - 1}$ as function of $w_k$. This norm is clearly
just the tail of the Taylor series of $S_{2 n - 1}$ expanded around the point
$\tmop{Re} p$, and evaluated at $(\overline{w_{n - 1}}, \ldots,
\overline{w_1}, \log 2 t, w_1, \ldots, w_{n - 1})$. (We are not expanding in
$t$.) Since $S_{2 n - 1}$ is holomorphic in the polydisk $P$ centered at
$\tmop{Re} p$, this Taylor series converges in $P$ and thus this remainder
tends to $0$ there.

Since $p = \tmmathbf{} (\overline{w_{n - 1}^0} \ldots \overline{w_1^0}, \log
2 t, w_1^0 \ldots w_{n - 1}^0) \in P$, the remainder tends to 0 at $p$, and
thus {\eqref{stateext}} converges at $\tmmathbf{} (t ; w_1^0 \ldots w_{n -
	1}^0)$. Furthermore, since $P$ is open, it follows that {\eqref{stateext}}
converges in some neighborhood of $(t ; w_1^0 \ldots w_{n - 1}^0)$, defining
$| \Psi_n (t ; w_1 \ldots w_{n - 1}) \rangle$ as a holomorphic
$\mathcal{H}^{\tmop{OS}}$-valued function in that neighborhood. Since the
choice of $\tmmathbf{} (t ; w_1^0 \ldots w_{n - 1}^0) \in D$ was arbitrary, we
have defined $| \Psi_n (t ; w_1 \ldots w_{n - 1}) \rangle$ as a holomorphic
function of $w_i$ for all points in $D$.

\subsection{Temperedness bound}

Now that the correlation functions $S_n (y_1 \ldots .y_n)$ have been
analytically continued from $y_k > 0$ to $\tmop{Re} y_k > 0$, we only need to
establish a bound on their growth as $\tmop{Re} y_i \rightarrow 0$ in order to
construct tempered Wightman distributions by an application of Vladimirov's
theorem. The logic proceeds by establishing a bound on $S_n (y_1 \ldots y_n)$
for real $y_k$, and then repeating the analytic continuation described above,
while keeping track of this bound. We will only sketch this rather technical
argument in very general terms.

The final temperedness bound that we want to establish is
\begin{equation}
	| S_n (y_1 \ldots y_n) | \leqslant c_n \left( \left( 1 + \sum_k | y_k |
	\right) \left( 1 + \sum_k (\tmop{Re} y_k)^{- 1} \right) \right)^{p_n},
	\label{tempbound}
\end{equation}
for some sequences $c_n$ and $p_n$.\footnote{Here, for simplicity, we again
	ignore spatial arguments of the correlation functions, although they need to
	be taken care of at this step in order to establish ``temperedness in spatial
	directions.'' Furthermore, note that Osterwalder and Schrader establish
	additional bounds on $c_k$, etc., which are not important for the application
	of Vladimirov's theorem.} We would like {\eqref{tempbound}} to hold for all
$y_k,$ $\tmop{Re} y_k > 0.$ For real positive $y_k$ (i.e.\ in the Euclidean)
this holds as a consequence of {\eqref{OSmod}}. As discussed in Remark
\ref{OSnewVSold}, the original OS axioms did not include {\eqref{OSmod}}, so
their first step was to derive {\eqref{tempbound}} for $y_k > 0$ using
{\eqref{J0distr}}.

In principle at fixed $n$, {\eqref{tempbound}} looks reasonable given
{\eqref{J0distr}}: both say, intuitively, that the correlation functions
cannot be too singular at coincident points or grow too fast at infinity.
However, {\eqref{tempbound}} imposes this in a much more direct way. It turns
out that in general one cannot derive direct bounds such as
{\eqref{tempbound}} from averaged statements such as {\eqref{J0distr}}, even
if we know that $S_n$ is real analytic.

Consider the real-analytic function $\sin (e^x)$, $x \in \mathbb{R}$. It is a
bounded function, hence a tempered distribution. Thus its first derivative
$e^x \cos (e^x)$ is also a tempered distribution. This is an example of a
real-analytic tempered distribution which is not polynomially bounded. So some
further assumptions are needed beyond real analyticity.\footnote{Incidentally,
	our example shows that the Corollary of Lemma 1 in {\cite{Glaser1974}} is
	wrong.} \

In our case, the functions $S_n (y_1 \ldots y_n)$ are real-analytic and
satisfy {\eqref{J0distr}}. In addition, they satisfy OS positivity. We already
used OS positivity to show real-analyticity, and we will now have to invoke it
again to prove {\eqref{tempbound}} for $y_k > 0$. The full argument is rather
technical; we will explain the main idea on the example of $S_1 (y)$. Since
we know that $S_1$ is holomorphic, in particular harmonic, by the mean value
theorem for harmonic functions we can write it as a radially symmetric average
\begin{eqnarray}
	S_1 (y) & = & \int d x\, d t\, S_1 (y + x + i t) k_{\rho}
	(x, t) \nonumber\\
	& = & \int_{| t |, | t' | < \rho} d t\, d t'\, T (t | \nobracket g_{\rho}
	(\cdot, t + t'), g_{\rho} (\cdot, t')),  \label{SfromT}\\
	T (t | \nobracket \varphi_1, \varphi_2) & \assign & \int d x\, d x'\, S_1 (y
	+ x + i t) \varphi_1 (x + x') \varphi_2 (x'), \nonumber
\end{eqnarray}
where $k_{\rho}$ is a $C_0^{\infty}$ radial function supported in a ball of
radius $\rho$ and of integral 1, and we choose $\rho$ sufficiently small so
that all points under the integral sign are where $S_2$ is analytic. We also
chose
\begin{equation}
	k_{\rho} (x, t) = \int d x'\, d t'\, g_{\rho} (x + x', t + t') g_{\rho} (x',
	t'),
\end{equation}
a convolution of another radial $C_0^{\infty}$ function with itself (and hence
a radial function). The point of this construction is that, for generic
$\varphi_1, \varphi_2$, $T (0 | \nobracket \varphi_1, \varphi_2)$ is an inner
product $\langle \Psi_1 | \Psi_2 \rangle$ of two OS states:
\begin{equation}
	\langle \Psi_1 | \nobracket = \int d x\, \mathcal{O} (y / 2 + x) \varphi_1
	(x), \qquad | \nobracket \Psi_2 \rangle = \int d x\, \mathcal{O} (- y / 2 + x)
	\varphi_2 (x) .
\end{equation}
The norm of these states, and hence their inner product, can be bounded using
{\eqref{J0distr}}. Furthermore $T (t | \nobracket \varphi_1, \varphi_2) =
\langle \Psi_1 | e^{- i H t} | \nobracket \Psi_2 \rangle$ satisfies the same
bound. Using this bound for $\varphi_1 = g_{\rho} (\cdot, t + t'), \varphi_2 =
g_{\rho} (\cdot, t')$, Eq.\ {\eqref{SfromT}} gives a bound on $S_1 (y)$. The
same idea works for higher point functions. We first have to estimate the norm
of some states using {\eqref{J0distr}}.\footnote{Note that the linear growth
	condition is not needed at this point: Eq.\ {\eqref{J0distr}} with some
	$\sigma_n$ and $q_n$ suffices to establish {\eqref{tempbound}} with some $c_n$
	and $p_n$. The linear growth condition gives in addition $c_n$ of factorial
	growth and $p_n$ growing at most linearly. This turns out important later in
	the proof, see below.} We then analytically continue separately in each time,
and then use Malgrange-Zerner theorem to extend the bound on $T$ to an open
set. A single use of Malgrange-Zerner theorem suffices here, like in Fig.\ \ref{extend1}. We refer the reader to Sec.\ VI.1 of {\cite{osterwalder1975}}
for full details.

Once {\eqref{tempbound}} is established for $y_k > 0$, one repeats the
analytic continuation procedure that we described above, keeping track of the
implications of {\eqref{tempbound}}. The analytic continuation used three
tricks: (1) analytically continuing $S_n$ by representing it in the form
{\eqref{trick1}} (as $e^{- H \tau}$ inserted between two states), (2)
expressing the norms of these states in terms of higher-point $S_n$ as in
{\eqref{trick2}}, and (3) analytic completion.

The bound {\eqref{tempbound}} propagates through the tricks (1) and (2) by the
use of Cauchy-Schwartz inequality, as well as by using the fact that the norm
of $e^{- H \tau}$ is bounded from above by 1 (i.e.\ Eq.
{\eqref{contractive}}).

To propagate the bound through trick (3), the following simple idea is used.
Suppose we have domains $\mathcal{D}' \supset \mathcal{D}$ such that any
holomorphic function $f$ on $\mathcal{D}$ can be extended to a holomorphic
function on $\mathcal{D}'$. Then we have the equality of images
\begin{equation}
	f (\mathcal{D}') = f (\mathcal{D}),
\end{equation}
and in particular
\begin{equation}
	\sup_{z \in \mathcal{D}'} | f (z) | = \sup_{z \in \mathcal{D}} | f (z) | .
\end{equation}
To see this, suppose $a \in \mathbb{C}$ is a value which $f$ assumes in
$\mathcal{D}'$ but not in $\mathcal{D}$. Then the function $(f (z) - a)^{- 1}$
is holomorphic in $\mathcal{D}$ but has a singularity in $\mathcal{D}'$, which
is a contradiction. This shows that if we have a bound on $f$ in
$\mathcal{D}$, it is also valid in $\mathcal{D}'$.

Finally, recall that in order to construct the analytic continuation of
$S_{n_0}$ for some fixed $n_0$, we had to use $S_n$ with arbitrarily high $n$
in the process. This means that in order to establish the bound
{\eqref{tempbound}} on $S_{n_0}$ for all $\tmop{Re} y_k$, we have to use
{\eqref{tempbound}} for $y_k > 0$ for $S_n$ with arbitrarily high $n$. These
bounds need to combine in a way that is strong enough to establish
{\eqref{tempbound}} for $S_{n_0}$. For this, it is important that $c_n$ is of
factorial growth and $p_n$ grows at most linearly. This requires the same of
the sequence $\sigma_n$ and the index of the seminorm in {\eqref{J0distr}},
explaining the need for the linear growth condition.

\chapter{Conclusions}\label{chap:conclusions}

In this part of the thesis we studied the relationship between the modern Euclidean CFT
axioms (which we formulated in Sec.\ \ref{ECFTax}) and the more traditional
Osterwalder-Schrader and Wightman axioms. We showed that at least for $(n
\leqslant 4)$-point functions, both OS and Wightman axioms follow from the
Euclidean CFT axioms. Our Euclidean CFT axioms are quite modest. In
particular, beyond the minimal assumptions of regularity of correlators and
the standard constraints of unitarity, we assumed only a very weak form of the
convergent OPE.

Our derivation of Wightman axioms is of particular importance: it shows that
the conformal Wightman 4-point functions are well-defined tempered
distributions for arbitrary configurations of the 4 points, even when no OPE
channel is convergent in the sense of functions. We have furthermore shown
that these tempered distributions can always be computed by a conformal block
expansion which is convergent in the sense of distributions, generalizing
our previous results in part \ref{part:crossratio}, and giving a
derivation of Mack's results {\cite{Mack:1976pa}} from Euclidean CFT axioms.

\begin{figure}[h]\centering
	\raisebox{-0.481498738926148\height}{\includegraphics[width=7.92247146792601cm,height=5.45475862521317cm]{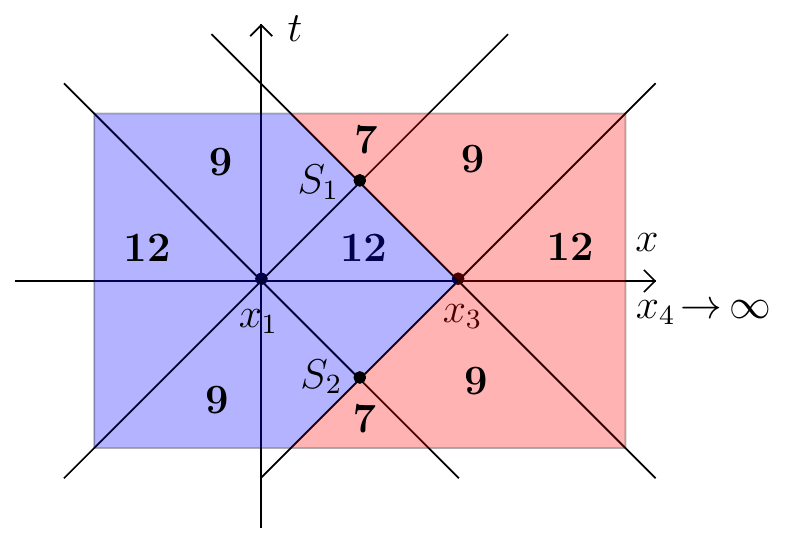}}
	\caption{\label{134}Minkowski configurations with $x_1 = 0, x_3 = \hat{e}_1,
		x_4 = \infty \hat{e}_1$ and $x_2 = t \hat{e}_0 + x \hat{e}_1$. Blue:
		configurations where $| \rho |, | \bar{\rho} | < 1$ and the 4-point
		functions is analytic. Red: configurations where $| \rho |$ and/or $|
		\bar{\rho} | =1$. Boldface numbers $X =\tmmathbf{7}, \tmmathbf{9},
		\tmmathbf{1}\tmmathbf{2}$ denote the causal type of the configuration
		according to part \ref{part:ope} (excluding lightlike separations). $S_{1,
			2}$ are double light-cone singularities.}
\end{figure}

For example, consider the configuration in Fig.\ \ref{134}, where the
operators in a 4-point function are inserted at $x_1 = 0, x_3 = \hat{e}_1,
x_4 = \infty \hat{e}_1$, while $x_2 = t \hat{e}_0 + x \hat{e}_1$ is allowed to
move in a plane parametrized by $(t, x) .$ The cross-ratios for this
configuration are $z, \bar{z} = x \pm t$. It is then easy to see that for
$x_2$ in the blue region of Fig.\ \ref{134} $| \rho |, | \bar{\rho} | < 1$
and the s-channel OPE converges in the sense of functions. Our results imply
that the s-channel OPE also converges in the red region where $| \rho |$ and/or
$| \bar{\rho} | = 1$, but now the convergence is in the sense of
distributions. In particular, the 4-point function is at least a
distribution for all values of $x_2$. Of course, in some regions of Fig.\ \ref{134} this was obviously true -- for example, in the red part of the
regions $\tmmathbf{9}, \tmmathbf{1}\tmmathbf{2}$ (labeling according to the
classification in part \ref{part:ope}), one can show that the 4-point
function is real-analytic using the convergent t-channel OPE. One may hope
to establish real-analyticity also in the region \tmtextbf{7} using
u-channel OPE. {This would indeed be the case for the ordering
	$\langle \mathcal{O} (x_2) \mathcal{O} (x_1) \mathcal{O} (x_3) \mathcal{O}
	(x_4) \rangle$. However, for the ordering \ $\langle \mathcal{O} (x_1)
	\mathcal{O} (x_2) \mathcal{O} (x_3) \mathcal{O} (x_4) \rangle$ that we are
	discussing here}, it turns out that no OPE channel converges in region
\tmtextbf{7} in the sense of functions.\footnote{For a reader comfortable with
	cuts in $z, \bar{z}$ plane the intuitive argument is simple: we have $z < 0$
	(on what we'll call s-channel cut), $\bar{z} > 1$ (on t-channel cut).
	Furthermore according to the operator ordering, when $z$ crosses $0$ we need
	to make $\tmop{Im} t$ slightly negative (and thus $\tmop{Im} z$ slightly
	negative), because $x_2$ at this point crosses the null cone of $x_1$, and
	when $\bar{z}$ crosses 1 we need to make $\tmop{Im} t$ slightly positive (and
	thus $\tmop{Im} \bar{z}$ slightly negative) because it corresponds to $x_2$
	crossing the null cone of $x_3$. Thus both $z$ and $\bar{z}$ end on lower
	sides of their respective cuts, and so one of them must have crossed the
	$u$-channel cut at $(0, 1)$ when analytically continuing from a Euclidean
	configuration. We conclude that s- and t- channel OPEs are only
	distributionally convergent, while $u$-channel is badly divergent.} Therefore,
before our work it was not at all clear whether this correlator makes any
sense in region \tmtextbf{7} if we assume only the Euclidean CFT axioms.

While we have shown that the correlator is \tmtextit{at least }distributional
in region \tmtextbf{7}, we have not excluded the possibility of it being
real-analytic there. For example, in 2 dimensions Virasoro symmetry implies
that the 4-point function is analytic everywhere away from light-cone
singularities {\cite{Maldacena:2015iua}}. This is perhaps too much to expect
in higher dimensions, but one can still ask whether analyticity can be
established in a larger domain. One approach is to ask for the envelope of
holomorphy of the known domain of analyticity. {Since the 4-point function is
	essentially only a function of two cross-ratios, this might be a tractable
	question \cite{PetrEnvelope}. We leave working out the full consequences of this idea for future work.\footnote{Another approach could be via alternative representations of the 4-point function having an extended region of analyticity, e.g.~\cite{Caron-Huot:2020nem}.}}


{In an upcoming paper {\cite{paper2a}}, we will generalize our
	results to external operators with spins.} In addition, there are many other
fundamental open questions which we believe are important to understand. First
of all, this part of the thesis is concerned with properties of CFT Wightman functions in
Minkowski space. However, it is expected that Lorentzian CFTs should be
naturally defined on Minkowski cylinder {\cite{Luscher:1974ez}}, which is the
smallest physically-sensible space on which finite conformal transformations
can act. Yet, it is not known whether CFT Wightman functions can be defined as
tempered distributions on Minkowski cylinder (see note \ref{LMcomplain}).
Answering this question in the positive for CFT $(n \leqslant 4)$-point
functions is the main goal of our forthcoming paper {\cite{paper3}}.

An important problem is to extend our results to $(n > 4)$-point functions. As
we discuss in App.\ \ref{OShigher}, even deriving the OS
axioms might require some strengthening of the OPE axiom. Another interesting
possibility is to formulate Euclidean CFT axioms as OS axioms supplemented
with a very weak form of the OPE (for example, asymptotic OPE in Euclidean
space). This is perhaps less attractive, since it is desirable to formulate
CFT axioms directly in terms of the CFT data (scaling dimensions and OPE
coefficients). However, it will still be interesting to establish an
equivalence between OS+(weak OPE axiom) and (possibly a stronger version of)
our Euclidean CFT axioms, perhaps using arguments similar to those of
{\cite{Mack:1976pa}}. Once OS axioms are established, it is likely that a
strategy similar to that of the present paper can be pursued to establish
Wightman axioms, using a comb-like OPE channel.

{In this part of the thesis we only considered Wightman functions, but in
	practice one often needs time-order Minkowski correlators. Textbook definition
	of time-ordered correlators involves multiplying Wightman functions by
	$\theta$-functions implementing time ordering. Since Wightman functions are in
	general distributions, this definition does not make rigorous sense at
	coincident points. As a matter of fact, time-ordered correlators have not
	been rigorously defined just from Wightman axioms alone (see e.g.\ {\cite{bogolubov2012general}}, p.505) in a general QFT. In a general QFT
	setting, it is known that defining time-ordered Minkowski correlators is
	closely related to defining Euclidean correlators at coincident points
	{\cite{Eckmann:1979vq}}. In the future, it would be interesting to construct
	time-ordered CFT Minkowski correlators as distributions just from Euclidean
	CFT axioms.}

A more ambitious goal is to understand the relationship of CFT axioms to
Haag-Kastler axioms. This appears to be considerably harder since these axioms
deal with operator algebras rather than local correlation functions, and some
qualitatively new ideas seem to be required.

\end{part}

\begin{part}{Classification of Convergent OPE Channels for Lorentzian CFT Four-Point Functions}\label{part:ope}
\thispagestyle{fancy}

\chapter{Introduction}\label{section:intro}
In this part of the thesis we study the convergence properties of operator product expansion (OPE) for Lorentzian four-point functions in conformal field theories (CFT). 

Historically, analyticity of correlation functions is an important bridge connecting Lorentzian quantum field theories (QFT) and Euclidean QFTs. Starting from a Lorentzian correlator, we can get a Euclidean correlator by analytically continuing the time variables onto the imaginary axis \cite{jost1979general}. Under certain conditions we can also do the reverse \cite{osterwalder1973,Glaser1974,osterwalder1975}. This procedure of analytic continuation, called Wick rotation, allows us to explore the Lorentzian nature of QFTs which may originate from statistical models in the Euclidean signature.

The Lorentzian correlators are not always genuine functions, instead they belong to a class of tempered distributions which are called Wightman distributions \cite{Streater:1989vi}. It is interesting to know at which Lorentzian configurations $(x_1,\ldots,x_n)$ the correlators $G_n(x_1,\ldots,x_n)$ are indeed functions. The Wightman distributions are known to be analytic functions in some regions $\mathcal{J}_n$ which are the sets of ``Jost points" \cite{jost1957bemerkung} ($G_n$ are called Wightman functions in their domains of analyticity).\footnote{It does not mean that the Lorentzian correlators cannot be functions in other regions. For example, the correlators of generalized free fields are functions aside from light-cone singularities. Here we are talking about the minimal domain of Lorentzian correlators which can be derived from general principles of QFT.} $\mathcal{J}_n$ corresponds to some (not all) Lorentzian configurations with totally space-like separations. By using the microscopic causality constraints, one can extend $G_n$ to a larger domain, including all configurations with totally space-like separations \cite{ruelle1959}. In Minkowski space $\bbR{d-1,1}$,\footnote{Often one uses Minkowski space to denote $\bbR{3,1}$ only. While in this part of the thesis, we use this terminology for $\bbR{d-1,1}$ and general $d$.} two points can also have time-like or light-like separation. The Lorentzian correlators usually diverge at configurations with light-like separations, and these configurations are called light-cone singularities \cite{dietz1973lightcone}. Except for some exactly solvable models, the Lorentzian correlators at configurations which contain time-like separations are not fully studied. 	 

There are more constraints in CFTs. In general QFTs, the domains of Wightman functions are Poincar\'e invariant, while in CFTs this Poincar\'e invariant domain can be further extended by using conformal symmetry. Furthermore, in CFTs we have better control on correlators with the help of OPE \cite{Pappadopulo:2012jk}. A successful example is the four-point functions in 2d local unitary CFTs, where the conformal algebra is infinite dimensional \cite{belavin1984infinite}. In this case, by using Al. Zamolodchikov's uniformizing variables $q,\bar{q}$ \cite{zamolodchikov1987conformal}, one can show that the four-point function is regular analytic at all possible Lorentzian configurations aside from light-cone singularities \cite{Maldacena:2015iua}. We are going to study a similar problem in $d\geq3$, for which the conformal group is finite dimensional and the radial coordinates $\rho,\bar{\rho}$ \cite{Hogervorst:2013sma} are used in our analysis.\footnote{The set of four-point configurations $(x_1,x_2,x_3,x_4)$ with $\abs{\rho},\abs{\bar{\rho}}<1$ is a subset of $(x_1,x_2,x_3,x_4)$ with $\abs{q},\abs{\bar{q}}<1$. Since the $q$-variable argument is based on the Virasoro symmetry which is only true in 2d \cite{francesco1997conformal}, we cannot apply it to the case of $d\geq3$.} In addition, in 2d there exists non-local unitary CFTs, which only have the global conformal symmetry. The analysis in this part of the thesis also applies to 2d non-local unitary CFTs.

Recently the conformal bootstrap approach has become a powerful tool in the study of strongly coupled systems \cite{Poland:2018epd}. On the numerical side, it gives precise predictions of experimentally measurable quantities, such as the critical exponents of the 3d Ising model \cite{ElShowk:2012ht,El-Showk:2014dwa,Kos:2014bka,Simmons-Duffin:2015qma}, $O(N)$ model \cite{Kos:2016ysd,Kos:2013tga,Kos:2015mba,Chester:2019ifh} and other critical systems. The functional methods, which are used in the numerical approach, can be realized analytically in low dimensions , and lead to insights into low dimensional CFTs and S-matrices \cite{Mazac:2016qev,Mazac:2018mdx,Mazac:2018ycv,Kaviraj:2018tfd,Paulos:2019gtx}. While the basic CFT assumptions are made in the Euclidean signature, many attempts have been made to study the bootstrap equations in the Lorentzian signature \cite{Komargodski:2012ek,Fitzpatrick:2012yx,Hartman:2015lfa,Hartman:2016dxc,Hartman:2016lgu,Caron-Huot:2017vep,Simmons-Duffin:2017nub}. In the conformal bootstrap approach, for crossing equations to be valid in the sense of functions, there should be at least two convergent OPE channels. To play the bootstrap game for four-point functions in the Lorentzian signature, it is important to know the convergent domains of various OPE channels. This provides an additional motivation for our work.

The main goal of this part of the thesis is to give complete tables of Lorentzian four-point configurations with the information about convergence in the sense of analytic functions in various OPE channels. In this part of the thesis we will mostly focus on four-point functions of identical scalar operators. Our techniques can be immediately generalized to the case of non-identical scalar operators. The four-point funcitons of spinning operators require extra work because of tensor structures. In this part of the thesis, we will only make some comments on the case of spinning operators. One may also be interested in the convergence of OPE in the sense of distributions \cite{Mack:1976pa}. We leave the discussions of distributional properties to the series of papers \cite{Kravchuk:2021kwe,paper3}.

The outline of this part of the thesis is as follows. In chapter \ref{section:main} we introduce the main problem and provide a quick summary of the main results in this part of the thesis. In chapter \ref{section:lorentz4pt} we give criteria of OPE convergence in s-, u- and t-channels. In chapter \ref{section:classifylorentzconfig}, we make a classification of the Lorentzian four-point configurations. All configurations in the same class have the same convergent OPE channels. All information on the OPE convergence properties can be looked up in appendix \ref{appendix:tableopeconvergence}. In appendix \ref{section:Wightman} we review some classical results from Wightman QFT, and compare them with CFT four-point functions. In chapter \ref{section:nonidscalar} we generalize our results to the case of non-identical scalar operators and make some comments on the case of spinning operators. In chapter \ref{section:conclusion} we make conclusions and point out some open questions related to this part of the thesis.

\chapter{Main problem and summary of results}\label{section:main}
\section{Main problem}\label{section:problem}
We start from CFT in the Euclidean signature. Let $x_k=(\tau_k,\mathbf{x}_k)$ denote the $k$-th point in the Euclidean space ($k=1,2,3,4$), where $\tau_k=x_k^0$ is the temporal variable and $\mathbf{x}_k=(x_k^1,x_k^2,\ldots,x_k^{d-1})\in\bbR{d-1}$ represents the vector of spatial variables. Lorentzian points are given by Wick rotating the temporal variables: $\tau=it$ where $t$ is a real number. To get Lorentzian four-point functions we need to analytically continue the Euclidean four-point functions to the Lorentzian regime. We define the Wick rotation of the four-point function as follows: \\ \\
\textbf{\large Step 1.} \\ \\
We construct the analytic continuation of the Euclidean CFT four-point function $G_4(c)$ to the forward tube $\mathcal{T}_4$, recall the definition
\begin{equation*}
	\begin{split}
		\mathcal{T}_4=\left\{c=(x_1,x_2,x_3,x_4)\in\mathbb{C}^{4d}\Big{|}x_k=(\tau_k,\mathbf{x}_k),\ \mathrm{Re}(\tau_k)-\mathrm{Re}(\tau_{k+1})>\abs{\mathrm{Im}(\mathbf{x}_k)-\mathrm{Im}(\mathbf{x}_{k+1})}\right\}
	\end{split}
\end{equation*}
\textbf{\large Step 2.}\\ \\
For a Lorentzian configuration $c_L=(x_1,x_2,x_3,x_4)$, where $x_k=(t_k,\mathbf{x}_k)$. Lorentzian CFT four-point function is defined by the limit
\begin{equation}\label{def:Lorentz4ptfct}
	\begin{split}
		G^L_4(c_L)\coloneqq\lim\limits_{\substack{\epsilon_k,\mathbf{y}_k\rightarrow0 \\\epsilon_k-\epsilon_{k+1}>\abs{\mathbf{y}_k-\mathbf{y}_{k+1}} \\}}G_4(\epsilon_k+it_k,\mathbf{x}_k+i\mathbf{y}_k).
	\end{split}
\end{equation}
The above definition is consistent with Wightman QFT, where the Lorentzian four-point function is the boundary value of the Wightman four-point function from its domain of complex coordinates \cite{jost1979general}. The domain of the four-point Wightman function includes $\mathcal{T}_4$, so the limit (\ref{def:Lorentz4ptfct}) gives the Wightman four-point distribution when such a limit exists. We review the properties of Wightman functions in appendix \ref{section:Wightman}.

Our starting point is Euclidean CFT axioms (see section \ref{ECFTax}) instead of Wightman axioms. However it is proved in part \ref{part:minkowski} that Euclidean CFT axioms imply Wightman axioms at the level of four-point functions. So step 1 is finished, and the limit in step 2 exists in the sense of tempered distributions.

The CFT four-point function can be computed via three OPE channels (s, t and u). The union of their domains of convergence determines the minimal domain of analyticity of the four-point function, including much more configurations than the general Wightman four-point function. In this part, we would like to study the following problem: 
\begin{itemize}
	\item In which Lorentzian regions does the Lorentzian CFT four-point function, defined by (\ref{def:Lorentz4ptfct}), have a convergent operator product expansion in the sense of \textbf{functions}?
\end{itemize}
The goal of this part of the thesis is to determine the OPE convergence properties of four-point functions at all possible Lorentzian configurations.

\subsection{Summary of results}
In this subsection, we provide a quick summary of the main results for readers who wish to know the general ideas of this part of the thesis before going into the technical details. Readers will find here:
\begin{itemize}
	\item The criteria of OPE convergence of the Lorentzian CFT four-point function $G_4^L$ in s-, t- and u-channels (chapter \ref{section:lorentz4pt}). $G_4^L$ is defined to be the boundary value of analytically continued Euclidean four-point function (see eq. (\ref{def:Lorentz4ptfct})). One can imagine that the OPE convergence properties of $G_4^L$ rely on the behavior of cross-ratio variables along the analytic continuation path. In the end we will see that for any fixed Lorentzian four-point configuration, one can check the criteria using any analytic continuation path in the forward tube $\mathcal{T}_4$ (starting from a Euclidean four-point configuration), and the conclusion does not depend on the choice of the path.
	
	\item A classification of the Lorentzian four-point configurations (chapter \ref{section:classifylorentzconfig}). The Lorentzian configurations are classified into a finite number of classes according to the range of cross-ratio variables $(z,\bar{z})$ (section \ref{section:classifyz}) and the causal orderings (section \ref{section:causal}). In each class, all configurations have the same OPE convergence properties (section \ref{section:classifyconfig}). 
	
	Then the problem is reduced to checking convergence properties in a finite number of cases. We use time-reversal symmetry to further reduce the problem to fewer cases (see section \ref{section:timereversal}). The conclusion of OPE convergence properties is lengthy because there are many cases (although finite) to check even after reduction, so we leave this part to appendix \ref{appendix:tableopeconvergence}. We share the Mathematica code for readers who wish to reproduce and check our results (see the auxiliary file on the arXiv webpage of \cite{Qiao:2020bcs}).
	
\end{itemize}
\color{black}

\chapter{Lorentzian CFT four-point function}\label{section:lorentz4pt}

\section{Some preparations}
In part \ref{part:minkowski}, we have shown that the s-channel expansion of the CFT four-point function is always convergent in the forward tube $\mathcal{T}_4$, thus it performs the analytic continuation of the Euclidean four-point function to $\mathcal{T}_4$ (step 1 done). Moreover, the s-channel expansion is convergent in the sense of distributions in Minkowski space. At some of the Minkowski four-point configurations, the CFT correlator is well-defined as an analytic function. In this part, we would like to study the OPE convergence properties in the sense of functions in Minkowski space.  We will call a configuration $c=(x_1,x_2,x_3,x_4)$ a \emph{Lorentzian (four-point) configuration} if all its points lie in Minkowski space.

Lemma \ref{xij2h} shows that for any four-point configuration $c=(x_1,x_2,x_3,x_4)$ in the forward tube $\mathcal{T}_4$, $x_{ij}^2$ is always non-zero. However if the four points of $c$ are in Minkowski space, $x_{ij}^2$ vanishes when $x_i$ and $x_j$ are light-like separated. The CFT four-point function may have singularities at these configurations.

By lemma \ref{xij2h}, \ref{bound} and the definition of $z,\bar{z}$ variables (see eqs.\,(\ref{uv}) and (\ref{zzbar})), we know that for all configurations in $\mathcal{T}_4$, the cross-ratio variables $z$ and $\bar{z}$ never belong to $\left\{0\right\}\cup[1,+\infty)$. However, since the Lorentzian four-point configurations are in the closure of $\mathcal{T}_4$, instead of $\mathcal{T}_4$ itself, some Lorentzian configurations will have $z$ or $\bar{z}$ in $\left\{0\right\}\cup[1,+\infty)$. These exceptional configurations are where s-channel expansion does not converge. We will see later that such configurations make up a large proportion of all Lorentzian configurations. 

A priori we can also use the t- and u-channel expansions to construct the analytic continuation of the four-point function, starting from the t- and u-channel versions of eqs.\,(\ref{def:Euclidean4-point}) and (\ref{uv}):
\begin{equation}\label{ope:tuchannel}
	\begin{split}
		G_4^E(c_E)=\dfrac{g(u_t,v_t)}{\fr{x_{14}^2\,x_{23}^2}^\Delta}=\dfrac{g(u_u,v_u)}{\fr{x_{13}^2\,x_{24}^2}^\Delta},
	\end{split}
\end{equation}
where
\begin{equation}\label{crossratios:tu}
	\begin{split}
		u_t=v,\quad v_t=u,\quad u_u=\dfrac{1}{u},\quad v_u=\dfrac{v}{u}.
	\end{split}
\end{equation}
In Euclidean space, these expansions should give the same result in their common domain of convergence. This consistency condition is called \emph{crossing symmetry} {\cite{ferrara1973tensor,polyakov1974nonhamiltonian}}. It can be written in the following graphical way:
\begin{equation*}
	\begin{tikzpicture}[baseline={(e.base)},circuit logic US]		
		\node (a) at (0,0) {$\mathcal{O}_1$};
		\node (b) at (0,-2) {$\mathcal{O}_2$};
		\node (c) at (5,0) {$\mathcal{O}_4$};
		\node (d) at (5,-2) {$\mathcal{O}_3$};
		\node (e) at (2.5,-0.7) {$\mathcal{O}$};
		\draw [thick] (0.3,0) -- (1.3,-1);
		\draw [thick] (0.3,-2) -- (1.3,-1);
		\draw [thick] (4.7,0) -- (3.7,-1);
		\draw [thick] (4.7,-2) -- (3.7,-1);
		\draw [thick] (1.3,-1) -- (3.7,-1);
		\node (f) at (2.5,-3.2) {s-channel};
		\node (g) at (-0.7,-1) {\Large$\sum\limits_{\mathcal{O}}$};
	\end{tikzpicture}\qquad=\qquad
	\begin{tikzpicture}[baseline={(e.base)},circuit logic US]		
		\node (a) at (0,0) {$\mathcal{O}_1$};
		\node (b) at (0,-4) {$\mathcal{O}_2$};
		\node (c) at (2,0) {$\mathcal{O}_4$};
		\node (d) at (2,-4) {$\mathcal{O}_3$};
		\node (e) at (0.7,-2) {$\mathcal{O}$};
		\draw [thick] (0,-0.3) -- (1,-1.3);
		\draw [thick] (0,-3.7) -- (1,-2.7);
		\draw [thick] (2,-0.3) -- (1,-1.3);
		\draw [thick] (2,-3.7) -- (1,-2.7);
		\draw [thick] (1,-1.3) -- (1,-2.7);
		\node (f) at (1,-4.5) {t-channel};
		\node (g) at (-0.7,-2.3) {\Large$\sum\limits_{\mathcal{O}}$};
	\end{tikzpicture}\qquad=\qquad
	\begin{tikzpicture}[baseline={(e.base)},circuit logic US]		
		\node (a) at (0,0) {$\mathcal{O}_1$};
		\node (b) at (0,-4) {$\mathcal{O}_2$};
		\node (c) at (2,0) {$\mathcal{O}_4$};
		\node (d) at (2,-4) {$\mathcal{O}_3$};
		\node (e) at (0.7,-2) {$\mathcal{O}$};
		\draw [thick] (0,-0.3) -- (1,-1.3);
		\draw [thick] (0,-3.7) -- (1,-2.7);
		\draw [thick] (2,-0.3) -- (1,-2.7);
		\draw [thick] (2,-3.7) -- (1,-1.3);
		\draw [thick] (1,-1.3) -- (1,-2.7);
		\node (f) at (1,-4.5) {u-channel};
		\node (g) at (-0.7,-2.3) {\Large$\sum\limits_{\mathcal{O}}$};
	\end{tikzpicture}.
\end{equation*}
We want to remark that only the s-channel expansion could be used to extend the Euclidean CFT four-point function to the whole forward tube $\mathcal{T}_4$, since lemma \ref{bound} holds only for the s-channel. We can use t- and u-channel expansion to analytically continue the four-point function to part of $\mathcal{T}_4$, but not to the whole $\mathcal{T}_4$. We will also consider t- and u-channel expansions because there are Lorentzian configurations where the s-channel expansion does not converge, but the t- or u-channel expansion converges (in fact such configurations make up a large part of the whole Lorentzian configuration space).

\section{Excluding light-cone singularities}
When $x_{ij}^2=0$ for some $x_i,x_j$ pair, since at least one of the scaling factors $(x_{ij}^2x_{kl}^2)^{-\Delta_\mathcal{O}}$ in eqs.\,(\ref{def:Euclidean4-point}) and (\ref{ope:tuchannel}) is infinity, we expect the four-point function to be infinity. The configurations which contain at least one light-like $x_i,x_j$ pair are called light-cone singularities. 

One example, for which the correlation functions are divergent at light-cone singularities, is the generalized free field (GFF)
\begin{equation}
	\begin{split}
	G_4^{\mathrm{GFF}}(c)=\dfrac{1}{\fr{x_{12}^2\,x_{34}^2}^\Delta}+\dfrac{1}{\fr{x_{14}^2\,x_{23}^2}^\Delta}+\dfrac{1}{\fr{x_{13}^2\,x_{24}^2}^\Delta}	.
	\end{split}
\end{equation}
Since we are interested in the Lorentzian configurations where the four-point functions are genuine functions for all unitary CFTs, we only consider the configurations which are not light-cone singularities. In other words, we will only consider the following set of Lorentzian configurations:
\begin{equation}\label{def:setlorconfig}
	\begin{split}
		\mathcal{D}_L\coloneqq\left\{(x_1,x_2,x_3,x_4)\Big{|}x_k=(it_k,\mathbf{x}_k),\ \forall k;\quad x_{ij}^2\neq0,\ \forall i\neq j\right\}.
	\end{split}
\end{equation}

\section{Criteria of OPE convergence}\label{section:criteria}
Now that $x_{ij}^2\neq0$ for all configurations in $\mathcal{D}_L$, all the cross-ratios defined in (\ref{uv}) and (\ref{crossratios:tu}) are finite and non-zero, which implies
\begin{equation}\label{z:DL}
	\begin{split}
		z,\bar{z}\neq0,1,\infty.
	\end{split}
\end{equation}
So the real axis in the $z,\bar{z}$-space is divided into three parts:
\begin{equation}\label{zaxis:split}
	\begin{split}
		(-\infty,0)\cup(0,1)\cup(1,+\infty).
	\end{split}
\end{equation}
In this section, we are going to establish criteria of OPE convergence in s-, t- and u-channels. The three intervals in (\ref{zaxis:split}) will play important roles because each of them is the place where one OPE channel stops being convergent.

\subsection{s-channel}\label{section:sconvergence}
The analytic continuation of $G_4$ (from Euclidean region to $\mathcal{T}_4$) was discussed in detail in section \ref{anal4-point}. So far, as already mentioned, we only used the s-channel expansion because of lemma \ref{bound}. Actually by using the s-channel expansion, we are able to extend $G_4$ to a larger domain $\mathcal{T}^s\supset\mathcal{T}_4$ (the label ``s" just means s-channel) according the constraint $0<\abs{\rho},\abs{\bar{\rho}}<1$ (or equivalently, $z,\bar{z}\neq\left\{0\right\}\cup[1,+\infty)$). $\mathcal{T}^s$ contains some but not all Lorentzian configurations. In other words, the Lorentzian four-point function has convergent s-channel OPE on the set $\mathcal{T}^s\cap\mathcal{D}_L$.

By eqs.\,(\ref{def:rho}) and (\ref{z:DL}), we have $\rho,\bar{\rho}\neq0,\pm1$ for all configurations in $\mathcal{D}_L$. Because of lemma \ref{bound} and the continuity, all configurations in $\mathcal{D}_L$ have $\abs{\rho},\abs{\bar{\rho}}\leq1$. To check the convergence of s-channel OPE, it suffices to check whether $\abs{\rho},\abs{\bar{\rho}}\neq1$ or not. Equivalently, it suffices to check whether $z,\bar{z}\notin(1,+\infty)$ or not. 

Therefore, given a Lorentzian configuration $c_L\in\mathcal{D}_L$, we have the following criterion of s-channel OPE convergence:
\begin{theorem}\label{theorem:schannel}
	(\emph{s-channel OPE convergence}) If neither $z$ nor $\bar{z}$ computed from $c_L$ belong to $(1,+\infty)$, then the Lorentzian four-point function $G_4$ is analytic at $c_L$ and is given by the formula (\ref{eq:gtilde}) in $d\geqslant3$ or (\ref{eq:gtilde2d}) in $d=2$.
\end{theorem}

\subsection{t-channel and u-channel}\label{section:tuconvergence}
We define the variables $z_t,\bar{z}_t$ and $z_u,\bar{z}_u$ by replacing $u,v$ with $u_t,v_t$ and $u_u,v_u$ in eq.\,(\ref{zzbar}). By eq.\,(\ref{crossratios:tu}), we choose proper solutions to the t- and u-channel versions of eq.\,(\ref{zzbar}), and get the following relations\footnote{The other solutions of $z_t,\bar{z}_t$ ($z_u,\bar{z}_u$) differ from (\ref{z:relation}) by interchanging $z_t$ and $\bar{z}_t$ ($z_u$ and $\bar{z}_u$), which will give the same conclusions of convergence properties in the t-channel (u-channel) expansion.}
\begin{equation}\label{z:relation}
	\begin{split}
		z_t=1-z,\quad\bar{z}_t=1-\bar{z},\quad z_u=1/z,\quad\bar{z}_u=1/\bar{z}.
	\end{split}
\end{equation}
Then we define the t- and u-channel versions of radial coordinates $\rho_t,\bar{\rho}_t,\rho_u,\bar{\rho}_u$ by replacing $z,\bar{z}$ with $z_t,\bar{z}_t$ and $z_u,\bar{z}_u$ in eq.\,(\ref{def:rho}). 

\textbf{\large t-channel}

Let us demonstrate how analytic continuation is performed using the t-channel expansion. The u-channel expansion argument will be similar. We first start from the Euclidean configuration space with time ordering:
\begin{equation}
	\begin{split}
		\mathcal{D}_E=\left\{c_E=(x_1,x_2,x_3,x_4)\in\mathbb{R}^{4d}\Big{|}x_1^0>x_2^0>x_3^0>x_4^0\right\}.
	\end{split}
\end{equation}
Since the cross-ratio variables $z$ and $\bar{z}$ are always complex conjugate for Euclidean configurations, they are not real as long as $z\neq\bar{z}$. Then recall eqs.\,(\ref{zuv}) and (\ref{Gamma}), $z$ and $\bar{z}$ are not real when $c_E\in\mathcal{D}_E\backslash\Gamma$. In particular, we have $z_t,\bar{z}_t\notin[1,+\infty)$ for all configurations in $\mathcal{D}_E\backslash\Gamma$, which allows us to choose $\abs{\rho_t}=\abs{\bar{\rho}_t}<1$ to start with convergent t-channel expansion. Analogously to the s-channel expansion, the t-channel expansion is given by sticking $\rho_t=re^{i\theta},\bar{\rho}_t=re^{-i\theta}$ into the series expansion (\ref{g:rhoexpansion}).

Now let us enter the forward tube $\mathcal{T}_4$ via the t-channel expansion. First of all, for the t-channel expansion to be convergent, the configuration must satisfy $z_t,\bar{z}_t\notin[1,+\infty)$. Otherwise we will have $\abs{\rho_t}\,\mathrm{or}\,\abs{\bar{\rho}_t}=1$ and the series expansion does not converge. Suppose we have a path $\gamma$ in $\mathcal{T}_4\backslash\Gamma$ such that $\gamma(0)\in\mathcal{D}_E\backslash\Gamma$, we can find a neighbourhood $U_\gamma\subset\mathcal{T}_4\backslash\Gamma$ of the set $\left\{\gamma(s)\ |\ 0\leq s\leq1\right\}$ and perform the analytic continuation of $z,\bar{z}$ in $U_\gamma$ via eqs.\,(\ref{uv}) and (\ref{zzbarsolved}).\footnote{As long as $\gamma(s)\notin\mathcal{D}\backslash\Gamma$ along the path $\gamma$, such a neighbourhood $U_\gamma$ always exists.} Then we get the analytic continuation of $z_t,\bar{z}_t$ in $U_\gamma$ by the relation in eq.\,(\ref{z:relation}). If $z_t,\bar{z}_t\notin[1,+\infty)$ in $U_\gamma$, or equivalently, $\abs{\rho_t},\abs{\bar{\rho}_t}<1$ in $U_\gamma$, then the t-channel expansion of $G_4$ is convergent in $U_\gamma$, and gives the analytic continuation to $U_\gamma$. Since the start point $\gamma(0)$ is a Euclidean configuration, $U_\gamma\cap\mathcal{D}_E$ is an open subset of $\mathcal{D}_E$, where the temporal variables $\tau_k$ are independent real numbers. According the crossing symmetry, the s- and t-channel expansions agree in $U_\gamma\cap\mathcal{D}_E$, so they also agree in $U_\gamma$, where $\tau_k$ are independent complex numbers. Furthermore, by taking the limit from $\mathcal{T}_4\backslash\Gamma$ to $\Gamma$, we can also use the t-channel expansion to compute the four-point function for configurations in $\Gamma$ with the constraint $\abs{\rho_t},\abs{\bar{\rho}_t}<1$, and the result also agrees with the s-channel expansion by continuity. So we conclude that
\begin{itemize}
	\item Given a configuration $c$ in $\mathcal{T}_4$, the t-channel expansion gives the same analytic continuation of $G_4$ as the s-channel expansion if there exists a path $\gamma$ in $\mathcal{D}$ such that $\gamma(0)\in\mathcal{D}_E\backslash\Gamma$, $\gamma(1)=c$ and $z_t,\bar{z}_t\notin[1,+\infty)$ along $\gamma$.
\end{itemize}
Analogously, by replacing $z_t,\bar{z}_t$ with $z_u,\bar{z}_u$, we have the similar conclusion for the u-channel expansion. 

While lemma \ref{bound} holds for $z,\bar{z}$, it does not hold for $z_t,\bar{z}_t$ or $z_u,\bar{z}_u$, which means that the t- and u-channel expansions may diverge in $\mathcal{T}_4$. To know the convergence properties of t- and u-channel expansions, we need to know not only the values of $z_t,\bar{z}_t,z_u,\bar{z}_u$ of a configuration, but also the values of these variables along a path. For convenience we use the relation (\ref{z:relation}) to translate $z_t,\bar{z}_t,z_u,\bar{z}_u\notin(1,+\infty)$ to equivalent conditions in $z,\bar{z}$:
\begin{equation}
	\begin{split}
		z_t,\bar{z}_t\notin(1,+\infty)\quad&\Leftrightarrow\quad z,\bar{z}\notin(-\infty,0) \\
		z_u,\bar{z}_u\notin(1,+\infty)\quad&\Leftrightarrow\quad z,\bar{z}\notin(0,1) \\
	\end{split}
\end{equation}
Since in each interval there is one OPE that stops converging, we call these three intervals \textbf{s-channel cut} ($(1,+\infty)$), \textbf{t-channel cut} ($(-\infty,0)$)and \textbf{u-channel cut} ($(0,1)$) respectively. Then it suffices to compute the $z,\bar{z}$-curves along the path and count how many times they cross the cuts.

To give the final criteria of convergence properties in t- channel expansions, we define some quantities which count how $z,\bar{z}$-curves cross the t-channel cut $(-\infty,0)$. Given a path $\gamma$ defined as follows
\begin{equation}\label{defpath:DL}
	\begin{split}
		\gamma&:\ [0,1]\ \longrightarrow\ \overline{\mathcal{T}}_4, \\
		\gamma&(0)\in\mathcal{D}_E\backslash\Gamma, \\
		\gamma&(s)\in\mathcal{T}_4\backslash\Gamma,\quad s<1, \\
	\end{split}
\end{equation}
if the variables $z,\bar{z}$ at the final point $\gamma(1)$ satisfy $z,\bar{z}\notin(-\infty,0)$, we define
\begin{equation}\label{def:nt}
	\begin{split}
		n_{t}\fr{\gamma}\coloneqq&\ \mathrm{number\ of\ times}\ z\ \mathrm{crosses}\ (-\infty,0)\ \mathrm{from\ above} \\
		&-\mathrm{number\ of\ times}\ z\ \mathrm{crosses}\ (-\infty,0)\ \mathrm{from\ below}, \\
		\bar{n}_{t}\fr{\gamma}\coloneqq&\ \mathrm{number\ of\ times}\ \bar{z}\ \mathrm{crosses}\ (-\infty,0)\ \mathrm{from\ above} \\
		&-\mathrm{number\ of\ times}\ \bar{z}\ \mathrm{crosses}\ (-\infty,0)\ \mathrm{from\ below}, \\
	\end{split}
\end{equation}
and
\begin{equation}
	\begin{split}
		N_t(\gamma)\coloneqq n_t(\gamma)+\bar{n}_t(\gamma). \\
	\end{split}
\end{equation}
Let us consider the t-channel expansion. We claim that $N_t$ is a path-independent quantity:
\begin{lemma}
	Given a configuration $c\in\overline{\mathcal{T}}_4$ with $z,\bar{z}\notin(-\infty,0]$, $N_t$ is independent of the choice of the path. Therefore, we can write $N_t(\gamma)$ as $N_t(c)$. 
\end{lemma}
\begin{proof}
	Without loss of generality, we assume that $\mathrm{Im}(z)>0$ and $\mathrm{Im}(\log\rho)\in(0,\pi)$ at the start point (resp. $\mathrm{Im}(\bar{z})<0$ and $\mathrm{Im}(\log\bar{\rho})\in(-\pi,0)$).\footnote{Choosing the convention $\mathrm{Im}(z)<0$ only exchanges $n_t$ and $\bar{n}_t$, which does not effect $N_t$.} 
	
	Let $\gamma$ be a path satisfying condition (\ref{defpath:DL}). By definition (\ref{def:nt}), we have
	\begin{equation}\label{imlogrhorhobar}
		\begin{split}
			\mathrm{Im}(\log\rho(c))=2\pi n_t(\gamma)+\varphi(\gamma), \\
			\mathrm{Im}(\log\bar{\rho}(c))=2\pi \bar{n}_t(\gamma)+\bar{\varphi}(\gamma), \\
		\end{split}
	\end{equation}
	where $\varphi(\gamma),\bar{\varphi}(\gamma)\in(-\pi,\pi)$. By eqs. (\ref{zzbar}) and (\ref{def:rho}), choosing a different path $\gamma$ can only exchange $\varphi(\gamma)$ and $\bar{\varphi}(\gamma)$. So $\varphi(\gamma)+\bar{\varphi}(\gamma)$ is path independent.
	
	By the analysis in section \ref{anal4-point}, we know that $R(c) = \rho (c)  \bar{\rho} (c)$ is an analytic function on the forward tube $\mathcal{T}_4$. Together with the fact that $\mathcal{T}_4$ is simply connected, we conclude that $\log(\rho(c))+\log(\bar{\rho}(c))$ is also an analytic function on $\mathcal{T}_4$. Therefore, by (\ref{imlogrhorhobar}) conclude that $N_t(\gamma)=n_t(\gamma)+\bar{n}_t(\gamma)$ is path independent.
\end{proof}
Suppose $c$ is a configuration in $\mathcal{T}_4$ with $z,\bar{z}\notin(-\infty,0)$ and $N_t=0$. By choosing an arbitrary path $\gamma$ with conditions (\ref{defpath:DL}) and $\gamma(1)=c$, we get the paths of $\log\rho$ and $\log\bar{\rho}$ along $\gamma$. By (\ref{imlogrhorhobar}) and $N_t=0$, (for each fixed path) there exists a unique integer $k$ such that
\begin{equation}
	\begin{split}
		\mathrm{Im}(\log\rho(c))+2k\pi,\,\mathrm{Im}(\log\bar{\rho}(c))-2k\pi\,\in\,(-\pi,\pi).
	\end{split}
\end{equation}
According to the s-channel expansion formula (\ref{g:rhoexpansion2}), the four-point function does not change if we replace $(\log\rho,\log\bar{\rho})$ with
\begin{equation}\label{deform:logrho}
	\begin{split}
		\log\tilde{\rho}=\log\rho+2i k\pi,\quad\log\tilde{\bar{\rho}}=\log\bar{\rho}-2i k\pi.
	\end{split}
\end{equation} 
In the $z,\bar{z}$ space, transformation (\ref{deform:logrho}) corresponds to that $z$ and $\bar{z}$ move around zero in opposite directions, without touching the s-channel cut. Therefore, we can deform the $z,\bar{z}$-curves by letting $z$ and $\bar{z}$ go around zero in opposite directions. The following plot shows an example of $k=1$. The blue and red solid curves on the left are the trajectories of $z$ and $\bar{z}$ along the path, which give $n_t=-1$ and $\bar{n}_t=1$, i.e. $N_t=0$. We add a clockwise loop (cyan dashed) to the $z$-curve and an anticlockwise loop (purple dashed) to the $\bar{z}$-curve. Then we continuously deform the curves without touching 0 and 1. In the end we arrive at the curves on the right.
\begin{equation*}
	\begin{split}
		\begin{tikzpicture}[baseline={((0,0).base)}]	
			\draw[->,thick] (-3,0)--(3,0) node[right]{};
			\filldraw[black] (0,0) circle (2pt) node[anchor=north]{0};
			\filldraw[black] (1,0) circle (2pt) node[anchor=north]{1};
			\draw [blue, thick] plot [smooth,tension=1] coordinates { (-1,1) (1,2) (2,1) (0.5,0.5) (0.2,-0.5) (-1,-2) (-2,2)};
			\draw [cyan, thick, dashed, ->] plot [smooth,tension=1] coordinates { (0.2,-0.5) (0.2,0.3)};
			\draw [cyan, thick, dashed, ->] plot [smooth,tension=1] coordinates { (0.2,0.3) (0,0.5) (-0.2,0.3) (-0.5,-0.2)};
			\draw [cyan, thick, dashed, ->] plot [smooth,tension=1] coordinates { (-0.5,-0.2) (-0.5,-0.8) (0.2,-0.5)};
			\filldraw[blue] (-1,1) circle (1pt) node[anchor=north]{$z(0)$};
			\filldraw[blue] (-2,2) circle (1pt) node[anchor=west]{$z(1)$};
			\draw [red, thick] plot [smooth,tension=1] coordinates { (-1,-1) (0.5,-1) (0.3,1.5) (-2,1) (-2.5,-1)};
			\draw [purple, thick,dashed,->] plot [smooth,tension=1] coordinates { (0.5,-1) (-0.6,-0.9)(-0.5,0.5)};
			\draw [purple, thick,dashed,->] plot [smooth,tension=1] coordinates {  (-0.5,0.5)(-0.2,0.7) (0.3,0.5)};
			\draw [purple, thick,dashed,->] plot [smooth,tension=1] coordinates {  (0.3,0.5)(0.6,0) (0.5,-1)};
			\filldraw[red] (-1,-1) circle (1pt) node[anchor=south]{$\bar{z}(0)$};
			\filldraw[red] (-2.5,-1) circle (1pt) node[anchor=north]{$\bar{z}(1)$};
		\end{tikzpicture}\quad\Longrightarrow\quad
	    \begin{tikzpicture}[baseline={((0,0).base)}]
	    	\draw[->,thick] (-3,0)--(3,0) node[right]{};
	    	\filldraw[black] (0,0) circle (2pt) node[anchor=north]{0};
	    	\filldraw[black] (1,0) circle (2pt) node[anchor=north]{1};
	    	\draw [blue, thick] plot [smooth,tension=1] coordinates { (-1,1) (0,0.5) (-1.5,0.5) (-2,2)};
	    	\filldraw[blue] (-1,1) circle (1pt) node[anchor=south]{$z(0)$};
	    	\filldraw[blue] (-2,2) circle (1pt) node[anchor=west]{$z(1)$};
	    	\draw [red, thick] plot [smooth,tension=1] coordinates { (-1,-1) (0.5,-1) (0.3,-0.5) (-1,-0.7) (-2.5,-1)};
	    	\filldraw[red] (-1,-1) circle (1pt) node[anchor=north]{$\bar{z}(0)$};
	    	\filldraw[red] (-2.5,-1) circle (1pt) node[anchor=north]{$\bar{z}(1)$};
	    \end{tikzpicture}
	\end{split}
\end{equation*}
Eq.\,(\ref{g:rhoexpansion2}) tells us that the left and the right curves give the same analytic continuation of the four-point function at the end point. Note that if $z$ nor $\bar{z}$ do not cross $(-\infty,0)$ at all, then $|\rho_t|, |\bar{\rho}_t|<1$ along the whole path, and the t-channel OPE is guaranteed to converge. So for the right $z,\bar{z}$-curves, we can safely use the t-channel expansion to compute the four-point function. Therefore, any $c\in\mathcal{T}_4$ with $z,\bar{z}\notin(-\infty,0)$ and $N_t=0$ has convergent t-channel expansion.

For $c\in\overline{\mathcal{T}}_4$ such that $z,\bar{z}\notin(-\infty,0)$ and $N_t=0$, it can always be approached by the configurations in $\mathcal{T}_4$ with the same condition. Then the t-channel expansion still converges in the end by continuity. So we conclude that
\begin{theorem}\label{theorem:tchannel}
	(\emph{t-channel OPE convergence}) Let $c\in\overline{\mathcal{T}}_4$ be a four-point configuration aside from light-cone singularities. If the cross-ratio variables $z(c),\bar{z}(c)$ do not belong to $(-\infty,0)$, and furthermore if $N_t(c)=0$, then the four-point function $G_4$ is analytic at $c$ and can be computed using the t-channel expansion
	\begin{equation}\label{G4:tchannel}
		\begin{split}
			G_4(c)=\dfrac{g_t(c)}{(x_{23}^2x_{14}^2)^\Delta}.
		\end{split}
	\end{equation}
	The function $g_t(c)$ is defined by replacing $\rho,\bar{\rho}$ with $\tilde{\rho}_t(c),\tilde{\bar{\rho}}_t(c)$ in the series expansion (\ref{g:rhoexpansion2}), and $\tilde{\rho}_t(c),\tilde{\bar{\rho}}_t(c)$ are the corresponding $t$-channel variables after transformation (\ref{deform:logrho}):
	\begin{equation}
		\begin{split}
			\rho\stackrel{(\ref{deform:logrho})}{\longrightarrow}\tilde{\rho}\rightarrow\tilde{z}\rightarrow\tilde{z}_t=1-z\stackrel{(\ref{def:rho})}{\longrightarrow}\tilde{\rho}_t=f(z_t).
		\end{split}
	\end{equation}
\end{theorem}

\textbf{\large u-channel}

The argument for the u-channel expansion is similar. First of all, for the u-channel expansion to be convergent, the configuration should satisfy $z,\bar{z}\notin(0,1)$, otherwise we will have $\abs{\rho_u}\,\mathrm{or}\,\abs{\bar{\rho}_u}=1$. Given a configuration $c$ with $z,\bar{z}\notin(0,1)$, we choose a path $\gamma$ satisfying condition (\ref{defpath:DL}) and $\gamma(1)=c$. We define
\begin{equation}
	\begin{split}
		n_{u}\fr{\gamma}\coloneqq&\ \mathrm{number\ of\ times}\ z\ \mathrm{crosses}\ (0,1)\ \mathrm{from\ above} \\
		&-\mathrm{number\ of\ times}\ z\ \mathrm{crosses}\ (0,1)\ \mathrm{from\ below}, \\
		\bar{n}_{u}\fr{\gamma}\coloneqq&\ \mathrm{number\ of\ times}\ \bar{z}\ \mathrm{crosses}\ (0,1)\ \mathrm{from\ above} \\
		&-\mathrm{number\ of\ times}\ \bar{z}\ \mathrm{crosses}\ (0,1)\ \mathrm{from\ below}, \\
	\end{split}
\end{equation}
and
\begin{equation}
	\begin{split}
		N_u(\gamma)\coloneqq n_u(\gamma)+\bar{n}_u(\gamma). \\
	\end{split}
\end{equation}
Similarly to the t-channel case, $N_u(\gamma)$ does not depend on the choice of $\gamma$, so we can write it as $N_u(c)$. When $N_u(c)=0$, we choose an arbitrary $\gamma$, compute the $z,\bar{z}$-curves, then deform the curves by moving $z$ and $\bar{z}$ around 0 in opposite directions.  The deformed curves give the same analytic continuation of the four-point function at the end point. We can find a proper integer $k$ in (\ref{deform:logrho}) such that the deformed curves do not touch the u-channel cut $(0,1)$. Then we can safely use the u-channel expansion. Therefore, we conclude that 
\begin{theorem}\label{theorem:uchannel}
	(\emph{u-channel OPE convergence}) Let $c\in\overline{\mathcal{T}}_4$ be a four-point configuration aside from light-cone singularities.  If the cross-ratio variables $z(c),\bar{z}(c)$ do not belong to $(0,1)$, and furthermore if $N_u(c)=0$, then the four-point function $G_4$ is analytic at $c$ and can be compute using the u-channel expansion
	\begin{equation}\label{G4:uchannel}
		\begin{split}
			G_4(c)=\dfrac{g_u(c)}{(x_{13}^2x_{24}^2)^\Delta}.
		\end{split}
	\end{equation}
	The function $g_u(c)$ is defined by replacing $\rho,\bar{\rho}$ with $\tilde{\rho}_u(c),\tilde{\bar{\rho}}_u(c)$ in the series expansion (\ref{g:rhoexpansion2}), and $\tilde{\rho}_u(c),\tilde{\bar{\rho}}_u(c)$ are the corresponding $u$-channel variables after transformation (\ref{deform:logrho}):\footnote{In the u-channel case, the integer $k$ in (\ref{deform:logrho}) should be such that the deformed $z,\bar{z}$-curves do not touch the u-channel cut $(0,1)$.} 
	\begin{equation}
		\begin{split}
			\rho\stackrel{(\ref{deform:logrho})}{\longrightarrow}\tilde{\rho}\rightarrow\tilde{z}\rightarrow\tilde{z}_u=1/z\stackrel{(\ref{def:rho})}{\longrightarrow}\tilde{\rho}_u.
		\end{split}
	\end{equation}
\end{theorem}
We would like to make four comments on the t-channel and u-channel expansions. 
\begin{remark}
	Theorems \ref{theorem:tchannel} and \ref{theorem:uchannel} cover the case of Lorentzian four-point functions because Lorentzian configurations lie on the boundary of $\mathcal{T}_4$. 
\end{remark}
\begin{remark}
	Theorem \ref{theorem:tchannel} and \ref{theorem:uchannel} cover the case when s-channel expansion is not convergent.
\end{remark}
\begin{remark}
	To compute the four-point function in t-channel or u-channel, it is important to know the arguments of $\tilde{\rho}_t$, $\tilde{\bar{\rho}}_t$, $\tilde{\rho}_u$ and $\tilde{\bar{\rho}}_u$. When $z,\bar{z}\notin(1,\infty)$, one can determine these arguments directly from the configuration itself, without looking at the path. The reason is that (up to homotopy of the path) there is only one possible way to connect start point $\tilde{z}(0)$ and end point $\tilde{\bar{z}}(1)$, without touching the s- and t-(u-)channel cuts (since $\mathbb{C}\backslash\left\{(-\infty,0]\cup[1,+\infty)\right\}$ and $\mathbb{C}\backslash[0,+\infty)$ are simply connected).  When $z\,\mathrm{or}\,\bar{z}\in(1,\infty)$, one can determine these arguments by looking at the behavior near the end of $z$-curve or $\bar{z}$-curve.
\end{remark}
\begin{remark}
	Unlike the s-channel case, even if we only want to check whether t- or u-channel expansion is convergent or not, we have to choose a path to compute $N_t$ or $N_u$. 
\end{remark}

Before finishing this section, we want to remark that in practice, the condition (\ref{defpath:DL}) of the path $\gamma$ can be relaxed in the way that $\gamma$ is allowed to touch $\Gamma$:
\begin{equation}\label{defpath:DLrelax}
	\begin{split}
		\gamma&:\ [0,1]\ \longrightarrow\ \overline{\mathcal{T}}_4, \\
		\gamma&(0)\in\mathcal{D}_E\backslash\Gamma, \\
		\gamma&(s)\in\mathcal{T}_4,\quad s<1. \\
	\end{split}
\end{equation}
Suppose we have a path $\gamma$ which intersects with $\Gamma$. Let $\gamma(s_*)\in\Gamma$ be the first intersection point. At $s_*$ we have $z(s_*)=\bar{z}(s_*)$, then $z(s),\bar{z}(s)$ become indistinguishable for $s>s_*$, so the quantities $n_t,\bar{n}_t,n_u,\bar{n}_u,\chi,\bar{\chi}$ are not well defined for $\gamma$. However, by manually choosing $z,\bar{z}$ after each intersection, we still get two curves $z(s),\bar{z}(s)$: they may not be smooth at intersection points, but they are still continuous. By this trick we get $n_t,\bar{n}_t,n_u,\bar{n}_u,\chi,\bar{\chi}$, so that we are able to compute $N_t,N_u$ and the four-point function. On the other hand, we can always deform $\gamma$ to a path $\gamma^\prime$, such that $\gamma^\prime$ has the same start and final points as $\gamma$ but $\gamma^\prime$ does not intersect with $\Gamma$. By doing proper deformation, we can make $\gamma^\prime$ have the same $n_t,\bar{n}_t,n_u,\bar{n}_u,\chi,\bar{\chi}$ as selected on $\gamma$. Therefore, our manual selection will give the correct OPE convergence properties and the correct value of the four-point function.

\section{Violating the conditions of theorem \ref{theorem:schannel}, \ref{theorem:tchannel} and \ref{theorem:uchannel}}
We would like to make a comment that theorem \ref{theorem:schannel}, \ref{theorem:tchannel} and \ref{theorem:uchannel} only give sufficient conditions for OPE convergence. For a Lorentzian configuration $c_L$ which is not a light-cone singularity and which does not satisfy the conditions in these theorems, it does not mean that $G_4$ cannot be a function at $c_L$. It just means that for general CFT, we are not able to use the radial coordinates $\rho,\bar{\rho}$ ($\rho_t,\bar{\rho}_t$, $\rho_u,\bar{\rho}_u$) and the expansion (\ref{g:rhoexpansion2}) to prove the analyticity of $G_4$ at $c_L$. The four-point function still has a chance to be analytic at $c_L$. For example, the Euclidean four-point function of generalized free fields has analytic continuation to the whole Lorentzian region except for the light-cone singularities.

An interesting related open question is: can we relax the conditions in theorem \ref{theorem:schannel}, \ref{theorem:tchannel} and \ref{theorem:uchannel}? \\

\textbf{\large The s-channel condition} \\  \\
In theorem \ref{theorem:schannel}, we only assume the condition $z,\bar{z}\notin(1,+\infty)$ (equivalently, $|\rho|,\abs{\bar{\rho}}<1$). The Lorentzian configurations which violate this condition has $|\rho|=1$ or $\abs{\bar{\rho}}=1$, then the proof of theorem \ref{theorem:schannel} fails because in the proof we used the fact that the series expansion (\ref{g:rhoexpansion2}) is absolutely convergent when $\abs{\rho},\abs{\bar{\rho}}<1$.

We are interested in the Lorentzian configurations where the s-channel expansion is convergent for all unitary CFTs. For configurations with $|\rho|=1$ or $\abs{\bar{\rho}}=1$, we may exhibit an explicit CFT four-point function, for which the s-channel expansion is divergent (then such configurations are ruled out). The generalized free field (GFF) theory is such an example. The GFF four-point function of identical scalar operators (with scaling dimension $\Delta$) is defined by
\begin{equation}
	\begin{split}
		\fr{G_4}_{GFF}(x_1,x_2,x_3,x_4)=\dfrac{1}{\left[x_{12}^2x_{34}^2\right]^\Delta}+\dfrac{1}{\left[x_{23}^2x_{14}^2\right]^\Delta}+\dfrac{1}{\left[x_{13}^2x_{24}^2\right]^\Delta}.
	\end{split}
\end{equation}
By (\ref{def:Euclidean4-point}), the conformal invariant part of $\fr{G_4}_{GFF}$ is given by
\begin{equation}
	\begin{split}
		g_{_{GFF}}(\rho,\bar{\rho})=1+\left(\dfrac{16\rho\bar{\rho}}{(1+\rho)^2(1+\bar{\rho})^2}\right)^\Delta+\left(\dfrac{16\rho\bar{\rho}}{(1-\rho)^2(1-\bar{\rho})^2}\right)^\Delta.
	\end{split}
\end{equation}
It has the series expansion
\begin{equation}
	\begin{split}
		g_{_{GFF}}(\rho,\bar{\rho})=1+\fr{16\rho\bar{\rho}}^\Delta\sumlim{m,n=0}^{\infty}\dfrac{\fr{1+(-1)^{m+n}}\Gamma(\Delta+m)\Gamma(\Delta+n)}{m!n!\Gamma(\Delta)^2}\rho^m\bar{\rho}^n
	\end{split}
\end{equation}
which diverges when $\abs{\rho}=1$ or $\abs{\bar{\rho}}=1$. It follows that theorem \ref{theorem:schannel} cannot be extended to configurations with $\abs{\rho}=1$ or $\abs{\bar{\rho}}=1$ without extra assumptions on the theory. One such extra assumption will be mentioned in section \ref{section:comments2d} (locality of 2d CFT).\\

\textbf{\large The t-channel and u-channel conditions} \\ \\
In theorem \ref{theorem:tchannel}, we assumed two conditions: $z,\bar{z}\notin(-\infty,0)$ and $N_t=0$. 

For Lorentzian configurations which violate the first condition, (analogously to the s-channel case) we can use GFF to conclude that these configurations do not have convergent t-channel expansion for some unitary CFTs.

When the second condition is violated, i.e. $N_t\neq0$, according to our discussion in section \ref{section:tuconvergence}, this is equivalent to putting $\bar{z}$ in the first sheet of the complex plane with the t-channel cut $(-\infty,0]$ and letting $z$ cross the t-channel cut from above for $N_t$ times. When $z$ cross the t-channel cut for the first time, the corresponding radial variable $\rho_t$ becomes greater than 1 in absolute value, so the t-channel expansion diverges. One may ask what happens if $z$ cross the t-channel cut for even number of times (in these cases one has $\abs{\rho_t}<1$). Our answer is we do not know since the analytic continuation has already been interrupted in the second sheet (i.e. after $z$ crossing the cut for the first time).

The arguments for theorem \ref{theorem:uchannel} (u-channel) are similar.

\chapter{Classifying the Lorentzian four-point configurations}\label{section:classifylorentzconfig}
In the previous chapter we gave the criteria of convergence properties of OPE in various channels for Lorentzian CFT four-point functions. These criteria say that given a Lorentzian configuration $c_L$, one can just start with an arbitrary Euclidean configuration in $\mathcal{D}_E\backslash\Gamma$ and choose an arbitrary path towards $c_L$, then decide if the conditions in theorem \ref{theorem:schannel}, \ref{theorem:tchannel} and \ref{theorem:uchannel} hold or not by watching the $z,\bar{z}$-curves (in theorem \ref{theorem:schannel} one does not even have to choose a path).

However, it would be frustrating if we have to check the analytic continuation curves for all Lorentzian configurations in $\mathcal{D}_L$ (recall definition (\ref{def:setlorconfig})). We expect that these Lorentzian configurations can be classified such that for each class it suffices to choose one representative configuration to see if various OPE channels converge or not. There are two natural classification methods, one according to the range of $z$ and $\bar{z}$, the other according to the causal orderings. We will show that combining these two methods leads to a complete classification for the convergence properties of Lorentzian CFT four-point functions.

\section{\texorpdfstring{$z,\bar{z}$}{z,zbar} of Lorentzian configurations}\label{section:classifyz}
For all Lorentzian configurations, since $x_{ij}^2$ are real, the cross-ratios $u,v$ are also real. By (\ref{zzbarsolved}), there are only two possibilities for $z,\bar{z}$:
\begin{enumerate}
	\item $z,\bar{z}$ are independent real variables.
	\item $z,\bar{z}$ are complex conjugate to each other.
\end{enumerate}
In addition, we have already excluded light-cone singularities in $\mathcal{D}_L$ (recall its definition in eq.\,(\ref{def:setlorconfig})), so the configurations in $\mathcal{D}_L$ have $z,\bar{z}\neq0,1,\infty$. According to the range of the $z,\bar{z}$ variables, we divide $\mathcal{D}_L$ into four classes:
\begin{equation}\label{division:class}
	\begin{split}
		\mathcal{D}_L=\mathrm{S}\sqcup \mathrm{T}\sqcup \mathrm{U}\sqcup \mathrm{E},
	\end{split}
\end{equation}
where the classes are defined as follows.
\begin{itemize}
	\item Class S: configurations with $0<z<1,\ \bar{z}<0\ \mathrm{or}\ z<0,\ 0<\bar{z}<1$.
	\item Class T: configurations with $z>1,\ 0<\bar{z}<1\ \mathrm{or}\ 0<z<1,\ \bar{z}>1$.
	\item Class U: configurations with $z>1,\ \bar{z}<0\ \mathrm{or}\ z<0,\ \bar{z}>1$.
	\item Class E: configurations with $z,\bar{z}<0\ \mathrm{or}\ 0<z,\bar{z}<1\ \mathrm{or}\ z,\bar{z}>1\ \mathrm{or}\ z^*=\bar{z}$.
\end{itemize}
We use the name ``S" (resp. ``T", ``U") because it corresponds to the configurations where only the s-channel (resp. t-channel, u-channel) expansion has a chance to converge. The name ``E" means ``Euclidean", since the variables $z,\bar{z}$ in class E can be realized by the configurations with totally space-like separation. In addition, we divide the class E into four subclasses:
\begin{equation}\label{division:subclass}
	\begin{split}
		\mathrm{E}=\mathrm{E_{su}}\sqcup\mathrm{E_{st}}\sqcup\mathrm{E_{tu}}\sqcup\mathrm{E_{stu}},
	\end{split}
\end{equation}
where the subclasses are defined as follows.
\begin{itemize}
	\item Subclass $\mathrm{E_{su}}$: configurations with $z,\bar{z}<0$.
	\item Subclass $\mathrm{E_{st}}$: configurations with $0<z,\bar{z}<1$.
	\item Subclass $\mathrm{E_{tu}}$: configurations with $z,\bar{z}>1$.
	\item Subclass $\mathrm{E_{stu}}$: configurations with $z^*=\bar{z}$ not real.
\end{itemize} 
The subscripts in the above names indicate the possible convergent channels. Figure \ref{fig:zzbarclass} shows the range of $(z,\bar{z})$ pair corresponding to each class/subclass. Let $P\fr{\mathcal{C}}$ denote the subset of $(z,\bar{z})$ pairs corresponding to class/subclass $\mathcal{C}$. Under identification $(z,\bar{z})\sim(\bar{z},z)$, $P(\mathcal{C})$ are connected subsets of $\mathbb{C}/\bbZ_2$. $P(\mathrm{S}),P(\mathrm{T}),P(\mathrm{U}),P(\mathrm{E})$ are disconnected from each other, but $P\fr{\mathrm{E_{su}}}$, $P\fr{\mathrm{E_{st}}}$ and $P\fr{\mathrm{E_{tu}}}$ are connected to $P\fr{\mathrm{E_{stu}}}$ (also note that $P\fr{\mathrm{E_{su}}}$, $P\fr{\mathrm{E_{st}}}$ and $P\fr{\mathrm{E_{tu}}}$ are disconnected from each other).
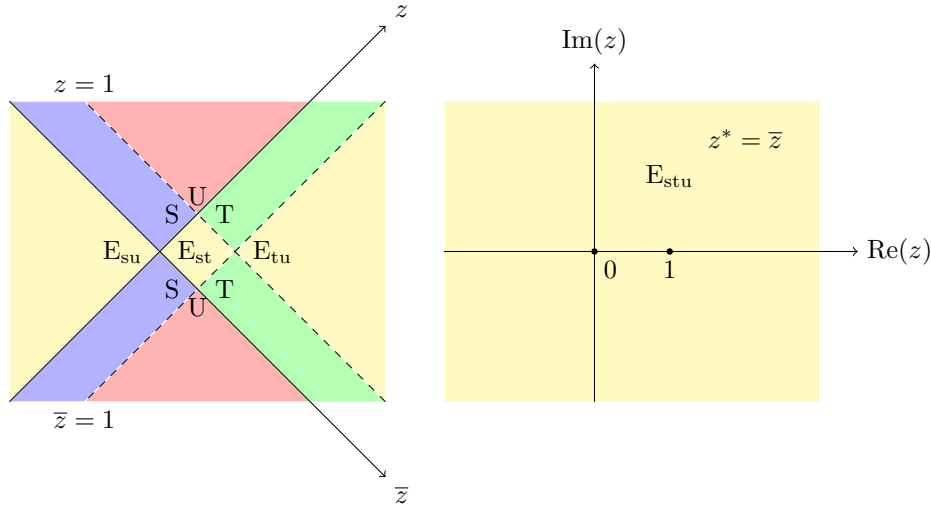
\begin{figure}[H]
	\centering
	\begin{tikzpicture}[baseline={(0,0)},circuit logic US]
		\draw[ultra thin, white, fill=blue!30] (0.5,0.5) -- (-1,2) -- (-2,2) -- (0,0) -- (0.5,0.5) node[anchor=east,black]{S$\ $};
		\draw[ultra thin, white, fill=blue!30] (0.5,-0.5) -- (-1,-2) -- (-2,-2) -- (0,0) -- (0.5,-0.5) node[anchor=east,black]{S$\ $};
		\draw[ultra thin, white, fill=green!30] (0.5,0.5) -- (1,0) -- (3,2) -- (2,2) -- (0.5,0.5) node[anchor=west,black]{$\ $T};
		\draw[ultra thin, white, fill=green!30] (0.5,-0.5) -- (1,0) -- (3,-2) -- (2,-2) -- (0.5,-0.5) node[anchor=west,black]{$\ $T};
		\draw[ultra thin, white, fill=red!30] (0.5,0.5) -- (2,2) -- (-1,2) -- (0.5,0.5) node[anchor=south,black]{U};
		\draw[ultra thin, white, fill=red!30] (0.5,-0.5) -- (2,-2) -- (-1,-2) -- (0.5,-0.5) node[anchor=north,black]{U};
		\draw[ultra thin, white, fill=yellow!30] (0,0) -- (-2,2) -- (-2,-2) -- (0,0) node[anchor=east,black]{$\mathrm{E_{su}}\ $};
		\draw[ultra thin, white, fill=yellow!30] (0,0) -- (0.5,0.5) -- (1,0) -- (0.5,-0.5) -- (0,0) node[anchor=west,black]{$\ \mathrm{E_{st}}$};
		\draw[ultra thin, white, fill=yellow!30] (1,0) -- (3,2) -- (3,-2) -- (1,0) node[anchor=west,black]{$\ \mathrm{E_{tu}}$};
		\draw[->] (-2,-2)--(3,3) node[anchor=south west]{$z$};
		\draw[->] (-2,2)--(3,-3) node[anchor=north west]{$\bar{z}$};
		\draw[dashed] (3,2)--(-1,-2) node[anchor=north]{$\bar{z}=1$};
		\draw[dashed] (3,-2)--(-1,2) node[anchor=south]{$z=1$};
	\end{tikzpicture}\quad
	\begin{tikzpicture}[baseline={(0,0)},circuit logic US]	
		\draw[ultra thin, white, fill=yellow!30] (-2,2) -- (3,2) -- (3,-2) -- (-2,-2) -- (-2,2);
		\draw[->] (-2,0)--(3.5,0) node[right]{Re$(z)$};
		\draw[->] (0,-2)--(0,2.5) node[above]{Im$(z)$};
		\filldraw[black] (0,0) circle (1pt) node[anchor=north west] {0};
		\filldraw[black] (1,0) circle (1pt) node[anchor=north] {1};	
		\draw (1,1) node {$\mathrm{E_{stu}}$};
		\draw (2,1.5) node {$z^*=\bar{z}$};
	\end{tikzpicture}  
	\caption{\label{fig:zzbarclass}The corresponding range of $(z,\bar{z})$ pair of each class/subclass.} 
\end{figure}
For each class/subclass, we immediately get some information about OPE convergence properties by theorem \ref{theorem:schannel}, \ref{theorem:tchannel} and \ref{theorem:uchannel} (see table \ref{table:class}).
\begin{table}[H]
	\setlength{\tabcolsep}{3mm} 
	\def\arraystretch{1.5} 
	\centering
	\begin{tabular}{|c|c|c|c|}
		\hline
		class/subclass  &  s-channel    &   t-channel   &   u-channel
		\\ \hline
		S  & \cmark  &  \xmark  &  \xmark
		\\ \hline
		T  & \xmark  &    &   \xmark  
		\\ \hline
		U  & \xmark  &  \xmark   &  
		\\ \hline
		$\mathrm{E_{st}}$  & \cmark  &    &  \xmark 
		\\ \hline
		$\mathrm{E_{su}}$  & \cmark  &  \xmark  &  
		\\ \hline
		$\mathrm{E_{tu}}$  & \xmark  &    &  
		\\ \hline
		$\mathrm{E_{stu}}$  & \cmark  &    &  
		\\ \hline
	\end{tabular}
	\caption{OPE convergence properties of classes/subclasses}
	\label{table:class}
\end{table}
In table \ref{table:class}, the check mark means that the sufficient conditions in theorem \ref{theorem:schannel} or \ref{theorem:tchannel} or \ref{theorem:uchannel} holds, hence the corresponding channel is convergent. The cross mark means that the sufficient conditions do not hold, we cannot conclude that the corresponding channel is convergent or not (basically because one or both $\rho,\bar{\rho}$ variables are on the unit circles). The blank means that there is room for convergence but we need to check $N_t,N_u$ conditions.

\section{Causal orderings}\label{section:causal}
In Minkowski space $\bbR{d-1,1}$, causal ordering is a binary relation between two arbitrary points. Let $x_1=(it_1,\mathbf{x}_1)$ and $x_2=(it_2,\mathbf{x}_2)$ be two points in $\bbR{d-1,1}$,\footnote{Since in this part of the thesis our discussions start from the Euclidean signature, we use the Euclidean coordinates $x=(\epsilon+it,\mathbf{x})$. The Euclidean points correspond to $t=0$ and the Lorentzian points correspond to $\epsilon=0$.} we say $x_1\rightarrow x_2$ if $x_2$ is in the open forward light-cone of $x_1$, or equivalently, $t_2-t_1>\abs{\mathbf{x}_1-\mathbf{x}_2}$. 

By the triangle inequality, the causal ordering is transitive: if $x_1\rightarrow x_2$ and $x_2\rightarrow x_3$, then $x_1\rightarrow x_3$. 

Causal orderings are preserved by translations, Lorentz transformations and dilatations. But special conformal transformations may violate causal orderings. Given a pair of time-like separated points $x_i,x_j$ in $\bbR{d-1,1}$, there exists a special transformation such that the images $x_i^\prime,x_j^\prime$ are space-like separated \cite{go1974properties}.

By ``the causal ordering of a configuration $C=(x_1,x_2,x_3,x_4)$", we will mean the directed graph $(V,E)$, where $V=\left\{1,2,3,4\right\}$ is the set of indices and $E=\left\{(ij)\right\}$ is the set of arrows $i\rightarrow j$ encoding the causal orderings $x_i\rightarrow x_j$. For example, the causal ordering of the configuration
\begin{equation}
	\begin{split}
		x_1=&(0,0,\ldots,0), \\
		x_2=&(i,0,\ldots,0), \\
		x_3=&(2i,0,\ldots,0), \\
		x_4=&(3i,0,\ldots,0), \\
	\end{split}
\end{equation}
is given by
\begin{equation}\label{graph:example}
	\begin{split}
		\begin{tikzpicture}
			\draw (0,0) node{$1$};
			\draw (2,0) node{$2$};
			\draw (2,-2) node{$3$};
			\draw (0,-2) node{$4$};
			\draw[-stealth] (0.5,0) -- (1.5,0);
			\draw[-stealth] (0,-0.5) -- (0,-1.5);
			\draw[-stealth] (0.5,-0.5) -- (1.5,-1.5);
			\draw[-stealth] (2,-0.5) -- (2,-1.5);
			\draw[-stealth] (1.5,-0.5) -- (0.5,-1.5);
			\draw[-stealth] (1.5,-2) -- (0.5,-2);
		\end{tikzpicture}
	\end{split}
\end{equation}
Since causal ordering is transitive, some arrows in the graph (\ref{graph:example}) are redundant and we will drop them. E.g. the graph
\begin{equation}\label{graph:simpleexaple}
	\begin{split}
		1\ \rightarrow\ 2\ \rightarrow\ 3\ \rightarrow\ 4
	\end{split}
\end{equation}
represents the same causal ordering as (\ref{graph:example}). For simplicity, we will use the graphic notation with the least number of arrows like (\ref{graph:simpleexaple}).

\section{Classifying convergent OPE channels}\label{section:classifyconfig}
We decompose the set $\mathcal{D}_L$ according to the causal orderings of the configurations:
\begin{equation}\label{Dl:decomp}
	\begin{split}
		\mathcal{D}_L=\bigsqcup_{\alpha}\mathcal{D}_L^{\alpha},
	\end{split}
\end{equation}
where each $\mathcal{D}_L^{\alpha}$ is the set of configurations with the same causal ordering, labelled by the index $\alpha$. 

\subsection{Case \texorpdfstring{$d\geq3$}{d>=3}}
In $d\geq3$, each $\mathcal{D}_L^{\alpha}$ in (\ref{Dl:decomp}) is a connected component of $\mathcal{D}_L$. It is not hard to see that different $\mathcal{D}_L^\alpha$ are disconnected to each other. The proof that each $\mathcal{D}_L^\alpha$ is connected is given in appendix \ref{appendix:connectedness}.

Since $\mathcal{D}_L^\alpha$ is connected, with the identification $(z,\bar{z})\sim(\bar{z},z)$, the set of corresponding $(z,\bar{z})$ pairs is a connected subset of $\mathbb{C}^2/\bbZ_2$. Recalling our classification in section \ref{section:classifyz}, we conclude that
\begin{lemma}\label{lemma:causaltoclass}
	For $d\geq3$, all configurations with the same causal ordering belong to the same class S, T, U, E (recall their definitions in section \ref{section:classifyz}).
\end{lemma}
By the lemma, we can assign class S, T, U and E to each causal ordering of the configurations. In addition, if $\mathcal{D}_L^\alpha$ is in class E, we subdivide $\mathcal{D}_L^\alpha$ according to the subclasses of class E. We summarize these relations in figure \ref{fig:class}.
\begin{figure}[H]
	\centering
	\begin{tikzpicture}
		\draw[thick,black] (0,0) ellipse (1 and 1.5);
		\draw[thick,black] (-0.866025,0.75) -- (0.866025,0.75); 
		\draw[black] (0,0.75) node[anchor=south]{$\mathcal{D}_L^{\alpha_1}$};
		\draw[thick,black] (-1,0) -- (1,0);
		\draw[black] (0,0) node[anchor=south]{$\mathcal{D}_L^{\alpha_2}$};
		\draw[black] (0,-0.4) node[circle,fill,inner sep=0.8pt]{};
		\draw[black] (0,-0.6) node[circle,fill,inner sep=0.8pt]{};
		\draw[black] (0,-0.8) node[circle,fill,inner sep=0.8pt]{};
		\draw[black] (0,-2) node{S};
		\draw[thick,black] (3,0) ellipse (1 and 1.5);
		\draw[thick,black] (2.13397,0.75) -- (3.866025,0.75); 
		\draw[black] (3,0.75) node[anchor=south]{$\mathcal{D}_L^{\alpha_3}$};
		\draw[thick,black] (2,0) -- (4,0);
		\draw[black] (3,0) node[anchor=south]{$\mathcal{D}_L^{\alpha_4}$};
		\draw[black] (3,-0.4) node[circle,fill,inner sep=0.8pt]{};
		\draw[black] (3,-0.6) node[circle,fill,inner sep=0.8pt]{};
		\draw[black] (3,-0.8) node[circle,fill,inner sep=0.8pt]{};
		\draw[black] (3,-2) node{T};
		\draw[thick,black] (6,0) ellipse (1 and 1.5);
		\draw[thick,black] (5.13397,0.75) -- (6.866025,0.75); 
		\draw[black] (6,0.75) node[anchor=south]{$\mathcal{D}_L^{\alpha_5}$};
		\draw[thick,black] (5,0) -- (7,0);
		\draw[black] (6,0) node[anchor=south]{$\mathcal{D}_L^{\alpha_6}$};
		\draw[black] (6,-0.4) node[circle,fill,inner sep=0.8pt]{};
		\draw[black] (6,-0.6) node[circle,fill,inner sep=0.8pt]{};
		\draw[black] (6,-0.8) node[circle,fill,inner sep=0.8pt]{};
		\draw[black] (6,-2) node{U};
		\draw[thick,black] (8,-1.5) rectangle (11,1.5);
		\draw[thick,black] (8,0) -- (11,0);
		\draw[thick,black] (9.5,-1.5) -- (9.5,1.5);
		\draw[thick,black] (9.5,0) circle (1) node[anchor=south east]{$\mathcal{D}_L^{\alpha_7}$};
		\draw[black] (8.5,1) node{$\mathrm{E_{st}}$};
		\draw[black] (10.5,1) node{$\mathrm{E_{su}}$};
		\draw[black] (8.5,-1) node{$\mathrm{E_{tu}}$};
		\draw[black] (10.5,-1) node{$\mathrm{E_{stu}}$};
		\draw[black] (9.5,-2) node{E};
	\end{tikzpicture}
	\caption{\label{fig:class}The class S, T, U and E are subdivided according causal orderings. For each $\mathcal{D}_L^\alpha$ in class E, $\mathcal{D}_L^\alpha$ is subdivided according to subclasses.}
\end{figure}
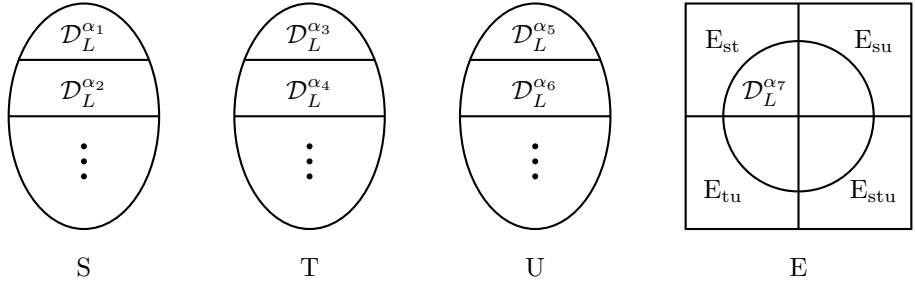

Now we are ready to state the classification of convergent OPE channels for Lorentzian CFT four-point functions. 
\begin{theorem}\label{theorem:classification}
	Let $G_4^L$ be the Lorentzian four-point function which is defined by the Wick rotation (\ref{def:Lorentz4ptfct}) from a Euclidean unitary CFT in $d\geq3$. Let $\alpha$ be a causal ordering and let $\mathcal{D}_L^\alpha$ be the set of all configurations with this causal ordering. 
	\begin{itemize}
		\item If $\mathcal{D}_L^\alpha$ is in class S, then all configurations in $\mathcal{D}_L^\alpha$ only have convergent s-channel expansion for $G_4^L$.
		\item If $\mathcal{D}_L^\alpha$ is in class T, then all configurations in $\mathcal{D}_L^\alpha$ have the same $N_t$.
		\item If $\mathcal{D}_L^\alpha$ is in class U, then all configurations in $\mathcal{D}_L^\alpha$ have the same $N_u$.
		\item If $\mathcal{D}_L^\alpha$ is in class E, then
		\begin{itemize}
			\item All configurations in $\mathcal{D}_L^\alpha\cap\mathrm{E_{st}}$ have the convergent s-channel expansion and the same $N_t$.
			\item All configurations in $\mathcal{D}_L^\alpha\cap\mathrm{E_{su}}$ have the convergent s-channel expansion and the same $N_u$.
			\item All configurations in $\mathcal{D}_L^\alpha\cap\mathrm{E_{tu}}$ have the same $N_t,N_u$.
			\item All configurations in $\mathcal{D}_L^\alpha\cap\mathrm{E_{stu}}$ have the convergent s-channel expansion and the same $N_t,N_u$.
		\end{itemize}
	\end{itemize}
\end{theorem}
\begin{proof}
	Let us check the conclusions case by case.
	
	Case 1: $\mathcal{D}_L^\alpha$ is in class S.
	
	The s-channel convergence follows from theorem \ref{theorem:schannel}. For other cases, the s-channel arguments are the same, and we will only focus on $N_t$ and $N_u$. 
	
	Case 2: $\mathcal{D}_L^\alpha$ is in class T.
	
	It remains to show that $N_t$ is a constant in $\mathcal{D}_L^\alpha$. For any $c_L,c_L^\prime\in\mathcal{D}_L^\alpha$, since $\mathcal{D}_L^\alpha$ is connected, there exists a path $\gamma_1$ which connects $c_L$ and $c_L^\prime$:
	\begin{equation}\label{path:DLalpha}
		\begin{split}
			\gamma_1&:\ [0,1]\ \longrightarrow\ \mathcal{D}_L^\alpha,\\
			\gamma_1&(0)=c_L,\quad\gamma_1(1)=c_L^\prime.\\
		\end{split}
	\end{equation}
	Since $\gamma_1(s)$ are always configurations in class T, the corresponding $z,\bar{z}$ never touch the interval $(-\infty,0)$. So $n_t(\gamma_1)=\bar{n}_t(\gamma_1)=0$, which implies $N_t(\gamma_1)=0$. On the other hand, given a path $\gamma_2$ from $\mathcal{D}_E\backslash\Gamma$ to $c_L$, we get a path from $\mathcal{D}_E\backslash\Gamma$ to $c_L^\prime$ by connecting $\gamma_1$ and $\gamma_2$. So we have
	\begin{equation}\label{proof:t}
		\begin{split}
			N_t(c_L^\prime)=N_t(\gamma_1)+N_t(\gamma_2)=N_t(c_L).
		\end{split}
	\end{equation}
	In other words, $N_t$ is a constant in $\mathcal{D}_L^\alpha$.
	
	Case 3: $\mathcal{D}_L^\alpha$ is in class U.
	
	It remains to show that $N_u$ is a constant in $\mathcal{D}_L^\alpha$. The argument is similar to case 2.
	
	Case 4: $\mathcal{D}_L^\alpha$ is in class E.
	
	Suppose $c_L,c_L^\prime$ are two configurations in $\mathcal{D}_L^\alpha\cap\mathrm{E_{st}}$. It remains to show that $N_t(c_L)=N_t(c_L^\prime)$. Analogously to case 2, there exists a path $\gamma_1$ satisfying the condition (\ref{path:DLalpha}), and it suffices to show that $N_t(\gamma_1)=0$. Here it is different from case 2 because $\gamma_1(s)$ may go through the other subclasses of the class E, and the curves of $z,\bar{z}$ may touch the interval $(-\infty,0)$. In class E, the curves $z(s),\bar{z}(s)$ touch the interval $(-\infty,0)$ only when $\gamma(s)$ enters the subclass $\mathrm{E_{su}}$. However, $\gamma(s)$, which starts from $\mathrm{E_{st}}$, must go through $\mathrm{E_{stu}}$ before entering $\mathrm{E_{su}}$. When $\gamma(s)$ leaves $\mathrm{E_{su}}$, it must go through $\mathrm{E_{stu}}$ again. Since in $\mathrm{E_{stu}}$, the variables $z,\bar{z}$ are complex conjugate to each other, the curves of $z,\bar{z}$ must cross $(-\infty,0)$ from opposite directions, e.g. see figure \ref{fig:case4}.
	\begin{figure}[H]
		\centering
		\includegraphics[scale=0.3]{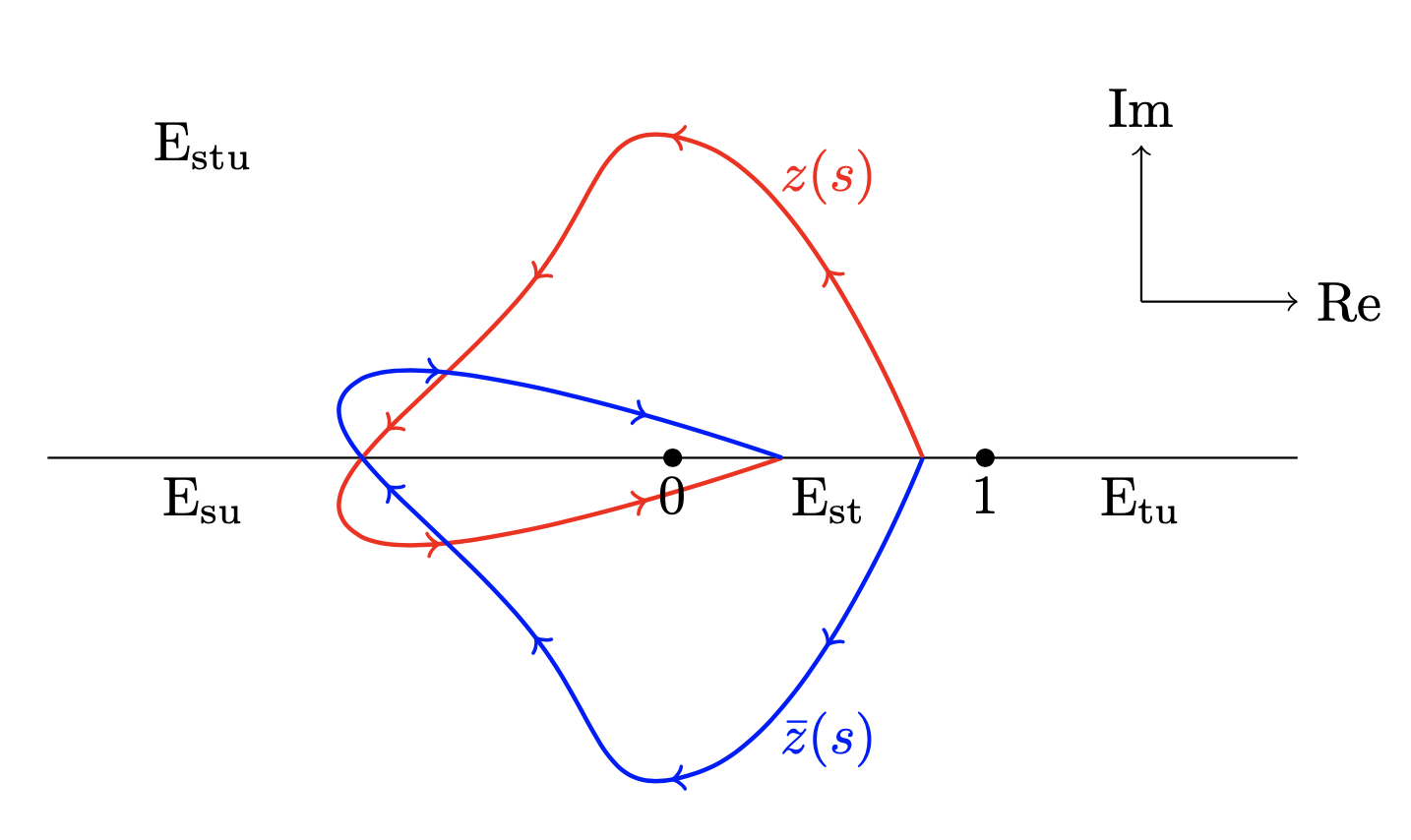}
		\caption{\label{fig:case4} An example of $z(s),\bar{z}(s)$ along $\gamma_1$ in case 4.}
	\end{figure}
	So we get
	\begin{equation}
		\begin{split}
			n_t(\gamma_1)=-\bar{n}_t(\gamma_1),
		\end{split}
	\end{equation}
	which implies $N_t(\gamma_1)=0$, hence $N_t(c_L)=N_t(c_L^\prime)$.
	
	The arguments for $\mathcal{D}_L^\alpha\cap\mathrm{E_{su}}$, $\mathcal{D}_L^\alpha\cap\mathrm{E_{tu}}$ and $\mathcal{D}_L^\alpha\cap\mathrm{E_{stu}}$ are similar.
\end{proof}
An immediate consequence of theorem \ref{theorem:classification} is that for a fixed causal ordering (say $\mathcal{D}_L^\alpha$), each blank space in table \ref{table:class} satisfy the all-or-none law: either check mark for all configurations in $\mathcal{D}_L^\alpha$ or cross mark for all configurations in $\mathcal{D}_L^\alpha$. Therefore, if $\mathcal{D}_L^\alpha$ is in class S or T or U, then all its configurations have the same OPE convergence properties; if $\mathcal{D}_L^\alpha$ is in class E, then all its configurations in the same subclass have the same OPE convergence properties.\footnote{By configurations having the same OPE convergence properties, we mean that in each OPE channel, all or none of these configurations have the convergent expansion for the four-point function.}

\subsection{Comments on the \texorpdfstring{2d}{} case}\label{section:comments2d}
In 2d unitary local CFTs, we have Al. Zamolodchikov's uniformizing variables $q,\bar{q}$ \cite{zamolodchikov1987conformal}. The function $g$ in eq.\,(\ref{def:Euclidean4-point}) has a convergent expansion in terms of Virasoro blocks, and Virasoro blocks have convergent series expansions in $q,\bar{q}$ if $0<\abs{q},\abs{\bar{q}}<1$, which includes the configurations with $0\leq\abs{\rho},\abs{\bar{\rho}}\leq1$ except for $\rho$ or $\bar{\rho}=\pm1$. However, $\rho$ or $\bar{\rho}=\pm1$ only happens at light-cone singularities.\footnote{If $\rho$ or $\bar{\rho}=1$, then $v=0$. If $\rho$ or $\bar{\rho}=-1$, then $u$ or $v=\infty$. Thus, for any configuration with $\rho\ or\ \bar{\rho}=\pm1$, there exists at least one $x_i,x_j$ pair such that $(x_i-x_j)^2=0$.} So we conclude that in the Lorentzian signature, the s-channel OPE is always convergent aside from light-cone singularities \cite{Maldacena:2015iua}.

The above CFT argument is valid only for 2d unitary local CFTs, where by local we mean there exists a stress tensor $T_{\mu\nu}(x)$, which has the mode expansion in Virasoro generators \cite{francesco1997conformal}. There are also non-local CFTs, e.g. the generalized free field theories. These non-local CFTs have only global conformal symmetry, for which we can only use $\rho,\bar{\rho}$ instead of $q,\bar{q}$. 

We claim that the conclusions in theorem \ref{theorem:classification} are still true for 2d unitary CFT (here we only assume global conformal symmetry). Unlike the case $d\geq3$, the sets $\mathcal{D}_L^\alpha$ are usually disconnected in 2d. This is because in 2d, there are two disconnected space-like separations. So we cannot copy the proof of theorem \ref{theorem:classification}. However, any 2d configuration can be embedded into $d\geq3$. Since our criteria of OPE convergence properties are based on counting how the analytic continuation curves of $z,\bar{z}$ cross the intervals $(-\infty,0)$, $(0,1)$ and $(1,+\infty)$, which is dimension independent, the 2d path gives the same counting of $N_t,N_u$ as in $d\geq3$. Therefore, theorem \ref{theorem:classification} also covers the 2d case.

The only little difference is that in the 2d case, the Lorentzian four-point configurations only have real $z,\bar{z}$. This follows from (\ref{def:coord2d}) and (\ref{zzbarglobal}). So the subclass $\mathrm{E_{stu}}$, where $z,\bar{z}$ are not real, does not exist in 2d.

\section{Time reversals}\label{section:timereversal}
In theorem \ref{theorem:classification}, we have classified the Lorentzian configurations in $\mathcal{D}_L$ into a finite number of cases. For each case, we will have to choose a representative configuration and a path from $\mathcal{D}_E\backslash\Gamma$, then check if conditions of theorem \ref{theorem:schannel}, \ref{theorem:tchannel} and \ref{theorem:uchannel} hold. Actually, there are some further simplifications which will reduce the number of checks to perform. We are going to show that different $\mathcal{D}_L^\alpha$ which are related by time reversals have the same convergent OPE channels. 

We define two time reversals:
\begin{equation}\label{def:timereversalpt}
	\begin{split}
		\theta_E:&\ (\epsilon+it,\mathbf{x}+i\mathbf{y})\mapsto(-\epsilon+it,\mathbf{x}-i\mathbf{y}) \\
		\theta_L:&\ (\epsilon+it,\mathbf{x}+i\mathbf{y})\mapsto(\epsilon-it,\mathbf{x}-i\mathbf{y}) \\
	\end{split}
\end{equation}
They correspond to the time reversals in Euclidean and Lorentzian space. Under time reversals, $x_{ij}^2$ takes its complex conjugate
\begin{equation}\label{prop:timereversal1}
	\begin{split}
		(\theta_E x_i-\theta_E x_j)^2=(\theta_L x_i-\theta_L x_j)^2=\left[(x_i-x_j)^2\right]^*
	\end{split}
\end{equation}
Given a configuration $c=(x_1,x_2,x_3,x_4)$, we define the time reversals of the configuration by (notice the change of order of points in $\theta_E c$)
\begin{equation}\label{def:timereversalconfig}
	\begin{split}
		\theta_Ec=&(\theta_Ex_4,\theta_Ex_3,\theta_Ex_2,\theta_Ex_1), \\
		\theta_Lc=&(\theta_Lx_1,\theta_Lx_2,\theta_Lx_3,\theta_Lx_4). \\
	\end{split}
\end{equation}
Then the following properties are easily checked:
\begin{itemize}
	\item The sets $\mathcal{T}_4$, $\mathcal{D}_E$ and $\mathcal{D}_L$ are preserved by $\theta_E$ and $\theta_L$.
	\item Under the transformation $c\mapsto\theta_E c$ or $c\mapsto\theta_L c$, the conformal invariants $u,v,z,\bar{z},\rho,\bar{\rho}$ become their complex conjugates.
\end{itemize}
Suppose we have a path $\gamma$ from $\mathcal{D}_E\backslash\Gamma$ to $\mathcal{D}_L$. Then $\theta_E\gamma$ and $\theta_L\gamma$ are still paths from $\mathcal{D}_E\backslash\Gamma$ to $\mathcal{D}_L$. The curves of $z,\bar{z}$ are reflected with respect to the real axis, which implies
\begin{equation}
	\begin{split}
		N_t(\theta_E\gamma),N_t(\theta_L\gamma)=-N_t(\gamma), \\
		N_u(\theta_E\gamma),N_u(\theta_L\gamma)=-N_u(\gamma). \\
	\end{split}
\end{equation}
By theorem \ref{theorem:schannel}, \ref{theorem:tchannel} and \ref{theorem:uchannel}, we conclude that 
\begin{itemize}
	\item Different Lorentzian configurations which are related by $\theta_E,\theta_L$ have the same convergent OPE channels.
\end{itemize}
By lemma \ref{lemma:causaltoclass} and theorem \ref{theorem:classification}, we translate the above results to the level of causal orderings:
\begin{itemize}
	\item If two different sets $\mathcal{D}_L^{\alpha},\mathcal{D}_L^{\beta}$ are related by $\theta_E,\theta_L$, then they belong to the same class (S, T, U, E).
	\item If two different sets $\mathcal{D}_L^{\alpha},\mathcal{D}_L^{\beta}$ are in class S or T or U and are related by $\theta_E,\theta_L$, then they have the same convergent OPE channels. 
	\item If two different sets $\mathcal{D}_L^{\alpha},\mathcal{D}_L^{\beta}$ are in class E and are related by $\theta_E,\theta_L$, then their intersections with each subclass have the same convergent OPE channels. 
	
\end{itemize}
Given a Lorentzian configuration $c=(x_1,x_2,x_3,x_4)$, $\theta_E$ interchanges $x_1\leftrightarrow x_4$ and $x_2\leftrightarrow x_3$. At the level of causal orderings, $\theta_E$ is the permutation of indices
\begin{equation}\label{action:thetaE}
	\begin{split}
		1\leftrightarrow4,\quad2\leftrightarrow3,    
	\end{split}
\end{equation}
with all the arrows kept fixed. For example, under $\theta_E$ we have
\begin{equation}
	\begin{split}
		\begin{tikzpicture}[baseline={(a.base)},circuit logic US]		
			\node (a) at (0,0) {$1$};
			\node (b) at (1,0) {$2$};
			\node (c) at (2,0.5) {$3$};
			\node (d) at (2,-0.5) {$4$};
			\draw[-stealth] (a) to (b);
			\draw[-stealth] (b) to (c);
			\draw[-stealth] (b) to (d);		
		\end{tikzpicture}\stackrel{\theta_E}{\Longrightarrow}
		\begin{tikzpicture}[baseline={(a.base)},circuit logic US]		
			\node (a) at (0,0) {$4$};
			\node (b) at (1,0) {$3$};
			\node (c) at (2,0.5) {$2$};
			\node (d) at (2,-0.5) {$1$};
			\draw[-stealth] (a) to (b);
			\draw[-stealth] (b) to (c);
			\draw[-stealth] (b) to (d);		
		\end{tikzpicture}.
	\end{split}
\end{equation}
Under $\theta_L$, the Lorentzian configuration $x_k=(it_k,\mathbf{x}_k)$ is mapped to $c^\prime=(x_1^\prime,x_2^\prime,x_3^\prime,x_4^\prime)$ with
\begin{equation}
	\begin{split}
		x_k^\prime=\theta_Lx_k=(-it_k,\mathbf{x}_k),\quad k=1,2,3,4.
	\end{split}
\end{equation}
So the operator ordering does not change but the causal ordering is reversed. For example, under $\theta_L$ we have
\begin{equation}
	\begin{split}
		\begin{tikzpicture}[baseline={(a.base)},circuit logic US]		
			\node (a) at (0,0) {$1$};
			\node (b) at (1,0) {$2$};
			\node (c) at (2,0.5) {$3$};
			\node (d) at (2,-0.5) {$4$};
			\draw[-stealth] (a) to (b);
			\draw[-stealth] (b) to (c);
			\draw[-stealth] (b) to (d);		
		\end{tikzpicture}\stackrel{\theta_L}{\Longrightarrow}
		\begin{tikzpicture}[baseline={(a.base)},circuit logic US]		
			\node (a) at (0,0) {$1$};
			\node (b) at (1,0) {$2$};
			\node (c) at (2,0.5) {$3$};
			\node (d) at (2,-0.5) {$4$};
			\draw[-stealth] (b) to (a);
			\draw[-stealth] (c) to (b);
			\draw[-stealth] (d) to (b);		
		\end{tikzpicture}.
	\end{split}
\end{equation}
By definitions (\ref{def:timereversalpt}) and (\ref{def:timereversalconfig}), we have the following properties for $\theta_E,\theta_L$:
\begin{equation}
	\begin{split}
		\theta_E^2=id,\quad\theta_L^2=id,\quad\theta_E\theta_L=\theta_L\theta_E. \\
	\end{split}
\end{equation}
So the group generated by $\theta_E,\theta_L$ is $\mathbb{Z}_2\times\mathbb{Z}_2$. Under the $\bbZ_2\times\bbZ_2$-actions, the orbit of a given causal ordering contains 1 or 2 or 4 causal orderings. In each orbit, it suffices to check the OPE convergence properties of only one causal ordering and make the same conclusions for other causal orderings. This simplifies our work.

\section{The table of four-point causal orderings}\label{section:tablecausal}
Given two Lorentzian configurations $(x_1,\ldots,x_n)$ and $(y_1,\ldots,y_n)$, we say that they are in the same causal type if there is a permutation $\sigma\in S_n$ such that $(x_{\sigma(1)},\ldots,x_{\sigma(n)})$ has the same causal ordering as $(y_1,\ldots,y_n)$ or $(\theta_Ly_1,\ldots,\theta_Ly_n)$. 

In table \ref{table:causalclassification}, we give a classification of four-point causal orderings according to the causal types. The vertices labelled by $a,b,c,d$ can be any permutation of $1,2,3,4$. In the end we will give one table about OPE convergence properties for each causal type in table \ref{table:causalclassification}.
\begin{table}[H]
	\caption{Classification of four-point causal orderings}
	\label{table:causalclassification}
	\setlength{\tabcolsep}{5mm} 
	\def\arraystretch{1.25} 
	\centering
	\begin{tabular}{|c|c|c|}
		\hline
		Type No.    &   causal ordering    &   $\theta_L$ time reversal
		\\ \hline
		1   &    \begin{tikzpicture}[baseline={(a.base)},circuit logic US]		
			\node (a) at (0,0) {$a$};
			\node (b) at (1,0) {$b$};
			\node (c) at (2,0) {$c$};
			\node (d) at (3,0) {$d$};
			\draw[-stealth] (a) to (b);
			\draw[-stealth] (b) to (c);
			\draw[-stealth] (c) to (d);	
		\end{tikzpicture}   & same
		\\ \hline
		2   &    \begin{tikzpicture}[baseline={(a.base)},circuit logic US]		
			\node (a) at (0,0) {$a$};
			\node (b) at (1,0) {$b$};
			\node (c) at (2,0.5) {$c$};
			\node (d) at (2,-0.5) {$d$};
			\draw[-stealth] (a) to (b);
			\draw[-stealth] (b) to (c);
			\draw[-stealth] (b) to (d);		
		\end{tikzpicture}   & \begin{tikzpicture}[baseline={(c.base)},circuit logic US]		
			\node (a) at (0,0.5) {$c$};
			\node (b) at (0,-0.5) {$d$};
			\node (c) at (1,0) {$b$};
			\node (d) at (2,0) {$a$};
			\draw[-stealth] (a) to (c);
			\draw[-stealth] (b) to (c);
			\draw[-stealth] (c) to (d);		
		\end{tikzpicture}
		\\ \hline
		3   &    \begin{tikzpicture}[baseline={(0,-0.35)},circuit logic US]		
			\node (a) at (0,0) {$a$};
			\node (b) at (1,0) {$b$};
			\node (c) at (2,0) {$c$};
			\node (d) at (1,-0.5) {$d$};
			\draw[-stealth] (a) to (b);
			\draw[-stealth] (b) to (c);
			\draw[-stealth] (a) to (d);		
		\end{tikzpicture}   & \begin{tikzpicture}[baseline={(0,-0.35)},circuit logic US]		
			\node (a) at (0,0) {$c$};
			\node (b) at (1,0) {$b$};
			\node (c) at (2,0) {$a$};
			\node (d) at (1,-0.5) {$d$};
			\draw[-stealth] (a) to (b);
			\draw[-stealth] (b) to (c);
			\draw[-stealth] (d) to (c);		
		\end{tikzpicture}
		\\ \hline
		4   &    \begin{tikzpicture}[baseline={(a.base)},circuit logic US]		
			\node (a) at (0,0) {$a$};
			\node (b) at (1,0.5) {$b$};
			\node (c) at (1,-0.5) {$c$};
			\node (d) at (2,0) {$d$};
			\draw[-stealth] (a) to (b);
			\draw[-stealth] (a) to (c);
			\draw[-stealth] (b) to (d);
			\draw[-stealth] (c) to (d);		
		\end{tikzpicture}   & same
		\\ \hline
		5   &    \begin{tikzpicture}[baseline={(a.base)},circuit logic US]		
			\node (a) at (0,0) {$a$};
			\node (b) at (1,0.5) {$b$};
			\node (c) at (1,0) {$c$};
			\node (d) at (1,-0.5) {$d$};
			\draw[-stealth] (a) to (b);
			\draw[-stealth] (a) to (c);
			\draw[-stealth] (a) to (d);		
		\end{tikzpicture}   & \begin{tikzpicture}[baseline={(b.base)},circuit logic US]		
			\node (a) at (0,0.5) {$b$};
			\node (b) at (0,0) {$c$};
			\node (c) at (0,-0.5) {$d$};
			\node (d) at (1,0) {$a$};
			\draw[-stealth] (a) to (d);
			\draw[-stealth] (b) to (d);
			\draw[-stealth] (c) to (d);		
		\end{tikzpicture}
		\\ \hline
		6   &    \begin{tikzpicture}[baseline={(0,-0.35)},circuit logic US]		
			\node (a) at (0,0) {$a$};
			\node (b) at (1,0) {$b$};
			\node (c) at (2,0) {$c$};
			\node (d) at (1,-0.5) {$d$};
			\draw[-stealth] (a) to (b);
			\draw[-stealth] (b) to (c);		
		\end{tikzpicture}   & same
		\\ \hline
		7   &    \begin{tikzpicture}[baseline={(0,-0.4)},circuit logic US]		
			\node (a) at (0,0) {$a$};
			\node (b) at (1,0.5) {$b$};
			\node (c) at (1,-0.5) {$c$};
			\node (d) at (0.5,-1) {$d$};
			\draw[-stealth] (a) to (b);
			\draw[-stealth] (a) to (c);		
		\end{tikzpicture}   & \begin{tikzpicture}[baseline={(0,-0.4)},circuit logic US]		
			\node (a) at (0,0.5) {$b$};
			\node (b) at (0,-0.5) {$c$};
			\node (c) at (1,0) {$a$};
			\node (d) at (0.5,-1) {$d$};
			\draw[-stealth] (a) to (c);
			\draw[-stealth] (b) to (c);		
		\end{tikzpicture}
		\\ \hline
		8   &    \begin{tikzpicture}[baseline={(0,-0.4)},circuit logic US]		
			\node (a) at (0,0) {$a$};
			\node (b) at (1,0.5) {$b$};
			\node (c) at (1,-0.5) {$d$};
			\node (d) at (0,-1) {$c$};
			\draw[-stealth] (a) to (b);
			\draw[-stealth] (a) to (c);
			\draw[-stealth] (d) to (c);		
		\end{tikzpicture}   & same
		\\ \hline
		9   &    \begin{tikzpicture}[baseline={(c.base)},circuit logic US]		
			\node (a) at (0,0) {$a$};
			\node (b) at (1,0) {$b$};
			\node (c) at (0.5,-0.5) {$c$};
			\node (d) at (0.5,-1) {$d$};
			\draw[-stealth] (a) to (b);		
		\end{tikzpicture}   & same
		\\ \hline
		10   &    \begin{tikzpicture}[baseline={(c.base)},circuit logic US]		
			\node (a) at (0,0) {$a$};
			\node (b) at (0,-1) {$b$};
			\node (c) at (1,-0.5) {$c$};
			\node (d) at (2,-0.5) {$d$};
			\draw[-stealth] (a) to (c);
			\draw[-stealth] (b) to (c);
			\draw[-stealth] (a) to (d);
			\draw[-stealth] (b) to (d);		
		\end{tikzpicture}   & same
		\\ \hline
		11   &    \begin{tikzpicture}[baseline={(0,-0.4)},circuit logic US]		
			\node (a) at (0,0) {$a$};
			\node (b) at (1,0) {$b$};
			\node (c) at (0,-0.5) {$c$};
			\node (d) at (1,-0.5) {$d$};
			\draw[-stealth] (a) to (b);
			\draw[-stealth] (c) to (d);		
		\end{tikzpicture}   & same
		\\ \hline
		12   &    \begin{tikzpicture}[baseline={(0,0)},circuit logic US]		
			\node (a) at (0,0) {$a$};
			\node (b) at (1,0) {$b$};
			\node (c) at (2,0) {$c$};
			\node (d) at (3,0) {$d$};	
		\end{tikzpicture}   & same
		\\ \hline
	\end{tabular}
\end{table}
Each causal type thus represents at most $4!\times2=48$ causal orderings (4! for possible assignments of $1,2,3,4\rightarrow a,b,c,d$ and $\times2$ for two columns). This maximal number is realized in type 3, while there are less causal orderings in other types because of symmetries of the graphs.\footnote{It may happen that the causal ordering of a configuration does not change under time reversals $\theta_E,\theta_L$ or permutations of the points. A typical example is the configuration whose points are totally space-like separated.}

often second column is equivalent to the first and because of little group (see appendix \ref{section:graphsymmetry}).

It makes sense to do this grouping of causal orderings into causal types for two reasons:
\begin{itemize}
	\item causal orderings related by $\theta_E$ and $\theta_L$ action (and which thus have same OPE convergence properties) belong to the same causal type.
	\item if we know class/subclass of $\mathcal{D}_L^\alpha$ for one $\alpha$ in a given causal type, it is easy to determine the class/subclass of any other $\mathcal{D}_L^\alpha$ in the same causal type (see appendix \ref{section:S4action}).
\end{itemize}

\section{Examples}\label{section:examples}
The tables which classify the OPE convergence properties will be particularly large, we leave them in appendix \ref{appendix:tableopeconvergence}. Readers can pick the cases they are interested in. To make it easy for readers to check, we also share the Mathematica code which contain the OPE convergence results of all causal orderings, see the auxiliary file on the arXiv webpage of \cite{Qiao:2020bcs}. In this section we only give some examples.

The Lorentzian four-point correlation functions defined in eq.\,(\ref{limit}) are either time-ordered ($t_1>t_2>t_4>t_4$) or out-of-time-order (not $t_1>t_2>t_4>t_4$). The time-ordered correlators have applications in scattering theories \cite{lehmann1955formulierung,peskin1995introduction}, and the out-of-time-order correlation functions have applications in the study of many-body systems  \cite{keldysh1965diagram,kitaev2014hidden,maldacena2016chaos,aleiner2016microscopic,bagrets2017power,garttner2017measuring,fan2017out,haehl2019classification}. An example in part \ref{part:crossratio} shows the existence of out-of-time-order correlators which do not have a convergent OPE channel (see appendix A in \cite{paper1}).\footnote{By ``a configuration do not have a convergent OPE channel" we mean the configuration do not satisfy the conditions of theorem \ref{theorem:schannel}, \ref{theorem:tchannel} and \ref{theorem:uchannel}.} Our first example is to show that not all time-ordered correlators have a convergent OPE channel. 

Then we will discuss two other examples from AdS/CFT. One is the Regge kinematics \cite{Cornalba:2007fs,Cornalba:2008qf}, the other is related to the bulk-point singularities \cite{Maldacena:2015iua}. 

\subsection{A time-orderd correlation function}
Let us consider the following two-dimensional configuration:
\begin{equation}\label{config:ex1}
	\begin{split}
		x_1=(0,0),\quad x_2=(-0.1i,1),\quad x_3=(-2i,-1.5),\quad x_4=(-2.1i,1.5).
	\end{split}
\end{equation}
The four-point function $G_4^L(x_1,x_2,x_3,x_4)$ at the configuration (\ref{config:ex1}) is time-ordered. The causal ordering of (\ref{config:ex1}) is given by
\begin{equation}\label{causal:ex1}
	\begin{split}
		\begin{tikzpicture}[baseline={(0,-0.4)},circuit logic US]		
			\node (a) at (0,0) {$4$};
			\node (b) at (1,0.5) {$2$};
			\node (c) at (1,-0.5) {$1$};
			\node (d) at (0,-1) {$3$};
			\draw[-stealth] (a) to (b);
			\draw[-stealth] (a) to (c);
			\draw[-stealth] (d) to (c);		
		\end{tikzpicture},
	\end{split}
\end{equation}
which is of causal type 8 in table \ref{table:causalclassification}. A quick way to know the OPE convergence property is to look up the table of OPE convergence in appendix \ref{appendix:type8}. The causal ordering (\ref{causal:ex1}) corresponds to the label ``(4231)" in table \ref{table:type8convergence}. We see from table \ref{table:type8convergence} that there is no convergent OPE channel for this causal ordering.

Let us also choose a start point in $\mathcal{D}_E\backslash\Gamma$ and a path to compute the $z,\bar{z}$-curves, and directly check the OPE convergence properties. Figure \ref{fig:timeorderedcurve} shows the $z,\bar{z}$-curves along the path.\footnote{We choose the start point $x_1^E=(0,-0.8)$, $x_2^E=(-1,-0.2)$, $x_3^E=(-2,-0.6)$ and $x_4^E=(-3,-0.3)$. The path is given by the straight line.}
\begin{figure}[H]
	\centering
	\includegraphics[scale=0.5]{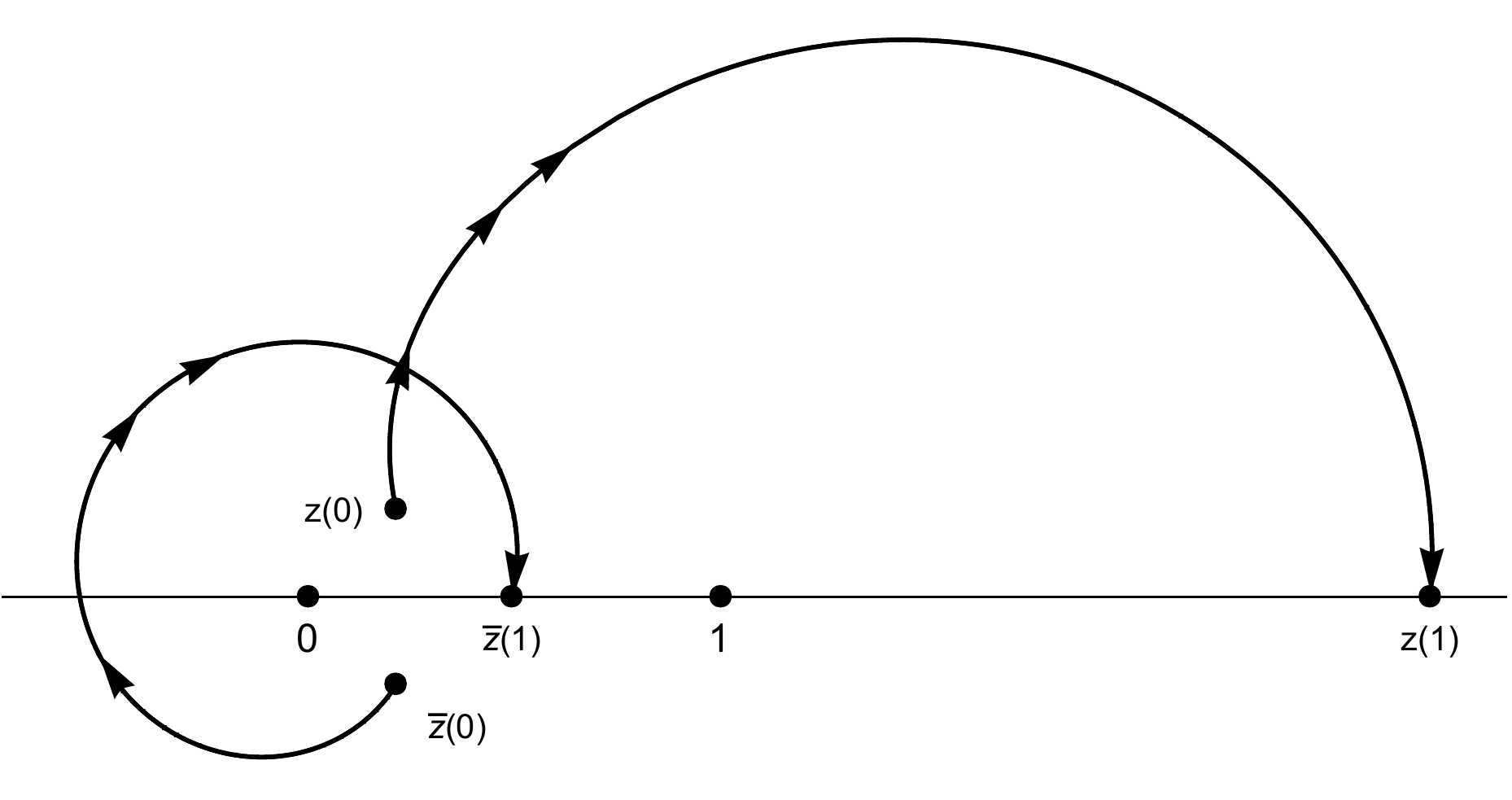}
	\caption{\label{fig:timeorderedcurve}$z,\bar{z}$-curves of the configuration (\ref{config:ex1}).}
\end{figure} 
We see from figure \ref{fig:timeorderedcurve} that $z>1$, $0<\bar{z}<1$ at the final point, which implies that the configuration (\ref{config:ex1}) is in class T. The curve of $z$ variable crosses $(-\infty,0)$ from below, which gives $N_t=-1$. So the t-channel OPE (the only undetermined case by table \ref{table:class}) is not convergent. Thus, as already mentioned, there is no convergent OPE channel for the four-point function at the configuration (\ref{config:ex1}). 

This example shows that not all time-ordered correlation functions have a convergent OPE channel.

\subsection{Regge kinematics} 
The second example is the Lorentzian four-point function in the Regge regime \cite{Cornalba:2007fs,Cornalba:2008qf,Costa:2012cb}. Let $x_1,x_4$ and $x_2,x_3$ pairs be time-like separated, while other pairs be space-like separated (see figure \ref{fig:regge}).
\begin{figure}[H]
	\centering
	\begin{tikzpicture}
		\draw (-2,-2) -- (2,2);
		\draw (-2,2) -- (2,-2);
		\filldraw[black] (-1.5,-1) circle (1pt) node[anchor=east] {$x_1$};
		\filldraw[black] (1,-0.8) circle (1pt) node[anchor=west] {$x_2$};
		\filldraw[black] (1.2,1) circle (1pt) node[anchor=west] {$x_3$};
		\filldraw[black] (-1.1,0.7) circle (1pt) node[anchor=east] {$x_4$};
		\draw[-stealth] (0,0) -- (0,1) node[anchor=south]{$t$};
		\draw[-stealth] (0,0) -- (1,0) node[anchor=west]{$x$};
	\end{tikzpicture}
	\caption{\label{fig:regge}Regge kinematics.}
\end{figure}
It is well known that the four-point function at Regge kinematics only has convergent t-channel expansion \cite{Cornalba:2007fs}. Here we just review this result. The causal ordering of the Regge kinematics is given by
\begin{equation}\label{causal:regge}
	\begin{split}
		\begin{tikzpicture}[baseline={(0,-0.4)},circuit logic US]		
			\node (a) at (0,0) {$1$};
			\node (b) at (1,0) {$4$};
			\node (c) at (0,-0.5) {$2$};
			\node (d) at (1,-0.5) {$3$};
			\draw[-stealth] (a) to (b);
			\draw[-stealth] (c) to (d);		
		\end{tikzpicture}
	\end{split}
\end{equation}
The Regge kinematics belongs to causal type 11 in table \ref{table:causalclassification}. Let us look up this causal ordering in appendix \ref{appendix:type11}. The causal ordering (\ref{causal:regge}) corresponds to the label ``(1423)" in table \ref{table:type11convergence}. We see that only t-channel OPE is convergent. 

We would like to also choose a representative configuration and a path to compute the curves of $z,\bar{z}$. The plot is given by figure \ref{fig:reggecurve}. \footnote{We choose the Euclidean configuration $x_1=0$, $x_2=(-1,0,0,0)$, $x_3=(-2,0.9,0,0)$, $x_4=(-4,0,0,0)$ and the representative Lorentzian configuration $y_1=0$, $y_2=(0,0,0.6,0)$, $y_3=(2i,0,0,0.7)$, $y_4=(2i,-0.05,0,-3)$. We choose the path to be the straight line between them.}
\begin{figure}[H]
	\centering
	\includegraphics[scale=0.7]{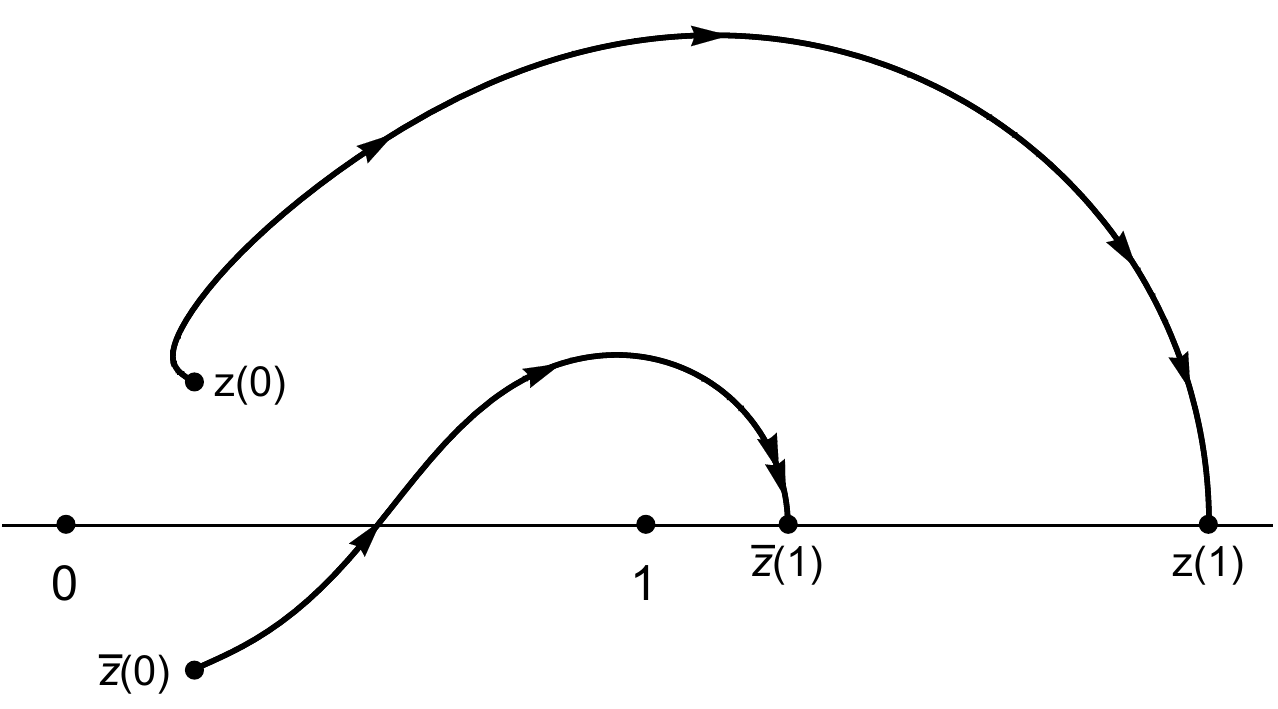}
	\caption{\label{fig:reggecurve}The plot of $z,\bar{z}$-curves of the Regge kinematics.}
\end{figure} 
We see from figure \ref{fig:reggecurve} that $z,\bar{z}>1$ at the final point,\footnote{The definition of $z,\bar{z}$ in \cite{Cornalba:2007fs} is different from this part of the thesis. In their work, $0<z,\bar{z}<1$ at Regge kinematics, while in this part, $z,\bar{z}>1$. One can compare the definitions and get the relation of $z,\bar{z}$ between \cite{Cornalba:2007fs} and our work: $z\rightarrow1/z$, $\bar{z}\rightarrow1/\bar{z}$.} which implies that the Regge kinematics is in class E. In fact the Regge kinematics can only be in the subclass $\mathrm{E_{tu}}$, where $z,\bar{z}>1$ \cite{Cornalba:2007fs}, so only t- and u-channel expansions have a chance to converge. We see from figure \ref{fig:reggecurve} that the $\bar{z}$-curve crosses $(0,1)$ from below, and $z,\bar{z}$-curves do not cross $(-\infty,0)$. So we get
\begin{equation}
	\begin{split}
		N_t=0,\quad N_u=-1,
	\end{split}
\end{equation}
which implies that only the t-channel expansion is convergent.

\subsection{Causal ordering of bulk-point singularities}\label{section:bulkptcausalorder}
The third example is as follows. Let $x_1,x_2$ and $x_3,x_4$ pairs be space-like separated. We put the $x_1,x_2$ pair in the open backward light-cone of some base point and $x_3,x_4$ pair in the open forward light-cone of the base point (see figure \ref{fig:lightconesing}).
\begin{figure}[H]
	\centering
	\begin{tikzpicture}
		\draw (-2,-2) -- (2,2);
		\draw (-2,2) -- (2,-2);
		\filldraw[black] (-1,-1.3) circle (1pt) node[anchor=north] {$x_1$};
		\filldraw[black] (1,-1.5) circle (1pt) node[anchor=north] {$x_2$};
		\filldraw[black] (-1,1.2) circle (1pt) node[anchor=south] {$x_3$};
		\filldraw[black] (1,1.3) circle (1pt) node[anchor=south] {$x_4$};
		\draw[-stealth] (0,0) -- (0,1) node[anchor=south]{$t$};
		\draw[-stealth] (0,0) -- (1,0) node[anchor=west]{$x$};
	\end{tikzpicture}
	\caption{\label{fig:lightconesing}Configuration of example 3.}
\end{figure}
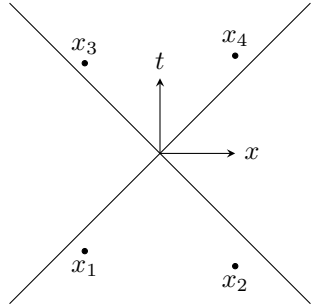
Such configurations have the causal ordering
\begin{equation}\label{causal:bulkpointsing}
	\begin{split}
		\begin{tikzpicture}[baseline={(c.base)},circuit logic US]		
			\node (a) at (0,0) {$1$};
			\node (b) at (0,-1) {$2$};
			\node (c) at (1,-0.5) {$3$};
			\node (d) at (2,-0.5) {$4$};
			\draw[-stealth] (a) to (c);
			\draw[-stealth] (b) to (c);
			\draw[-stealth] (a) to (d);
			\draw[-stealth] (b) to (d);		
		\end{tikzpicture}.
	\end{split}
\end{equation}
The causal ordering (\ref{causal:bulkpointsing}) is of causal type 10 in table \ref{table:causalclassification}. We look up the OPE convergence properties in appendix \ref{appendix:type10}. The causal ordering (\ref{causal:bulkpointsing}) corresponds to the label ``(1234)" in table \ref{table:type10convergence}. We see that this causal ordering is in class E, which has four subclasses. From table \ref{table:type10convergence} we also see that the configurations with the causal ordering (\ref{causal:bulkpointsing}) exist in each subclass. We wish to consider the subclass $\mathrm{E_{ut}}$, where $z,\bar{z}>1$. In table \ref{table:type10convergence}, we see that the configurations with causal ordering (\ref{causal:bulkpointsing}) and in subclass $\mathrm{E_{ut}}$ have no convergent OPE channels. 

Let us also choose a representative configuration to check this result. We want to remark that such case does not exist in 2d (see appendix \ref{appendix:type10} for the proof). We choose the following three-dimensional configuration
\begin{equation}\label{finalpoint:bulkptsing}
	\begin{split}
		x_1=(0,0,0),\quad x_2=(0,1,0),\quad x_3=(i,0.2,0.5),\quad x_4=(i,0.5,0.8).
	\end{split}
\end{equation}
Figure \ref{fig:bulkpointcurve} shows the plot of $z,\bar{z}$-curves.
\begin{figure}[H]
	\centering
	\includegraphics[scale=0.6]{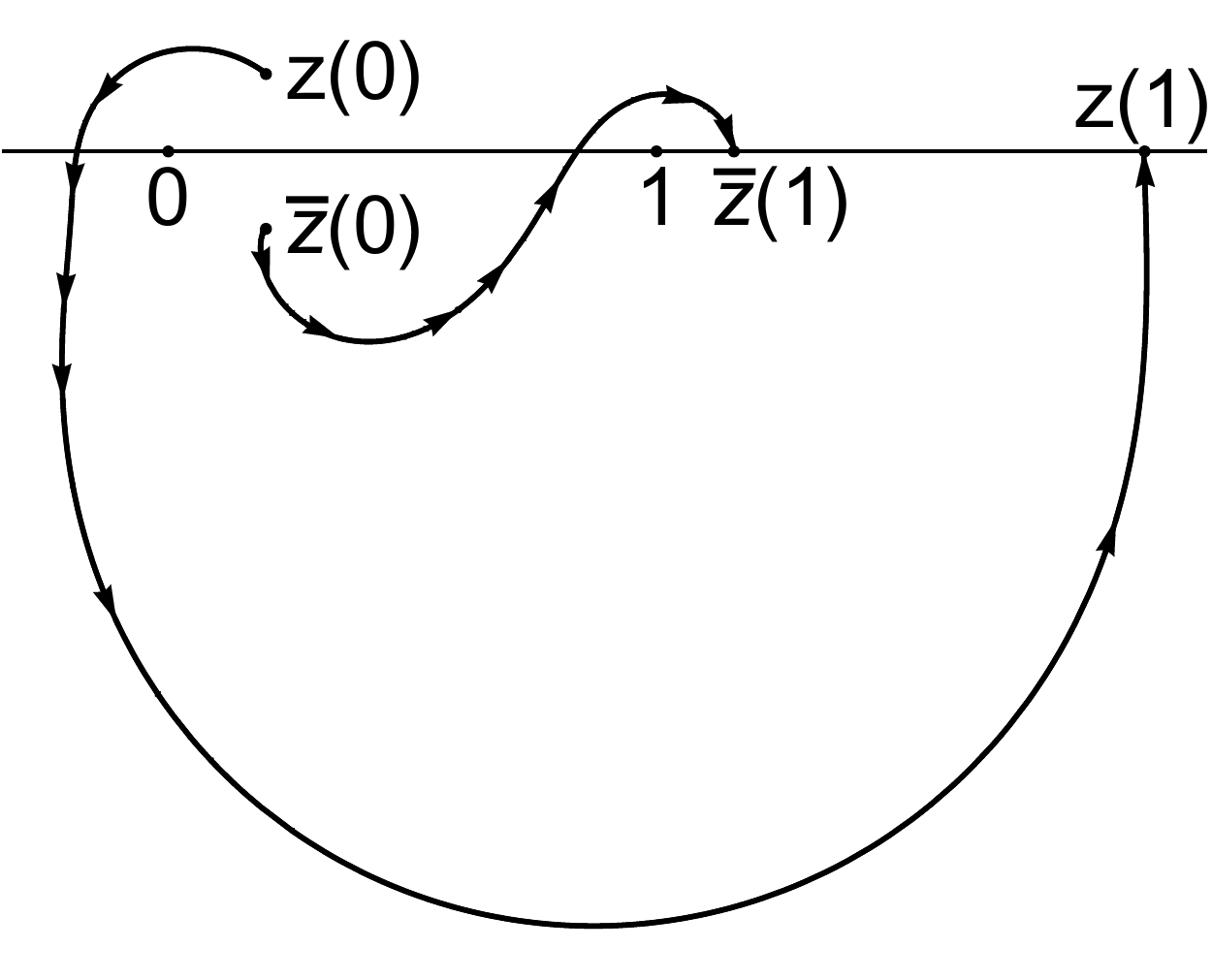}
	\caption{\label{fig:bulkpointcurve}The plot of $z,\bar{z}$-curves of the configuration (\ref{finalpoint:bulkptsing}).}
\end{figure} 
We see that along the path, $z$ crosses the interval $(-\infty,0)$ and $\bar{z}$ crosses the interval (0,1). We get
\begin{equation}\label{cylinder:NtNu}
	\begin{split}
		N_t=1,\quad N_u=-1.
	\end{split}
\end{equation}
which implies that the t- and u-channel expansions do not converge. 

We conclude that there is no convergent OPE channel for the causal ordering (\ref{causal:bulkpointsing}) with $z,\bar{z}>1$.

Here we give a hint why this example is related to the bulk-point singularities in AdS/CFT \cite{Maldacena:2015iua}. The bulk-point singularities are not exactly the configurations in Minkowski space $\bbR{d-1,1}$, instead they are configurations on the Minkowski cylinder $\bbR{}\times S^{d-1}$ \cite{Luscher:1974ez}. The Minkowski space can be embedded into a patch of the Minkowski cylinder in a Weyl equivalent way, this patch is called the Poincar\'e patch\cite{Aharony:1999ti}. The Minkowski cylinder also admits a causal ordering which is equivalent to the causal ordering of the Minkowski space in the Poincar\'e patch \cite{segal1971causally,todorov1973conformal}. One can show the following facts:
\begin{itemize}
	\item The bulk-point singularities have the causal ordering (\ref{causal:bulkpointsing}) and $z,\bar{z}>1$.
	\item One can find a path from an arbitrary bulk-point singularity to a configuration in the Poincar\'e patch, such that the path preserves the causal ordering (\ref{causal:bulkpointsing}) and dose not touch the light-cone singularities in the Minkowski cylinder.
	\item The CFT four-point function in the Poincar\'e patch is the same as the CFT four-point function in the Minkowski space up to a scaling factor.\footnote{The definition of the CFT four-point function on the Minkowski cylinder is similar to Minkowski space. We replace the planar time variables $\tau_k$ by the cylindrical time variables. Then do Wick rotations.} 
\end{itemize}
Based on the above facts, the OPE convergence properties of the bulk-point singularities are exactly the same as this example: there is no convergent OPE channel. We will revisit this case in section \ref{section:cylinderope}, where the similar classification results will be given for CFT four-point functions in the Minkowski cylinder. Our result does not contradict the two-dimensional result in \cite{Maldacena:2015iua} (see the beginning of section \ref{section:comments2d}) because here we only use the global conformal symmetry instead of the Virasoro symmetry.

\chapter{Generalization to non-identical scalar or spinning operators}\label{section:nonidscalar}
\section{Generalization to the case of non-identical scalar operators}
In this section we show that the arguments in section \ref{section:lorentz4pt} and \ref{section:classifylorentzconfig} are valid for CFT four-point functions of non-identical scalar operators. We start from a Euclidean CFT four-point function
\begin{equation}\label{def:4ptnonid}
	\begin{split}
		G_{1234}(c)=\braket{\mathcal{O}_1(x_1)\mathcal{O}_2(x_2)\mathcal{O}_3(x_3)\mathcal{O}_4(x_4)},\quad c=(x_1,x_2,x_3,x_4).
	\end{split}
\end{equation}
The scalar operators $\mathcal{O}_i$ in (\ref{def:4ptnonid}) have scaling dimensions $\Delta_i$. By conformal symmetry, (\ref{def:4ptnonid}) can be factorized as
\begin{equation}\label{nonidfactorize}
	\begin{split}
		G_{1234}(c)=&\dfrac{1}{\fr{x_{12}^2}^{\frac{\Delta_1+\Delta_2}{2}}\fr{x_{34}^2}^{\frac{\Delta_3+\Delta_4}{2}}}\fr{\dfrac{x_{24}^2}{x_{14}^2}}^{\frac{\Delta_{1}-\Delta_2}{2}}\fr{\dfrac{x_{14}^2}{x_{13}^2}}^{\frac{\Delta_{3}-\Delta_4}{2}}g_{1234}(\rho,\bar{\rho}) \\
	\end{split}
\end{equation}
As we discussed in chapter \ref{23-point}, the prefactor multiplying $g_{1234}(\rho,\bar{\rho})$ has analytic continuation to the forward tube $\mathcal{T}_4$. In section \ref{sec:generalcb} and \ref{sec:generalbounds}, we showed that $g_{1234}(\rho,\bar{\rho})$ is an analytic function (with branch cut) on the polydisc $\abs{\rho},\abs{\bar{\rho}}<1$. Using the same argument as in section \ref{anal4-point}, we composite $g_{1234}(\rho,\bar{\rho})$ and $(\rho(c),\bar{\rho}(c))$. This procedure performs the analytic continuation of $g_{1234}(c)$, as a function of $x_k$'s, to $\mathcal{T}_4$.

The remaining steps are the same as the case of identical scalar operators. Since our criteria of OPE convergence only concern the properties of cross-ratios, which are purely geometric, we conclude that for the case of four-point functions with non-identical scalar operators, the OPE convergence properties are the same as the case of identical scalar operators.

\section{Comments on the case of spinning operators}
Before finishing this section we want to make some comments on the case of four-point functions with spinning operators:
\begin{equation}\label{def:4ptnonidspinning}
	\begin{split}
		G_{1234}^{a_1a_2a_3a_4}(c)=\braket{\mathcal{O}_1^{a_1}(x_1)\mathcal{O}_2^{a_2}(x_2)\mathcal{O}_3^{a_3}(x_3)\mathcal{O}_4^{a_4}(x_4)},\quad c=(x_1,x_2,x_3,x_4).
	\end{split}
\end{equation}
where $\mathcal{O}_i^{a_i}$ are primary operators with scaling dimensions $\Delta_i$ and $SO(d)$-representation $\rho_i$. $a_i$ are the indices for the spin representations $\rho_i$. In the Euclidean signature, the Jacobian of any conformal transformation $f$ in $SO(1,d+1)$ can be factorized as
\begin{equation}
	\begin{split}
		J^\mu_\nu(x)\coloneqq\dfrac{\partial f^\mu(x)}{\partial x^\nu}=\Omega(x)\mathcal{R}^\mu_\nu(x),
	\end{split}
\end{equation}
where $\Omega(x)>0$ is a scaling factor and $\mathcal{R}$ is a rotation matrix. The four-point function $G_{1234}^{a_1a_2a_3a_4}$ is invariant if we replace all $\mathcal{O}_i^{a_i}(x)$ in (\ref{def:4ptnonidspinning}) with
\begin{equation}\label{conformaltransf:operator}
	\begin{split}
		\mathcal{O}_i^{a_i}(x)\rightarrow\Omega(x)^{\Delta_i}\left[\rho_i\fr{\mathcal{R}(x)^{-1}}\right]^{a_i}_{\ b_i}\mathcal{O}_i^{b_i}\fr{f(x)}.
	\end{split}
\end{equation}
If we choose the conformal transformation $f$ to be the one which maps $(x_1,x_2,x_3,x_4)$ to its $\rho,\bar{\rho}$-configuration ($x_1^\prime=\rho,x_2^\prime=-\rho,x_3^\prime=-1,x_4^\prime=1$ in the (01)-plane)
, then by conformal invariance we get
\begin{equation}\label{factorize:spinning}
	\begin{split}
		G_{1234}^{a_1a_2a_3a_4}(c)=&\dfrac{1}{\fr{x_{12}^2}^{\frac{\Delta_1+\Delta_2}{2}}\fr{x_{34}^2}^{\frac{\Delta_3+\Delta_4}{2}}}\fr{\dfrac{x_{24}^2}{x_{14}^2}}^{\frac{\Delta_{1}-\Delta_2}{2}}\fr{\dfrac{x_{14}^2}{x_{13}^2}}^{\frac{\Delta_{3}-\Delta_4}{2}} \\
		&\times T^{a_1a_2a_3a_4}_{b_1b_2b_3b_4}(\mathcal{R}_1,\mathcal{R}_2,\mathcal{R}_3,\mathcal{R}_4)g_{1234}^{b_1b_2b_3b_4}(\rho,\bar{\rho})
	\end{split}
\end{equation}
where $\mathcal{R}_k$ is the rotation matrix of $f$ at $x_k$, and $T$ is a function of rotation matrices, which is determined by the representations $\rho_i$ of the spinning operators $\mathcal{O}_i$.

The analytic continuation of the $x_{ij}^2$ prefactor is trivial. Analogously to the scalar case, the function $g_{1234}^{b_1b_2b_3b_4}(\rho,\bar{\rho})$ has series expansion (\ref{g:rhoexpansion}), which is convergent in the polydisc $\abs{\rho},\abs{\bar{\rho}}<1$. Then $g_{1234}^{b_1b_2b_3b_4}(\rho(c),\bar{\rho}(c))$, as a function of the four-point configuration $c$, has analytic continuation to the forward tube $\mathcal{T}_4$. 

The main difficulty is that there is the function $T\fr{\mathcal{R}_k}$ in (\ref{factorize:spinning}) because of the non-trivial representations of the spinning operators. If we think of the above conformal transformation $f$ as a conformal-group-valued function of $(x_1,x_2,x_3,x_4)$,\footnote{Since such a conformal transformation $f$ is not unique, the definition of this group-valued function depends on how we construct it.} then $\mathcal{R}_k$ are also functions of $(x_1,x_2,x_3,x_4)$. In general, the entries of $\mathcal{R}_k$ have singularities in $\mathcal{T}_4$, e.g. at configurations where $z=\bar{z}$ \cite{Kravchuk:2018htv,Karateev:2019pvw}. In a word, it is not obvious that $T(\mathcal{R}_k)$ in (\ref{factorize:spinning}) is under control. Some extra work is required for a good estimate on the object
\begin{equation*}
	\begin{split}
		T^{a_1a_2a_3a_4}_{b_1b_2b_3b_4}(\mathcal{R}_1,\mathcal{R}_2,\mathcal{R}_3,\mathcal{R}_4)g_{1234}^{b_1b_2b_3b_4}(\rho,\bar{\rho})
	\end{split}
\end{equation*}
In this part of the thesis we do not study the correlators of spinning operators. In chapter \ref{chap:spinning4pt} we will justify the analytic continuation of the spinning four-point functions for almost all four-point functions of spinning operators. Here we say ``almost all" because there will be a very small set of exceptional cases that we do not know how to deal with, see section \ref{section:spinexceptional}. For the spinning cases where we manage to justify the analytic continuation, the classification results will be the same as the scalar case. We leave this part of the generalization for future work.

\chapter{Conclusions and outlooks}\label{section:conclusion}
In this part of the thesis we studied the convergence properties of various OPE channels for Lorentzian CFT four-point functions of scalar operators in $d\geq2$, assuming global conformal symmetry. Our analysis is based on the convergence properties of OPE in the Euclidean unitary CFTs. We classified the Lorentzian four-point configurations according to their causal orderings and the range of the variables $z,\bar{z}$. The Lorentzian correlators are analytic functions in a neighbourhood of some Lorentzian configuration as long as there exists at least one convergent OPE channel in the sense of functions. We showed that the convergence properties of various OPE channels are fully determined by the causal orderings and the range of $z,\bar{z}$ of the four-point configurations. The CFT four-point functions are analytic in a very big domain, including configurations with totally space-like separations and configurations with some other causal orderings. All the results of OPE convergence properties are given in Appendix \ref{appendix:tableopeconvergence}.

Before ending, we would like to point out some related open questions. 
\begin{enumerate}
	\item We mainly used the radial coordinates $\rho,\bar{\rho}$ in our analysis. We have seen that by using the $q,\bar{q}$-variables in 2d, the domain of CFT four-point functions are larger than the domain derived by using the $\rho,\bar{\rho}$-variables. A natural question is
	\begin{itemize}
		\item For CFTs with only global conformal symmetry, are there any other coordinates which allow us to extend $G_4$ to some other domains which are not covered by using radial variables?
	\end{itemize}
	Our conjecture is that there are no such coordinates.
	\item Our results provide some safe Lorentzian regions where conformal bootstrap approach can be applied. One can use bootstrap equations to analyze the four-point functions at Lorentzian configurations with at least two convergent OPE channels. It is also interesting to play with crossing symmetry at Lorentzian configurations with
	\begin{itemize}
		\item One convergent OPE channel in the sense of functions, another one in the sense of distributions.
		\item Two convergent OPE channels in the sense of distributions.\footnote{The similar idea was proposed recently by Gillioz et al, see \cite{Gillioz:2019iye}, section 5.}
	\end{itemize}
	The above situations are closely related to the topics on analytic functional bootstrap when the functionals touch the boundaries of the regions with convergent OPE \cite{Mazac:2016qev,Mazac:2018mdx,Mazac:2018ycv,Paulos:2019gtx,Kaviraj:2018tfd,Mazac:2018biw,Mazac:2019shk}.
	\item There are also Lorentzian configurations with no convergent OPE channels. For these cases we do not know whether the general four-point correlators are genuine functions or not. We may need other techniques to handle these situations. For example, there are questions similar to appendix \ref{section:envelope}:
	\begin{itemize}
		\item One can derive a complex domain $\mathcal{D}^{stu}$ which is the union of the domains of three OPE channels. Then what is $\mathcal{D}^{stu}$ and what is its envelope of holomorphy $H\fr{\mathcal{D}^{stu}}$? Does $H\fr{\mathcal{D}^{stu}}$ contain more Lorentzian configurations than those provided by the results in this part of the thesis?
	\end{itemize}
	Once we are able to construct $H\fr{\mathcal{D}^{stu}}$, one can ask
	\begin{itemize}
		\item Given a Lorentzian configuration $c_L\in\mathcal{D}\backslash H\fr{\mathcal{D}^{stu}}$, can we find a CFT example such that the four-point function is divergent at $c_L$?
	\end{itemize}
	\item One can also consider higher-point correlation functions in CFTs. A natural question is:
	\begin{itemize}
		\item For $n\geq5$, what is the Lorentzian domain of $G_n$ in the sense of functions?
	\end{itemize}
\end{enumerate}

\end{part}

\begin{part}{Preview on some generalizations}\label{part:generalization}
\thispagestyle{fancy}

\chapter{Spinning four-point functions}\label{chap:spinning4pt}
This chapter is aiming for the generalization to four-point functions of general bosonic operators, i.e.\,operators with SO(d)-spins (fermionic operators are excluded since they carry half-integer Spin(d)-spins). We only focus on the temperedness property in the Lorentzian signature, because the derivation of the other Wightman axioms are similar to the scalar case. According to the strategy of the scalar case (see chapter \ref{strategy}), proving temperedness consists of two steps: analytically continuating the correlation function function to the forward tube and proving the power-law bound.

Let $\mathcal{O}^a$ denote a bosonic primary operator with SO(d)-spin indices $a$. Here we use the tensor indices $a = (\mu_1, \ldots, \mu_l)$, as described in section \ref{OSaxioms}. A general Euclidean CFT four-point function of bosonic primary operators has partial wave
expansion:
\begin{equation} \label{4pt:OPE}
	G_{1234}^{a_1 a_2 a_3 a_4} (c_E) \assign \langle \mathcal{O}_1^{a_1} (x_1)
	\mathcal{O}_2^{a_2} (x_2) \mathcal{O}_3^{a_3} (x_3) \mathcal{O}_4^{a_4}
	(x_4) \rangle = \underset{\mathcal{O}}{\sum} G_{1234, \mathcal{O}}^{a_1 a_2
		a_3 a_4} (c_E),
\end{equation}
where $c_E = (x_1, x_2, x_3, x_4)$ is the Euclidean configuration and the sum over the primary operators that appear in the OPE. According to the OPE convergence assumption in the Euclidean CFT axioms, the expansion (\ref{4pt:OPE}) is convergent as long as there exists a sphere $S^{d - 1}$ which separates the $x_1, x_2$ pair from the $x_3, x_4$ pair, e.g.\,see figure \ref{fig:sphereseparation}.
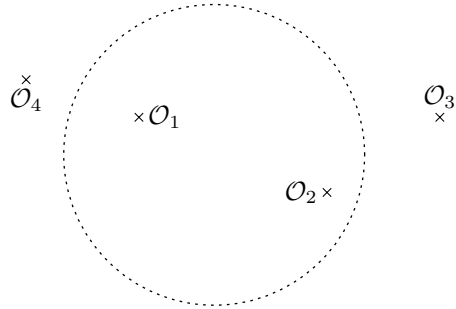
\begin{figure}[ht]
	\centering
	\begin{tikzpicture}
		\draw [line width=.5pt,dash pattern=on 1pt off 2pt]circle(2 cm);
		\coordinate (x1) at (-1,0.5);
		\coordinate (x2) at (1.5,-0.5);
		\coordinate (x3) at (3,0.5);
		\coordinate (x4) at (-2.5,1);
		\draw (x1) node[cross] {};
		\draw (x2) node[cross] {};
		\draw (x3) node[cross] {};
		\draw (x4) node[cross] {};
		\draw (x1) node[right] {$\mathcal{O}_1$};
		\draw (x2) node[left] {$\mathcal{O}_2$};
		\draw (x3) node[above] {$\mathcal{O}_3$};
		\draw (x4) node[below] {$\mathcal{O}_4$};
	\end{tikzpicture}
    \caption{\label{fig:sphereseparation} An example of a sphere separating $\mathcal{O}_1,\mathcal{O}_2$ from $\mathcal{O}_3,\mathcal{O}_4$.}
\end{figure}

Consider the configurations with complex coordinates: $x_k =
(\varepsilon_k + i t_k, \mathbf{x}_k + i\mathbf{y}_k)$. We would like to
analytically continue the four-point function to the forward tube $\mathcal{T}_4$ (i.e.\,$\varepsilon_k-\varepsilon_{k+1}>\abs{\mathbf{y}_k-\mathbf{y}_{k+1}}$, $k=1,2,3$).
The analytic continuation and the power-law bound for the scalar case were done in section \ref{anal4-point} and \ref{power4-point}. We would like to briefly review the key strategies in the scalar
case, and then see the extra subtleties that arise in the case of spinning
operators.

\section{Revisit the scalar case}\label{section:scalarrevisit}
Given any configuration $c = (x_1, x_2, x_3, x_4)$ in $\mathcal{T}_4$, we can
always find a conformal transformation $f$ which maps $c$ to the
``$\rho$-configuration'' $c_{\rho} = (x_1', x_2', x_3', x_4')$ with
\begin{equation}\label{config:rho}
	x_1' = - x_2' = \left( \frac{\rho + \bar{\rho}}{2}, \frac{\rho -
		\bar{\rho}}{2 i}, 0, \ldots, 0 \right), \quad x_3' = - x_4' = (- 1, 0,
	\ldots, 0) .
\end{equation}
For scalar 4pt functions, conformal invariance implies
\begin{equation}
	G_{1234} (c) = \Omega_f (x_1)^{\Delta_1} \Omega_f (x_2)^{\Delta_2} \Omega_f
	(x_3)^{\Delta_3} \Omega_f (x_4)^{\Delta_4} G_{1234} (c_{\rho}),
	\label{confinv:scalar}
\end{equation}
where $\Omega_f (x) = \det (\partial f^{\mu} / \partial x^{\nu})^{1 / d}$.  This formula is just a writting of eq.\,(\ref{def:Eucl4-pointgeneral}) in section \ref{nonId}. These two equations are matched by
\begin{equation}\label{nonidscalar:matching}
	\begin{split}
		G_{1234}(c_\rho)=\dfrac{1}{2^{\Delta_1+\Delta_2+\Delta_3+\Delta_4}(\rho\bar{\rho})^{\frac{\Delta_1+\Delta_{2}}{2}}}\fr{\dfrac{(1+\rho)(1+\bar{\rho})}{(1-\rho)(1-\bar{\rho})}}^{\frac{\Delta_1-\Delta_2-\Delta_3+\Delta_4}{2}}g_{1234}(c_E).
	\end{split}
\end{equation}
In section \ref{nonId}, the analytic continuation of $G_{1234}(c)$ relies on the fact that both the $x_{ij}^2$-prefactor and $g_{1234}(c)$ has the analytic continuation as a function of $c$. We would like to reinterpret the analytic continuation procedure, using the following facts: under the constraints $f (x_k) = x_k'$ ($k = 1, 2, 3,
4$), the scale factors $\Omega_f (x_k)$ do not depend on the choice of $f$ and
they are analytic in Cartesian coordinates of $c$. To see this we use the
formula
\begin{equation}
	{x_{i j}'}^2 = \Omega_f (x_i) \Omega_f (x_j) x_{i j}^2,
\end{equation}
which implies
\begin{equation}\label{eq:Omega}
	\begin{split}
		\Omega_f (x_1)^2 & = \frac{x_{34}^2}{x_{13}^2 x_{14}^2} \times \frac{(1 -
			\rho^2) (1 - \bar{\rho}^2)}{4}, \\
		\Omega_f (x_2)^2 & = \frac{x_{34}^2}{x_{23}^2 x_{24}^2} \times \frac{(1 -
			\rho^2) (1 - \bar{\rho}^2)}{4},  \\
		\Omega_f (x_3)^2 & = \frac{x_{12}^2}{x_{13}^2 x_{23}^2} \times \frac{(1 -
			\rho^2) (1 - \bar{\rho}^2)}{4 \rho \bar{\rho}},  \\
		\Omega_f (x_4)^2 & = \frac{x_{12}^2}{x_{14}^2 x_{24}^2} \times \frac{(1 -
			\rho^2) (1 - \bar{\rho}^2)}{4 \rho \bar{\rho}} . 
	\end{split}
\end{equation}
We see from (\ref{eq:Omega}) that $\Omega (x_k)$'s only depend on $c$ and they are analytic functions on
$\mathcal{T}_4$.\footnote{The argument for analyticity is as follows.
	First, it is obvious that the $x_{ij}^2$ factors are analytic
	functions on $\mathcal{T}_4$. The analyticity of the $\rho,
	\bar{\rho}$ factors follows from the facts that $\rho, \bar{\rho}$ always appear symmetrical and $0 < | \rho |, | \bar{\rho} | < 1$ on
	$\mathcal{T}_4$.} So we will use the notation
\begin{equation}
	\Omega_k (c) \assign \Omega_f (x_k) .
\end{equation}
On the other hand, $G_{1234} (c_{\rho})$ has convergent series expansion in
$\rho, \bar{\rho}$, which defines an analytic function on $\mathcal{T}_4$.
Therefore, the RHS of (\ref{confinv:scalar}) is a product of five analytic functions on $\mathcal{T}_4$, so it performs the analytic
continuation of $G_{1234}$ to $\mathcal{T}_4$.

Furthermore, we can show that $\Omega_k (c)$ and $G_{1234} (c_{\rho})$ have
power-law bounds when $c \in \mathcal{T}_4$. So $G_{1234} (c)$, as a product of these factors, also has a power-law bound on $\mathcal{T}_4$. Then by Vladimirov's theorem (theorem \ref{ThVlad}), $G_{1234}$ is a tempered distribution in the Lorentzian regime.

\section{Subtlety in the spinning case, strategy}\label{section:difficulty}
Now let us turn to the spinning case. The spinning-field analog of eq.\ (\ref{confinv:scalar}) is given by
\begin{equation}\label{confinv:spinning}
	G_{1234}^{a_1 a_2 a_3 a_4} (c) = \left( \underset{i = 1}{\overset{4}{\prod}}
	\Omega_i (c)^{\Delta_i} \rho_i (R_f (x_i)^{- 1})^{a_i}_{\quad b_i} \right)
	G^{b_1 b_2 b_3 b_4}_{1234} (c_{\rho}), 
\end{equation}
where $\Omega_i (c)$ is the scale factor and $(R_f)^{\mu}_{\ \nu} (x)
\assign \Omega_f (x_i)^{- 1} \partial f^{\mu} / \partial x^{\nu}$ is the
rotation matrix. For $a_i = (\mu_1, \ldots, \mu_{l_i})$ and $b_i = (\nu_1,
\ldots, \nu_{l_i})$, the matrix $\rho_i (R)^{a_i}_{\quad b_i}$ is the tensor
product of rotation matrices
\begin{equation}\label{rho:Rproduct}
	\rho_i (R)^{a_i}_{\quad b_i} = R^{\mu_1}_{\quad \nu_1} R^{\mu_2}_{\quad
		\nu_2} \ldots R^{\mu_{l_i}}_{\quad \nu_{l_i}} .
\end{equation}
As reviewed in the previous section, $\Omega_k (c)^{\Delta_k}$ and $G^{b_1 b_2 b_3
	b_4}_{1234} (c_{\rho} (c))$ in eq.\,(\ref{confinv:spinning}) are
analytic functions on $\mathcal{T}_4$. However, in $d\geqslant3$, the rotation matrices $R_f(x_i)$'s are not analytic in $c$, mainly because the conformal transformation $f$ has SO$(d-2)$ or SO$(d-1)$ ambiguity.\footnote{By eq.\,(\ref{config:rho}), $c_\rho$ stays in the two-dimensional plane when $\rho\neq\bar{\rho}$, or in the one-dimensional line when $\rho=\bar{\rho}$. It is invariant under rotations $g$ of the other $d-2$ or $d-1$ dimensions. Therefore, any such $g\circ f$ maps $c$ to $c_\rho$.} So we do not know how to perform the analytic continuation using the
RHS of eq.\,(\ref{confinv:spinning}).

In $d=2$, both SO$(d-2)$ and SO$(d-1)$ are trivial groups. So given a four-point configuration $c$, there is a unique conformal transformation $f$ which maps $c$ to its corresponding $\rho$-configuration $c_\rho$. Using proper coordinates to characterize $f$, one can show that $f$ is an analytic function of $c$ in the forward tube, so the rotation matrices induced by $f$ are also analytic functions of $c$. Then we are able to use eq.\ (\ref{confinv:spinning}) to construct the analytic continuation of the four-point function, and show its power-law bound in the two-dimensional case (see appendix \ref{appendix:2dconformal} for details).

In summary, eq.\,(\ref{confinv:spinning}) is still available for the analytic continuation in $d=2$, but no longer useful in $d\geqslant3$. For the analytic continuation in $d\geqslant3$, we would like to take a different strategy:
\begin{itemize}
	\item[]\textbf{Step 1.} Show directly that each conformal partial wave $G_{1234,
		\mathcal{O}}^{a_1 a_2 a_3 a_4} (c)$ on the RHS of (\ref{4pt:OPE}) has
	analytic continuation to $\mathcal{T}_4$.
	
	\item[]\textbf{Step 2.} Show that the partial wave expansion (\ref{4pt:OPE}) is convergent in $\mathcal{D}_4
	\subset \mathcal{T}_4$. Recall that $\mathcal{D}_4$ is defined to be the
	$\mathcal{T}_4$ configurations with real spatial components.\footnote{We consider $\mathcal{D}_4$ instead of $\mathcal{T}_4$ because the reflection-symmetric configurations in $\mathcal{D}_4$ can be mapped into the two-dimensional subspace in a nice way, which will be useful to prove convergence of the partial wave expansion (\ref{4pt:OPE}).}
\end{itemize}
The main point here is that the non-analyticity of the rotation matrices are packaged into the conformal partial waves, so we can just bypass this difficulty. We will see that once step 1 is done, we can reduce step 2 to the case of $d=2$, where everything is already well-established.

\section{Step 1: analytic continuation of the conformal partial wave}
We would like to use the following integral representation to perform the analytic continuation of the conformal partial wave in $x_i$'s, inspired by Mack (\cite{Mack:1976pa}, eq.\ (8.7)):
\begin{equation}\label{partialwave:integral}
	\begin{split}
			G_{1234, \mathcal{O}}^{a_1 a_2 a_3 a_4} (c) = \int \frac{d^d p}{(2 \pi)^d}
			(\hat{G}_{\mathcal{O}})_{a b} (p) \hat{B}_{\mathcal{O},12}^{a\, a_1 a_2} (p,x_1, x_2) \overline{\hat{B}_{\mathcal{O},\bar{4}\, \bar{3}}^{b\, a_4 a_3} (p,x_4^{\theta}, x_3^{\theta})}.
	\end{split}
\end{equation}
Here $(\hat{G}_{\mathcal{O}})_{a b} (p)$ is the Fourier transform of the two-point function
\begin{equation}\label{2ptmomentum}
	\begin{split}
		(\hat{G}_{\mathcal{O}})_{a b} (p) = \int d^d x \langle
		\mathcal{O}_a^{\dag} (0) \mathcal{O}_b (x) \rangle e^{i p
			\cdummy x} , \quad \fr{x=(it,\mathbf{x}),\,p=(E,\mathbf{p}),\,x\cdot p=-Et+\mathbf{p}\cdot\mathbf{x},}\\
	\end{split}
\end{equation}
the function $\hat{B}_{\mathcal{O},ij}^{a\,a_i a_j} (p,x_i, x_j)$ is the OPE kernel in the momentum space:\footnote{In \cite{Mack:1976pa}, Mack formulated the operator product expansion in the Lorentzian position space:
\begin{equation}
	\begin{split}
		\mathcal{O}_i(x_i)\mathcal{O}_j(x_j)|0\rangle=\sum\limits_{\mathcal{O}}\int\dfrac{d^dx}{(2\pi)^d}\,B_{\mathcal{O},ij}^{a\,a_i a_j} (x,x_i, x_j)\mathcal{O}^\dagger_a(x)|0\rangle.
	\end{split}
\end{equation}
This is the reason why we use the notation $\hat{B}$ in the momentum space. In fact we have already mentioned it in section \ref{MackComp}.}
\begin{equation}\label{OPEkernal:O}
	\begin{split}
		[\mathcal{O}_i(x_i)\mathcal{O}_j(x_j)]_\mathcal{O}=\int\dfrac{d^dp}{(2\pi)^d}\,\hat{B}_{\mathcal{O},ij}^{a\,a_i a_j} (p,x_i, x_j)\hat{\mathcal{O}}^\dagger_a(-p),
	\end{split}
\end{equation}
and the complex version of the reflection operator $\theta$ is defined by
\begin{equation}
	\begin{split}
		x=(\varepsilon+i t,\mathbf{x}+i\mathbf{y})\rightarrow x^\theta=(-\varepsilon+i t,\mathbf{x}-i\mathbf{y}).
	\end{split}
\end{equation}
Regardless of the convergence issue, one can formally check that (\ref{partialwave:integral}) should be correct:
\begin{equation}\label{partialwave:formalopecheck}
	\begin{split}
		G_{1234, \mathcal{O}}^{a_1 a_2 a_3 a_4} (c)
		=&\int\dfrac{d^dp}{(2\pi)^d}\,\hat{B}_{\mathcal{O},12}^{a\,a_1 a_2} (p,x_1, x_2)\langle\hat{\mathcal{O}}^\dagger_a(-p)\mathcal{O}_3^{a_3}(x_3)\mathcal{O}_4^{a_4}(x_4)\rangle \qquad\mathrm{by}\,(\ref{OPEkernal:O}) \\
		=&\int\dfrac{d^dp}{(2\pi)^d}\,\hat{B}_{\mathcal{O},12}^{a\,a_1 a_2} (p,x_1, x_2)\overline{\langle\hat{\mathcal{O}}_a(p)\mathcal{O}_3^{\dagger a_3}(x_3^\theta)\mathcal{O}_4^{\dagger a_4}(x_4^\theta)\rangle} \qquad \mathrm{by}\,(\ref{CFThermiticity})\\
		=&\int\dfrac{d^dp\,d^dq}{(2\pi)^{2d}}\,\hat{B}_{\mathcal{O},12}^{a\,a_1 a_2} (p,x_1, x_2)\overline{\hat{B}_{\mathcal{O},\bar{4}\, \bar{3}}^{b\,a_4 a_3} (q,x_4^{\theta}, x_3^{\theta})}\overline{\langle\hat{\mathcal{O}}_a(p)\hat{\mathcal{O}}^\dagger_b(-q)\rangle} \qquad\mathrm{by}\,(\ref{OPEkernal:O}) \\
		=&\int\dfrac{d^dp\,d^dq}{(2\pi)^{2d}}\,\hat{B}_{\mathcal{O},12}^{a\,a_1 a_2} (p,x_1, x_2)\overline{\hat{B}_{\mathcal{O},\bar{4}\, \bar{3}}^{b\,a_4 a_3} (q,x_4^{\theta}, x_3^{\theta})}\langle\hat{\mathcal{O}}^\dagger_a(-p)\hat{\mathcal{O}}_b(q)\rangle \qquad \mathrm{by}\,(\ref{hermicity2-point})\\
		=&\int\dfrac{d^dp\,d^dq}{(2\pi)^{2d}}\,\hat{B}_{\mathcal{O},12}^{a\,a_1 a_2} (p,x_1, x_2)\overline{\hat{B}_{\mathcal{O},\bar{4}\, \bar{3}}^{b\,a_4 a_3} (q,x_4^{\theta}, x_3^{\theta})}(\hat{G}_{\mathcal{O}})_{a b} (q) (2\pi)^d\delta(p-q) \qquad\mathrm{by}\,(\ref{2ptmomentum})\\
		=&\int \frac{d^d p}{(2 \pi)^d}
		(\hat{G}_{\mathcal{O}})_{a b} (p) \hat{B}_{\mathcal{O},12}^{a\,a_1 a_2} (p,x_1, x_2) \overline{\hat{B}_{\mathcal{O},\bar{4}\, \bar{3}}^{b\,a_4 a_3} (p,x_4^{\theta}, x_3^{\theta})}.
	\end{split}
\end{equation}
To make the integral representation (\ref{partialwave:integral}) valid in rigorous sense, we would like to show that it can be viewed as an inner product in an $L^2$-space, where the inner product is defined by
\begin{equation}
	\langle f, g \rangle_{\mathcal{O}} \assign \int \frac{d^d p}{(2 \pi)^d}
	(\hat{G}_{\mathcal{O}})_{a b} (p) f^a (p) \overline{g^b (p)} .
	\label{def:innerO}
\end{equation}
In addition, to show that eq.\,(\ref{partialwave:integral}) performs the analytic continuation, one needs to show that (a) the OPE kernel $\hat{B}(p,x_i,x_j)$ is analytic in $x_i$ and $x_j$ in the regime
\begin{equation}
	\begin{split}
		\varepsilon_i-\epsilon_j>\abs{\mathbf{y}_i-\mathbf{y}_j},\quad\epsilon_j>\abs{\mathbf{y}_j},\quad(x=(\varepsilon+it,\mathbf{x}+i\mathbf{y}));
	\end{split}
\end{equation}
(b) $\hat{B}(p,x_i,x_j)$ and its derivatives in $x$ have sufficiently good upper bound, such that the derivatives commute with the integration. (c) the r.h.s. of eq.\,(\ref{partialwave:integral}) is indeed the conformal partial wave in the Euclidean regime.

Based on the above ideas, we would like to show the followings things:
\begin{itemize}
	\item The momentum-space two-point function $\hat{G}_\mathcal{O}$ is a good measure. \\
	For $\mathcal{O}^a$ above unitarity bound,
	i.e. $\Delta \geqslant \Delta_{\ast} (\rho)$, $(\hat{G}_{\mathcal{O}})_{a b}
	(p)$ is a positive tempered measure, means that (a) $(\hat{G}_{\mathcal{O}})_{a b} (p) \xi^a
	\overline{\eta^b} d^dp$ is a tempered measure for any constant tensors $\xi,\eta$, (b)$(\hat{G}_{\mathcal{O}})_{a b} (p) \xi^a
	\overline{\xi^b} d^dp$ is a positive tempered measure for any constant tensor $\xi$. (In fact (b) implies (a).)
	
	\item The OPE kernel $\hat{B}_{\mathcal{O},12}$ is well-defined. \\
	Define the Fourier-transform of the three-point function with respect to the third point:
    \begin{equation}\label{def:3ptmomentum}
    	\hat{G}^{a_1 a_2 a}_{12\mathcal{O}} (x_1, x_2, p) = \int d^d x \langle
    	\mathcal{O}_1^{a_1} (x_1) \mathcal{O}_2^{a_2} (x_2) \mathcal{O}^a (i t,
    	\mathbf{x}) \rangle e^{i p \cdummy x},
    \end{equation}
	The OPE kernel $\hat{B}_{\mathcal{O},12}$ is defined by the decomposition 
	\begin{equation}\label{Ghat:decomp}
		\begin{split}
			\hat{G}^{a_1 a_2 a}_{12\mathcal{O}} (x_1, x_2,
			p) = \hat{B}_{\mathcal{O},12}^{b\, a_1 a_2} (p,x_1, x_2)
			(\hat{G}_{\mathcal{O}})_b^{\ a} (p),
		\end{split}
	\end{equation}
    $\hat{B}_{\mathcal{O},12}$ has a crucial property: it is in the form of $
	(x_{12}^2)^{- \sigma} \times$entire function of $x_1, x_2, p$.
	
	\item Some upper bounds of the OPE kernel and its derivatives. \\
	For $x_1 = ( \varepsilon_1+i t_1,
	\mathbf{x}_1+i\mathbf{y}_1), x_2 = (\varepsilon_2+i t_2, \mathbf{x}_2+i\mathbf{y}_2)$ under conditions
	$\varepsilon_1 - \varepsilon_2 > | \mathbf{y}_1 -\mathbf{y}_2 |, \nospace
	\varepsilon_2 > | \mathbf{y}_2 | \nobracket$, in a small neighbourhood
	$\mathcal{U}$ of $(x_1, x_2)$ we have the following bound
	\begin{equation}\label{bound:Bhat}
		\left| \frac{\partial^{| \alpha | + | \beta |}}{\partial x_2^{(\alpha)}
			\partial x_{12}^{(\beta)}}  \hat{B}_{\mathcal{O},12}^{a\, a_1 a_2} (p,x_1, x_2) \right| \leqslant K (\mathcal{U}) e^{- E
			\varepsilon_2 +\mathbf{p} \cdummy \mathbf{y}_2} [1 + | p |]^{\gamma}
	\end{equation}
	for $p=(E,\mathbf{p})$ in the closed forward light cone $\bar{V}_+$. Here $K (\mathcal{U}),
	\gamma < + \infty$.
\end{itemize}
Let us first see how the above three properties lead to the analytic continuation of the conformal partial wave. By translation invariance, we can always shift the configuration $c=(x_1,x_2,x_3,x_4)$ from the forward tube to the regime where
\begin{equation}\label{domain:forwardtuberestricted}
	\begin{split}
		\varepsilon_1 - \varepsilon_2 > | \mathbf{y}_1 -\mathbf{y}_2 |, \quad \varepsilon_2 > | \mathbf{y}_2 |,\\
		\varepsilon_3 - \varepsilon_4 > | \mathbf{y}_3 -\mathbf{y}_4 |, \quad\varepsilon_3 <- | \mathbf{y}_3 |,\\
	\end{split}
\end{equation}
so the OPE kernels $\hat{B}_{\mathcal{O},12}^{a\, a_1 a_2} (p,x_1, x_2)$ and $\hat{B}^{b\,a_4 a_3}_{\mathcal{O},\bar{4}\,\bar{3}} (p,x_4^\theta, x_3^\theta))$ satisfy the condition for the upper bound (\ref{bound:Bhat}).

By Vladimirov's theorem and Lorentz invariance, we know that $\hat{G}_{ab}(p)$ is a tempered distribution which is supported in the forward light-cone. Now since $\hat{G}_{ab}(p)$ is also a measure, together with the exponential-decay bounds (\ref{bound:Bhat}), we see that the r.h.s. of eq.\,(\ref{partialwave:integral}) is an integral of a tempered measure against two exponential-decay functions over the forward light-cone, so it is convergent. Then since the derivatives of the OPE kernels also satisfy the exponential-decay bounds, uniform in a small neighbourhood of the configuration $c$, we can interchange the derivatives (in $x_i$'s) and the integration (in $p$) according to Lebesgue's dominated convergence theorem. So the r.h.s. of eq.\,(\ref{partialwave:integral}) defines an analytic function on the complex domain (\ref{domain:forwardtuberestricted}). Then by translation invariance, its domain is easily extended to the forward tube.

Now we have to argue that this analytic function agrees with the Euclidean conformal partial wave in the Euclidean regime. This is actually done formally in eq.\,(\ref{partialwave:formalopecheck}). We claim that using the assumption of OPE convergence in Euclidean, it can be done in a rigorous way but the idea is basically the same as eq.\ (\ref{partialwave:formalopecheck}). We leave the details to \cite{paper2a}.

We conclude that the conformal partial wave $G_{1234,\mathcal{O}}(c)$ has analytic continuation to $c\in\mathcal{T}_4$, performed by the integral representation (\ref{partialwave:integral}). If the configuration $c$ satisfies the condition (\ref{domain:forwardtuberestricted}), then by positivity of the measure $\hat{G}_\mathcal{O}(p)$, we have the Cauchy-Schwarz inequality:
\begin{equation}
	| G_{1234, \mathcal{O}}^{a_1 a_2 a_3 a_4} (c) |^2 \leqslant G_{12 \bar{2}\,
		\bar{1}, \mathcal{O}}^{a_1 a_2 a_2 a_1} (c_{12}) G_{\bar{4}\, \bar{3} 34,
		\mathcal{O}}^{a_4 a_3 a_3 a_4} (c_{34}), \label{CB:CSineq}
\end{equation}
where $c_{12} = (x_1, x_2, x_2^{\theta},
x_1^{\theta})$ and $c_{34} = (x_4^{\theta}, x_3^{\theta}, x_3, x_4)$.

It remains to prove the required properties of $(\hat{G}_{\mathcal{O}})_{a b} (p)$ and $\hat{B}_{\mathcal{O},12}(p,x_1,x_2)$. The following subsections present a sketch of the proof. We claim that the properties of $\hat{G}_\mathcal{O}$ is true. But for the properties of $\hat{B}_{\mathcal{O},12}$, we only manage to show that they are true for generic $\Delta_{\mathcal{O}}$, and there a discrete set of exceptional $\Delta_{\mathcal{O}}$ for which we do not have a proof.
\subsection{Properties of $(\hat{G}_{\mathcal{O}})_{a b}$}
We would like to show that $\frac{d^d p}{(2 \pi)^d}
(\hat{G}_{\mathcal{O}})_{a b} (p)$ is a positive tempered measure. This is actually true for general QFT two-point functions. The claim is that given any QFT two-point function satisfying Wightman axioms (W0) - (W2) in Lorentzian or OS axioms (OS0) - (OS2) in Euclidean,\footnote{At the level of two-point functions, without any extra assumptions, one can use the argument in \cite{osterwalder1973} to show that Wightman axioms (W0) - (W2) are equivalent to OS axioms (OS0) - (OS2). } the corresponding momentum-space two-point function is a positive tempered measure.\footnote{The conclusion consists of three ingredients: (a) measure; (b) positive; (c) temperedness. The concept of ``measure" is in the usual mathematical sense. By ``positive" we means that for any constant tensor $\xi^a$ in $V_{\rho_\mathcal{O}}$, the SO(d) representation space of $\mathcal{O}$, the measure $(\hat{G}_{\mathcal{O}})_{a b} (p)\xi^a\overline{\xi^b}\,d^dp$ is positive. By ``tempered" we mean that the integral $\int (\hat{G}_{\mathcal{O}})_{a b} (p)\varphi(p)\,d^dp$ defines a tempered distribution, where $\varphi$ is a Schwartz test function.} This claim is a corollary of \emph{Bochner-Schwartz theorem}:
\begin{theorem}
	\label{theorem:BS}(Bochner-Schwartz, see {\cite{Vladimirov2}}, p125.) For a generalized function $f$ taken from $\mathcal{D}'
	(\mathbb{R}^d)$ to be positive definite, which means that
	\begin{equation}\label{positivedistribution}
		\begin{split}
		\int d^dx f(x-y)\varphi(x)\overline{\varphi(y)}\geqslant0,\qquad\forall\,\varphi\in\mathcal{D},	
		\end{split}
	\end{equation}
    it is necessary and sufficient that
	it be a Fourier transform of a nonnegative tempered measure: $f =\mathcal{F}
	[\mu]$, $\mu \in \mathcal{S}' (\mathbb{R}^d)$, $\mu \geqslant 0$. \\
\end{theorem}
\begin{remark}
	Recall that $\mathcal{S}$ is the Schwarz space and $\mathcal{D}$ is the space of compactly supported Schwartz test functions, we always have $\mathcal{D}\subset\mathcal{S}$, thus $\mathcal{S}^\prime\subset\mathcal{D}^\prime$. In the case of Wightman QFT, $f$ is the two-point function, which is a tempered distribution, so $f\in\mathcal{D}^\prime$ is satisfied.
\end{remark}
\begin{remark}
	To show that the diagonal components $\hat{G}_{aa}$ are tempered measures, one just need to set $f(p)=(\hat{G}_{\mathcal{O}})_{a b} (p)\xi^a\overline{\xi^b}$ with proper $\xi$ in eq.\,(\ref{positivedistribution}). For the off-diagonal components, one need to apply the Cauchy-Schwarz inequality
	\begin{equation}
		\begin{split}
		\abs{	\int\dfrac{d^dp}{(2\pi)^d}(\hat{G}_{\mathcal{O}})_{a b} (p)\varphi(p)\overline{\eta}(p)}^2\leqslant\int\dfrac{d^dp}{(2\pi)^d}(\hat{G}_{\mathcal{O}})_{a a} (p)\abs{\varphi(p)}^2\times\int\dfrac{d^dp}{(2\pi)^d}(\hat{G}_{\mathcal{O}})_{b b} (p)\abs{\eta(p)}^2
		\end{split}
	\end{equation}
   and choose proper test functions $\varphi,\eta\in\mathcal{S}(\mathbb{R}^d)$.
\end{remark}

\subsection{Properties of $\hat{B}_{\mathcal{O},12}$.}
We would like to explain why $\hat{B}_{\mathcal{O},12}$ is an entire function of $p$, and why it has an exponentially decaying upper bound in the light cone, leaving the technical details to the upcoming paper \cite{paper2a}.

\subsubsection{Analyticity}
Let us first consider the scalar case, i.e.\,no spin indices. The three-point function $G_{12\mathcal{O}}(x_1,x_2,x)$ is fixed up to a constant factor
\begin{equation}
	\begin{split}
		G_{12\mathcal{O}}(x_1,x_2,x)=\dfrac{1}{(x_{12}^2)^{\frac{\Delta_1+\Delta_2-\Delta}{2}}\fr{(x_1-x)^2}^{\frac{\Delta_1-\Delta_2+\Delta}{2}}\fr{(x_2-x)^2}^{\frac{\Delta_2-\Delta_1+\Delta}{2}}},
	\end{split}
\end{equation}
where $\Delta_{1},\Delta_{2},\Delta$ are the scaling dimensions of $\mathcal{O}_1,\mathcal{O}_2,\mathcal{O}$. We put $\mathcal{O}(x)$ in Minkowski space and take Fourier transform with respect to $x$. This gives $\hat{G}_{12\mathcal{O}}(x_1,x_2,p)$, the scalar version of eq. (\ref{def:3ptmomentum}). By explicit computation \cite{Mack:1976pa}, one can show that $\hat{G}_{12\mathcal{O}}$ has the following integral representation
\begin{equation}\label{G:integralrepr}
	\begin{split}
		 \hat{G}_{12\mathcal{O}} (x_1, x_2, p) = & \frac{2 \pi^{d / 2 + 1}}{\Gamma
			\left( \frac{\Delta + \Delta_1 - \Delta_2}{2} \right) \Gamma \left(
			\frac{\Delta - \Delta_1 + \Delta_2}{2} \right)} \frac{(- p^2 /
			4)^{\frac{\Delta - d / 2}{2}}}{(x_{12}^2)^{\frac{\Delta_1 + \Delta_2 - d /
					2}{2}}} \theta_{V_+} (p) \\
		& \times \int_0^1 d u \left( \frac{u}{1 - u} \right)^{\frac{\Delta_1 -
				\Delta_2}{2}} [u (1 - u)]^{\frac{d - 4}{4}} e^{i p \cdummy (x_2+ux_{12})} J_{\Delta
			- d / 2} ([- u (1 - u) x_{12}^2 p^2]^{1 / 2}),  
	\end{split}
\end{equation}
where $J_\nu$ is the Bessel function of the first kind. Under normalization $\braket{\mathcal{O}(0)\mathcal{O}(x)}=\frac{1}{(x^2)^\Delta}$, the momentum-space two-point function of $\mathcal{O}$ (defined in eq. (\ref{2ptmomentum})) is given by
\begin{equation}\label{G2scalarFT:result}
	\hat{G}_{\mathcal{O}} (p) = \frac{2 \pi^{d / 2 + 1}}{\Gamma (\Delta + 1 - d
		/ 2) \Gamma (\Delta)} \left( - \frac{1}{4} p^2 \right)^{\Delta - d / 2}
	\theta_{V_+} (p) . 
\end{equation}
By comparing (\ref{G:integralrepr}) and (\ref{G2scalarFT:result}), we see that
$\hat{G}_{12\mathcal{O}}$ is factorized into
\begin{equation}
	\hat{G}_{12\mathcal{O}} (x_1, x_2, p) = \hat{G}_{\mathcal{O}} (p)
	\hat{B}_{\mathcal{O},12} (p,x_1, x_2) . \label{G3:decompscalar}
\end{equation}
By expanding the Bessel function in eq.\,(\ref{G:integralrepr}) into power series and integrating over $u$ term by term, we will get a power series for $\hat{G}_{12\mathcal{O}}$. Then after removing the $\hat{G}_\mathcal{O}$-factor, we get the power series for $\hat{B}_{\mathcal{O},12}$ :
\begin{equation}\label{B3:serieshigherd}
	\begin{split}
		\hat{B}_{\mathcal{O},12} (p,x_1, x_2) =& \frac{e^{i p \cdummy
				x_2}}{(x_{12}^2)^{\frac{\Delta_1 + \Delta_2 - \Delta}{2}}} F(\Delta_1,\Delta_2,\Delta;i p\cdot x_{12},x_{12}^2 p^2 /
		4), \\ 
		F(\Delta_1,\Delta_2,\Delta;x,y)=&\underset{m, n =
			0}{\overset{\infty}{\sum}} \frac{\left( \frac{\Delta + \Delta_1 - \Delta_2}{2} \right)_{m +
				n} \left( \frac{\Delta - \Delta_1 + \Delta_2}{2} \right)_n}{(\Delta)_{m + 2
				n} (\Delta - d / 2 + 1)_n}\,\frac{x^m y^n}{m!n!} . \\
	\end{split}
\end{equation}
When $\Delta, \Delta - \frac{d - 2}{2} \nin -\mathbb{N}$, the power series
above is absolutely convergent for all $x,y\in
\mathbb{C}$, hence $\hat{B}_{\mathcal{O},12} (p,x_1, x_2)$ is
$(x_{12}^2)^{\frac{\Delta - \Delta_1 - \Delta_2}{2}} \times$an entire function
of $x_1, x_2, p$. 

Now let us consider the three-point function of spinning operators, $\braket{\mathcal{O}_1^{a_1}\mathcal{O}_2^{a_2}\mathcal{O}_3^{a_3}}$, which is kinematically determined by the scaling dimensions $\Delta_k$ and the SO$(d)$ representations $\rho_k$:  
\begin{equation}\label{spinning3pt}
	\begin{split}
		G_{123}^{a_1a_2a_3}(x_1,x_2,x_3)=\sum\limits_{\ell_1,\ell_2,\ell_3}T_{\ell_1,\ell_2,\ell_3}^{a_1a_2a_3}(x_k)G_3(x_k;\Delta_k+\ell_k),
	\end{split}
\end{equation}
where $T_{\ell_1,\ell_2,\ell_3}^{a_1a_2a_3}(x_k)$'s are homogeneous polynomials of $x_k$'s, $\ell_k$'s are integers and $G_3(x_k;\Delta_k+\ell_k)$'s are the scalar three-point functions with the scaling dimensions $\Delta_k+\ell_k$. This sum is finite, and each set of $\ell_k$'s is determined by a totally symmetric traceless representation that appears in $\rho_1\otimes\rho_2\otimes\rho_3$. Since $T$'s are polynomials, we have
\begin{equation}
	\begin{split}
		\hat{G}_{123}^{a_1a_2a_3}(x_1,x_2,p)=\sum\limits_{\ell_1,\ell_2,\ell_3}T_{\ell_1,\ell_2,\ell_3}^{a_1a_2a_3}(x_1,x_2,i\partial_{p})\hat{G}_3(x_1,x_2,p;\Delta_k+\ell_k).
	\end{split}
\end{equation}
We claim that $\hat{B}_{3,12}^{a_3 a_1 a_2}(p,x_1,x_2)$ has the form\footnote{Here we remind readers that in $\hat{B}$, the scaling dimension $\Delta_3$ is shifted by $-\ell_3$ instead of $+\ell_3$. }
\begin{equation}\label{Bspinning:basis}
	\begin{split}
		\hat{B}_{3,12}^{a_3 a_1 a_2}(p,x_1,x_2)=\sum\limits_{\ell_1,\ell_2,\ell_3}c_{\ell_1,\ell_2,\ell_3}T_{\ell_1,\ell_2,\ell_3}^{a_1a_2a_3}(x_1,x_2,i\partial_{p})\hat{B}_3(p,x_1,x_2;\Delta_3-\ell_3,\Delta_1+\ell_1,\Delta_2+\ell_2),
	\end{split}
\end{equation}
where $\hat{B}_3$ is the scalar OPE kernel. This claim is based on three facts:
\begin{enumerate}
	\item For fixed SO(d) spins $(\rho_1,\rho_2,\rho_3)$, there are only finitely many independent tensor structures $T_{\ell_1,\ell_2,\ell_3}$ allowed by conformal invariance.
	\item $T_{\ell_1,\ell_2,\ell_3}$ only depends on $(\rho_1,\rho_2,\rho_3,\ell_1,\ell_2,\ell_3)$. In particular, it does not depend on scaling dimensions.
	\item If $\hat{B}_{3,12}^{a_3 a_1 a_2}(p,x_1,x_2)$ has the form (\ref{Bspinning:basis}), then  $\hat{G}^{a_1 a_2 a_3}_{123} (x_1, x_2,
	p)$($=\hat{B}^{b_3 a_1 a_2}_{3,12} (p,x_1, x_2)
	(\hat{G}_{\mathcal{O}_3})_{b_3}^{\ a_3} (p)$) satisfies the conformal Ward identities for scaling dimensions $(\Delta_1,\Delta_2,\Delta_3)$ and SO(d) spins $(\rho_1,\rho_2,\rho_3)$.
\end{enumerate}
The above three facts imply that under (\ref{Ghat:decomp}), the space spanned by the basis $\{T_{\ell_1,\ell_2,\ell_3}^{a_1a_2a_3}(x_1,x_2,i\partial_{p})\hat{B}_3(x_1,x_2,p;\Delta_1+\ell_1,\Delta_2+\ell_2,\Delta_3-\ell_3)\}$ the space of $\hat{G}_{123}^{a_1 a_2 a_3}(x_1,x_2,p)$ (the space of tensor structures allowed by conformal invariance).

Since $\hat{B}_{123}(x_1,x_2,p)$ is an entire function of $p$ in the scalar case except when $\Delta_3\in(-\mathbb{N})\cup(\frac{d-2}{2}-\mathbb{N})$, by eq. (\ref{Bspinning:basis}), it is also an entire function of $p$ in the spinning case because it is the derivatives of entire functions, except when $\Delta_3-\ell_3\in(-\mathbb{N})\cup(\frac{d-2}{2}-\mathbb{N})$. We will comment on these exceptional cases in section \ref{section:spinexceptional}.

\subsubsection{Exponential decay bound}
In this section, we would like to give a sketch of the proof of the upper bound (\ref{bound:Bhat}). 

For convenience, we think of $\hat{B}_{\mathcal{O},12}^{a\,a_1a_2}(p,x_1,x_2)$ as a function of $x_{12},x_2$ and $p$. By eqs.\,(\ref{bound:Bhat}) and (\ref{Bspinning:basis}), $\partial^{(\mu)}_{x_2}\partial^{(\nu)}_{x_{12}}\hat{B}_{\mathcal{O},12}^{a\,a_1a_2}(p,x_1,x_2)$ is a finite sum of terms in the following form:
\begin{equation}\label{B3:derivativeform}
	\begin{split}
		\dfrac{e^{ip\cdot x_2}}{(x_{12}^2)^{\frac{\Delta_1+\Delta_2-\Delta+\ell_1+\ell_2+\ell}{2}+|\nu|}}\times(\mathrm{polynomial}\ \mathrm{of}\ x_2,x_{12}\ \mathrm{and}\ p)\times F^{(k,l)}(\Delta_1+\ell_1,\Delta_2+\ell_2,\Delta-\ell;x,y),
	\end{split}
\end{equation}
where $F^{(k,l)}(x,y)=\partial^k_x\partial^l_yF(x,y)$. So we see that the exponential decay factor in eq. (\ref{bound:Bhat}) comes from $e^{ip\cdot x_2}$. To prove the upper bound (\ref{bound:Bhat}), it remains to show that the extra factors in eq.\,(\ref{B3:derivativeform}) has a power-law bound in $p$ for $p$ in the forward light cone.

When $x_1$ and $x_2$ are Lorentzian points, $e^{ip\cdot x_2}$ becomes a phase factor, then the upper bound (\ref{bound:Bhat}) becomes a power-law bound. Mack claimed this upper bound in his paper \cite{Mack:1976pa}, where it says that the temperedness of $G_{12\mathcal{O}}$ and the decomposition $\hat{G}_{12\mathcal{O}}=\hat{B}_{\mathcal{O},12}\hat{G}_\mathcal{O}$ imply the power-law bound of $\hat{B}_{\mathcal{O},12}$ in $p$ for $p$ in the forward light cone (see \cite{Mack:1976pa}, p181, the last paragraph). From my own understanding, the underlying argument of this claim is that if an analytic function is also a tempered distribution, then it should be power-law bounded. However this is not true. A toy counterexample is the function
\begin{equation}
	\begin{split}
		f(t)=\dfrac{d}{dt}sin(e^t).
	\end{split}
\end{equation}
We see that $f(t)$ is an analytic function as well as a tempered distribution, but it does not have a power-law bound. For this reason, we regard the upper bound (\ref{bound:Bhat}) as an unknown result, and we need some careful computation to justify it.

For the technical reason, we divide the forward light-cone into two regimes: $0\leqslant-p^2\leqslant m^2$ and $-p^2\geqslant m^2$, where $m^2>0$ is an arbitrary positive cutoff. \\ \\
\textbf{Case $0\leqslant-p^2\leqslant m^2$} \\ \\
We take the $(k,l)$-th derivative and sum over $m$ in the series expansion (\ref{B3:serieshigherd}), which gives
\begin{equation}\label{Fkl:series}
	\begin{split}
		F^{(k,l)}(x,y)=&\sum\limits_{n=0}^{\infty}\dfrac{(\alpha)_{n+k+l}(\beta)_{n+l}}{n!(\alpha+\beta)_{2n+k+2l}(\alpha+\beta-\frac{d-2}{2})_{n+l}}{}_{1}F_{1}(\alpha+n+k+l;\alpha+\beta+2n+k+2l;x)y^n.
	\end{split}
\end{equation}
When $x\rightarrow\infty$, each term in series (\ref{Fkl:series}) has the following asymptotic behavior
\begin{equation}\label{Fkl:nasymp}
	\begin{split}
		{}[F^{(k, l)} (x, y)]_n = & \frac{(\beta)_{n + l} \Gamma (\alpha +
			\beta)}{n! \Gamma (\alpha) (\alpha+\beta)_{n + l}} e^x x^{- \beta - n - l}y^n \left[
		1 + O \left( \frac{1}{x} \right) \right] \\
		& + \frac{(\alpha)_{n + k + l} \Gamma (\alpha + \beta)}{n! \Gamma
			(\beta) (\alpha+\beta)_{n + l}} (- x)^{- \alpha - n - k - l} y^n \left[ 1 + O
		\left( \frac{1}{x} \right) \right] . 
	\end{split}
\end{equation}
For $x_{12}=(\varepsilon+it,\mathbf{x}+i\mathbf{y}))$ such that $\varepsilon>|\mathbf{y}|$ and $p \in \bar{V}_+$, one can show that $x=ip\cdot x_{12}$ is in the regime where
\begin{equation}\label{x:range}
	\tmop{Re} (x) \leqslant 0, \quad \left| \dfrac{\tmop{Im} (x)}{\tmop{Re} (x)}
	\right| \leqslant \dfrac{\sqrt{2 (t^2 +\mathbf{x}^2)}}{\varepsilon - |
		\mathbf{y} |}. 
\end{equation}
Then $e^x$ in the first term of (\ref{Fkl:nasymp}) decays exponentially in $p$ when $p$ goes to $\infty$ in $\overline{V}_+$. In addition, since $0\leqslant-p^2\leqslant m^2$, $y=\frac{x_{12}^2p^2}{4}$ is bounded. Therefore, each term in series (\ref{Fkl:series}) has the following power-law bound in $p$:
\begin{equation}
	\begin{split}
		{}[F^{(k, l)} (x, y)]_n\leqslant C_n^{(k,l)}(x_{12},m)(1+|p|)^{-\alpha-n-k-l},\qquad p^0 \geqslant 0, \tmxspace 0 \leqslant
		- p^2 \leqslant m^2,
	\end{split}
\end{equation}
where $C_n^{(k,l)}(x_{12},m)$ is some constant.

We split above sum into two parts:
\begin{equation}
	\begin{split}
		F^{(k, l)} (x, y) = \left( \underset{n = 0}{\overset{N}{\sum}} +
		\underset{n = N + 1}{\overset{\infty}{\sum}} \right) [\ldots] \assign
		F^{(k, l)}_N + F^{(k, l)}_{> N} .
	\end{split}
\end{equation}
$F_N^{(k,l)}$ has a power-law bound in $p$ as described above. For $F^{(k, l)}_{> N}$, we choose $N$ to be sufficiently large such that $\alpha + N
+ k + l, \beta + N + l> 0$. Then using the following property of the hypergeometric function:
\begin{equation}\label{1F1:bound}
	\begin{split}
		| {}_1 F_1 (\alpha ; \alpha + \beta ; x) | \leqslant \frac{\Gamma (\alpha
			+ \beta)}{\Gamma (\beta)} \frac{1}{[- \tmop{Re} (x)]^{\alpha}},\quad\alpha,\beta>0,\ \mathrm{Re}(x)<0,
	\end{split}
\end{equation}
one can show that
\begin{equation}\label{Fkl>N:bound}
	| F^{(k, l)}_{> N} (x, y) | \leqslant C_N^{(k, l)} (x_{12}, m) [1 + | p
	|]^{- \alpha - k - l - N - 1}, \qquad p^0 \geqslant 0, \tmxspace 0 \leqslant
	- p^2 \leqslant m^2, 
\end{equation}
where $C_N^{(k, l)} (x_{12}, m)$ is some constant. Combining the upper bounds of $F_N^{(k,l)}$ and $F^{(k, l)}_{> N}$, we get an estimate
\begin{equation}\label{Fkl:bound}
	| F^{(k, l)} (x, y) | \leqslant C^{(k, l)} (x_{12}, m) [1 + | p |]^{- \alpha
		- k - l}, \qquad p^0 \geqslant 0, \tmxspace 0 \leqslant - p^2 \leqslant m^2.
\end{equation}
The constant $C^{(k, l)} (x_{12}, m)$ is uniformly bounded for $x_{12}$ in a small neighbourhood. 

We would like make a comment on why the above estimate of $F^{(k,l)}$ is not available for $p$ in the whole forward light-cone. In the estimate (\ref{Fkl>N:bound}) of $F_{>N}^{k,l}(x,y)$ for $0\leqslant-p^2\leqslant m^2$, we implicitly used the condition $|y|\leqslant|x_{12}^2|m^2/4$, and the $y$-dependence of this estimate is absorbed into the constant $C_n^{(k,l)}(x_{12},m)$. If there is no restriction $0\leqslant-p^2\leqslant m^2$, then $y\sim p^2$ is not bounded, and the estimate of $F_{>N}$ will contain the factor $e^{|y/\mathrm{Re}(x)|}$, which grows exponentially in $p$ towards some direction in the forward light-cone. This is the reason why we have to introduce the cutoff $m^2$. \\

\textbf{Case $-p^2\geqslant m^2$} \\ \\
In this case, we would like to use the inversion of (\ref{G3:decompscalar}):
\begin{equation}\label{Bscalar:inversion}
	\begin{split}
		 \hat{B}_{\mathcal{O},12}(p,x_1, x_2)=\hat{G}_{12\mathcal{O}} (x_1, x_2,
		 p)
		\hat{G}_{\mathcal{O}}(p)^{-1},
	\end{split}
\end{equation}
and use eq.\,(\ref{Bspinning:basis}) to produce $\hat{B}$ of spinning operators. Then the estimate of $\partial_{x_2}^{(\alpha)} \partial_{x_{12}}^{(\beta)}\hat{B}_{\mathcal{O},12}^{a\, a_1 a_2} (p,x_1, x_2)$ is reduced to the estimate of $\hat{G}_{12\mathcal{O}} (x_1, x_2,
p)$, $\hat{G}_{\mathcal{O}}(p)^{-1}$ and their derivatives in the scalar case.

By (\ref{G2scalarFT:result}), $\hat{G}_{\mathcal{O}}(p)^{-1}$ is well-defined except for $\Delta\in(-\mathbb{N})\cup\fr{\frac{d-2}{2}-\mathbb{N}}$, which are poles of the Gamma functions. Since the singularities at $p^2=0$ are excluded by the condition $-p^2\geqslant m^2$, $\hat{G}(p)^{-1}$ and its derivatives have power-law upper bounds in $p$. Then it suffices to show that $\hat{G}_{12\mathcal{O}} (x_1, x_2,
p)$ and its derivatives have an exponential decay bounds in $p$ for $-p^2\geqslant m^2$.

For convenience we introduce the notation $\alpha = \frac{\Delta + \Delta_1 -
	\Delta_2}{2}, \beta = \frac{\Delta - \Delta_1 + \Delta_2}{2}$. Since $G_{12\mathcal{O}}(x_1,x_2,x)$ (as a function of $x$) is a product of two two-point functons, $\hat{G}_{12\mathcal{O}} (x_1, x_2, p)$ is
the convolution of two 2pt functions (up to $x_{12}^{\#}$)
\begin{equation}\label{G3FT:convolution}
	\hat{G}_{12\mathcal{O}} (x_1, x_2, p) = \frac{C_{\alpha}
		C_{\beta}}{(x_{12}^2)^{\sigma}} \int \frac{d^d q}{(2 \pi)^d} (- q^2)^{\alpha
		- d / 2} (- (p - q)^2)^{\beta - d / 2} e^{i q \cdummy x_1 + i (p - q)
		\cdummy x_2} \theta_{V_+} (q) \theta_{V_+} (p - q), 
\end{equation}
where
\begin{equation}
	C_{\alpha} = \frac{\pi^{d / 2 + 1}}{2^{2 \alpha - d - 1} \Gamma (\alpha + 1
		- d / 2) \Gamma (\alpha)} . \label{def:Calpha}
\end{equation}
The above integral is over a finite domain because of the $\theta$-functions. When $\alpha, \beta > d / 2 - 1$, the integral is absolutely convergent and is bounded from above by
some polynomial of $p$.

When $\alpha\ \infixor\ \beta \leqslant d / 2 - 1$, $\hat{G}_{12\mathcal{O}}$ is
regularized by the analytic continuation in $\alpha$ and $\beta$. To show the general feature of this regulation, we would like to introduce the following toy example (from \cite{gelfandshilov}, chapter 3):
\begin{equation}\label{I:toyexample}
	\begin{split}
		I(\alpha)=\int_0^1 x^{\alpha}\varphi(x),
	\end{split}
\end{equation}
where $\varphi$ is a smooth function. This integral is convergent and analytic in $\alpha$ when Re$(\alpha)>-1$. To analytically continue $I(\alpha)$ to Re$(\alpha)\leqslant-1$, we split the integral into two parts:
\begin{equation}
	\begin{split}
		I(\alpha)=\int_0^1 x^{\alpha}\left[\varphi(x)-\sum\limits_{n=0}^{N}\dfrac{\varphi^{(n)}(0)}{n!}x^n\right]+\sum\limits_{n=0}^{N}\dfrac{\varphi^{(n)}(0)}{n!}\int_0^1 x^{\alpha+n}.
	\end{split}
\end{equation}
The first term defines an analytic function of $\alpha$ in the region Re$(\alpha)>-1-N$. The second term can be evaluated explicitly when Re$(\alpha)>-1$
\begin{equation}
	\begin{split}
		\sum\limits_{n=0}^{N}\dfrac{\varphi^{(n)}(0)}{n!}\int_0^1 x^{\alpha+n}=\sum\limits_{n=0}^{N}\dfrac{\varphi^{(n)}(0)}{n!}\dfrac{1}{\alpha+n+1}.
	\end{split}
\end{equation}
Each term has analytic continuation to the whole complex plane, with a pole at some special point. By increasing $N$, we see that $I(\alpha)$ has analytic continuation to $\alpha\in\mathbb{C}\backslash\left\{-1,-2,\ldots\right\}$.

Back to the three-point function, the regulation of the integral (\ref{G3FT:convolution}) is quite similar to the above toy example. The only complexity is that we need to deal with more integral variables and more singularities (including single and double light-cone singularities). Here we only summarize the main results, and leave the technical details to \cite{paper2a}: 
\begin{itemize}
	\item For $-p^2>0$ fixed, the integral in (\ref{G3FT:convolution}) has analytic continuation to $\alpha,\beta\in\mathbb{C}\backslash(-\mathbb{N}) \cup \left( \frac{d - 2}{2} -\mathbb{N} \right)$. It has poles at these exceptional points, but they are canceled out by the zeros of the prefactor $C_\alpha C_\beta$. Therefore, $\hat{G}_{12\mathcal{O}} (x_1, x_2, p)$ is an entire function of $\alpha$ and $\beta$.\footnote{This conclusion agrees with the series expansion (\ref{B3:serieshigherd}). If we multiply $\hat{B}_3$ by $(\Gamma(\Delta)\Gamma(\Delta-\frac{d-2}{2}))^{-1}$, which comes from the two-point function $\hat{G}_\mathcal{O}$ in eq.\,(\ref{G3:decompscalar}), we will get an entire function of $\Delta_1$, $\Delta_2$ and $\Delta$.}
	\item When $-p^2\geqslant m^2$, $\hat{G}_{12\mathcal{O}} (x_1, x_2, p)$ is bounded from above by $const\times\abs{e^{ip\cdot x_2}}\times$(polynomial of $p$ and $x_{12}$). The constant coefficient is finite and only depends on $x_{12}$ and $m$.
	\item The same is true for the derivatives of $\hat{G}_{12\mathcal{O}} (x_1, x_2, p)$.
\end{itemize}
For $\hat{B}_{\mathcal{O},12}$, there is an extra factor $\Gamma(\Delta)\Gamma(\Delta-\frac{d-2}{2})$ which comes from the two-point function $\hat{G}_\mathcal{O}$ in eq.\,(\ref{G3:decompscalar}). So $\hat{B}_{\mathcal{O},12}$ may have poles at $\Delta\in (-\mathbb{N}) \cup \left( \frac{d - 2}{2} -\mathbb{N} \right)$, which agrees with our computation result (\ref{B3:serieshigherd}). Therefore, the conclusions in the case $-p^2\geqslant m^2$ are the same as $0\leqslant-p^2\leqslant m^2$. 

We identify $\Delta$ in the scalar case to $\Delta_3-\ell_3$ in the spinning case, according to eq.\,(\ref{Bspinning:basis}). We conclude that
\begin{proposition}
	\label{prop:B3spinning} Let $\hat{B}^{a_3 a_1 a_2}_{3,12} (p,x_1, x_2)$ be the OPE kernel defined by eq.\,(\ref{Bspinning:basis}). Then 
	\begin{itemize}
		\item[](a) As a
		function of $\Delta_i$, $\hat{B}^{a_3 a_1 a_2}_{3,12} (p,x_1, x_2)$ has analytic continuation to
		\[ \left\{ (\Delta_1, \Delta_2, \Delta_3) \in \mathbb{C}^3 \left| \Delta_3
		- \ell_3 \nin (-\mathbb{N}) \cup \left( \frac{d - 2}{2} -\mathbb{N} \right)	\right. \right\}, \]
		and may have simple or double poles at $\Delta_3 -
		\ell_3 \in (-\mathbb{N}) \cup \left( \frac{d - 2}{2} -\mathbb{N} \right)$. Here $\ell_3$ is the same as in eq.\,(\ref{Bspinning:basis}).
		\item[](b) For
		each regular $(\Delta_1, \Delta_2, \Delta_3)$, the function
		$\hat{B}_{123}^{a_1 a_2 a_3} (x_1, x_2, p)$ and its derivatives have the form
		\begin{equation}
			\begin{split}
				(x_{12})^{\#}\times\mathrm{entire\ function\ of}\ x_1,\ x_2\ \mathrm{and}\ p.
			\end{split}
		\end{equation}  
	    They have the following bound in the light cone:
		\begin{equation}
			\left| \frac{\partial^{| \lambda | + | \mu |} \hat{B}_{123}^{a_1 a_2
					a_3}}{\partial x_2^{(\lambda)} \partial x_{12}^{(\mu)}} \right| \leqslant
			C_{(\lambda), (\mu)} (x_{12}) | \tmop{polynomial} \tmop{of} p | \times |
			e^{i p \cdummy x_2} | . \label{B3:spinningupperbound}
		\end{equation}
		The constant $C_{(\lambda), (\mu)} (x_{12})$ is finite as long as $x_{12} =
		(t - i \varepsilon, \mathbf{x}- i\mathbf{y})$ satisfies $\varepsilon > |
		\mathbf{y} |$. For each $x_{12}$ we can choose a small neighbourhood of it
		such that $C_{(\lambda), (\mu)} (x_{12})$ is uniformly bounded in the
		neighbourhood.
	\end{itemize}
\end{proposition}

\subsection{Comments on the exceptional cases}\label{section:spinexceptional}
The OPE kernel $\hat{B}_{3,12}$ may be singular when $\Delta_3-\ell_3\in(-\mathbb{N}) \cup \left( \frac{d - 2}{2} -\mathbb{N} \right)$, because of poles of $\Delta=\Delta_3-\ell_3$ in the power-series (\ref{B3:serieshigherd}). So far we do not have a proof of analyticity and the exponential-decay bound for these cases. Here we just makes some comments on why we believe such a proof exist.

In section \ref{section:relateB}, we related the OPE kernel $\hat{B}_{3,12}(p,x_1,x_2)$ to the Euclidean OPE kernel $C_{3}(x_1,x_2,x_0,\partial_0)$ (see eq.\,(\ref{ope:relation}) with $``3"$ replaced by $``\chi"$). Since we assume OPE convergence in the Euclidean signature, we expect that $\hat{B}_{3,12}(p,x_1,x_2)$ is finite when $x_1,x_2$ are Euclidean points.

By Euclidean OPE convergence, we also expect that the integral representation (\ref{partialwave:integral}) converges when $c$ is a Euclidean four-point configuration. This gives us a hint that $\hat{B}$'s should have good upper bound to make (\ref{partialwave:integral}) convergent. We hope that this upper bound does not change much when $x_k$'s are moved to the complex regime.

For the above reasons, we expect that the conclusions of proposition \ref{prop:B3spinning} is also true for these exceptional values of $\Delta_3$.

\section{Step 2: convergence of the partial wave expansion in $\mathcal{D}_4$}
We would like to show that the partial wave expansion (\ref{4pt:OPE}), as a sum of analytic functions, converges to an analytic function on the domain $\mathcal{D}_4$ (the subset of $\mathcal{T}_4$ with real spatial components). By translation invariance, we can always act with a time translation on $c\in\mathcal{D}_4$, such that the temporal components of the new configuration satisfies
\begin{equation}
	\begin{split}
		\epsilon_1>\epsilon_2>0>\epsilon_3>\epsilon_4.
	\end{split}
\end{equation}
Then using the Cauchy-Schwarz inequality (\ref{CB:CSineq}), we bound the sum in the r.h.s. of eq.\,(\ref{4pt:OPE}) as follows:
\begin{equation}\label{spinning:CSestimate}
	\begin{split}
		| G_{1234}^{a_1 a_2 a_3 a_4} (c) | & \leqslant 
		\underset{\mathcal{O}}{\sum} | G_{1234, \mathcal{O}}^{a_1 a_2 a_3 a_4} (c)
		|\\
		& \leqslant  \underset{\mathcal{O}}{\sum} \sqrt{G_{12 \bar{2} \bar{1},
				\mathcal{O}}^{a_1 a_2 a_2 a_1} (c_{12}) G_{\bar{4} \bar{3} 34,
				\mathcal{O}}^{a_4 a_3 a_3 a_4} (c_{34})}\\
		& \leqslant  \sqrt{\left( \underset{\mathcal{O}}{\sum} G_{12 \bar{2}
				\bar{1}, \mathcal{O}}^{a_1 a_2 a_2 a_1} (c_{12}) \right) \left(
			\underset{\mathcal{O}'}{\sum} G_{\bar{4} \bar{3} 34, \mathcal{O}'}^{a_4 a_3
				a_3 a_4} (c_{34}) \right)}.
	\end{split}
\end{equation}
So it suffices to show that the partial wave expansion is convergent at reflection-symmetric configurations in $\mathcal{D}_4$. A nice property of such configurations is that they are mapped to two-dimensional configurations by a composition of translation and real rotation, preserving the reflection symmetry:
\begin{equation*}
	\begin{split}
		&x_1=(\varepsilon_1+i t_1,\mathbf{x}_1),\quad x_2=(\varepsilon_2+i t_2,\mathbf{x}_2),\quad x_3=(-\varepsilon_2+i t_2,\mathbf{x}_2),\quad x_4=(-\varepsilon_1+i t_1,\mathbf{x}_1); \\
		\Longrightarrow\quad&x_1^{\prime}=(\varepsilon_1,0),\quad x_2^{\prime}=(\varepsilon_2+i t_2-it_1,\abs{\mathbf{x}_2-\mathbf{x}_1},0),\quad x_3^{\prime}=(-\varepsilon_2+i t_2-it_1,\abs{\mathbf{x}_2-\mathbf{x}_1},0),\quad x_4^{\prime}=(-\varepsilon_1,0).\\
	\end{split}
\end{equation*}
By Poincaré invariance, the partial waves at configuration $c_{12}=(x_1,x_2,x_2^\theta,x_1^\theta)$ is equal to a finite linear combination of the partial waves at $c_{12}^{\prime}=(x_1^{\prime},x_2^{\prime},x_2^{\prime\theta},x_1^{\prime\theta})$:
\begin{equation}
	\begin{split}
		G_{1234, \mathcal{O}}^{a_1 a_2 a_3 a_4} (c_{12})=\rho_1(R)^{a_1}_{b_1}\rho_2(R)^{a_2}_{b_2}\rho_3(R)^{a_3}_{b_3}\rho_4(R)^{a_4}_{b_4}G_{1234, \mathcal{O}}^{b_1 b_2 b_3 b_4} (c_{12}^{\prime}),
	\end{split}
\end{equation}
where the coefficients $\rho_1(R)^{a_1}_{b_1}\rho_2(R)^{a_2}_{b_2}\rho_3(R)^{a_3}_{b_3}\rho_4(R)^{a_4}_{b_4}$ are less than or equal to 1 since $R$ is a real rotation matrix. 

We claim that the conformal partial wave expansion (\ref{4pt:OPE}) is absolutely convergent if $c$ is a two-dimensional configuration in the forward tube. The main reason for this claim is that in the two-dimensional case, we can explicitly write down the conformal transformation which maps $c$ to $c_\rho$, and estimate the corresponding rotation matrices. We leave the proof to appendix \ref{appendix:2dconformal}.

Continue the estimate (\ref{spinning:CSestimate}):
\begin{equation}
	\begin{split}
		\underset{\mathcal{O}}{\sum}\abs{G_{1234, \mathcal{O}}^{a_1 a_2 a_3 a_4} (c_{12})}\leqslant&\sum\limits_{b_1,b_2,b_3,b_4}\abs{\rho_1(R)^{a_1}_{b_1}\rho_2(R)^{a_2}_{b_2}\rho_3(R)^{a_3}_{b_3}\rho_4(R)^{a_4}_{b_4}}\underset{\mathcal{O}}{\sum}\abs{G_{1234, \mathcal{O}}^{b_1 b_2 b_3 b_4} (c_{12}^{\prime})} \\
		\leqslant&\sum\limits_{b_1,b_2,b_3,b_4}\underset{\mathcal{O}}{\sum}\abs{G_{1234, \mathcal{O}}^{b_1 b_2 b_3 b_4} (c_{12}^{\prime})}.
	\end{split}
\end{equation}
The last line is a finite sum of absolutely convergent series, thus it is convergent. The estimate of the $c_{34}$-part is similar. This finishes the proof of the convergence of  the partial wave expansion (\ref{4pt:OPE}) for $c\in\mathcal{D}_4$. Thus $G_{1234, \mathcal{O}}^{a_1 a_2 a_3 a_4} (c)$ is an analytic function for $c\in\mathcal{D}_4$.

In appendix \ref{appendix:2dconformal}, we also show that the four-point function $G_{1234, \mathcal{O}}^{a_1 a_2 a_3 a_4} (c)$ has a power-law bound if $c$ is in the two-dimensional subspace of $\mathcal{T}_4$. Then the power-law bound follows using the above Cauchy-Schwarz argument.

\chapter{CFT four-point functions in the Minkowski cylinder}\label{chap:Minkcylinder}
In this chapter we discuss the generalization to the scalar four-point functions in the Minkowski cylinder, which has the topology $\mathbb{R}\times S^{d-1}$. The Minkowski cylinder is known to be the boundary of Lorentzian AdS${}_{d+1}$ in the framework of AdS${}_{d+1}$/CFT${}_{d}$ correspondence \cite{Aharony:1999ti}. We use coordinates $(t,\Omega)$ to describe its points, where $t$ is a real number and $\Omega=(\Omega^1,\ldots,\Omega^d)$ is a $d$-dimensional unit vector (i.e.\,$\sum_{i}(\Omega^i)^2=1$). 

The Minkowski cylinder can be viewed as an infinite-sheet version of the compactified Minkowski space,\footnote{Here we include the points at infinity in Minkowski space, which correspond to $\cos t+\Omega^d=0$.} with covering map
\begin{equation}\label{cylinder:cyltoflat}
	\begin{split}
		x_M^0=&\dfrac{\sin{t}}{\cos{t}+\Omega^d}, \\
		x_M^i=&\dfrac{\Omega^i}{\cos{t}+\Omega^d},\quad i=1,2,\ldots,d-1. \\
	\end{split}
\end{equation}
This covering map is conformal. One can show that the Minkowski metric is pulled back to the cylinder by
\begin{equation}\label{Mmetric:confequiv}
	\begin{split}
		ds^2_M=&-(dx_M^0)^2+\sumlim{i=1}^{d-1}(dx_M^i)^2=\dfrac{1}{\fr{\cos t+\Omega^d}^2}\fr{-d t^2+d\Omega^2}. \\
	\end{split}
\end{equation}
Therefore, these two Lorentzian spaces share the same conformal algebra. But the topologies of their conformal groups are different.

The Minkowski space is conformally embedded into the \emph{Poincaré patch} of the Minkowski cylinder\cite{Luscher:1974ez}:
\begin{equation}\label{cylinder:cond:poincare}
	\begin{split}
		-\pi<t<\pi, \quad\cos t+\Omega^d>0.
	\end{split}
\end{equation}
The Minkowski cylinder admits a causal ordering
\begin{equation}\label{def:cylcausalordering}
	\begin{split}
		(t_1,\Omega_1)>(t_2,\Omega_2)\quad\Leftrightarrow\quad t_1-t_2>\arccos(\Omega_1\cdot\Omega_2),
	\end{split}
\end{equation}
which agrees with the Minkowski-space causal ordering in the Poincaré patch and is preserved by any finite conformal transformation \cite{todorov1973conformal}. In contrast, in Minkowski space, the special conformal transformations can violate the causal ordering. Because of this subtlety, in the early days, people realized that one cannot naively postulate global conformal invariance in the framework of Wightman QFT because it will violate the causality condition.\footnote{By global conformal invariance we mean that the correlators are invariant under any finite conformal transformation. Then given any two local operators $\mathcal{O}_1(x_1),\mathcal{O}_2(x_2)$ which are time-like separated, one can find a special conformal transformation which maps $x_1,x_2$ to a space-like separated pair. As a consequence of the global conformal invariance and the causality condition, $[\mathcal{O}_1(x_1),\mathcal{O}_2(x_2)]=0$ for all possible $x_1,x_2$.} There are two resolutions to the puzzle. The first one is to assume Wightman axioms in Lorentzian but global conformal invariance in the regime where Wightman distributions are well-defined analytic functions, e.g. in the Euclidean space. Then after Wick rotation, the Lorentzian conformal invariance only holds in the infinitesimal way (the same as we have justified in part \ref{part:minkowski}, see section \ref{ConfMink}). This way of formulating conformal invariance is called the hypothesis of ``weak conformal invariance" \cite{Hortacsu:1972bw}. The second one is to put CFT in the Minkowski cylinder, where the conformal group action is globally well-defined and does not contradict anything \cite{Luscher:1974ez}.

Assuming Wightman axioms and Euclidean conformal invariance, Lüscher and Mack tried to show that the correlators in Minkowski space (viewed as the Poincaré patch) can be extended to the whole Minkowski cylinder. Similarly to the Osterwalder-Schrader theorem, L\&M started from the Euclidean CFT correlators (whose existence is easily justified using Wightman axioms \cite{jost1957bemerkung}) and tried to perform the analytic continuation in cylinder temporal variables. However, they did not manage to show the temperedness in the whole Lorentzian regime. Therefore, we still consider this as an open problem.

In the following sections, we will show that the Euclidean CFT axioms imply Wightman axioms in the Minkowski cylinder. We will only demonstrate the proof of temperedness, since the proof of other Wightman axioms are the same as part \ref{part:minkowski}.

\section{Euclidean cylinder and its Wick rotation}
We define the hyperbolic coordinates $(\sigma,\Omega)\in\bbR{}\times S^{d-1}$, where $\Omega=(\Omega^1,...,\Omega^d)$ is under constraint $\sumlim{i=1}^d\fr{\Omega^i}^2=1$. The relation of the Cartesian coordinates $(x^0,x^1,...,x^{d-1})$ and the hyperbolic coordinates is given in \cite{Luscher:1974ez}: 
\begin{equation}\label{relation:cartesianhyperbolic}
	\begin{split}
		x^0=&\dfrac{\sinh\sigma}{\cosh\sigma+\Omega^d}, \\
		x^i=&\dfrac{\Omega^i}{\cosh\sigma+\Omega^d},\quad i=1,2,...,d-1. \\
	\end{split}
\end{equation}
By eq.\,(\ref{relation:cartesianhyperbolic}), one can show that the Euclidean flat metric is conformally equivalent to a cylinder metric:
\begin{equation}\label{metric:confequiv}
	\begin{split}
		ds^2=&\sumlim{i=0}^{d-1}(dx^i)^2=\dfrac{1}{\fr{\cosh\sigma+\Omega^d}^2}\fr{d\sigma^2+d\Omega^2}. \\
	\end{split}
\end{equation}
Also, the Euclidean distance between two points $x_1$, $x_2$ is given by
\begin{equation}\label{cylinder:2ptdistance}
	\begin{split}
		(x_1-x_2)^2=\dfrac{2\fr{\cosh\fr{\sigma_1-\sigma_2}-\Omega_1\cdot\Omega_2}}{\fr{\cosh\sigma_1+\Omega_1^d}\fr{\cosh\sigma_2+\Omega_2^d}}
	\end{split}
\end{equation}
In the context of CFT, the hyperbolic coordinates correspond to the so-called \emph{N-S (North-South pole) quantization picture} \cite{EPFL}. In the N-S picture, the Euclidean flat space is foliated into slices of constant-$\sigma$ spheres, characterized by
\begin{equation}
	\begin{split}
		(x^0-\coth\sigma)^2+\mathbf{x}^2=\dfrac{1}{\sinh^2\sigma},
	\end{split}
\end{equation}
see fig. \ref{fig:NS}.
\begin{figure}[h]
	\centering
	\includegraphics[scale=0.3]{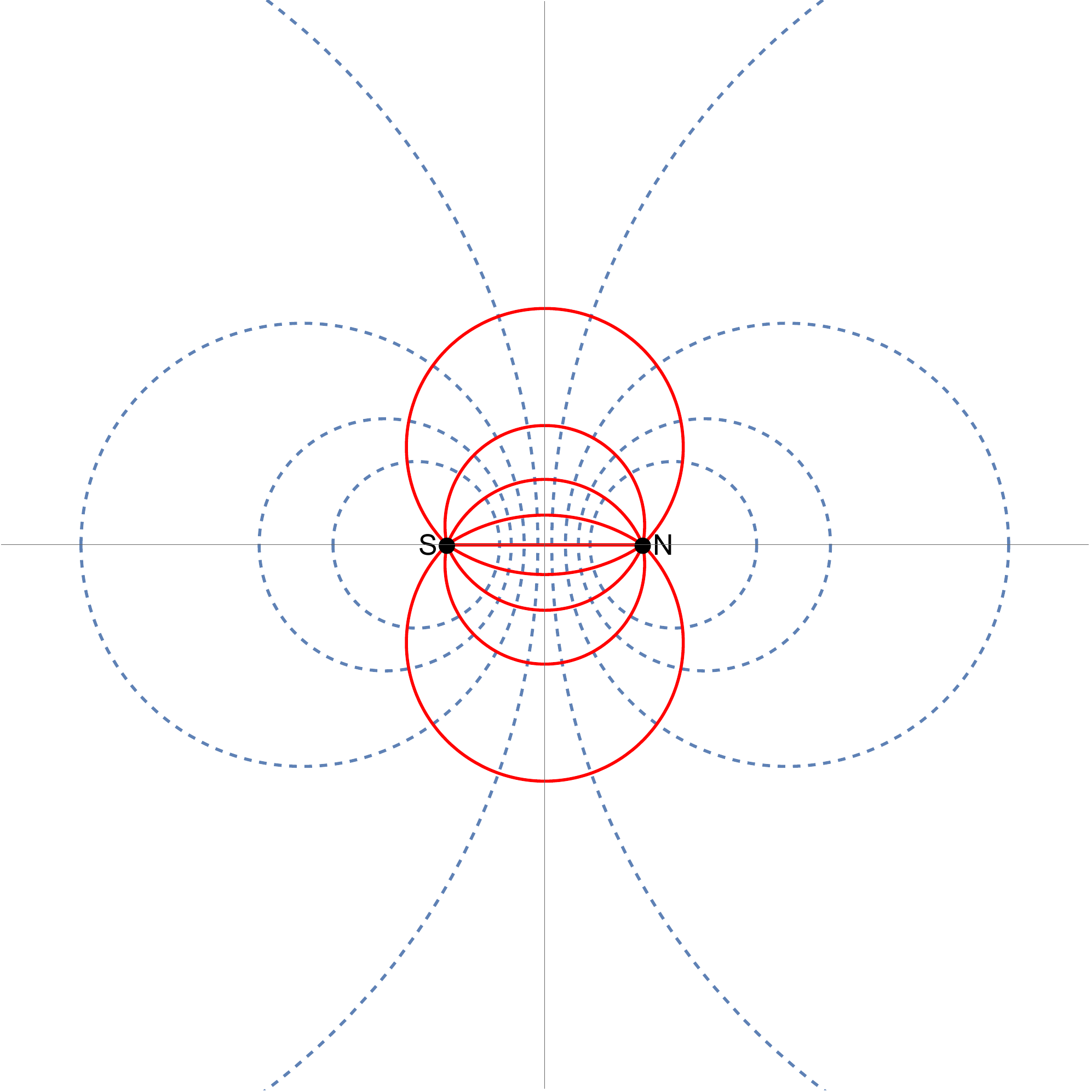}
	\caption{\label{fig:NS}N-S quantization picture. The dashed blue circles are the constant-$\sigma$ slices, and the red curves from the south pole to the north pole are the flows induced by $\partial_\sigma$.}
\end{figure}

We would like to make some remarks on why hyperbolic coordinates are good. \\ \\
\textbf{Good reflection operator} \\
In the Euclidean flat space, the reflection of a point is defined by $x=(x^0,\mathbf{x})\mapsto x^\theta=\fr{-\overline{x^0},\overline{\mathbf{x}}}$. By eq. (\ref{relation:cartesianhyperbolic}), the reflection is written in terms of hyperbolic coordinates as follows
\begin{equation}\label{def:cylreflection}
	\begin{split}
		x=(\sigma,\Omega)\mapsto x^\theta=(-\bar{\sigma},\Omega)
	\end{split}
\end{equation}
So it turns out that the flat space reflection corresponds to the cylinder reflection. Based on this observation, the formulations of the reflection positivity condition are exactly the same for the Euclidean flat space and Euclidean cylinder.\\ \\
\textbf{Good Wick rotation} \\
If we take imaginary cylinder time, i.e.\,$\sigma=it$, then by eq.\,(\ref{relation:cartesianhyperbolic}), the corresponding flat space coordinates are given by
\begin{equation}\label{cylinder:coordtransf:im}
	\begin{split}
		x^0=&i\dfrac{\sin{t}}{\cos{t}+\Omega^d}, \\
		x^i=&\dfrac{\Omega^i}{\cos{t}+\Omega^d},\quad i=1,2,\ldots,d-1. \\
	\end{split}
\end{equation}
We see that the imaginary cylinder time corresponds to the imaginary temporal variable and real spatial variables in the flat space. This is exactly eq.\,(\ref{cylinder:cyltoflat}) under substitution $x^0=ix^0_M,\,x^i=x^i_M$. So the Wick rotation of under the hyperbolic coordinates agrees perfectly with the Wick rotation under the Cartesian coordinates.

\section{CFT in the Euclidean cylinder}
For a scalar primary operator with scaling dimension $\Delta$, we establish the correspondence between the flat-space operator and the cylinder operator:
\begin{equation}\label{cylinder:def:op}
	\begin{split}
		\mathcal{O}_{cyl}(\sigma,\Omega)=\fr{\cosh\sigma+\Omega^d}^{-\Delta}\mathcal{O}_{flat}(x). \\
	\end{split}
\end{equation}
The scale factor comes from eq.\ (\ref{metric:confequiv}). Under this definition, the correlation functions of $\mathcal{O}_{cyl}$ are conformally invariant with respect to the cylinder metric $\fr{ds^2}_{cyl}=d\sigma^2+d\Omega^2$.

For the correlators in the flat space, the Wick rotation is defined in two steps: analytic continuation + taking the limit to the Lorentzian regime (see step 1 and step 2 in chapter \ref{strategy}). For cylinder correlators, we define the following sets of configurations:
\begin{equation}\label{cylinder:def:D}
	\begin{split}
		(\mathcal{D}_n)_{cyl}=&\left\{(x_1,\ldots,x_n)\in \fr{\mathbb{C}\times S^{d-1}}^n\,\Big{|}\,x_i=(\sigma_i,\Omega_i),\quad \mathrm{Re}(\sigma_i)>\mathrm{Re}(\sigma_{i+1})\,\mathrm{for}\,i=1,2,\ldots,n-1\right\}. \\
		(\mathcal{D}_n^E)_{cyl}=&\left\{c\in(\mathcal{D}_n)_{cyl}\,\Big{|}\,\mathrm{Im}(\sigma_i)=0\ \mathrm{for}\,i=1,2,\ldots,n\right\}.
	\end{split}
\end{equation} 
Then the Wick rotation of cylinder correlators is defined similarly by:

\tmtextbf{Step 1.} One finds an extension $G_n^E$ from $(\mathcal{D}^E_n)_{cyl}$ to a
function $G_n$ such that the following condition is
satisfied:
\begin{equation}
	\text{$G_n$ is defined on $(\mathcal{D}_n)_{cyl}$, is analytic in variables $\sigma_k$
		and is continuous in $\Omega_k$.} \label{Gasscylinder}
\end{equation}

\tmtextbf{Step 2.} Let $\sigma_k=\epsilon_k+it_k$ be the complex (Euclidean) cylinder temporal variable. Lorentzian correlators are defined as the limits of $G_n$
from inside $(\mathcal{D}_n)_{cyl}$ by sending $\epsilon_i \rightarrow 0$:
\begin{equation}
	G_n^L (x_1^L, \ldots, x_n^L) = \lim_{\epsilon_i \rightarrow 0} G_n (x_1,
	\ldots, x_n), \qquad x_k^L = \left( t_k, \Omega_k \right), \quad k = 1
	\ldots n \label{limitcylinder} .
\end{equation}
Now let us focus on the four-pont function. By conformal invariance, the four-point function of $\mathcal{O}_{cyl}$ can be written as
\begin{equation}\label{cylinder:4pt}
	\begin{split}
		\braket{\mathcal{O}_{cyl}(x_1)\mathcal{O}_{cyl}(x_2)\mathcal{O}_{cyl}(x_3)\mathcal{O}_{cyl}(x_4)}=\dfrac{g\fr{\rho,\bar{\rho}}}{2^{2\Delta}\left[\cosh\fr{\sigma_1-\sigma_2}-\Omega_1\cdot\Omega_2\right]^\Delta\left[\cosh\fr{\sigma_3-\sigma_4}-\Omega_3\cdot\Omega_4\right]^\Delta},
	\end{split}
\end{equation}
where the conformally invariant factor $g(\rho,\bar{\rho})$ is the same as eq.\,(\ref{def:Euclidean4-point}) in the Euclidean flat space. 

The analytic continuation procedure of the cylinder four-point function is analogous to section \ref{anal4-point}. Firstly, one can show that in the complex regime, $\cosh\fr{\sigma_1-\sigma_2}-\Omega_1\cdot\Omega_2\neq0$ as long as $\mathrm{Re}(\sigma_1)>\mathrm{Re}(\sigma_2)$.\footnote{Using the identity $\cosh(\epsilon+it)=\cosh\epsilon\cos t+i\sinh\epsilon\sin t$, we see that $\cosh(\epsilon+it)\notin[-1,1]$ as long as $\epsilon\neq0$.} So the prefactor in eq.\,(\ref{cylinder:4pt}) has analytic continuation to $(\mathcal{D}_n)_{cyl}$. Secondly, we have the series expansion (\ref{g:rhoexpansion}) (or (\ref{g:rhoexpansion2})) for $g(\rho,\bar{\rho})$. Using the same argument in section \ref{anal4-point}, the analytic continuation problem is reduced to showing that $z,\bar{z}\neq[1,+\infty)$, or equivalently, $\abs{\rho},\abs{\bar{\rho}}<1$ for configurations in $(\mathcal{D}_n)_{cyl}$. Then we need power-law bounds of the prefactor $[\cosh(\sigma_i-\sigma_j)-\Omega_i\cdot\Omega_j]^{-\Delta}$ and the cross-ratio quantities $(1-\abs{\rho})^{-1},(1-\abs{\bar{\rho}})^{-1}$ to show that the limit to the Lorentzian regime (step 2) defines a tempered distribution.

The estimate of $\fr{\cosh\fr{\sigma_1-\sigma_2}-\Omega_1\cdot\Omega_2}^{-1}$ is trivial:
\begin{equation}\label{cylinder:2ptbound}
	\begin{split}
		\abs{\dfrac{1}{\cosh\fr{\epsilon+i t}-\Omega_1\cdot\Omega_2}}\leqslant const\fr{1+\dfrac{1}{\epsilon}}^2\,e^{-\epsilon}.
	\end{split}
\end{equation}

We will give a proof of $\abs{\rho},\abs{\bar{\rho}}<1$ and the power-law bound of $\rho,\bar{\rho}$ in the next section. Our strategy will be the same as chapter \ref{secondpass}.

\section{Cauchy-Schwarz type inequality, power-law bound of $\rho,\bar{\rho}$} 
In chapter \ref{secondpass}, we derived Cauchy-Schwarz type inequality (\ref{maxrhoineq}) of the radial cross-ratio variables, using the Cauchy-Schwarz inequality of conformal blocks that appear in the generalized free theory. The same argument applies to the cylinder case. We let $(\mathcal{D}_4^{(0)})_{cyl}$ denote the set of cylinder configurations such that
\begin{equation}
	\begin{split}
		c=&(x_1,x_2,x_3,x_4), \\
		x_j=&(\sigma_j,\Omega_j),\quad\sigma_j=\epsilon_j+it_j, \\
		\epsilon_1>&\epsilon_2>0>\epsilon_3>\epsilon_4. \\
	\end{split}
\end{equation}
Each $c\in(\mathcal{D}_4^{(0)})_{cyl}$ has its corresponding reflection-symmetric configurations, given by
\begin{equation}
	\begin{split}
		c_{12}=(x_1,x_2,x_2^\theta,x_1^\theta),\quad
		c_{34}=(x_4^\theta,x_3^\theta,x_3,x_4).
	\end{split}
\end{equation}
Using the conformal block argument in section (\ref{rhorhobarProof}), we get
\begin{equation}\label{cylinder:maxrhoineq}
	\begin{split}
		\max\left\{\abs{\rho(c)},\abs{\bar{\rho}(c)}\right\}^2\leq\max\left\{\abs{\rho(c_{12})},\abs{\bar{\rho}(c_{12})}\right\}\times\max\left\{\abs{\rho(c_{34})},\abs{\bar{\rho}(c_{34})}\right\}
	\end{split}
\end{equation}
in the region where $\abs{\rho(c_{12})},\abs{\bar{\rho}(c_{12})},\abs{\rho(c_{34})},\abs{\bar{\rho}(c_{34})}<1$. To make eq.\,(\ref{cylinder:maxrhoineq}) true for all $c\in(\mathcal{D}_4^{(0)})_{cyl}$, we need an estimate on the $\rho,\bar{\rho}$ of reflection-symmetric configurations.

Let us consider the configuration $c_{12}$. We compute $u,v$ for $c_{12}$:
\begin{equation}\label{cylinder:uvsym}
	\begin{split}
		u=\dfrac{\abs{\cosh(\epsilon_1-\epsilon_2+i(t_1-t_2))-\Omega_1\cdot\Omega_2}^2}{\abs{\cosh(\epsilon_1+\epsilon_2+i(t_1-t_2))-\Omega_1\cdot\Omega_2}^2},\quad v=\dfrac{(\cosh(2\epsilon_1)-1)(\cosh(2\epsilon_2)-1)}{\abs{\cosh(\epsilon_1+\epsilon_2+i(t_1-t_2))-\Omega_1\cdot\Omega_2}^2}.
	\end{split}
\end{equation}
We see that (a) $u,v$ are real, (b) $0<u<1$, (c) $v>0$. Since $u=z\bar{z},\,v=(1-z)(1-\bar{z})$, we have $0<z,\bar{z}<1$.\footnote{By $u,v>0$ there are only three possibilties (a) $z,\bar{z}<0$, (b) $0<z,\bar{z}<1$, (c) $z,\bar{z}>1$. The last one is ruled out by $u<1$. By continuity we set $t_1=t_2$ and $\Omega_1=\Omega_2$, which gives $v<1$, so the first possibility is ruled out.} So we conclude that $\abs{\rho},\abs{\rho}<1$ for all reflection-symmetric configurations in $(\mathcal{D}_4^{(0)})_{cyl}$, thus eq.\,(\ref{cylinder:maxrhoineq}) holds for all $c\in(\mathcal{D}_4^{(0)})_{cyl}$. One can also check this by explicit computation of $z,\bar{z}$, using eq. (\ref{zzbarsolved}):
\begin{equation}
	\begin{split}
		z,\bar{z}=&\dfrac{\cosh^2\epsilon_1+\cosh^2\epsilon_2+\cos^2t+\cos^2\theta-2\cosh\epsilon_1\cosh\epsilon_2\cos t\cos\theta\pm2\sinh\epsilon_1\sinh\epsilon_2\sin t\sin\theta-2}{\fr{\cosh(\epsilon_1+\epsilon_2)\cos t-\cos\theta}^2+\sinh^2(\epsilon_1+\epsilon_2)\sin^2t}, \\
		(t=&t_1-t_2,\quad\cos\theta=\Omega_1\cdot\Omega_2.) \\
	\end{split}
\end{equation}
From this we get
\begin{equation}\label{cylinder:1-z}
	\begin{split}
		1-z,1-\bar{z}=\dfrac{2\sinh\epsilon_1\sinh\epsilon_2\fr{\cosh(\epsilon_1+\epsilon_2)-\cos(t\pm\theta)}}{\fr{\cosh(\epsilon_1+\epsilon_2)\cos t-\cos\theta}^2+\sinh^2(\epsilon_1+\epsilon_2)\sin^2t},
	\end{split}
\end{equation}
which implies the power-law bound\footnote{A naive estimate may give $\fr{1+{1}/{\epsilon_2}}^4$. By a careful estimate, one can get an upper bound
	\begin{equation*}
		\begin{split}
			\dfrac{\fr{\cosh(\epsilon_1+\epsilon_2)\cos t-\cos\theta}^2+\sinh^2(\epsilon_1+\epsilon_2)\sin^2t}{\cosh(\epsilon_1+\epsilon_2)-\cos(t\pm\theta)}\leqslant const\, e^{\epsilon_1+\epsilon_2},
		\end{split}
	\end{equation*}
which is regular at small $\epsilon$. So the only power-law singularities come from $\fr{\sinh\epsilon_1\sinh\epsilon_2}^{-1}$.}
\begin{equation}
	\begin{split}
		\dfrac{1}{1-z(c_{12})},\dfrac{1}{1-\bar{z}(c_{12})}\leq const\fr{1+\dfrac{1}{\epsilon_1}}\fr{1+\dfrac{1}{\epsilon_2}}\leqslant const\fr{1+\dfrac{1}{\epsilon_2}}^2.
	\end{split}
\end{equation}
Then by eq. (\ref{rhozest}) we get an estimate for $\rho$-variables
\begin{equation}\label{cylinder:rhosymest}
	\begin{split}
		\dfrac{1}{1-\rho(c_{12})},\dfrac{1}{1-\bar{\rho}(c_{12})}\leq const\fr{1+\dfrac{1}{\epsilon_2}}.
	\end{split}
\end{equation}
For a general configuration $c\in(\mathcal{D}_4)_{cyl}$, we shift the coordinates by time translation:
\begin{equation}
	\begin{split}
		x_k=(\epsilon_k+it_k,\Omega_k)\rightarrow x_k^\prime=(\epsilon_k-\dfrac{\epsilon_2+\epsilon_3}{2}+it_k,\Omega_k),
	\end{split}
\end{equation}
so the new configuration $c^\prime=(x_1^\prime,x_2^\prime,x_3^\prime,x_4^\prime)$ belongs to $(\mathcal{D}_4^{(0)})_{cyl}$ and it has the same $\rho,\bar{\rho}$ as $c$. Combining eqs. (\ref{cylinder:maxrhoineq}) and (\ref{cylinder:rhosymest}), we get an estimate of $\rho$-variables:
\begin{equation}\label{cylinder:rhobound}
	\begin{split}
		\dfrac{1}{1-\rho(c)},\dfrac{1}{1-\bar{\rho}(c)}\leq const\fr{1+\dfrac{1}{\epsilon_2-\epsilon_3}}\qquad\fr{c\in(\mathcal{D}_4)_{cyl}}.
	\end{split}
\end{equation}
This is the power-law bound we want. 

\section{Wightman axioms in the Minkowski cylinder}
Now we briefly demonstrate the Wightman axioms in the Minkowski cylinder, using the same argument as chapter \ref{sec:4-point}. In the cylinder case, the spherical coordinates $\Omega_k$'s are treated as function variables because these directions are compactified. \\ \\
\textbf{Temperedness} \\ \\
By eqs.\,(\ref{cylinder:2ptbound}), (\ref{cylinder:rhobound}) and (\ref{g:rhoexpansion}), the cylinder four-point function $G_4(c)$ has analytic continuation to the complex domain $(\mathcal{D}_4)_{cyl}$ (defined in eq.\,(\ref{cylinder:def:D})), and satisfies a power-law bound in $\epsilon_k-\epsilon_{k+1}$ (recall that the temporal variables are $\sigma_k=\epsilon_k+it_k$). Then by the first part of theorem \ref{ThVlad}, $G_4^L(t_k,\Omega_k)$ (defined by the limit (\ref{limitcylinder})) is a tempered distribution in the temporal variables $t_k$, and is continuous in spherical coordinates $\Omega_k$. 

Since the power-law bound is uniform for truncated conformal block expansions, the conformal block expansion converges in the sense of tempered distributions (by the same argument as theorem \ref{theorem:districonverge}). \\

\textbf{Conformal invariance} \\ \\
Using the same argument as section \ref{ConfMink}, one can show that $G_4^L(t_k,\Omega_k)$ satisfies the conformal Ward identities, i.e.\, the infinitesimal cylinder conformal symmetry holds. Since the conformal group action is globally well-defined on the Minkowski cylinder, the conformal Ward identities imply the global conformal symmetry of $G_4^L(t_k,\Omega_k)$. In other words, for any finite conformal transformation $x^\prime=f(x)$, the following identity holds:
\begin{equation}
	J(x_1)^{\Delta_{\mathcal{O}}} J(x_2)^{\Delta_{\mathcal{O}}} J(x_3)^{\Delta_{\mathcal{O}}} J(x_4)^{\Delta_{\mathcal{O}}} G^L_4 (x'_1, x'_2, x'_3, x'_4)
	= G^L_4 (x_1, x_2, x_3, x_4), \label{cylinder:finconfinv},
\end{equation}
where $J(x)=\mathrm{det}(\partial f^\mu/\partial x^\nu)^{1/d}$ is the local scale factor. \\

\textbf{Unitarity} \\ \\
Since the Euclidean flat-space reflection coincides with the Euclidean cylinder reflection (\ref{def:cylreflection}), the conjugation of the Euclidean cylinder operator is given by
\begin{equation}
	\begin{split}
		\fr{\mathcal{O}_{cyl}(\sigma,\Omega)}^\dagger=\mathcal{O}_{cyl}(-\bar{\sigma},\Omega).
	\end{split}
\end{equation}
We define the Minkowski cylinder operator by $\mathcal{O}^L(t,\Omega)=\mathcal{O}_{cyl}(it,\Omega)$. Then we have
\begin{equation}
	\begin{split}
		\fr{\mathcal{O}^L(t,\Omega)}^\dagger=\mathcal{O}^L(t,\Omega).
	\end{split}
\end{equation}
Since the reflection positivity condition in the Euclidean cylinder is formulated in the same way as in the Euclidean space, the unitarity (Wightman positivity) condition in the Minkowski cylinder is also formulated in the same way as in the Minkowski space, i.e.\, eqs.\,(\ref{Wightman:inner}) and (\ref{Wightman:positivity}). \\

\textbf{Mircocausality} \\ \\
The microcausality condition in the Minkowski cylinder is with respect to the causal ordering (\ref{def:cylcausalordering}). Inside the four-point function,
\begin{equation}\label{cylinder:microcausality}
	\begin{split}
		[\mathcal{O}^L(t_k,\Omega_k),\mathcal{O}^L(t_{k+1},\Omega_{k+1})]=0\quad\mathrm{if}\quad\abs{t_k-t_{k+1}}<\arccos(\Omega_k\cdot\Omega_{k+1}).
	\end{split}
\end{equation}
Similarly to the argument in section \ref{localCFT}, it suffices to show that $\abs{\rho},\abs{\bar{\rho}}<1$ for the following three types of configurations:
\begin{itemize}
	\item[]$k=1$: $\epsilon_1=\epsilon_2>\epsilon_3>\epsilon_4$ and $\abs{t_1-t_2}<\arccos(\Omega_1\cdot\Omega_2)$;
	\item[]$k=2$: $\epsilon_1>\epsilon_2=\epsilon_3>\epsilon_4$ and $\abs{t_2-t_3}<\arccos(\Omega_2\cdot\Omega_3)$;
	\item[]$k=3$: $\epsilon_1>\epsilon_2>\epsilon_3=\epsilon_4$ and $\abs{t_2-t_3}<\arccos(\Omega_2\cdot\Omega_3)$.
\end{itemize}
Using eq. (\ref{cylinder:1-z}), we see that the condition for $\abs{\rho},\abs{\bar{\rho}}<1$ can actually be relaxed to $\epsilon_1,\epsilon_2>\epsilon_3,\epsilon_4$. This applies to the cases of $k=1$ and $k=3$. 

For $k=2$, we shift the configuration to $\epsilon_2=\epsilon_3=0$, i.e.\,$\epsilon_1>0>\epsilon_4$. To analyze the $\rho$-variables, we map the cylinder configuration to the flat space by eq.\,(\ref{relation:cartesianhyperbolic}) (the cross-ratios do not change). Then by eqs. (\ref{cylinder:2ptdistance}) and (\ref{cylinder:coordtransf:im}), $x_2$ and $x_3$ are mapped to two space-like separated points in Minkowski space. According to the $k=2$ argument in section \ref{localCFT}, it suffices to show that the flat-space images of $x_1$ and $x_4$ satisfy
\begin{equation}\label{cylinder:condx1x4}
	\begin{split}
		\mathrm{Re}(x_1^0)>\abs{\mathrm{Im}(\mathbf{x}_1)},\quad \mathrm{Re}(x_4^0)<-\abs{\mathrm{Im}(\mathbf{x}_4)}.
	\end{split}
\end{equation} 
By (\ref{relation:cartesianhyperbolic}), we have
\begin{equation}
	\begin{split}
		\mathrm{Re}(x_1^0)=&\dfrac{\sinh\epsilon_1(\cosh\epsilon_1+\cos t_1\Omega_1^d)}{\fr{\cosh\epsilon_1\cos t_1+\Omega_1^d}^2+\sinh^2\epsilon_1\sin^2t_1}, \\
		\mathrm{Im}(x_1^i)=&-\dfrac{\sinh\epsilon_1\sin t_1\Omega_1^i}{\fr{\cosh\epsilon_1\cos t_1+\Omega_1^d}^2+\sinh^2\epsilon_1\sin^2t_1}, \quad i=1,\ldots,d-1,\\
	\end{split}
\end{equation}
which implies that
\begin{equation}
	\begin{split}
		\mathrm{Re}(x_1^0)-\abs{\mathrm{Im}(\mathbf{x}_1)}=&\dfrac{\sinh\epsilon_1(\cosh\epsilon_1+\cos t_1\Omega_1^d-\abs{\sin t_1}\sqrt{1-(\Omega_1^d)^2})}{\fr{\cosh\epsilon_1\cos t_1+\Omega_1^d}^2+\sinh^2\epsilon_1\sin^2t_1} \\
		\geqslant&\dfrac{\sinh\epsilon_1(\cosh\epsilon_1-1)}{\fr{\cosh\epsilon_1\cos t_1+\Omega_1^d}^2+\sinh^2\epsilon_1\sin^2t_1}>0. \\
	\end{split}
\end{equation}
The computation for $x_4$ is similar. So the condition (\ref{cylinder:condx1x4}) holds when $\epsilon_1>0>\epsilon_4$. Now we are in the same situation as the $k=2$ case in section \ref{localCFT}, so we conclude that $\abs{\rho},\abs{\bar{\rho}}<1$ in this case. This finishes the proof of eq. (\ref{cylinder:microcausality}). \\

\textbf{Cluster property} \\ \\
Recall that the cluster property talks about the asymptotic behavior of the correlator when the two clusters of operators are separated very far from each other in the space-like direction, there is no cluster property for the cylinder CFT since the spatial directions are compactified. \\

\textbf{Spectral condition} \\ \\
As a consequence of the power-law bound and the second part of theorem \ref{ThVlad}, $G_4^L(t_k,\Omega_k)$ has the Fourier transform (in the sense of distributions) in temporal variables:
\begin{equation}
	\begin{split}
		G_4^L(t_k,\Omega_k)=\int\dfrac{d E_1\,d E_2\, d E_3}{(2\pi)^3}\,F(E_1,E_2,E_3;\Omega_k)e^{-i(E_1t_1+E_2t_2+E_3t_3)},
	\end{split}
\end{equation}
where $F$ is a tempered distribution in $E_k$'s and continuous in $\Omega_k$'s. $F$ is supported in the region $E_1,E_2,E_3\geqslant0$.

\section{Matching the four-point functions in the flat space and in the cylinder}\label{section:cylinder:matching}
When we define the correlators in the Minkowski space and in the Minkowski cylinder, the Wick rotations start from the same correlation function but they are performed in different coordinates. As we mentioned at the beginning of this chapter, the Minkowski space is conformally embedded into the Poincaré patch of the Minkowski cylinder. Then a question naturally arises: do the correlators in the Poincaré patch agree with the ones in the Minkowski space, up to the scale factors in eq. (\ref{cylinder:def:op})? In this section, we claim that the answer is yes.

To show this claim, we would like to introduce the following lemma:
\begin{lemma}\label{lemma:pathmatching}
	For any $n$-point configuration $c=(x_1,x_2,\ldots,x_n)$, if all the points of $c$ are in the Poincaré patch, i.e. $x_k=(it_k,\Omega_k)$ is under condition (\ref{cylinder:cond:poincare}), then there exists a collection of $\epsilon_k$'s such that
	\begin{enumerate}
		\item$\epsilon_1>\epsilon_2>\ldots>\epsilon_n>0$.
		\item The the cylinder path $c(s)$ with
		\begin{equation}
			\begin{split}
				x_k(s)=((1-s)\epsilon_k+ist_k,\Omega_k),\quad 0\leqslant s<1,\quad k=1,2,\ldots,n
			\end{split}
		\end{equation}
	is mapped into the forward tube $\mathcal{T}_n$ in the flat space, i.e. $\mathrm{Re}(x_k^0(s))-\mathrm{Re}(x_{k+1}^0(s))>\abs{\mathrm{\mathbf{x}_k(s)-\mathbf{x}_{k+1}(s)}}$ for $k=1,2,\ldots,n-1$.
	\end{enumerate}
\end{lemma}
\begin{proof}
	Let $\eta_1>\eta_2>\ldots>\eta_n>0$ be a collection of positive numbers, we choose $\epsilon_k$'s to be
	\begin{equation}
		\begin{split}
			\epsilon_k=\lambda\eta_k
		\end{split}
	\end{equation}
    with a positive real number $\lambda$. For sufficiently small $\lambda$, we have
    \begin{equation}\label{cylinder:path}
    	\begin{split}
    		\mathrm{Re}(x_k^0(s))=&\dfrac{\sinh((1-s)\lambda\eta_k)[\cosh((1-s)\lambda\eta_k)+\cos (st_k)\Omega_k^d]}{\fr{\cosh((1-s)\lambda\eta_k)\cos (st_k)+\Omega_k^d}^2+\sinh^2((1-s)\lambda\eta_k)\sin^2(st_k)} \\
    		=&\dfrac{(1-s)\lambda\eta_k[1+\Omega_k^d\cos(st_k)]}{[\cos(st_k)+\Omega_k^d]^2}\fr{1+O(\lambda^2\eta_k^2)}, \\
    		\mathrm{Im}(x_k^i(s))=&-\dfrac{\sinh((1-s)\lambda\eta_k)\sin (st_k)\Omega_k^i}{\fr{\cosh((1-s)\lambda\eta_k)\cos (st_k)+\Omega_k^d}^2+\sinh^2((1-s)\lambda\eta_k)\sin^2(st_k)} \\
    		=&-\dfrac{(1-s)\lambda\eta_k\Omega_k^i\sin(st_k)}{[\cos(st_k)+\Omega_k^d]^2}\fr{1+O(\lambda^2\eta_k^2)}. \\
    	\end{split}
    \end{equation}
   As long as $x_k=(it_k,\Omega_k)$ is in the Poincaré patch, the denominators in eq. (\ref{cylinder:path}) are always bounded from below by $\cos(t_k)+\Omega_k^d$, which is strictly positive. To make the path mapped into the forward tube, we require that
   \begin{equation}
   	\begin{split}
   		\fr{\dfrac{\eta_k[1+\Omega_k^d\cos(st_k)]}{[\cos(st_k)+\Omega_k^d]^2},-\dfrac{\eta_k\Omega_k^i\sin(st_k)}{[\cos(st_k)+\Omega_k^d]^2}}\succ\fr{\dfrac{\eta_{k+1}[1+\Omega_{k+1}^d\cos(st_{k+1})]}{[\cos(st_{k+1})+\Omega_{k+1}^d]^2},-\dfrac{\eta_{k+1}\Omega_{k+1}^i\sin(st_{k+1})}{[\cos(st_{k+1})+\Omega_{k+1}^d]^2}}
   	\end{split}
   \end{equation}
for all $s\in[0,1]$ and $k=1,2,\ldots,n-1$. Here $``x\succ y"$ means that $x$ is in the future light-cone of $y$. The above condition is easily satisfied by choosing $\eta_1\gg\eta_2\gg\ldots\gg\eta_n$.
\end{proof}
Since the Wick rotations of the flat-space correlator and the cylinder correlator start from the same Euclidean correlator (up to the scale factor), by choosing the analytic continuation path constructed in lemma \ref{lemma:pathmatching} and by the uniqueness of analytic continuation, we see that the CFT correlator in the Poincaré patch is the same as in the Minkowski space, up to the scale factors.

\begin{remark}
	The result in this section is valid for all $n$-point correlation functions. In the case of four-point functions, one can check it explicitly by computing the phase factors of $\rho$-variables.
\end{remark}

\section{OPE convergence in the sense of functions}\label{section:cylinderope}
In part \ref{part:ope}, we classify the four-point configurations in the Minkowski space and show the OPE convergence properties of each configuration class in appendix \ref{appendix:tableopeconvergence}. Our analysis was basically computing the range of cross-ratio $z,\bar{z}$ and the numbers $N_t,N_t$ that count how many times the analytic continuation path crosses the t-channel and u-channel cuts.

The same analysis works for the four-point configurations in the Minkowski cylinder. However, there are infinitely many such classes in the cylinder case because each configuration in the Minkowski space corresponds to infinitely many preimages in the Minkowski cylinder, and the four-point function may take different values at different preimages. For this reason, we cannot finish this work directly. 

In this section, we would like to argue that the problem of classifying cylinder four-point configurations can be reduced to the cases of configurations the Poincaré patch. 

For convenience, we use $x=(t,\Omega)$, instead of $x=(it,\Omega)$, to denote Minkowski cylinder points. The cross-ratios are given by the formula
\begin{equation}
	u = \frac{x_{12}^2 x_{34}^2}{x_{13}^2 x_{24}^2}, \qquad v = \frac{x_{14}^2
		x_{23}^2}{x_{13}^2 x_{24}^2}, \label{def:uv}
\end{equation}
where
\begin{equation}
	x_{i j}^2 = \underset{\varepsilon \rightarrow 0^+}{\lim}  [\tmop{ch}
	[\varepsilon + i (t_i - t_j)] - \Omega_i \cdummy \Omega_j], \qquad (i < j) .
	\label{def:xij}
\end{equation}
From above we can compute the corresponding radial variables $\rho,
\bar{\rho}$ and their phases.

By eq. (\ref{cylinder:4pt}), the four-point function in the Minkowski cylinder has the following s-channel expansion
\begin{equation}\label{cylinder:schannel}
	\begin{split}
		G^L_4(c)=\dfrac{g\fr{\rho,\bar{\rho}}}{2^{2\Delta}\left[\cos\fr{t_1-t_2}-\Omega_1\cdot\Omega_2\right]^\Delta\left[\cos\fr{t_3-t_4}-\Omega_3\cdot\Omega_4\right]^\Delta},
	\end{split}
\end{equation}
where $g(\rho,\bar{\rho})$ given by the power series (\ref{g:rhoexpansion}). Analogous to eq. (\ref{ope:tuchannel}), we also have the t-channel and u-channel expansions. The above formula and its t-/u-channel version imply that in the Minkowski cylinder, the light-cone singularities are the configurations where
\begin{equation}
	\begin{split}
		\cos(t_i-t_j)=\Omega_i\cdot\Omega_j
	\end{split}
\end{equation}
for some $x_i,x_j$ pair. In the following discussion we exclude the light-cone singularities.

Given a four-point configuration $c = (x_1, x_2, x_3, x_4)$, if all $x_k$'s are in
the Poincaré patch, then according to section \ref{section:cylinder:matching}, the OPE convergence properties are the same as in the Minkowski space.

Now we consider the general case, when $x_k$'s are not necessarily in the Poincaré
patch. Without loss of generality we assume that
\begin{equation}\label{cylinder:timecondition}
	\begin{split}
		\cos(t_k)+\Omega_k^d\neq0,\qquad(k=1,2,3,4.)
	\end{split}
\end{equation}
Otherwise we can always shift $c$ by a small time translation, which does not change the four-point function. 

The cylinder conformal group has a $\mathbb{Z}$-center, generated by
the following map
\begin{equation}
	x = (t, \Omega) \mapsto T x \assign (t + \pi, - \Omega) .
\end{equation}
\begin{lemma}
	\label{lemma:uvtransform}Let $c = (x_1, x_2, x_3, x_4)$ be a four-point
	configuration in the Minkowski cylinder, $n = (n_1, n_2, n_3, n_4)$ and $T^n c =
	(T^{n_1} x_1, T^{n_2} x_2, T^{n_3} x_3, T^{n_4} x_4)$. Assuming that $x_{i
		j}^2 \neq 0$ for all $i < j$, we have
	\begin{equation}
		\log (u (T^n c)) = \log (u (c)) + 2 \pi i (n_3 - n_2), \qquad \log (v (T^n
		c)) = \log (v (c)) . \label{transform:uv}
	\end{equation}
\end{lemma}

Given a (Euclidean) cylinder point $x = (\sigma, \Omega) \in \mathbb{R} \times S^{d - 1}$, we define. The cross-ratios are given by the formula
\begin{proof}
	By eq.\,(\ref{def:xij}) we have
	\begin{equation}
		\log [(T^{n_i} x_i - T^{n_j} x_j)^2] = \log (x_{i j}^2) + \pi i (n_i -
		n_j), \qquad (i < j) .
	\end{equation}
	Then summing over all the $x_{ij}^2$ contributions in eq.\,(\ref{def:uv}), we get (\ref{transform:uv}).
\end{proof}

\begin{remark}
	The causal ordering of $T^n c$ may be different from the causal ordering of
	$c$.
\end{remark}

\begin{remark}
	In eq. (\ref{transform:uv}), there is no contribution from $n_1$ and $n_4$.
	Therefore, shifting the first or the last operator by the center of the conformal group does not change $g(\rho,\bar{\rho})$ in eq.\ (\ref{cylinder:schannel}). However it produces some overall phase factor via the
	prefactor $(x_{12}^2 x_{34}^2)^{-\Delta}$. 
\end{remark}

Let $N_t$ and $N_u$ count the total (signed) number that $z$ and $\bar{z}$
cross the t- and u-channel cuts from above during the analytic continuation
from Euclidean. Then lemma \ref{lemma:uvtransform} implies
\begin{equation}
	N_t (T^n c) = N_t (c) + n_3 - n_2, \quad N_u (T^n c) = N_u (c) - n_3 + n_2 .
	\label{transform:NtNu}
\end{equation}
Therefore, the OPE classification problem in Minkowski cylinder is reduced to
the same problem in Minkowski space, via the following procedure:
\begin{itemize}
	\item Find $n = (n_1, n_2, n_3, n_4)$ such that $T^{- n} c$ is in the
	Poincar{\'e} patch.
	
	\item Find the causal ordering of $T^{- n} c$, compute $N_t (c), N_u (c)$ by
	eq. (\ref{transform:NtNu}).  \\
	(Here we assume that $N_t (T^{- n} c), N_u (T^{-
		n} c)$ are already known as the Minkowski space results.)
	
	\item Determine the OPE convergence properties in s-, t- and u-channels.
\end{itemize}
\tmtextbf{Example: bulk-point singularity}

A bulk-point singularity is the configuration $c = (x_1, x_2, x_3, x_4)$ with
\begin{equation}
	x_1 = (0, \Omega), \quad x_2 = (0, - \Omega), \quad x_3 = (\pi, \Omega'),
	\quad x_4 = (\pi, - \Omega') . \label{bulkpt}
\end{equation}
According to the above described scheme, we first find the proper $n$ such that
$T^{- n} c$ is in the Poincar{\'e} patch. The result is $n = (0, 0, 1, 1)$,
and $T^{- n} c$ is a totally space-like configuration because all its points
are in the time slice $t = 0$. By explicit computation we get
\[ u (c) = \frac{4}{(1 + \Omega \cdummy \Omega')^2}, \quad v (c) = \left(
\frac{1 - \Omega \cdummy \Omega'}{1 + \Omega \cdummy \Omega'} \right)^2, \]
which implies that $z (c) = \bar{z} (c) = \frac{2}{1 + \Omega \cdummy \Omega'}
> 1$. So the s-channel OPE does not converge. For t- and u-channel OPE, since
we know that in the Minkowski space, the totally space-like configuration with
$z = \bar{z} > 1$ has $N_t = N_u = 0$, by (\ref{transform:NtNu}) we get
\begin{equation}\label{cylinder:Nbulkpt}
	\begin{split}
		N_t (c) = 1, \quad N_u (c) = - 1. 
	\end{split}
\end{equation}
So neither t- nor u-channel OPE converges. We conclude that there is no convergent OPE channel for bulk-point singularities.

As we hinted in section \ref{section:bulkptcausalorder} that in the Minkowski space, the causal ordering (\ref{causal:bulkpointsing}) with $z,\bar{z}>1$ has the same OPE convergence properties as the bulk-point singularities. Here we confirmed that indeed $z,\bar{z}>1$ for bulk-point singularities. Also, the results of $N_t$ and $N_u$ here are the same as eq.\,(\ref{cylinder:NtNu}) there.

\end{part}

\nnchapter{Conclusions and Outlooks}\label{conclusions}
\thispagestyle{empty}

In this thesis we have presented studies of functional and distributional properties of CFT four-point functions. An important feature of our analysis is that we only rely on Euclidean assumptions made in the modern conformal bootstrap approach, without assuming Wightman nor Osterwalder-Schrader axioms. 

In part \ref{part:crossratio}, we studied the CFT four-point functions in the cross-ratio space. We start with the well-known conformal block expansion of the CFT four-point function, which is convergent when the radial cross-ratios $\rho,\bar{\rho}$ are in the open unit disk. Using Vladimirov's theorem, we showed that both the correlation functions and conformal blocks are tempered distributions on the boundary $\abs{\rho}=\abs{\bar{\rho}}=1$, i.e. they make sense when being smeared with a smooth test function. We showed that the conformal block expansion converges in the sense of distributions. In other words, the pairing of the correlation function with a test function can be evaluated by summing over the contributions from conformal blocks. 

In the Lorentzian signature, there are four-point configurations where the correlation functions only converge in the distributional sense. In these cases, we can only compute the correlation function by pairing it with various test functions. Therefore, these results are important for understanding the nature of the correlation functions in the Lorentzian signature.

The results of part \ref{part:crossratio} also gives us a hint at a uniform description of the bootstrap functionals that in the conformal bootstrap. From the distribution point of view, the bootstrap functionals correspond to test functions and the swapping property corresponds to OPE convergence in the sense of distributions. We managed to rewrite some bootstrap functionals (e.g.\, derivatives at $z=\bar{z}=1/2$) in terms of smooth test functions. However, the space of smooth test functions is too small to enough to contain all the bootstrap functionals. In the future research, it is would be interesting to find a sufficiently large space of test functions, which contains all the bootstrap functionals. A better understanding of this space will be helpful in deepening the analytical understanding of conformal bootstrap.

In part \ref{part:minkowski}, we studied the relation of the modern Euclidean CFT axioms to the more traditional Osterwalder-Schrader and Wightman axioms. In our version of Euclidean CFT axioms, we only assumed a minimal set of assumptions, including real analyticity, conformal invariance, unitarity condition of two-point functions, reality constraints of three-point functions and a very weak form of convergent OPE. We showed that at least for two-, three- and four-point functions, the Euclidean CFT axioms imply both OS and Wightman axioms. In particular, an important conclusion is that the CFT four-point functions are tempered distributions in Lorentzian signature, and the conformal invariance holds in the infinitesimal sense. We also showed that the Lorentzian CFT four-point functions have convergent s-channel OPE in the sense of distributions, generalizing our results in part \ref{part:crossratio} and giving a derivation of classic Lorentzian CFT assumptions (by Mack \cite{Luscher:1974ez}) from Euclidean axioms. 

We noticed that the radial cross-ratios satisfy $\abs{\rho},\abs{\bar{\rho}}<1$ for four-point configurations in the forward tube, and $\abs{\rho},\abs{\bar{\rho}}\leqslant1$ in the Lorentzian regime. This agrees with the range of $\rho,\bar{\rho}$ we considered in part \ref{part:crossratio}.

In part \ref{part:ope}, we revisited the classic problem about the domain of analyticity of QFT correlators. We focus on CFT four-point functions which assume the Euclidean CFT axioms as in the previous parts. We established criteria of OPE convergence in s-, t- and u-channels. In Lorentzian signature, we gave a complete classification of the four-point configurations according to their causal orderings and range of cross-ratio variables $z,\bar{z}$. We checked the OPE convergence properties case by case according to the classification and gave the tables of results in appendix \ref{appendix:tableopeconvergence}. Our results show that the domain of a general CFT four-point function is much bigger than general QFT four-point functions. On the other hand, there are still many regions of Lorentzian four-point configurations where there is no convergent OPE channel.

In our analysis, we used variables $\rho,\bar{\rho}$ to probe OPE convergence properties. It would be interesting if there are some other coordinates that can be used to discover more domain of CFT four-point functions. It is also an open question whether the old results in multi-variable complex analysis, such as envelope of analyticity, can be applied to this problem. 

In part \ref{part:generalization}, we briefly previewed on the generalizations to two directions. The first generalization is about the CFT four-point functions with arbitrary bosonic spinning operators. We showed that for generic internal scaling dimensions, the conformal partial wave has analytic continuation to the forward tube, and it satisfies a Cauchy-Schwarz type inequality. There is a discrete set of exceptional internal scaling dimensions, for which we do not know whether the same conclusions hold. Based on the existence of conformal partial waves in the Euclidean signature and its continuity as a function of internal scaling dimension, we conjecture that the answer is yes if the scaling dimension is above the unitarity bound. The analyticity and power-law bound of the spinning four-point functions follow from these two properties of the conformal partial waves. The remaining derivation of the Wightman axioms are the same as the scalar case.

The second generalization is about the scalar CFT four-point functions in the Minkowski cylinder. Using the same technique as the scalar four-point function in the flat space, we proved Wightman axioms, conformal invariance and OPE convergence in the sense of distributions. Since the conformal transformations are globally well-defined in the Minkowski cylinder, the conformal invariance of the cylinder four-point functions hold not only in the infinitesimal form, but also in the finite form. We showed that up to scaling prefactors, the $n$-point functions in the Poincaré patch agrees with $n$-point functions in the Minkowski space. 

We also generalized the OPE classification results in part \ref{part:ope} to the cylinder case. Using the center elements of the cylinder conformal group, we can map a generic cylinder four-point configuration into the Poincaré patch. Using such maps, we show that for the OPE convergence properties in the sense of functions, the classification of the four-point configurations in the Minkowski cylinder can be reduced to the case of the Poincaré patch, i.e.\,the results in part \ref{part:ope}.



\appendix
\chapter{Appendices of Part \ref{part:crossratio}}
\section{Lorentzian 4pt correlator with no convergent OPE channel}
\label{app:lorentz}
In this section we will give an example of a Lorentzian 4pt configuration in which there's no convergent OPE channel. For simplicity let's consider the correlators of identical scalar operators. Recall that, in a general QFT, Lorentzian correlators can be recovered from Euclidean correlators by analytic continuation. Starting from a configuration of Euclidean points $x_i=(\tau_i,\mathbf{x}_i)$ with ordered times 
\be
\label{eq:or0}
\tau_1>&\tau_2>\ldots>\tau_n\,,
\ee
we analytically continue each time variable as $\tau_i = \eps_i+i t_i$ and take the limit $\eps_i\to0$, preserving the ordering of real parts. The result is interpreted as the Lorentzian correlator at (Lorentzian) points $y_i=(t_i,\mathbf{x}_i)$. Schematically:
\begin{equation}
	\begin{split}
		\<0|\phi(t_1,\mathbf{x}_1)\ldots\phi(t_n,\mathbf{x}_n)|0\>:=\lim\limits_{\substack{\eps_i\rightarrow0 \\\eps_1>\ldots>\eps_n}}\<\phi(\eps_1+it_1,\mathbf{x}_1)\ldots\phi(\eps_n+it_n,\mathbf{x}_n)\>
	\end{split}
\end{equation}
Now we will apply this to a 4pt function in a CFT. In a CFT, this analytic continuation can be performed starting from Eq.~(\ref{eq:4pt}). We just complexify all Euclidean times as described above, and then take the limit. It is easy to see (exercise) that the distances $x_{ij}^2$ do not vanish in this process, except perhaps at the very end if the Lorentzian points $y_i$ are lightlike separated. We will be interested in the case when all points are spacelike or timelike separated. So the prefactor in Eq.~(\ref{eq:4pt}) is thus analytically continued (notice that there is an interesting phase for timelike separation).

In order to analytically continue the factor $g(u,v)$, we will use the existence of the conformal block expansion (\ref{eq:cbexpansion}) which as mentioned there is convergent for $|\rho|,|\bar{\rho}|<1$ (``OPE convergence region''). Concretely, we are instructed to compute $u,v$ corresponding to complexified Euclidean times, then evaluate $z,\bar{z}$ defined by~\eqref{eq:zzbar0}, which gives 
\be
z,\bar{z}=\frac{1}{2}(1+u-v\pm\sqrt{(1+u-v)^2-4u}),
\ee
then evaluate the corresponding $\rho,\bar\rho$ via~\eqref{eq:rho}, and finally stick these into the expansion (\ref{eq:cbexpansion}). This procedure defines an analytic function of $\tau_i$ as long as $|\rho|,|\bar{\rho}|<1$.\footnote{Note that even though $z,\bar{z}$ will have a branch point when $(1+u-v)^2-4u=0$, the function $g(u,v)$ is symmetric under the intercharge of $z,\bar{z}$ and will remain analytic as a function of complexified Euclidean times.} The question then is if this condition will hold all along the analytic continuation curve needed to recover the Lorentzian correlator, including the endpoint. If this happens, Lorentzian correlator can be computed by summing up a convergent expansion, in particular it is non-singular. 

Above we describe how to use the $s$-channel expansion for the analytic continuation. A priori we can also use the $t$- and $u$-channels for this purpose, starting from the $t$- and $u$-channel versions of Eq.~(\ref{eq:4pt}): 
\begin{equation}
	\label{eq:4ptt}
	\<\f(x_1)\f(x_2)\f(x_3)\f(x_4)\>=\frac{1}{(x_{23}^2)^{\De_\f}(x_{14}^2)^{\De_\f}}g(u_t,v_t)  =\frac{1}{(x_{13}^2)^{\De_\f}(x_{24}^2)^{\De_\f}}g(u_u,v_u).
\end{equation}
The cross ratios $u_t,v_t$ are obtained from $u,v$ via $x_1\leftrightarrow x_3$, and $u_u,v_u$ via $x_2\leftrightarrow x_3$. The functions $g(u_t,v_t)$, $g(u_u,v_u)$ can be computed via the corresponding conformal block expansions with their own regions of analyticity set by the conditions $|\rho_t|,|\bar{\rho}_t|<1$ and $|\rho_u|,|\bar{\rho}_u|<1$. 

It is not a priori clear and requires a separate analysis, which OPE channel, if any, is convergent for a given Lorentzian configuration. The answer turns out to depend, generically, only on the causal structure of the configuration (who is timelike, who is spacelike). 
The OPE can stop converging in two ways: either at the end point of the analytic continuation, or somewhere along the way. As we will show in~\cite{Kravchuk:2021kwe}, for the $s$-channel we always have $|\rho|,|\bar\rho|\le 1$, so OPE converges along the way but may diverge at the end point. For other channels the OPE may start diverging already along the way.

We will give an exhaustive discussion of these phenomena, for all possible causal structures, in a later publication~\cite{Qiao:2020bcs}. Here we will just give an extreme example of a configuration where all channels diverge. 

Consider the causal ordering 
\be
\label{eq:ord}
y_3\rightarrow y_1\rightarrow y_4\rightarrow y_2\,,
\ee
where $y_i\to y_j$ means that $y_i$ is in the past open lightcone of $y_j$. We pick some points $(i t_i, \mathbf{x}_i)$ corresponding to this causal ordering, as well as some initial Euclidean times $\eps_i$ satisfying the ordering $\eps_1>\eps_2>\eps_3>\eps_4$, and consider a curve of complexified points corresponding to these initial and final positions. E.g.~we can use linear interpolation:
\be
x_i(\theta) = ( (1-\theta)\eps_i + \theta it_i , \mathbf{x}_i),\qquad \theta\in[0,1]\,.
\ee
We choose the initial point with $|\rho|,|\bar\rho|<1$, and we would like to see if this condition stays true along this curve. For this it is enough to evaluate $z,\bar z$ and see if they cross the cut $[1,+\infty)$ which corresponds to $|\rho|=1$. This is how the check is carried out in practice for the $s$-channel. For the $t$- and $u$-channel, we have the same check in terms of $z_t,\bar{z}_t$ and $z_u,\bar{z}_u$. But in fact we have relations 
\be
z_t= 1-z,\qquad z_u= 1/z
\ee
and similarly for $\bar z$. These relations map the $[1,+\infty)$ cut on $(-\infty,0]$ and $[0,1]$, respectively.
Thus we don't have to redo the analysis for $z_t,\bar{z}_t$ and $z_u,\bar{z}_u$ separately, we just have to watch if the $s$-channel $z,\bar z$ crosses these additional cuts to conclude about the convergence of the $t$- and $u$-channel OPEs.

\begin{figure}[H]
	\centering
	\includegraphics[scale=0.6]{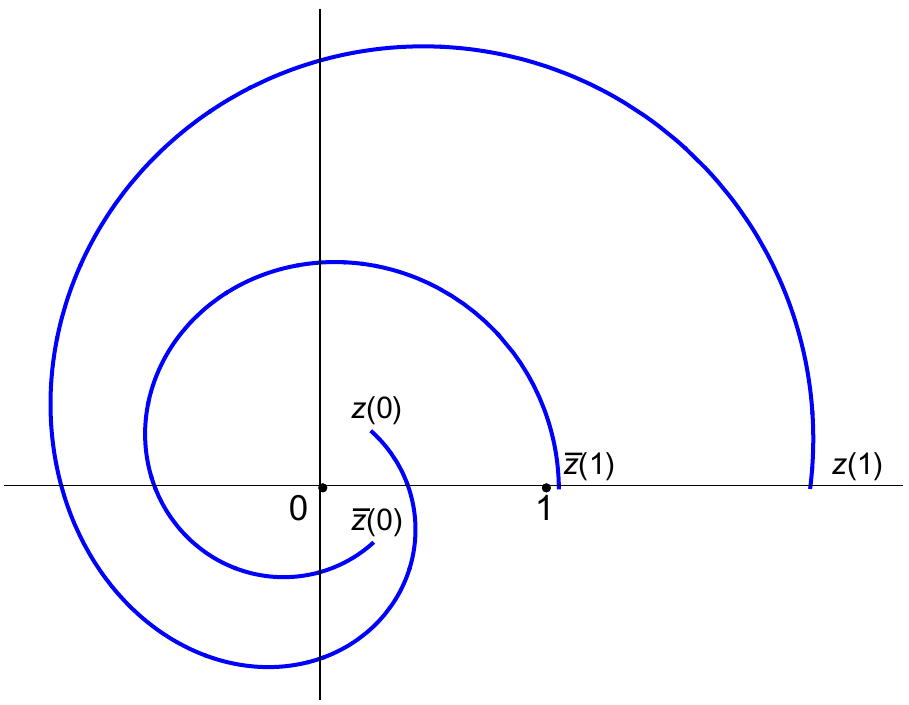}
	\caption{\label{plot-orbit3142}The curves $z(\theta)$ and $\bar{z}(\theta)$ for the causal ordering $y_3\rightarrow y_1\rightarrow y_4\rightarrow y_2$.}
\end{figure}

In practice, we just pick some numerical values for the initial and final points (respecting the orderings), plot the curves $z(\theta)$, $\bar{z}(\theta)$ and see what they do. For the causal ordering~\eqref{eq:ord}, we get the plot shown in Fig.~\ref{plot-orbit3142}. To draw the plot we picked numerical values:
\be
\begin{array}{llll}
	\eps_1=4\,, &\eps_2=3\,, &\eps_3=2\,, &\eps_4=0\,,\\ 
	y_1 =(2,0,0,0)\,,\ &y_2 =(20,0,0,0)\,,\ &y_3 =(0,0.9,0,0)\,,\ &y_4 =(3,0,0,0)\,,
\end{array}
\ee
where $y_i=(t_i,{\bf x}_i)$. Any other initial point $\eps_1>\eps_2>\eps_3>\eps_4$ and the final point corresponding to the ordering~\eqref{eq:ord} gives rise to a topologically equivalent configuration of curves.
We see that the curves $z(\theta)$, $\bar z(\theta)$ touch the $[1,\infty)$ cut at $\theta=1$ but do not cross it at the intermediate values of $\theta$. This means that $|\rho|=|\bar{\rho}|=1$ at the corresponding Lorentzian configuration. Furthermore, both curves cross the $t$-channel cut $(-\infty,0]$, which according to the above discussion means $|\rho_t|>1$, $|\bar\rho_t|>1$. One of the two curves also crosses the $u$-channel cut $[0,1]$, which means $|\rho_u|>1$. We conclude that the Lorentzian configuration under study is outside the region of OPE convergence of any of the three channels.

The given recipe to determine which channels diverge would require some care in situations when a curve crosses a cut and then goes back, or when the $z(\theta)$ and $\bar{z}(\theta)$ curves cross the same cut in opposite directions. We will discuss these subtleties and their interpretation in~\cite{Qiao:2020bcs}. In the given example they do not occur, so our conclusion that all three channels diverge is robust.

Another comment is in order concerning the 2d CFT case. In this case, the region of analyticity of 4pt functions is larger than $|\rho|, |\bar{\rho}|<1$, being instead given by the condition $|q|, |\bar{q}|<1$~\cite{Hogervorst:2013sma} where $q$ is Al.~Zamolodchikov's uniformizing variable.  Using this variable, one can show that the Lorentzian 4pt function in a 2d CFT is analytic for all possible causal orderings away from null cone singularities~\cite{Maldacena:2015iua}.

\section{Proof of lemma~\ref{lemma:functions}}
\label{app:lemma}

To prove the first inequality,\footnote{See~\cite{Schwarz}, Exercise 6.3 for similar arguments. For this result it's only important that $|\phi(w)|\le 1$. That it's one-to-one and avoids the cut does not matter.} we start by constructing a map $\tl\varphi(w)$ from $\mathbb{D}$ into $\mathbb{D}$ which satisfies $\tl\varphi(0)=0$. This is achieved by a fractional linear transformation as follows
\be
\label{eq:FLT}
\tl\varphi(w)=\frac{\varphi(w)-\varphi(0)}{1-\varphi(w)\overline{\varphi(0)}}.
\ee
Now, Schwarz lemma implies that $|\tl\varphi(w)|\leq |w|$ and so $1-|\tl\varphi(w)|\geq 1-|w|$. At the same time, we find 
\be
1-|\tl\varphi(w)|^2=\frac{(1-|\varphi(w)|^2)(1-|\varphi(0)|^2)}{(1-\varphi(w)\overline{\varphi(0)})(1-\varphi(0)\overline{\varphi(w)})}\leq C(1-|\varphi(w)|)\,,\qquad C=2\frac{1+|\varphi(0)|}{1-|\varphi(0)|}
\ee
where the first equality follows by a short computation from~\eqref{eq:FLT}, and to get the inequality we bounded some factors using $|\varphi(w)|\le 1$. Furthermore, since $1-|\tl\varphi(w)|^2=(1-|\tl\varphi(w)|)(1+|\tl\varphi(w)|)\geq 1-|\tl\varphi(w)|$, we find
\be
1-|w|\leq 1-|\tl\varphi(w)|\leq1-|\tl\varphi(w)|^2\leq  C(1-|\varphi(w)|).
\ee

To prove the second inequality, it will be important that $\varphi(w)$ is one-to-one and that $\varphi(w)\ne 0$.\footnote{It won't be important that it avoids the rest of the cut.} Under these conditions the function $\frac{1}{\varphi(w)}$ is holomorphic and one-to-one. Such functions from $\mathbb{D}$ onto a subset of $\C$ are called univalent, or schlicht~\cite{Duren}. The shifted and rescaled function
\be
h(w)=-\frac{\varphi(0)^2}{\varphi'(0)}\p{\frac{1}{\varphi(w)}-\frac{1}{\varphi(0)}},
\ee
is then also univalent, and in addition satisfies normalization conditions $h(0)=0$ and $h'(0)=1$. A basic result about normalized univalent functions is the Growth Theorem (\cite{Duren}, Theorem 2.6)
\be
|h(w)|\leq \frac{|w|}{(1-|w|)^2}.
\ee
This immediately implies the second bound in~\eqref{eq:Lind-bound}.

\section{Comments on the proof of theorem~\ref{thm:vlad2}}
\label{app:proofVlad2}

Compared to theorem~\ref{thm:vlad}, theorem~\ref{thm:vlad2} has only two essentially new ingredients. First, we now have the freedom of choosing $v\in V$ so we want to show that this choice doesn't matter, and second, we have to prove that the boundary value is holomorphic in $w$. Without these two ingredients, the proof of section~\ref{sec:proof} goes through without any essential modifications. 

Let us briefly recall the main steps of that proof, but now in the context of theorem~\ref{thm:vlad2}.\footnote{Our proof is an adaptation of the proof of theorem 7.2.6 in~\cite{RealSubmanifolds}.} First, for a Schwartz test function $f(x)$ we define
\be
L_v(w,\e) = \int d^dx g(w,x+iv\e) f(x).
\ee
Using integration by parts, we show that
\be\label{eq:integrationbypartsidentity}
\ptl_\e^k L_v(w,\e) = (-i)^k\int d^dx g(w,x+iv\e)v^{\mu_1}\cdots v^{\mu_k} \ptl_{\mu_1}\cdots\ptl_{\mu_k}f(x).
\ee
We then use this identity and the slow-growth condition on $g$~\eqref{eq:slow-growth2} to bound
\be
|\ptl_{\e}^k L_v(w,\e)| \leq \frac{C_k}{\e^{2K}}.
\ee
for some $C_k>0$ that is proportional to some semi-norm of $f$. In what follows, it will be important to us how $C_k$ depends on $v$. It is easy to see that
\be
|\ptl_{\e}^k L_v(w,\e)| \leq \frac{C_k'||v||_\infty^k ||v||_2^{-2K}}{\e^{2K}}.
\ee
for some $C_k'>0$ that is independent of $v$. Furthermore, since the bound~\eqref{eq:slow-growth2} is independent of $w$, $C'_k$ is also independent of $w$.\footnote{This holds on compact subsets $\cK\subset U$, see footnote~\ref{ft:compactfootnote}.} Then we use the obvious analogue of~\eqref{eq:NLformula} starting from sufficiently large $k$ to conclude 
\be\label{eq:improvedbound}
|\ptl_{\e} L_v(w,\e)| \leq {C||v||_\infty^k ||v||_2^{-2K}}
\ee
for some $C>0$ proportional to a semi-norm of $f$. This immediately implies that 
\be
L_v(w,\e)=-\int_\e^{\e_0}\ptl_{\e} L_v(w,\e)+ L_v(w,\e_0)
\ee
is continuous down to $\e=0$ and that thus defined $L_v(w,0)$ depends continuously on $f$ in $\cS(\R^d)$. The slight refinements that we made to the bound~\eqref{eq:improvedbound}, i.e.\ observing that it holds uniformly in $w$ and exhibiting its dependence on $v$, allow us to make the following statement: the limit $L_v(w,\e)\to L_v(w,0)$ is reached uniformly on compact sets $\cK\subset U$ in $w$ and on compact sets $\cV\subset V$ in $v$ (recall that $V$ doesn't contain $0$). This statement is the key in proving that the limit is independent of $v\in V$ and is holomorphic in $w$.

The fact that $L_v(w,0)$ is holomorphic in $w$ is now indeed straightforward, since $L_v(w,\e)$ is holomorphic in $w$ for $\e>0$.\footnote{The standard argument is as follows. Suppose holomorphic functions $h_n$ converge uniformly to some function $h$. Then, first of all, $h$ is continuous because $h_n$ are and the limit is uniform. Second, the uniform limit can be exchanged with contour integration. Since integrals of $h_n$ over closed curves are $0$, so are the integrals of $h$. By Morera's theorem, this implies holomorphicity of $h$.} To prove that it is independent of $v$ requires a bit more work. Take $v_1,v_2\in V$ and write
\be
L_{v_1}(w,\e)-L_{v_2}(w, \e)&=\int d^dx (g(w,x+iv_1\e)-g(w,x+iv_2\e)) f(x)\nn\\
& = \int d^dx \int_0^1 dt \,\ptl_t g(w,x+iv(t)\e) f(x), \qquad v(t)=t v_1+(1-t)v_2\nn\\
& = -i\e \int_0^1 dt \int d^dx g(w,x+iv(t)\e) \,\,(v_1-v_2)\cdot\ptl f(x)\nn\\
& = -i\e \int_0^1 dt \,\tl L_{v(t)}(w,\e).
\ee
where $\tl L_v(w,\e)$ is defined as $L_v(w,\e)$ but with $(v_1-v_2)\cdot\ptl f(x)$ instead of $f(x)$. Since $(v_1-v_2)\cdot\ptl f(x)$ is also a test function, we have that by the same arguments as the above,
$\tl L_v(w,\e)$ converges to a finite limit $\tl L_v(w,0)$ uniformly in $v$ on compacts of $V$. This implies that the integral
\be
\int_0^1 dt \,\tl L_{v(t)}(w,\e)
\ee
has a finite limit as $\e\to 0$, and thus
\be
L_{v_1}(w,\e)-L_{v_2}(w, \e)= -i\e \int_0^1 dt \,\tl L_{v(t)}(w,\e)\to 0.
\ee
\chapter{Appendices of Part \ref{part:minkowski}}
\section{Lorentzian CFT literature}\label{literature}

Recent years have seen an explosion of the uses of Lorentzian CFT, motivated
in particular by the conformal bootstrap applications. In this appendix we
will mention some of these works, and comment on their underlying assumptions.
{See also \cite{DSDLorentzian} for a modern pedagogical introduction to Lorentzian CFT.}

\tmtextbf{Conformal collider bounds.} One of the first ``modern'' Lorentzian
CFT results was obtained in {\cite{Hofman:2008ar}}. This work considered a
thought experiment, creating a CFT state via a (smeared) local operator and
measuring energy coming out at null infinity in a particular direction,
integrated over time. On physical grounds, one expects $\<\Psi| \int d x^-\, T_{- -} | \Psi \> \geqslant 0$ for any
state (``averaged null energy condition'' - ANEC). One interesting case is of
3-point functions $\langle \mathcal{O}^{\dagger} T_{\mu \nu} \mathcal{O} \rangle$
where $\mathcal{O}$ has nontrivial spin, when there are several independent
OPE coefficients multiplying different tensor structures allowed by conformal
symmetry. In this case ANEC implies that certain linear combinations of these
OPE coefficients must be non-negative (``conformal collider bounds''). Interference effects can be used to strengthen conformal collider bounds to provide
explicit lower bounds~\cite{Cordova:2017zej}, while combining conformal collider bounds
with stress-tensor Ward identities leads to 
constraints on operator dimensions which are sometimes stronger than standard unitarity bounds~\cite{Cordova:2017dhq}. See below for work aiming to justify ANEC, or to derive conformal conformal bounds
directly without using ANEC.

\tmtextbf{Light-cone bootstrap.} Refs.\ {\cite{Fitzpatrick:2012yx,Komargodski:2012ek}} were the first to notice that
some bootstrap constraints become more visible in the Lorentzian signature.
These references pioneered the ``analytic light-cone bootstrap'' which studies
conformal four point functions in the regime of $0 < z, \bar{z} < 1$ real,
i.e.\ in the kinematics of Fig.\ \ref{134} when the point $x_2$ is spacelike
with respect to $x_1, x_3$. By studying the light cone limit $z \rightarrow 0$
at fixed $\bar{z}$ of one OPE channel and requiring that it should be
reproduced by the crossed channel, they argued that, in any CFT for $d > 2$,
the OPE should contain a series of operators of arbitrarily large spin and
twist asymptoting to a particular value. The original argument has some
caveats (see the discussion in {\cite{Qiao:2017xif}}, App.\ F) and a
mathematically rigorous proof is lacking. It would be nice to provide such a
proof, given the extreme importance of the light-cone bootstrap in the modern
bootstrap literature. There is little doubt that the light-cone bootstrap
results are correct. Numerical bootstrap studies of the critical 3d Ising and
the $O (2)$ models {\cite{Simmons-Duffin:2016wlq,Liu:2020tpf}} have found the
series of operators predicted by the light-cone bootstrap {see also \cite{Caron-Huot:2020ouj})}.
Ref.\ {\cite{Hofman:2016awc}} used the light-cone bootstrap to derive the conformal
collider bounds of {\cite{Hofman:2008ar}} without using ANEC.

\tmtextbf{Causality constraints.} Refs.\ {\cite{Hartman:2015lfa,Hartman:2016dxc,Hartman:2016lgu}} pioneered the study
of causality constraints for CFT 4-point functions. In particular Ref.\ {\cite{Hartman:2015lfa}} pointed out that the $z, \bar{z}$ and $\rho,
\bar{\rho}$ expansions are sufficient to construct Lorentzian 4-point functions
for many kinematic configurations and show local commutativity (i.e.\ that
spacelike-separated operators commute). See also note \ref{noteShock}. These
techniques led to a proof of ANEC {\cite{Hartman:2016lgu}}. As mentioned in
footnote \ref{caveats}, some steps in these papers are not completely
rigorous. See App.\ \ref{Tom} below for a more detailed review of
{\cite{Hartman:2015lfa}}.

\tmtextbf{Bulk point singularity.} Ref.\ {\cite{Maldacena:2015iua}} studied the
CFT 4-point function on the Lorentzian cylinder focusing on ``bulk-point''
configurations which correspond to scattering events in AdS/CFT
{\cite{Polchinski:1999yd,Gary:2009ae,Heemskerk:2009pn,Penedones:2010ue,Okuda:2010ym}}.
Using a local AdS dual description, one may suspect that the 4-point
function should be singular at such configurations. However, on the boundary
CFT side, one does not see this singularity in perturbation theory in $d = 2$
and $d = 3$ dimensions {\cite{Maldacena:2015iua}}. In $d = 2$, Ref.\ {\cite{Maldacena:2015iua}} showed non-perturbatively that the CFT 4-point
function is analytic everywhere away from light cones (in particular regular at
bulk-point configurations). This assumes Virasoro symmetry and unitarity and
uses Zamolodchikov's $q$-variables {\cite{zamolodchikov1987conformal}}. What
happens non-perturbatively in $d \geqslant 3$ (or in $d = 2$ in the absence of
the local stress tensor) is still an open problem. Note that at bulk-point
configurations, the $\rho$-expansion of the CFT 4-point function does not
absolutely converge in s-channel (as $| \rho | = | \bar{\rho} | = 1$ there)
and diverges in t-,u-channels {\cite{Qiao:2020bcs}}. In this paper we only
considered the CFT 4-point functions in flat space, but by the same
strategy we will show in {\cite{paper3}} that the Wightman axioms also hold
for CFT 4-point functions on Lorentzian cylinder. In particular, this will
show that the CFT 4-point functions are well defined at bulk-point
configurations in the sense of tempered distributions (but it will not settle
the question of their analyticity there).

\tmtextbf{Lorentzian inversion formula.} Ref.\ {\cite{Caron-Huot:2017vep}}
introduced an analogue of Froissart-Gribov formula in the context of conformal
field theory, which is now known as the Lorentzian inversion formula (LIF).
This formula computes the OPE data of a scalar 4-point function in terms of
a Lorentzian integral of this 4-point function. The OPE data is extracted
in the form of a function $C (\Delta, \ell)$. For integer $\ell$, the function
$C (\Delta, \ell)$ encodes the scaling dimensions of exchanged primary
operators of spin $\ell$ in the positions of poles in $\Delta$, and the
corresponding OPE coefficients are encoded in residues. LIF has many
interesting properties, such as analyticity in $\ell$, and suppression of
double-twist operators when a cross-channel conformal block expansion is used
under the integral. The original derivation of {\cite{Caron-Huot:2017vep}} was
done in cross-ratio space. The formula was re-derived in position space in
{\cite{Simmons-Duffin:2017nub}}. The derivation was further simplified and
generalized in {\cite{Kravchuk:2018htv}}.

Among other applications, LIF has been used to systematize and extend many of
the results of light-cone bootstrap (see, e.g.,
{\cite{Liu:2018jhs,Cardona:2018dov,Albayrak:2019gnz,Cardona:2018qrt,Liu:2020tpf,Iliesiu:2018zlz,Iliesiu:2018fao,Albayrak:2020rxh}}).
Similarly to light-cone bootstrap, this application is not completely rigorous
simply due to the fact that LIF expresses $C (\Delta, \ell)$ in terms of an
integral, and the local operators correspond to singularities of $C (\Delta,
\ell)$. In other words, the integral has no chance of converging near the
values of $\Delta, \ell$ relevant to local operators, except perhaps for
leading-twist operators {(see~\cite{Caron-Huot:2020ouj} for steps in this direction)}. This necessarily makes any conclusions about
anomalous dimensions of local operators reliant on additional assumptions.
These are easy to justify in some perturbative expansions, but in
non-perturbative setting do not appear to have been solidly understood.

\tmtextbf{Light-ray operators.} Ref.\ {\cite{Kravchuk:2018htv}} generalized LIF
to external operators with spin and uncovered an interesting relation to
Knapp-Stein intertwining operators, especially to what they called the light
transform. They interpreted the analyticity of LIF in $\ell$ in terms of
families of non-local non-integer-spin operators, the light-ray operators. These
operators are defined for generic complex $\ell$ and reduce to
light-transforms (null integrals) of local operators for integer spins. More
recently, light-ray operators have been used to understand an OPE for
event-shape observables such as energy-energy correlators in CFT
{\cite{Kologlu:2019mfz,Kologlu:2019bco,Chang:2020qpj}} (see also
{\cite{Dixon:2019uzg,Korchemsky:2019nzm}}). The light-ray operators correspond
to poles in $\Delta$ of $C (\Delta, \ell),$ and the issues with convergence of
LIF described above prevent a simple rigorous proof of their non-perturbative
existence. (E.g., for generic $\ell$, $C (\Delta, \ell)$ could have cuts or a
natural boundary of analyticity in $\Delta$.) It would be interesting to find
such a proof. In addition to clarifying the nature of light-ray operators, it
would probably also have a bearing on the light-cone bootstrap results
discussed above.

\tmtextbf{Conformal Regge theory} provides a way to understand Minkowski
correlators in Regge limit, and was developed in Refs.\ {\cite{Brower:2006ea,Cornalba:2007fs,Cornalba:2008qf,Costa:2012cb}}. Regge
limit in CFT is a limit of a 4-point function in Lorentzian signature in
which $\mathcal{O}_2$ approaches the ``image of $\mathcal{O}_3$ in the next
Poincar\'e patch,'' in 4-point function with the ordering
\begin{equation}
	\langle \mathcal{O}_4 \mathcal{O}_3 \mathcal{O}_2 \mathcal{O}_1 \rangle .
\end{equation}
The operators $\mathcal{O}_1$ and $\mathcal{O}_4$ are kept spacelike
separated, with $\mathcal{O}_1$ in past of $\mathcal{O}_2$ and $\mathcal{O}_4$
in the future of $\mathcal{O}_3$.\footnote{In a symmetric version of the
	limit, which is related to the one described here by a conformal
	transformation, the operators $\mathcal{O}_1$ and $\mathcal{O}_4$ approach
	each other's images in the same way as $\mathcal{O}_2$ and $\mathcal{O}_3$
	do.} The image of $\mathcal{O}_3$ in the next Poincar\'e patch is the first
point on Minkowski cylinder where all future-directed null geodesics from
$\mathcal{O}_3$ meet. A lot of interest in Regge limit comes from its
interpretation as bulk high-energy scattering through AdS/CFT. Kinematically,
this limit is somewhat similar to the $\mathcal{O}_2 \rightarrow
\mathcal{O}_3$ limit because $\mathcal{O}_3$ and its image in the next
Poincar\'e patch transform in the same way under conformal group. For example,
the cross-ratios $z_t, \bar{z}_t \rightarrow 0$ in Regge limit. (Here by $z_t,
\bar{z}_t$ we mean the cross-ratios for t-channel $\mathcal{O}_2 \times
\mathcal{O}_3$.) However, they do so after $\bar{z}_t$ crosses the cut $[1,
\infty)$, and so in terms of $\rho_t, \bar{\rho}_t$ we have $\rho_t
\rightarrow 0$ and $\bar{\rho}_t \rightarrow \infty$. Therefore, the
$\mathcal{O}_2 \times \mathcal{O}_3$ OPE is divergent. Conformal Regge theory
gives a way to resum the $\mathcal{O}_2 \times \mathcal{O}_3$ OPE in a way
that exhibits a dominant contribution from a ``Reggeon'' exchange, which is an
example of a light-ray operator. Justification for this resummation, which
involves analytic continuation of OPE data in spin, comes from LIF (which
historically was understood after Conformal Regge theory was established). In
the context of our paper, it would be interesting to understand whether such
resummations can be made rigorous enough (in axiomatic sense) and used to
prove that Minkowski correlators are functions in regions where so far only
temperedness has been proven.\footnote{In the classic Regge limit there is a
	channel in which the OPE converges regularly, but it is possible that some
	causal orderings can be relaxed while keeping the resummation procedure
	valid.} For this it might not be necessary to understand the Reggeon or more
general light-ray operators, since the resummation procedure can be stopped at
a point where the correlator is expressed as an integral of $C (\Delta, \ell)$
over a region where LIF converges. See~\cite{Caron-Huot:2020nem} for progress on these questions.

\tmtextbf{Works of Gillioz, Luty et al.} Papers by this group of authors are
characterized by the systematic use of momentum space in Lorentzian CFT. So,
Refs.\ {\cite{Gillioz:2019lgs,Gillioz:2020wgw}} computed Lorentzian momentum
space 3-point functions (3 scalars and scalar-scalar-spin $\ell$) by solving the
conformal Ward identities. In momentum space, it's also possible to form
conformal blocks by gluing 3-point functions {\cite{Gillioz:2020wgw}}. See also
notes \ref{noteMarc1}, \ref{noteMarc2}.

Ref.\ {\cite{Gillioz:2019iye}} carried out this program quite explicitly in 2d
CFT, with an eye towards eventual conformal bootstrap applications. They
stressed that the momentum conformal block expansion generally converges only
in the sense of distributions---one of the first mentions of distributional
convergence in the modern CFT literature. For some momenta configurations,
they argued that the momentum conformal blocks can be pointwise bounded by the
position conformal blocks with an appropriately chosen real $z \in (0, 1)$.
For such configurations the momentum expansion converges in the ordinary sense
of functions. The same work also proposes a bootstrap equation in the momentum
space, obtained by transforming the local commutativity constraint multiplied
by a test function selecting configurations with a spacelike pair of points
(however, examples of test functions chosen in {\cite{Gillioz:2019iye}} may be
too singular).

Refs.\ {\cite{Gillioz:2016jnn,Gillioz:2018kwh,Gillioz:2020mdd} studied the Fourier transform of the time-ordered Minkowski 4-point function in
	relation to various interesting physics questions. Note that, as mentioned in
	the conclusions, time-ordered Minkowski CFT 4-point functions have not yet been
	rigorously defined as a distributions. The Fourier transform depends on 4
	momenta $p_i$, and to reduce functional complexity it is interesting to take
	some or all of these momenta lightlike, $p_i^2 \rightarrow 0$. \
	
	So, Ref.\ {\cite{Gillioz:2016jnn}} considered the Fourier transform of the
	connected time-ordered 4-point function $\langle \mathcal{T} \{
	\mathcal{O}_1 \mathcal{O}_2 \mathcal{O}_3 \mathcal{O}_4 \} \rangle_c$. Here
	they worked with operators of scaling dimension $\Delta_i > d / 2$ for which
	the Fourier transform is expected to have a finite limit as $p_i^2 \rightarrow
	0$.\footnote{We thank Marc Gillioz for explanations of his work and in
		particular of the distinction between the high dimension case discussed here
		and the low dimension case below.} Ref.\ {\cite{Gillioz:2016jnn}} proposed a
	Lorentzian CFT analogue of the optical theorem:
	\begin{equation}
		\tmop{Im} \mathcal{M}_{1234} (s, t) = \underset{\mathcal{O} \neq 1}{\sum}
		f_{12\mathcal{O}} f_{\bar{4} \bar{3} \mathcal{O}}^{\ast}
		\mathcal{N}_{\mathcal{O}} (q) \langle \bar{\mathcal{T}} \{
		\hat{\mathcal{O}}_1 (p_1) \hat{\mathcal{O}}_2 (p_2) \} \mathcal{O}^{\dag}
		(0) \rangle \langle \mathcal{O} (0) \mathcal{T} \{ \hat{\mathcal{O}}_3 (p_3)
		\hat{\mathcal{O}}_4 (p_4) \} \rangle, \label{CFToptical}
	\end{equation}
	where $\mathcal{M}_{1234}$ is proportional to the Fourier transform of
	$\langle \mathcal{T} \{ \mathcal{O}_1 \mathcal{O}_2 \mathcal{O}_3
	\mathcal{O}_4 \} \rangle_c$, $f_{i j k}$ is the same as in (\ref{3-pointGeneral}),
	$\mathcal{N}_{\mathcal{O}} (q)$ is some normalization factor at $q = p_1 + p_2
	(= - p_3 - p_4)$, and $\hat{\mathcal{O}}$ denotes the Fourier transform of
	$\mathcal{O}$. Eq.\ (\ref{CFToptical}) is supposed to hold in the following
	kinematic region in the momentum space:
	\[ p_i^2 = 0, \qquad s = (p_1 + p_2)^2 > 0, \qquad t = (p_1 + p_3)^2 \leqslant
	0, \]
	and was derived from a combinatorial operator identity
	\begin{equation}
		\underset{k = 0}{\overset{n}{\sum}} (- 1)^k \underset{\sigma \in S_n}{\sum}
		\frac{1}{k! (n - k) !} \bar{\mathcal{T}} \{ \mathcal{O}_{\sigma_1}
		(x_{\sigma_1}) \ldots \mathcal{O}_{\sigma_k} (x_{\sigma_k}) \} \mathcal{T}
		\{ \mathcal{O}_{\sigma_{k + 1}} (x_{\sigma_{k + 1}}) \ldots
		\mathcal{O}_{\sigma_n} (x_{\sigma_n}) \} = 0, \label{identity:comb}
	\end{equation}
	summing over all permutations, with $\mathcal{T}$($\bar{\mathcal{T}}$) time
	ordering (anti-time ordering). Note that the use of this identity may be not
	fully safe in the distributional context, as it arises from a non-smooth
	partition of unity.
	
	The CFT optical theorem (\ref{CFToptical}) was used in Ref.\ {\cite{Gillioz:2016jnn}} to study the scale anomalies that appear in a
	specific class of CFT correlation functions. In fact, unlike Wightman
	functions which are conformally invariant distributions, time-ordered
	correlator distributions may, for certain scaling dimensions, contain pieces
	which violate scale invariance. Thus the scale anomaly describes the violation
	of dilatation Ward identities, and in position space it is an ultralocal term,
	located at coincident points. In Fourier space, scale anomaly translates into
	a nonzero imaginary part of $\mathcal{M}_{1234} (s, t)$ at $t = 0$. Eq.
	(\ref{CFToptical}) then computes the scale anomaly coefficient through a
	positive definite sum rule (in particular predicts that it is positive). These
	scale anomalies also appear in the Euclidean signature, and a similar sum rule
	for anomaly coefficients can also be found in Ref.\ {\cite{Gillioz:2016jnn}}.
	However, the Euclidean sum rule is not positive definite unlike the Lorentzian
	case.
	
	Ref.\ {\cite{Gillioz:2016jnn}} tested the above ideas for the scalar 4-point
	function of external dimensions $\Delta = 3 d / 4$. Ref.\ {\cite{Gillioz:2018kwh}} then studied the more interesting case of the stress
	tensor 4-point function $\langle \mathcal{T} \{ T_{\mu_1 \nu_1} T_{\mu_2
		\nu_2} T_{\mu_3 \nu_3} T_{\mu_4 \nu_4} \} \rangle$ whose scale anomaly is
	proportional to the stress-tensor 2-point function coefficient $c_T$.
	Assuming that the $t \rightarrow 0$ limit is finite, the CFT optical theorem
	expresses $c_T$ as a sum of positive contributions of all operators in the $T
	\times T$ OPE apart from the identity (the stress tensor contribution is
	known, proportional to $c_T$, and can be moved to the l.h.s.). The
	contributions from the scalars and the spin-2 operators are computed
	explicitly in {\cite{Gillioz:2018kwh}}.
	
	The more recent Ref.\ {\cite{Gillioz:2020mdd}} studied instead the Fourier
	transform of {{the time-ordered 4-point function}} (or Euclidean
	4-point function) in the opposite case of the low external dimensions
	$\Delta_{\phi} < d / 2$. Unlike in {\cite{Gillioz:2016jnn,Gillioz:2018kwh}},
	in this low dimension case the Fourier transform is singular as $p_i^2
	\rightarrow 0$, and one obtains a finite quantity multiplying it by
	$(p_i^2)^{d / 2 - \Delta_{\phi}}$ before taking the limit, a CFT analogue of
	LSZ reduction. Doing so, they defined a ``CFT scattering amplitude'' $A (s, t,
	u)$ ($p_i^2 \rightarrow 0$ for all $i$) and a closely related ``form factor''
	$F (s, t, u)$ where $p_i^2 \rightarrow 0$ only for $i = 1, 2, 3$. Because the
	limit $p_i^2 \rightarrow 0$ has to be taken one momentum at a time, crossing
	symmetry is not obvious. Ref.\ {\cite{Gillioz:2020mdd}} also gave an
	alternative derivation, starting from the Mellin representation of the CFT
	4-point function, where crossing symmetry of $F (s, t, u)$ and $A (s, t,
	u)$ follows from the crossing symmetry of the Mellin amplitude. In the future,
	crossing symmetric quantities $A (s, t, u)$ and $F (s, t, u)$ may turn out
	useful in a bootstrap analysis. It should be stressed that the results of
	{\cite{Gillioz:2020mdd}} in no way contradict the usual lore that there are no
	S-matrices in interacting CFTs. In spite of the name adopted in
	{\cite{Gillioz:2020mdd}}, the existence of the quantity $A (s, t, u)$ does not
	imply that we can set up a wave-packet scattering experiment in a CFT.
	Wave-packets would quickly diffuse before reaching the interaction region, the
	singularity of $(p_i^2)^{\Delta_{\phi} - d / 2}$ being a cut rather than a
	pole.}

\subsection{Review of Hartman et al {\cite{Hartman:2015lfa}}}\label{Tom}

\tmtextbf{Relating different orderings via analytic continuations.} Here we
will comment on some of the results of {\cite{Hartman:2015lfa}} in more
details. The first part of this paper considers the Lorentzian CFT 4-point
functions with operators $\mathcal{O}_1$, $\mathcal{O}_3$, $\mathcal{O}_4$ fixed
at zero time and the spatial positions 0, $\hat{e}_1$ and $\infty$, while the operator $\mathcal{O}_2$ is inserted at Minkowski position $t_2\hat e_0+y_2\hat e_1$. 
They consider
four different operator orderings
\begin{equation}
	\langle \mathcal{O}_2 \mathcal{O}_1 \mathcal{O}_3 \mathcal{O}_4 \rangle,
	\langle \mathcal{O}_1 \mathcal{O}_2 \mathcal{O}_3 \mathcal{O}_4 \rangle,
	\langle \mathcal{O}_3 \mathcal{O}_2 \mathcal{O}_1 \mathcal{O}_4 \rangle,
	\langle \mathcal{O}_1 \mathcal{O}_3 \mathcal{O}_2 \mathcal{O}_4 \rangle
\end{equation}
in the region of $0 < y_2 < 1 / 2$ and $t_2$ positive. As
$t_2$ is increased from zero, the operator $\mathcal{O}_2$, initially
spacelike with respect to all other insertions, crosses the light cone first of
$\mathcal{O}_1$ and then of $\mathcal{O}_3$. With $z, \bar{z} = y_2 \pm t_2$,
denoting $G (z, \bar{z}) = \langle \mathcal{O}_2 \mathcal{O}_1 \mathcal{O}_3
\mathcal{O}_4 \rangle$, they give the following prescription to compute the
correlators for the other orderings (Ref.~\cite{Hartman:2015lfa}, Eq.\ (3.22)):
\begin{eqnarray}
	\langle \mathcal{O}_1 \mathcal{O}_2 \mathcal{O}_3 \mathcal{O}_4 \rangle & =
	& G (z, \bar{z})_{z \rightarrow e^{- 2 \pi i} z}, \label{Tomrecipe} \\
	\langle \mathcal{O}_3 \mathcal{O}_2 \mathcal{O}_1 \mathcal{O}_4 \rangle & =
	& G (z, \bar{z})_{(\bar{z} - 1) \rightarrow e^{- 2 \pi i} (\bar{z} - 1)},
	\nonumber\\
	\langle \mathcal{O}_1 \mathcal{O}_3 \mathcal{O}_2 \mathcal{O}_4 \rangle & =
	& G (z, \bar{z})_{z \rightarrow e^{- 2 \pi i} z, (\bar{z} - \bar{z}_0)
		\rightarrow e^{- 2 \pi i} (\bar{z} - \bar{z}_0)} . \nonumber
\end{eqnarray}
Their justification of this prescription relied on some presumed analyticity
properties of $G (z, \bar{z})$ which, to our knowledge, have never been shown
in a general QFT context. Nevertheless we will see below that for CFTs Eq.
{\eqref{Tomrecipe}} turns out to be true (with $\bar{z}_0 = 1$).

The real parameter $\bar{z}_0$ was introduced in {\cite{Hartman:2015lfa}} as
the position of the first $\bar{z}$ singularity of $G (z, \bar{z})_{z
	\rightarrow e^{- 2 \pi i} z}$. Their goal was to show that $\bar{z}_0
\geqslant 1$, which using {\eqref{Tomrecipe}} then implies local commutativity
$\langle \mathcal{O}_1 [\mathcal{O}_2, \mathcal{O}_3] \mathcal{O}_4 \rangle =
0$ for $\bar{z} < 1$ i.e.\ when $\mathcal{O}_2$ is spacelike to
$\mathcal{O}_3$. In our paper (Sec.\ \ref{local-comm}) we presented a
different way to understand and derive local commutativity which is closer to
the classic literature: as we have reviewed there, it is a robust consequence
of the existence of the analytic continuation to the forward tube, which we
constructed.

Let's see what it would take to justify {\eqref{Tomrecipe}}. We define the
function $G (z, \bar{z})$ by the $\mathcal{O}_2 \times \mathcal{O}_1$ OPE
expansions, which converges for $| \rho |, | \bar{\rho} | < 1$, i.e.\ as long
as $z, \bar{z}$ stay away from $z, \bar{z} \in (+ 1, \infty)$. The points $z =
0$ and $\bar{z} = 0$ are branch points singularities with cuts which we put
along the negative real axis. Note that the contours in {\eqref{Tomrecipe}}
are all within the analyticity domain of $G (z, \bar{z})$. So the prescription
{\eqref{Tomrecipe}} is meaningful. We still have to see if it agrees with the
rigorous definition which computes the Minkowski 4-point function by analytically
continuing from the Euclidean region staying in the forward tube corresponding
to the chosen operator ordering. We will see that it will indeed agree, but
showing it for the last ordering will be subtle.

\begin{figure}[h]\centering
	\raisebox{-0.437612502733711\height}{\includegraphics[width=16.2870097074643cm,height=4.87365538501902cm]{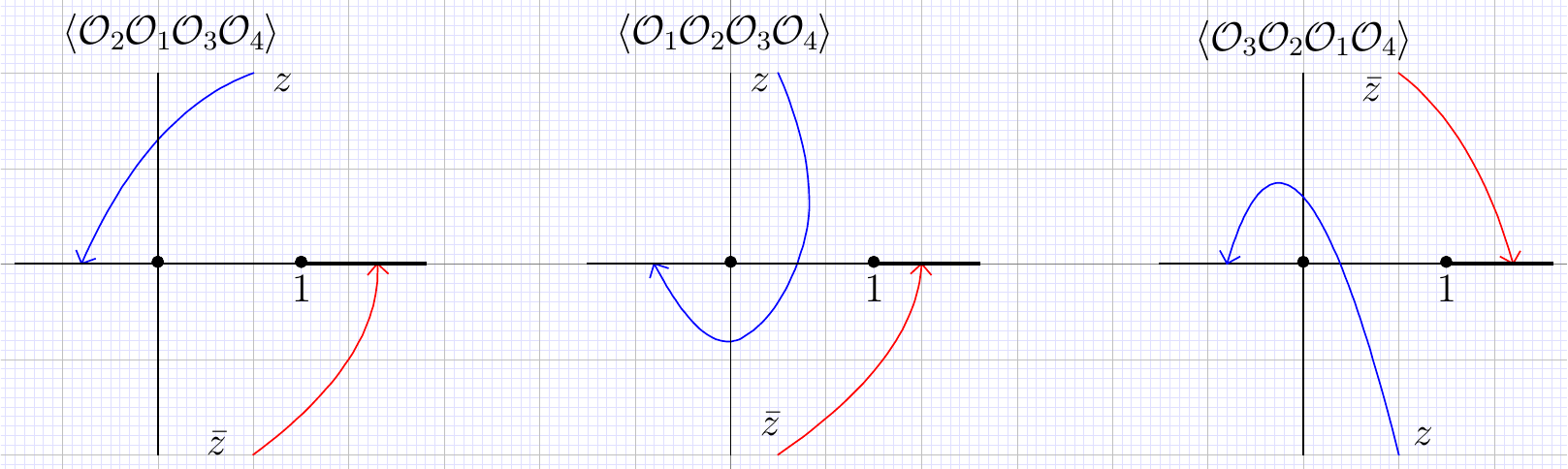}}
	\caption{\label{first3orderings}$z, \bar{z}$ curves for the analytic
		continuation from the Euclidean; the first 3 orderings.}
\end{figure}

For definiteness we will focus on the region $t_2 > 1 - y_2$ \ i.e.\ $\bar{z}
> 1$, where $\mathcal{O}_2$ crossed both light cones. The end point of the
analytic continuation contour is always the same while the initial point
depends on the operator ordering. E.g.\ for the ordering $\langle \mathcal{O}_2
\mathcal{O}_1 \mathcal{O}_3 \mathcal{O}_4 \rangle$ we have to pick initial
Euclidean times $\varepsilon_2 > \varepsilon_1 > \varepsilon_3$, while for
$\langle \mathcal{O}_1 \mathcal{O}_2 \mathcal{O}_3 \mathcal{O}_4 \rangle$ we
have $\varepsilon_1 > \varepsilon_2 > \varepsilon_3$ etc. For any of these
orderings, we denote
\begin{equation}
	z_1, \bar{z}_1 = \pm i \varepsilon_1, \quad z_2, \bar{z}_2 = y_2 \pm i
	(\varepsilon_2 + i t_2), \quad z_3, \bar{z}_3 = 1 \pm i \varepsilon_3
\end{equation}
and compute (in the limit $z_4, \bar{z}_4 = \infty$)
\begin{equation}
	z = \frac{z_1 - z_2}{z_1 - z_3}, \quad \bar{z} = \frac{\bar{z}_1 -
		\bar{z}_2}{\bar{z}_1 - \bar{z}_3} .
\end{equation}
We are interested in the curves which $z, \bar{z}$ trace as the Euclidean
times are scaled to zero and the Minkowski time $t_2$ from 0 to its final
value. For the first three orderings the resulting curves are shown in Fig.\ \ref{first3orderings}.

We see that in all these cases, the curves lie in the analyticity domain of $G
(z, \bar{z})$ i.e.\ they don't cross $(1, + \infty)$. For the first two cases
this was guaranteed by our results that $| \rho |, | \bar{\rho} | < 1$ for the
s-channel OPE expansion. For the third case it was not guaranteed but it also
turns out to be true, by inspection. We also see that in all these 3 cases,
the curves go around $z = 0$ and $\bar{z} = 1$ in agreement with
{\eqref{Tomrecipe}}. \footnote{To see this more clearly in the third case, deform the
	curves continuously moving the initial $z$ into the upper half plane and the
	initial $\bar{z}$ into the lower half plane.}

\begin{figure}[h]\centering
	\raisebox{-0.4738733350422\height}{\includegraphics[width=5.97432113341204cm,height=3.45444706808343cm]{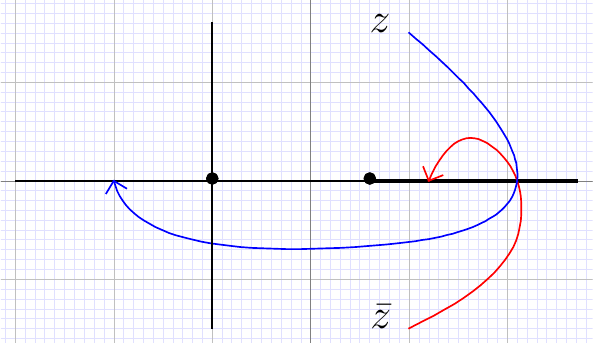}}
	\caption{\label{last0}$z, \bar{z}$ curves for the analytic continuation from
		the Euclidean for the $\langle \mathcal{O}_1 \mathcal{O}_3 \mathcal{O}_2
		\mathcal{O}_4 \rangle$ ordering.}
\end{figure}

\begin{figure}[h]\centering
	\raisebox{-0.4738733350422\height}{\includegraphics[width=5.97432113341204cm,height=3.45444706808343cm]{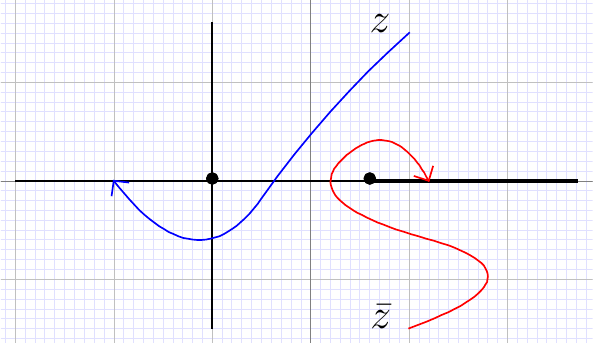}}
	\caption{\label{last1}$z, \bar{z}$ curves for computing the $\langle
		\mathcal{O}_1 \mathcal{O}_3 \mathcal{O}_2 \mathcal{O}_4 \rangle$ ordering
		via {\eqref{Tomrecipe}}.}
\end{figure}

For the fourth ordering $\langle \mathcal{O}_1 \mathcal{O}_3 \mathcal{O}_2
\mathcal{O}_4 \rangle$ when we have to assign $\varepsilon_1 > \varepsilon_3 >
\varepsilon_2$, the analytic continuation inside the forward tube gives the
$z, \bar{z}$ curves shown in Fig.\ \ref{last0}, while prescription
{\eqref{Tomrecipe}} would correspond to Fig.\ \ref{last1}.

\

The two figures are clearly not the same. Moreover the curves in the first
figure cross $(+ 1, \infty)$ where the definition of $G (z, \bar{z})$ via the
$\mathcal{O}_1 \times \mathcal{O}_2$ channel OPE expansion stops converging.
Can we show that the analytic continuation in Fig.\ \ref{last0} exists and that
it agrees with the one in Fig.\ \ref{last1}?

\begin{figure}[h]\centering
	\raisebox{-0.4738733350422\height}{\includegraphics[width=5.97432113341204cm,height=3.45444706808343cm]{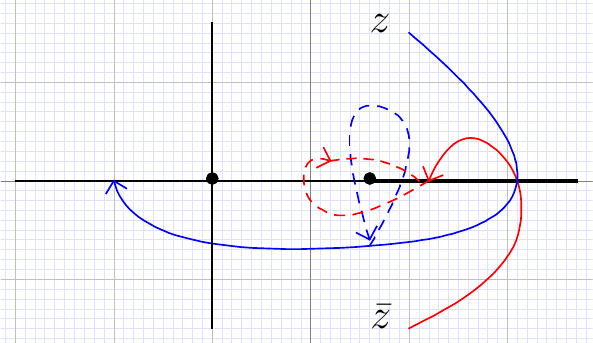}}$\Longrightarrow$\raisebox{-0.4738733350422\height}{\includegraphics[width=5.97432113341204cm,height=3.45444706808343cm]{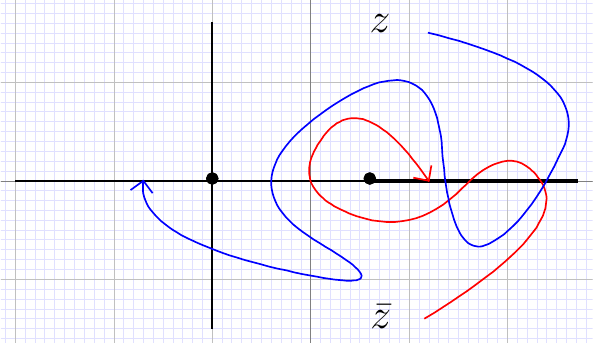}}
	\caption{\label{last2}Deforming the $z, \bar{z}$ curves for the $\langle
		\mathcal{O}_1 \mathcal{O}_3 \mathcal{O}_2 \mathcal{O}_4 \rangle$ ordering.}
\end{figure}

For this, let us bring in the $\mathcal{O}_2 \times \mathcal{O}_3$ OPE
expansion, which correspond to expanding in $z_t = 1 - z$, $\bar{z}_t = 1 -
\bar{z}$ or in the corresponding $\rho_t$, $\bar{\rho}_t$. In the Euclidean
region the two expansions agree. The $\mathcal{O}_2 \times \mathcal{O}_3$
expansion converges away from $z_t, \bar{z}_t \in (+ 1, \infty)$ i.e.\ $z,
\bar{z} \in (- \infty, 0)$. Thus the curves in both Figs.\ \ref{last0},
\ref{last1} lie within the range of analyticity of the $\mathcal{O}_2 \times
\mathcal{O}_3$ expansion, so we can compare the analytic continuations. Since
only integer spins occur in the expansion, the analytic continuation does not
change under $\rho_t \rightarrow e^{2 \pi i} \rho_t$, $\bar{\rho}_t
\rightarrow e^{- 2 \pi i} \bar{\rho}_t$ (such arguments were systematically
exploited in {\cite{Qiao:2020bcs}}).\footnote{Sometimes this property is called ``Euclidean single-valuedness''.} So let us add extra loops to the blue and
the red curves in the opposite directions around $1$, see Fig.\ \ref{last2}.
Adding the loops and deforming the curves continuously (the first step is
shown in Fig.\ \ref{last2}) we can bring them to those in Fig.\ \ref{last1}.
This finishes the proof that the prescription {\eqref{Tomrecipe}} is correct
also for the fourth ordering.

\begin{figure}[h]\centering
	\raisebox{-0.485281603306646\height}{\includegraphics[width=5.02620359438541cm,height=2.83259543486816cm]{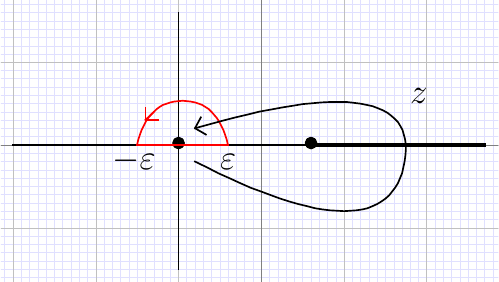}}
	\caption{\label{Gtilde}Definition of $\hat{G} (z, \bar{z})$. Red:
		integration contour in Eq.\ {\eqref{contour}}.}
\end{figure}

\tmtextbf{Positivity constraints.} We wish to comment on another result of
{\cite{Hartman:2015lfa}}: an argument for positivity of certain conformal
block expansion coefficients. We present the argument exchanging the role of s
an t channels w.r.t. {\cite{Hartman:2015lfa}}. Let $G (z, \bar{z}) = 1 +
\ldots$ be the holomorphic function defined by the s-channel OPE expansion (i.e.\ $(z \bar{z})^{\Delta_1 + \Delta_2}$ times the $G (z, \bar{z})$ discussed
above). We will define a certain analytic continuation of the function $G (z,
\bar{z})$. Let us start with $\bar{z}$ and $z$ close to zero, $\bar{z}$ in the
upper half plane and $z$ in the lower half plane. In this range all three
channels s,t,u converge. We wish to analytically continue $G (z, \bar{z})$ by
taking $z$ through $(1, + \infty)$ and bring it back close to zero, in the
upper half plane (see Fig.\ \ref{Gtilde}), while we don't touch $\bar{z}$. This
analytic continuation can be performed using the t-channel or u-channel
expansions, with the same result (but not the s-channel since it stops
converging on $(1, + \infty)$). We denote the result of this analytic
continuation by $\hat{G} (z, \bar{z})$, with $\tmop{Im} z, \tmop{Im} \bar{z} >
0$.

Although $\hat{G} (z, \bar{z})$ is so defined with both $z, \bar{z}$ in the
upper half plane, it has continuous limits when they both approach positive
real axis, or both approach negative real axis, since in the first case the
t-channel and in the second case the u-channel remains convergent. We will be
interested in the situation when \ $\mathcal{O}_1 =\mathcal{O}_2$,
$\mathcal{O}_3 =\mathcal{O}_4$. In this case all expansion coefficients in the
t and u channels are positive. This implies that the Euclidean correlator $G_E
(z, \bar{z})$ will be positive for real $z, \bar{z} > 0$ (using t-channel) and
for real $z, \bar{z} < 0$ (using u-channel). The difference between $G_E (z,
\bar{z})$ and $\hat{G} (z, \bar{z})$, for real $z, \bar{z} > 0$ or $z, \bar{z}
< 0$ is that in the first case $z, \bar{z}$ approach the real axis from the
opposite sides while in the second case from the same one. When we take $z$
through $(1, + \infty)$ cut, this only brings in some phases in the t and
u-channel expansion of $\hat{G} (z, \bar{z})$ with respect to $G_E (z,
\bar{z})$. This implies that we have a bound for real $z, \bar{z} > 0$ or $z,
\bar{z} < 0$:
\begin{equation}
	| \hat{G} (z, \bar{z}) | \leqslant G_E (z, \bar{z}) \label{GtildeG}
\end{equation}
In what follows $\hat{G} (z, \bar{z})$ will be used as a holomorphic function
with $z, \bar{z}$ in the upper half plane satisfying the bound
{\eqref{GtildeG}} on its boundary, while $G_E (z, \bar{z})$ will be used only
with real $z, \bar{z}$.

In particular, since $G_E \approx 1$ for $z, \bar{z}$ near zero up to small
corrections, Eq.\ {\eqref{GtildeG}} says that $\hat{G}$ is bounded, for small
real $z, \bar{z} > 0$ or $z, \bar{z} < 0$, by 1 up to small corrections. This
argument can be generalized to show that $\hat{G} (z, \eta z)$ for $\eta > 0$
real and $z$ near zero in the upper half plane is bounded by a
constant.\footnote{Let $z = r e^{i \varphi}$, $r \ll 1$. We consider $0
	\leqslant \varphi \leqslant \pi / 2$, when the argument uses the t-channel,
	the case $\pi / 2 \leqslant \varphi \leqslant \pi$ is analogous using the
	u-channel. The key point is that the $\rho$ variable in the t-channel $\rho_t
	\approx 1 - \sqrt{r} e^{i \varphi / 2}$, $| \rho_t | \approx 1 - \sqrt{r} \cos
	(\varphi / 2)$. This allows to compare the function $\hat{G} (z, \eta z)$ to
	$G_E (z', \eta z')$ with real $z' = r \cos^2 (\varphi / 2)$, times a factor
	$\sim (z \bar{z})^{\Delta_1} / (z' \bar{z}')^{\Delta_1}$ from the crossing
	kernel, which is bounded by a constant.}

We now pass to the non-rigorous part of the argument. Although the s-channel
stops converging when crossing $(1, + \infty)$, Ref.\ {\cite{Hartman:2015lfa}}
proposed that, in the regime $| \bar{z} | \ll | z | \ll 1$, the behavior
$\hat{G} (z, \bar{z})$ can nevertheless be predicted from the s-channel
expansion, by organizing it in $\bar{z}^{\tau / 2}$ where $\tau = \frac{1}{2}
(\Delta - \ell)$ is the twist. The typical term is
\begin{equation}
	\bar{z}^{\tau / 2} k_{\frac{1}{2} (\Delta + \ell)} (z),
\end{equation}
where $k_h (x) =_2 F_1 (h, h, 2 h, x)$ is the collinear conformal block. This
is the same expansion as used in the light-cone bootstrap, which has its own
problems of rigor, but here it is proposed to use it after $z - 1 \rightarrow
e^{2 \pi i} (z - 1)$. Under this continuation the collinear conformal block,
which has a $\log (1 - z)$ behavior near $z = 1$, picks up an imaginary piece
which, for $z$ small, behaves as $\sim i z^{1 - \frac{1}{2} (\Delta + \ell)}$
(see {\cite{Hartman:2015lfa}}, Eq.\ (4.28)). Considering $\bar{z} = \eta z$,
$\eta \ll 1$, $| z | \ll 1$, we then have, according to the proposal of Ref.\ {\cite{Hartman:2015lfa}},
\begin{equation}
	\hat{G} (z, \eta z) \approx 1 - B (\Delta, \ell) p_{\Delta, \ell} \times i
	\frac{\eta^{\tau / 2}}{z^{\ell - 1}}, \label{pred4}
\end{equation}
where $\Delta, \ell$ are the dimension and spin of the leading twist operator
(which may e.g.\ be the stress tensor), $p_{\Delta, \ell}$ its conformal block
coefficient, and $B (\Delta, \ell) \geqslant 0$ some explicitly known
constant. The spin $\ell$ is even since we are assuming $\mathcal{O}_1
=\mathcal{O}_2$. Eq.\ {\eqref{pred4}} assumes that the limit $\eta \rightarrow
0$ is taken before $z \rightarrow 0$.\footnote{In fact, in the opposite limit
	$z \rightarrow 0$ for fixed $\eta$, Eq.\ {\eqref{pred4}} would violate the
	discussed above rigorous bound that $\hat{G} (z, \eta z)$ is bounded by a
	constant. There is no paradox because that's not the limit we are supposed to
	be taking.}

Now, let us consider the holomorphic function $f (z) = 1 - \hat{G} (z, \eta z)$,
and integrate $z^{\ell - 2} f (z)$ along the contour shown in Fig.\ \ref{Gtilde}.\footnote{We can also take an intermediate step of adding a small
	semicircle of radius $\varepsilon'$ around zero, but since $\hat{G}$ is
	bounded for small $z$, the limit $\varepsilon' \rightarrow 0$ is not
	problematic.} We have
\begin{equation}
	\int_{\tmop{arc}} z^{\ell - 2} f (z)\, d z + \int_{-
		\varepsilon}^{\varepsilon} x^{\ell - 2} f (x)\, d x = 0 . \label{contour}
\end{equation}
Using {\eqref{pred4}} and that the integral over the arc of $1 / z$ is $\pi
i$, we get in particular:\footnote{Note that
	the quantity $\tmop{Re} [G_E (z, \eta z) - \hat{G} (z, \eta z)]$ appearing in~\eqref{result4} is essentially
	the double discontinuity considered in~\cite{Caron-Huot:2017vep}. Similarly, Eq.~\eqref{result4} can
	be formally obtained from the Lorentzian inversion formula of~\cite{Caron-Huot:2017vep} by expanding the integrand
	in a light-cone limit. We thank Tom Hartman for pointing this out.
}
\begin{equation}
	\pi B (\Delta, \ell) p_{\Delta, \ell} \approx \eta^{- \tau / 2} \int_{-
		\varepsilon}^{\varepsilon} x^{\ell - 2} \tmop{Re} [1 - \hat{G} (z, \eta z)]\,
	d x \approx \eta^{- \tau / 2} \int_{- \varepsilon}^{\varepsilon} x^{\ell
		- 2} \tmop{Re} [G_E (z, \eta z) - \hat{G} (z, \eta z)]\, d x, \label{result4}
\end{equation}
where in the final step we replaced 1 by $G_E (z, \eta z)$. Since $G_E (z,
\eta z)$ has a rigorously convergent expansion for small $z$, it satisfies the
bound:
\begin{equation}
	G_E (z, \bar{z}) = 1 + O (\eta^{\tau / 2}) .
\end{equation}
So the last replacement was legitimate if e.g.\ $\ell \geqslant 2$. By
{\eqref{GtildeG}}, the r.h.s.\ of {\eqref{result4}} is a positive quantity. \
This equation then implies that $p_{\Delta, \ell}$ must be positive as well.

As already mentioned, the weak point of this argument is that the s-channel
expansion stops converging when we cross $(1, + \infty)$. It is therefore not
at all obvious that analytic continuations of the individual conformal block
expansion terms have anything to do with the asymptotics of $\hat{G} (z,
\bar{z})$. Ref.\ {\cite{Hartman:2015lfa}} was of course aware of this, and
provided some arguments, inspired by the light-cone bootstrap, why nevertheless
the asymptotics from the leading twist terms can be trusted. We don't know how
to make those arguments rigorous. It would be interesting to understand if
asymptotics {\eqref{pred4}} can be justified using just Euclidean CFT axioms
of Sec.\ \ref{ECFTax} or requires additional assumptions. The same question
also looms over the proofs of ANEC {\cite{Hartman:2016lgu}} {and ANEC commutativity \cite{Kologlu:2019bco} which involved similar ``light-cone limit
	on the second sheet'' considerations.}

\section{OS axioms for higher-point functions}\label{OShigher}

In this appendix we discuss the modifications necessary to derive from the
Euclidean CFT axioms the OS axioms (positivity and cluster property) for
$n$-point functions with $n > 4$, compared to the $n \leqslant 4$ case
considered in Sec.\ \ref{OSfromCFT}. As we explain below, it appears that
there is no simple proof of OS positivity for $n > 4$ from the Euclidean CFT
axioms of Sec.\ \ref{ECFTax}. Since the reason for this is rather technical,
let us first discuss the conceptual implications of this.

Ideally, one would like to have a set of Euclidean CFT axioms that would imply
Wightman axioms (and therefore also OS axioms) and also be powerful enough to
derive all the usual CFT lore such as OPEs, radial quantization,
operator-state correspondence, crossing symmetry, etc. These statements, as
we saw in the main text, make sense and can be non-trivial even when we
restrict our attention to $n$-point functions with bounded $n$.

In particular, we have found that the axioms we formulated in Sec.\ \ref{ECFTax} achieve the above goal for $n \leqslant 4$. Extending our results
to $n > 4$ using the same strategy would require a solution to two problems:
first, we need to derive OS axioms (specifically, positivity and cluster
property) for $n > 4$, and, second, we need to prove that OS axioms together
with the OPE imply Wightman axioms.

Conceptually, it seems plausible that OS axioms + OPE imply Wightman axioms
for $n > 4$ because we expect that for $n > 4$ there is again an OPE channel
which is convergent in the entire forward tube (i.e.\ the one given by taking
the OPE in the same order as the operators appear in the Wightman ordering).
This question clearly merits further study but is beyond the scope of this
paper.

However, it is less clear to us how to even attempt a derivation of OS
positivity for $n > 4$ from Euclidean CFT axioms of Sec.\ \ref{ECFTax}. Let
us first explain why this is the case, and then we will discuss the possible
modifications to these CFT axioms.

Suppose we want to prove the positivity
\begin{equation}
	\langle \Psi | \nobracket \Psi \rangle \geqslant 0,
\end{equation}
where $\Psi$ is a state created by a product of three local operators, $| \Psi
\rangle = | \varphi_1 (x_1) \varphi_2 (x_2) \varphi_3 (x_3) \rangle$.

To prove this positivity the natural idea would be to use the OPE expansion
repeatedly for the two copies of $\Psi$ and then use the positivity of the
2-point function. However, for this we need our OPE approximation for
$\langle \nobracket \Psi | \nobracket$ to be conjugate to our approximation
for $| \Psi \rangle$. This is non-trivial to achieve because we have to
perform the OPEs one at a time. For example, we can first construct an
approximation of $| \Psi \rangle$ in terms of a state $| \Psi' \rangle$,
created by single operator insertions, such that
\begin{equation}
	| \langle \Psi | \nobracket \Psi \rangle - \langle \Psi | \nobracket \Psi'
	\rangle | < \varepsilon .
\end{equation}
Similarly, we can construct a state $\langle \Psi'' |$ such that
\begin{equation}
	| \langle \Psi'' | \nobracket \Psi' \rangle - \langle \Psi | \nobracket
	\Psi' \rangle | < \varepsilon
\end{equation}
and thus
\begin{equation}
	| \langle \Psi'' | \nobracket \Psi' \rangle - \langle \Psi | \nobracket \Psi
	\rangle | < 2 \varepsilon .
\end{equation}
These approximations are possible by the repeated use of the OPE
{\eqref{OPEplanar}}. Note, however, that since the OPE axiom is formulated for
correlation functions, the number of terms we have to include in the OPE for a
given $\varepsilon$ depends on the correlation function in which the OPE is
performed. It then follows that the state $\langle \Psi'' |$ depends on $|
\Psi' \rangle$ (because in order to construct it we use the OPE in the
correlation function $\langle \Psi | \nobracket \Psi' \rangle$) and is in
general different from it. It is therefore not obvious that $\langle \Psi'' |
\nobracket \Psi' \rangle \geqslant 0$, which is what we would like to use in
order to prove $\langle \Psi | \nobracket \Psi \rangle \geqslant 0$ with the
help of the above inequalities.

In the case when $n = 4$ and $| \Psi \rangle$ is created by 2 operators we
were able to solve this difficulty. This was because in this case the only
difference between $| \Psi' \rangle$ and $\langle \Psi'' |$ can be in the
number of OPE terms included in the approximation, and we were able to use an
orthogonality property of the 2-point function to show $\langle \Psi'' |
\nobracket \Psi' \rangle = \langle \Psi' | \nobracket \Psi' \rangle$ by
arguing that we can assume that $\langle \Psi'' |$ contains more terms than $|
\Psi' \rangle$ and that those terms which are in $\langle \Psi'' |$ but not in
$| \Psi' \rangle$ do not contribute to the product $\langle \Psi'' |
\nobracket \Psi' \rangle$.

This argument does not work in the case at hand, $| \Psi \rangle = | \varphi_1
(x_1) \varphi_2 (x_2) \varphi_3 (x_3) \rangle$. The reason for this is that in
order to construct $| \Psi' \rangle$ or $\langle \Psi'' |$ we need to perform
two OPE's in each case. For example, the first one can be $\varphi_1 \times
\varphi_2 = \sum_k \mathcal{O}_k$ and the second one can be $\varphi_3 \times
\mathcal{O}_k$. Both OPE's have to be truncated at some point, and while the
truncation of the second OPE affects only the set of terms that are present in
$| \Psi' \rangle$ or $\langle \Psi'' |$, where we truncate the first
$\varphi_1 \times \varphi_2$ OPE affects the \tmtextit{coefficients} of these
terms. Since now $| \Psi' \rangle$ and $\langle \Psi'' |$ contain terms with
differing coefficients, we cannot use orthogonality to argue $\langle \Psi'' |
\Psi' \rangle = \langle \Psi' | \Psi' \rangle $anymore.
There is no way to ensure that $\varphi_1 \times \varphi_2$ OPEs are truncated
in the same way in the construction of both states because the truncation in
$\langle \Psi'' |$ depends, through our OPE axiom, on $| \Psi' \rangle,$ and
thus might happen to be always at a higher order than the truncation used for
$| \Psi' \rangle .$

This all is to say that due to a rather technical reason it appears that there
is no \tmtextit{simple} proof of OS positivity of higher-point functions from
the Euclidean CFT axioms as stated in Sec.\ \ref{ECFTax}. Importantly, this
doesn't mean that there is no proof at all. Indeed, the Euclidean CFT axioms
are sufficient to derive the standard crossing-symmetry equations for
4-point functions. It could happen that in all solutions to these
crossing-symmetry equations the OPE coefficients have such asymptotics that a
stronger form of the OPE axiom holds and allows us to prove the OS positivity
for $n > 4$. However, it is not clear how to implement this line of reasoning
in practice.

It is therefore interesting to look for a stronger version of Euclidean CFT
axioms. We discuss below some simple modifications of the OPE axiom which
avoid the above problem and allow to prove OS positivity for higher-point
functions.

Morally, we want some kind of statement of uniformity for the convergence rate of the OPE: 
it should make $|\Psi''\>$ above independent of the truncation made in $|\Psi'\>$, as long as this truncation
is done at a sufficiently high order. This would allow us to make both truncations at a high order and
ensure $\langle \Psi'' |\Psi' \rangle = \langle \Psi' | \Psi' \rangle \geq 0$.

One option is to assume a stronger form of the OPE, which allows us to perform
two OPE's simultaneously Specifically, we can assume that the double sum
\begin{equation}
	\langle \mathcal{O}_1 \mathcal{O}_2 \mathcal{O}_3 \mathcal{O}_4 \ldots
	\rangle = \sum_{k, l} \langle \mathcal{O}_k \mathcal{O}_l \ldots \rangle,
\end{equation}
is convergent, where we wrote the two OPEs schematically as $\mathcal{O}_1
\mathcal{O}_2 = \sum_k \mathcal{O}_k$ and $\mathcal{O}_3 \mathcal{O}_4 =
\sum_l \mathcal{O}_l$. Convergence of the double sum means that
\begin{equation}
	\left| \langle \mathcal{O}_1 \mathcal{O}_2 \mathcal{O}_3 \mathcal{O}_4
	\ldots \rangle - \sum_{k, l} \langle \mathcal{O}_k \mathcal{O}_l \ldots
	\rangle \right| < \varepsilon,
\end{equation}
when the sums are truncated in a way that includes some
$\varepsilon$-dependent finite set of terms, but is otherwise arbitrary. In
particular, both sums can be truncated in the same way, and this solves the
problem that we encountered above. A disadvantage of this approach is that it
is unclear how to derive this axiom from OS axioms and the usual single OPE axiom (however,
a heuristic argument based on cutting the Euclidean path integral can be
made). This is somewhat subtle and is related to the question of whether the path integral over a spherical layer $(r_1<r<r_2)$ with operator insertions in the interior represents a bounded operator. We can't say with confidence whether or not this is the case.

Another option is to assume resummed repeated OPE, i.e.\ that the following sum
converges, schematically,
\begin{equation}
	\langle \mathcal{O}_1 \ldots \mathcal{O}_m \mathcal{O}_{m + 1} \ldots
	\mathcal{O}_n \rangle = \sum_k c_k \langle \mathcal{O}_k \mathcal{O}_{m + 1}
	\ldots \mathcal{O}_n \rangle,
\end{equation}
where the coefficients $c_k$ are chosen so that
\begin{equation}
	\langle \mathcal{O}_1 \ldots \mathcal{O}_m \mathcal{O}_k^{\theta} \rangle =
	c_k \langle \mathcal{O}_k \mathcal{O}_k^{\theta} \rangle,
\end{equation}
assuming $\langle \mathcal{O}_k \mathcal{O}_l^{\theta} \rangle \propto
\delta_{k l} .$ This version of the axiom is essentially the statement that
one-operator states form a basis of the CFT Hilbert space, formulated without
explicitly introducing the Hilbert space. In other words, above we are
approximating the state $\langle \mathcal{O}_1 \ldots \mathcal{O}_m |$ in
terms of an orthonormal basis of states $\langle \mathcal{O}_k |$, and the
coefficients are computed by inner products. This form of the axiom is easy to
derive from OS + convergent OPE, and also easily allows us to solve our
problem by using the same strategy as in the case $n = 4$. However, it does
appear to be an overly strong assumption, making our axioms not very different
from assuming OS + convergent OPE outright.

Finally, an interesting prospect might be, instead of formulating an entirely
new set of axioms, to add an {\tmem{asymptotic}} OPE axiom (and conformal
invariance) to OS axioms. It is likely that using logic very similar to that
of Mack {\cite{Mack:1976pa}}, which we reviewed in Sec.\ \ref{MackComp}, one
can prove that (OS axioms)+(asymptotic OPE)+(conformal invariance) imply
convergent OPE.

\section{Details on Vladimirov's theorem}\label{Vlad}

\subsection{Limit in the sense of distributions}

Let us start with a reminder of what the limit in the sense of tempered
distributions means. Let $f (u)$, $u = (t_k, \mathbf{x}_k) \equiv (t_1,
\mathbf{x}_1, \ldots, t_n, \mathbf{x}_n) \in \mathbb{R}^{n d}$, be a Schwartz
test function, i.e.\ an infinitely differentiable function decreasing at
infinity faster than any power together with all its derivatives. This can be
also stated as finiteness of all Schwartz norms:
\begin{equation}
	| f |_N = \sup_{u \in \mathbb{R}^{n d}, | \alpha | \leqslant N} (1 + u^2)^{N
		/ 2} | \partial^{\alpha}_u f | < \infty \qquad \forall N \geqslant 0 .
	\label{semin}
\end{equation}
That the limit {\eqref{limit}} exists in the sense of distributions means two
requirements. First, that the r.h.s.\ of {\eqref{limit}} has a finite limit
integrated against any $f$ as above:
\begin{equation}
	(G_n^M, f) \assign \lim_{\epsilon_k \rightarrow 0} \int d t\, d\mathbf{x}\, G_n
	(\epsilon_k + i t_k, \mathbf{x}_k) f (t_k, \mathbf{x}_k) \quad \text{exists
		for any Schwartz} f. \label{intGn}
\end{equation}
The $G_n^M$ defined by this equation is a linear functional on the Schwartz
space. The second requirement is that this functional should be continuous
(and thus is itself a tempered distribution). Continuity means that it should
be bounded by one of the norms {\eqref{semin}} with a sufficiently large $N$,
i.e.:
\begin{equation}
	| (G_n^M, f) | \leqslant C | f |_{N_{\ast}}, \label{GMcont}
\end{equation}
with $f$-independent $C$ and $N_{\ast}$.

Note that by Eq.\ {\eqref{GMcont}}, $G_n^M$ can be extended from the
Schwartz space to a larger space of test functions, which are required to be
differentiable only $N_{\ast}$ times and have a finite $| f
|_{N_{\ast}}$. Parameter $N_{\ast}$ thus characterizes regularity of the
distribution $G_n^M$. The proof of Theorem \ref{ThVlad} will determine
$N_{\ast}$ in terms of $A_n$ and $B_n$, see Eq.\ {\eqref{GMreg}}.

\subsection{Proof of Theorem \ref{ThVlad}}\label{Proof1}

Unfortunately, we do not know a reference where Theorem \ref{ThVlad} is stated
and proved succinctly in the form we need. Such results are considered
standard in the theory of several complex variables. For similar statements
see {\cite{Vladimirov}}, Chapter 5, and {\cite{Streater:1989vi}}, Theorem
2-10. For the convenience of the reader, we present here a proof based on
these sources.

The usefulness of Vladimirov's theorems for establishing distributional
properties of CFT correlators was recognized in our recent work
part \ref{part:crossratio}. There, we considered expansions of the CFT 4-point function $g
(\rho, \bar{\rho})$ in terms of conformally invariant cross-ratios $\rho$,
$\bar{\rho}$. It is well known that such expansions converge in the interior
of the unit disk $| \rho |, | \bar{\rho} | < 1$. Using Vladimirov's theorems,
we showed in part \ref{part:crossratio} that they also converge on the boundary of this
disk, in the sense of distributions. In this paper we are interested in CFT
correlators as functions of positions $x_k$, not of cross-ratios, but the
basic principle is the same as in part \ref{part:crossratio}: a powerlaw bound on an
holomorphic function near a boundary implies temperedness of the limiting
distribution.

By translation invariance it's enough to study the function $G_n$ expressed
in terms of the differences $y_k = x_k - x_{k + 1}$ which we denote by
$\mathcal{G} (y)$, $y = (y_1, \ldots, y_{n - 1})$. We also denote $y_k =
(y_k^0, \mathbf{y}_k),$ $y_k^0 = \varepsilon_k + i s_k$, $\varepsilon_k > 0$,
$\mathbf{y}_k \in \mathbb{R}^{d - 1}$.

Consider first the case when all $\varepsilon_k$ go to zero together along a
fixed direction: $\varepsilon_k = r v_k$ where $r \rightarrow 0$ and $v =
(v_k)$ is a vector with positive components. Later on we will show that the
limit continues to exist if $\varepsilon_k \rightarrow 0$ independently (as
well as the more general statement about the limit from inside the forward
tube).

So, let us prove that $\mathcal{G} (y)$ has a limit as $r \rightarrow 0$ which
is a tempered distribution in variables $s_k$, $\mathbf{y}_k$. As in
{\eqref{intGn}}, we fix a Schwartz test function $f$ and consider the integral
(we will omit index $k$ on $\varepsilon, s, v, \mathbf{y}$ if no confusion may
arise)
\begin{equation}
	h (r) = \int d s\, d\mathbf{y}\,\mathcal{G} (r v + i s, \mathbf{y}) f (s,
	\mathbf{y}) .
\end{equation}
The problem is analogous to theorems used in part \ref{part:crossratio}, so we will be
brief. As in section \ref{ThVlad} and appendix \ref{app:proofVlad2}, using analyticity in
$y^0$, integration by parts, and the powerlaw bound one can show that
derivatives of $h$ in $r$ satisfy the bound:
\begin{equation}
	| \partial_r^j h (r) | \leqslant \frac{C}{r^{A_n}}  | f |_N,
	\label{jbound}
\end{equation}
where $| f |_N$ is a Schwartz norm {\eqref{semin}} of a sufficiently large
order $N$ depending on $j$ and $B_n$. The constant $A_n$ is the same as in
{\eqref{powerlawbound}}, in particular the same $A_n$ works for all $j$. In
what follows we only need this equation for finitely many $j$ (up to $[A_n] +
1$). Using the Newton-Leibniz formula in the $r$ direction several times, one
then proves that the same bound as {\eqref{jbound}} holds in fact without $1 /
r^{A_n}$ singularity in the r.h.s. It then follows that, first of all,
$\lim_{r \rightarrow 0} h (r)$ exists, and second, it is a continuous linear
functional of $f$, that is, a distribution. The limit holds uniformly when the
components $v_k$ vary on any fixed compact interval contained in $(0, +
\infty)$. Its $v$-independence is shown exactly as in part \ref{part:crossratio}, Eq.
(C.7). Let us denote the limiting distribution $\mathcal{G} (i s, \mathbf{y})
\equiv \mathcal{G}^M (s, \mathbf{y})$.

It is of some interest to know the precise regularity of the distribution
$\mathcal{G}^M$ (i.e.\ how many derivatives the test function must have to be
pairable with $\mathcal{G}^M$) and the rate of its growth at infinity.
Following the above argument in detail, one can show the following bound which
contains this information:
\begin{equation}
	| (\mathcal{G}^M, f) | \leqslant \tmop{Const} . \int d s\, d\mathbf{y}\, (1 +
	| s | + | \mathbf{y} |)^{B_n} \max_{| \alpha | \leqslant [A_n] + 1} |
	\partial^{\alpha}_s f (s, \mathbf{y}) | . \label{GMreg}
\end{equation}
This in particular implies {\eqref{GMcont}} with $N_{\ast} = \max ([A_n] + 1,
B_n + n d + 1)$.

Parts 2,3 of Theorem \ref{ThVlad} are new compared to part \ref{part:crossratio}, since
such questions do not arise in the cross-ratio space.

Lorentz invariance is easy to show, as follows. Rotation invariance of
$G^E_n$ implies that $\mathcal{G} (y)$ satisfies for real $y$ the
differential equations
\begin{equation}
	\{ y^a \partial_{y^b} - y^b \partial_{y^a} \} \mathcal{G} (y) = 0,
	\qquad a, b \in \{ 0, 1, \ldots, d - 1 \} \label{DErot}
\end{equation}
(as usual $y = (y_k)$, summation in $k$ understood). By the uniqueness of
analytic continuation, these equations continue to hold for complex $y_0$.
That's the only place where we use real-analyticity in the spatial
direction.\footnote{With some extra tricks, it's possible to replace it by the
	assumption of mere continuity in $\mathbf{y}$, as in {\cite{osterwalder1975}},
	Theorem 4.3. In the CFT applications we have in mind, real analyticity appears
	a more natural assumption.} By taking the limit $\varepsilon \rightarrow 0$ in
{\eqref{DErot}}, we recover precisely the differential equations expressing
the Lorentz invariance of $\mathcal{G}^M$. Let us explain in more detail how
the limit is taken and why it exists. Consider for definiteness $a = 0$, $b =
1$, other cases being similar. Eq.\ {\eqref{DErot}} then says \ $\{
(\varepsilon + i t) \partial_{y^1} + i y^1 \partial_t \} \mathcal{G}
(\varepsilon + i t, \mathbf{y}) = 0$, in the sense of functions, and hence
integrating by parts in the sense of distributions acting on test functions
$\varphi (t, \mathbf{y})$:
\begin{equation}
	(\mathcal{G}_{\varepsilon}, \{ (\varepsilon + i t) \partial_{y^1} + i y^1
	\partial_t \} \varphi) = 0, \label{Gepsdistr}
\end{equation}
where we denoted $\mathcal{G}_{\varepsilon} (t, \mathbf{y}) =\mathcal{G}
(\varepsilon + i t, \mathbf{y})$. Now we take the limit $\varepsilon
\rightarrow 0$. We know that (a) $\mathcal{G}_{\varepsilon} \rightarrow
\mathcal{G}^M$ in the sense of distributions, and also that (b) $|
(\mathcal{G}_{\varepsilon}, \varphi) |$ is uniformly bounded as $\varepsilon
\rightarrow 0$ by some Schwartz norm of $\varphi$. By (b) the term
$(\mathcal{G}_{\varepsilon}, \varepsilon \varphi)$ in {\eqref{Gepsdistr}}
drops out when $\varepsilon \rightarrow 0$, and by (a) the rest tends to
$(\mathcal{G}^M, \{ i t \partial_{y^1} + i y^1 \partial_t \} \varphi)$.
So we conclude that \ $(\mathcal{G}^M, \{ i t \partial_{y^1} + i y^1
\partial_t \} \varphi) = 0$ which expresses invariance of $\mathcal{G}^M$
under the $01$ Lorentz transformation.

Let us proceed to show the rest of Parts 2,3. It will be crucial that
$\mathcal{G}$ can be written as a ``Fourier-Laplace transform'':
\begin{equation}
	\mathcal{G} (\varepsilon + i s, \tmmathbf{\mathbf{y}}) = \int \frac{d E\,
		d\mathbf{p}}{(2 \pi)^{d (n - 1)}}\, g (E, \mathbf{p}) e^{- (\varepsilon + i s)
		E - i\mathbf{p}\mathbf{y}} \label{FL},
\end{equation}
where $g (E, \mathbf{p}), E \in \mathbb{R}^{n - 1}, \mathbf{p} \in
(\mathbb{R}^{d - 1})^{n - 1}$ is a tempered distribution, called ``spectral
function'', supported at $E \geqslant 0$ (by which we mean all $E_k \geqslant
0$) [later this will be improved to $E \geqslant | \mathbf{p} |$]. We are
omitting the indices, thus $\varepsilon E = \sum_k \varepsilon_k E_k$, etc.
The equality in {\eqref{FL}} is understood in the sense of distributions, with
the r.h.s.\ being the inverse Fourier transform of the tempered distribution $g
(E, \mathbf{p}) e^{- \varepsilon E}$. In other words, what this means is that
\begin{equation}
	\int d s\, d\mathbf{y} \, \mathcal{G} (\varepsilon + i s,
	\tmmathbf{\mathbf{y}}) f (- s, -\mathbf{y}) = \int \frac{d E\, d\mathbf{p}}{(2
		\pi)^{d (n - 1)}}  \, g (E, \mathbf{p}) e^{- \varepsilon E} \hat{f} (E,
	\mathbf{p}), \label{FTmeans}
\end{equation}
for any Schwartz test function $f$, and $\hat{f}$ its Fourier transform.

Let us show {\eqref{FL}}. Notice first that for every $\varepsilon > 0$ we can
write
\begin{equation}
	\mathcal{G} (\varepsilon + i s, \tmmathbf{\mathbf{y}}) = \int \frac{d E\,
		d\mathbf{p}}{(2 \pi)^{d (n - 1)}}\, g_{\varepsilon} (E, \mathbf{p}) e^{- i s E
		- i\mathbf{p}\mathbf{y}} \label{gx},
\end{equation}
where $g_{\varepsilon}$ is the Fourier transform of $\mathcal{G} (\varepsilon
+ i s, \tmmathbf{\mathbf{y}})$ with respect to $s, \mathbf{y}$. This Fourier
transform exists as a tempered distribution, since $\mathcal{G} (\varepsilon +
i s, \tmmathbf{\mathbf{y}})$ is itself a tempered distribution in $s,
\mathbf{y}$ (being a real-analytic function, bounded by a power at infinity).
In addition, $\mathcal{G} (\varepsilon + i s, \tmmathbf{\mathbf{y}})$ is
differentiable in $\varepsilon$ and $s$ and satisfies the Cauchy-Riemann
equations. From here it's easy to show that $g_{\varepsilon}$ as a
distribution is differentiable in $\varepsilon$ and satisfies the differential
equations:
\begin{equation}
	\frac{\partial g_{\varepsilon}}{\partial \varepsilon_k} + E_k
	g_{\varepsilon} = 0 \qquad (k = 1, \ldots, d - 1) .
\end{equation}
From here we conclude that
\begin{equation}
	g (E, \mathbf{p}) : = g_{\varepsilon} (E, \mathbf{p}) e^{\varepsilon E}
	\label{gge}
\end{equation}
is an $\varepsilon$-independent distribution. Substituting $g_{\varepsilon}
(E, \mathbf{p}) = g (E, \mathbf{p}) e^{- \varepsilon E}$ into {\eqref{gx}},
we obtain {\eqref{FL}}. Note that since $g$ and $g_{\varepsilon}$ are related
by an exponential factor, we can so far only claim that $g$ is defined as a
distribution on test functions of compact support. Let us show next that it is
in fact tempered (i.e.\ extends to Schwartz test functions).

To this end, consider the inverse of {\eqref{gx}}:
\begin{equation}
	g_{\varepsilon} (E, \mathbf{p}) = g (E, \mathbf{p}) e^{- \varepsilon E} =
	\int d s\, d\mathbf{y}\, \mathcal{G} (\varepsilon + i s, \tmmathbf{\mathbf{y}})
	e^{i s E + i\mathbf{p}\mathbf{y}},
\end{equation}
and integrate it against a compactly supported test function $\varphi (E,
\mathbf{p})$. We get (compare {\eqref{FTmeans}}):
\begin{equation}
	\int d E\, d\mathbf{p}\, g (E, \mathbf{p}) e^{- \varepsilon E} \varphi (E,
	\mathbf{p}) = \int d s\, d\mathbf{y}\, \mathcal{G} (\varepsilon + i s,
	\tmmathbf{\mathbf{y}}) \hat{\varphi} (- s, -\mathbf{y}) .
\end{equation}
As $\varepsilon \rightarrow 0$, the l.h.s.\ tends to the pairing $(g,
\varphi)$. Using Part 1 of the theorem, the r.h.s.\ tends in the same limit to
$\int d s\, d\mathbf{y}\, \mathcal{G}^M (s, \tmmathbf{\mathbf{y}}) \hat{\varphi} (-
s, -\mathbf{y})$ which exists in the sense of tempered distributions and so is
bounded by some Schwartz-space norm $| \hat{\varphi} |_N$. We get
\begin{equation}
	| (g, \varphi) | \leqslant \tmop{const} . | \hat{\varphi} |_N \leqslant
	\tmop{const} . | \varphi |_{N'},
\end{equation}
where in the second inequality we used that the Fourier transform is
continuous in the Schwartz space. This inequality, valid for any compactly
supported $\varphi$, means that $g$ extends to a tempered distribution on the
whole Schwartz space. The representation {\eqref{FL}} is thus established.

Next let us show that $g$ is supported at $E \geqslant 0$. For this we will
pass to the large $\varepsilon$ limit in {\eqref{gge}}. Supposing that $E_k <
0$ for some $k$, the factor $e^{E \varepsilon}$ in {\eqref{gge}} decreases
exponentially as the corresponding $\varepsilon_k \rightarrow + \infty$. On
the other hand $g_{\varepsilon} (E, \mathbf{p})$ is bounded in the same limit
by a power of $\varepsilon$, because it's the Fourier transform of
$\mathcal{G} (\varepsilon + i s, \tmmathbf{\mathbf{y}})$ which satisfies a
powerlaw bound.\footnote{This is the only place where we use the powerlaw
	bound on $\mathcal{G} (\varepsilon + i s, \tmmathbf{\mathbf{y}})$ for large
	rather than small $\varepsilon$.} This implies that $g (E, \mathbf{p}) = 0$
unless $E \geqslant 0$.\footnote{If unhappy with this intuitive reasoning, the
	argument may be made more rigorous in its integrated version: show that $g$
	vanishes on test functions supported in the complement of $E \geqslant 0$.}

Consider then the following lemma, proven analogously to, and easier than,
Lemma \ref{tubeLemma} below.

\begin{lemma}
	\label{gEp}Let $g (E, \mathbf{p})$ be a tempered distribution supported at
	$E \geqslant 0$, and consider the distribution $g (E, \mathbf{p}) e^{-
		\varepsilon E}$ ($\varepsilon > 0$). This distribution, being initially
	defined by this formula on compactly supported test functions, extends to a
	tempered distribution, and moreover $g (E, \mathbf{p}) e^{- \varepsilon E}
	\rightarrow g (E, \mathbf{p})$ as $\e\to 0$, in the sense of tempered distributions.
\end{lemma}

Let us now take the $\varepsilon \rightarrow 0$ limit on both sides of
{\eqref{FL}} (or, which is the same, {\eqref{FTmeans}}). The l.h.s.\ has a
limit by Part 1, while the r.h.s.\ has a limit by Lemma \ref{gEp}. We obtain
that $\mathcal{G}^M (s, \tmmathbf{\mathbf{y}})$ and $g (E,
\mathbf{p})$ are related by the Fourier transform:
\begin{equation}
	\mathcal{G}^M (s, \tmmathbf{\mathbf{y}}) = \int \frac{d E\, d\mathbf{p}}{(2
		\pi)^{d (n - 1)}}\, g (E, \mathbf{p}) e^{- i s E - i\mathbf{p}\mathbf{y}} .
	\label{gxM}
\end{equation}
We can now complete the proof of Part 2, namely to show the spectral
condition. Above we proved that $\mathcal{G}^M (s, \tmmathbf{\mathbf{y}})$ is
Lorentz invariant. Since $g (E, \mathbf{p})$ is its Fourier transform, it is
also Lorentz invariant, and in particular its support must be a
Lorentz-invariant set. We also know that $\tmop{supp} g \subset \{ E \geqslant
0 \}$. These two facts together imply that $\tmop{supp} g$ must be contained
in the product of the forward null cones, i.e.\ $g (E, \mathbf{p}) = 0$ unless
each $E_k \geqslant | \mathbf{p}_k |$, which is the spectral condition.

Part 3 follows by standard Wightman theory arguments. Namely, let us write $(i
y_k^0, \mathbf{y}_k) = \xi_k + i \eta_k $ where $\xi_k, \eta_k \in
\mathbb{R}^{1, d - 1}$ and $\eta_k = (\tmop{Re} y_k^0, \tmop{Im}
\mathbf{y}_k) \succ 0$. The extension to the forward tube is given by the
equation (with $p=(E,\mathbf{p})$)
\begin{equation}
	\int d p\, g (p) e^{i (p, \xi)} e^{(p, \eta)}, \label{FText}
\end{equation}
which reduces to {\eqref{FL}} for real $\mathbf{y}_k$. It is holomorphic by Part
(c) of the following lemma, while Parts (a,b) imply that this extension has
the same limit as {\eqref{FL}}.

\begin{lemma}
	\label{tubeLemma}Let $g (p)$ be a tempered supported at $p \succeq 0$
	(closed forward light cone). Consider the distribution $g_{\eta} (p) = g (p)
	e^{- (p, \eta)}$, initially defined by this formula on compactly supported
	test functions. Then
	
	(a) $g_{\eta}$ for $\eta \succ 0$ extends to a tempered distribution;
	
	(b) $g_{\eta} \rightarrow g$ as $\eta \rightarrow 0$ from inside the forward
	light cone $\eta \succ 0$, in the sense of tempered distributions;
	
	(c) The Fourier transform $\widehat{g_{\eta}} (\xi)$ of the distribution
	$g_{\eta} (p)$ is a holomorphic function of $\xi + i \eta$ {for $\eta\succ 0$}.
\end{lemma}

\begin{proof}
	Let $\omega (p)$ be a $C^{\infty}$ function which is identically 1 on the
	forward light cone $\overline{V_+}$, and zero as soon as $\tmop{dist} (p,
	\overline{V_+}) \geqslant 1$ where $\tmop{dist}$ is the Euclidean distance.
	We can choose this function so that all its derivatives are uniformly
	bounded by a constant depending only on the derivative order: $|
	\omega^{(\alpha)} (p) | \leqslant C_{\alpha}$ for any $p$.
	
	Consider the family of $C^{\infty}$ functions parametrized by $\xi, \eta \in
	\mathbb{R}^{1, d - 1}$:
	\begin{equation}
		\Omega_{\xi, \eta} (p) = e^{i (p, \xi)} e^{(p, \eta)} \omega (p) .
	\end{equation}
	It is not hard to check that $\Omega_{\xi, \eta}$ is a Schwartz function for
	$\eta \succ 0$ and any $\xi$.
	
	Let us define $g_{\eta}$ paired with a Schwartz function $\varphi (p)$ via
	\begin{equation}
		(g_{\eta}, \varphi) = (g, \Omega_{0, \eta} \varphi) .
	\end{equation}
	We know that $\Omega_{0, \eta} \varphi$ is a Schwartz function for $\eta
	\succ 0$, so this definition makes sense. Furthermore it is not hard to
	check that $\Omega_{0, \eta} \varphi \rightarrow \omega \varphi$ in the
	Schwartz space topology as $\eta \rightarrow 0, \eta \succ 0$. This proves
	Parts (a),(b).
	
	Next, let us define
	\begin{equation}
		F (\xi, \eta) = (g, \Omega_{\xi, \eta}), \qquad \xi, \eta \in
		\mathbb{R}^{1, d - 1} .
	\end{equation}
	We know that $\Omega_{\xi, \eta}$ is a Schwartz function for $\eta \succ 0$,
	so $F (\xi, \eta)$ is a function. Moreover it is not hard to show that the
	family $\Omega_{\xi, \eta}$ is continuous and continuously differentiable in
	the Schwartz space topology. It also obviously satisfies the Cauchy-Riemann
	equations: $(\partial_{\xi} + i \partial_{\eta}) \Omega_{\xi, \eta} = 0$.
	This implies that $F (\xi, \eta)$ is a holomorphic function in $\xi + i \eta$.
	It remains to show that $F (\xi, \eta) = \widehat{g_{\eta}} (\xi)$. It's
	enough to check this integrated against a compactly supported test function
	$\chi (\xi)$:
	\begin{equation}
		\int F (\xi, \eta) \chi (\xi)\, d \xi = \int (g, \Omega_{\xi, \eta}) \chi
		(\xi)\, d \xi = \left( g, \int d \xi\, \chi (\xi) \Omega_{\xi, \eta} \right) =
		(g, \Omega_{0, \eta} \hat{\chi} ) = (g_{\eta}, \hat{\chi} ) =
		(\widehat{g_{\eta}}, \chi) .
	\end{equation}
	The proof is complete.
\end{proof}

\section{Intuition about Lemma \ref{lemma:fcheckdense}}\label{IntLem1}

The proof of Lemma \ref{lemma:fcheckdense} in Sec.\ \ref{MinkFromEucl} was
by contradiction. To help intuition, we will give here a constructive argument
of a special case of Lemma \ref{lemma:fcheckdense}, namely $d = 1$ and $n =
2$. I.e.\ we will show how any Schwartz function $f \in \mathcal{S}
(\mathbb{R})$ can be approximated by Schwartz functions $g$ which for $E
\geqslant 0$ agree with Laplace transform:
\begin{equation}
	\mathcal{L} (\varphi) (E) = \int_0^{\infty} d t\, \varphi (t) e^{- E t},
\end{equation}
$\varphi \in C_0^{\infty} (\mathbb{R}_+)$ (compactly supported with support
strictly inside $(0, + \infty)$), while for $E < 0$, $g (E)$ is extended
arbitrarily. Recall that the Schwartz space topology is given by the family of
norms
\begin{equation}
	| f |_n = \sup_{E \in \mathbb{R}, m \leqslant n} (1 + E^2)^{n / 2} | f^{(m)}
	(E) |, \label{fnnorm}
\end{equation}
and we need to find a sequence $\{ g_r \}_{r = 1}^{\infty}$ such that $| f -
g_r |_n \rightarrow 0$ as $r \rightarrow \infty$ for any $n$ (we stress that
one sequence $g_r$ should work for any $n$).

We will also consider the Schwartz space $\mathcal{S}
(\overline{\mathbb{R}_+})$, consisting of $C^{\infty}$ functions on $E
\geqslant 0$ (not necessarily vanishing at $E = 0$) with topology given by the
family of norms $| f |_{n, +}$ defined by the same equations as
{\eqref{fnnorm}} but with sup taken over $E \geqslant 0$. It will be
sufficient to arrange that for any $n$
\begin{equation}
	| f -\mathcal{L} (\varphi_r) |_{n, +} \rightarrow 0 \qquad (r \rightarrow
	\infty) \label{suffSRp} .
\end{equation}
This is because there exists an extension operator which takes a function $h
\in \mathcal{S} (\overline{\mathbb{R}_+})$ and provides a function
$\mathcal{E} (h) \in \mathcal{S} (\mathbb{R})$ such that $\mathcal{E}
(h) = h$ for $E \geqslant 0$ (which is why it called an extension operator),
and in addition
\begin{equation}
	| \mathcal{E} (h) |_n \leqslant C_n | h |_{n, +} \label{Seeley}
\end{equation}
for all $n$ with some finite constants $C_n$ independent of $h$. E.g.,
Seeley's linear extension operator
{\cite{seeley_1964,wiki:Whitney_extension_theorem}} has this property. Then,
given {\eqref{suffSRp}}, we put
\begin{equation}
	g_r = f +\mathcal{E} (\mathcal{L} (\varphi_r) - f),
\end{equation}
which, on the one hand satisfies $g_r (E) =\mathcal{L} (\varphi_r) (E)$ for $E
\geqslant 0$ and on the other hand by {\eqref{suffSRp}} and {\eqref{Seeley}}
has $| g_r - f |_n \leqslant C_n | \mathcal{L} (\varphi_r) - f |_{n, +}
\rightarrow 0$ which is what we need.

So let us focus on satisfying {\eqref{suffSRp}}. By a map $x = \frac{1}{1 +
	E}$ the half-line $[0, + \infty)$ is mapped to the interval $(0, 1]$ and the
function $f (E)$ is mapped to a function $F (x) = f \left( \frac{1}{x} - 1
\right)$ which is a $C^{\infty}$ function vanishing at $x = 0$ faster than any
power of $x$. For any $\varepsilon$ and any $N$ we can find, by the
Weierstrass theorem, a polynomial $Q (x)$ such that
\begin{equation}
	| F^{(N)} (x) - Q (x) | \leqslant \varepsilon \qquad (0 \leqslant x
	\leqslant 1) .
\end{equation}
Let $P (x)$ be the polynomial such that $P^{(N)} (x) = Q (x)$ and $P (0) =
\cdots = P^{(N - 1)} (0) = 0$. Then $P (x) = O (x^N)$ and it is not hard to
see that
\begin{equation}
	| F^{(n)} (x) - P^{(n)} (x) | \leqslant \varepsilon x^{N - n} \qquad (0
	\leqslant x \leqslant 1) . \label{FPbound}
\end{equation}
We also put $p (E) = P \left( \frac{1}{1 + E} \right)$. From \ $f (E) = F
\left( \frac{1}{1 + E} \right)$ we know that
\begin{equation}
	| f^{(n)} (E) | \leqslant B_n \max_{m \leqslant n} \left| F^{(m)} \left(
	\frac{1}{1 + E} \right) \right| .
\end{equation}
So combining this with Eq.\ {\eqref{FPbound}}, and going up to $n = N / 2$ we
may conclude that
\begin{equation}
	| f - p |_{N / 2, +} \leqslant B'_N \varepsilon . \label{f-p}
\end{equation}
Now, by construction $p$ has the form
\begin{equation}
	p (E) = \underset{N \leqslant n \leqslant M}{\sum} a_n \frac{1}{(1 + E)^n} .
	\label{def:g}
\end{equation}
Since $\frac{1}{(1 + E)^n} = \frac{1}{(n - 1) !} \int_0^{\infty} t^{n - 1}
e^{- (1 + E) t}\, d t$, we see that $p (E)$ is the Laplace transform of a
function $\psi (t)$:
\begin{equation}
	p =\mathcal{L} (\psi), \quad \psi (t) = \underset{N \leqslant n \leqslant
		M}{\sum} \frac{a_n}{(n - 1) !} t^{n - 1} e^{- t} .
\end{equation}

Now we can finish the argument as follows. For $r = 1, 2, 3, \ldots$ we apply
the above argument with $N = 2 r$ and $\varepsilon = 1 / (B_N' r)$ to find
$\psi_r$ such that, by {\eqref{f-p}},
\begin{equation}
	| f -\mathcal{L} (\psi_r) |_{r, +} \leqslant 1 / r . \label{fpsir}
\end{equation}
The function $\psi_r$ is not in $C_0^{\infty} (0, \infty)$ although
$\psi_r^{(k)} = 0$ for $k = 0 \ldots 2 r - 2$, and it vanishes at $\infty$
exponentially. We can therefore approximate $\psi_r$ by a $C_0^{\infty} (0,
\infty)$ function $\varphi_r$ so that $| \psi_r - \varphi_r |_{2 r - 2, +}$ is
arbitrarily small, where the order $2 r - 2$ of the norm is related to the
order of the vanishing of $\psi_r$ at $t = 0$. Furthermore we have the
following lemma:

\begin{lemma}
	Let $\chi$ be a $C^{\infty}$ function on $[0, + \infty)$ which exponentially
	vanishes at infinity and
	\begin{equation}
		\chi^{(k)} (0) = 0, \quad k = 0 \ldots n - 1 \label{chik} .
	\end{equation}
	Then, with some constant $D_n$ independent of $\chi$,
	\begin{equation}
		| \mathcal{L} (\chi) |_{n, +} \leqslant D_n | \chi |_{n + 2, +} .
		\label{further}
	\end{equation}
\end{lemma}

\begin{proof}
	We use the following elementary properties of Laplace transform:
	\begin{eqnarray}
		& \left( \frac{d}{d E} \right)^m \mathcal{L} (\chi) (E) =\mathcal{L}
		[\chi (t) (- t)^m] (E), &  \nonumber\\
		& E^n \mathcal{L} (\chi) (E) =\mathcal{L} [\chi^{(n)} (t)] (E), & 
	\end{eqnarray}
	where the second equation is derived by integration by parts and is valid
	under {\eqref{chik}} and exponential decay. So we have (where $\lesssim$
	denotes $\leqslant$ with some $n$-dependent but function-independent
	constant)
	\begin{equation}
		| \mathcal{L} (\chi) |_{n, +} \lesssim \sum_{m = 0}^n \sup_{E \geqslant 0}
		(1 + E^n) | \mathcal{L} (\chi)^{(m)} (E) | \leqslant \sum_{m = 0}^n
		\sup_{E \geqslant 0} | \mathcal{L} [\chi (t) t^m] (E) | + | \mathcal{L}
		[(\chi (t) t^m)^{(n)}] (E) |,
	\end{equation}
	Using further the elementary bound $| \mathcal{L} (f) (E) | \lesssim \sup_{t
		\geqslant 0} | (1 + t^2) f (t) |$ we deduce {\eqref{further}}.
\end{proof}

We use this lemma with $n = r$ and $\chi = \psi_r - \varphi_r$, which
satisfies $\chi^{(k)} = 0$ up to $k = 2 r - 2 \geqslant r - 1$, so
{\eqref{chik}} is satisfied. By {\eqref{further}}, we have
\begin{equation}
	| \mathcal{L} (\psi_r) -\mathcal{L} (\varphi_r) |_{r, +} \leqslant D_r |
	\psi_r - \varphi_r |_{r + 2, +} \leqslant D_r | \psi_r - \varphi_r |_{2 r -
		2, +}
\end{equation}
as long as $r \geqslant 4$ so that $2 r - 2 \geqslant r + 2$. As mentioned
above $| \psi_r - \varphi_r |_{2 r - 2, +}$ can be made arbitrarily small.
Combining with {\eqref{fpsir}}, we can arrange so that $| f -\mathcal{L}
(\varphi_r) |_{r, +} \leqslant 2 / r \rightarrow 0$ as $r \rightarrow \infty$,
which in particular implies {\eqref{suffSRp}}.
\chapter{Appendices of Part \ref{part:ope}}
\section{Connectedness of \texorpdfstring{$\mathcal{D}_L^\alpha$}{DLalpha} in \texorpdfstring{$d\geq3$}{d>=3}}\label{appendix:connectedness}
In this section we are going to show that in $d\geq3$, each $\mathcal{D}_L^\alpha$ in (\ref{Dl:decomp}) is connected.

Observe first of all that all $\mathcal{D}_L^\alpha$ of the same causal type in table \ref{table:causalclassification} have the same connectedness property (this is obvious because they are related by renumbering points), so it suffices to prove the connectedness property for one $\mathcal{D}_L^\alpha$ in each causal type.

Given a $\mathcal{D}_L^\alpha$, we define $(\mathcal{D}_L^\alpha)_3$ to be the set of all three-point Lorentzian configurations which have the causal ordering of the first three points of the configurations in $\mathcal{D}_L^\alpha$. Then there is a natural projection from $\mathcal{D}_L^\alpha$ to $(\mathcal{D}_L^\alpha)_3$:
\begin{equation}\label{def:piprojection}
	\begin{split}
		\pi: \mathcal{D}_L^\alpha&\ \longrightarrow\ \fr{\mathcal{D}_L^\alpha}_3, \\
		(x_1,x_2,x_3,x_4)&\ \mapsto\ (x_1,x_2,x_3), \\
	\end{split}
\end{equation}
Then $\mathcal{D}_L^\alpha$ has the following decomposition
\begin{equation}
	\begin{split}
		\mathcal{D}_L^\alpha=\bigcup_{(x_1,x_2,x_3)\in\fr{\mathcal{D}_L^\alpha}_3}\left\{x_4\ \big{|}\ (x_1,x_2,x_3,x_4)\in\mathcal{D}_L^\alpha\right\}.
	\end{split}
\end{equation}
For each causal type in table \ref{table:causalclassification}, we want to show that there exists a $\mathcal{D}_L^\alpha$ in this causal type such that
\begin{enumerate}
	\item For fixed $(x_1,x_2,x_3)\in\fr{\mathcal{D}_L^\alpha}_3$, the set $F^\alpha_{x_1,x_2,x_3}=\left\{x_4\ \big{|}\ (x_1,x_2,x_3,x_4)\in\mathcal{D}_L^\alpha\right\}$ is non-empty and connected.
	\item $\fr{\mathcal{D}_L^\alpha}_3$ is connected.
\end{enumerate} 

\subsection{Step 1}\label{section:step1}
For a fixed three-point configuration $(x_1,x_2,x_3)$, the set $F^\alpha_{x_1,x_2,x_3}=\left\{x_4\ \big{|}\ (x_1,x_2,x_3,x_4)\in\mathcal{D}_L^\alpha\right\}$ is non-empty and connected if one of the following conditions holds as a consequence of causal ordering imposed by $\mathcal{D}_L^\alpha$:
\begin{enumerate}
	\item $x_i\rightarrow x_4$ for $i=1,2,3$.
	\item $x_4\rightarrow x_i$ for $i=1,2,3$.
	\item $x_4$ is space-like separated from all of $x_1,x_2,x_3$.
\end{enumerate}
For the first case, $F^\alpha_{x_1,x_2,x_3}$ is given by the intersection of the open forward light-cones of $x_1,x_2,x_3$, which is non-empty. Since cones are convex, $F^\alpha_{x_1,x_2,x_3}$ is also convex, thus connected. The connectedness for the second case follows from a similar argument. For the third case, we use the fact that the connectedness property does not change under Poincar\'e transformations, which allows us to move $x_1$ to 0 by translation
\begin{equation}
	\begin{split}
		x_k\mapsto x_k-x_1,\quad k=1,2,3,4,
	\end{split}
\end{equation}
and then move $x_2,x_3$ onto a 2d subspace by a Lorentz transformation. We enumerate all possible three-point causal orderings
\begin{equation}\label{3ptordering}
	\begin{split}
		\begin{tikzpicture}[baseline={(a.base)},circuit logic US]		
			\node (a) at (0,0) {$a$};
			\node (b) at (1,0) {$b$};
			\node (c) at (2,0) {$c$};
			\draw[-stealth] (a) to (b);
			\draw[-stealth] (b) to (c);	
		\end{tikzpicture},\quad
		\begin{tikzpicture}[baseline={(a.base)},circuit logic US]		
			\node (a) at (0,0) {$a$};
			\node (b) at (1,0.5) {$b$};
			\node (c) at (1,-0.5) {$c$};
			\draw[-stealth] (a) to (b);
			\draw[-stealth] (a) to (c);	
		\end{tikzpicture},\quad
		\begin{tikzpicture}[baseline={(a.base)},circuit logic US]		
			\node (b) at (1,0.5) {$b$};
			\node (c) at (1,-0.5) {$c$};
			\node (a) at (2,0) {$a$};
			\draw[-stealth] (b) to (a);
			\draw[-stealth] (c) to (a);		
		\end{tikzpicture},\quad
		\begin{tikzpicture}[baseline={(a.base)},circuit logic US]		
			\node (a) at (0,0) {$b$};
			\node (b) at (1,0) {$c$};
			\node (c) at (0.5,-0.5) {$a$};
			\draw[-stealth] (a) to (b);		
		\end{tikzpicture},\quad
		\begin{tikzpicture}[baseline={(b.base)},circuit logic US]		
			\node (a) at (0,0.5) {$a$};
			\node (b) at (0,0) {$b$};
			\node (c) at (0,-0.5) {$c$};	
		\end{tikzpicture},
	\end{split}
\end{equation}
and check case by case that in $d\geq3$ we can always move the extra point $x_4$ from any position to $\infty$, preserving the constraint that $x_4$ is space-like to $a,b,c$. This observation implies that $F^\alpha_{x_1,x_2,x_3}$ is connected for the third case. 

In table \ref{table:causalclassification}, we find a $\mathcal{D}_L^\alpha$ satisfying one of the above conditions for some but not all causal types. That's why the connectedness of $\mathcal{D}_L^\alpha$ is not so obvious. The exceptional cases are causal type 8, 10 and 11, for which we need to discuss case by case. Without loss of generality we set $a=1,\ b=2,\ c=3,\ d=4$ (comparing with table \ref{table:causalclassification}) in the following discussion.
\newline
\newline$\textbf{Type 8.}$
By translations, Lorentz transformations and dilatations we fix the configurations to
\begin{equation}
	\begin{split}
		x_1=0,\quad x_2=(x^0,x^1,0,\ldots,0),\quad x_3=(0,1,0\ldots,0).
	\end{split}
\end{equation}
\begin{figure}[H]
	\centering
	\begin{tikzpicture}
		\draw[ultra thin, white, fill=gray!50] (0,0) -- (0.5,0.5) -- (-1,2) -- (-2,2) -- (0,0);
		\draw[ultra thin, white, fill=red!50] (0.5,0.5) -- (-1,2) -- (2,2) -- (0.5,0.5);
		\draw[ultra thin, white, pattern=north west lines,pattern color=red] (0,1) -- (0.5,0.5) -- (2,2) -- (1,2) -- (0,1);
		\draw[dashed] (-1,-1)--(2,2) ;
		\draw[dashed] (1,-1)--(-2,2) ;
		\draw[dashed] (0,-1)--(3,2) ;
		\draw[dashed] (2,-1)--(-1,2) ;
		\draw[dashed] (-0.4,0.6)--(1,2) ;
		\draw[dashed] (-0.4,0.6)--(-1.8,2) ;
		\draw[->] (-1,0)--(3,0) node[right]{$x^1$};
		\draw[->] (0,-1)--(0,3) node[above]{$x^0$};
		\filldraw[black] (0,0) circle (1pt) node[anchor=north west] {$x_1$};
		\filldraw[black] (1,0) circle (1pt) node[anchor=north west] {$x_3$};
		\filldraw[black] (-0.4,0.6) circle (1pt) node[anchor=north east] {$x_2$};	
	\end{tikzpicture}
	\caption{\label{connectedness:type8}Type 8}  
\end{figure}
Then $x_2$ is in the open forward light-cone of $x_1$, but out of the light-cones of $x_3$ (see the grey region in figure \ref{connectedness:type8}), and $x_4$ is in the intersection of open forward light-cones of $x_1$ and $x_3$ (see the red region in figure \ref{connectedness:type8}). Once $x_2$ is fixed somewhere in the grey region, the space of allowed positions for $x_4$ is given by the red region minus the forward light-cone of $x_2$, so the remaining region for $x_4$, which is $F^\alpha_{x_1,x_2,x_3}$, is the red dashed region in figure \ref{connectedness:type8}. Figure \ref{connectedness:type8} shows the 2d situation but a similar 3d figure shows that $F^\alpha_{x_1,x_2,x_3}$ is non-empty and connected in 3d, thus also non-empty and connected in higher d (because we can always find a spatial rotation which preserves $x_1,x_2,x_3$ and maps $x_4$ to $(x,y,z,0,\ldots,0)$).
\newline
\newline\textbf{Type 10.} By translations, Lorentz transformations and dilatations we fix the configurations to
\begin{equation}
	\begin{split}
		x_1=0,\quad x_2=(0,1,0\ldots,0),\quad x_3=(x^0,x^1,0,\ldots,0).
	\end{split}
\end{equation}
\begin{figure}[H]
	\centering
	\caption{\label{connectedness:type10}Type 10}
	\begin{tikzpicture}
		\draw[ultra thin, white, fill=gray!50] (0.5,0.5) -- (-1,2) -- (2,2) -- (0.5,0.5);
		\draw[ultra thin, white, pattern=north west lines,pattern color=gray] (0.6,1.3) -- (0.95,0.95) -- (2,2) -- (1.3,2) -- (0.6,1.3);
		\draw[ultra thin, white, pattern=north east lines,pattern color=gray] (0.6,1.3) -- (0.15,0.85) -- (-1,2) -- (-0.1,2) -- (0.6,1.3);
		\draw[dashed] (-1,-1)--(2,2) ;
		\draw[dashed] (1,-1)--(-2,2) ;
		\draw[dashed] (0,-1)--(3,2) ;
		\draw[dashed] (2,-1)--(-1,2) ;
		\draw[dashed] (-0.1,2)--(3,-1.05) ;
		\draw[dashed] (1.3,2)--(-1.7,-1) ;
		\draw[->] (-1,0)--(3,0) node[right]{$x^1$};
		\draw[->] (0,-1)--(0,3) node[above]{$x^0$};
		\filldraw[black] (0,0) circle (1pt) node[anchor=north west] {$x_1$};
		\filldraw[black] (1,0) circle (1pt) node[anchor=north] {$x_2$};
		\filldraw[black] (0.6,1.3) circle (1pt) node[anchor=north] {$x_3$};	
	\end{tikzpicture}  
\end{figure}
The remaining $x_3,x_4$ pair are in the intersection of the open forward light-cones of $x_1,x_2$, i.e. the grey region in figure \ref{connectedness:type10}. Once $x_3$ is fixed, by the constraint that $x_3,x_4$ are space-like separated, $F^\alpha_{x_1,x_2,x_3}$ is given by the grey dashed region in figure \ref{connectedness:type10}, which is obviously non-empty. This region is topologically the same as $\bbR{d}$ minus the light-cones of $x_3$, thus connected when $d\geq3$.
\newline
\newline\textbf{Type 11.} By translations, Lorentz transformations and dilatations we fix the configurations to
\begin{equation}
	\begin{split}
		x_1=0,\quad x_2=(i,0,\ldots,0),\quad x_3=(x^0,x^1,0,\ldots,0)
	\end{split}
\end{equation}
\begin{figure}[H]
	\centering
	\begin{tikzpicture}
		\draw[ultra thin, white, fill=gray!50] (0.5,0.5) -- (2,2) -- (2,-1) -- (0.5,0.5);
		\draw[ultra thin, white, fill=gray!50] (-0.5,0.5) -- (-2,2) -- (-2,-1) -- (-0.5,0.5);
		\draw[ultra thin, white, pattern=horizontal lines,pattern color=gray] (1.5,0.5) -- (2,1) -- (2,2) -- (1,1) -- (1.5,0.5);
		\draw[dashed] (-2,-2)--(2,2) ;
		\draw[dashed] (2,-2)--(-2,2) ;
		\draw[dashed] (-2,-1)--(2,3) ;
		\draw[dashed] (2,-1)--(-2,3) ;
		\draw[dashed] (1.5,0.5)--(2,1) ;
		\draw[dashed] (1.5,0.5)--(1,1) ;
		\draw[->] (-2,0)--(2,0) node[right]{$x^1$};
		\draw[->] (0,-2)--(0,3) node[above]{$x^0$};
		\filldraw[black] (0,0) circle (1pt) node[anchor=north west] {$x_1$};
		\filldraw[black] (0,1) circle (1pt) node[anchor=west] {$x_2$};
		\filldraw[black] (1.5,0.5) circle (1pt) node[anchor=north west] {$x_3$};	
	\end{tikzpicture}  
	\caption{\label{connectedness:type11}Type 11}
\end{figure}
The remaining $x_3,x_4$ pair are in the grey region in figure \ref{connectedness:type11}. One can see that in $d\geq3$ the grey region is topologically the same as the triangle slice (see one of the grey triangle in figure \ref{connectedness:type11}) times $S^{d-2}$, which is connected. Once $x_3$ is fixed, $F^\alpha_{x_1,x_2,x_3}$ is given by the forward light-cone of $x_3$ in the grey region (see the grey dashed region in figure \ref{connectedness:type11}), which is connected.

\subsection{Step 2}
By slightly improving the argument in step 1, we claim that for the representative set $\mathcal{D}_L^\alpha$ (which we chose in step 1) of each causal type in table \ref{table:causalclassification}, the map (\ref{def:piprojection}) is surjective. In other words, for each three-point configuration in $(\mathcal{D}_L^\alpha)_3$, its preimage in $\mathcal{D}_L^\alpha$ is non-empty.

This claim is true for the cases which satisfy one of the conditions at the beginning of section \ref{section:step1} because for these cases we can always find an $x_4$ which is very far away from $x_1$, $x_2$ and $x_3$. This claim is also true for the exceptional cases because from the figure \ref{connectedness:type8}, \ref{connectedness:type10} and \ref{connectedness:type11} we see that the remaining region for $x_4$ is always non-empty.

Now it remains to show that $(\mathcal{D}_L^\alpha)_3$, which is the set of all three-point configurations with a fixed causal ordering, is connected. For each causal type in (\ref{3ptordering}), we choose 
\begin{equation}\label{choice:3pt}
	\begin{split}
		x_1=b,\quad x_2=c,\quad x_3=a.
	\end{split}
\end{equation}
Analogously to the four-point case we define a projection
\begin{equation}
	\begin{split}
		\pi:\ \fr{\mathcal{D}_L^\alpha}_3\ \longrightarrow\ \fr{\mathcal{D}_L^\alpha}_2, \\
		(x_1,x_2,x_3)\mapsto(x_1,x_2). \\
	\end{split}
\end{equation}
Then we decompose $\fr{\mathcal{D}_L^\alpha}_3$ into
\begin{equation}
	\begin{split}
		\fr{\mathcal{D}_L^\alpha}_3=\bigcup_{(x_1,x_2)\in\fr{\mathcal{D}_L^\alpha}_2}\left\{x_3\ \big{|}\ (x_1,x_2,x_3)\in\fr{\mathcal{D}_L^\alpha}_3\right\}.
	\end{split}
\end{equation}
By comparing (\ref{3ptordering}) and (\ref{choice:3pt}), we find each $\fr{\mathcal{D}_L^\alpha}_3$ satisfies one of the following conditions:
\begin{enumerate}
	\item $x_i\rightarrow x_3$ for $i=1,2$.
	\item $x_3\rightarrow x_i$ for $i=1,2$.
	\item $x_3$ is space-like separated from both of $x_1,x_2$.
\end{enumerate}
This observation implies that for each $\fr{\mathcal{D}_L^\alpha}_3$:
\begin{itemize}
	\item For any fixed $(x_1,x_2)\in\fr{\mathcal{D}_L^\alpha}_2$, the set $\left\{x_3\ \big{|}\ (x_1,x_2,x_3)\in\fr{\mathcal{D}_L^\alpha}_3\right\}$ is connected.
	\item $\fr{\mathcal{D}_L^\alpha}_2=\pi\fr{\fr{\mathcal{D}_L^\alpha}_3}$ contains all two-point configurations with the corresponding causal ordering.
\end{itemize}
It remains to show that in $d\geq3$, the set of two-point configurations with a given causal ordering is connected. This is trivial.

\section{Tables of OPE convergence}\label{appendix:tableopeconvergence}
In this appendix we will give 12 tables of the results about convergence properties of three OPE channels: one table for one causal type. For each causal type we will give a template graph with points $a,b,c,d$. Given a Lorentzian configuration $c_L=(x_1,x_2,x_3,x_4)\in\mathcal{D}_L$, the way to look up the tables is as follows.
\begin{enumerate}
	\item Compute the causal ordering of $c_L$, draw the graph of this causal ordering. Find the corresponding type number (say type X) in table \ref{table:causalclassification}.
	\item Go to the section of causal type X. Compare the causal ordering of $c_L$ with the template causal ordering of causal type X at the beginning of appendix C.X. Match the points $i_1,i_2,i_3,i_4$ with $(abcd)$. We will get a sequence $(i_1i_2i_3i_4)$.
	\item Look up the convergence properties of $(i_1i_2i_3i_4)$ in the table of causal type X. 
\end{enumerate}
For example, consider the following template causal ordering 
\begin{equation*}
	\begin{split}
		\begin{tikzpicture}[baseline={(0,-0.4)},circuit logic US]		
			\node (a) at (0,0) {$a$};
			\node (b) at (1,0.5) {$b$};
			\node (c) at (1,-0.5) {$c$};
			\node (d) at (0.5,-1) {$d$};
			\draw[-stealth] (a) to (b);
			\draw[-stealth] (a) to (c);		
		\end{tikzpicture},\quad\mathrm{or}\quad
		\begin{tikzpicture}[baseline={(0,-0.4)},circuit logic US]		
			\node (a) at (0,0.5) {$b$};
			\node (b) at (0,-0.5) {$c$};
			\node (c) at (1,0) {$a$};
			\node (d) at (0.5,-1) {$d$};
			\draw[-stealth] (a) to (c);
			\draw[-stealth] (b) to (c);		
		\end{tikzpicture}.
	\end{split}
\end{equation*}
Then $(i_1i_2i_3i_4)$ means
\begin{equation*}
	\begin{split}
		\begin{tikzpicture}[baseline={(0,-0.4)},circuit logic US]		
			\node (a) at (0,0) {${i_1}$};
			\node (b) at (1,0.5) {${i_2}$};
			\node (c) at (1,-0.5) {${i_3}$};
			\node (d) at (0.5,-1) {${i_4}$};
			\draw[-stealth] (a) to (b);
			\draw[-stealth] (a) to (c);		
		\end{tikzpicture},\quad\mathrm{or}\quad
		\begin{tikzpicture}[baseline={(0,-0.4)},circuit logic US]		
			\node (a) at (0,0.5) {${i_2}$};
			\node (b) at (0,-0.5) {${i_3}$};
			\node (c) at (1,0) {${i_1}$};
			\node (d) at (0.5,-1) {${i_4}$};
			\draw[-stealth] (a) to (c);
			\draw[-stealth] (b) to (c);		
		\end{tikzpicture}.
	\end{split}
\end{equation*}
In appendix \ref{appendix:type1} we will explain in detail how to make the table of OPE convergence for type 1 causal ordering. The procedure is similar for the other causal types, so we will only give the results for them. Before we start, we would like to introduce some tricks in appendix \ref{section:S4action}, \ref{section:lorentzconfframe} and \ref{section:graphsymmetry}. They will be helpful in making the tables. 

\subsubsection{\texorpdfstring{$S_4$}{S4}-action}\label{section:S4action}
There is a natural $S_4$-action on the space of four-point configurations. Let $\sigma\in S_4$ be a symmetry group element:
\begin{equation}
	\begin{split}
		\sigma=\left(\begin{array}{cccc}
			1 & 2 & 3 & 4 \\
			\sigma(1) & \sigma(2) & \sigma(3) & \sigma(4) \\
		\end{array}\right).
	\end{split}
\end{equation}
Let $C=(x_1,x_2,x_3,x_4)$ be a four-point configuration such that $x_{ij}^2\neq0$ for all $x_i,x_j$ pairs. We define the action
\begin{equation}\label{def:permutation}
	\begin{split}
		\sigma\cdot C=(x_1^\prime,x_2^\prime,x_3^\prime,x_4^\prime),\quad x_k^\prime=x_{\sigma^{-1}(k)}.
	\end{split}
\end{equation}
By computing $z,\bar{z}$ of $\sigma\cdot C$ and comparing with $z,\bar{z}$ of $C$, we get a natural $S_4$-action on $z,\bar{z}$:
\begin{equation}
	\begin{split}
		w_\sigma:\ \mathbb{C}\backslash\left\{0,1\right\}&\ \longrightarrow\ \mathbb{C}\backslash\left\{0,1\right\}, \\
		z&\quad\mapsto\quad w_\sigma(z), \\
		\bar{z}&\quad\mapsto\quad w_\sigma(\bar{z}), \\
	\end{split}
\end{equation}
where $w_\sigma(z),w_\sigma(\bar{z})$ are the variables $z,\bar{z}$ computed from $\sigma\cdot C$. We have the following properties:
\begin{itemize}
	\item $\left\{w_\sigma\right\}_{\sigma\in S_4}$ belong to a set of 6 fractional linear transformation forming a group which is isomorphic to $S_3$. The map $\sigma\mapsto w_\sigma$ is a group homomorphism from $S_4$ to $S_3$ (i.e. $w_{\sigma_1}\circ w_{\sigma_2}=w_{\sigma_1\sigma_2}$).
	\item The $S_4$-action on $\mathcal{D}_L$ permutes classes S,T,U among themselves.
	\item The $S_4$-action on $\mathcal{D}_L$ permutes subclasses $\mathrm{E_{su}}$,$\mathrm{E_{st}}$,$\mathrm{E_{tu}}$ among themselves.
	\item The $S_4$-action on $\mathcal{D}_L$ preserves the subclass $\mathrm{E_{stu}}$.
\end{itemize}
Let us denote $\sigma$ by $\left[\sigma(1)\sigma(2)\sigma(3)\sigma(4)\right]$. We summarize the above properties in table \ref{table:wsigma}. 
\begin{table}[H]
	\caption{The list of $w_\sigma$ and the $S_4$ transformation between classes and subclasses.}
	\label{table:wsigma}
	\setlength{\tabcolsep}{3mm} 
	\def\arraystretch{1.5} 
	\centering
	\begin{tabular}{|c|c|c|c|c|c|c|c|c|}
		\hline
		$\sigma$    &   $w_\sigma(z)$  &  S  &  T  &  U  &  $\mathrm{E_{su}}$  &  $\mathrm{E_{st}}$  &  $\mathrm{E_{tu}}$  &  $\mathrm{E_{stu}}$
		\\ \hline
		[1234], [2143], [3412], [4321]  &  $z$  &  S  &  T  &  U  &  $\mathrm{E_{su}}$  &  $\mathrm{E_{st}}$  &  $\mathrm{E_{tu}}$  &  $\mathrm{E_{stu}}$ 
		\\ \hline
		[2134], [1243], [4312], [3421]  &  $\frac{z}{z-1}$  &  S  &  U  &  T  &  $\mathrm{E_{st}}$  &  $\mathrm{E_{su}}$  &  $\mathrm{E_{tu}}$  &  $\mathrm{E_{stu}}$
		\\ \hline
		[3214], [4123], [1432], [2341]  &  $1-z$  &  T  &  S  &  U  &  $\mathrm{E_{tu}}$  &  $\mathrm{E_{st}}$  &  $\mathrm{E_{su}}$  &  $\mathrm{E_{stu}}$
		\\ \hline
		[1324], [2413], [3142], [4231]  &  $\frac{1}{z}$  &  U  &  T  &  S  &  $\mathrm{E_{su}}$  &  $\mathrm{E_{tu}}$  &  $\mathrm{E_{st}}$  &  $\mathrm{E_{stu}}$
		\\ \hline
		[2314], [1423], [4132], [3241]  &  $\frac{1}{1-z}$  &  T  &  U  &  S  &  $\mathrm{E_{st}}$  &  $\mathrm{E_{tu}}$  &  $\mathrm{E_{su}}$  &  $\mathrm{E_{stu}}$
		\\ \hline
		[3124], [4213], [1342], [2431]  &  $1-\frac{1}{z}$  &  U  &  S  &  T  &  $\mathrm{E_{tu}}$  &  $\mathrm{E_{su}}$  &  $\mathrm{E_{st}}$  &  $\mathrm{E_{stu}}$
		\\ \hline
	\end{tabular}
\end{table}
Suppose a configuration $c_L$ gives the template causal ordering of a causal type, which means that $c_L$ corresponds to the sequence (1234). For $\sigma=[i_1i_2i_3i_4]$, we get a configuration $c_L^\prime=\sigma\cdot C$ by eq. (\ref{def:permutation}). The causal ordering of $c_L^\prime$ is in the same causal type as $c_L$. By comparing the causal orderings of $c_L$ and $c_L^\prime$, we see that the sequence of $c_L^\prime$ is exactly $(i_1i_2i_3i_4)$. Therefore, given a causal type, if we know the class/subclass of the template causal ordering, by looking up table \ref{table:wsigma} we decide the classes/subclasses of the other causal ordering in the same causal type. Then by looking up table \ref{table:class}, we immediately get a part of the OPE convergence properties for each causal ordering.

By using the above trick, the problem of determining the classes/subclasses of causal orderings belong is reduced to determining the class/subclass of the template causal ordering in each causal type. In appendix \ref{section:lorentzconfframe}, we will introduce a trick to determine the classes/subclasses of the template causal orderings.

\subsubsection{Lorentzian conformal frame}\label{section:lorentzconfframe}
Our goal in this subsection is to give a systematic way to determine the class/subclass of $\mathcal{D}_L^\alpha$, where $\alpha$ is a fixed causal ordering.

Recalling lemma \ref{lemma:causaltoclass}, all configurations in $\mathcal{D}_L^\alpha$ belong to the same class. We can choose a particular configuration $c_L\in\mathcal{D}_L^\alpha$ and compute $z,\bar{z}$ of $c_L$, then we immediately know the class (not subclass) of $\mathcal{D}_L^\alpha$. 

If $\mathcal{D}_L^\alpha$ belongs to class S/T/U, then we are done. The rest of this subsection is for the case that $\mathcal{D}_L^\alpha$ belongs to class E. If $\mathcal{D}_L^\alpha$ belongs to class E, then according to theorem \ref{theorem:classification}, we need to check the OPE convergence properties for the intersection of $\mathcal{D}_L^\alpha$ and each subclass of class E as long as the intersection is non-empty. We will find that only the type 1, 5, 6, 10, 11, 12 causal orderings in table \ref{table:causalclassification} belong to class E.\footnote{This can be easily done by choosing one particular configuration and compute $z,\bar{z}$ for each template causal ordering, and by the fact that the $S_4$-action preserves class E (as discussed in appendix \ref{section:S4action}).} In the tables of OPE convergence properties of causal type 5, 10 and 12 , we give the results of all subclasses for each causal ordering (see table \ref{table:type5convergence}, \ref{table:type10convergence} and \ref{table:type12convergence}); while in the tables of causal type 1 ,6 and 11, we only give the results of one subclass for each causal ordering (see table \ref{table:type1convergence}, \ref{table:type6convergence} and \ref{table:type11convergence}). We claim that our tables are complete, based on the following lemma.
\begin{lemma}\label{lemma:subclass}
	Given a fixed causal ordering $\alpha$, if $\alpha$ is in causal type 1/6/11, then $\mathcal{D}_L^\alpha$ only belongs to one of the three subclasses $\mathrm{E_{st}},\mathrm{E_{su}},\mathrm{E_{tu}}$. 
\end{lemma}
The basic tool we use to prove the above lemma is the Lorentzian conformal frame. The Lorentzian conformal frame is similar to the Euclidean conformal frame (\ref{frameE}). Given a Lorentzian configuration $c_L$, its conformal frame configuration $c_L^\prime$ is a Lorentzian configuration which has one of the following forms
\begin{equation}\label{conformalframe:lorentz}
	\begin{split}
		&1.\ x_1^\prime=0,\  x_2^\prime=(ia,b,0,\ldots,0),\ x_3^\prime=(i,0,\ldots,0),\ x_4^\prime=\infty. \\
		&2.\ x_1^\prime=0,\ x_2^\prime=(ib,a,0,\ldots,0),\ x_3^\prime=(0,1,0,\ldots,0),\ x_4^\prime=\infty. \\
		&3.\ x_1^\prime=0,\  x_2^\prime=(0,a,b,0,\ldots,0),\ x_3^\prime=(0,1,0,\ldots,0),\ x_4^\prime=\infty. \\
	\end{split}
\end{equation}
$c_L^\prime$ and $c_L$ are related by a Lorentzian conformal transformation. Computing the cross-ratios from (\ref{conformalframe:lorentz}), we see that: for the first and second cases $z=a+b$, $\bar{z}=a-b$; for the third case $z=a+ib$, $\bar{z}=a-ib$. Analogously to the Euclidean conformal frame, the Lorentzian conformal frame configuration is unique up to a reflection $b\mapsto-b$, which corresponds to interchanging $z$ and $\bar{z}$.

\begin{figure}[t]
	\centering
	\begin{tikzpicture}
		\draw[ultra thin, white, fill=gray!50] (0,0) -- (0.5,0.5) -- (0,1) -- (-0.5,0.5) -- (0,0);
		\draw[ultra thin, white, fill=gray!50] (0,1) -- (1,2) -- (-1,2) -- (0,1);
		\draw[ultra thin, white, fill=gray!50] (0,0) -- (1,-1) -- (-1,-1) -- (0,0);
		\draw[dashed] (-1,-1)--(2,2) ;
		\draw[dashed] (1,-1)--(-2,2) ;
		\draw[dashed] (-2,-1)--(1,2) ;
		\draw[dashed] (2,-1)--(-1,2) ;
		\draw[->] (-2,0)--(2,0) node[right]{$x^1$};
		\draw[->] (0,-1)--(0,2) node[above]{$x^0$};
		\filldraw[black] (0,0) circle (1.5pt) node[anchor=north west] {$x_1^\prime$};
		\filldraw[black] (0,1) circle (1.5pt) node[anchor=west] {$x_3^\prime$};
		\draw (0,0.5) node[red]{$\mathrm{E_{st}}$};	
		\draw (0,1.5) node[red]{$\mathrm{E_{tu}}$};
		\draw (0,-0.5) node[red]{$\mathrm{E_{su}}$};
	\end{tikzpicture}
	\begin{tikzpicture}
		\draw[ultra thin, white, fill=gray!50] (0,0) -- (-1,1) -- (-1,-1) -- (0,0);
		\draw[ultra thin, white, fill=gray!50] (0,0) -- (0.5,0.5) -- (1,0) -- (0.5,-0.5) -- (0,0);
		\draw[ultra thin, white, fill=gray!50] (1,0) -- (2,1) -- (2,-1) -- (1,0);
		\draw[dashed] (-1,-1)--(2,2);
		\draw[dashed] (2,-2)--(-1,1);
		\draw[dashed] (-1,-2)--(2,1);
		\draw[dashed] (2,-1)--(-1,2);
		\draw[->] (-1,0)--(2,0) node[right]{$x^1$};
		\draw[->] (0,-2)--(0,2) node[above]{$x^0$};
		\filldraw[black] (0,0) circle (1.5pt) node[anchor=north west] {$x_1^\prime$};
		\filldraw[black] (1,0) circle (1.5pt) node[anchor=north] {$x_3^\prime$};
		\draw (0.5,0) node[red]{$\mathrm{E_{st}}$};	
		\draw (1.5,0) node[red]{$\mathrm{E_{tu}}$};
		\draw (-0.5,0) node[red]{$\mathrm{E_{su}}$};
	\end{tikzpicture}
	\begin{tikzpicture}
		\draw[ultra thin, white, fill=gray!50] (-2,-2) -- (-2,2) -- (3,2) -- (3,-2) -- (-2,-2);
		\draw[->] (-2,0)--(3,0) node[right]{$x^1$};
		\draw[->] (0,-2)--(0,2) node[above]{$x^2$};
		\filldraw[black] (0,0) circle (1.5pt) node[anchor=north west] {$x_1^\prime$};
		\filldraw[black] (1,0) circle (1.5pt) node[anchor=north] {$x_3^\prime$};
		\draw (1,1) node[red]{$\mathrm{E_{stu}}$};	
	\end{tikzpicture}
	\caption{\label{fig:conformalframetoclass}The conformal frame configurations realized by configurations in class E.}  
\end{figure}
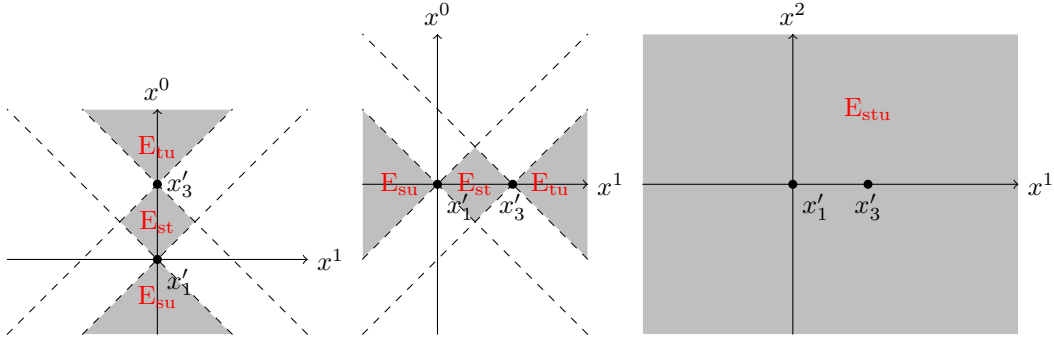

Let us describe how we map a four-point configuration to the conformal frame by conformal transformations. Let $c_L=(x_1,x_2,x_3,x_4)$ be a Lorentzian configuration. We will go from $c_L$ to $c_L^\prime$ in a few steps, and each step is a conformal transformation. The configuration after the k-th step is denoted by $c_L^{(k)}$.
\newline\textbf{Step 1.} We move $x_1$ to 0 by translation. The configuration $c_L^{(1)}$ after the first step is given by
\begin{equation}
	\begin{split}
		c_L^{(1)}=\fr{x_1^{(1)},x_2^{(1)},x_3^{(1)},x_4^{(1)}}=(0,x_2-x_1,x_3-x_1,x_4-x_1).
	\end{split}
\end{equation}
This step preserves the causal ordering. 
\newline\textbf{Step 2.} We move $x_4$ to $\infty$ by special conformal transformation
\begin{equation}\label{conformalframe:sct}
	\begin{split}
		{x^\prime}^\mu=\dfrac{x^\mu-x^2b^\mu}{1-2x\cdot b+x^2b^2},\quad b^\mu=\dfrac{(x_4-x_1)^\mu}{(x_4-x_1)^2}.
	\end{split}
\end{equation}
$x_1=0$ is preserved by special conformal transformation. This step may change the causal ordering. Under general conformal transformations, $x_{ij}^2$ transforms as
\begin{equation}\label{sct:squaretransform}
	\begin{split}
		(x_i^\prime-x_j^\prime)^2=&\Omega(x_i)\Omega(x_j)(x_i-x_j)^2, \\
		(ds^{\prime2}=&\Omega(x)^2ds^2), \\
	\end{split}
\end{equation}
and for special conformal transformation (\ref{conformalframe:sct}), the scaling factor at $x_k^{(1)}=x_k-x_1$ is given by
\begin{equation}\label{scaling:sct}
	\begin{split}
		\Omega\fr{x_k^{(1)}}=\dfrac{(x_4-x_1)^2}{(x_4-x_k)^2},\quad k=1,2,3,4.
	\end{split}
\end{equation}
Let $c_L^{(2)}=\fr{x_1^{(2)},x_2^{(2)},x_3^{(2)},x_4^{(2)}}$ be the configuration after step 2. For any configuration $c_L$ in $\mathcal{D}_L$ (where the light-cone singularities are excluded), by (\ref{sct:squaretransform}) and (\ref{scaling:sct}) we have
\begin{equation}\label{lorentzconformal:step2}
	\begin{split}
		\fr{x_i^{(2)}-x_j^{(2)}}^2\neq0,\quad i,j=1,2,3,
	\end{split}
\end{equation}
Furthermore, the sign of $\fr{x_i^{(2)}-x_j^{(2)}}^2$ is determined by the causal ordering of $c_L$. So we know if each $x_i^{(2)},x_j^{(2)}$ pair of $c_L^{(2)}$ is space-like or time-like. The information we do not know a priori from (\ref{sct:squaretransform}) and (\ref{scaling:sct}) is the causal orderings of time-like $x_i^{(2)},x_j^{(2)}$ pairs (who is in the future of whom).\footnote{Of course for any particular configuration we can just compute $c_L^{(2)}$ and then determine its causal ordering.}  
\newline\textbf{Step 3.} We move $x_3$ to its final position by some composition of Lorentz transformations, dilatations and time reversal $\theta_L$ (these conformal transformations preserve $x_1=0$ and $x_4=\infty$). Lorentz transformations and dilatations preserve causal orderings, and time reversal only reverse causal orderings (i.e. $x_i,x_j$ pairs change from time-like to time-like, or from space-like to space-like). There are two possibilities after step 2: $x_3^{(2)}$ could be space-like or time-like to $x_1^{(2)}$. If $x_1^{(2)},x_3^{(2)}$ are time-like, then $x_3^{(3)}$ is put at $(i,0,\ldots,0)$. If $x_1^{(2)},x_3^{(2)}$ are space-like, then $x_3^{(3)}$ is put at $(0,1,0,\ldots,0)$. Therefore, $c_L^{(3)}$ is in one of the following forms:
\begin{enumerate}
	\item $x_1^{(3)}=0,\ x_3^{(3)}=(i,0,\ldots,0),\ x_4^{(3)}=\infty$.
	\item $x_1^{(3)}=0,\ x_3^{(3)}=(0,1,0,\ldots,0),\ x_4^{(3)}=\infty$.
\end{enumerate} 
\textbf{Step 4.} We move $x_2$ to somewhere in the $(01)$-plane or $(12)$-plane by Lorentz transformations in the little group of $x_3^{(3)}$. If $x_3^{(3)}=(i,0,\ldots,0)$, then we move $x_2$ to the $(01)$-plane by rotation, i.e. $x_2^{(4)}=(ia,b,0,\ldots,0)$. If $x_3^{(3)}=(0,1,0,\ldots,0)$, then we move $x_2$ onto the $(01)$-plane or the $(12)$-plane, determined as follows:
\begin{itemize}
	\item If $x_2^{(3)}=(i\beta_1,a,\beta_2,\ldots,\beta_{d-1})$ with $(\beta_1)^2\geq(\beta_2)^2+\ldots(\beta_{d-1})^2$, then $x_2$ is put in the (01)-plane, i.e. $x_2^{(4)}=(ib,a,0,\ldots,0)$ and $b^2=(\beta_1)^2-(\beta_2)^2-\ldots-(\beta_{d-1})^2$. 
	\item If $x_2^{(3)}=(i\beta_1,a,\beta_2,\ldots,\beta_{d-1})$ with $(\beta_1)^2\leq(\beta_2)^2+\ldots(\beta_{d-1})^2$, then $x_2$ is put in the (01)-plane, i.e. $x_2^{(4)}=(0,a,b,0,\ldots,0)$ and $b^2=(\beta_2)^2+\ldots+(\beta_{d-1})^2-(\beta_1)^2$. 
\end{itemize}
In the end, $c_L^\prime=c_L^{(4)}$ has one of the forms in eq. (\ref{conformalframe:lorentz}). Moreover, the above discussion provides us with the following fact:
\begin{itemize}
	\item the sign of $\fr{x_{ij}^\prime}^2$ of $c_L^\prime$ is going to be the same for all configurations $c_L\in\mathcal{D}_L^\alpha$ in each causal ordering $\alpha$. 
\end{itemize}
Now back to our question. Suppose $\mathcal{D}_L^\alpha$ is in class E. Let $c_L^\prime=(x_1^\prime,x_2^\prime,x_3^\prime,x_4^\prime)$ be the conformal frame configuration of $c_L\in\mathcal{D}_L^\alpha$. Comparing the range of the $(a,b)$ pair in (\ref{conformalframe:lorentz}) with the range of the $(z,\bar{z})$ pair in class E (see section \ref{section:classifyz}) and using the above fact, we see that there are only two possibilities for $c_L^\prime$:
\begin{enumerate}
	\item $\fr{x_{13}^\prime}^2,\fr{x_{12}^\prime}^2,\fr{x_{23}^\prime}^2<0$ for all $c_L\in\mathcal{D}_L^\alpha$. All possible $c_L^\prime$ are given by the grey region in the first picture of figure \ref{fig:conformalframetoclass}. In this case $z,\bar{z}$ are real, so we have
	\begin{equation}
		\begin{split}
			\mathcal{D}_L^\alpha=\fr{\mathcal{D}_L^\alpha\cap\mathrm{E_{st}}}\sqcup\fr{\mathcal{D}_L^\alpha\cap\mathrm{E_{su}}}\sqcup\fr{\mathcal{D}_L^\alpha\cap\mathrm{E_{tu}}}.
		\end{split}
	\end{equation}
	Because the $z,\bar{z}$ ranges corresponding to $\mathrm{E_{st}},\mathrm{E_{su}},\mathrm{E_{tu}}$ are disconnected from each other in figure \ref{fig:conformalframetoclass} and because $\mathcal{D}_L^\alpha$ is connected in $d\geq3$, only one of the above intersections is non-empty. We conclude that such $\mathcal{D}_L^\alpha$ only belongs to one of the three subclasses $\mathrm{E_{st}},\mathrm{E_{su}},\mathrm{E_{tu}}$. This conclusion remains valid also in 2d, because 2d configurations can be embedded into higher d.
	\item $\fr{x_{13}^\prime}^2,\fr{x_{12}^\prime}^2,\fr{x_{23}^\prime}^2>0$ for all $c_L\in\mathcal{D}_L^\alpha$. All possible conformal frame configurations are given by the grey region in the second and third pictures of figure \ref{fig:conformalframetoclass}. In this case we have
	\begin{equation}
		\begin{split}
			\mathcal{D}_L^\alpha=\fr{\mathcal{D}_L^\alpha\cap\mathrm{E_{st}}}\sqcup\fr{\mathcal{D}_L^\alpha\cap\mathrm{E_{su}}}\sqcup\fr{\mathcal{D}_L^\alpha\cap\mathrm{E_{tu}}}\sqcup\fr{\mathcal{D}_L^\alpha\cap\mathrm{E_{stu}}},
		\end{split}
	\end{equation}
	which means that the configurations in $\mathcal{D}_L^\alpha$ may appear in all subclasses.
\end{enumerate}
To see which of these possibilities is realized, it is enough to know the sign of $\fr{x_{13}^\prime}^2$.  To finish the proof of lemma \ref{lemma:subclass}, it remains to check that $\fr{x_{13}^\prime}^2<0$ for all type 1, 6, 11 causal orderings (so that possibility 1 is realized for $c_L^\prime$). Since all causal orderings of a fixed causal type can be realized by permuting the indices of the template causal ordering and such $S_4$-action permutes subclasses $\mathrm{E_{st}},\mathrm{E_{su}},\mathrm{E_{tu}}$ among themselves (see section \ref{section:S4action}), it suffices to check that $\fr{x_{13}^\prime}^2<0$ for the template causal orderings of causal type 1, 6, 11. Moreover, since we are only interested in the sign of $\fr{x_{13}^\prime}^2$, we can use the following formula:
\begin{equation}\label{formula:x13prime}
	\begin{split}
		\mathrm{Sign}\fr{\fr{x_{13}^\prime}^2}=\mathrm{Sign}\fr{x_{13}^2x_{14}^2x_{34}^2}.
	\end{split}
\end{equation}
(\ref{formula:x13prime}) follows from (\ref{sct:squaretransform}), (\ref{scaling:sct}) and the fact that step 3,4 (of constructing the conformal frame) preserve the sign of $\fr{x_{ij}^{(k)}}^2$. Now let us do the check.
\newline\textbf{Type 1.} The template causal ordering is given by 
\begin{equation}
	\begin{split}
		1\rightarrow 2\rightarrow 3\rightarrow 4,
	\end{split}
\end{equation}
which gives $x_{13}^2,x_{14}^2,x_{34}^2<0$, hence $\fr{x_{13}^\prime}^2<0$.
\newline\textbf{Type 6.} The template causal ordering is given by 
\begin{equation}
	\begin{split}
		\begin{tikzpicture}[baseline={(0,-0.35)},circuit logic US]		
			\node (a) at (0,0) {$1$};
			\node (b) at (1,0) {$2$};
			\node (c) at (2,0) {$3$};
			\node (d) at (1,-0.5) {$4$};
			\draw[-stealth] (a) to (b);
			\draw[-stealth] (b) to (c);		
		\end{tikzpicture},
	\end{split}
\end{equation}
which gives $x_{13}^2<0$ and $x_{14}^2,x_{34}^2>0$, hence $\fr{x_{13}^\prime}^2<0$.
\newline\textbf{Type 11.} The template causal ordering is given by 
\begin{equation}
	\begin{split}
		\begin{tikzpicture}[baseline={(0,-0.4)},circuit logic US]		
			\node (a) at (0,0) {$1$};
			\node (b) at (1,0) {$2$};
			\node (c) at (0,-0.5) {$3$};
			\node (d) at (1,-0.5) {$4$};
			\draw[-stealth] (a) to (b);
			\draw[-stealth] (c) to (d);		
		\end{tikzpicture}
	\end{split}
\end{equation}
which gives $x_{13}^2,x_{14}^2>0$ and $x_{34}^2<0$, hence $\fr{x_{13}^\prime}^2<0$.

So we finished the proof of lemma \ref{lemma:subclass}. As a consistency check one can also compute the sign of $\fr{x_{13}^\prime}^2$ for type 5,10,12 causal orderings and find that $\fr{x_{13}^\prime}^2>0$ for these cases.

For type 1, 6, 11 causal orderings, to determine the subclasses, it remains to determine the subclass of one particular configuration of the template causal ordering in each type, then determine the subclasses of other causal orderings by looking up table \ref{table:wsigma}. 

We would like to remark that Lorentzian conformal frame is just a way to figure out the range of $z,\bar{z}$. There is no claim that correlation functions at $c_L,c_L^\prime$ agree. As mentioned in section \ref{section:wightmanconformal}, the global conformal invariance does not hold in a general Lorentzian CFT.

\subsubsection{Symmetry of the graph}\label{section:graphsymmetry} 
Usually, we go from one causal ordering to another by permuting the indices. For example, given the causal ordering $1\rightarrow2\rightarrow3\rightarrow4$, by permuting $1,4$ we get another causal ordering $4\rightarrow2\rightarrow3\rightarrow1$.

However, a causal ordering may have a non-trivial little group.\footnote{By little group of a causal ordering, we mean the permutations of the indices which do not change this causal ordering.} For example, consider the following causal ordering
\begin{equation}\label{ex:graphsymmetry}
	\begin{split}
		\begin{tikzpicture}[baseline={(a.base)},circuit logic US]		
			\node (a) at (0,0) {$a$};
			\node (b) at (1,0) {$b$};
			\node (c) at (2,0.5) {$c$};
			\node (d) at (2,-0.5) {$d$};
			\draw[-stealth] (a) to (b);
			\draw[-stealth] (b) to (c);
			\draw[-stealth] (b) to (d);		
		\end{tikzpicture}.
	\end{split}
\end{equation}
The causal ordering (\ref{ex:graphsymmetry}) does not change if we interchange $c$ and $d$, so it has a non-trivial little group $\bbZ_2$.

The little group is unique up to an isomorphism for all causal orderings in the same causal type. Let $G$ be the little group of one causal type and let $|G|$ be the order of $G$. Then the total number of causal orderings in this type is given by $24/|G|$. So in the table of each causal type, we will only list $24/|G|$ sequences $(i_1i_2i_3i_4)$. Below the tables, we will point out the sequences which give the same causal ordering.

\subsection{Type 1}\label{appendix:type1}
The type 1 causal ordering is given by
\begin{equation}\label{template:type1}
	\begin{split}
		a\rightarrow b\rightarrow c\rightarrow d.
	\end{split}
\end{equation}
We let (\ref{template:type1}) be the template causal ordering, then the causal ordering $i_1\rightarrow i_2\rightarrow i_3\rightarrow i_4$ is labelled by the sequence $(i_1i_2i_3i_4)$. Any permutation of the indices in (\ref{template:type1}) will change the causal ordering, so the little group of the type 1 causal orderings is trivial. We have to list 24 causal orderings in the table.

Under time reversal $\theta_L$, $a\rightarrow b\rightarrow c\rightarrow d$ is mapped to $d\rightarrow c\rightarrow b\rightarrow a$, which is equivalent to the permutation $\theta_L:\ a\leftrightarrow d$, $b\leftrightarrow c$. This action is causal-type specific. In addition, we have $\theta_E$ action which is always given by $\theta_E:\ 1\leftrightarrow4,2\leftrightarrow3$ (see eq. (\ref{action:thetaE})). Using $\bbZ_2\times\bbZ_2$ generated by these permutations, we divide 24 type 1 causal orderings into 8 orbits:
\begin{enumerate}
	\item (1234), (4321).
	\item (1243), (4312), (3421), (2134).
	\item (1324), (4231).
	\item (1423), (3241), (4132), (2314).
	\item (1342), (2431), (4213), (3124).
	\item (1432), (2341), (4123), (3214).
	\item (2143), (3412).
	\item (2413), (3142).
\end{enumerate}
As discussed in section \ref{section:timereversal}, all the causal orderings in each orbit have the same convergent OPE channels.

Let us consider (1234), or equivalently the causal ordering $1\rightarrow2\rightarrow3\rightarrow4$. We first pick a particular configuration and compute $z,\bar{z}$. Here we choose 
\begin{equation}\label{type1:example}
	\begin{split}
		x_1=0,\ x_2=(i,0,\ldots,0),\ x_3=(2i,0,\ldots,0),\ x_4=(3i,0,\ldots,0)
	\end{split}
\end{equation}
and get $z=\bar{z}=\frac{1}{4}$, which is in the range corresponding to subclass $\mathrm{E_{st}}$. By lemma \ref{lemma:subclass}, the whole (1234) is in subclass $\mathrm{E_{st}}$.

All other $(i_1i_2i_3i_4)$ causal orderings can be obtained by applying permutation $\sigma=[i_1i_2i_3i_4]$ to the template ordering (1234). Using table \ref{table:wsigma}, we can easily determine the subclasses of all other $(i_1i_2i_3i_4)$ sequences (look at the column having $\mathrm{E_{st}}$ on top). Then by looking up table \ref{table:class}, we get some OPE convergence properties of type 1 causal orderings, which are summarized in table \ref{table:type1step1}.
\begin{table}[H]
	\setlength{\tabcolsep}{3mm} 
	\def\arraystretch{1.25} 
	\centering
	\begin{tabular}{|c|c|c|c|c|}
		\hline
		causal ordering    &   class/subclass  &  s-channel    &   t-channel   &   u-channel
		\\ \hline
		(1234),\ (4321)   &  $\mathrm{E_{st}}$  & \cmark  &    &  \xmark
		\\ \hline
		(1243),\ (3421),\ (4312),\ (2134)   &  $\mathrm{E_{su}}$  & \cmark  &  \xmark  &  
		\\ \hline
		(1324),\ (4231)   & $\mathrm{E_{tu}}$  & \xmark  &    &  
		\\ \hline
		(1423),\ (3241),\ (4132),\ (2314)   & $\mathrm{E_{tu}}$  & \xmark  &    &  
		\\ \hline
		(1342),\ (2431),\ (4213),\ (3124)   & $\mathrm{E_{su}}$  & \cmark  &  \xmark  &  
		\\ \hline
		(1432),\ (2341),\ (4123),\ (3214)   & $\mathrm{E_{st}}$  & \cmark  &    &  \xmark
		\\ \hline
		(2143),\ (3412)   & $\mathrm{E_{st}}$  & \cmark  &    &  \xmark
		\\ \hline
		(2413),\ (3142)   & $\mathrm{E_{tu}}$  & \xmark  &    &  
		\\ \hline
	\end{tabular}
	\caption{\label{table:type1step1} The classes/subclasses of type 1 causal orderings}
\end{table}
It remains to complete the rest of table \ref{table:type1step1}. For this we choose a representative configuration for each orbit of causal orderings and choose a path to compute the curves of $z,\bar{z}$, then decide the convergence properties of t- and u-channel expansions. In practice this is done numerically, by plotting the curves of $z,\bar{z}$ and staring at the plots to determine $N_t,N_u$ (as in examples in section \ref{section:examples}). The final results of OPE convergence properties in the three channels are shown in table \ref{table:type1convergence}, where we use red to indicate the new marks. 
\begin{table}[H]
	\setlength{\tabcolsep}{3mm} 
	\def\arraystretch{1.25} 
	\centering
	\begin{tabular}{|c|c|c|c|c|}
		\hline
		causal ordering    &   class/subclass      &   s-channel    &   t-channel   &   u-channel
		\\ \hline
		(1234),\ (4321)   &  $\mathrm{E_{st}}$     & \cmark  &  \textcolor{red}{\cmark}  &  \xmark
		\\ \hline
		(2143),\ (3412)   & $\mathrm{E_{st}}$     & \cmark  &  \textcolor{red}{\xmark}  &  \xmark
		\\ \hline
		(1432),\ (2341),\ (4123),\ (3214)   & $\mathrm{E_{st}}$     & \cmark  &  \textcolor{red}{\cmark}  &  \xmark
		\\ \hline
		(1324),\ (4231)   & $\mathrm{E_{tu}}$     & \xmark  &  \textcolor{red}{\cmark}  &  \textcolor{red}{\xmark}
		\\ \hline
		(2413),\ (3142)   & $\mathrm{E_{tu}}$     & \xmark  &  \textcolor{red}{\xmark}  &  \textcolor{red}{\xmark}
		\\ \hline
		(1423),\ (3241),\ (4132),\ (2314)   & $\mathrm{E_{tu}}$     & \xmark  &  \textcolor{red}{\cmark}  &  \textcolor{red}{\xmark}
		\\ \hline
		(1342),\ (2431),\ (4213),\ (3124)   & $\mathrm{E_{su}}$     & \cmark  &  \xmark  &  \textcolor{red}{\xmark}
		\\ \hline
		(1243),\ (3421),\ (4312),\ (2134)   &  $\mathrm{E_{su}}$     & \cmark  &  \xmark  &  \textcolor{red}{\xmark}
		\\ \hline
	\end{tabular}
	\caption{\label{table:type1convergence}OPE convergence properties of type 1 causal orderings}
\end{table}

\subsection{Type 2}\label{appendix:type2}
The type 2 causal ordering is given by
\begin{equation}
	\begin{split}
		\begin{tikzpicture}[baseline={(a.base)},circuit logic US]		
			\node (a) at (0,0) {$a$};
			\node (b) at (1,0) {$b$};
			\node (c) at (2,0.5) {$c$};
			\node (d) at (2,-0.5) {$d$};
			\draw[-stealth] (a) to (b);
			\draw[-stealth] (b) to (c);
			\draw[-stealth] (b) to (d);		
		\end{tikzpicture},\quad \begin{tikzpicture}[baseline={(c.base)},circuit logic US]		
			\node (a) at (0,0.5) {$c$};
			\node (b) at (0,-0.5) {$d$};
			\node (c) at (1,0) {$b$};
			\node (d) at (2,0) {$a$};
			\draw[-stealth] (a) to (c);
			\draw[-stealth] (b) to (c);
			\draw[-stealth] (c) to (d);		
		\end{tikzpicture}.
	\end{split}
\end{equation}
We choose a particular configuration for (1234):
\begin{equation}
	\begin{split}
		x_1=0,\ x_2=(i,0,\ldots,0),\ x_3=(2i,0.5,0,\ldots,0),\ x_4=(2i,-0.5,0,\ldots,0),
	\end{split}
\end{equation}
and get $z=-\frac{8}{5},\ \bar{z}=\frac{4}{9}$, which is in the range corresponding to class S. So we conclude that (1234) is in class S. The remaining steps are the same as appendix \ref{appendix:type1}. The results of OPE convergence properties in the three channels are shown in table \ref{table:type2convergence}.
\begin{table}[H]
	\caption{OPE convergence properties of type 2 causal orderings}
	\label{table:type2convergence}
	\setlength{\tabcolsep}{3mm} 
	\def\arraystretch{1.25} 
	\centering
	\begin{tabular}{|c|c|c|c|c|}
		\hline
		causal ordering    &  class/subclass  &   s-channel    &   t-channel   &   u-channel
		\\ \hline
		(1234),\ (4321)   &  S  & \cmark  &  \xmark  &  \xmark
		\\ \hline
		(1324),\ (4231)   &  U  & \xmark  &  \xmark  &  \xmark
		\\ \hline
		(1423),\ (4132)   &  T  & \xmark  &  \cmark  &  \xmark
		\\ \hline
		(2134),\ (3421)   &  S  & \cmark  &  \xmark  &  \xmark
		\\ \hline
		(3124),\ (2431)   &  U  & \xmark  &  \xmark  &  \xmark
		\\ \hline
		(2314),\ (3241)   &  T  & \xmark  &  \cmark  &  \xmark
		\\ \hline
	\end{tabular}
\end{table}
There are only 12 causal orderings because $(i_1i_2i_3i_4)$ and $(i_1i_2i_4i_3)$ are the same causal ordering (the little group is $\bbZ_2$).

\subsection{Type 3}\label{appendix:type3}
The type 3 causal ordering is given by
\begin{equation}
	\begin{split}
		\begin{tikzpicture}[baseline={(0,-0.35)},circuit logic US]		
			\node (a) at (0,0) {$a$};
			\node (b) at (1,0) {$b$};
			\node (c) at (2,0) {$c$};
			\node (d) at (1,-0.5) {$d$};
			\draw[-stealth] (a) to (b);
			\draw[-stealth] (b) to (c);
			\draw[-stealth] (a) to (d);		
		\end{tikzpicture},\quad \begin{tikzpicture}[baseline={(0,-0.35)},circuit logic US]		
			\node (a) at (0,0) {$c$};
			\node (b) at (1,0) {$b$};
			\node (c) at (2,0) {$a$};
			\node (d) at (1,-0.5) {$d$};
			\draw[-stealth] (a) to (b);
			\draw[-stealth] (b) to (c);
			\draw[-stealth] (d) to (c);		
		\end{tikzpicture}.
	\end{split}
\end{equation}
We choose a particular configuration for (1234):
\begin{equation}
	\begin{split}
		x_1=0,\ x_2=(i,0,\ldots,0),\ x_3=(2i,0,\ldots,0),\ x_4=(1.5i,1,0,\ldots,0),
	\end{split}
\end{equation}
and get $z=\frac{1}{6},\ \bar{z}=\frac{3}{2}$, which is in the range corresponding to class T. So we conclude that (1234) is in class T. The results of OPE convergence properties in the three channels are shown in table \ref{table:type3convergence}:
\begin{table}[H]
	\caption{OPE convergence properties of type 3 causal orderings}
	\label{table:type3convergence}
	\setlength{\tabcolsep}{3mm} 
	\def\arraystretch{1.25} 
	\centering
	\begin{tabular}{|c|c|c|c|c|}
		\hline
		causal ordering    &   class/subclass  &  s-channel    &   t-channel   &   u-channel
		\\ \hline
		(1234),\ (4321)   & T & \xmark  &  \cmark  &  \xmark
		\\ \hline
		(1243),\ (4312)   & U & \xmark  &  \xmark  &  \cmark
		\\ \hline
		(1324),\ (4231)   & T & \xmark  &  \cmark  &  \xmark
		\\ \hline
		(1342),\ (4213)   & S & \cmark  &  \xmark  &  \xmark
		\\ \hline
		(1423),\ (4132)   & U & \xmark  &  \xmark  &  \cmark
		\\ \hline
		(1432),\ (4123)   & S & \cmark  &  \xmark  &  \xmark
		\\ \hline
		(2134),\ (3421)   & U & \xmark  &  \xmark  &  \xmark
		\\ \hline
		(2143),\ (3412)   & T & \xmark  &  \xmark  &  \xmark
		\\ \hline
		(2314),\ (3241)  & U & \xmark  &  \xmark  &  \xmark
		\\ \hline
		(2341),\ (3214)   & S & \cmark  &  \xmark  &  \xmark
		\\ \hline
		(2413),\ (3142)   & T & \xmark  &  \xmark  &  \xmark
		\\ \hline
		(2431),\ (3124)   & S & \cmark  &  \xmark  &  \xmark
		\\ \hline
	\end{tabular}
\end{table}

\subsection{Type 4}\label{appendix:type4}
The type 4 causal ordering is given by
\begin{equation}
	\begin{split}
		\begin{tikzpicture}[baseline={(a.base)},circuit logic US]		
			\node (a) at (0,0) {$a$};
			\node (b) at (1,0.5) {$b$};
			\node (c) at (1,-0.5) {$c$};
			\node (d) at (2,0) {$d$};
			\draw[-stealth] (a) to (b);
			\draw[-stealth] (a) to (c);
			\draw[-stealth] (b) to (d);
			\draw[-stealth] (c) to (d);		
		\end{tikzpicture}.
	\end{split}
\end{equation}
We choose a particular configuration for (1234):
\begin{equation}
	\begin{split}
		x_1=0,\ x_2=(i,0.5,0,\ldots,0),\ x_3=(i,-0.5,0,\ldots,0),\ x_4=(2i,0,\ldots,0),
	\end{split}
\end{equation}
and get $z=\frac{1}{9},\ \bar{z}=9$, which is in the range corresponding to class T. So we conclude that (1234) is in class T. The results of OPE convergence properties in the three channels are shown in table \ref{table:type4convergence}:
\begin{table}[H]
	\caption{OPE convergence properties of type 4 causal orderings}
	\label{table:type4convergence}
	\setlength{\tabcolsep}{3mm} 
	\def\arraystretch{1.25} 
	\centering
	\begin{tabular}{|c|c|c|c|c|}
		\hline
		causal ordering    &   class/subclass  &  s-channel    &   t-channel   &   u-channel
		\\ \hline
		(1234),\ (4321)   & T  &  \xmark  &  \cmark  &  \xmark
		\\ \hline
		(1243),\ (4312),\ (2134),\ (3421)   & U  &  \xmark  &  \xmark  &  \xmark
		\\ \hline
		(1342),\ (4213),\ (2341),\ (3214)   & S  &  \cmark  &  \xmark  &  \xmark
		\\ \hline
		(2143),\ (3412)   & T  &  \xmark  &  \xmark  &  \xmark
		\\ \hline
	\end{tabular}
\end{table}
Here we use the fact that $(i_1i_2i_3i_4)$ and $(i_1i_3i_2i_4)$ are the same causal ordering (the little group is $\bbZ_2$).

\subsection{Type 5}\label{appendix:type5}
The type 5 causal ordering is given by
\begin{equation}\label{type5:template}
	\begin{split}
		\begin{tikzpicture}[baseline={(a.base)},circuit logic US]		
			\node (a) at (0,0) {$a$};
			\node (b) at (1,0.5) {$b$};
			\node (c) at (1,0) {$c$};
			\node (d) at (1,-0.5) {$d$};
			\draw[-stealth] (a) to (b);
			\draw[-stealth] (a) to (c);
			\draw[-stealth] (a) to (d);		
		\end{tikzpicture},\quad \begin{tikzpicture}[baseline={(b.base)},circuit logic US]		
			\node (a) at (0,0.5) {$b$};
			\node (b) at (0,0) {$c$};
			\node (c) at (0,-0.5) {$d$};
			\node (d) at (1,0) {$a$};
			\draw[-stealth] (a) to (d);
			\draw[-stealth] (b) to (d);
			\draw[-stealth] (c) to (d);		
		\end{tikzpicture}.
	\end{split}
\end{equation}
We choose a particular configuration for (1234):
\begin{equation}\label{type5:particularconfig1}
	\begin{split}
		x_1=0,\ x_2=(i,0.5,0,\ldots,0),\ x_3=(i,0,\ldots,0),\ x_4=(i,-0.5,\ldots,0),
	\end{split}
\end{equation}
and get $z=\frac{1}{4},\ \bar{z}=\frac{3}{4}$, which is in the range corresponding to subclass $\mathrm{E_{st}}$. So (1234) is in class E. We would like to find a particular configuration in each subclass of class E. The little group of this causal type is $S_3$, which corresponds to permutations among $b,c,d$ in (\ref{type5:template}). By looking up table \ref{table:wsigma}, we see that permuting $x_2,x_3$ in (\ref{type5:particularconfig1}) gives $\mathrm{E_{tu}}$ and permuting $x_3,x_4$ gives $\mathrm{E_{su}}$. To realize $\mathrm{E_{stu}}$ we choose the following configuration in (1234):
\begin{equation}\label{type5:particularconfig2}
	\begin{split}
		x_1=0,\ x_2=(i,0.5,0,\ldots,0),\ x_3=(i,-0.5,0,\ldots,0),\ x_4=(i,0,0.5,0,\ldots,0),
	\end{split}
\end{equation}
and get $z=i,\ \bar{z}=-i$, which is indeed in the range corresponding to subclass $\mathrm{E_{stu}}$. So we conclude that the configurations of (1234) do appear in all subclasses of class E in $d\geq3$, while they only appear in subclasses $\mathrm{E_{st}},\mathrm{E_{su}},\mathrm{E_{tu}}$ in 2d.\footnote{We used two dimensions in (\ref{type5:particularconfig1}) and three dimensions in (\ref{type5:particularconfig2}). On the other hand, as mentioned at the end of section \ref{section:comments2d}, subclass $\mathrm{E_{stu}}$ does not exist in 2d because $z,\bar{z}$ can only be real.}

The results of OPE convergence properties in the three channels are shown in table \ref{table:type5convergence}: 
\begin{table}[H]
	\caption{OPE convergence properties of type 5 causal orderings}
	\label{table:type5convergence}
	\setlength{\tabcolsep}{3mm} 
	\def\arraystretch{1.25} 
	\centering
	\begin{tabular}{|c|c|c|c|c|}
		\hline
		causal ordering   & class/subclass &   s-channel    &   t-channel   &   u-channel
		\\ \hline
		(1234),\ (4321)   &
		\begin{tikzpicture}[baseline={(0,-1)},circuit logic US]
			\node at (0,0){$\mathrm{E_{st}}$};
			\node at (0,-0.5){$\mathrm{E_{su}}$};
			\node at (0,-1){$\mathrm{E_{tu}}$};
			\node at (0,-1.5){$\mathrm{E_{stu}}$};
		\end{tikzpicture}&
		\begin{tikzpicture}[baseline={(0,-1)},circuit logic US]
			\node at (0,0){\cmark};
			\node at (0,-0.5){\cmark};
			\node at (0,-1){\xmark};
			\node at (0,-1.5){\cmark};
		\end{tikzpicture}&
		\begin{tikzpicture}[baseline={(0,-1)},circuit logic US]
			\node at (0,0){\cmark};
			\node at (0,-0.5){\xmark};
			\node at (0,-1){\cmark};
			\node at (0,-1.5){\cmark};
		\end{tikzpicture}&
		\begin{tikzpicture}[baseline={(0,-1)},circuit logic US]
			\node at (0,0){\xmark};
			\node at (0,-0.5){\cmark};
			\node at (0,-1){\cmark};
			\node at (0,-1.5){\cmark};
		\end{tikzpicture}
		\\ \hline
		(2134),\ (3124)   &
		\begin{tikzpicture}[baseline={(0,-1)},circuit logic US]
			\node at (0,0){$\mathrm{E_{st}}$};
			\node at (0,-0.5){$\mathrm{E_{su}}$};
			\node at (0,-1){$\mathrm{E_{tu}}$};
			\node at (0,-1.5){$\mathrm{E_{stu}}$};
		\end{tikzpicture}&
		\begin{tikzpicture}[baseline={(0,-1)},circuit logic US]
			\node at (0,0){\cmark};
			\node at (0,-0.5){\cmark};
			\node at (0,-1){\xmark};
			\node at (0,-1.5){\cmark};
		\end{tikzpicture}&
		\begin{tikzpicture}[baseline={(0,-1)},circuit logic US]
			\node at (0,0){\xmark};
			\node at (0,-0.5){\xmark};
			\node at (0,-1){\xmark};
			\node at (0,-1.5){\xmark};
		\end{tikzpicture}&
		\begin{tikzpicture}[baseline={(0,-1)},circuit logic US]
			\node at (0,0){\xmark};
			\node at (0,-0.5){\xmark};
			\node at (0,-1){\xmark};
			\node at (0,-1.5){\xmark};
		\end{tikzpicture}
		\\ \hline
	\end{tabular}
\end{table}
Here we use the fact that for $(i_1i_2i_3i_4)$, any permutation of $2,3,4$ does not change the causal ordering (the little group is $S_3$).

\subsection{Type 6}\label{appendix:type6}
The type 6 causal ordering is given by
\begin{equation}
	\begin{split}
		\begin{tikzpicture}[baseline={(0,-0.35)},circuit logic US]		
			\node (a) at (0,0) {$a$};
			\node (b) at (1,0) {$b$};
			\node (c) at (2,0) {$c$};
			\node (d) at (1,-0.5) {$d$};
			\draw[-stealth] (a) to (b);
			\draw[-stealth] (b) to (c);		
		\end{tikzpicture}.
	\end{split}
\end{equation}
We choose a particular configuration for (1234):
\begin{equation}
	\begin{split}
		x_1=0,\ x_2=(i,0,\ldots,0),\ x_3=(2i,0,\ldots,0),\ x_4=(i,2,\ldots,0),
	\end{split}
\end{equation}
and get $z=\frac{1}{4},\ \bar{z}=\frac{3}{4}$, which is in the range corresponding to subclass $\mathrm{E_{st}}$. By lemma \ref{lemma:subclass}, the whole (1234) is in subclass $\mathrm{E_{st}}$. The results of OPE convergence properties in the three channels are shown in table \ref{table:type6convergence}.
\begin{table}[H]
	\caption{OPE convergence properties of type 6 causal orderings}
	\label{table:type6convergence}
	\setlength{\tabcolsep}{3mm} 
	\def\arraystretch{1.25} 
	\centering
	\begin{tabular}{|c|c|c|c|c|}
		\hline
		causal ordering    &   class/subclass  &  s-channel    &   t-channel   &   u-channel
		\\ \hline
		(1234),\ (4321),\ (3214),\ (2341)   & $\mathrm{E_{st}}$  &  \cmark  &  \cmark  &  \xmark
		\\ \hline
		(1243),\ (4312),\ (4213),\ (1342)   & $\mathrm{E_{su}}$  &  \cmark  &  \xmark  &  \cmark
		\\ \hline
		(1324),\ (4231),\ (2314),\ (3241)   & $\mathrm{E_{tu}}$  &  \xmark  &  \cmark  &  \xmark
		\\ \hline
		(1423),\ (4132),\ (2413),\ (3142)   & $\mathrm{E_{tu}}$  &  \xmark  &  \xmark  &  \cmark
		\\ \hline
		(1432),\ (4123),\ (3412),\ (2143)   & $\mathrm{E_{st}}$  &  \cmark  &  \xmark  &  \xmark
		\\ \hline
		(2134),\ (3421),\ (3124),\ (2431)   & $\mathrm{E_{su}}$  &  \cmark  &  \xmark  &  \xmark
		\\ \hline
	\end{tabular}
\end{table}

\subsection{Type 7}\label{appendix:type7}
The type 7 causal ordering is given by
\begin{equation}
	\begin{split}
		\begin{tikzpicture}[baseline={(0,-0.4)},circuit logic US]		
			\node (a) at (0,0) {$a$};
			\node (b) at (1,0.5) {$b$};
			\node (c) at (1,-0.5) {$c$};
			\node (d) at (0.5,-1) {$d$};
			\draw[-stealth] (a) to (b);
			\draw[-stealth] (a) to (c);		
		\end{tikzpicture},\quad \begin{tikzpicture}[baseline={(0,-0.4)},circuit logic US]		
			\node (a) at (0,0.5) {$b$};
			\node (b) at (0,-0.5) {$c$};
			\node (c) at (1,0) {$a$};
			\node (d) at (0.5,-1) {$d$};
			\draw[-stealth] (a) to (c);
			\draw[-stealth] (b) to (c);		
		\end{tikzpicture}.
	\end{split}
\end{equation}
We choose a particular configuration for (1234):
\begin{equation}
	\begin{split}
		x_1=0,\ x_2=(i,0.5,0,\ldots,0),\ x_3=(i,-0.5,0,\ldots,0),\ x_4=(0,2,0,\ldots,0),
	\end{split}
\end{equation}
and get $z=\frac{7}{15},\ \bar{z}=9$, which is in the range corresponding to class T. So we conclude that (1234) is in class T. The results of OPE convergence properties in the three channels are shown in table \ref{table:type7convergence}.
\begin{table}[H]
	\caption{OPE convergence properties of type 7 causal orderings}
	\label{table:type7convergence}
	\setlength{\tabcolsep}{3mm} 
	\def\arraystretch{1.25} 
	\centering
	\begin{tabular}{|c|c|c|c|c|}
		\hline
		causal ordering    &   class/subclass  &  s-channel    &   t-channel   &   u-channel
		\\ \hline
		(1234),\ (4321)   & T  &  \xmark  &  \cmark  &  \xmark
		\\ \hline
		(1243),\ (4312)   & U  &  \xmark  &  \xmark  &  \cmark
		\\ \hline
		(1342),\ (4213)   & S  &  \cmark  &  \xmark  &  \xmark
		\\ \hline
		(2134),\ (3421)   & U  &  \xmark  &  \xmark  &  \xmark
		\\ \hline
		(2143),\ (3412)   & T  &  \xmark  &  \xmark  &  \xmark
		\\ \hline
		(2341),\ (3124)   & S  &  \cmark  &  \xmark  &  \xmark
		\\ \hline
	\end{tabular}
\end{table}	
Here we use the fact that $(i_1i_2i_3i_4)$ and $(i_1i_3i_2i_4)$ are the same causal ordering (the little group is $\bbZ_2$).

\subsection{Type 8}\label{appendix:type8}
The type 8 causal ordering is given by
\begin{equation}
	\begin{split}
		\begin{tikzpicture}[baseline={(0,-0.4)},circuit logic US]		
			\node (a) at (0,0) {$a$};
			\node (b) at (1,0.5) {$b$};
			\node (c) at (1,-0.5) {$d$};
			\node (d) at (0,-1) {$c$};
			\draw[-stealth] (a) to (b);
			\draw[-stealth] (a) to (c);
			\draw[-stealth] (d) to (c);		
		\end{tikzpicture}.
	\end{split}
\end{equation}
We choose a particular configuration for (1234):
\begin{equation}
	\begin{split}
		x_1=0,\ x_2=(i,-0.5,0,\ldots,0),\ x_3=(0,1,0,\ldots,0),\ x_4=(i,0.5,0,\ldots,0),
	\end{split}
\end{equation}
and get $z=\frac{1}{4},\ \bar{z}=\frac{9}{4}$, which is in the range corresponding to class T. So we conclude that (1234) is in class T. The results of OPE convergence properties in the three channels are shown in table \ref{table:type8convergence}:
\begin{table}[H]
	\caption{OPE convergence properties of type 8 causal orderings}
	\label{table:type8convergence}
	\setlength{\tabcolsep}{3mm} 
	\def\arraystretch{1.25} 
	\centering
	\begin{tabular}{|c|c|c|c|c|}
		\hline
		causal ordering    &   class/subclass  &  s-channel    &   t-channel   &   u-channel
		\\ \hline
		(1234),\ (4321)   & T  &  \xmark  &  \xmark  &  \xmark
		\\ \hline
		(1243),\ (4312),\ (3421),\ (2134)   & U  &  \xmark  &  \xmark  &  \cmark
		\\ \hline
		(1342),\ (4213),\ (2431),\ (3124)   & S  &  \cmark  &  \xmark  &  \xmark
		\\ \hline
		(2143),\ (3412)   & T  &  \xmark  &  \xmark  &  \xmark
		\\ \hline
		(1324),\ (4231)   & T  &  \xmark  &  \xmark  &  \xmark
		\\ \hline
		(1432),\ (4123),\ (2341),\ (3214)   & S  &  \cmark  &  \xmark  &  \xmark
		\\ \hline
		(1423),\ (4132),\ (3241),\ (2314)   & U  &  \xmark  &  \xmark  &  \xmark
		\\ \hline
		(2413),\ (3142)   & T  &  \xmark  &  \xmark  &  \xmark
		\\ \hline
	\end{tabular}
\end{table}

\subsection{Type 9}\label{appendix:type9}
The type 9 causal ordering is given by
\begin{equation}
	\begin{split}
		\begin{tikzpicture}[baseline={(c.base)},circuit logic US]		
			\node (a) at (0,0) {$a$};
			\node (b) at (1,0) {$b$};
			\node (c) at (0.5,-0.5) {$c$};
			\node (d) at (0.5,-1) {$d$};
			\draw[-stealth] (a) to (b);		
		\end{tikzpicture}.
	\end{split}
\end{equation}
We choose a particular configuration for (1234):
\begin{equation}
	\begin{split}
		x_1=0,\ x_2=(i,0,\ldots,0),\ x_3=(0,2,0,\ldots,0),\ x_4=(0,3,0,\ldots,0),
	\end{split}
\end{equation}
and get $z=-\frac{1}{8},\ \bar{z}=\frac{1}{4}$, which is in the range corresponding to class S. So we conclude that (1234) is in class S. The results of OPE convergence properties in the three channels are shown in table \ref{table:type9convergence}:
\begin{table}[H]
	\caption{OPE convergence properties of type 9 causal orderings}
	\label{table:type9convergence}
	\setlength{\tabcolsep}{3mm} 
	\def\arraystretch{1.25} 
	\centering
	\begin{tabular}{|c|c|c|c|c|}
		\hline
		causal ordering    &   class/subclass  &  s-channel    &   t-channel   &   u-channel
		\\ \hline
		(1234),\ (4312),\ (2134),\ (3412)   & S  &  \cmark  &  \xmark  &  \xmark
		\\ \hline
		(1324),\ (4213),\ (3124),\ (2413)   & U  &  \xmark  &  \xmark  &  \cmark
		\\ \hline
		(1423),\ (4123)   & T  &  \xmark  &  \cmark  &  \xmark
		\\ \hline
		(2314),\ (3214)   & T  &  \xmark  &  \cmark  &  \xmark
		\\ \hline
	\end{tabular}
\end{table}	
Here we use the fact that $(i_1i_2i_3i_4)$ and $(i_1i_2i_4i_3)$ are the same causal ordering (the little group is $\bbZ_2$).

\subsection{Type 10}\label{appendix:type10}
The type 10 causal ordering is given by
\begin{equation}\label{template:type10}
	\begin{split}
		\begin{tikzpicture}[baseline={(c.base)},circuit logic US]		
			\node (a) at (0,0) {$a$};
			\node (b) at (0,-1) {$b$};
			\node (c) at (1,-0.5) {$c$};
			\node (d) at (2,-0.5) {$d$};
			\draw[-stealth] (a) to (c);
			\draw[-stealth] (b) to (c);
			\draw[-stealth] (a) to (d);
			\draw[-stealth] (b) to (d);		
		\end{tikzpicture}.
	\end{split}
\end{equation}
We choose a particular configuration for (1234):
\begin{equation}\label{type10:particularconfig1}
	\begin{split}
		x_1=0,\ x_2=(0,1,0,\ldots,0),\ x_3=(2i,0,\ldots,0),\ x_4=(2i,1,\ldots,0),
	\end{split}
\end{equation}
and get $z=\bar{z}=\frac{1}{4}$, which is in the range corresponding to subclass $\mathrm{E_{st}}$. So (1234) is in class E. We would like to find a particular configuration in each subclass of class E. The little group of this causal type is $\bbZ_2\times\bbZ_2$, which is generated by $a\leftrightarrow b$ and $c\leftrightarrow d$ in (\ref{template:type10}). By looking up table \ref{table:wsigma}, we see that by acting with the little group on configuration (\ref{type10:particularconfig1}), we can get $\mathrm{E_{su}}$, but we cannot get $\mathrm{E_{tu}}$. The underlying fact is that the 2d configurations of (1234) do not appear in $\mathrm{E_{tu}}$ (it is obvious that 2d configurations do not appear in $\mathrm{E_{stu}}$.). Let us show this fact. In 2d Minkowski space we can use the light-cone coordinates:
\begin{equation}
	\begin{split}
		z_k=t_k+\mathbf{x}_k,\quad\bar{z}_k=t_k-\mathbf{x}_k,\quad \fr{x_k=(it_k,\mathbf{x}_k)}. 
	\end{split}
\end{equation}
The causal ordering (\ref{template:type10}) implies
\begin{equation}
	\begin{split}
		&z_3,z_4>z_1,z_2,\quad\bar{z}_3,\bar{z}_4>\bar{z}_1,\bar{z}_2, \\
		&(z_1-z_2)(\bar{z}_1-\bar{z}_2)<0,\quad(z_3-z_4)(\bar{z}_3-\bar{z}_4)<0. \\
	\end{split}
\end{equation}
Since the little group $\bbZ_2\times\bbZ_2$ of (1234) preserves $\mathrm{E_{tu}}$ (see table \ref{table:wsigma}), by the $\bbZ_2\times\bbZ_2$-action, it suffices to show that $\mathrm{E_{tu}}$ configurations do not exist when
\begin{equation}
	\begin{split}
		&z_3,z_4>z_1,z_2,\quad\bar{z}_3,\bar{z}_4>\bar{z}_1,\bar{z}_2, \\
		&z_1-z_2<0,\quad\bar{z}_1-\bar{z}_2>0, \\
		&z_3-z_4<0,\quad\bar{z}_3-\bar{z}_4>0. \\
	\end{split}
\end{equation}
In this case the computation is straightforward:
\begin{equation}
	\begin{split}
		z=&\dfrac{(z_2-z_1)(z_4-z_3)}{(z_3-z_1)(z_4-z_2)}<\dfrac{(z_3-z_1)(z_4-z_2)}{(z_3-z_1)(z_4-z_2)}=1, \\
		\bar{z}=&\dfrac{(\bar{z}_1-\bar{z}_2)(\bar{z}_3-\bar{z}_4)}{(\bar{z}_3-\bar{z}_1)(\bar{z}_4-\bar{z}_2)}<\dfrac{(\bar{z}_4-\bar{z}_2)(\bar{z}_3-\bar{z}_1)}{(\bar{z}_3-\bar{z}_1)(\bar{z}_4-\bar{z}_2)}=1. \\
	\end{split}
\end{equation}
So we conclude that the 2d configurations in (1234) have $z,\bar{z}<1$, i.e. (1234) does not intersect with subclass $\mathrm{E_{tu}}$ in 2d. To find a $\mathrm{E_{tu}}$ configuration in (1234) we need to construct it in $d\geq3$. We choose the 3d configuration (\ref{finalpoint:bulkptsing}) and get $z\approx1.1,\ \bar{z}\approx6.3$, which is in the range corresponding to subclass $\mathrm{E_{tu}}$.

To realize $\mathrm{E_{stu}}$ we choose the following configuration in (1234):
\begin{equation}\label{type10:particularconfig3}
	\begin{split}
		x_1=0,\ x_2=(0,0.5,0,\ldots,0),\ x_3=(2i,0,0.5,0,\ldots,0),\ x_4=(i,0.5,0,\ldots,0),
	\end{split}
\end{equation}
and get $z\approx0.33+0.24i,\ \bar{z}=0.33-0.24i$, which is in the range corresponding to subclass $\mathrm{E_{stu}}$. So we conclude that the configurations of (1234) do appear in all subclasses of class E in $d\geq3$, while they only appear in subclasses $\mathrm{E_{st}},\mathrm{E_{su}}$ in 2d.

The results of OPE convergence properties in the three channels are shown in table \ref{table:type10convergence}: 
\begin{table}[H]
	\caption{OPE convergence properties of type 10 causal orderings}
	\label{table:type10convergence}
	\setlength{\tabcolsep}{3mm} 
	\def\arraystretch{1.25} 
	\centering
	\begin{tabular}{|c|c|c|c|c|}
		\hline
		causal ordering   & class/subclass &   s-channel    &   t-channel   &   u-channel
		\\ \hline
		(1234),\ (3412)   &
		\begin{tikzpicture}[baseline={(0,-0.75)},circuit logic US]
			\node at (0,0){$\mathrm{E_{st}}$};
			\node at (0,-0.5){$\mathrm{E_{su}}$};
			\node at (0,-1){$\mathrm{E_{tu}}$};
			\node at (0,-1.5){$\mathrm{E_{stu}}$};
		\end{tikzpicture}&
		\begin{tikzpicture}[baseline={(0,-0.75)},circuit logic US]
			\node at (0,0){\cmark};
			\node at (0,-0.5){\cmark};
			\node at (0,-1){\xmark};
			\node at (0,-1.5){\cmark};
		\end{tikzpicture}&
		\begin{tikzpicture}[baseline={(0,-0.75)},circuit logic US]
			\node at (0,0){\xmark};
			\node at (0,-0.5){\xmark};
			\node at (0,-1){\xmark};
			\node at (0,-1.5){\xmark};
		\end{tikzpicture}&
		\begin{tikzpicture}[baseline={(0,-0.75)},circuit logic US]
			\node at (0,0){\xmark};
			\node at (0,-0.5){\xmark};
			\node at (0,-1){\xmark};
			\node at (0,-1.5){\xmark};
		\end{tikzpicture}
		\\ \hline
		(1324),\ (2413)   &
		\begin{tikzpicture}[baseline={(0,-0.75)},circuit logic US]
			\node at (0,0){$\mathrm{E_{st}}$};
			\node at (0,-0.5){$\mathrm{E_{su}}$};
			\node at (0,-1){$\mathrm{E_{tu}}$};
			\node at (0,-1.5){$\mathrm{E_{stu}}$};
		\end{tikzpicture}&
		\begin{tikzpicture}[baseline={(0,-0.75)},circuit logic US]
			\node at (0,0){\cmark};
			\node at (0,-0.5){\cmark};
			\node at (0,-1){\xmark};
			\node at (0,-1.5){\cmark};
		\end{tikzpicture}&
		\begin{tikzpicture}[baseline={(0,-0.75)},circuit logic US]
			\node at (0,0){\xmark};
			\node at (0,-0.5){\xmark};
			\node at (0,-1){\xmark};
			\node at (0,-1.5){\xmark};
		\end{tikzpicture}&
		\begin{tikzpicture}[baseline={(0,-0.75)},circuit logic US]
			\node at (0,0){\xmark};
			\node at (0,-0.5){\xmark};
			\node at (0,-1){\xmark};
			\node at (0,-1.5){\xmark};
		\end{tikzpicture}
		\\ \hline
		(1423),\ (2314)   &
		\begin{tikzpicture}[baseline={(0,-0.75)},circuit logic US]
			\node at (0,0){$\mathrm{E_{st}}$};
			\node at (0,-0.5){$\mathrm{E_{su}}$};
			\node at (0,-1){$\mathrm{E_{tu}}$};
			\node at (0,-1.5){$\mathrm{E_{stu}}$};
		\end{tikzpicture}&
		\begin{tikzpicture}[baseline={(0,-0.75)},circuit logic US]
			\node at (0,0){\cmark};
			\node at (0,-0.5){\cmark};
			\node at (0,-1){\xmark};
			\node at (0,-1.5){\cmark};
		\end{tikzpicture}&
		\begin{tikzpicture}[baseline={(0,-0.75)},circuit logic US]
			\node at (0,0){\cmark};
			\node at (0,-0.5){\xmark};
			\node at (0,-1){\cmark};
			\node at (0,-1.5){\cmark};
		\end{tikzpicture}&
		\begin{tikzpicture}[baseline={(0,-0.75)},circuit logic US]
			\node at (0,0){\xmark};
			\node at (0,-0.5){\cmark};
			\node at (0,-1){\cmark};
			\node at (0,-1.5){\cmark};
		\end{tikzpicture}
		\\ \hline
	\end{tabular}
\end{table}
Here we use the fact $(i_1i_2i_3i_4)$, $(i_2i_1i_3i_4)$, $(i_1i_2i_4i_3)$ and $(i_2i_1i_4i_3)$ are the same causal ordering (the little group is $\bbZ_2\times\bbZ_2$).

\subsection{Type 11}\label{appendix:type11}
The type 11 causal ordering is given by
\begin{equation}
	\begin{split}
		\begin{tikzpicture}[baseline={(0,-0.4)},circuit logic US]		
			\node (a) at (0,0) {$a$};
			\node (b) at (1,0) {$b$};
			\node (c) at (0,-0.5) {$c$};
			\node (d) at (1,-0.5) {$d$};
			\draw[-stealth] (a) to (b);
			\draw[-stealth] (c) to (d);		
		\end{tikzpicture}.
	\end{split}
\end{equation}
We choose a particular configuration for (1234):
\begin{equation}
	\begin{split}
		x_1=0,\ x_2=(i,0,\ldots,0),\ x_3=(0,2,\ldots,0),\ x_4=(i,2,\ldots,0),
	\end{split}
\end{equation}
and get $z=\bar{z}=\frac{1}{4}$, which is in the range corresponding to subclass $\mathrm{E_{st}}$. By lemma \ref{lemma:subclass}, the whole (1234) is in subclass $\mathrm{E_{st}}$. The results of OPE convergence properties in the three channels are shown in table \ref{table:type11convergence}.
\begin{table}[H]
	\caption{OPE convergence properties of type 11 causal orderings}
	\label{table:type11convergence}
	\setlength{\tabcolsep}{3mm} 
	\def\arraystretch{1.25} 
	\centering
	\begin{tabular}{|c|c|c|c|c|}
		\hline
		causal ordering    &   class/subclass  &  s-channel    &   t-channel   &   u-channel
		\\ \hline
		(1234),\ (2143)   & $\mathrm{E_{st}}$  &  \cmark  &  \xmark  &  \xmark
		\\ \hline
		(1243),\ (2134)   & $\mathrm{E_{su}}$  &  \cmark  &  \xmark  &  \cmark
		\\ \hline
		(1324),\ (3142)   & $\mathrm{E_{tu}}$  &  \xmark  &  \xmark  &  \cmark
		\\ \hline
		(1342),\ (3124)   & $\mathrm{E_{su}}$  &  \cmark  &  \xmark  &  \cmark
		\\ \hline
		(1423),\ (3241)   & $\mathrm{E_{tu}}$  &  \xmark  &  \cmark  &  \xmark
		\\ \hline
		(1432),\ (2341)   & $\mathrm{E_{st}}$  &  \cmark  &  \cmark  &  \xmark
		\\ \hline
	\end{tabular}
\end{table}
Here we use the fact that $(i_1i_2i_3i_4)$ and $(i_3i_4i_1i_2)$ are the same causal ordering (the little group is $\bbZ_2$).

\subsection{Type 12}\label{app:type12}
The type 12 causal ordering is given by
\begin{equation}
	\begin{split}
		a\quad b\quad c\quad d.
	\end{split}
\end{equation}
One can show that this causal type belongs to class E. In fact this is the ``Euclidean" case (that's why this class is called class E), and there is no need to check numerically for this type. Suppose we have a totally space-like separated configuration $(x_1,x_2,x_3,x_4)$, where $x_k=(it_k,\mathbf{x}_k)$. We can reach this configuration via the path
\begin{equation}
	\begin{split}
		x_k(s)=\left\{
		\begin{array}{rcl}
			\fr{(1-2s)\epsilon_k,2s \mathbf{x}_k}& &s\in[0,1/2] \\
			\fr{(2s-1)it_k,\mathbf{x}_k}& &s\in[1/2,1] \\
		\end{array}
		\right.
	\end{split}
\end{equation}
Along the path all the $x_i,x_j$ pairs are space-like separated. As a result, the totally space-like separated configurations always have $N_t=N_u=0$ (as long as $N_t,N_u$ are well-defined). On the other hand, there is no doubt that all subclasses of class E can be realized in $d\geq3$.\footnote{One can realize one of the three subclasses $\mathrm{E_{st}},\mathrm{E_{su}},\mathrm{E_{tu}}$ in 2d, and then realize the other two subclasses by $S_4$-action. For subclass $\mathrm{E_{stu}}$, one can put four points in the hyperplane with $t=0$.} We summarize the OPE convergence properties of this type in table \ref{table:type12convergence}.
\begin{table}[H]
	\caption{OPE convergence properties of type 12 causal orderings}
	\label{table:type12convergence}
	\setlength{\tabcolsep}{3mm} 
	\def\arraystretch{1.25} 
	\centering
	\begin{tabular}{|c|c|c|c|c|}
		\hline
		causal ordering   & class/subclass &   s-channel    &   t-channel   &   u-channel
		\\ \hline
		(1234)   &
		\begin{tikzpicture}[baseline={(0,-0.75)},circuit logic US]
			\node at (0,0){$\mathrm{E_{st}}$};
			\node at (0,-0.5){$\mathrm{E_{su}}$};
			\node at (0,-1){$\mathrm{E_{tu}}$};
			\node at (0,-1.5){$\mathrm{E_{stu}}$};
		\end{tikzpicture}&
		\begin{tikzpicture}[baseline={(0,-0.75)},circuit logic US]
			\node at (0,0){\cmark};
			\node at (0,-0.5){\cmark};
			\node at (0,-1){\xmark};
			\node at (0,-1.5){\cmark};
		\end{tikzpicture}&
		\begin{tikzpicture}[baseline={(0,-0.75)},circuit logic US]
			\node at (0,0){\cmark};
			\node at (0,-0.5){\xmark};
			\node at (0,-1){\cmark};
			\node at (0,-1.5){\cmark};
		\end{tikzpicture}&
		\begin{tikzpicture}[baseline={(0,-0.75)},circuit logic US]
			\node at (0,0){\xmark};
			\node at (0,-0.5){\cmark};
			\node at (0,-1){\cmark};
			\node at (0,-1.5){\cmark};
		\end{tikzpicture}
		\\ \hline
	\end{tabular}
\end{table}
Here we use the fact that there is only 1 causal ordering in this type (the little group is $S_4$).

\section{Wightman functions: a brief review}\label{section:Wightman}
In this section we will review some classical results about regular points (points where Wightman distributions are genuine functions) in a general QFT \cite{jost1957bemerkung,Streater:1989vi,bogolubov2012general}. For simplicity let us still consider a scalar theory in the Minkowski space, which is characterized by a collection of Lorentzian correlators:
\begin{equation}
	\begin{split}
		\mathcal{W}_{n}(x_1,x_2,\ldots,x_n)=\bra{0}\phi(x_1)\phi(x_2)...\phi(x_n)\ket{0}
	\end{split}
\end{equation}
where $x_i$ are Lorentzian coordinates.\footnote{\label{footnote:coord}In the rest of part \ref{part:ope}, we the Lorentzian points were denoted by $(it_k,\mathbf{x}_k)$. Only in this section we use the notation $x_k=(t_k,\mathbf{x}_k)$.} We will introduce the Wightman axioms for QFTs, and then review the domain of correlation functions which can be derived from Wightman axioms. In the end, we will compare these classical results with our results for CFT four-point functions.

This section is logically independent from the rest of part \ref{part:ope}. Here we assume Wightman axioms while in the rest we did not. The only connection is to justify the definition of Wick rotation (steps 1 and 2 in chapter \ref{strategy}).

\subsection{Wightman axioms for Lorentzian correlators}
We assume the Wightman axioms for correlators $\left\{\mathcal{W}_n\right\}$:

$(W1)$ Temperedness.

$\mathcal{W}_n$ is a tempered distribution (called Wightman distribution). It becomes a complex number after being smeared with rapidly decreasing test functions $f_n$:
\begin{equation}
	\begin{split}
		\mathcal{W}_n(f_n)=\int f(x_1,\ldots,x_n)\mathcal{W}_n(x_1,\ldots,x_n)dx_1\ldots dx_n
	\end{split}
\end{equation}
The Fourier transform $\hat{\mathcal{W}}_n$ of $\mathcal{W}_n$ is well defined since the space of rapidly decreasing test functions (Schwartz space) is closed under Fourier transform \cite{vsilov1969generalized}. One has the definition $\hat{\mathcal{W}}_n\fr{f_n}\coloneqq\mathcal{W}_n\fr{\hat{f}_n}$.

$(W2)$ Poincar\'e invariance.

The correlators transform invariantly under action of the Poincar\'e group:
\begin{equation}
	\begin{split}
		\mathcal{W}_n(g\cdot x_1,\ldots,g\cdot x_n)=\mathcal{W}_n(x_1,\ldots,x_n)
	\end{split}
\end{equation}
for all $n\geq0$ and $g$ in the Poincar\'e group.  

$(W3)$ Unitarity.

The vector space generated by the states of the form
\begin{equation}
	\begin{split}
		\Psi\left(\underline{f}\right)=\sumlim{n\geq0}\int f_n(x_1,\ldots,x_n)\phi(x_1)\ldots\phi(x_n)\ket{0}dx_1\ldots dx_n
	\end{split}
\end{equation}
has a non-negative inner product. Here $\underline{f}$ is an arbitrary finite sequence of complex valued Schwartz functions: $\underline{f}=(f_0,f_1,f_2,\ldots)$ and $f_n$ denotes the Schwartz function with $n$ Lorentzian points as variables. If we assume that $\phi(x)$ are Hermitian operators, i.e. $\left[\phi(x)\right]^\dagger=\phi(x)$, then the unitarity condition is written as
\begin{equation}\label{wightman:unitarity}
	\begin{split}
		\sumlim{n,m}\int \overline{f_n(x_1\ldots,x_n)}f_m(y_1,\ldots,y_m)\mathcal{W}_{n+m}(x_n,\ldots,x_1,y_1,\ldots,y_m)dxdy\geq0
	\end{split}
\end{equation}

$(W4)$ Spectral condition.

The open forward light-cone $V_+$ is defined by the collection of vectors $x\in\bbR{d}$ such that
\begin{equation}
	\begin{split}
		x^0&>\sqrt{\sumlim{\mu\geq1}(x^\mu)^2}.
	\end{split}
\end{equation}
In a general QFT we have self-adjoint momentum operators $P^\mu$. The spectral condition says that the spectrum of $P=(P^0,P^1,\ldots,P^{d-1})$ is inside the closed forward light-cone $\overline{V_+}$, and the normalized eigenvector of $P=0$ is unique (up to a phase factor), denoted by $\ket{0}$.

We define the reduced correlators $W_{n-1}$ by
\begin{equation}\label{def:reducedW}
	\begin{split}
		W_{n}(x_2-x_1,\ldots,x_{n+1}-x_{n})=\mathcal{W}_{n+1}(x_1,\ldots,x_{n+1})
	\end{split}
\end{equation}
Since $\mathcal{W}_{n+1}$ is a translational invariant tempered distribution, $W_n$ is well defined and is also a tempered distribution. The spectral condition implies that the Fourier transforms $\hat{W}_{n}$ of $W_n$ is supported in the forward light-cone. That is to say, $\hat{W}_n(p_1,\ldots,p_n)\neq0$ only if all the momentum variables $p_k$ are inside $\overline{V}_+$.

$(W5)$ Microscopic causality.

$\mathcal{W}_n(x_1,\ldots,x_k,x_{k+1},\ldots,x_n)=\mathcal{W}_n(x_1,\ldots,x_{k+1},x_k,\ldots,x_n)$ if $x_k$ and $x_{k+1}$ are space-like separated.

\subsection{Wightman functions and their domains}\label{section:wightmandomain}
\subsubsection{Forward tube}
Let us consider the ``reduced correlator" $W_{n}$ defined in eq. (\ref{def:reducedW}). $W_n$ has Fourier transform
\begin{equation}\label{Wn:fourier}
	\begin{split}
		W_n(x_1,...,x_n)=\int\dfrac{dp_1}{(2\pi)^d}\ldots\dfrac{dp_n}{(2\pi)^d}\hat{W}_n(p_1,\ldots,p_n)e^{-i(p_1\cdot x_1+\ldots+p_n\cdot x_n)},
	\end{split}
\end{equation}
where $\hat{W}_n$ is also a tempered distribution, and the Lorentzian inner product is defined by $p\cdot x=-p^0x^0+p^1x^1+\ldots+p^{d-1}x^{d-1}$. In general, $W_n$ is not a function if $x_k$ are real. However, if we replace $x_k$ with complex coordinates $x_k\rightarrow z_k=x_k+iy_k$, because of the spectral condition $(W4)$, $W_n(z_1,\ldots,z_n)$ is indeed a function if the imaginary parts of $z_k$ belong to $V_+$. The argument is as follows. Suppose $y_k\in V_+$ for all $k=1,2,\ldots,n$, then there exists a Schwartz function $f(p_1,\ldots,p_n)$ in the momentum space such that $f(p_1,\ldots,p_n)=e^{-i(p_1\cdot z_1+\ldots+p_n\cdot z_n)}$ when all the momentum variables $p_k$ are inside the closure of forward light-cone.\footnote{The crucial point is that if all $y_k$ are inside the forward light-cone, then $f(p_1,...,p_n)$ decays exponentially fast when some $p_k$ goes to infinity inside the forward light-cone.} Since $\hat{W}_n$ is supported in $\overline{V}_+^n$, $\hat{W}_n(f)$ is exactly in the form of (\ref{Wn:fourier}) with $x_k$ replaced by $z_k$. So $W_n(z_1,\ldots,z_n)$ is a well-defined complex number when $Im(z_k)\in V_+$ for all $k$.

Furthermore, since $-i\fr{p_k}_\mu f(p_1,\ldots,p_n)$ is also a Schwartz test function, we have
\begin{equation}
	\begin{split}
		\dfrac{\partial}{\partial z_k^\mu}W_n(z_1,\ldots,z_n)=\hat{W}_n[-i\fr{p_k}_\mu f], \\
		k=1,\ldots,n\ \mathrm{and}\ \mu=0,\ldots,d-1.\\
	\end{split}
\end{equation}
As a result $W_n(z_1,\ldots,z_n)$ is an analytic function inside the ``forward tube", denoted as $I_n$
\begin{equation}\label{def:forwardtube}
	\begin{split}
		I_n=\left\{(z_1,\ldots,z_n)\in\mathbb{C}^{nd}\Big{|}Im(z_k)\in V_+,\ k=1,2,\ldots,n\right\},
	\end{split}
\end{equation}
The distribution $W_n(x_1,\ldots,x_n)$ is the boundary value of the analytic function $W_n(z_1,\ldots,z_n)$ on the forward tube $I_n$:
\begin{equation}
	\begin{split}
		W_n(x_1,\ldots,x_n)=\limlim{\substack{y\rightarrow0 \\ y\in V_+^n}}W_n(x_1+iy_1,\ldots,x_n+iy_n).
	\end{split}
\end{equation}

\subsubsection{Bargmann-Hall-Wightman theorem, extended tube}
Now let us use the Lorentz invariance $(W2)$ to analytically continue $W_n$ to a larger domain. By $(W2)$, $W_n$ is invariant under the action of real Lorentz group $SO^+(1,d-1)$:\footnote{By $SO^+(1,d-1)$ we mean the connected component of the identity element in $O(1,d-1)$.}
\begin{equation}\label{Wn:lorentz}
	\begin{split}
		W_n(g\cdot x_1,\ldots,g\cdot x_n)=W_n(x_1,\ldots,x_n),\quad\forall g\in SO^+(1,d-1).
	\end{split}
\end{equation}
The Lorentz transformations preserve the inner product $p_k\cdot x_k$ and the measure $dx_k$, so the Fourier transform $\hat{W}_n$ is also Lorentz invariant
\begin{equation}
	\begin{split}
		W_n(g\cdot p_1,\ldots,g\cdot p_n)=W_n(p_1,\ldots,p_n),\quad\forall g\in SO^+(1,d-1).
	\end{split}
\end{equation}
Since $W_n(z_1,\ldots,z_n)$ is defined by the Fourier transform (\ref{Wn:fourier}) (replace $x_k$ with $z_k$), we have
\begin{equation}\label{Wn:lorentz2}
	\begin{split}
		W_n(g\cdot z_1,\ldots,g\cdot z_n)=W_n(z_1,\ldots,z_n),\quad\forall g\in SO^+(1,d-1).
	\end{split}
\end{equation}
Here we remark that the real Lorentz group actions preserve the forward tube $I_n$. 

An important observation is that (\ref{Wn:lorentz2}) remains true if we replace the real Lorentz group $SO^+(1,d-1)$ by the proper complex Lorentz group $L_+\fr{\mathbb{C}}$.\footnote{Let $d$ be the spacetime dimension. The complex Lorentz group $L\fr{\mathbb{C}}$ is defined by the set of all $d\times d$ complex matrices $M$ such that $M^t\eta M=\eta$. Here $\eta=\mathrm{diag}(-1,1,\ldots,1)$ is the matrix of Lorentzian inner product, and $M^t$ is the transpose of $M$. $L_+\fr{\mathbb{C}}$ is the subgroup of $L\fr{\mathbb{C}}$ with constraint det$M$=1. $L_+\fr{\mathbb{C}}$ is connected, unlike the real case where we need to introduce the constraints ``proper", ``orthochronous" for connectedness.} Given an arbitrary $g\in L_+\fr{\mathbb{C}}$, we define
\begin{equation}\label{def:complexextension}
	\begin{split}
		W_n^g(z_1,\ldots,z_n)\coloneqq W_n\fr{g^{-1}\cdot z_n,\ldots,g^{-1}\cdot z_n}
	\end{split}
\end{equation}
for $(z_1,\ldots,z_n)\in gI_n$. The Bargmann-Hall-Wightman theorem \cite{hall1957theorem} tells us that if we choose different complex Lorentz group elements $g_1,g_2$, the functions $W^{g_1}_n$ and $W^{g_2}_n$ coincide in the domain $g_1I_n\cap g_2I_n$. So $W_n(z_1,\ldots,z_n)$ has analytic continuation to the ``extended forward tube", denoted by $\tilde{I}_n$:
\begin{equation}\label{def:extendedtube}
	\begin{split}
		\tilde{I}_n\coloneqq\left\{(z_1,\ldots,z_n)\in\mathbb{C}^{nd}\Big{|}(g\cdot z_1,\ldots,g\cdot z_n)\in I_n\ \mathrm{for}\ \mathrm{some}\ g\in L_+\fr{\mathbb{C}}\right\}.
	\end{split}
\end{equation}
Here we only give the idea of the proof. It suffices to show that for any $g\in L_+\fr{\mathbb{C}}$, the function $W^g_n$ coincides with $W_n$ in the domain $gI_n\cap I_n$. Since $I_n$ and $gI_n$ are two convex sets, their intersection $gI_n\cap I_n$ is also convex, thus connected. So it suffices to show that $W^g_n$ coincides with $W_n$ in the neighbourhood of one point. This is obvious for $g$ near the identity element, but the proof for an arbitrary $g$ is based on the fact that the set $\left\{g\in L_+\fr{\mathbb{C}}\ |\ gI_n\cap I_n\neq\emptyset\right\}$ is connected, which follows from the group structure of the complex Lorentz group $L_+\fr{\mathbb{C}}$ (for more details, see \cite{jost1979general}). 

\subsubsection{Jost points}
While $I_n$ does not contain Lorentzian points (i.e. points with $\mathrm{Im}(z_k)=0$ for all $k$), $\tilde{I}_n$ contains a region of Lorentzian points. These points are called Jost points \cite{jost1957bemerkung}, and they are defined by the configurations $(x_1,\ldots,x_n)$ such that the following cone
\begin{equation}\label{jostpts}
	\begin{split}
		\left\{\lambda_1 x_1+\ldots+\lambda_n x_n\Big{|}\lambda_k\geq0\ \mathrm{for\ all}\ k,\quad\sumlim{k=1}^n\lambda_k>0\right\}
	\end{split}
\end{equation}
contains only space-like points (see \cite{jost1979general}, the theorem on page 81 and the corollary on page 82). 

\subsubsection{Microscopic causality, envelope of holomorphy}\label{section:envelope}
Now let us go back to $\mathcal{W}_{n}(x_1,\ldots,x_{n})$ via (\ref{def:reducedW}). We define $\mathcal{J}_{n}$ as the set of $(x_1,\ldots,x_{n})$ such that $(x_2-x_1,\ldots,x_n-x_{n-1})$ are Jost points. The configurations in $\mathcal{J}_n$ have totally space-like separations. To see this we rewrite $x_i-x_j$ ($i>j$) as
\begin{equation}
	\begin{split}
		x_i-x_j=(x_i-x_{i-1})+(x_{i-1}-x_{i-2})+\ldots+(x_{j+1}-x_j),
	\end{split}
\end{equation}
which is in the form of (\ref{jostpts}). By the definition of Jost points, we have $(x_i-x_j)^2>0$ for $i\neq j$. It is obvious that $\mathcal{J}_2$ contains all totally space-like configurations. For $n\geq3$, $\mathcal{J}_n$ does not contain all totally space-like configurations.

Since Jost points are the configurations with totally space-like separations, by the microscopic causality condition (W5), $\mathcal{W}_{n}(x_1,\ldots,x_{n})$ is also regular at $(x_1,\ldots,x_n)$ if there exists a permutation $\sigma\in S_{n}$ such that
\begin{equation}
	\begin{split}
		\fr{x_{\sigma\fr{1}},x_{\sigma\fr{2}},\ldots,x_{\sigma\fr{n}}}\in\mathcal{J}_n.
	\end{split}
\end{equation}
Then the equation $\mathcal{W}_n\fr{x_1,\ldots,x_n}=\mathcal{W}_n\fr{x_{\sigma(1)},\ldots,x_{\sigma(n)}}$ in $\mathcal{J}_n$ can be analytically continued to
\begin{equation}
	\begin{split}
		\mathcal{W}_n\fr{z_1,\ldots,z_n}=\mathcal{W}_n\fr{z_{\sigma(1)},\ldots,z_{\sigma(n)}},\quad (z_2-z_1,\ldots,z_{n}-z_{n-1})\in\tilde{I}_{n-1}.
	\end{split}
\end{equation}
Therefore, $\mathcal{W}_n$ has analytic continuation to the following domain of complex coordinates $(z_1,\ldots,z_n)$:
\begin{equation}\label{def:domainpermutation}
	\begin{split}
		\mathcal{U}_n=\left\{(z_1,\ldots,z_n)\in\mathbb{C}^{nd}\Big{|}\exists\sigma\in S_n\ \mathrm{s.t.}\ (z_{\sigma\fr{2}}-z_{\sigma\fr{1}},\ldots,z_{\sigma\fr{n}}-z_{\sigma\fr{n-1}})\in\tilde{I}_{n-1}\right\}.
	\end{split}
\end{equation}
$\mathcal{U}_n$ have the following properties:\footnote{We were unable to track properties 1,2 in prior literatures. Readers are welcome to provide us with references.}
\begin{enumerate}
	\item In 2d, $\mathcal{U}_n$ contains all totally space-like configurations.
	\item $\mathcal{U}_3$ contains all totally space-like configurations.
	\item In $d\geq3$ and $n\geq4$, $\mathcal{U}_n$ does not contain all totally space-like  configurations \cite{jost1979general}.
\end{enumerate}
To show the first property, we use an analogous version of the complex coordinates (\ref{zzbarglobal}):
\begin{equation}
	\begin{split}
		w_k=\mathbf{x}_k+t_k,\quad\bar{w}_k=\mathbf{x}_k-t_k.
	\end{split}
\end{equation}
Given an arbitrary totally space-like configuration, we have
\begin{equation}
	\begin{split}
		(w_j-w_k)(\bar{w}_j-\bar{w}_k)>0,\quad(j\neq k)
	\end{split}
\end{equation}
which implies $w_j>w_k,\bar{w}_j>\bar{w}_k$ or $w_j<w_k,\bar{w}_j<\bar{w}_k$. So we can find a permutation $\sigma$ such that
\begin{equation}\label{jostpt:2d}
	\begin{split}
		w_{\sigma(k)}<w_{\sigma(k+1)},\quad\bar{w}_{\sigma(k)}<\bar{w}_{\sigma(k+1)},\quad k=1,2,\ldots,n-1.
	\end{split}
\end{equation}
We see from (\ref{jostpt:2d}) that $w_{\sigma(k+1)}-w_{\sigma(k)},\bar{w}_{\sigma(k+1)}-\bar{w}_{\sigma(k)}$ are positive, so the cone (\ref{jostpts}) generated from these vectors only contain points with positive components. Thus the configuration $(x_{\sigma(1)},\ldots,x_{\sigma(n)})$ is in $\mathcal{J}_n$, or equivalently, $(x_1,\ldots,x_n)$ is in $\mathcal{U}_n$.

To show the second property, we consider the totally space-like three-point configurations in the following form:
\begin{equation}\label{jostpt:3d3n}
	\begin{split}
		x_1=0,\quad x_2=(0,1,0),\quad x_3=(a,b,c).
	\end{split}
\end{equation}
To check that all totally space-like configurations are in $\mathcal{U}_3$, it suffices to check the totally space-like configurations in the form of (\ref{jostpt:3d3n}) because $\mathcal{U}_n$ is Poincar\'e invariant and scale invariant, and any totally space-like configuration can be mapped to a configuration in the form of (\ref{jostpt:3d3n}) by Poincar\'e transformations and dilatation. If $(x_1,x_2,x_3)\in\mathcal{J}_3$ then we are done. Suppose $(x_1,x_2,x_3)\notin\mathcal{J}_3$, which means that there exists a positive $\lambda$ such that $\lambda(x_2-x_1)+(x_3-x_2)$ is a null vector:
\begin{equation}
	\begin{split}
		a^2=(b-1+\lambda)^2+c^2.
	\end{split}
\end{equation}
The above equation implies $a^2\geq c^2$, so there exists a Lorentz transformation which maps the configuration (\ref{jostpt:3d3n}) to
\begin{equation}\label{lorentz:abc}
	\begin{split}
		x_1^\prime=0,\quad x_2^\prime=(0,1,0),\quad x_3^\prime=(a^\prime,b,0).
	\end{split}
\end{equation}
We see that the problem is reduced to the 2d case. According second property, the configuration (\ref{lorentz:abc}) is in $\mathcal{U}_3$. So we conclude that all totally space-like configurations are in $\mathcal{U}_3$.

To show the third property, it suffices to give a counterexample for $d=3$ and $n=4$:
\begin{equation}\label{envelophol:example}
	\begin{split}
		x_1=&(1-\epsilon,1,1),\quad x_2=(1-\epsilon,-1,-1), \\
		x_3=&(\epsilon-1,1,-1),\quad x_4=(\epsilon-1,-1,1). \\
	\end{split}
\end{equation}
where $\epsilon>0$ is small. (\ref{envelophol:example}) is a totally space-like configuration but it does not belong to $\mathcal{U}_4$ (see \cite{jost1979general}, p. 89). 

$\mathcal{W}_n$ has analytic continuation from $\mathcal{U}_n$ to its envelope of holomorphy $H(\mathcal{U}_n)$ \cite{vladimirov1966methods}, which is defined by the following property:
\begin{itemize}
	\item Any holomorphic function on $\mathcal{U}_n$ has analytic continuation to $H\fr{\mathcal{U}_n}$.
\end{itemize}
A theorem proved by Ruelle \cite{ruelle1959} shows that $H\fr{\mathcal{U}_n}$ contains all totally space-like configurations.

We conclude that Wightman distributions $\mathcal{W}_n(x_1,\ldots,x_n)$ are analytic functions at all totally space-like configurations.

\subsection{Comparison with CFT}
\subsubsection{Justifying the definition of Wick rotation}
In section \ref{anal4-point}, we showed that the CFT four-point function is analytic in the forward tube $\mathcal{T}_4$.\footnote{Here we are abusing terminology ``forward tube". The forward tube $(x_1,x_2,x_3,x_4)\in\mathcal{T}_4$ corresponds to the forward tube $(x_2-x_1,x_3-x_2,x_4-x_3)\in I_3$.} Consider the points $(x_2-x_1,x_3-x_2,x_4-x_3)$ where $(x_1,\ldots,x_4)\in\mathcal{T}_4$. The notation we use in the rest of part \ref{part:ope} is $x_k=(\epsilon_k+it_k,\mathbf{x}_k)$, so we have
\begin{equation}\label{3pt:difference1}
	\begin{split}
		x_{k+1}-x_k=(\epsilon_{k+1}-\epsilon_k+it_{k+1}-it_k,\mathbf{x}_{k+1}-\mathbf{x}_k+i\mathbf{y}_{k+1}-i\mathbf{y}_k).
	\end{split}
\end{equation}
By translating (\ref{3pt:difference1}) to the notation in this section (see footnote \ref{footnote:coord}), we have
\begin{equation}\label{3pt:difference2}
	\begin{split}
		x_{k+1}-x_k=(t_{k+1}-t_k,\mathbf{x}_{k+1}-\mathbf{x}_k)+i(\epsilon_k-\epsilon_{k+1},0).
	\end{split}
\end{equation}
We see from (\ref{3pt:difference2}) that the points $(x_2-x_1,x_3-x_2,x_4-x_3)$ are in the forward tube $I_3$ if $(x_1,x_2,x_3,x_4)\in\mathcal{D}$ (because $\epsilon_1>\epsilon_2>\epsilon_3>\epsilon_4$). Recalling that the  Wightman distributions $\mathcal{W}_n(x_1,\ldots,x_n)$ can be obtained by taking the limit of the analytic functions $\mathcal{W}_n(z_1,\ldots,z_n)$ from the domain $\left\{\fr{z_2-z_1,\ldots,z_n-z_{n-1}}\in I_{n-1}\right\}$, we see that our definition of Wick rotation (\ref{def:Lorentz4ptfct}) is consistent with Wightman QFT. 

\subsubsection{Osterwalder-Schrader theorem}
In fact we use the same analytic continuation path as in the Osterwalder-Schrader (OS) theorem \cite{osterwalder1973}. The OS theorem shows that under certain conditions, a Euclidean QFT can be Wick rotated to a Wightman QFT. 

In part \ref{part:ope} we focus on the domain of analyticity of the four-point function, and we do not explore the distributional properties of it. To show that the limit (\ref{def:Lorentz4ptfct}) of the CFT four-point functions defines a Wightman four-point distribution, one needs to deal with the four-point function not only at regular points (where $|\rho|,|\bar{\rho}|<1$ in s- or t- or u-channel), but also at all the other Lorentzian configurations where $\rho$ and/or $\bar{\rho}$ is 1 in absolute value (this needs a lot of extra work). Readers can go to \cite{Kravchuk:2021kwe}, where we show that Wick rotating a Euclidean CFT four-point function indeed results in a Wightman four-point distribution.

Let us contrast our construction and the OS construction. Our construction extends $G_4$ in a CFT to domain $\mathcal{D}$ using only information about the four-point function itself. The OS construction can extend $G_4$ (in fact any $G_n$) to domain $\mathcal{D}$ in a general QFT. But the price to pay is that analytic continuation involves infinitely many steps and needs information about higher-point functions \cite{Glaser1974,osterwalder1975}.

\subsubsection{Domain of analyticity: Wightman axioms + conformal invariance}\label{section:wightmanconformal}
Let us summarize how the domains of Wightman functions are derived in Wightman QFT. We first use the temperedness property (W1), translational invariance (W2) and the spectral condition (W4) to show that the reduced  Wightman distribution $W_n$ is an analytic function in the forward tube $I_n$. Then we use the Lorentz invariance (W2) to show that $W_n$ has analytic continuation to the ``extended tube" $\tilde{I}_n$, which includes the set of Jost points. Finally we use the microscopic causality condition (W5) to show that $\mathcal{W}_{n+1}$ has analytic continuation to all configurations with totally space-like separations.

The unitarity condition (W3) is not involved in the above argument, only the conditions (W1), (W2), (W4) and (W5) are used.  

In the rest of thesis we explored the domain of CFT four-point functions by assuming unitarity, Euclidean conformal invariance and OPE (not assuming Wightman axioms). Here we would like to discuss a related but different situation:
\begin{itemize}
	\item What is the domain of the four-point function if we only assume Wightman axioms and conformal invariance (not assuming OPE)?
\end{itemize}
We want to emphasize that global conformal invariance does not hold for general CFT in $\bbR{d-1,1}$ because Lorentzian special conformal transformations may violate causal orderings. The precise meaning of conformal invariance here is the Euclidean global conformal invariance: we Wick rotate Wightman functions to the Euclidean signature, then the corresponding Euclidean correlation functions are invariant under Euclidean global conformal transformations. This assumption is called \emph{weak conformal invariance}\cite{Luscher:1974ez}.

It is obvious that the Wightman function $\mathcal{W}_4$ is analytic in $\mathcal{U}_4$ (as discussed in section \ref{section:wightmandomain}). In section \ref{section:wightmandomain}, a crucial step is to extend the real Poincar\'e invariance to complex Poincar\'e invariance. Then the reduced Wightman function $W_3$ has analytic continuation from the forward tube $I_3$ to the extended forward tube $\tilde{I}_3$. Now that we assume weak conformal invariance, given any Euclidean conformal transformation $g$ and Euclidean configuration $C=(x_1,x_2,x_3,x_4)$, we have
\begin{equation}\label{euclconformal}
	\begin{split}
		\mathcal{W}_4(C)=&\Omega(x_1)^\Delta\Omega(x_2)^\Delta\Omega(x_3)^\Delta\Omega(x_4)^\Delta \mathcal{W}_4(g\cdot C), \\
		g\cdot C=&(g\cdot x_1,g\cdot x_2,g\cdot x_3,g\cdot x_4), \\
	\end{split}
\end{equation}
where $\Omega(x)$ is the scaling factor of the conformal transformation. It is not hard to show that given any configuration $C$ (which can be non-Euclidean) in the domain of $\mathcal{W}_4$, eq. (\ref{euclconformal}) holds for $g$ in a neighbourhood of the identity element in the Euclidean conformal group (this neighbourhood depends on configuration). Then we can show that for a fixed $C$, eq. (\ref{euclconformal}) holds not only in a neighbourhood of identity element in the Euclidean conformal group, but also in a neighourhood in the complex conformal group.\footnote{The complex conformal group is generated by translations $x\rightarrow x+a$, rotations $x\rightarrow \exp\left[\alpha^{\mu\nu} M_{\mu\nu}\right]x$, dilatations $x\rightarrow e^{\tau}x$, special conformal translations $x\rightarrow {x^\prime}^\mu=\frac{x^\mu-x^2b^\mu}{1-2b\cdot x+b^2x^2}$ with complex parameters.} Therefore, by using (\ref{euclconformal}) with $g$ in the complex conformal group, we can extend $\mathcal{W}_4$ to a bigger domain than $\mathcal{U}_4$ (recall definition (\ref{def:domainpermutation})).\footnote{This can be called conformal extension of Bargmann-Hall-Wightman theory. We were unable to find this idea in prior literature.} We say that configurations $C=(x_1,x_2,x_3,x_4),C^\prime=(x_1^\prime,x_2^\prime,x_3^\prime,x_4^\prime)$ are conformally equivalent if there exists a path $g(s)$ in the complex conformal group such that
\begin{itemize}
	\item $g(0)=$id, and $g(1)\cdot C_1=C_2$.
	\item The scaling factors $\Omega(x_1),\Omega(x_2),\Omega(x_3),\Omega(x_4)$ along $g(s)$ do not go to $0$ or $\infty$.
\end{itemize} 
We define
\begin{equation}
	\begin{split}
		\mathcal{U}_4^c=\left\{C=(x_1,x_2,x_3,x_4)\in\mathbb{C}^{4d}\ \big{|}\ C\ \mathrm{is\ conformally\ equivalent\ to\ some\ C^\prime\in\mathcal{U}_4}\right\},
	\end{split}
\end{equation}
where the superscript ``c" in $\mathcal{U}_4^c$ means ``conformal". Naively, one may expect that $\mathcal{W}_4$ has analytic continuation to $\mathcal{U}_4^c$ by (\ref{euclconformal}).\footnote{As a next step, one could consider the Lorentzian configurations in envelope of holomorphy $H(\mathcal{U}_4^c)$.} However, for each conformally equivalent $C,C^\prime$ pair in $\mathcal{U}_4^c$, the path $g(s)$ in the complex conformal group is not unique, which means choosing different $g(s)$ may give different analytic continuation. In other words, $\mathcal{W}_4$ may not be a single-valued function on $\mathcal{U}_4^c$.

There is one very simple partial case which is guaranteed not to lead to multi-valuedness. It is the case when the whole curve $C(s)=g(s)\cdot C=\fr{x_1(s),x_2(s),x_3(s),x_4(s)}$ has point differences $\fr{x_2(s)-x_1(s),x_3(s)-x_2(s),x_4(s)-x_3(s)}$ in the forward tube $I_3$, except for the end point $g(1)\cdot C$.

In this thesis we do not fully explore the Lorentzian domain of $\mathcal{W}_4$ by using the above method. We left it for future work. Here we only give a simple example which shows that by assuming weak conformal invariance in Wightman QFT, the domain of the Wightman function contains more Lorentzian configurations than totally space-like configurations. 

We would like to show that the following Lorentzian configurations are in the domain of (conformally invariant) $\mathcal{W}_4$:
\begin{equation}\label{ex:totallytimelike}
	\begin{split}
		x_k=(t_k,\mathbf{x}_k),\quad1\rightarrow2\rightarrow3\rightarrow4. \\
	\end{split}
\end{equation}
We act with complex dilatation on (\ref{ex:totallytimelike}):
\begin{equation}\label{totallytimelike:complexdil}
	\begin{split}
		x_k^\prime=e^{i\alpha}x_k.
	\end{split}
\end{equation}
Then the point differences are given by $x_{jk}^\prime=x_{jk}\cos\alpha+ix_{jk}\sin\alpha$. By the causal ordering in (\ref{ex:totallytimelike}), we have $(x_2^\prime-x_1^\prime,x_3^\prime-x_2^\prime,x_4^\prime-x_3^\prime)\in I_3$ when $0<\alpha\leq1$. On the other hand, the scaling factors of complex dilatations (\ref{totallytimelike:complexdil}) are constants: $\Omega(x)=e^{i\alpha}$, which are finite and non-zero. Thus, we can use the above-mentioned simple partial case. We conclude that the Lorentzian configurations in the form of (\ref{ex:totallytimelike}) are in the domain of (conformally invariant) $\mathcal{W}_4$.

\subsubsection{Domain of analyticity: unitarity + conformal invariance + OPE}
Now back to our CFT construction in this thesis. We assumed unitarity, conformal invariance and OPE, but did not assume Wightman axioms. 

With the help of OPE, we are able to control the upper bound of the CFT four-point function more efficiently \cite{Pappadopulo:2012jk}. It seems to be rather difficult to apply the unitarity condition (\ref{wightman:unitarity}) in a general Wightman QFT because it is a non-linear constraint. While, in the expansion (\ref{g:rhoexpansion2}), we are able to use the unitarity condition for CFT four-point functions.

The domain of the Lorentzian CFT four-point function $G_4$ contains all configurations where there exists at least one convergent OPE channel. The results in appendix \ref{appendix:tableopeconvergence} show that the domain of $G_4$ contains much richer set of causal orderings than just the totally space-like causal ordering obtainable from Wightman axioms alone.

One interesting point is that if we act with conformal transformations on the configurations which have point differences in the forward tube $I_3$ (let us call this set $\mathcal{I}_4$), we can get at most Lorentzian configurations whose cross-ratios can be realized by configurations in $\mathcal{I}_4$ because conformal transformations do not change cross-ratios.\footnote{The similar idea has been used to look for the domain of analyticity of the Wightman functions in a general QFT. E.g. G. K\"{a}ll{\'e}n explored the domain of the Wightman four-point function by studying configurations $(x_1,x_2,x_3,x_4)$ such that the Poincar\'e invariants $x_{ij}\cdot x_{kl}$ can be realized by configurations in $\mathcal{I}_4$ \cite{kallen1961analyticitydomain}.} However, if we additionally assume OPE, then it is not necessary that the cross-ratios of the Lorentzian configurations can be realized by configurations in $\mathcal{I}_4$ (the only requirement is that $\abs{\rho},\abs{\bar{\rho}}<1$ in the corresponding OPE channel). It would be interesting to figure out:
\begin{itemize}
	\item can we get more Lorentzian configurations by assuming unitarity + conformal invariance + OPE than by assuming Wightman axioms + conformal invariance?
\end{itemize}
We left it for future work.
\thispagestyle{fancy}
\chapter{Appendices of Part \ref{part:generalization}}
\section{Some properties of 2d conformal
	transformations}\label{appendix:2dconformal}

In this appendix we consider conformal transformations which preserves the 2d
subspace $\mathbb{R}^2 \subset \mathbb{R}^d$.

The 2d Euclidean conformal group is $\tmop{SO} (1, 3) \simeq \tmop{PSL} (2 ;
\mathbb{C})$. To construct the conformal transformation we will use complex
coordinates $(z_k, \bar{z}_k)$:
\begin{equation}
	z_k = x_k^0 + i x_k^1, \qquad \bar{z}_k = x_k^0 + i x_k^1 .
	\label{def:complexcoord}
\end{equation}
The 2d conformal transformations are given by
\begin{equation}
	z_k' = f (z_k), \qquad \bar{z}_k' = \bar{f} (\bar{z}_k),
	\label{2dconftransf}
\end{equation}
where $f$ and $\bar{f}$ are M{\"o}bius transformations. In the Euclidean
region, $f$ and $\bar{f}$ are complex conjugate to each other, which means
\begin{equation}
	[f (z)]^{\ast} = \bar{f} (z^{\ast}) .
\end{equation}
When we analytically continue the correlation functions to complex $x_k$'s,
the conformal transformations, which map $(x_1, x_2, x_3, x_4)$ to
$\rho$-configurations (\ref{config:rho}), belong to the complex conformal
group $\tmop{SO} (1, 3)_{\mathbb{C}}$, where $f$ and $\bar{f}$ are independent
M{\"o}bius transformations.

By (\ref{def:complexcoord}) and (\ref{2dconftransf}) we have $\frac{{\partial
		x'}^{\mu}}{\partial x^{\nu}} = \Omega (x) R^{\mu}_{\quad \nu} (x)$, where
\begin{equation}\label{Jacobian:2d}
	\begin{split}
		\Omega (x_k) = & \sqrt{f' (z_k) \bar{f}' (\bar{z}_k)},\\
		R (x_k)= & \left(\begin{array}{c}
			\cos \theta_k \quad - \sin \theta_k\\
			\sin \theta_k \qquad \cos \theta_k
		\end{array}\right), \qquad e^{i \theta_k} = \sqrt{\frac{f' (z_k)}{\bar{f}'
				(\bar{z}_k)}} .  
	\end{split}
\end{equation}

\subsection{Mapping $(z_1, z_2, z_3, z_4)$ to $(\rho, - \rho, - 1, 1)$ }\label{section:2dconformalmap}

Let us first consider the M{\"o}bius transformation which maps $(z_1, z_2,
z_3, z_4)$ to $(\rho, - \rho, - 1, 1)$. Such an $f$ can be constructed via
composing two intermediate maps:
\begin{equation}
	f = f_2 \circ f_1 .
\end{equation}
$f_1$ maps $(z_1, z_2, z_3, z_4)$ to $(0, z, 1, \infty)$, where $z =
\frac{(z_1 - z_2) (z_3 - z_4)}{(z_1 - z_3) (z_2 - z_4)}$:
\begin{equation}
	f_1 (w) = \dfrac{z_3 - z_4}{z_3 - z_1} \dfrac{w - z_1}{w - z_4} .
\end{equation}
$f_2$ maps $(0, z, 1, \infty)$ to $(\rho, - \rho, - 1, 1)$, where $\rho =
\frac{z}{\left( 1 + \sqrt{1 - z} \right)^2}$:
\begin{equation}
	f_2 (w) = \dfrac{aw + b}{cw + d},
\end{equation}
with
\begin{equation}
	\begin{split}
		a, c = & \dfrac{1}{\sqrt{2}  (1 - z)^{1 / 4}},\\
		b = & - \dfrac{1}{\sqrt{2}} [(1 - z)^{- 1 / 4} - (1 - z)^{1 / 4}],\\
		d = & - \dfrac{1}{\sqrt{2}} [(1 - z)^{- 1 / 4} + (1 - z)^{1 / 4}] . 
	\end{split}
\end{equation}
Then the Jacobian of $f$ is given by
\begin{equation}
	f' (w) = - \dfrac{(z_3 - z_4)  (z_1 - z_4)}{(z_3 - z_1)} \dfrac{2 \sqrt{1 -
			z}}{\left[ \dfrac{z_3 - z_4}{z_3 - z_1} (w - z_1) - \left( 1 + \sqrt{1 - z}
		\right) (w - z_4) \right]^2} .
\end{equation}
In particular
\begin{equation}\label{2d:fprime}
	\begin{split}
		f' (z_1) = & - \dfrac{z_3 - z_4}{(z_3 - z_1)  (z_1 - z_4)} \dfrac{2
			\sqrt{1 - z}}{\left( 1 + \sqrt{1 - z} \right)^2},\\
		f' (z_2) = & - \dfrac{z_3 - z_4}{(z_3 - z_1)  (z_1 - z_4)  (z_2 - z_4)^2}
		\dfrac{2}{\sqrt{1 - z} \left( 1 + \sqrt{1 - z} \right)^2},\\
		f' (z_3)  = & - \dfrac{z_1 - z_4}{(z_3 - z_1)  (z_3 - z_4)}
		\dfrac{2}{\sqrt{1 - z}},\\
		f' (z_4) = & - 2 \dfrac{z_3 - z_1}{(z_3 - z_4)  (z_1 - z_4)} \sqrt{1 - z}
		.  
	\end{split}
\end{equation}
Analogously we can map $(\bar{z}_1, \bar{z}_2, \bar{z}_3, \bar{z}_4)$ to
$(\bar{\rho}, - \bar{\rho}, - 1, 1)$ by a conformal transformation $\bar{f}$
and compute the Jacobian. The results are similar:
\begin{equation}\label{2d:fbarprime}
	\begin{split}
		\bar{f}' (\bar{z}_1) = & - \dfrac{\bar{z}_3 - \bar{z}_4}{(\bar{z}_3 -
			\bar{z}_1)  (\bar{z}_1 - \bar{z}_4)} \dfrac{2 \sqrt{1 - \bar{z}}}{\left( 1 +
			\sqrt{1 - \bar{z}} \right)^2},\\
		\bar{f}' (\bar{z}_2) = & - \dfrac{\bar{z}_3 - \bar{z}_4}{(\bar{z}_3 -
			\bar{z}_1)  (\bar{z}_1 - \bar{z}_4)  (\bar{z}_2 - \bar{z}_4)^2}
		\dfrac{2}{\sqrt{1 - \bar{z}} \left( 1 + \sqrt{1 - \bar{z}} \right)^2},\\
		\bar{f}' (\bar{z}_3) = & - \dfrac{\bar{z}_1 - \bar{z}_4}{(\bar{z}_3 -
			\bar{z}_1)  (\bar{z}_3 - \bar{z}_4)} \dfrac{2}{\sqrt{1 - \bar{z}}},\\
		\bar{f}' (\bar{z}_4) = & - 2 \dfrac{\bar{z}_3 - \bar{z}_1}{(\bar{z}_3 -
			\bar{z}_4)  (\bar{z}_1 - \bar{z}_4)} \sqrt{1 - \bar{z}} . 		
	\end{split}
\end{equation}
Eqs.\,(\ref{def:complexcoord}), (\ref{Jacobian:2d}), (\ref{2d:fprime}) and
(\ref{2d:fbarprime}) imply that in 2d, $\Omega (x_k)$ and $R (x_k)$ are
analytic functions of the complex coordinates $x_i^{\mu}$ (where $i \in \{ 1,
2, 3, 4 \}$ and $\mu \in \{ 0, 1 \}$) in the region
\begin{equation}\label{forwardtube:2d}
	\tmop{Re} (z_1) > \tmop{Re} (z_2) > \tmop{Re} (z_3) > \tmop{Re} (z_4),
	\qquad \tmop{Re} (\bar{z}_1) > \tmop{Re} (\bar{z}_2) > \tmop{Re} (\bar{z}_3)
	> \tmop{Re} (\bar{z}_4) .
\end{equation}

\subsection{Extension to higher dimensions}

We consider 01-plane which is a 2d subspace in $\mathbb{R}^d$. Any 2d
M{\"o}bius transformation $f : \mathbb{R}^2 \longrightarrow \mathbb{R}^2$ can
be naturally extended to a higher dimensional conformal transformation
$\tilde{f} : \mathbb{R}^d \longrightarrow \mathbb{R}^d$ as follows:
\begin{itemize}
	\item $f$ has a unique decomposition $f = \exp (a \cdummy P) \exp (b \cdummy
	K) \exp (\omega^{01} M_{01}) \exp (\lambda D)$ in 2d. Here
	\begin{eqnarray*}
		a & = & (a^0, a^1),\\
		b & = & (b^0, b^1) .
	\end{eqnarray*}
	\item Then $\tilde{f} = \exp (a \cdummy P) \exp (b \cdummy K) \exp
	(\omega^{01} M_{01}) \exp (\lambda D)$ in higher d.
\end{itemize}
An important observation is that

\begin{lemma}
	\label{lemma:extension}Let $f : \mathbb{R}^2 \longrightarrow \mathbb{R}^2$
	be a 2d M{\"o}bius transformation and $\tilde{f} : \mathbb{R}^d
	\longrightarrow \mathbb{R}^d$ be its higher dimensional extension as
	decribed above. Then in the 2d subspace, the scaling factors $\Omega_f (x),
	\tilde{\Omega}_f (x)$ and the rotation matrices $R_f (x), \tilde{R}_f (x)$ induced
	by $f, \tilde{f}$ satisfy the following relations:
	\begin{equation}
		\tilde{\Omega} (x)_f = \Omega_f (x), \qquad \tilde{R}_f (x) =
		\left(\begin{array}{cc}
			\tmxspace R_f (x) & 0\\
			0 & \tmop{Id}_{d - 2}
		\end{array}\right), \qquad \forall x \in \mathbb{R}^2 .
		\label{prop:extension}
	\end{equation}
	where $\tmop{Id}_{d - 2}$ is the $(d - 2) - \tmop{by} - (d - 2)$ identity
	matrix.
\end{lemma}

\begin{proof}
	(\ref{prop:extension}) is trivial for the case when $f$ is generated by
	translations, rotations and dilatations in $\mathbb{R}^2$. So it remains to
	check special conformal transformations, for which it suffices to check
	inversion:
	\begin{equation}
		{x'}^{\mu} = \frac{x^{\mu}}{x^2} . \label{def:inversion}
	\end{equation}
	The Jacobian of (\ref{def:inversion}) is given by
	\begin{equation}
		\frac{{\partial x'}^{\mu}}{\partial x^{\nu}} = \frac{\delta^{\mu}_{\quad
				\nu}}{x^2} - \frac{2 x^{\mu} x_{\nu}}{(x^2)^2} = \frac{1}{x^2} \left(
		\delta^{\mu}_{\quad \nu} - \frac{2 x^{\mu} x_{\nu}}{x^2} \right),
		\label{inversion:jacobian}
	\end{equation}
	which implies that $\Omega_{\tmop{inv}} (x) = \frac{1}{x^2}$ and
	$(R_{\tmop{inv}})^{\mu}_{\quad \nu} (x) = \delta^{\mu}_{\quad \nu} - \frac{2
		x^{\mu} x_{\nu}}{x^2}$. Here $R_{\tmop{inv}}$ belongs to $O (d)$ instead of
	$\tmop{SO} (d)$ because inversion does not belong to the connected conformal
	group $\tmop{SO} (1, d + 1)$.
	
	When $x$ is in the $(01)$-plane, i.e. $x^{\mu} = 0$ for $\mu \nin \{ 0, 1 \}$, we
	see that
	\begin{itemize}
		\item $\Omega_{\tmop{inv}} (x) = \frac{1}{x^2}$ agrees with the scaling
		factor in 2d.
		
		\item When $\mu, \nu \in \{ 0, 1 \}$, $(R_{\tmop{inv}})^{\mu}_{\quad \nu}
		(x) = \delta^{\mu}_{\quad \nu} - \frac{2 x^{\mu} x_{\nu}}{x^2}$ which
		agrees with the rotation matrix in 2d.
		
		\item When $\mu \in \{ 0, 1 \}, \nu \nin \{ 0, 1 \}$ or $\mu \nin \{ 0, 1
		\}, \nu \in \{ 0, 1 \}$, $(R_{\tmop{inv}})^{\mu}_{\quad \nu} (x) = 0.$
		
		\item When $\mu, \nu \in \{ 0, 1 \}$, $(R_{\tmop{inv}})^{\mu}_{\quad \nu}
		= \delta^{\mu}_{\quad \nu}$.
	\end{itemize}
	Therefore the inversion map satisfies (\ref{prop:extension}).
\end{proof}

\begin{remark}
	Lemma \ref{lemma:extension} can be easily generalized to complex conformal
	transformations.
\end{remark}

\subsection{CFT four-point function in the two-dimensional subspace}
We consider the four-point function $G_{1234}^{a_1a_2a_3a_4}(c)=\braket{\mathcal{O}_1^{a_1}(x_1)\mathcal{O}_2^{a_2}(x_2)\mathcal{O}_3^{a_3}(x_3)\mathcal{O}_4^{a_4}(x_4)}$ with $c=(x_1,x_2,x_3,x_4)$ staying in the (01)-plane (i.e.\,$x_k^\mu=0$ for $\mu=2,3,\ldots,d-1$).

By the arguments in the previous two subsections, we can find a $d$-dimensional complex conformal transformation $\tilde{f}$, which extends the two-dimensional conformal transformation $f$ acting on the (01)-plane and maps $c$ to the $\rho$-configuration $c_\rho$, defined in eq.\,(\ref{config:rho}). Then the relation of $G_{1234}^{a_1a_2a_3a_4}(c)$ to $G_{1234}^{b_1b_2b_3b_4}(c_\rho)$ is given in eq.\,(\ref{confinv:spinning}).

By lemma \ref{lemma:extension}, the rotation matrices induced by $\tilde{f}$ is block diagonal: $R_{\tilde{f}}(x_k)=\mathrm{diag}(R_f(x_k)_{2\times2},Id_{d-2})$. By the discussions in section \ref{section:2dconformalmap}, the entries of $R_f(x_k)$ are analytic functions of the (01)-plane coordinates. By eqs.\,(\ref{Jacobian:2d}), (\ref{2d:fprime}) and (\ref{2d:fbarprime}), these entries have power-law bounds in the (01)-plane coordinates. The extra entries of the rotation matrices are either 0 or 1.

The analyticity and power-law bound of $\Omega_{\tilde{f}}(x_k)$ and $G_{1234} (c_{\rho})$ have been reviewed in section \ref{section:scalarrevisit}. Together with the results of the rotation matrices here, we conclude that
\begin{theorem}
In the unitary CFT, given a four-point function $G_{1234}^{a_1a_2a_3a_4}(c)=\braket{\mathcal{O}_1^{a_1}(x_1)\mathcal{O}_2^{a_2}(x_2)\mathcal{O}_3^{a_3}(x_3)\mathcal{O}_4^{a_4}(x_4)}$ with $c=(x_1,x_2,x_3,x_4)$ staying in the (01)-plane, then
\begin{itemize}
	\item[]$\mathrm{(a)}$ $G_{1234}^{a_1a_2a_3a_4}(c)$ (as a function of the (01)-plane coordinates) has analytic continuation to the forward tube $\mathcal{T}_4$, with $x_k$'s in the complex (01)-plane;
	\item[]$\mathrm{(b)}$ $G_{1234}^{a_1a_2a_3a_4}(c)$ satisfies power-law bound
	\begin{equation}
		\begin{split}
			\abs{G_{1234}^{a_1a_2a_3a_4}(c)}\leqslant C\fr{1+\max\limits_{k}\dfrac{1}{\mathrm{Re}(z_k-z_{k+1})}}^\alpha\fr{1+\max\limits_{k}\dfrac{1}{\mathrm{Re}(\bar{z}_k-\bar{z}_{k+1})}}^\alpha\fr{1+\max\limits_{k}|x_k-x_{k+1}|}^\beta,
		\end{split}
	\end{equation}
	for $c$ in this region. Here the coordinates $(z_k,\bar{z}_k)$ are the same as in eq.\,(\ref{def:complexcoord}).
\end{itemize}
\end{theorem}

\backmatter
\pagestyle{empty}
\bibliographystyle{utphys}
\bibliography{bibliography}

\providecommand{\href}[2]{#2}\begingroup\raggedright\begin{thebibliography}{100}

\bibitem{osterwalder1973}
K.~Osterwalder and R.~Schrader, ``{Axioms for Euclidean Green's functions},''
\href{http://dx.doi.org/10.1007/BF01645738}{{\em Commun. Math. Phys.}
  {\bfseries 31} (1973) 83--112}.

\bibitem{osterwalder1975}
K.~{Osterwalder} and R.~{Schrader}, ``{Axioms for Euclidean Green's functions
  II},''
\href{http://dx.doi.org/10.1007/BF01608978}{{\em Commun. Math. Phys.}
  {\bfseries 42} (1975) 281}.

\bibitem{Hartman:2015lfa}
T.~Hartman, S.~Jain, and S.~Kundu, ``{Causality Constraints in Conformal Field
  Theory},'' \href{http://dx.doi.org/10.1007/JHEP05(2016)099}{{\em JHEP}
  {\bfseries 05} (2016) 099},
\href{http://arxiv.org/abs/1509.00014}{{\ttfamily arXiv:1509.00014 [hep-th]}}.

\bibitem{belavin1984infinite}
A.~A. Belavin, A.~M. Polyakov, and A.~B. Zamolodchikov, ``{Infinite Conformal
  Symmetry in Two-Dimensional Quantum Field Theory},''
\href{http://dx.doi.org/10.1016/0550-3213(84)90052-X}{{\em Nucl. Phys.}
  {\bfseries B241} (1984) 333--380}.

\bibitem{francesco1997conformal}
P.~Di~Francesco, P.~Mathieu, and D.~Senechal,
  \href{http://dx.doi.org/10.1007/978-1-4612-2256-9}{{\em {Conformal Field
  Theory}}}.
\newblock Graduate Texts in Contemporary Physics. Springer-Verlag, New York,
1997.
\newblock

\bibitem{Rattazzi:2008pe}
R.~Rattazzi, V.~S. Rychkov, E.~Tonni, and A.~Vichi, ``{Bounding scalar operator
  dimensions in 4D CFT},''
  \href{http://dx.doi.org/10.1088/1126-6708/2008/12/031}{{\em JHEP} {\bfseries
  12} (2008) 031},
\href{http://arxiv.org/abs/0807.0004}{{\ttfamily arXiv:0807.0004 [hep-th]}}.

\bibitem{ElShowk:2012ht}
S.~El-Showk, M.~F. Paulos, D.~Poland, S.~Rychkov, D.~Simmons-Duffin, and
  A.~Vichi, ``{Solving the 3D Ising Model with the Conformal Bootstrap},''
  \href{http://dx.doi.org/10.1103/PhysRevD.86.025022}{{\em Phys. Rev.}
  {\bfseries D86} (2012) 025022},
\href{http://arxiv.org/abs/1203.6064}{{\ttfamily arXiv:1203.6064 [hep-th]}}.

\bibitem{El-Showk:2014dwa}
S.~El-Showk, M.~F. Paulos, D.~Poland, S.~Rychkov, D.~Simmons-Duffin, and
  A.~Vichi, ``{Solving the 3d Ising Model with the Conformal Bootstrap II.
  c-Minimization and Precise Critical Exponents},''
  \href{http://dx.doi.org/10.1007/s10955-014-1042-7}{{\em J. Stat. Phys.}
  {\bfseries 157} (2014) 869},
\href{http://arxiv.org/abs/1403.4545}{{\ttfamily arXiv:1403.4545 [hep-th]}}.

\bibitem{Kos:2014bka}
F.~Kos, D.~Poland, and D.~Simmons-Duffin, ``{Bootstrapping Mixed Correlators in
  the 3D Ising Model},'' \href{http://dx.doi.org/10.1007/JHEP11(2014)109}{{\em
  JHEP} {\bfseries 11} (2014) 109},
\href{http://arxiv.org/abs/1406.4858}{{\ttfamily arXiv:1406.4858 [hep-th]}}.

\bibitem{Simmons-Duffin:2015qma}
D.~Simmons-Duffin, ``{A Semidefinite Program Solver for the Conformal
  Bootstrap},'' \href{http://dx.doi.org/10.1007/JHEP06(2015)174}{{\em JHEP}
  {\bfseries 06} (2015) 174},
\href{http://arxiv.org/abs/1502.02033}{{\ttfamily arXiv:1502.02033 [hep-th]}}.

\bibitem{Kos:2016ysd}
F.~Kos, D.~Poland, D.~Simmons-Duffin, and A.~Vichi, ``{Precision Islands in the
  Ising and $O(N)$ Models},''
  \href{http://dx.doi.org/10.1007/JHEP08(2016)036}{{\em JHEP} {\bfseries 08}
  (2016) 036},
\href{http://arxiv.org/abs/1603.04436}{{\ttfamily arXiv:1603.04436 [hep-th]}}.

\bibitem{Kos:2013tga}
F.~Kos, D.~Poland, and D.~Simmons-Duffin, ``{Bootstrapping the $O(N)$ vector
  models},'' \href{http://dx.doi.org/10.1007/JHEP06(2014)091}{{\em JHEP}
  {\bfseries 06} (2014) 091},
\href{http://arxiv.org/abs/1307.6856}{{\ttfamily arXiv:1307.6856 [hep-th]}}.

\bibitem{Kos:2015mba}
F.~Kos, D.~Poland, D.~Simmons-Duffin, and A.~Vichi, ``{Bootstrapping the O(N)
  Archipelago},'' \href{http://dx.doi.org/10.1007/JHEP11(2015)106}{{\em JHEP}
  {\bfseries 11} (2015) 106},
\href{http://arxiv.org/abs/1504.07997}{{\ttfamily arXiv:1504.07997 [hep-th]}}.

\bibitem{Chester:2019ifh}
S.~M. Chester, W.~Landry, J.~Liu, D.~Poland, D.~Simmons-Duffin, N.~Su, and
  A.~Vichi, ``{Carving out OPE space and precise $O(2)$ model critical
  exponents},''
\href{http://arxiv.org/abs/1912.03324}{{\ttfamily arXiv:1912.03324 [hep-th]}}.

\bibitem{Poland:2018epd}
D.~Poland, S.~Rychkov, and A.~Vichi, ``{The Conformal Bootstrap: Theory,
  Numerical Techniques, and Applications},''
  \href{http://dx.doi.org/10.1103/RevModPhys.91.015002}{{\em Rev. Mod. Phys.}
  {\bfseries 91} (2019) 015002},
\href{http://arxiv.org/abs/1805.04405}{{\ttfamily arXiv:1805.04405 [hep-th]}}.

\bibitem{Hofman:2008ar}
D.~M. Hofman and J.~Maldacena, ``{Conformal collider physics: Energy and charge
  correlations},'' \href{http://dx.doi.org/10.1088/1126-6708/2008/05/012}{{\em
  JHEP} {\bfseries 05} (2008) 012},
\href{http://arxiv.org/abs/0803.1467}{{\ttfamily arXiv:0803.1467 [hep-th]}}.

\bibitem{Komargodski:2012ek}
Z.~Komargodski and A.~Zhiboedov, ``{Convexity and Liberation at Large Spin},''
  \href{http://dx.doi.org/10.1007/JHEP11(2013)140}{{\em JHEP} {\bfseries 11}
  (2013) 140},
\href{http://arxiv.org/abs/1212.4103}{{\ttfamily arXiv:1212.4103 [hep-th]}}.

\bibitem{Fitzpatrick:2012yx}
A.~L. Fitzpatrick, J.~Kaplan, D.~Poland, and D.~Simmons-Duffin, ``{The Analytic
  Bootstrap and AdS Superhorizon Locality},''
  \href{http://dx.doi.org/10.1007/JHEP12(2013)004}{{\em JHEP} {\bfseries 12}
  (2013) 004}, \href{http://arxiv.org/abs/1212.3616}{{\ttfamily arXiv:1212.3616
  [hep-th]}}.

\bibitem{Hartman:2016dxc}
T.~Hartman, S.~Jain, and S.~Kundu, ``{A New Spin on Causality Constraints},''
  \href{http://dx.doi.org/10.1007/JHEP10(2016)141}{{\em JHEP} {\bfseries 10}
  (2016) 141}, \href{http://arxiv.org/abs/1601.07904}{{\ttfamily
  arXiv:1601.07904 [hep-th]}}.

\bibitem{Hartman:2016lgu}
T.~Hartman, S.~Kundu, and A.~Tajdini, ``{Averaged Null Energy Condition from
  Causality},'' \href{http://dx.doi.org/10.1007/JHEP07(2017)066}{{\em JHEP}
  {\bfseries 07} (2017) 066}, \href{http://arxiv.org/abs/1610.05308}{{\ttfamily
  arXiv:1610.05308 [hep-th]}}.

\bibitem{Kologlu:2019bco}
M.~Kologlu, P.~Kravchuk, D.~Simmons-Duffin, and A.~Zhiboedov, ``{Shocks,
  Superconvergence, and a Stringy Equivalence Principle},''
  \href{http://dx.doi.org/10.1007/JHEP11(2020)096}{{\em JHEP} {\bfseries 11}
  (2020) 096}, \href{http://arxiv.org/abs/1904.05905}{{\ttfamily
  arXiv:1904.05905 [hep-th]}}.

\bibitem{Caron-Huot:2017vep}
S.~Caron-Huot, ``{Analyticity in Spin in Conformal Theories},''
  \href{http://dx.doi.org/10.1007/JHEP09(2017)078}{{\em JHEP} {\bfseries 09}
  (2017) 078},
\href{http://arxiv.org/abs/1703.00278}{{\ttfamily arXiv:1703.00278 [hep-th]}}.

\bibitem{Simmons-Duffin:2017nub}
D.~Simmons-Duffin, D.~Stanford, and E.~Witten, ``{A spacetime derivation of the
  Lorentzian OPE inversion formula},''
  \href{http://dx.doi.org/10.1007/JHEP07(2018)085}{{\em JHEP} {\bfseries 07}
  (2018) 085}, \href{http://arxiv.org/abs/1711.03816}{{\ttfamily
  arXiv:1711.03816 [hep-th]}}.

\bibitem{Streater:1989vi}
R.~F. Streater and A.~S. Wightman, {\em {PCT, spin and statistics, and all
  that}}.
\newblock Benjamin, New York,
1964.
\newblock

\bibitem{Luscher:1974ez}
M.~{L\"uscher} and G.~Mack, ``{Global Conformal Invariance in Quantum Field
  Theory},''
\href{http://dx.doi.org/10.1007/BF01608988}{{\em Commun. Math. Phys.}
  {\bfseries 41} (1975) 203--234}.

\bibitem{Mack:1976pa}
G.~Mack, ``{Convergence of Operator Product Expansions on the Vacuum in
  Conformal Invariant Quantum Field Theory},''
\href{http://dx.doi.org/10.1007/BF01609130}{{\em Commun. Math. Phys.}
  {\bfseries 53} (1977) 155}.

\bibitem{Vladimirov}
V.~S. Vladimirov, {\em {Methods of the theory of functions of many complex
  variables}}.
\newblock MIT Press: Cambridge, Massachusetts, 1966.

\bibitem{Rychkov:2017tpc}
J.~Qiao and S.~Rychkov, ``{Cut-touching linear functionals in the conformal
  bootstrap},'' \href{http://dx.doi.org/10.1007/JHEP06(2017)076}{{\em JHEP}
  {\bfseries 06} (2017) 076},
\href{http://arxiv.org/abs/1705.01357}{{\ttfamily arXiv:1705.01357 [hep-th]}}.

\bibitem{ruelle1959}
D.~{Ruelle}, ``Analyticity of wightman functions at completely space-like
  points,'' \href{http://dx.doi.org/10.5169/seals-113000}{{\em Helv. Phys.
  Acta} {\bfseries 32} (1959) 135--137}.

\bibitem{Aharony:1999ti}
O.~Aharony, S.~S. Gubser, J.~M. Maldacena, H.~Ooguri, and Y.~Oz, ``{Large N
  field theories, string theory and gravity},''
  \href{http://dx.doi.org/10.1016/S0370-1573(99)00083-6}{{\em Phys. Rept.}
  {\bfseries 323} (2000) 183--386},
\href{http://arxiv.org/abs/hep-th/9905111}{{\ttfamily arXiv:hep-th/9905111
  [hep-th]}}.

\bibitem{Glaser1974}
V.~Glaser, ``{On the equivalence of the Euclidean and Wightman formulation of
  field theory},'' \href{http://dx.doi.org/10.1007/BF01645941}{{\em Comm. Math.
  Phys.} {\bfseries 37} no.~4, (1974) 257--272}.

\bibitem{Mazac:2016qev}
D.~Mazac, ``{Analytic bounds and emergence of AdS$_{2}$ physics from the
  conformal bootstrap},'' \href{http://dx.doi.org/10.1007/JHEP04(2017)146}{{\em
  JHEP} {\bfseries 04} (2017) 146},
\href{http://arxiv.org/abs/1611.10060}{{\ttfamily arXiv:1611.10060 [hep-th]}}.

\bibitem{Mazac:2018mdx}
D.~Mazac and M.~F. Paulos, ``{The analytic functional bootstrap. Part I: 1D
  CFTs and 2D S-matrices},''
  \href{http://dx.doi.org/10.1007/JHEP02(2019)162}{{\em JHEP} {\bfseries 02}
  (2019) 162},
\href{http://arxiv.org/abs/1803.10233}{{\ttfamily arXiv:1803.10233 [hep-th]}}.

\bibitem{Mazac:2018ycv}
D.~Mazac and M.~F. Paulos, ``{The analytic functional bootstrap. Part II.
  Natural bases for the crossing equation},''
  \href{http://dx.doi.org/10.1007/JHEP02(2019)163}{{\em JHEP} {\bfseries 02}
  (2019) 163},
\href{http://arxiv.org/abs/1811.10646}{{\ttfamily arXiv:1811.10646 [hep-th]}}.

\bibitem{Kaviraj:2018tfd}
A.~Kaviraj and M.~F. Paulos, ``{The Functional Bootstrap for Boundary CFT},''
\href{http://arxiv.org/abs/1812.04034}{{\ttfamily arXiv:1812.04034 [hep-th]}}.

\bibitem{Mazac:2018biw}
D.~Mazáč, L.~Rastelli, and X.~Zhou, ``{An Analytic Approach to BCFT$_d$},''
  \href{http://dx.doi.org/10.1007/JHEP12(2019)004}{{\em JHEP} {\bfseries 12}
  (2019) 004},
\href{http://arxiv.org/abs/1812.09314}{{\ttfamily arXiv:1812.09314 [hep-th]}}.

\bibitem{Hartman:2019pcd}
T.~Hartman, D.~Mazáč, and L.~Rastelli, ``{Sphere Packing and Quantum
  Gravity},''
\href{http://arxiv.org/abs/1905.01319}{{\ttfamily arXiv:1905.01319 [hep-th]}}.

\bibitem{Paulos:2019gtx}
M.~F. Paulos, ``{Analytic Functional Bootstrap for CFTs in $d\ge 1$},''
\href{http://arxiv.org/abs/1910.08563}{{\ttfamily arXiv:1910.08563 [hep-th]}}.

\bibitem{Mazac:2019shk}
D.~Mazáč, L.~Rastelli, and X.~Zhou, ``{A Basis of Analytic Functionals for
  CFTs in General Dimension},''
\href{http://arxiv.org/abs/1910.12855}{{\ttfamily arXiv:1910.12855 [hep-th]}}.

\bibitem{Kravchuk:2021kwe}
P.~Kravchuk, J.~Qiao, and S.~Rychkov, ``{Distributions in CFT II. Minkowski
  Space},'' \href{http://arxiv.org/abs/2104.02090}{{\ttfamily arXiv:2104.02090
  [hep-th]}}.

\bibitem{paper3}
P.~Kravchuk, J.~Qiao, and S.~Rychkov, ``{Distributions in CFT IV. Lorentzian
  Cylinder},''. work in progress.

\bibitem{Bissi:2019kkx}
A.~Bissi, P.~Dey, and T.~Hansen, ``{Dispersion Relation for CFT Four-Point
  Functions},''
\href{http://arxiv.org/abs/1910.04661}{{\ttfamily arXiv:1910.04661 [hep-th]}}.

\bibitem{Pappadopulo:2012jk}
D.~Pappadopulo, S.~Rychkov, J.~Espin, and R.~Rattazzi, ``{OPE Convergence in
  Conformal Field Theory},''
  \href{http://dx.doi.org/10.1103/PhysRevD.86.105043}{{\em Phys. Rev.}
  {\bfseries D86} (2012) 105043},
\href{http://arxiv.org/abs/1208.6449}{{\ttfamily arXiv:1208.6449 [hep-th]}}.

\bibitem{Hogervorst:2013sma}
M.~Hogervorst and S.~Rychkov, ``{Radial Coordinates for Conformal Blocks},''
  \href{http://dx.doi.org/10.1103/PhysRevD.87.106004}{{\em Phys. Rev.}
  {\bfseries D87} (2013) 106004},
\href{http://arxiv.org/abs/1303.1111}{{\ttfamily arXiv:1303.1111 [hep-th]}}.

\bibitem{Maldacena:2015iua}
J.~Maldacena, D.~Simmons-Duffin, and A.~Zhiboedov, ``{Looking for a bulk
  point},'' \href{http://dx.doi.org/10.1007/JHEP01(2017)013}{{\em JHEP}
  {\bfseries 01} (2017) 013},
\href{http://arxiv.org/abs/1509.03612}{{\ttfamily arXiv:1509.03612 [hep-th]}}.

\bibitem{Cornalba:2007fs}
L.~Cornalba, ``{Eikonal methods in AdS/CFT: Regge theory and multi-reggeon
  exchange},''
\href{http://arxiv.org/abs/0710.5480}{{\ttfamily arXiv:0710.5480 [hep-th]}}.

\bibitem{Cornalba:2008qf}
L.~Cornalba, M.~S. Costa, and J.~Penedones, ``{Eikonal Methods in AdS/CFT: BFKL
  Pomeron at Weak Coupling},''
  \href{http://dx.doi.org/10.1088/1126-6708/2008/06/048}{{\em JHEP} {\bfseries
  06} (2008) 048},
\href{http://arxiv.org/abs/0801.3002}{{\ttfamily arXiv:0801.3002 [hep-th]}}.

\bibitem{Costa:2012cb}
M.~S. Costa, V.~Goncalves, and J.~Penedones, ``{Conformal Regge theory},''
  \href{http://dx.doi.org/10.1007/JHEP12(2012)091}{{\em JHEP} {\bfseries 12}
  (2012) 091},
\href{http://arxiv.org/abs/1209.4355}{{\ttfamily arXiv:1209.4355 [hep-th]}}.

\bibitem{Tillmann1961}
H.~G. Tillmann, ``{Darstellung der Schwartzschen Distributionen durch
  analytische Funktionen},'' \href{http://dx.doi.org/10.1007/BF01180167}{{\em
  Mathematische Zeitschrift} {\bfseries 77} no.~1, (1961) 106--124}.

\bibitem{RealSubmanifolds}
M.~S. Baouendi, P.~Ebenfelt, and L.~P. Rothschild,
  \href{http://dx.doi.org/10.1515/9781400883967}{{\em Real submanifolds in
  complex space and their mappings}}, vol.~47 of {\em Princeton Mathematical
  Series}.
\newblock Princeton University Press, Princeton, NJ, 1999.

\bibitem{Kravchuk:2016qvl}
P.~Kravchuk and D.~Simmons-Duffin, ``{Counting Conformal Correlators},''
  \href{http://dx.doi.org/10.1007/JHEP02(2018)096}{{\em JHEP} {\bfseries 02}
  (2018) 096},
\href{http://arxiv.org/abs/1612.08987}{{\ttfamily arXiv:1612.08987 [hep-th]}}.

\bibitem{Karateev:2019pvw}
D.~Karateev, P.~Kravchuk, M.~Serone, and A.~Vichi, ``{Fermion Conformal
  Bootstrap in 4d},'' \href{http://dx.doi.org/10.1007/JHEP06(2019)088}{{\em
  JHEP} {\bfseries 06} (2019) 088},
\href{http://arxiv.org/abs/1902.05969}{{\ttfamily arXiv:1902.05969 [hep-th]}}.

\bibitem{Simmons-Duffin:2016wlq}
D.~Simmons-Duffin, ``{The Lightcone Bootstrap and the Spectrum of the 3d Ising
  CFT},'' \href{http://dx.doi.org/10.1007/JHEP03(2017)086}{{\em JHEP}
  {\bfseries 03} (2017) 086},
\href{http://arxiv.org/abs/1612.08471}{{\ttfamily arXiv:1612.08471 [hep-th]}}.

\bibitem{Haag:1963dh}
R.~Haag and D.~Kastler, ``{An Algebraic approach to quantum field theory},''
  \href{http://dx.doi.org/10.1063/1.1704187}{{\em J. Math. Phys.} {\bfseries 5}
  (1964) 848--861}.

\bibitem{Haag:1992hx}
R.~Haag, {\em {Local quantum physics: Fields, particles, algebras}}.
\newblock Springer, 1992.

\bibitem{Zamolodchikov:1978xm}
A.~B. Zamolodchikov and A.~B. Zamolodchikov, ``{Factorized S-Matrices in
  Two-Dimensions as the Exact Solutions of Certain Relativistic Quantum Field
  Models},'' \href{http://dx.doi.org/10.1016/0003-4916(79)90391-9}{{\em Annals
  Phys.} {\bfseries 120} (1979) 253--291}.

\bibitem{Chester:2020iyt}
S.~M. Chester, W.~Landry, J.~Liu, D.~Poland, D.~Simmons-Duffin, N.~Su, and
  A.~Vichi, ``{Bootstrapping Heisenberg Magnets and their Cubic Instability},''
  \href{http://arxiv.org/abs/2011.14647}{{\ttfamily arXiv:2011.14647
  [hep-th]}}.

\bibitem{Paulos:2016fap}
M.~F. Paulos, J.~Penedones, J.~Toledo, B.~C. van Rees, and P.~Vieira, ``{The
  S-matrix bootstrap. Part I: QFT in AdS},''
  \href{http://dx.doi.org/10.1007/JHEP11(2017)133}{{\em JHEP} {\bfseries 11}
  (2017) 133},
\href{http://arxiv.org/abs/1607.06109}{{\ttfamily arXiv:1607.06109 [hep-th]}}.

\bibitem{Paulos:2016but}
M.~F. Paulos, J.~Penedones, J.~Toledo, B.~C. van Rees, and P.~Vieira, ``{The
  S-matrix bootstrap II: two dimensional amplitudes},''
  \href{http://dx.doi.org/10.1007/JHEP11(2017)143}{{\em JHEP} {\bfseries 11}
  (2017) 143},
\href{http://arxiv.org/abs/1607.06110}{{\ttfamily arXiv:1607.06110 [hep-th]}}.

\bibitem{Paulos:2017fhb}
M.~F. Paulos, J.~Penedones, J.~Toledo, B.~C. van Rees, and P.~Vieira, ``{The
  S-matrix bootstrap. Part III: higher dimensional amplitudes},''
  \href{http://dx.doi.org/10.1007/JHEP12(2019)040}{{\em JHEP} {\bfseries 12}
  (2019) 040},
\href{http://arxiv.org/abs/1708.06765}{{\ttfamily arXiv:1708.06765 [hep-th]}}.

\bibitem{Cordova:2018uop}
L.~Cordova and P.~Vieira, ``{Adding flavour to the S-matrix bootstrap},''
  \href{http://dx.doi.org/10.1007/JHEP12(2018)063}{{\em JHEP} {\bfseries 12}
  (2018) 063},
\href{http://arxiv.org/abs/1805.11143}{{\ttfamily arXiv:1805.11143 [hep-th]}}.

\bibitem{Guerrieri:2019rwp}
A.~L. Guerrieri, J.~Penedones, and P.~Vieira, ``{Bootstrapping QCD Using Pion
  Scattering Amplitudes},''
  \href{http://dx.doi.org/10.1103/PhysRevLett.122.241604}{{\em Phys. Rev.
  Lett.} {\bfseries 122} no.~24, (2019) 241604},
\href{http://arxiv.org/abs/1810.12849}{{\ttfamily arXiv:1810.12849 [hep-th]}}.

\bibitem{EliasMiro:2019kyf}
J.~Elias~Miro, A.~L. Guerrieri, A.~Hebbar, J.~Penedones, and P.~Vieira, ``{Flux
  Tube S-matrix Bootstrap},''
  \href{http://dx.doi.org/10.1103/PhysRevLett.123.221602}{{\em Phys. Rev.
  Lett.} {\bfseries 123} no.~22, (2019) 221602},
\href{http://arxiv.org/abs/1906.08098}{{\ttfamily arXiv:1906.08098 [hep-th]}}.

\bibitem{Cordova:2019lot}
L.~Cordova, Y.~He, M.~Kruczenski, and P.~Vieira, ``{The O(N) S-matrix
  Monolith},'' \href{http://dx.doi.org/10.1007/JHEP04(2020)142}{{\em JHEP}
  {\bfseries 04} (2020) 142},
\href{http://arxiv.org/abs/1909.06495}{{\ttfamily arXiv:1909.06495 [hep-th]}}.

\bibitem{Karateev:2019ymz}
D.~Karateev, S.~Kuhn, and J.~Penedones, ``{Bootstrapping Massive Quantum Field
  Theories},'' \href{http://dx.doi.org/10.1007/JHEP07(2020)035}{{\em JHEP}
  {\bfseries 07} (2020) 035},
\href{http://arxiv.org/abs/1912.08940}{{\ttfamily arXiv:1912.08940 [hep-th]}}.

\bibitem{Correia:2020xtr}
M.~Correia, A.~Sever, and A.~Zhiboedov, ``{An Analytical Toolkit for the
  S-matrix Bootstrap},''
\href{http://arxiv.org/abs/2006.08221}{{\ttfamily arXiv:2006.08221 [hep-th]}}.

\bibitem{Guerrieri:2020bto}
A.~Guerrieri, J.~Penedones, and P.~Vieira, ``{S-matrix Bootstrap for Effective
  Field Theories: Massless Pions},''
  \href{http://arxiv.org/abs/2011.02802}{{\ttfamily arXiv:2011.02802
  [hep-th]}}.

\bibitem{Hebbar:2020ukp}
A.~Hebbar, D.~Karateev, and J.~Penedones, ``{Spinning S-matrix Bootstrap in
  4d},'' \href{http://arxiv.org/abs/2011.11708}{{\ttfamily arXiv:2011.11708
  [hep-th]}}.

\bibitem{Tourkine:2021fqh}
P.~Tourkine and A.~Zhiboedov, ``{Scattering from production in 2d},''
  \href{http://arxiv.org/abs/2101.05211}{{\ttfamily arXiv:2101.05211
  [hep-th]}}.

\bibitem{Sinha:2020win}
A.~Sinha and A.~Zahed, ``{Crossing Symmetric Dispersion Relations in QFTs},''
  \href{http://arxiv.org/abs/2012.04877}{{\ttfamily arXiv:2012.04877
  [hep-th]}}.

\bibitem{Haldar:2021rri}
P.~Haldar, A.~Sinha, and A.~Zahed, ``{Quantum field theory and the Bieberbach
  conjecture},'' \href{http://arxiv.org/abs/2103.12108}{{\ttfamily
  arXiv:2103.12108 [hep-th]}}.

\bibitem{He:2021eqn}
Y.~He and M.~Kruczenski, ``{S-matrix bootstrap in 3+1 dimensions:
  regularization and dual convex problem},''
  \href{http://arxiv.org/abs/2103.11484}{{\ttfamily arXiv:2103.11484
  [hep-th]}}.

\bibitem{paper1}
P.~Kravchuk, J.~Qiao, and S.~Rychkov, ``{Distributions in CFT. Part I.
  Cross-ratio space},'' \href{http://dx.doi.org/10.1007/JHEP05(2020)137}{{\em
  JHEP} {\bfseries 05} (2020) 137},
\href{http://arxiv.org/abs/2001.08778}{{\ttfamily arXiv:2001.08778 [hep-th]}}.

\bibitem{Qiao:2020bcs}
J.~Qiao, ``{Classification of Convergent OPE Channels for Lorentzian CFT
  Four-Point Functions},''
\href{http://arxiv.org/abs/2005.09105}{{\ttfamily arXiv:2005.09105 [hep-th]}}.

\bibitem{Rychkov:2020rcd}
S.~Rychkov, ``{3D Ising Model: a view from the Conformal Bootstrap Island},''
  \href{http://dx.doi.org/10.5802/crphys.23}{{\em Comptes Rendus Physique}
  {\bfseries 21} no.~2, (2020) 185--198},
  \href{http://arxiv.org/abs/2007.14315}{{\ttfamily arXiv:2007.14315
  [math-ph]}}.

\bibitem{Schwarz:2015fva}
A.~Schwarz, ``{Axiomatic conformal theory in dimensions $> 2$ and AdS/CT
  correspondence},'' \href{http://dx.doi.org/10.1007/s11005-016-0866-2}{{\em
  Lett. Math. Phys.} {\bfseries 106} no.~9, (2016) 1181--1197},
\href{http://arxiv.org/abs/1509.08064}{{\ttfamily arXiv:1509.08064 [hep-th]}}.

\bibitem{tillmann2004}
G.~Segal, {\em The definition of CFT},
  \href{http://dx.doi.org/10.1017/CBO9780511526398.019}{p.~432–575}.
\newblock Cambridge Univ. Press, 2004.

\bibitem{paper2a}
P.~Kravchuk, J.~Qiao, and S.~Rychkov, ``{Distributions in CFT III. Spinning
  Fields in Minkowski space},''. work in progress.

\bibitem{simon1974}
B.~Simon, {\em {The $P(\phi)_2$ Euclidean (quantum) Theory}}.
\newblock Princeton University Press, 1974.

\bibitem{Casini:2010bf}
H.~Casini, ``{Wedge reflection positivity},''
  \href{http://dx.doi.org/10.1088/1751-8113/44/43/435202}{{\em J. Phys. A}
  {\bfseries 44} (2011) 435202},
  \href{http://arxiv.org/abs/1009.3832}{{\ttfamily arXiv:1009.3832 [hep-th]}}.

\bibitem{Minwalla:1997ka}
S.~Minwalla, ``{Restrictions Imposed by Superconformal Invariance on Quantum
  Field Theories},'' {\em Adv. Theor. Math. Phys.} {\bfseries 2} (1998)
  781--846,
\href{http://arxiv.org/abs/hep-th/9712074}{{\ttfamily arXiv:hep-th/9712074}}.

\bibitem{EPFL}
S.~Rychkov, \href{http://dx.doi.org/10.1007/978-3-319-43626-5}{{\em {EPFL
  Lectures on Conformal Field Theory in $D\ge 3$ Dimensions}}}.
\newblock SpringerBriefs in Physics. Springer, 2016.
\newblock
\href{http://arxiv.org/abs/1601.05000}{{\ttfamily arXiv:1601.05000 [hep-th]}}.
\newblock

\bibitem{Yamazaki:2016vqi}
M.~Yamazaki, ``{Comments on Determinant Formulas for General CFTs},''
  \href{http://dx.doi.org/10.1007/JHEP10(2016)035}{{\em JHEP} {\bfseries 10}
  (2016) 035},
\href{http://arxiv.org/abs/1601.04072}{{\ttfamily arXiv:1601.04072 [hep-th]}}.

\bibitem{Penedones:2015aga}
J.~Penedones, E.~Trevisani, and M.~Yamazaki, ``{Recursion Relations for
  Conformal Blocks},'' \href{http://dx.doi.org/10.1007/JHEP09(2016)070}{{\em
  JHEP} {\bfseries 09} (2016) 070},
\href{http://arxiv.org/abs/1509.00428}{{\ttfamily arXiv:1509.00428 [hep-th]}}.

\bibitem{jantzen_kontravariante_1977}
J.~C. Jantzen, ``Kontravariante {Formen} auf induzierten {Darstellungen}
  halbeinfacher {Lie}-{Algebren},''
  \href{http://dx.doi.org/10.1007/BF01391218}{{\em Mathematische Annalen}
  {\bfseries 226} (1977) 53--65}.

\bibitem{Dymarsky:2017xzb}
A.~Dymarsky, J.~Penedones, E.~Trevisani, and A.~Vichi, ``{Charting the space of
  3D CFTs with a continuous global symmetry},''
  \href{http://dx.doi.org/10.1007/JHEP05(2019)098}{{\em JHEP} {\bfseries 05}
  (2019) 098},
\href{http://arxiv.org/abs/1705.04278}{{\ttfamily arXiv:1705.04278 [hep-th]}}.

\bibitem{Dymarsky:2017yzx}
A.~Dymarsky, F.~Kos, P.~Kravchuk, D.~Poland, and D.~Simmons-Duffin, ``{The 3d
  Stress-Tensor Bootstrap},''
  \href{http://dx.doi.org/10.1007/JHEP02(2018)164}{{\em JHEP} {\bfseries 02}
  (2018) 164},
\href{http://arxiv.org/abs/1708.05718}{{\ttfamily arXiv:1708.05718 [hep-th]}}.

\bibitem{Faulkner:2016mzt}
T.~Faulkner, R.~G. Leigh, O.~Parrikar, and H.~Wang, ``{Modular Hamiltonians for
  Deformed Half-Spaces and the Averaged Null Energy Condition},''
  \href{http://dx.doi.org/10.1007/JHEP09(2016)038}{{\em JHEP} {\bfseries 09}
  (2016) 038}, \href{http://arxiv.org/abs/1605.08072}{{\ttfamily
  arXiv:1605.08072 [hep-th]}}.

\bibitem{Kravchuk:2018htv}
P.~Kravchuk and D.~Simmons-Duffin, ``{Light-ray operators in conformal field
  theory},'' \href{http://dx.doi.org/10.1007/JHEP11(2018)102}{{\em JHEP}
  {\bfseries 11} (2018) 102}, \href{http://arxiv.org/abs/1805.00098}{{\ttfamily
  arXiv:1805.00098 [hep-th]}}.

\bibitem{brezis}
H.~Brezis, {\em {Functional Analysis, Sobolev Spaces and Partial Differential
  Equations}}.
\newblock Springer, 2010.

\bibitem{Gillioz:2018mto}
M.~Gillioz, ``{Momentum-space conformal blocks on the light cone},''
  \href{http://dx.doi.org/10.1007/JHEP10(2018)125}{{\em JHEP} {\bfseries 10}
  (2018) 125}, \href{http://arxiv.org/abs/1807.07003}{{\ttfamily
  arXiv:1807.07003 [hep-th]}}.

\bibitem{Gillioz:2019lgs}
M.~Gillioz, ``{Conformal 3-point functions and the Lorentzian OPE in momentum
  space},'' \href{http://dx.doi.org/10.1007/s00220-020-03836-8}{{\em Commun.
  Math. Phys.} {\bfseries 379} no.~1, (2020) 227--259},
  \href{http://arxiv.org/abs/1909.00878}{{\ttfamily arXiv:1909.00878
  [hep-th]}}.

\bibitem{gelfandshilov}
I.~Gelfand and G.~Shilov, {\em Generalized Functions Vol.1: Properties and
  Operations}.
\newblock Academic Press, 1964.

\bibitem{Jaffe}
A.~M. Jaffee, ``High-energy behavior in quantum field theory. i. strictly
  localizable fields,'' \href{http://dx.doi.org/10.1103/PhysRev.158.1454}{{\em
  Phys. Rev.} {\bfseries 158} (1967) 1454--1461}.

\bibitem{Dolan:2000ut}
F.~A. Dolan and H.~Osborn, ``{Conformal four point functions and the operator
  product expansion},''
  \href{http://dx.doi.org/10.1016/S0550-3213(01)00013-X}{{\em Nucl. Phys.}
  {\bfseries B599} (2001) 459--496},
\href{http://arxiv.org/abs/hep-th/0011040}{{\ttfamily arXiv:hep-th/0011040
  [hep-th]}}.

\bibitem{Dolan:2003hv}
F.~Dolan and H.~Osborn, ``{Conformal partial waves and the operator product
  expansion},'' \href{http://dx.doi.org/10.1016/j.nuclphysb.2003.11.016}{{\em
  Nucl. Phys. B} {\bfseries 678} (2004) 491--507},
  \href{http://arxiv.org/abs/hep-th/0309180}{{\ttfamily arXiv:hep-th/0309180}}.

\bibitem{Osgood}
Wikipedia, ``\href{https://en.wikipedia.org/wiki/Osgood's_lemma}{Osgood's
  lemma}.''.

\bibitem{lecturesSaclay}
S.~Rychkov, ``\href{https://courses.ipht.fr/?q=en/node/226}{Lorentzian methods
  in conformal field theory}.''. Lectures at IPHT Saclay, September-October
  2019 (videos and lecture notes).

\bibitem{KravchukSchwarz-Pick}
P.~Kravchuk, ``{A Schwarz-Pick lemma for analytic functions on the forward
  tube}.''. unpublished.

\bibitem{Costa:2011mg}
M.~S. Costa, J.~Penedones, D.~Poland, and S.~Rychkov, ``{Spinning Conformal
  Correlators},'' \href{http://dx.doi.org/10.1007/JHEP11(2011)071}{{\em JHEP}
  {\bfseries 11} (2011) 071}, \href{http://arxiv.org/abs/1107.3554}{{\ttfamily
  arXiv:1107.3554 [hep-th]}}.

\bibitem{Borchers1964}
H.~J. Borchers, ``{Field operators as $C^\infty$ functions in spacelike
  directions},'' \href{http://dx.doi.org/10.1007/BF02749678}{{\em Il Nuovo
  Cimento} {\bfseries 33} no.~6, (1964) 1600--1613}.

\bibitem{jost1979general}
R.~Jost, {\em {The General Theory of Quantized Fields}}.
\newblock American Mathematical Society, 1979.

\bibitem{Tomozawa}
Y.~Tomozawa, ``{Local Commutativity and the Analytic Continuation of the
  Wightman Function},'' \href{http://dx.doi.org/10.1063/1.1703896}{{\em J.
  Math. Phys.} {\bfseries 4} no.~10, (1963) 1240--1252}.

\bibitem{bogolubov2012general}
N.~Bogolubov, A.~Logunov, A.~Oksak, and I.~Todorov, {\em General Principles of
  Quantum Field Theory}.
\newblock Springer, 2012.

\bibitem{Hogervorst:2013kva}
M.~Hogervorst, H.~Osborn, and S.~Rychkov, ``{Diagonal Limit for Conformal
  Blocks in $d$ Dimensions},''
  \href{http://dx.doi.org/10.1007/JHEP08(2013)014}{{\em JHEP} {\bfseries 08}
  (2013) 014}, \href{http://arxiv.org/abs/1305.1321}{{\ttfamily arXiv:1305.1321
  [hep-th]}}.

\bibitem{KravchukCB}
P.~Kravchuk.
\newblock unpublished.

\bibitem{Gillioz:2020wgw}
M.~Gillioz, ``{Conformal partial waves in momentum space},''
  \href{http://arxiv.org/abs/2012.09825}{{\ttfamily arXiv:2012.09825
  [hep-th]}}.

\bibitem{Mack:1975je}
G.~Mack, ``{All unitary ray representations of the conformal group SU(2,2) with
  positive energy},'' \href{http://dx.doi.org/10.1007/BF01613145}{{\em Commun.
  Math. Phys.} {\bfseries 55} (1977) 1}.

\bibitem{Mackey1}
G.~W. Mackey, {\em {Unitary Group Representations in Physics, Probability and
  Number Theory}}.
\newblock Benjamin-Cummings, 1979.

\bibitem{Vladimirov2}
V.~S. Vladimirov, {\em {Methods of the theory of generalized functions}}.
\newblock CRC Press, 2002.

\bibitem{funcan}
F.~Riesz and B.~Sz.-Nagy, {\em Functional analysis}.
\newblock Frederick Ungar Publishing Co., New York, 1955.
\newblock Translated by Leo F. Boron.

\bibitem{Zinoviev1995}
Y.~M. Zinoviev, ``{Equivalence of Euclidean and Wightman field theories},''
  \href{http://dx.doi.org/10.1007/BF02099461}{{\em Comm. Math. Physics}
  {\bfseries 174} no.~1, (1995) 1--27}.

\bibitem{Post}
Wikipedia,
  ``\href{https://en.wikipedia.org/wiki/Inverse_Laplace_transform\#Post's_inversion_formula}{Post's
  inversion formula}.''.

\bibitem{GJS_os_axiom}
J.~Glimm, A.~Jaffe, and T.~Spencer, ``{The Wightman Axioms and Particle
  Structure in the $P(\phi)_2$ Quantum Field Model},'' {\em Annals of
  Mathematics} {\bfseries 100} no.~3, (1974) 585--632.

\bibitem{Glimm:1981xz}
J.~Glimm and A.~M. Jaffe, {\em {Quantum Physics. A Functional Integral Point of
  View, 2nd ed.}}
\newblock New York, Usa: Springer,
1987.
\newblock

\bibitem{Abdesselam:2016npc}
A.~Abdesselam, ``{A Second-Quantized Kolmogorov-Chentsov Theorem via the
  Operator Product Expansion},''
  \href{http://dx.doi.org/10.1007/s00220-019-03665-4}{{\em Commun. Math. Phys.}
  {\bfseries 376} no.~1, (2020) 555--608},
  \href{http://arxiv.org/abs/1604.05259}{{\ttfamily arXiv:1604.05259
  [math.PR]}}.

\bibitem{Epstein:1966yea}
H.~Epstein, ``{Some analytic properties of scattering amplitudes in quantum
  field theory},'' in {\em {8th Brandeis University Summer Institute in
  Theoretical Physics}: {Particle symmetries and axiomatic field theory}},
  M.~Chretien and S.~Deser, eds., pp.~1--128.
\newblock Gordon and Breach, 1966.

\bibitem{PetrEnvelope}
P.~Kravchuk, ``{Analytic completion of 4pt functions in CFT}.''. unpublished.

\bibitem{Caron-Huot:2020nem}
S.~Caron-Huot and J.~Sandor, ``{Conformal Regge Theory at Finite Boost},''
  \href{http://arxiv.org/abs/2008.11759}{{\ttfamily arXiv:2008.11759
  [hep-th]}}.

\bibitem{Eckmann:1979vq}
J.~P. Eckmann and H.~Epstein, ``{Time Ordered Products and Schwinger
  Functions},'' \href{http://dx.doi.org/10.1007/BF01197509}{{\em Commun. Math.
  Phys.} {\bfseries 64} (1979) 95--130}.

\bibitem{jost1957bemerkung}
R.~Jost, ``A remark on the c.t.p. theorem,''
\href{http://dx.doi.org/10.5169/seals-112824}{{\em Helv. Phys. Acta} {\bfseries
  30} (1957) 409--416}.

\bibitem{dietz1973lightcone}
K.~Dietz, ``{Light-cone singularities in quantum field theory},''
\href{http://dx.doi.org/10.1007/BF02820833}{{\em Nuovo Cim.} {\bfseries A18}
  (1973) 1--16}.

\bibitem{zamolodchikov1987conformal}
A.~B. {Zamolodchikov}, ``{Conformal symmetry in two-dimensional space:
  Recursion representation of conformal block},''
  \href{http://dx.doi.org/10.1007/BF01022967}{{\em Theor. Math. Phys}
  {\bfseries 73} no.~1, (Oct., 1987) 1088--1093}.

\bibitem{ferrara1973tensor}
S.~Ferrara, A.~F. Grillo, and R.~Gatto, ``{Tensor representations of conformal
  algebra and conformally covariant operator product expansion},''
\href{http://dx.doi.org/10.1016/0003-4916(73)90446-6}{{\em Annals Phys.}
  {\bfseries 76} (1973) 161--188}.

\bibitem{polyakov1974nonhamiltonian}
A.~M. Polyakov, ``{Nonhamiltonian approach to conformal quantum field
  theory},'' {\em Zh. Eksp. Teor. Fiz.} {\bfseries 66} (1974) 23--42.
[Sov. Phys. JETP39,9(1974)].

\bibitem{go1974properties}
T.~H. Go, H.~A. Kastrup, and D.~H. Mayer, ``{Properties of dilatations and
  conformal transformations in Minkowski space},''
\href{http://dx.doi.org/10.1016/S0034-4877(74)80006-6}{{\em Rept. Math. Phys.}
  {\bfseries 6} (1974) 395--430}.

\bibitem{lehmann1955formulierung}
H.~Lehmann, K.~Symanzik, and W.~Zimmermann, ``{On the formulation of quantized
  field theories},''
\href{http://dx.doi.org/10.1007/BF02731765}{{\em Nuovo Cim.} {\bfseries 1}
  (1955) 205--225}.

\bibitem{peskin1995introduction}
M.~E. Peskin and D.~V. Schroeder, {\em {An Introduction to quantum field
  theory}}.
\newblock Addison-Wesley, Reading, USA, 1995.
\newblock
\url{http://www.slac.stanford.edu/~mpeskin/QFT.html}.
\newblock

\bibitem{keldysh1965diagram}
L.~V. Keldysh, ``{Diagram technique for nonequilibrium processes},'' {\em Zh.
  Eksp. Teor. Fiz.} {\bfseries 47} (1964) 1515--1527.
[Sov. Phys. JETP20,1018(1965)].

\bibitem{kitaev2014hidden}
A.~Kitaev, ``{Hidden Correlations in the Hawking Radiation and Thermal Noise}
  \href{http://online.itp.ucsb.edu/online/joint98/kitaev/}{$\mathrm{KITP\
  Theory\ Seminar}$},'' 12 feb, 2015.

\bibitem{maldacena2016chaos}
J.~Maldacena, S.~H. Shenker, and D.~Stanford, ``{A bound on chaos},''
  \href{http://dx.doi.org/10.1007/JHEP08(2016)106}{{\em JHEP} {\bfseries 08}
  (2016) 106},
\href{http://arxiv.org/abs/1503.01409}{{\ttfamily arXiv:1503.01409 [hep-th]}}.

\bibitem{aleiner2016microscopic}
I.~L. Aleiner, L.~Faoro, and L.~B. Ioffe, ``{Microscopic model of quantum
  butterfly effect: out-of-time-order correlators and traveling combustion
  waves},'' \href{http://dx.doi.org/10.1016/j.aop.2016.09.006}{{\em Annals
  Phys.} {\bfseries 375} (2016) 378--406},
\href{http://arxiv.org/abs/1609.01251}{{\ttfamily arXiv:1609.01251
  [cond-mat.stat-mech]}}.

\bibitem{bagrets2017power}
D.~Bagrets, A.~Altland, and A.~Kamenev, ``{Power-law out of time order
  correlation functions in the SYK model},''
  \href{http://dx.doi.org/10.1016/j.nuclphysb.2017.06.012}{{\em Nucl. Phys.}
  {\bfseries B921} (2017) 727--752},
\href{http://arxiv.org/abs/1702.08902}{{\ttfamily arXiv:1702.08902
  [cond-mat.str-el]}}.

\bibitem{garttner2017measuring}
M.~Gärttner, J.~G. Bohnet, A.~Safavi-Naini, M.~L. Wall, J.~J. Bollinger, and
  A.~M. Rey, ``{Measuring out-of-time-order correlations and multiple quantum
  spectra in a trapped ion quantum magnet},''
  \href{http://dx.doi.org/10.1038/nphys4119}{{\em Nature Phys.} {\bfseries 13}
  (2017) 781},
\href{http://arxiv.org/abs/1608.08938}{{\ttfamily arXiv:1608.08938
  [quant-ph]}}.

\bibitem{fan2017out}
R.~Fan, P.~Zhang, H.~Shen, and H.~Zhai, ``{Out-of-Time-Order Correlation for
  Many-Body Localization},''
\href{http://arxiv.org/abs/1608.01914}{{\ttfamily arXiv:1608.01914
  [cond-mat.str-el]}}.

\bibitem{haehl2019classification}
F.~M. Haehl, R.~Loganayagam, P.~Narayan, and M.~Rangamani, ``{Classification of
  out-of-time-order correlators},''
  \href{http://dx.doi.org/10.21468/SciPostPhys.6.1.001}{{\em SciPost Phys.}
  {\bfseries 6} no.~1, (2019) 001},
\href{http://arxiv.org/abs/1701.02820}{{\ttfamily arXiv:1701.02820 [hep-th]}}.

\bibitem{segal1971causally}
I.~{Segal}, ``Causally oriented manifolds and groups,''
  \href{http://dx.doi.org/10.1090/S0002-9904-1971-12815-X}{{\em Bull. Amer.
  Math. Soc.} {\bfseries 77} no.~6, (1971) 958--959}.

\bibitem{todorov1973conformal}
I.~T. Todorov, ``{Conformal invariant quantum field theory with anomalous
  dimensions},'' Tech. Rep.
  \href{https://cds.cern.ch/record/967138/files/CM-P00060449.pdf}{CERN-TH-1697},
  CERN, Geneva, 1973.

\bibitem{Gillioz:2019iye}
M.~Gillioz, X.~Lu, M.~A. Luty, and G.~Mikaberidze, ``{Convergent Momentum-Space
  OPE and Bootstrap Equations in Conformal Field Theory},''
  \href{http://dx.doi.org/10.1007/JHEP03(2020)102}{{\em JHEP} {\bfseries 03}
  (2020) 102}, \href{http://arxiv.org/abs/1912.05550}{{\ttfamily
  arXiv:1912.05550 [hep-th]}}.

\bibitem{Hortacsu:1972bw}
M.~Hortacsu, R.~Seiler, and B.~Schroer, ``{Conformal symmetry and
  reverberations},'' \href{http://dx.doi.org/10.1103/PhysRevD.5.2519}{{\em
  Phys. Rev. D} {\bfseries 5} (1972) 2519--2534}.

\bibitem{Schwarz}
\href{http://dx.doi.org/https://doi.org/10.1016/S0079-8169(08)61286-6}{``{Schwarz'
  Lemma and its Many Applications (Chapter VI)},''} in {\em An Introduction to
  Classical Complex Analysis}, R.~B. Burckel, ed., vol.~82 of {\em Pure and
  Applied Mathematics}, pp.~191 -- 217.
\newblock Elsevier, 1979.

\bibitem{Duren}
P.~L. Duren, {\em Univalent functions}.
\newblock Grundlehren der mathematischen Wissenschaften 259. Springer-Verlag,
  1983.

\bibitem{DSDLorentzian}
D.~Simmons-Duffin,
  ``\href{https://gitlab.com/davidsd/lorentzian-cft-notes}{Conformal Field
  Theory in Lorentzian Signature},''. Lectures at TASI 2019.

\bibitem{Cordova:2017zej}
C.~Cordova, J.~Maldacena, and G.~J. Turiaci, ``{Bounds on OPE Coefficients from
  Interference Effects in the Conformal Collider},''
  \href{http://dx.doi.org/10.1007/JHEP11(2017)032}{{\em JHEP} {\bfseries 11}
  (2017) 032}, \href{http://arxiv.org/abs/1710.03199}{{\ttfamily
  arXiv:1710.03199 [hep-th]}}.

\bibitem{Cordova:2017dhq}
C.~Cordova and K.~Diab, ``{Universal Bounds on Operator Dimensions from the
  Average Null Energy Condition},''
  \href{http://dx.doi.org/10.1007/JHEP02(2018)131}{{\em JHEP} {\bfseries 02}
  (2018) 131}, \href{http://arxiv.org/abs/1712.01089}{{\ttfamily
  arXiv:1712.01089 [hep-th]}}.

\bibitem{Qiao:2017xif}
J.~Qiao and S.~Rychkov, ``{A tauberian theorem for the conformal bootstrap},''
  \href{http://dx.doi.org/10.1007/JHEP12(2017)119}{{\em JHEP} {\bfseries 12}
  (2017) 119}, \href{http://arxiv.org/abs/1709.00008}{{\ttfamily
  arXiv:1709.00008 [hep-th]}}.

\bibitem{Liu:2020tpf}
J.~Liu, D.~Meltzer, D.~Poland, and D.~Simmons-Duffin, ``{The Lorentzian
  inversion formula and the spectrum of the 3d O(2) CFT},''
  \href{http://dx.doi.org/10.1007/JHEP09(2020)115}{{\em JHEP} {\bfseries 09}
  (2020) 115}, \href{http://arxiv.org/abs/2007.07914}{{\ttfamily
  arXiv:2007.07914 [hep-th]}}.

\bibitem{Caron-Huot:2020ouj}
S.~Caron-Huot, Y.~Gobeil, and Z.~Zahraee, ``{The leading trajectory in the 2+1D
  Ising CFT},'' \href{http://arxiv.org/abs/2007.11647}{{\ttfamily
  arXiv:2007.11647 [hep-th]}}.

\bibitem{Hofman:2016awc}
D.~M. Hofman, D.~Li, D.~Meltzer, D.~Poland, and F.~Rejon-Barrera, ``{A Proof of
  the Conformal Collider Bounds},''
  \href{http://dx.doi.org/10.1007/JHEP06(2016)111}{{\em JHEP} {\bfseries 06}
  (2016) 111}, \href{http://arxiv.org/abs/1603.03771}{{\ttfamily
  arXiv:1603.03771 [hep-th]}}.

\bibitem{Polchinski:1999yd}
J.~Polchinski, L.~Susskind, and N.~Toumbas, ``{Negative energy, superluminosity
  and holography},'' \href{http://dx.doi.org/10.1103/PhysRevD.60.084006}{{\em
  Phys. Rev. D} {\bfseries 60} (1999) 084006},
  \href{http://arxiv.org/abs/hep-th/9903228}{{\ttfamily arXiv:hep-th/9903228}}.

\bibitem{Gary:2009ae}
M.~Gary, S.~B. Giddings, and J.~Penedones, ``{Local bulk S-matrix elements and
  CFT singularities},''
  \href{http://dx.doi.org/10.1103/PhysRevD.80.085005}{{\em Phys. Rev. D}
  {\bfseries 80} (2009) 085005},
  \href{http://arxiv.org/abs/0903.4437}{{\ttfamily arXiv:0903.4437 [hep-th]}}.

\bibitem{Heemskerk:2009pn}
I.~Heemskerk, J.~Penedones, J.~Polchinski, and J.~Sully, ``{Holography from
  Conformal Field Theory},''
  \href{http://dx.doi.org/10.1088/1126-6708/2009/10/079}{{\em JHEP} {\bfseries
  10} (2009) 079}, \href{http://arxiv.org/abs/0907.0151}{{\ttfamily
  arXiv:0907.0151 [hep-th]}}.

\bibitem{Penedones:2010ue}
J.~Penedones, ``{Writing CFT correlation functions as AdS scattering
  amplitudes},'' \href{http://dx.doi.org/10.1007/JHEP03(2011)025}{{\em JHEP}
  {\bfseries 03} (2011) 025}, \href{http://arxiv.org/abs/1011.1485}{{\ttfamily
  arXiv:1011.1485 [hep-th]}}.

\bibitem{Okuda:2010ym}
T.~Okuda and J.~Penedones, ``{String scattering in flat space and a scaling
  limit of Yang-Mills correlators},''
  \href{http://dx.doi.org/10.1103/PhysRevD.83.086001}{{\em Phys. Rev. D}
  {\bfseries 83} (2011) 086001},
  \href{http://arxiv.org/abs/1002.2641}{{\ttfamily arXiv:1002.2641 [hep-th]}}.

\bibitem{Liu:2018jhs}
J.~Liu, E.~Perlmutter, V.~Rosenhaus, and D.~Simmons-Duffin, ``{$d$-dimensional
  SYK, AdS Loops, and $6j$ Symbols},''
  \href{http://dx.doi.org/10.1007/JHEP03(2019)052}{{\em JHEP} {\bfseries 03}
  (2019) 052}, \href{http://arxiv.org/abs/1808.00612}{{\ttfamily
  arXiv:1808.00612 [hep-th]}}.

\bibitem{Cardona:2018dov}
C.~Cardona and K.~Sen, ``{Anomalous dimensions at finite conformal spin from
  OPE inversion},'' \href{http://dx.doi.org/10.1007/JHEP11(2018)052}{{\em JHEP}
  {\bfseries 11} (2018) 052}, \href{http://arxiv.org/abs/1806.10919}{{\ttfamily
  arXiv:1806.10919 [hep-th]}}.

\bibitem{Albayrak:2019gnz}
S.~Albayrak, D.~Meltzer, and D.~Poland, ``{More Analytic Bootstrap:
  Nonperturbative Effects and Fermions},''
  \href{http://dx.doi.org/10.1007/JHEP08(2019)040}{{\em JHEP} {\bfseries 08}
  (2019) 040}, \href{http://arxiv.org/abs/1904.00032}{{\ttfamily
  arXiv:1904.00032 [hep-th]}}.

\bibitem{Cardona:2018qrt}
C.~Cardona, S.~Guha, S.~K. Kanumilli, and K.~Sen, ``{Resummation at finite
  conformal spin},'' \href{http://dx.doi.org/10.1007/JHEP01(2019)077}{{\em
  JHEP} {\bfseries 01} (2019) 077},
  \href{http://arxiv.org/abs/1811.00213}{{\ttfamily arXiv:1811.00213
  [hep-th]}}.

\bibitem{Iliesiu:2018zlz}
L.~Iliesiu, M.~Kolo\u{g}lu, and D.~Simmons-Duffin, ``{Bootstrapping the 3d
  Ising model at finite temperature},''
  \href{http://dx.doi.org/10.1007/JHEP12(2019)072}{{\em JHEP} {\bfseries 12}
  (2019) 072}, \href{http://arxiv.org/abs/1811.05451}{{\ttfamily
  arXiv:1811.05451 [hep-th]}}.

\bibitem{Iliesiu:2018fao}
L.~Iliesiu, M.~Kolo\u{g}lu, R.~Mahajan, E.~Perlmutter, and D.~Simmons-Duffin,
  ``{The Conformal Bootstrap at Finite Temperature},''
  \href{http://dx.doi.org/10.1007/JHEP10(2018)070}{{\em JHEP} {\bfseries 10}
  (2018) 070}, \href{http://arxiv.org/abs/1802.10266}{{\ttfamily
  arXiv:1802.10266 [hep-th]}}.

\bibitem{Albayrak:2020rxh}
S.~Albayrak, D.~Meltzer, and D.~Poland, ``{The Inversion Formula and 6j Symbol
  for 3d Fermions},'' \href{http://dx.doi.org/10.1007/JHEP09(2020)148}{{\em
  JHEP} {\bfseries 09} (2020) 148},
  \href{http://arxiv.org/abs/2006.07374}{{\ttfamily arXiv:2006.07374
  [hep-th]}}.

\bibitem{Kologlu:2019mfz}
M.~Kologlu, P.~Kravchuk, D.~Simmons-Duffin, and A.~Zhiboedov, ``{The light-ray
  OPE and conformal colliders},''
  \href{http://dx.doi.org/10.1007/JHEP01(2021)128}{{\em JHEP} {\bfseries 01}
  (2021) 128}, \href{http://arxiv.org/abs/1905.01311}{{\ttfamily
  arXiv:1905.01311 [hep-th]}}.

\bibitem{Chang:2020qpj}
C.-H. Chang, M.~Kologlu, P.~Kravchuk, D.~Simmons-Duffin, and A.~Zhiboedov,
  ``{Transverse spin in the light-ray OPE},''
  \href{http://arxiv.org/abs/2010.04726}{{\ttfamily arXiv:2010.04726
  [hep-th]}}.

\bibitem{Dixon:2019uzg}
L.~J. Dixon, I.~Moult, and H.~X. Zhu, ``{Collinear limit of the energy-energy
  correlator},'' \href{http://dx.doi.org/10.1103/PhysRevD.100.014009}{{\em
  Phys. Rev. D} {\bfseries 100} no.~1, (2019) 014009},
  \href{http://arxiv.org/abs/1905.01310}{{\ttfamily arXiv:1905.01310
  [hep-ph]}}.

\bibitem{Korchemsky:2019nzm}
G.~P. Korchemsky, ``{Energy correlations in the end-point region},''
  \href{http://dx.doi.org/10.1007/JHEP01(2020)008}{{\em JHEP} {\bfseries 01}
  (2020) 008}, \href{http://arxiv.org/abs/1905.01444}{{\ttfamily
  arXiv:1905.01444 [hep-th]}}.

\bibitem{Brower:2006ea}
R.~C. Brower, J.~Polchinski, M.~J. Strassler, and C.-I. Tan, ``{The Pomeron and
  gauge/string duality},''
  \href{http://dx.doi.org/10.1088/1126-6708/2007/12/005}{{\em JHEP} {\bfseries
  12} (2007) 005},
\href{http://arxiv.org/abs/hep-th/0603115}{{\ttfamily arXiv:hep-th/0603115
  [hep-th]}}.

\bibitem{Gillioz:2016jnn}
M.~Gillioz, X.~Lu, and M.~A. Luty, ``{Scale Anomalies, States, and Rates in
  Conformal Field Theory},''
  \href{http://dx.doi.org/10.1007/JHEP04(2017)171}{{\em JHEP} {\bfseries 04}
  (2017) 171}, \href{http://arxiv.org/abs/1612.07800}{{\ttfamily
  arXiv:1612.07800 [hep-th]}}.

\bibitem{Gillioz:2018kwh}
M.~Gillioz, X.~Lu, and M.~A. Luty, ``{Graviton Scattering and a Sum Rule for
  the c Anomaly in 4D CFT},''
  \href{http://dx.doi.org/10.1007/JHEP09(2018)025}{{\em JHEP} {\bfseries 09}
  (2018) 025}, \href{http://arxiv.org/abs/1801.05807}{{\ttfamily
  arXiv:1801.05807 [hep-th]}}.

\bibitem{Gillioz:2020mdd}
M.~Gillioz, M.~Meineri, and J.~Penedones, ``{A scattering amplitude in
  Conformal Field Theory},''
  \href{http://dx.doi.org/10.1007/JHEP11(2020)139}{{\em JHEP} {\bfseries 11}
  (2020) 139}, \href{http://arxiv.org/abs/2003.07361}{{\ttfamily
  arXiv:2003.07361 [hep-th]}}.

\bibitem{seeley_1964}
R.~T. Seeley, ``{Extension of $C^\infty$ Functions Defined in a Half Space},''
  \href{http://dx.doi.org/10.2307/2034761}{{\em Proc.~Am.~Math.~Soc.}
  {\bfseries 15} no.~4, (1964) 625}.

\bibitem{wiki:Whitney_extension_theorem}
Wikipedia,
  ``{\href{https://en.wikipedia.org/wiki/Whitney_extension_theorem\#Extension_in_a_half_space}{Whitney
  extension theorem}}.''.

\bibitem{vsilov1969generalized}
G.~E. {{\v{S}}ilov} and I.~M. {Gelfand}, {\em Generalized functions. 1.
  Properties and operations}.
\newblock Academic Press, 1969.

\bibitem{hall1957theorem}
D.~Hall and A.~S. Wightman, {\em A theorem on invariant analytic functions with
  applications to relativistic quantum field theory}.
\newblock København, I kommission hos Munksgaard, 1957.

\bibitem{vladimirov1966methods}
V.~S. {Vladimirov}, {\em Methods of the theory of functions of several complex
  variables}.
\newblock MIT Press, 1966.

\bibitem{kallen1961analyticitydomain}
G.~{K\"{a}ll{\'e}n}, ``{The analyticity domain of the four point function},''
\href{http://dx.doi.org/10.1016/0029-5582(61)90184-5}{{\em Nucl. Phys.}
  {\bfseries 25} (1961) 568--603}.

\end{thebibliography}\endgroup
\pagestyle{empty}

\includepdf[pages={1}]{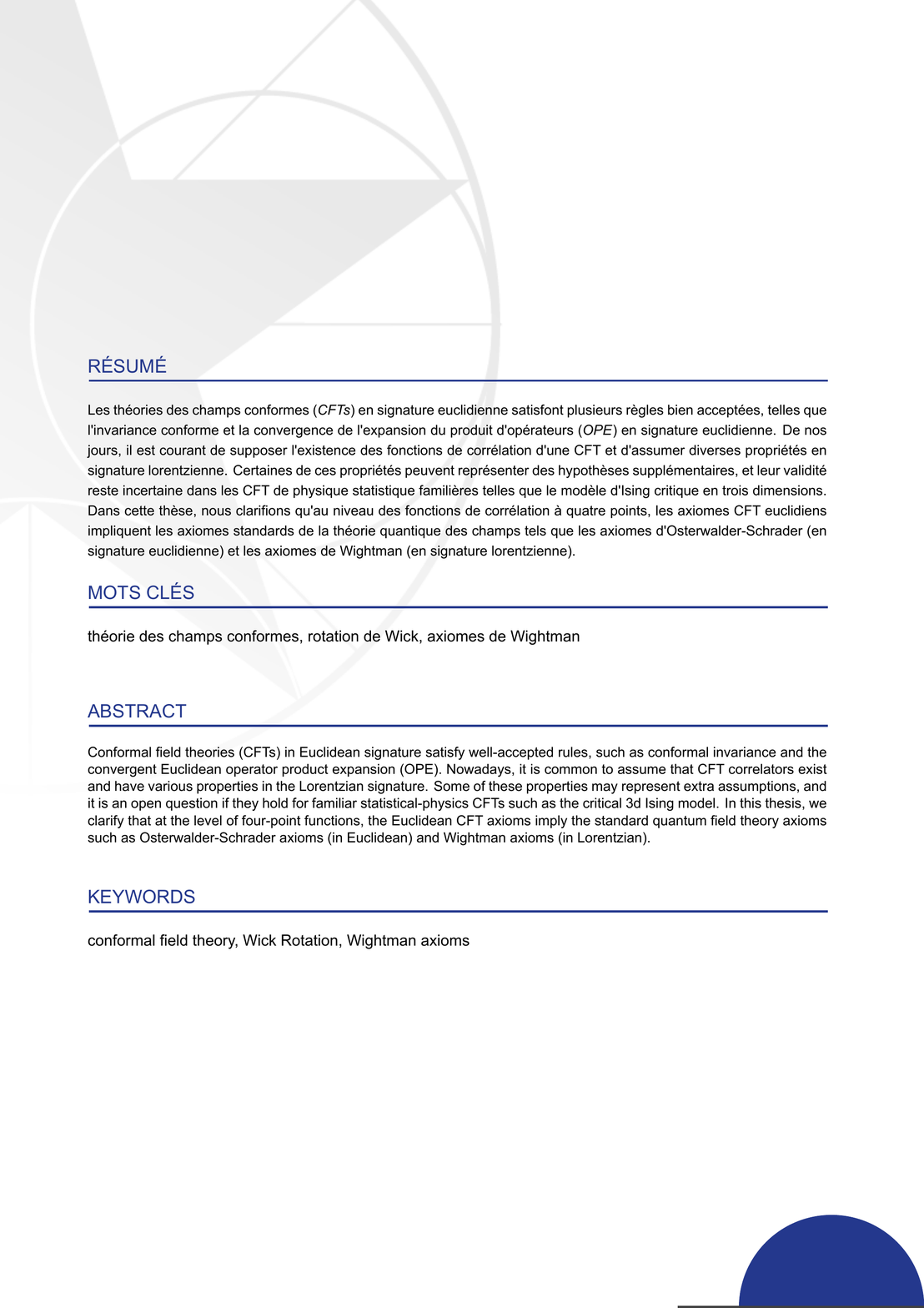}

\end{document}